\documentclass[twoside,openright,a4paper,11pt]{book}
\usepackage{amssymb,amsfonts,amsmath}   
\usepackage{graphicx,subfigure,float}   
\usepackage{palatino}          
\usepackage{physics}
\usepackage{dsfont}            		
\usepackage{mathrsfs}          		
\usepackage{tcolorbox}                  
\usepackage[normalem]{ulem}             
\usepackage{todonotes}					
\usepackage{xcolor}
\usepackage[toc,page]{appendix}
\usepackage[square,sort,comma,numbers,sort&compress]{natbib}
\usepackage[pagebackref=false,pdftex,pdfborderstyle={/S/U/W 1},hyperfootnotes=true,colorlinks,
linkcolor={darkred},
citecolor={darkviolet},
urlcolor={darkred}]{hyperref}
\usepackage{setspace}
\usepackage{wrapfig,sidecap,,fullpage,morefloats}
\usepackage{grffile,float,xspace}
\usepackage[sort&compress]{natbib}     

\newcounter{daggerfootnote}
\newcommand*{\daggerfootnote}[1]{%
	\setcounter{daggerfootnote}{\value{footnote}}%
	\renewcommand*{\thefootnote}{\fnsymbol{footnote}}%
	\footnote[2]{#1}%
	\setcounter{footnote}{\value{daggerfootnote}}%
	\renewcommand*{\thefootnote}{\arabic{footnote}}%
}

\usepackage{caption}                     
\newcommand{\TR}{\mathcal{T}}             	
\newcommand{\K}{\mathcal{K}}             	

\definecolor{darkred}{rgb}{0.55, 0.0, 0.0}
\definecolor{darkviolet}{rgb}{0.58, 0.0, 0.83}	
\definecolor{midnightblue}{rgb}{0.1, 0.1, 0.44}	
\definecolor{oxfordblue}{rgb}{0.0, 0.13, 0.28}
\definecolor{prussianblue}{rgb}{0.0, 0.19, 0.33}
\definecolor{goldenrod}{rgb}{0.85, 0.65, 0.13}
\definecolor{darkgreen}{rgb}{0.0, 0.5, 0.0}
\definecolor{Ultramarine}{rgb}{18,10,143}

\newcommand{\thesistitle}{Geometric phase and its applications: topological phases, quantum walks and non-inertial quantum systems}
 
\newcommand{\degree}{Doctor of Philosophy}
\newcommand{\candidate}{Vikash Mittal}

\newcommand{\department}{\href{https://web.iisermohali.ac.in/dept/physics/}{Department of Physical Sciences}}
\newcommand{\university}{\href{https://www.iisermohali.ac.in/}{Indian Institute of Science Education \& Research (IISER) Mohali,\\ Sector 81 SAS Nagar, Manauli PO 140306 Punjab, India}}

\begin{document}
\pagenumbering{roman}
\begin{titlepage}
	
	\vfill
	\begin{center}
	\vfill
	{\huge {\sc \textbf{\thesistitle}} \par}\vspace{0.01cm}
	\vfill
	
	{\LARGE \sc \candidate \vskip 0.3cm} 
	
	\vfill
	\textit{\large A thesis submitted for the partial fulfillment of }\\[0.1cm] 
	\textit{\large the degree of \degree}\\[0.4cm]
	\vfill
	
	\centering
	\includegraphics[width=3.5cm]{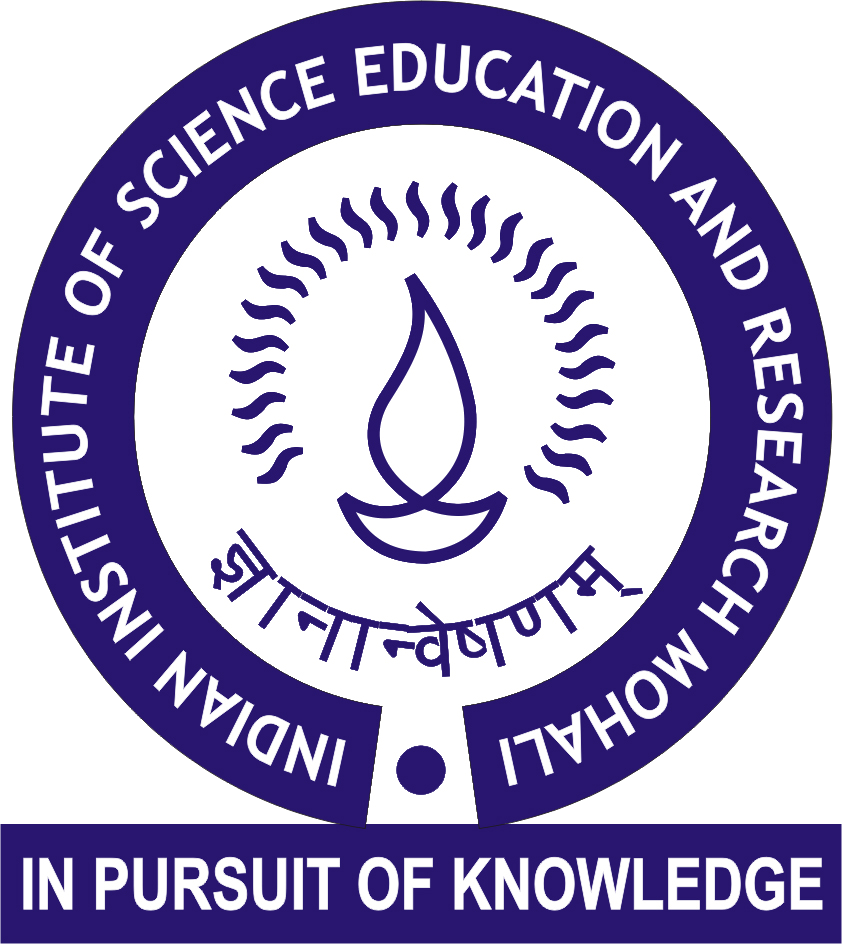}
	\vfill
	\sc \department
	\\
	\vspace{0.5cm}
	\university
	\vfill
	{August 2022}
	\end{center}
	\vfill
	
\end{titlepage}

\clearpage\mbox{}\clearpage


\vspace*{\fill}
\begin{center}
	\Large \emph{dedicated to my family and \\ 
		to all the people \\
		to whom I owe more than I can express \dots}
\end{center}
\vspace*{\fill}

\clearpage\mbox{}\clearpage
\clearpage\mbox{}\clearpage
\clearpage\mbox{}\clearpage
\begin{center}
	{\Large  \textbf{Abstract}}
\end{center}

Geometric phase plays a fundamental role in quantum theory and accounts for wide phenomena ranging from the Aharanov-Bohm effect, the integer and fractional quantum hall effects, and topological phases of matter, including topological insulators, to name a few. In this thesis, we have proposed a fresh perspective of geodesics and null phase curves, which are key ingredients in understanding the geometric phase. We have also looked at a number of applications of geometric phases in topological phases, quantum walks, and non-inertial quantum systems. 

The shortest curve between any two points on a given surface is a (minimal) geodesic. They are also the curves along which a system does not acquire any geometric phase. In the same context, we can generalize geodesics to define a larger class of curves, known as null phase curves (NPCs), along which also the acquired geometric phase is zero; however, they need not be the shortest curves between the two points. We have proposed a geometrical decomposition of geodesics and null phase curves on the Bloch sphere, which is crucial in improving our understanding of the geometry of the state space and the intrinsic symmetries of geodesics and NPCs.

We have also investigated the persistence of topological phases in quantum walks in the presence of an external (lossy) environment. We show that the topological order in one and two-dimensional quantum walks persist against moderate losses. Further, we use the geometric phase to detect the non-inertial modifications to the field correlators perceived by a circularly rotating two-level atom placed inside a cavity.  

\clearpage
\clearpage\mbox{}\clearpage
\listoffigures
\tableofcontents
\clearpage\mbox{}\clearpage
\pagenumbering{arabic}
	
\chapter{Introduction}
Berry showed that a quantum system that is taken around a closed path by varying the parameters $\vb{R}$ of its Hamiltonian $H(\vb{R})$ would acquire an additional phase factor in addition to the standard dynamical phase~\cite{Berry1984}. This extra phase factor, known as \emph{geometric phase} is purely geometric in nature and depends only on the geometry of the path traversed by the system in the parameter space. The geometric phase accounts for wide phenomena in physics, such as the Aharonov-Bohm effect~\cite{Bohm1959}, the integer and fractional hall effect~\cite{Klitzing1980,Stormer1982}, the topological phases of matter~\cite{Kane2010} and is instrumental in quantum information processing~\cite{Jones2000,Alicea2011}. In this thesis, we focus on the fundamental structure of the geometric phase and study its applications.

In this thesis, we introduce a Bloch sphere decomposition of geodesics and null phases curves, which are the key ingredients to understand the fundamental structure of the geometric phases~\cite{Mukunda1993, Rabei1999}. The shortest curve between any two points on a given surface is a (minimal) geodesic. Geodesics in the state space of quantum systems play an important role in the theory of geometric phases, as these are also the curves along which the acquired geometric phase is zero. Null phase curves (NPCs) are the generalization of the geodesics, which are defined as the curves along which the acquired geometric phase is zero even though they need not be the shortest curves between two points. We proposed a geometrical way to construct geodesics and a class of null phase curves in $n$-dimensional systems using Majorana star (MS) representation~\cite{Majorana1932}. This work might be useful in improving the understanding of the state-space structure of higher-dimensional systems. We have further looked at several applications of geometric phases.

One direct application of geometric phase is to define topological invariants which are used to characterize different topological phases of matter~\cite{Zak1989,Stanescu2016}. The geometric phases originated due to the underlying topology of the system are robust against local perturbations. Quantum walks are the quantum analog of classical random walks and are known to exhibit exotic topological phases~\cite{Kitagawa2010,Kitagawa2012}. We have investigated the behavior of topological phases in quantum walks in the presence of a lossy environment. We show that the topological phases of the quantum walks are robust against moderate losses. The topological order in one-dimensional (1D) split-step quantum walk persists as long as the Hamiltonian respects exact $\mathcal{PT}$-symmetry. We also observe that the topological nature persists in two-dimensional (2D) quantum walks as well; however, the $\mathcal{PT}$-symmetry has little role to play there.

We further use the geometric phase to detect the effects of the non-inertial motion of a rotating atom placed inside an electromagnetic cavity. The cavity enables the isolation or strengthening of the non-inertial response relative to the inertial one. The accumulative nature of the geometric phase may facilitate the experimental observation of the resulting, otherwise feeble, non-inertial contribution to the modified field correlations. We have also indicated the possibility of an experimental observation of the modified vacuum fluctuations using the geometric phase acquired by a circularly accelerated atom interacting with the field vacuum inside an electromagnetic cavity. We have shown that the atom acquires an experimental observable phase at accelerations as low as $10^7$m/s$^2$.  

\chapter{Geometric Phase}\label{chap:gp}
A quantum system taken around a closed path by varying the parameters $\vb{R}$ of its Hamiltonian $H(\vb{R})$ acquires a geometrical phase. This phase is different from the standard dynamical phase of quantum systems in the sense that it depends only on the geometry of the path traversed by the system in the parameter space~\cite{Berry1984,Shapere1989,Jamiolkowski2012,Bohm2013,Cheng2017}.
Aharonov-Bohm effect~\cite{Bohm1959} and Pancharatnam's phase~\cite{Pancharatnam1956} are the few manifestations of this phase. In the same year, immediately after Berry's paper, Barry Simon interpreted the geometric phase as the holonomy of the fiber bundle~\cite{Simon}. This geometric phase has wide applicability in the fields of quantum computation~\cite{Vedral2003}, condensed matter-physics~\cite{Zak1989}, optics~\cite{Bhandari1997} and high-energy physics~\cite{Sonoda1986}. Berry's derivation of the geometric phase (also known as the Berry phase) requires the assumption of quantum adiabatic approximation and cyclic evolution. It further made use of the eigenstates of the Hamiltonian of the qunatum system under consideration. In the following, the geometric phase was introduced for general unitary cyclic evolutions by Aharonov and Anandan~\cite{Anandan} in 1987 and subsequently generalized to arbitrary (not necessarily unitary or cyclic) evolutions by Samuel and Bhandari~\cite{Samuel1988}. The final generalization was introduced by Mukunda \emph{et al.}~\cite{Mukunda1993} using a kinematical approach. In this approach, the connection between the geometric phase and the Bargmann invaraint~\cite{Bargmann1964} was established which proved to be very important in order to find the geometric phase for a given set of discrete states. The definition of geometric phase in terms of Bargmann invariant can be taken very well as the fundamental definition of Berry's phase.

\noindent In this chapter we will first talk about geometric phase in the context of pure states starting with the original derivation due to Berry and then discuss it's subsequent generalizations by Aharanov and Anandan~\cite{Anandan}, Samuel and Bhandari~\cite{Samuel} and finally by Simon and Mukunda~\cite{Mukunda}. We will further extend the definition of the geometric phase for more general states, referred to as mixed states, which was developed by Uhlmann~\cite{Uhlmann1986} first in the context of parallel transport and then by Sjoqovist \emph{et. al}~\cite{Sjoqvist} in the context of interferometry. We discuss the most general form of the geometric phase by Tong \emph{et. al}~\cite{Tong2004}. We look at the idea of geometric phase from the perspective of weak measurements and finally conclude the chapter by looking at some applications of geometric phase in characterizing topological phases of matter.

\section{Berry's derivation of Geometric phase}
Consider a quantum mechanical system whose Hamiltonian $H(\vb{R})$ depends on real parameters $\vb{R} = (X, Y, \dots)$. The time evolution of the system (in the parameter space) is generated by slowly changing the parameters over time. The time dependence in Hamiltonian comes indirectly from $\vb{R}(t)$ such that $H(\vb{R}) \rightarrow H(\vb{R}(t))$.
The parameters are changed slowly enough to stay within the limits of the adiabatic theorem. The \textit{adiabatic theorem} in quantum mechanics~\cite{Kato} (or~\cite{Messiah}) states that if the Hamiltonian $H(\vb{R}(t))$ is slowly varying, then at any instant of time, the system will be in an eigenstate of the instantaneous $H(\vb{R}(t))$. 
\[ \vb{R} \rightarrow \vb{R}(t) = (X(t), Y(t), \dots ) \in \text{multidimensional parameter space} \]
The system evolves between times $t = 0 $ and $t = T$ which can be seen as a transport around a closed path $\vb{R}(t)$ in the parameter space such that $\vb{R}(T) = \vb{R}(0)$. This closed path is denoted by $\mathscr{C}$. Here, $T$ is the time it takes the system to return to the initial state (with an overall phase factor).
For the applicability of the adiabatic theorem, the time scale $T$ over which $H(\vb{R}(t))$ varies must be large.
For a given time-dependent Hamiltonian $H(\vb{R}(t))$, the state vector $\ket{\psi (t)}$ of the system evolves according to the time dependent Schr\"{o}dinger equation, 
\begin{equation} \label{eq:schroedingeeq}
	i \hbar \dfrac{d}{dt} \ket{\psi (t)} =  H(\vb{R}(t)) \ket{\psi (t)} 
\end{equation}
At a given instant of time $t$ for $\vb{R} = \vb{R}(t)$,  we have an orthonormal set of eigenstates $\{\ket{n; \vb{R}(t)}\}$ and the corresponding eigenvalues $E_n(\vb{R}(t))$ of $H(\vb{R}(t))$ such that
\begin{equation} \label{eq:eigenvalueHR}
	H(\vb{R}(t)) \ket{n; \vb{R}(t)} = E_n(\vb{R}(t)) \ket{n; \vb{R}(t)}
\end{equation}
Initially, the system is prepared in one of the eigenstate of $H(\vb{R}(0))$, $\ket{\psi (t =0)} = \ket{n; \vb{R}(0)}$ which will evolve adiabatically with $H(\vb{R}(t))$. The state at time $t$ is written as
\begin{equation} \label{eq:ketpsigp}
	\ket{\psi (t)} = \text{exp} \left \{ - \dfrac{i}{\hbar} \int_{0}^{t} dt' E_n (\vb{R}(t')) \right\} \text{exp} \left(i \gamma_n(t) \right) \ket{n; \vb{R}(t)}
\end{equation}
NOTE: This is the form of $\ket{\psi(t)}$ which Berry assumed and wanted to derive an explicit expression for the extra phase factor $\gamma_n(t)$.

\noindent The first exponential phase factor $\exp \left(- \tfrac{i}{\hbar} \int_{0}^{t} dt' E_n (\vb{R}(t')) \right)$ is well known in the theory of quantum mechanics.The other phase factor $\exp \left( i \gamma_n(t) \right)$ is the subject of interest here.

\noindent This extra phase factor $\text{exp}\left( i \gamma_n(t) \right)$ is non-integrable (because all the contribution from dynamics is embedded in the first phase factor). The expression for $\gamma_n(t)$ is written by substituting $\ket{\psi(t)}$ given by Eq. \eqref{eq:ketpsigp} in Eq. \eqref{eq:schroedingeeq} which gives 
\begin{equation}
	\dot{\gamma}_n (t) = i \expval{\dfrac{d}{dt}}{n; \vb{R}(t)} 
\end{equation} 
which can also be expressed in terms of gradient in parameter space,
\begin{equation}
	\dot{\gamma}_n (t) = i \expval{\nabla_{\vb{R}}}{n; \vb{R}(t)} \cdot \dot{\vb{R}}(t).
\end{equation}
So, the state at the end of the closed loop $\mathscr{C}$ when $t = T$, reads 
\begin{equation}
	\ket{\psi (T)} = \text{exp} \left \{ - \dfrac{i}{\hbar} \int_{0}^{T} dt' E_n (\vb{R}(t')) \right\} \text{exp} \left(i \gamma_n(T) \right) \ket{\psi(0)}
\end{equation} 
where
\begin{equation} \label{eq:beryyphase}
	\gamma_n(T) \equiv \gamma_n(\mathscr{C}) = i \oint_{\mathscr{C}} \expval{\nabla_{\vb{R}}}{n; \vb{R}(t)} \cdot \dot{\vb{R}}(t).
\end{equation} 
The integrand in the above integral is pure imaginary, which makes $\gamma_n(\mathscr{C})$ real and a physically observable quantity. It is very straightforward to see it using the fact that $\ket{n; \vb{R}(t)}$ are normalized. The Eq. \eqref{eq:beryyphase} can be written further as
\begin{equation}
	\gamma_n(\mathscr{C}) = -\text{Im} \oint_{\mathscr{C}}  \expval{\nabla_{\vb{R}}}{n;\vb{R}(t)} \cdot d \vb{R}(t) \qquad \mod 2\pi.
\end{equation}
If we consider the parameter space $\vb{R}(t)$ to be three-dimensional, then we can easily transform the line integral in the last equation to a surface integral using Stokes' theorem, which leads to
\begin{equation} \label{eq:stokestheorem}
	\begin{aligned}
		\gamma_n(\mathscr{C}) &= -\text{Im} \iint_{\mathscr{S}} d\vb{S} \cdot \nabla \times \expval{\nabla_{\vb{R}}}{n;\vb{R}(t)}, \qquad \nabla \equiv \nabla_{\vb{R}}\\
		&= -\text{Im} \iint_{\mathscr{S}} d\vb{S} \cdot \left( \nabla  \bra{n; \vb{R}(t)} \right) \times \left( \nabla  \ket{n; \vb{R}(t)}  \right)  \\
		&=-\text{Im} \iint_{\mathscr{S}} d\vb{S} \cdot \sum_{m \ne n} \left( \nabla  \bra{n; \vb{R}(t)}  \right) \dyad{m; \vb{R}(t)} \left( \nabla  \ket{n; \vb{R}(t)}  \right) \;\; \mod 2\pi                     
	\end{aligned}
\end{equation}
where $\mathscr{S}$ is the surface bounded by the curve $\mathscr{C}$ and $d\vb{S}$ is the surface element. In deriving the last expression, we have made use of the complete set of eigenstates 
\begin{equation}
	\sum _m \dyad{m; \vb{R}(t)}= \mathds{1}
\end{equation}
and the vector identity~\cite{Sakurai1985}
\begin{equation}
	\nabla \times [f (x) \nabla g(x)] = (\nabla f(x)) \times (\nabla g(x)).
\end{equation}
Note that the expression in Eq.~\eqref{eq:stokestheorem} does not depend on the choice of the surface $ \mathscr{S} $. We note that terms for which $m = n$ does not contribute in Eq. \eqref{eq:stokestheorem} by virtue of the fact that $\expval{\nabla_{\vb{R}}}{n;\vb{R}(t)}$ is a pure imaginary. For $m \ne n$, from Eq.~\eqref{eq:eigenvalueHR} we can deduce
\begin{equation}
	\mel*{m;\vb{R}(t)}{\nabla}{n;\vb{R}(t)} = \dfrac{ \mel*{m;\vb{R}(t)}{\nabla H(\vb{R}(t))}{n;\vb{R}(t)} }{E_n(\vb{R}(t)) - E_m(\vb{R}(t))}.
\end{equation}
By substituting this back in Eq. \eqref{eq:stokestheorem}, we have
\begin{equation}
	\gamma_n(\mathscr{C}) = -\text{Im} \iint_{\mathscr{S}} d\vb{S} \cdot \sum_{m \ne n} \dfrac{ \mel*{n;\vb{R}(t)}{\nabla H(\vb{R}(t))}{m;\vb{R}(t)} \times \mel*{m;\vb{R}(t)}{\nabla H(\vb{R}(t))}{n;\vb{R}(t)}}{(E_n(\vb{R}(t)) - E_m(\vb{R}(t)))^2}.
\end{equation}
and hence $\gamma_n(\mathscr{C})$ can be expressed in a more compact notation as,
\begin{equation}
	\gamma_n(\mathscr{C}) = -\iint_{\mathscr{S}} d\vb{S} \cdot \vb{V}_n(\vb{R})
\end{equation}
with
\begin{equation} \label{eq:berryconnection}
	\vb{V}_n(\vb{R}) = \text{Im} \sum_{m \ne n} \dfrac{ \mel*{n;\vb{R}(t)}{\nabla H(\vb{R}(t))}{m;\vb{R}(t)} \times \mel*{m;\vb{R}(t)}{\nabla H(\vb{R}(t))}{n;\vb{R}(t)}}{(E_n(\vb{R}(t)) - E_m(\vb{R}(t)))^2}.
\end{equation}
We make a few observations here. The $\vb{V}_n(\vb{R})$ is the curl of $\expval{\nabla_{\vb{R}}}{n;\vb{R}(t)}$ [from Eq.~\eqref{eq:stokestheorem}] which makes it invariant under gauge transformations of the type
\begin{equation}
	\ket{n; \vb{R}(t)} \rightarrow \exp(i \phi(\vb{R})) \ket{n; \vb{R}(t)}.
\end{equation}
In the later chapter [Chapter~\ref{chap:sym}], where we will discuss topological phases, we will identify $\vb{V}_n(\vb{R})$ with the Berry curvature and the quantity $i\expval{\nabla_{\vb{R}}}{n;\vb{R}(t)}$ is called Berry connection.

\section{An example of a spin - 1/2 particle in an external magnetic field} 
Let's consider a spin - $1/2$ particle in an external magnetic field with Hamiltonian,
\begin{equation}
	H(\vb{R}(t)) = - \dfrac{\mu}{2} \boldsymbol{\sigma} \vdot \vb{R}(t)
\end{equation}
where $\vb{R}(t) = (X(t), Y(t), Z(t))$ is the external magnetic field and $\boldsymbol{\sigma}$ is the vector operator whose components are Pauli operators,
\begin{equation}
	\sigma_X = \begin{pmatrix}
		0 & 1 \\ 
		1 & 0
	\end{pmatrix}, 
	\;\;\; \sigma_Y = \begin{pmatrix}
		0 & -i \\ 
		i & 0
	\end{pmatrix}, 
	\;\;\; \sigma_Z = \begin{pmatrix}
		1 & 0 \\ 
		0 & -1
	\end{pmatrix}.
\end{equation}
Thus,
\begin{equation}
	H(\vb{R}(t)) = - \dfrac{\mu}{2} \left(\begin{array}{cc} Z(t) & X(t) - iY(t) \\ X(t) + iY(t) & Z(t) \end{array}\right)
\end{equation}
By diagonalizing $H(\vb{R}(t))$ we will have eigenstates ${\ket{+; \vb{R}(t)}, \ket{-; \vb{R}(t)}}$ with corresponding eigenvalues, 
\begin{equation}
	\begin{aligned}
		E_{\pm} (\vb{R}(t)) &= \pm \dfrac{\mu}{2} \sqrt{X^2(t) + Y^2(t) + Z^2(t)} = \pm \dfrac{\mu}{2} R(t).
	\end{aligned}
\end{equation}
where $R(t) = \vert \vb{R}(t) \vert = \sqrt{X^2(t) + Y^2(t) + Z^2(t)} $. 
Also, 
\begin{equation}
	\nabla H(\vb{R}(t)) = - \dfrac{\mu}{2} \boldsymbol{\sigma}
\end{equation}
Now, if we choose $\vb{R}(t)$ to be along the $z$-axis, i.e. $\vb{R}(t) = R(t) \vb{\hat{k}}$, without any loss of generality and by using Eq.~\eqref{eq:berryconnection}, we get
\begin{equation} \label{eq:berrycspinhalf}
	\begin{aligned}
		\vb{V}_{+}(R(t)\vb{\hat{k}}) &= \dfrac{1}{4 R^2(t)} \text{Im}  \bra{+} \boldsymbol{\sigma} \ket{-} \times \bra{-} \boldsymbol{\sigma} \ket{+} \\
		&= \left[ \dfrac{\vb{\hat{k}}}{4 R^2(t)}\right] \text{Im} \left( \bra{+} \sigma_X \ket{-} \bra{-} \sigma_Y \ket{+} - \bra{+} \sigma_Y \ket{-} \bra{-} \sigma_X \ket{+}\right)
	\end{aligned}
\end{equation}
From Pauli's algebra, we have the following relations;
\begin{equation}
	\sigma_X \ket{\pm} = \ket{\mp}, \;\; \sigma_Y \ket{\pm} = \pm i \ket{\mp}, \;\; \sigma_Z \ket{\pm} = \pm \ket{\pm}
\end{equation}
where $\ket{\pm}$ are the eigenstates of $\sigma_Z$. Using these relations in Eq.~\eqref{eq:berrycspinhalf} we will get the following
\begin{equation}
	\vb{V}_{+}(R(t)\vb{\hat{k}}) = \dfrac{\vb{\hat{k}}}{2 R^2(t)}
\end{equation}
or, generally,
\begin{equation}
	\vb{V}_n(\vb{R}) = \dfrac{\vb{\hat{R}}}{2 R^2(t)} = \dfrac{\vb{R}}{2 R^3(t)}.
\end{equation}
Therefore, the Berry phase will be
\begin{equation}
	\begin{aligned}
		\gamma_{\pm}(\mathscr{C}) &= \mp \iint_{\mathscr{S}} d\vb{S}.\dfrac{\vb{R}}{2 R^3(t)},\\ 
		&= \mp \dfrac{1}{2} \iint_{\mathscr{S}} d\vb{S}.\dfrac{\vb{R}}{R^3(t)},\\
		&= \mp \dfrac{1}{2} \Omega (\mathscr{C})
	\end{aligned}
\end{equation}
where $\Omega(\mathscr{C})$ is the solid angle subtended by the closed path $\mathscr{C}$ at the origin.
\begin{tcolorbox}
	The solid angle is defined as
	\begin{equation}
		\Omega = \iint_{\mathscr{S}} \dfrac{\vb{\hat{r}}.\vb{\hat{n}}}{r^2} dS = \iint_{\mathscr{S}} \sin \theta d \theta d \varphi
	\end{equation}
	Here we have a cone for which we have
	\begin{equation}
		\Omega = \int_{0}^{\theta} \int_{0}^{2 \pi} \sin \theta d \theta d \varphi = 2 \pi (1 - \cos \theta)
	\end{equation}
\end{tcolorbox}
\section{Non-adiabatic, cyclic phase by Aharanov \& Anandan}
The first generalization of the Berry phase was given by Aharanov \& Anandan~\cite{Anandan} for nonadiabatic evolution under the assumption of a cyclic state. They used \textit{projective space}, $\mathcal{P}$ to derive the geometric phase expression. In this space, all normalized states in the Hilbert space $\mathcal{H}$, which differ only by a phase factor, are projected onto one point (state) in $\mathcal{P}$.
\noindent Let us consider a normalized state $\ket{\psi(t)} \in \mathcal{H}$ that evolves according to the Schr\"{o}dinger equation
\begin{equation}
	i \hbar \dfrac{d \ket{\psi(t)}}{dt} = H(t) \ket{\psi(t)} \implies \ket*{\dot{\psi}(t)} = -\dfrac{i}{\hbar} H(t) \ket{\psi(t)}.
\end{equation}
Since we are considering a cyclic evolution, the initial and final states can be related by a phase factor such that
\begin{equation}
	\ket{\psi(\tau)} = e^{i \phi}\ket{\psi(0)}.
\end{equation}
Now, we define a map $\Pi: \mathcal{H} \rightarrow \mathcal{P}$ as
\begin{equation}
	\Pi(\ket{\psi(t)}) = \{ \ket*{\psi'(t)} = e^{i \alpha} \ket{\psi(t)}| \alpha \in [0, 2\pi) \}
\end{equation}
which takes all states that differ only by a phase factor to a single point in $\mathcal{P}$. Then $\ket{\psi(t)}$ defines a curve
\begin{equation}
	\mathcal{C}: [0, \tau] \rightarrow \mathcal{H}
\end{equation}
such that $C \equiv \Pi(\mathcal{C})$ is the projection of $\mathcal{C}$ and is a closed curve in $\mathcal{P}$.
Now, define 
\begin{equation}
	\ket*{\tilde{\psi}(t)} = e^{- i f(t)} \ket{\psi(t)}
\end{equation}
such that $f(\tau) - f(0) = \phi$. The state $\ket*{\tilde{\psi}(t)}$ belongs to $\mathcal{P}$ where it forms a closed loop $C : [0, \tau]$ which is evident by the fact that
\begin{align}
	\ket*{\tilde{\psi}(\tau)} &= e^{- i f(\tau)} \ket{\psi(\tau)} = e^{- i f(\tau)} e^{i \phi}\ket{\psi(0)} \nonumber \\
	&= e^{-i (f(\tau) - \phi)} \ket{\psi(0)} = e^{-i f(0)} \ket{\psi(0)} \nonumber \\
	&= \ket*{\tilde{\psi}(0)}
\end{align} 
Further, by solving Schr\"{o}edinger equation for $\ket*{\tilde{\psi}(t)}$ we get
\begin{align}
	i \hbar \bra*{\tilde{\psi}(t)}\dfrac{d}{dt} \ket*{\tilde{\psi}(t)} = \hbar \dot{f}(t) + \bra{\psi(t)} H(t) \ket{\psi(t)}
\end{align}
By rearranging the last expression, we get
\begin{align*}
	-\dfrac{df(t)}{dt} &= \dfrac{1}{\hbar}\bra{\psi(t)} H(t) \ket{\psi(t)} - i \bra*{\tilde{\psi}(t)}\dfrac{d}{dt} \ket*{\tilde{\psi}(t)} 
\end{align*}
and by integrating the above equation over an interval $t \in [0, \tau]$ we have
\begin{align}
	\phi &= \int_{0}^{\tau}\dfrac{df(t)}{dt} dt = -\dfrac{1}{\hbar} \int_{0}^{\tau} dt \bra{\psi(t)} H(t) \ket{\psi(t)} + \int_{0}^{\tau} dt \bra*{\tilde{\psi}(t)}i\dfrac{d}{dt} \ket*{\tilde{\psi}(t)} \label{eq:AAPhase}
\end{align}
Here, we note that the total phase is naturally decomposed into two parts, one which depends on $H(t)$ and is the same in $\mathcal{H}$ and $\mathcal{P}$. Identifying the first part of LHS of Eq.~\eqref{eq:AAPhase} as the dynamical phase 
\begin{equation}
	\Phi_{dyn} = -\dfrac{1}{\hbar} \int_{0}^{\tau} dt \bra{\psi(t)} H(t) \ket{\psi(t)}
\end{equation}
leaves the other part to be \textit{geometric phase} given by
\begin{equation}
	\Phi_g \equiv \int_{0}^{\tau} dt \bra*{\tilde{\psi}(t)}i\dfrac{d}{dt} \ket*{\tilde{\psi}(t)} = \phi + \dfrac{1}{\hbar}\bra{\psi(t)} H(t) \ket{\psi(t)} 
\end{equation}
Since we can find many curves $\mathcal{C}$ in $\mathcal{H}$ such that they all have the same projection in $\mathcal{P}$ and these different curves are generated by different Hamiltonians $H$. We can find the same $\ket*{\tilde{\psi}(t)}$ for each $H$ by an appropriate choice of $f(t)$, the phase factor $\Phi_g$ is independent of $H$ for a given closed curve $C$. Thus, the \textit{geometric phase} defined in this manner does not depend on $H(t)$ and is purely a property of the projective space $\mathcal{P}$. We also note that $\Phi_g$ is uniquely defined up to $2\pi n$ with $n$ being an integer. Another difference to note in the Aharanov Anandan phase is that $\ket{\psi(t)}$ need not to be an eigenstate of $H(t)$ unlike in the case of Berry.

\section{Mathematical interlude} \label{sec:mathematical}
\begin{figure}
	\centering
	\subfigure[]{
		\includegraphics[width=6.5cm]{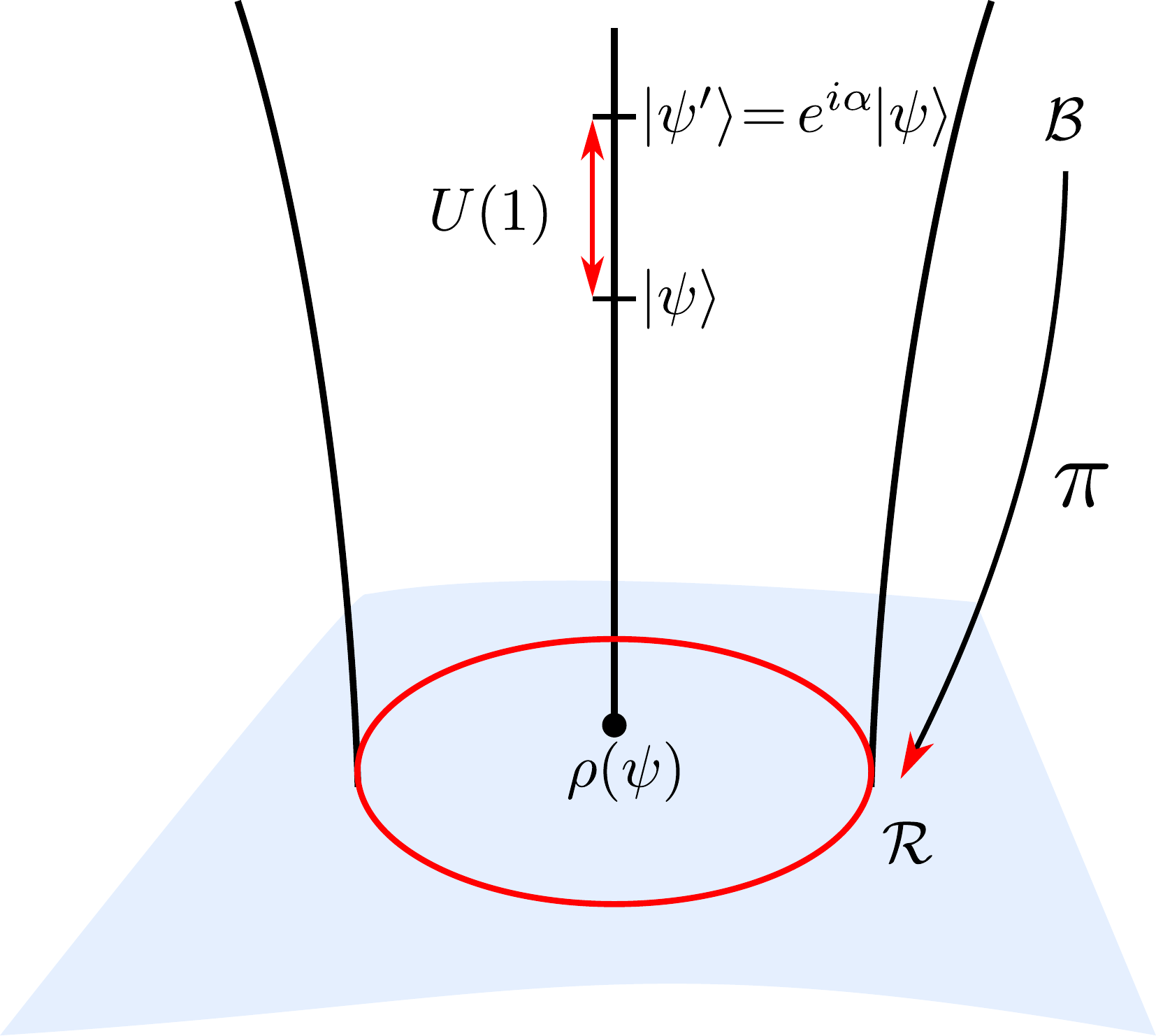}
		\label{fig:HBR}}
	\subfigure[]{
		\includegraphics[width=6.5cm]{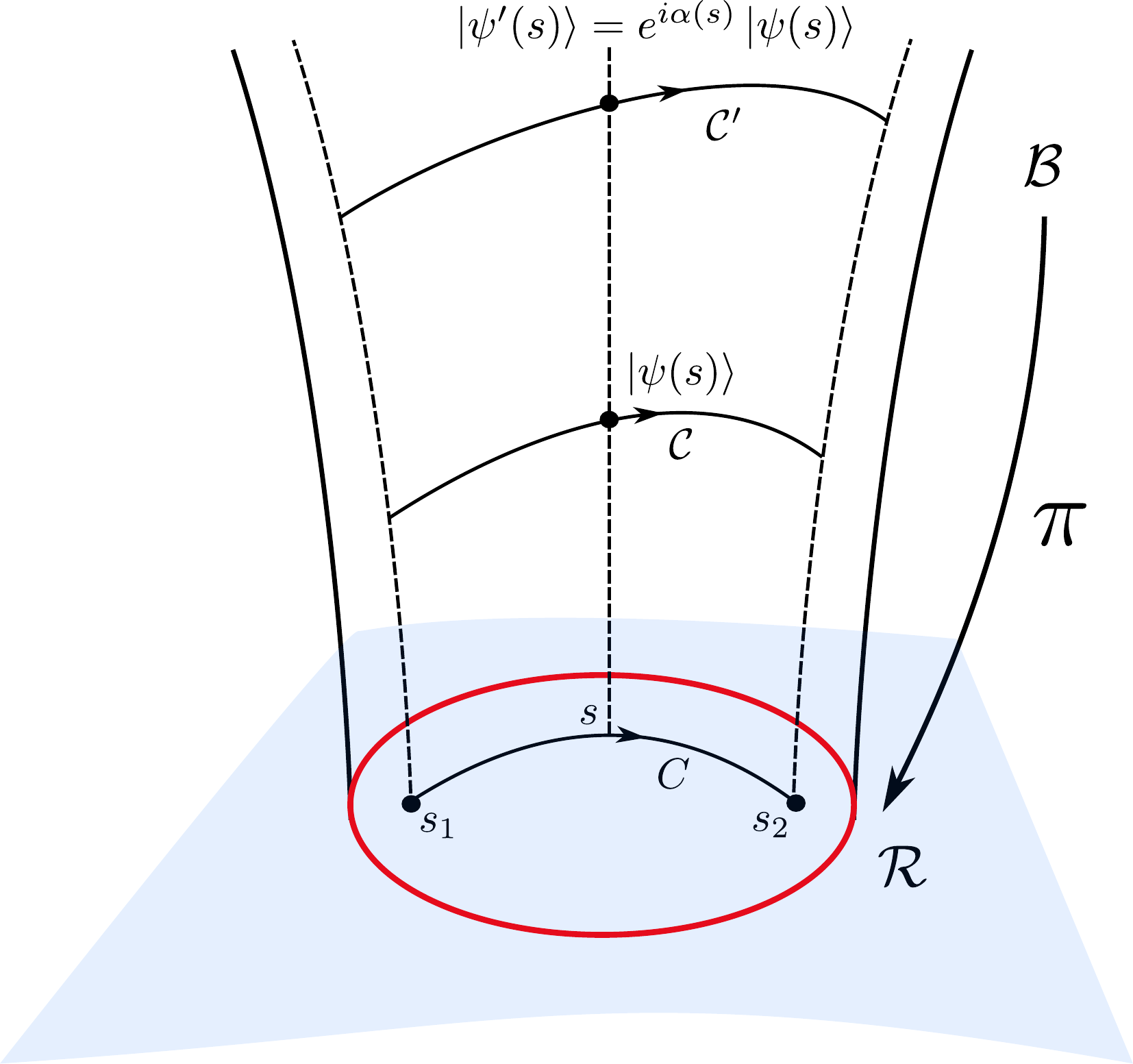}
		\label{fig:projective}}
	\caption{Structure of Hilbert space $\mathcal{H}$ and ray or projective space $\mathcal{R}$.}
\end{figure}

Before moving on to talk about the next generalization, we set up some mathematical definitions and notation and explain the necessary concepts which will help to grasp the subtlety of the topic. Consider a complex Hilbert space $\mathcal{H}$ representing a quantum system with complex dimension $N$. The states in the system are written as $\{\ket{\phi},\ket{\psi},\dots\}$ with an inner product structure $\ip{\psi}{\phi}$. The unit sphere $\mathcal{B}$  corresponding to the complex Hilbert space $\mathcal{H}$ is defined as
\begin{equation} \label{eq:unitsphere}
	\mathcal{B}=\{ \ket{\psi}\in \mathcal{H} \;|\; \ip{\psi}=1\}\subset \mathcal{H}
\end{equation}  
and has a real dimension $2N-1$. The group $U(1)$ acts on $\mathcal{B}$ as
\begin{equation}
	\ket{\psi} \in \mathcal{B} \implies \ket{\psi'} = e^{i \alpha} \ket{\psi} \in \mathcal{B}, \;\;\; 0 \le \alpha \le 2 \pi.
\end{equation}
Now, we define the space of unit rays $\mathcal{R}$ as
\begin{equation} \label{eq:rayspace}
	\mathcal{R}=\{\rho(\psi)=\dyad{\psi} \;|\;\ket{\psi} \in \mathcal{B}\}
\end{equation}                                           
which is the quotient of unit sphere $\mathcal{B}$ over U(1) actions. We can define a projection map $\pi:\mathcal{B}\rightarrow\mathcal{R}$ between $\mathcal{B}$ and $\mathcal{R}$ that takes $\ket{\psi} \in \mathcal{B}$ to a point $\rho(\psi)=\dyad{\psi} \in \mathcal{R}$ in the ray space, as illustrated in Fig.~\ref{fig:HBR}. Since $\mathcal{R}$ contains all the projections, it is also referred to as the projective space of $\mathcal{H}$. $\mathcal{R}$ is not a linear vector space, and it has a real dimension $ 2(N-1) $ and it is $ CP^{N-1} $ complex manifold. Note, that the dimension of the ray space $\mathcal{R}$ is always even. The ray space for pure quantum states is very special mathematical object and it happen to be Riemannian manifold that carries a natural Riemannian metric---the Fubini-Study metric~\cite{Page1987} and a natural symplectic structure~\cite{Arnold1997}. The projection map $\pi$ is defined as
\begin{equation}
	\pi: \mathcal{B} \rightarrow \mathcal{R} \;:\;\; \ket{\psi} \in \mathcal{B} \rightarrow \rho(\psi) = \dyad{\psi} \; \text{or} \; \psi \psi^{\dagger} \in \mathcal{B}.
\end{equation}
The one-dimensional $ U(1) $ \emph{fibre} sitting on top of $\rho(\psi) \in \mathcal{R}$ is an entire equivalence class of state vectors which are related by one another just by a phase factor and all project down to a single point in $\mathcal{R}$ i.e.
\begin{equation}
	\rho(\psi) \in \mathcal{R} \rightarrow \pi^{-1}(\rho(\psi)) = \{ \ket*{\psi'} = e^{i \alpha} \ket{\psi} \in \mathcal{B} \; | \; 0 \le \alpha < 2 \pi \} \subset \mathcal{B}
\end{equation}
Now, we define a smooth curve $\mathcal{C}$ of unit vectors in $\mathcal{B}$ as
\begin{equation}
	\mathcal{C} = \{\ket{\psi(s)} \in \mathcal{B} \; | \; s_1 \le s \le s_2 \} \subset \mathcal{B}
\end{equation} 
parameterized by a real variable $s$. The action of the map $\pi$ on $\mathcal{C}$ results in a curve $C \in \mathcal{R}$ that reads
\begin{equation}
	C = \pi[\mathcal{C}] = \{ \rho(\psi(s)) = \dyad{\psi(s)} \in \mathcal{R} \; | \; s_1 \le s \le s_2 \} \subset \mathcal{R}
\end{equation} 
Every parameterized curve $\mathcal{C} \subset \mathcal{R}$ as shown in Fig.~\ref{fig:projective} projecting onto the same curve $C \subset \mathcal{R}$ is called a \emph{lift} of $C$ from the ray space $\mathcal{R}$ to $\mathcal{B}$. Any other lift $\mathcal{C'}$ is related to $\mathcal{C}$ by a smooth pointwise phase shift known as \emph{gauge transformation} and is written as
\begin{equation}
	\mathcal{C'} = \{\ket*{\psi'(s)} = e^{i \alpha(s)} \ket{\psi(s)} \;|\; \ket{\psi(s)} \in \mathcal{C}, s_1 \le s \le s_2 \} \subset \mathcal{B}
\end{equation}
and
\begin{equation}
	\pi[\mathcal{C}] = \pi[\mathcal{C'}] = C \subset \mathcal{B}.
\end{equation}
This is all we need for the time being and we will discuss another mathematical interlude when we discuss the kinematic approach to the geometric phase.

\section{Non-adiabatic, non-cyclic geometric phase}
The next development in generalization is due to Samuel and Bhandari \cite{Samuel}. They generalized the definition for the cases when the curve is open in the projective Hilbert space $\mathcal{P}$ and the initial and final state does not belong to the same ray. They make use of the early work of Pancharatnam \cite{Pancharatnam} on the interference of polarized light beams and some concepts of differential geometry. We thus first go through Pancharatnam's phase and then move on to Samuel and Bhandari's work. 

\subsection{Pancharatnam phase}
If we have a pair of vectors (or states) $\ket{\psi_1}$ and $\ket{\psi_2}$ such that $\ket{\psi_2} = e^{i \Phi} \ket{\psi_1}$ then they are the same quantum states and represent the same quantum system due to the fact that they belong to the same \emph{ray}. A ray is defined as an equivalence class of states differing only by a phase factor. Furthermore, the two states $\ket{\psi_1}$ and $\ket{\psi_2}$ map to one point in the projective Hilbert space $\mathcal{R}$, that is, $\pi(\ket{\psi_1}) = \dyad{\psi_1} = \dyad{\psi_2} = \pi(\ket{\psi_2})$. In this case, the relative phase between $\ket{\psi_1}$ and $\ket{\psi_2}$ is naturally $\Phi$. However, when $\ket{\psi_1}$ and $\ket{\psi_2}$ represent two different quantum states, it is not trivial to define a relative phase between two and this question was first addressed by Pancharatnam. He gave a geometrical/physical interpretation of the relative phase between distinct polarization states of light. It is as follows: Given two non-orthogonal vectors / states $\ket{\psi_1}$ and $\ket{\psi_2}$, the relative phase difference $\Phi$ between them is given by
\begin{equation}
	\ip*{\psi_1}{\psi_2} = \abs{\ip*{\psi_1}{\psi_2}} e^{i \Phi}
\end{equation}
i.e. $\Phi$ is the phase of their inner product $\arg\ip*{\psi_1}{\psi_2}$. The $\ket{\psi_1}$ and $\ket{\psi_2}$ are said to be \emph{in phase} when 
\begin{equation}
	\ip*{\psi_1}{\psi_2} \text{is real and positive.}
\end{equation}
In the literature, it is commonly referred to as $\ket{\psi_1}$ and $\ket{\psi_2}$ are \emph{in phase} in Pancharatnam sense and is known as \emph{Pancharatnam connection}. An interesting point to note here is that this relation is not transitive, that is, if $ \ket{\psi_1} $ is in phase with $ \ket{\psi_2} $, and $ \ket{\psi_2} $ is in phase with another state $ \ket{\psi_3} $, then $ \ket{\psi_3} $ need not to be in phase with $ \ket{\psi_1} $. It can be illustrated by considering three normalized vectors
\begin{equation}
	\ket{\psi_1} = \dfrac{1}{\sqrt{2}} \begin{pmatrix}
		1 \\
		1
	\end{pmatrix}, \;\;\; \ket{\psi_2} = \begin{pmatrix}
		1 \\
		0
	\end{pmatrix},  \;\;\; \ket{\psi_3} = \dfrac{1}{\sqrt{2}} \begin{pmatrix}
		1 \\
		i
	\end{pmatrix}
\end{equation}
which are the eigenvectors of Pauli matrices. We can clearly see that
\begin{align*}
	\ip*{\psi_1}{\psi_2} = \dfrac{1}{\sqrt{2}} \implies \text{real and positive} \implies \ket{\psi_1} \text{is in phase with} \ket{\psi_2} \\ \noalign{\vskip5pt}
	\ip*{\psi_2}{\psi_3} = \dfrac{1}{\sqrt{2}} \implies \text{real and positive} \implies \ket{\psi_2} \text{is in phase with} \ket{\psi_3} \\ \noalign{\vskip5pt}
	\ip*{\psi_3}{\psi_1} = \dfrac{1 + i}{2} \implies \text{not real} \implies \ket{\psi_2} \text{is not in phase with} \ket{\psi_3} 
\end{align*}
\begin{figure}
	\centering
	\includegraphics[width=6.5cm]{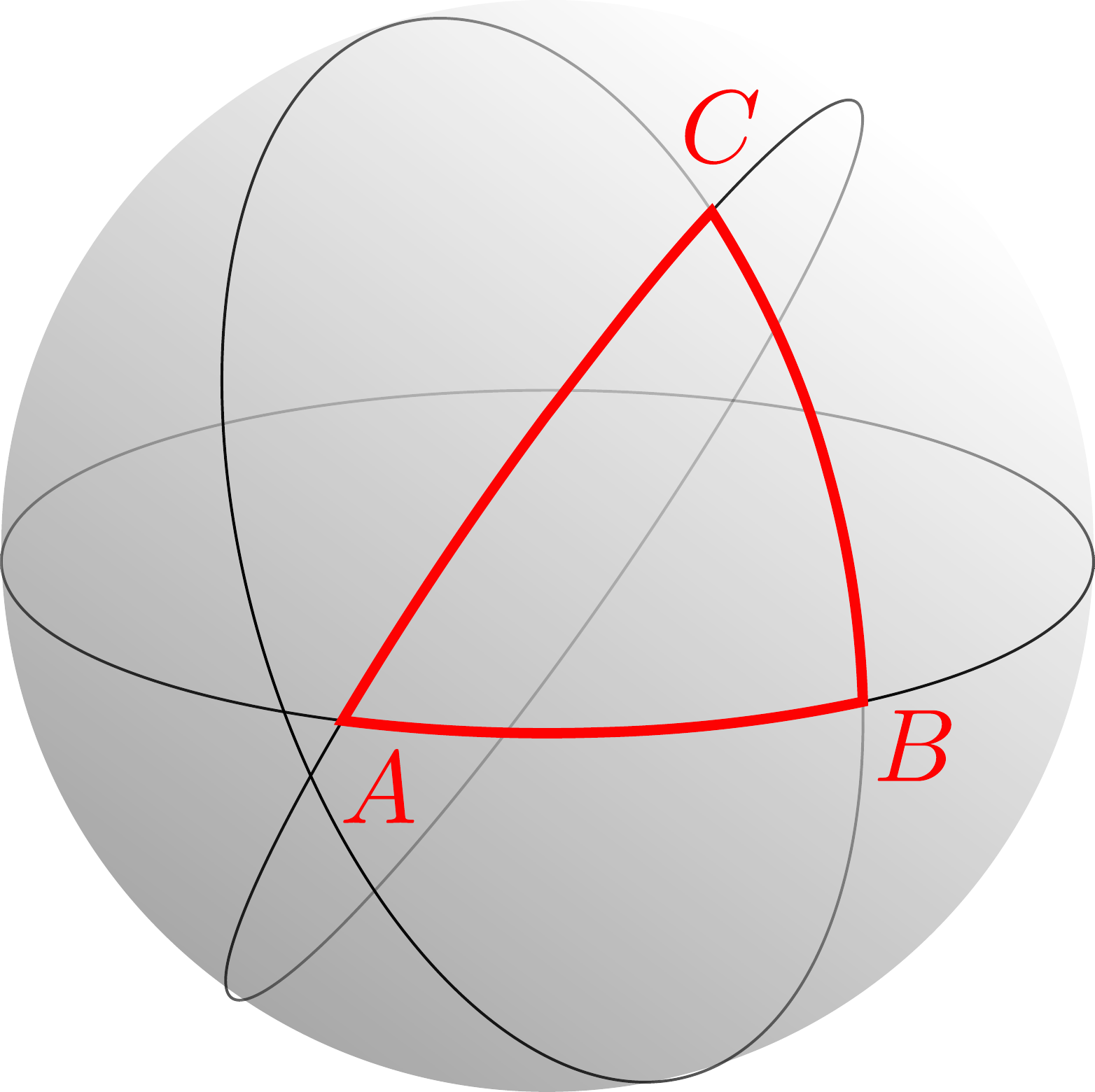}
	\caption{The Bargmann Invariant and the geometric phase.}
	\label{fig:BTriangle}
\end{figure}
The Pancharatnam phase $\Phi$ depends on the geometry of the state space. For two-level systems, let us take three states $ \{ \ket{A}, \ket{B}, \ket{C}\} $ such that $\ket{A}$ is in phase with $\ket{B}$, $\ket{B}$ is with $ \ket{C} $, then, in general, $\ket{C}$ is not in phase with $\ket{A}$. Let $ \ket{A'} = e^{i \Phi} \ket{A}$ be a state vector such that $ \ket{C} $ is in phase with $ \ket{A'}$, then this ``excess'' phase is expressed as 
\begin{equation}
	\Phi = - \dfrac{1}{2} \Omega_{ABC}
\end{equation} 
where $\Omega_{ABC}$ is the solid angle of triangle ABC subtended at the center as shown in Fig.~\ref{fig:BTriangle}. Pancharatnam arrives at this result by considering two polarization states $ \ket{A} $ and $ \ket{C}$ which are in phase with each other. Now, projecting both states onto a third state $ \ket{C}$, that is, $\ket{A} \rightarrow \ket*{\tilde{A}} = (\dyad{B}) \ket{A}$ and $\ket{C} \rightarrow \ket*{\tilde{C}} = (\dyad{B}) \ket{C}$ so that the relative phase between $ \ket*{\tilde{A}} $ and $ \ket*{\tilde{C}}$ becomes 
\begin{align*}
	\arg\ip*{\tilde{A}}{\tilde{C}} &= \arg\ip*{A}{B} \ip{B} \ip*{B}{C} \\ 
	&= \arg\ip*{A}{B} \ip{B}{C} \ip*{C}{A} \;\;\;\;\; \because \arg\ip*{C}{A} = 0 \\
	&= \Delta(A,B,C).
\end{align*}
The quantity $\Delta(A,B,C)$ is invariant under gauge transformations of the kind $ \ket{A} \rightarrow e^{i \beta} \ket{A} $ and therefore is a characteristic of a projective Hilbert space $\mathcal{R}$. Considering $ \{ \ket{A}, \ket{B}, \ket{C}\} $ to be polarization states of the light beam that are represented by a point on the \emph{Poincare} sphere, we find
\begin{equation}
	\Delta(A,B,C) = \dfrac{1}{2} \Omega_{ABC}.
\end{equation} 
The quantity $\Delta(A,B,C)$ is closely related to the geometric phase and is identified as \emph{Bargmann invariant}. This connection was made by Simon and Mukunda~\cite{Mukunda} which is the topic for the next section. We will come again to the Pancharatnam phase when we discuss the geometric phase for mixed states.  

\subsection{Parallel transport}
In the previous section, we identified $\Phi_{dyn}$ which is given by 
\begin{equation} \label{eq:dynamicalAA}
	\Phi_{dyn} = -\dfrac{1}{\hbar} \int_{0}^{\tau} dt \bra{\psi(t)} H(t) \ket{\psi(t)}
\end{equation}
and talked about the lifts of a closed curve $C$ in the projective Hilbert space $\mathcal{P}$ to $\mathcal{C}$ in the normalized Hilbert space $\mathcal{H}$. There exist special lifts of $C$ in Hilbert space, along which the dynamical phase vanishes and the whatever phase is accumulated during an evolution is purely geometrical. To achieve that, the integrand in Eq.~\eqref{eq:dynamicalAA} has to vanish i.e.,
\begin{equation}
	\expval{H(t)}{\psi(t)} = 0
\end{equation}
for all time $t$. Further, using the Schr\"{o}dinger equation, we can write
\begin{equation}
	\expval{\dfrac{d}{dt}}{\psi(t)} = 0
\end{equation}
and this is precisely the condition for \emph{parallel transport}. The parallel transport refers to an evolution in which the state vector $ \ket{\psi(t)} $ at time $t$ remains in \textit{in phase} with the adjacent state vector $ \ket{\psi(t + dt)} $. 

\noindent With the definition of Pancharatnam phase $\Phi$ and parallel transport at our disposal, we move to the results of Samuel and Bhandari. Consider a quantum system with Hilbert space $\mathcal{H}$. A state vector $ \ket{\psi(t)} \in \mathcal{H}$ evolves according to the Schr\"{o}dinger equation as
\begin{equation}
	i \hbar \dfrac{d \ket{\psi(t)}}{dt} = H(t) \ket{\psi(t)}.
\end{equation}
Here, $H$ is a linear operator and does not need to be Hermitian. We choose a lift, known as \emph{horizontal lift} of the curve $C$ in projective space $\mathcal{P}$ in such a way that the dynamical phase vanishes. That can be done by defining a new state vector $ \ket{\phi(t)} $ 
\begin{equation}
	\ket{\phi(t)} = e^{\tfrac{i}{\hbar} \int_{0}^{t} h (t') dt'} \ket{\psi(t)}
\end{equation}   
with
\begin{equation}
	h(t') = \dfrac{\Re\expval{H(t)}{\psi(t')}}{\ip*{\psi(t')}}.
\end{equation}
The difference between $\ket{\phi(t)}$ and $\ket{\psi(t)}$ is that we have removed the dynamical phase factor from $ \ket{\psi(t)} $ and left with only the geometric contribution. By substituting $ \ket{\phi(t)} $ into the Schr\"{o}dinger equation, we get
\begin{align*}
	i \hbar \dfrac{d \ket{\phi(t)}}{dt} &= i \hbar \left( \dfrac{i}{\hbar} h(t) \ket{\phi(t)} + e^{i \int_{0}^{t} h (t') dt'} \dfrac{d \ket{\psi(t)}}{dt} \right) \\
	&= i \left( i h(t) \ket{\phi(t)} - \dfrac{i}{\hbar} e^{\tfrac{i}{\hbar} \int_{0}^{t} h (t') dt'} H(t) \psi(t) \right) \\
	&= \left(H(t) - h(t)\right) \ket{\phi(t)}.
\end{align*} 
Further, taking the inner product with $ \ket{\phi(t)} $ from left, we get
\begin{align*}
	i \hbar \expval{\dfrac{d}{dt}}{\phi(t)} &= \expval{H(t) - h(t)}{\phi(t)} \\
	&= \expval{H(t)}{\phi(t)} - \expval{h(t)}{\phi(t)} \\
	&= \expval{H(t)}{\psi(t)} - \Re\expval{H(t)}{\psi(t)} \\
	&= \Im\expval{H(t)}{\psi(t)}
\end{align*}
which gives us the condition for parallel transport as
\begin{equation} \label{eq:parallel-transport}
	\Im \expval{\dfrac{d}{dt}}{\phi(t)} = 0
\end{equation}
which is valid for any general evolution governed by $H(t)$. In the case when $ H(t) $ is hermitian, the above condition reduces to 
\begin{equation}
	\expval{\dfrac{d}{dt}}{\phi(t)} = 0.
\end{equation}
It can be obtained independently using the normalization of $ \ket{\phi(t)} $. The parallel transport condition can be seen in an alternate way. We demand $\ket{\phi(t)}$ and $\ket{\phi(t + dt)}$ to be in phase, i.e. $\ip*{\phi(t)}{\phi(t + dt)}$ has to be real and positive during evolution. Using the series expansion for $\ip*{\phi(t)}{\phi(t + dt)}$ we get
\begin{equation}
	\ip*{\phi(t)}{\phi(t + dt)} = \ip*{\phi(t)} + \expval{\dfrac{d}{dt}}{\phi(t)} dt + \mathcal{O}(dt^2) 
\end{equation}
and $\expval*{\dfrac{d}{dt}}{\phi(t)}$ is pure imaginary. Therefore, $\ip*{\phi(t)}{\phi(t + dt)}$ is real and positive for any arbitrary small time interval $dt$ only when the second term $\expval*{\dfrac{d}{dt}}{\phi(t)}$ vanishes. The triplet $ (\mathcal{B}, \mathcal{R}, \pi) $ forms a principle bundle over the base space $\mathcal{R}$ (with structure group $U(1)$) and the parallel-transport law defines a \emph{natural connection} on this fibre bundle. \emph{A connection is an assignment of a ``horizontal subspace'' in the tangent space of each point in $\mathcal{B}$.} 

\subsection{Cyclic evolution}
\begin{figure}[H]
	\centering
	\includegraphics[width=9cm]{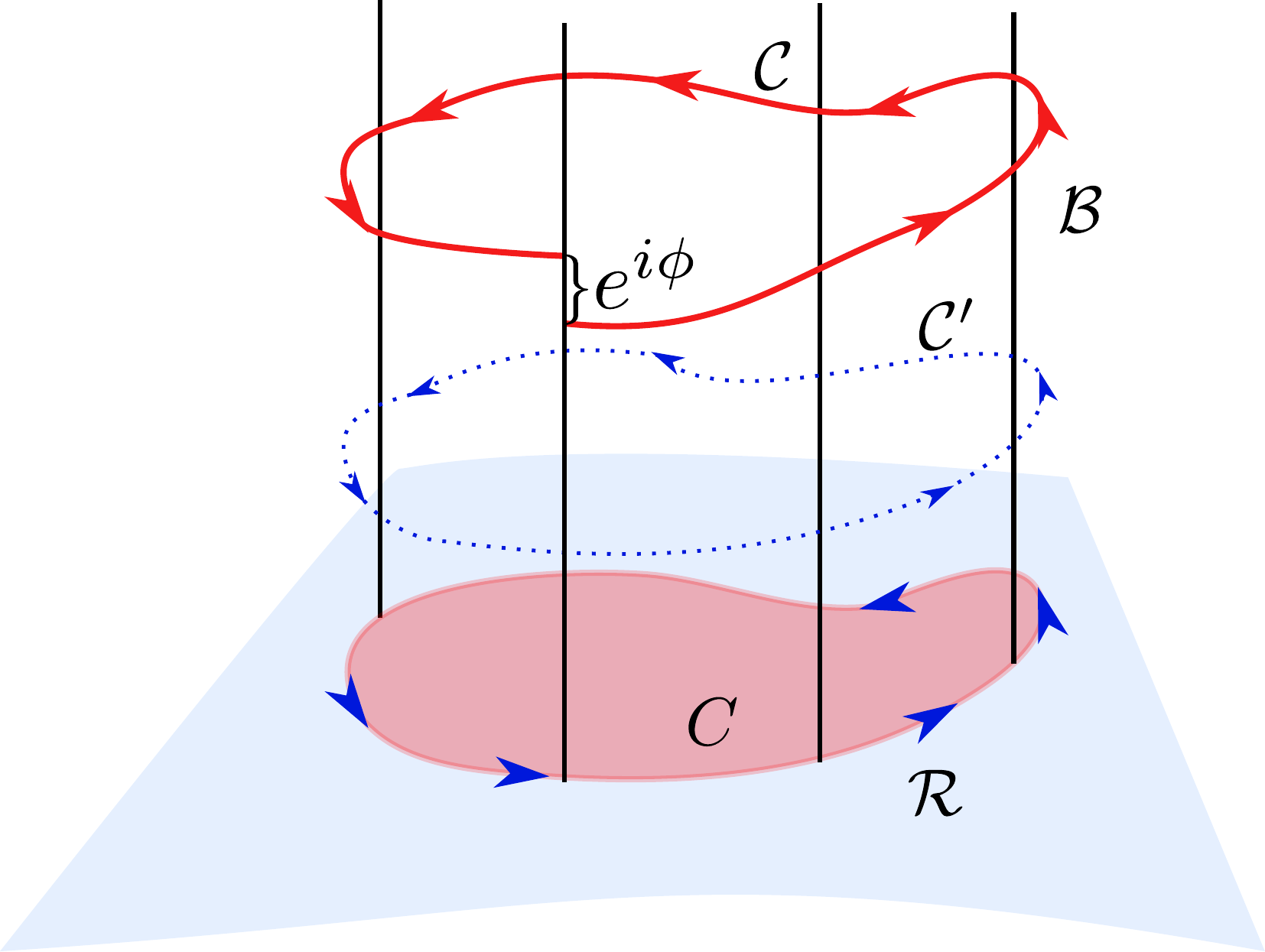}
	\caption{Schematic for cyclic geometric phase in Samuel and Bhandari's setting.}
	\label{fig:samuelbhandari}
\end{figure}
We now consider the cyclic evolution and derive the geometric phase with this setting. Let $ \mathcal{C} = \{\ket{\phi(s)} \;|\; s \in [s_1, s_2] \} $ be a curve in the Hilbert space of unit vectors $\mathcal{B}$. When $ \ket{\phi(s_2)} =  e^{i \phi} \ket{\phi(s_1)}$ for real $\phi$ as shown in Fig.~\ref{fig:samuelbhandari}, then its projection $C \in \mathcal{R}$ will be a closed curve. For this curve in $\mathcal{B}$ the tangent vector is given by
\begin{equation}
	\ket{u(s)} = \dfrac{d}{ds} \ket{u(s)}
\end{equation} 
and we define a quantity $ A_s = \Im\ip*{\phi(s)}{u(s)} $. This quantity $A_s$ resembles the vector potential in electrodynamics because, under gauge transformations of the kind $ \ket{\phi(s)} \rightarrow e^{i \alpha(s)} \ket{\phi(s)} $ it transforms like $ A_s \rightarrow A_s + d\alpha/ds $. The motivation to define this quantity will become clear in a moment.

\noindent Let's consider our original state vector $ \ket{\psi(s)} $ and a curve traced by it in $\mathcal{B}$. Now, if $ \ket{\psi(s)} $ is a cyclic solution of the Schr\"{o}dinger equation i.e. it returns to the same \emph{ray} after a time period $\tau$ (red curve $\mathcal{C}$ in Fig.~\ref{fig:samuelbhandari}), then the projection of this curve (under the projection map $\pi$) is a closed curve in $\mathcal{R}$. However, we are not actually interested in $ \ket{\psi(s)} $ because it also has a dynamical contribution in the total phase after evolving for time $\tau$. So, given a closed curve $C \in \mathcal{R}$, we need to find a lift $\mathcal{C} = \{\ket{\phi(s)}\:|\; s \in [0, \tau]\} \in \mathcal{B}$ which is traced by a state vector with the removed dynamical phase. Such lifts are called ``horizontal" lifts and are determined by the parallel-transport law Eq.~\eqref{eq:parallel-transport} which implies $A_s = 0$ along the curve. Consider the integral
\begin{equation} \label{eq:loopintegral}
	\phi_g = \oint A_s ds
\end{equation}
along the curve $\mathcal{C} \in \mathcal{B}$, which is closed by a vertical line that joins $ \ket{\phi(s_1)} $ and $ \ket{\phi(s_2)} $ as shown in Fig.~\ref{fig:samuelbhandari}. The part $ \ket{\phi(s)} $ shows the true evolution of the system, and $A_s$ vanishes along this curve. The only contribution to the integral Eq.~\eqref{eq:loopintegral} is due to the vertical line, and consequently $\phi_g$ is given by the Pancharatnam phase between $ \ket{\phi(s_1)} $ and $ \ket{\phi(s_2)} $ i.e. $\phi_g = \arg\ip*{\phi(s_1)}{\phi(s_2)}$. From this we see that the Pancharatnam phase is, in fact, an early example of Berry phase~\cite{Nityananda1986}. 
\begin{tcolorbox}
	Given a closed curve in $C \in \mathcal{R}$, we can lift this curve so that $A_s$ vanishes along that lift, i.e. a horizontal lift. However, the horizontal lift of a closed curve in $\mathcal{R}$ need not to be closed in $\mathcal{B}$. This is called holonomy of the connection \cite{Barry1983}. Furthermore, in the language of the fibre bundles, $A_s$ is referred to as \emph{one-form} which defines a horizontal lift.
\end{tcolorbox}
\noindent By using the gauge invariance of Eq.~\eqref{eq:loopintegral}, this integral can be identified as a characteristic of the projective Hilbert space $\mathcal{R}$. We can further use Stokes' theorem to write $\phi_g$ as
\begin{equation}
	\phi_g = \int_S \mathcal{F}
\end{equation} 
where $S$ denotes the surface bounded by a closed curve $C$ in $\mathcal{R}$ and $\mathcal{F}$ is the gauge-invariant \emph{two-form}. It denotes the exterior derivative of $A_s$ that is equivalent to $\nabla \times A_s$ in three dimensions. Hence, $\phi_g$ depends only on the curve $C \in \mathcal{R}$ and not on the rate at which it is transversed in the parameter space.
\subsection{Non-cyclic evolution}
\begin{figure}[H]
	\centering
	\includegraphics[width=9cm]{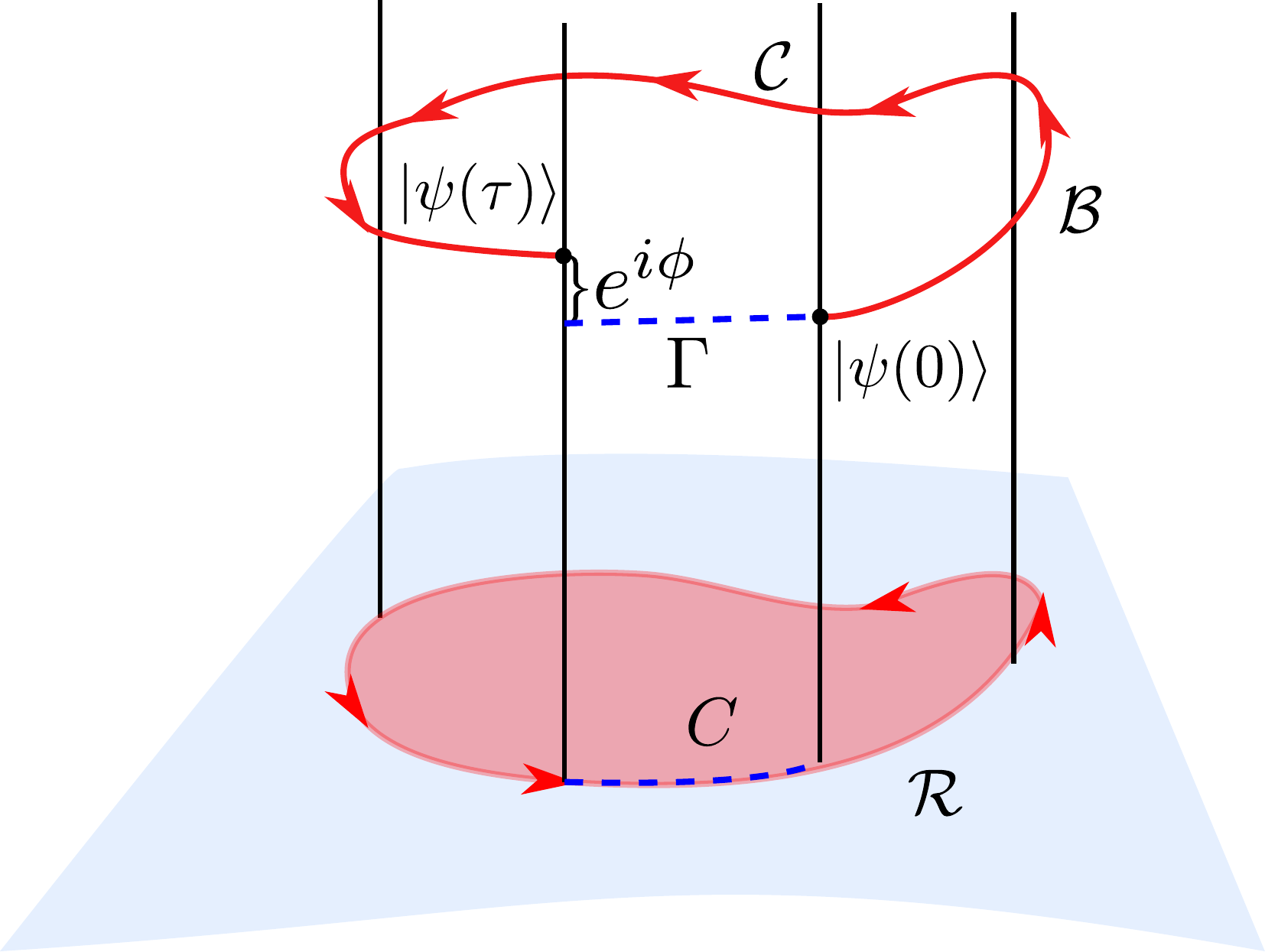}
	\caption{Schematic for non-cyclic geometric phase in Samuel and Bhandari's setting.}
	\label{fig:samuelbhandari2}
\end{figure}
Now, we come to the main results of Samuel \& Bhandari. Let us consider a quantum system undergoing a non-cyclic evolution and the initial $ \ket{\psi(0)} $ and final state vector $ \ket{\psi(\tau)} $ does not belong to the same ray. Also, the curve $ C \in \mathcal{R} $ is not a closed curve. At this point Pancharatnam comes to our rescue and shows us a way to proceed. 

\noindent The most important result proved by Samuel and Bhandari is that the Pancharatnam phase difference $\beta$ between any two nonorthogonal states $ \ket{\phi_1} $ and $ \ket{\phi_2} $ is written as the integral of one-form $A_s$ along a \emph{geodesic} i.e.
\begin{equation}
	\beta = \arg\ip{\phi_1}{\phi_2} = \int_{\Gamma} A_s ds
\end{equation}   
where $\Gamma$ is the geodesic curve $ \ket{\phi(s)} $ that connects $ \ket{\phi_1} $ and $ \ket{\phi_2} $. We are not going through the proof here, although we will discuss geodesics in great detail in the next section. The structure of geodesics is very important for one of our problem that we dealt with in this thesis. 

\noindent Therefore, the geometric phase acquired by a state vector $ \ket{\phi(s)} $ as it evolves from $ \ket{\phi(0)} $ to $ \ket{\phi(\tau)} $ (they are not orthogonal) is expressed as
\begin{equation}
	\phi_g = \int_{\mathcal{C} + \Gamma} A_s ds = \int_{\mathcal{C}} A_s ds + \int_{\Gamma} A_s ds
\end{equation}
where $\mathcal{C}$ is the horizontal curve along which $A_s$ and hence the first integral vanishes. Thus, the geometric phase between $ \ket{\phi(0)} $ and $ \ket{\phi(\tau)} $ is given by the integral Eq.~\eqref{eq:loopintegral} where the curve $\mathcal{C}$ is given by the actual evolution of the system from $ \ket{\phi(0)} $ to $ \ket{\phi(\tau)} $ and \emph{back along a geodesic curve} connecting $ \ket{\phi(\tau)} $ and $ \ket{\phi(0)} $. This can also be expressed as a surface integral of two-form $\mathcal{F}$ over a surface bounded by the curve $\pi(\mathcal{C})$ in the projective Hilbert space $\mathcal{R}$. This geodesic rule proposition for the non-cyclic geometric phase has recently been experimentally tested by Folman \emph{et al.} \cite{Folman2020}. They also discussed the possible applications of geodesic rule in general relativity to obtain the red-shifts. This might be the first step towards probing/measuring gravitational effects using a quantum system. 
\begin{tcolorbox}
	In the original paper of Samuel \& Bhandari, they worked with a subset of Hilbert space $\mathcal{H}$ with non-zero state vectors i.e. $ \ip{\psi} \ne 0$  and not with unit vectors.
\end{tcolorbox}

\section{Kinematic Approach to geometric phase}
The final generalization for the geometric phase has been proposed by Simon \& Mukunda~\cite{Mukunda} based entirely on the kinematic approach. In this approach, there is no need for dynamics (or Hamiltonian), and the whole idea is based on the characteristics of curves connecting two states in projective Hilbert space $\mathcal{R}$.

\noindent For any given two state vectors $ \ket{\psi_1}, \ket{\psi_2} \in \mathcal{B}$, we start by looking for the invariant under the $U(1)$ phase transformation given by 
\begin{equation} \label{eq:modulus}
	\ket{\psi'_1} = e^{i \alpha_1} \ket{\psi_1}, \;\;\;\; \ket{\psi'_2} = e^{i \alpha_2} \ket{\psi_2}.
\end{equation}
The one possible nontrivial invariant can be modulus of the inner product of the two vectors, which reads
\begin{equation}
	\abs{\ip*{\psi_1}{\psi_2}} = \abs{\ip*{\psi'_1}{\psi'_2}}
\end{equation}
that is $U(1) \times U(1)$-invariant. We can extend it to three given states $ \{\ket{\psi_1}, \ket{\psi_2},\ket{\psi_3}\} \in \mathcal{B}$ using \emph{Bargmann invariants}~\cite{Bargmann1964} of the form 
\begin{equation} \label{eq:BargmannInvaraint}
	\ip{\psi_1}{\psi_2}\ip{\psi_2}{\psi_3}\ip{\psi_3}{\psi_1} 
\end{equation} 
that are $U(1) \times U(1) \times U(1)$- invariant and subsequently for $ \{\ket{\psi_1}, \ket{\psi_2}, \dots \ket{\psi_N}\} \in \mathcal{B}$ we can write a cyclic quantity as
\begin{equation} \label{eq:BargmannInvaraintN}
	\ip{\psi_1}{\psi_2}\ip{\psi_2}{\psi_3} \dots \ip{\psi_N}{\psi_1}
\end{equation} 
that is $U(1) \times U(1) \times \dots \times U(1)$ - invariant (there are $N$ numbers of $U(1)$'s).
The modulus of the inner product of the two vectors is the simplest example of a Bargmann invariant
\begin{equation}
	\abs{\ip*{\psi_1}{\psi_2}}^2 = \ip{\psi_1}{\psi_2} \ip{\psi_2}{\psi_1}.
\end{equation}
The expression Eq.~\eqref{eq:BargmannInvaraint} is a $complex$ invariant in contrast with the expression Eq.~\eqref{eq:modulus} which are limited to real and positive values. We will soon explore the importance of Bargmann invariants in the study of geometric phases. 

\subsection{Geometric phase for smooth parametrized curve}
Consider a one-parameter smooth curve $\mathcal{C} \in \mathcal{B}$ that consists of state vectors $\ket{\psi(s)}$ and reads
\begin{equation}
	\mathcal{C} = \{\ket{\psi(s)} \in \mathcal{B} \;|\; s \in [s_1, s_2] \subset \mathds{R}\}.
\end{equation}
By invoking the normalization of $ \ket{\psi(s)} $, it is easy to show that the quantity $ \ip*{\psi(s)}{\dot{\psi}(s)} $ is pure imaginary i.e.
\begin{equation} \label{eq:oneform}
	\ip*{\psi(s)}{\dot{\psi}(s)} = i \Im\ip*{\psi(s)}{\dot{\psi}(s)}
\end{equation}
where the dot represents the derivative with respect to $s$. Now, we perform phase changes on $ \ket{\psi(s)} $ at each value of $s$, i.e., local phase transformation which we call \emph{gauge transformation}. These transformations are characterized by a smooth parameter $\alpha(s)$, which takes $\mathcal{C}$ to $\mathcal{C'}$ as
\begin{equation} \label{eq:gaugetransformation}
	\mathcal{C} \rightarrow \mathcal{C'} = \{ \ket*{\psi'(s)} = e^{i \alpha(s)} \ket{\psi(s)} \;|\; s \in [s_1, s_2] \}.
\end{equation}
Under this gauge transformation, the quantity in Eq.~\eqref{eq:oneform} transforms as 
\begin{equation}
	\expval{\dfrac{d}{ds}}{\psi(s)} \rightarrow \expval*{\dfrac{d}{ds}}{\psi'(s)} = \expval{\dfrac{d}{ds}}{\psi(s)} + i \dot{\alpha}(s)
\end{equation}
and
\begin{equation}
	\Im\ip*{\psi'(s)}{\dot{\psi'}(s)} = \ip*{\psi(s)}{\dot{\psi}(s)} + \dot{\alpha}(s).
\end{equation}
Note that the quantity in Eq.~\eqref{eq:oneform} is invariant under \emph{global transformation} of the kind $ \ket{\psi(s)} \rightarrow \ket{\psi'(s)} = e^{i \alpha} \ket{\psi(s)}$. Now, our goal is to construct a functional of $\mathcal{C}$ that is invariant under the gauge transformation given in Eq.~\eqref{eq:gaugetransformation} i.e. which should be same for $\mathcal{C}$ and $\mathcal{C'}$. Such a functional is given by
\begin{align}
	\phi_g[\mathcal{C}] &= \arg\ip*{\psi(s_1)}{\psi(s_2)} - \Im\int_{s_1}^{s_2} ds \ip*{\psi(s)}{\dot{\psi}(s)} \nonumber \\
	&= \arg\ip*{\psi'(s_1)}{\psi'(s_2)} - \Im\int_{s_1}^{s_2} ds \ip*{\psi'(s)}{\dot{\psi}'(s)} \label{eq:geometricphasemukunda} \\
	&= \text{gauge invariant} \nonumber.
\end{align}
Choosing different curves $\mathcal{C}$ or $\mathcal{C'}$ corresponds to choosing different Hamiltonians, and the invariance of the functional under the gauge transformations means that this is a functional of the image $C \in \mathcal{R}$ of $\mathcal{C}$ in the projective Hilbert space. In addition to the invariance under gauge transformation, the function $ \phi_g[\mathcal{C}] $ is also \emph{reparametrisation} invariant. The reparametrisation transformations are defined as
\begin{equation} \label{eq:reparameterisation}
	\mathcal{C} \rightarrow \mathcal{C'} = \{ \ket{\psi'(s')} \in \mathcal{B} \;|\; s' = f(s), \dfrac{df(s)}{ds} \ge 0  \}
\end{equation}
Under such a transformation, we have $ds = \tfrac{ds}{df} df$
\begin{align*}
	&\arg\ip*{\psi'(f(s_1))}{\psi'(f(s_2))} - \Im\int_{f(s_1)}^{f(s_2)}  \expval*{\dfrac{d}{df} \dfrac{df}{ds}}{\psi'(f(s))} \dfrac{ds}{df} df\\
	&=\arg\ip*{\psi'(f(s_1))}{\psi'(f(s_2))} - \Im\int_{f(s_1)}^{f(s_2)}  \expval*{\dfrac{d}{df} }{\psi'(f(s))} df
\end{align*}
and 
\begin{equation}
	\ket{\psi[f(s)]} = \ket{\psi(s)} \implies \ket{\psi'[f(s_1)]} = \ket{\psi(s_1)}, \;\;\; \ket{\psi'[f(s_2)]} = \ket{\psi(s_2)}
\end{equation}
for $ f(s_1) = s_1 $ and $ f(s_2) = s_2 $. The reparameterization transformation takes a curve $\mathcal{C} \in \mathcal{B}$ to $\mathcal{C'} \in \mathcal{B}$ that is transversed at a different rate. Thus, the functional $\phi_g[\mathcal{C}]$ gives the geometric phase associated with a smooth curve $C \in \mathcal{R}$. The individual terms in the functional depend on a particular lift $\mathcal{C} \in \mathcal{B}$ of the smooth curve $\mathcal{C} \in \mathcal{R}$, whereas the functional itself depends only on the smooth curve $C = \pi[\mathcal{C}] = \pi[\mathcal{C'}]$ in the projective Hilbert space $\mathcal{R}$. This implies the possibility that the geometric phase is the property of the projective Hilbert space $\mathcal{R}$ and we write it as 
\begin{equation} \label{eq:geometricphasefinal}
	\phi_g[C] \equiv \arg\ip*{\psi(s_1)}{\psi(s_2)} - \Im\int_{s_1}^{s_2} ds \ip*{\psi(s)}{\dot{\psi}(s)}.
\end{equation}
Also, this definition is only valid when $ \ket{\psi(s_1)} $ and $ \ket{\psi(s_2)} $ are \emph{non-orthogonal} because the first inner product between the initial and final states vanishes, and hence $\arg\ip*{\psi(s_1)}{\psi(s_2)}$ is not defined in case two states are orthogonal. The two terms on the RHS of the above equation are identified as 
\begin{align}
	\arg\ip*{\psi(s_1)}{\psi(s_2)} &\equiv \phi_t[\mathcal{C}] = \text{total phase of} \; \mathcal{C}, \\
	\Im\int_{s_1}^{s_2} ds \ip*{\psi(s)}{\dot{\psi}(s)} &\equiv \phi_{\text{dyn}}[\mathcal{C}] = \text{dynamical phase of} \; \mathcal{C}
\end{align} 
It is very important to note that these two functionals depend on $\mathcal{C}$ individually and it is only their difference that is a functional of $C$.

\begin{figure}
	\centering
	\includegraphics[width=6cm]{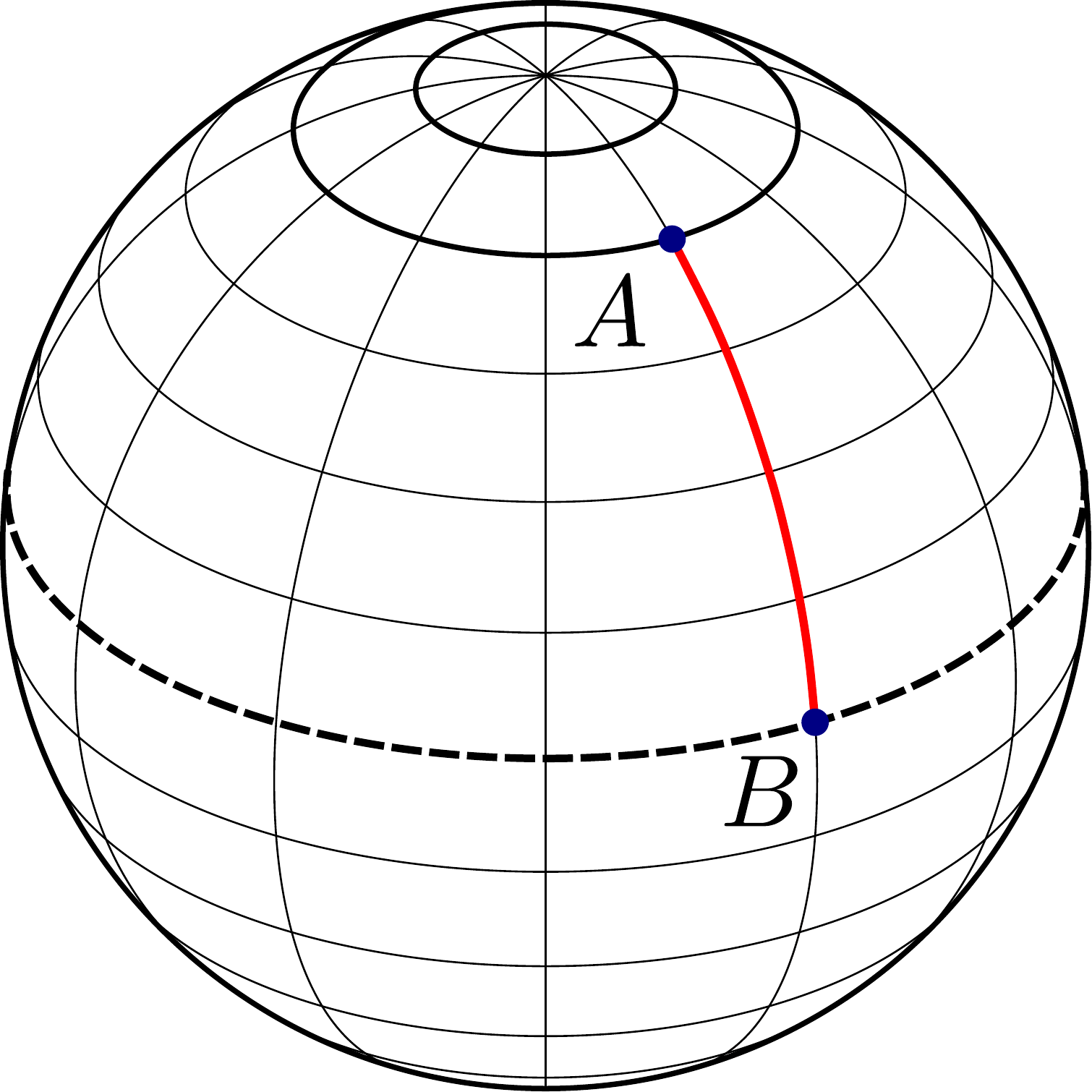}
	\caption{A geodesic on the surface of a sphere.}
	\label{fig:geodesic1}
\end{figure}

\noindent Now, for a given smooth curve $C$ in the projective Hilbert space $\mathcal{R}$. There are several options to choose a lift $\mathcal{C} \in \mathcal{B}$ to calculate $\phi_g[C]$. We point out two such lifts, which are interesting:
\begin{itemize}
	\item[(i)] we can choose a lift such that the total phase $\phi_t[\mathcal{C}] = \arg\ip*{\psi(s_1)}{\psi(s_2)} $ vanishes along the curve and the two state vectors $ \ket{\psi(s_1)} $ and $\ket{\psi(s_2)}$ are said to be \emph{in phase} i.e 
	\[ \phi_t[\mathcal{C}] = 0  \implies \phi_g[C] = -\phi_{\text{dyn}}[\mathcal{C}].\]
	\item[(ii)] or we can choose a lift such that the dynamical phase $\phi_{\text{dyn}}[C]$ vanishes. This lift is called a \emph{horizontal lift} and it hold the parallel-transport law i.e
	\[ \mathcal{C} \;\text{horizontal} \implies \ip*{\psi(s)}{\dot{\psi}(s)} = 0 \implies \phi_{\text{dyn}}[\mathcal{C}] = 0 \implies \phi_g[C] = \phi_t[\mathcal{C}].\] 
\end{itemize}
\subsection\protect{Geodesics in ray space $\mathcal{R}$}
We started by constructing gauge invariant quantities with a given finite set of vectors and then showed how can one approach to define a gauge invariant, reparametrization invariant functional to define the geometric phase. We will now derive an expression for geometric phase in terms of the gauge invariant quantities, very well known as Bargmann invariants. To do that, we first need to introduce the notion of a "geodesic". The shortest curve between any given two points on a surface is a geodesic\footnote{Geodesics are locally shortest paths. The minimal length path between two points is referred to as minimal geodesic connecting those points. Here, it is essential to note that a curve being locally the shortest does not necessarily imply it is globally the shortest~\cite{Do2016,Kimmel1995}. We will frequently use the term geodesic; by this, we shall always mean minimal geodesic, which is unique provided that the given pair of points correspond to nonorthogonal states.}. For example, a geodesic between two points on a sphere is a curve along the great circle passing through the two points, as shown in Fig.~\ref{fig:geodesic1}.  
In this section we will introduce the \emph{geodesics} and how they are important in the context of geometric phase. We will first derive a differential equation by minimizing the distance and using the variational approach. Let us start by considering a smooth parametrized curve $\mathcal{C} = \{ \ket{\psi(s)}\} \subset \mathcal{B}$. The tangent at the point $\ket{\psi(s)}$ on $\mathcal{C}$, gives the velocity that we write as
\begin{equation}
	\ket{u(s)} = \dfrac{d}{ds} \ket{\psi(s)}
\end{equation}
and the component of $ \ket{u(s)} $ that is orthogonal to $ \ket{\psi(s)} $ reads
\begin{equation}
	\ket{u_{\perp}(s)} = \ket{u(s)} - (\ip*{\psi(s)}{u(s)}) \ket{\psi(s)}
\end{equation}
The motivation to choose the orthogonal component of the tangent vector $ \ket{u_{\perp}(s)} $ is because under gauge transformation of the kind $ \ket{\psi(s)} \rightarrow \ket*{\psi'(s)} = e^{i \alpha(s)} \ket{\psi(s)} $ it is $ \ket{u_{\perp}(s)} $ that transforms linearly and homogeneously as same as $ \ket{\psi(s)} $ i.e. $ \ket*{u_{\perp}'(s)} = e^{i \alpha(s)} \ket{u_{\perp}(s)} $ ~\cite{Mukunda1993,Carollo2020}. We can now define a functional that is a ray space quantity called length as
\begin{equation} \label{eq:lengthfuncitional}
	\mathcal{L}[C] = \int_{s_1}^{s_2} ds \ip*{u_{\perp}(s)}^{1/2}.
\end{equation}
which is gauge and reparametrization invariant. The quantity 
\begin{equation}
	\norm{u_{\perp}} =  \ip*{u_{\perp}(s)}^{1/2} = \left( \ip*{u(s)} - \ip*{u(s)}{\psi(s)}\ip*{\psi(s)}{u(s)} \right)
\end{equation}
is the norm of the tangent vector $ \ket{u(s)} $. A class of curves in $\mathcal{R}$ for which this length is stationary ($\delta \mathcal{L} = 0 $) is called \textit{geodesic curves}. The differential equation for the geodesic curves can be obtained using the variational principle. By making an infinitesimal change in $ \ket{\psi(s)} $ we obtain the following
\begin{align}
	\delta \mathcal{L}[C] = &\dfrac{1}{2} \int_{s_1}^{s_2} ds \dfrac{1}{\norm{u_{\perp}}} \delta \left(\ip*{u} - \ip*{u}{\psi}\ip*{\psi}{u}\right),  \nonumber \\ \noalign{\vskip10pt}
	= &\dfrac{1}{2} \int_{s_1}^{s_2} ds \dfrac{1}{\norm{u_{\perp}}} \left(\ip*{\delta u}{u} + \ip*{ u}{\delta u} - \ip*{\delta u}{\psi}\ip*{\psi}{u} - \ip*{u}{\delta  \psi}\ip*{\psi}{u} \right.\nonumber \\ \noalign{\vskip10pt}
	&\qquad\qquad\qquad\qquad \left.- \ip*{u}{\psi}\ip*{\delta \psi}{u} - \ip*{u}{\psi}\ip*{\psi}{\delta u} \right),  \nonumber \\ \noalign{\vskip10pt}
	=&\int_{s_1}^{s_2} ds \dfrac{1}{\norm{u_{\perp}}} \Re\left(\ip*{\delta u}{u} - \ip*{\delta u}{\psi}\ip*{\psi}{u} - \ip*{\delta  \psi}{u}\ip*{u}{\psi} \right), \nonumber \\ \noalign{\vskip10pt}
	=&\int_{s_1}^{s_2} ds \dfrac{1}{\norm{u_{\perp}}} \Re\left(\ip*{\delta u}{u_{\perp}} - \ip*{\delta  \psi}{u}\ip*{u}{\psi} \right)
\end{align} 
where in the third step we collected (1) and (2), (3) and (6), (4) and (5) terms from inside the brackets which are complex conjugates of each other. If we look at the term, 
\begin{align*}
	\ip*{\delta\psi}{u}\ip*{u_{\perp}}{\psi} = &\ip*{\delta\psi}{u_{\perp}}\ip*{u}{\psi} \nonumber \\
	=&\ip*{\delta\psi}{(u - \ip*{\psi}{u} \psi)}\ip*{u}{\psi} \nonumber \\
	=&\ip*{\delta\psi}{u}\ip*{u}{\psi} - \ip*{\delta\psi}{\psi} \abs{\ip*{u}{\psi}}^2 \nonumber \\
	\implies \Re\ip*{\delta\psi}{u}\ip*{u_{\perp}}{\psi} = &\Re\ip*{\delta\psi}{u}\ip*{u}{\psi}
\end{align*}
where we used the fact that $ \ip*{u}{\psi} $ and $ \ip*{\delta\psi}{\psi} $, both are pure imaginary. Therefore,, we can write
\begin{equation*}
	\delta \mathcal{L}[C] = \int_{s_1}^{s_2} ds \dfrac{1}{\norm{u_{\perp}}} \Re\left(\ip*{\delta u}{u_{\perp}} + \ip*{\delta\psi}{u_{\perp}}\ip*{\psi}{u} \right)
\end{equation*}
Now, by integrating the first term with $\delta u$ by parts, we get
\begin{align*}
	\int_{s_1}^{s_2} ds \Re\ip{\dfrac{d}{ds} \delta \psi}{\dfrac{u_{\perp}}{\norm{u_{\perp}}}} = \ip{\delta \psi}{\dfrac{u_{\perp}}{\norm{u_{\perp}}}} \bigg \vert_{s_1}^{s_2}
	- \int ds \ip{\delta \psi}{\dfrac{d}{ds} \dfrac{u_{\perp}}{\norm{u_{\perp}}}}
\end{align*}
and by discarding the boundary terms, we finally have
\begin{equation}
	\delta \mathcal{L}[C] = -\int_{s_1}^{s_2} ds \Re\ip{\delta \psi}{\left(\dfrac{d}{ds} \dfrac{u_{\perp}}{\norm{u_{\perp}}} - \ip*{\psi}{u} \dfrac{u_{\perp}}{\norm{u_{\perp}}}\right)}
\end{equation}
By demanding that this integral vanishes for any arbitrary variations $\delta \psi$ and subjected to the condition $ \ip*{\delta \psi}{\psi} $ is pure imaginary, we get a differential equation for a geodesic which reads 
\begin{equation} \label{eq:diffgeodesic1}
	\left(\dfrac{d}{ds}-\ip*{\psi(s)}{u(s)}\right)\dfrac{\ket{u_{\perp}(s)}}{\norm{u_{\perp}(s)}}=f(s)\ket{\psi(s)}
\end{equation} 
where $f(s)$ is some real function. We note that since $\mathcal{L}[C]$ is gauge and reparametrization invariant, the differential equation for a geodesic is gauge and reparametrization covariant and yields the same result for any lift of a given smooth curve $C \in \mathcal{R}$. Also, if $C$ is a geodesic in $\mathcal{R}$, then any lift of $C$ in $\mathcal{B}$ will also be a geodesic. One such lift is the horizontal one along which $ \ip*{\psi(s)}{u(s)} $ vanishes and Eq.~\eqref{eq:diffgeodesic1} reduces to a simpler form
\begin{equation} \label{eq:geodesicsimple1}
	\frac{d}{ds} \dfrac{\ket{u(s)}}{\norm{u(s)}}=f(s)\ket{\psi(s)}
\end{equation}
with some real $f(s)$. Further, we can exploit the reparameterization freedom to demand $ \norm{u(s)} $ to be constant. This freedom makes our life easy and we can further write Eq.~\eqref{eq:geodesicsimple1} as
\begin{equation} \label{eq:geodesicsimple2}
	\frac{d^2}{ds^2} \ket{\psi(s)} = f(s) \ket{\psi(s)}
\end{equation}
with 
\begin{equation} \label{eq:conds}
	\ip{\psi(s)} = 1, \qquad \ip*{\psi(s)}{u(s)} = 0, \qquad \ip*{u(s)}{u(s)} = \text{constant}.
\end{equation}
Using the conditions given in the above expression, we can evaluate the real function $f(s)$ as
\begin{gather}
	\ip{\psi(s)} = 1 \implies \ip*{\psi(s)}{\dot{\psi}(s)} + \ip*{\dot{\psi}(s)}{\psi(s)} = 0 \nonumber \\
	\ip*{\psi(s)}{\ddot{\psi}(s)} + \ip*{\ddot{\psi}(s)}{\psi(s)} + 2\ip*{\dot{\psi}(s)}{\dot{\psi}(s)}  = 0 \nonumber \\
	\implies f(s) = - \ip*{\dot{\psi}(s)}{\dot{\psi}(s)}
\end{gather}
which further reduces Eq.~\eqref{eq:geodesicsimple2} to
\begin{equation} \label{eq:geodesicsimple3}
	\frac{d^2}{ds^2} \ket{\psi(s)} = - \ip*{\dot{\psi}(s)}{\dot{\psi}(s)} \ket{\psi(s)}
\end{equation}
Since the above differential equation resembles the simple harmonic motion. The general solution of the reads
\begin{equation}
	\ket{\psi(s)} = \ket{\psi(0)} \cos(\omega s) + \ket*{\dot{\psi}(0)} \dfrac{\sin(\omega s)}{\omega}
\end{equation} 
with
\begin{equation}
	\ip*{\psi(0)} = 1, \;\;\;\; \ip*{\psi(0)}{\dot{\psi}(0)} = 0, \;\;\;\; \ip*{\dot{\psi}(0)} = \omega^2
\end{equation}
We note that everything is reduced to the real domain here. For a horizontal lift, we have $\phi_g[C] = \arg\ip*{\psi(s_1)}{\psi(s_2)}$ which will be identically zero. Therefore, we have a very important result here, which becomes handy at many places in which the geometric phase vanishes along a geodesic curve. 
\begin{equation} \label{eq:geodesiczero}
	C = \text{geodesic} \implies \phi_g[C] = 0.
\end{equation}
Given any two non-orthogonal vectors in $\mathcal{B}$, we can connect them by a unique geodesic curve. We consider two nonorthogonal vectors $\ket{\psi_1},\ket{\psi_2}$ such that the inner product is
\begin{equation} \label{eq:inner-product1}
	\ip*{\psi_1}{\psi_2}=\cos(\theta) e^{i\beta}.
\end{equation}
The geodesic curve connecting $\ket{\psi_1}$ and $\ket{\psi_2}$ which is a unique solution of Eq.~\eqref{eq:diffgeodesic} reads \cite{Mukunda}
\begin{align} 
	\ket{\psi(s)} &= e^{i \beta s/ \theta}\left[\cos(s)\ket{\psi_1}+\left(\frac{e^{-i \beta}\ket{\psi_2}-\ket{\psi_1}\cos\theta}{\sin\theta}\right)\sin(s)\right], \nonumber \\
	&= e^{i \beta s/ \theta} ( \sin(\theta - s) \ket{\psi_1} + e^{-i \beta}\sin(s) \ket{\psi_2} )/\sin\theta \label{eq:geodesicequation}
\end{align}
where $s$ varies from $0$ to $\theta$. In this way, one can construct a unique geodesic between any given two nonorthogonal states~\cite{Mukunda1993}. Here we note that, a geodesic connecting the two orthogonal states may not be unique. For example, there exist infinite choices to connect two antipodals points, on the Bloch sphere, via a geodesic. 
\subsection{Bargmann invariant and geometric phase}
\begin{figure}[H]
	\centering
	\includegraphics[width=10cm]{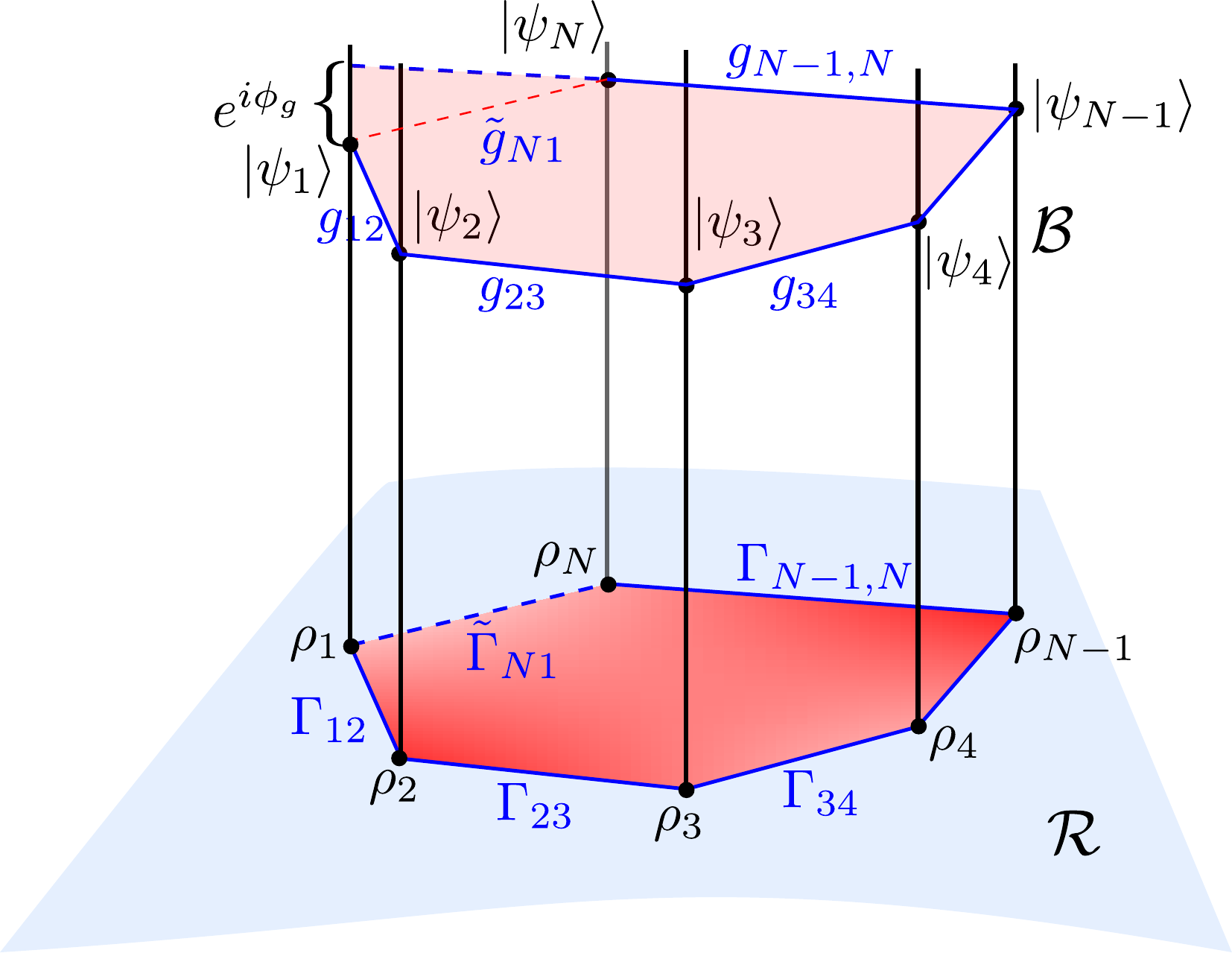}
	\caption{Bargmann invariant}
	\label{fig:bi}
\end{figure}
We now move on to second part of their result, which is about the Bargmann invariant and its connection with the geometric phase. Let us consider as open polygonal curve $C$ in projective Hilbert space $\mathcal{R}$ which consists of $N$-$1$ geodesics ($\Gamma_{12}, \Gamma_{23}$, \dots) connecting $\rho_1$ to $\rho_2$, $\rho_2$ to $\rho_3$, $\dots$, $\rho_{N-1}$ to $\rho_N$ as shown in Fig.~\ref{fig:bi}. The $\rho_1$, $\rho_2$, $\dots$, $\rho_N$ are the given $N$ points on the curve $C \in \mathcal{R}$. Now, we construct a lift $\mathcal{C} \in \mathcal{B}$ of the curve $C$ such that the state vectors $\ket{\psi_1}, \ket{\psi_2}, \ket{\psi_3}, \dots, \ket{\psi_N}$ are mutually nonorthogonal. We connect the successive pair of state vectors with a geodesic $g_{12}, g_{23}, \dots, g_{N-1,N}$ as shown in Fig.~\ref{fig:bi}. Then, the geometric phase for the open curve $C$ is given by
\begin{align}
	\phi_g[C] = &\phi_t[\mathcal{C}] - \phi_{\text{dyn}}[\mathcal{C}] \nonumber \\
	=& \arg \ip*{\psi_1}{\psi_N} - \sum_{i = 1}^{N-1} \phi_{\text{dyn}}[\text{geodesic} (g_{i, i+1}) \ket{\psi_i} \text{to} \ket{\psi_{i + 1}}] 
\end{align}
By making use of the result in Eq.~\eqref{eq:geodesiczero} and the dynamical part in the above expression can be further written as 
\begin{equation}
	\sum_{i = 1}^{N-1} \phi_{\text{dyn}}[\text{geodesic} (g_{i, i+1}) \ket{\psi_i} \text{to} \ket{\psi_{i + 1}}] = \sum_{i = 1}^{N-1} \left\{ \phi_t[g_{i, i+1}] - \phi_g[\Gamma_{i, i+1}] \right\} =  \sum_{i = 1}^{N-1} \phi_t[g_{i, i+1}].
\end{equation}
Therefore, $ \phi_g[C] $ becomes
\begin{align} \label{eq:BargmannIn}
	\phi_g[C] =& \arg \ip*{\psi_1}{\psi_N} - \sum_{i = 1}^{N-1} \phi_t[g_{i, i+1}] \nonumber \\
	=& \arg \ip*{\psi_1}{\psi_N} - \sum_{i = 1}^{N-1} \arg\ip*{\psi_i}{\psi_{i + 1}} \nonumber \\
	=& -\arg \ip*{\psi_1}{\psi_2}\ip*{\psi_2}{\psi_3}\dots\ip*{\psi_{N-1}}{\psi_N}\ip*{\psi_N}{\psi_1}.
\end{align}
Here, we see how beautifully Bargmann invariants appear in the discussion of geometric phase and finally the geometric phase is expressed as the negative of the argument of Bargmann invariant associated with a polygonal (discrete) path $C$. Just for completeness, we can go one step further to close the curve $C$ by connecting $\rho_N$ to $\rho_1$ with a geodesic as shown by the blue dotted line in Fig.~\ref{fig:bi} and call the new curve $\tilde{C}$. The same thing we do with the lift and connect $ \ket{\psi_N} $ to $ \ket{\psi_1} $ with a geodesic and close the curve as shown by the dotted red line in Fig.~\ref{fig:bi}. With these considerations, we shall have 
\begin{align}
	\phi_g[\tilde{C}] =&\phi_t[\mathcal{\tilde{C}}] - \phi_{\text{dyn}}[\mathcal{\tilde{C}}] \nonumber \\
	=&- \sum_{i = 1}^{N} \phi_{\text{dyn}}[g_{i, i+1}] \nonumber \\
	=& -\sum_{i = 1}^{N} \left\{ \phi_t[g_{i, i+1}] - \phi_g[\Gamma_{i, i+1}] \right\} \nonumber \\
	=& -\sum_{i = 1}^{N} \phi_t[g_{i, i+1}] = \sum_{i = 1}^{N} \arg\ip*{\psi_i}{\psi_{i + 1}} \nonumber \\
	=& -\arg \ip*{\psi_1}{\psi_2}\ip*{\psi_2}{\psi_3}\dots\ip*{\psi_{N-1}}{\psi_N}\ip*{\psi_N}{\psi_1} \nonumber \\
	=& \phi_g[C]
\end{align}
with $\ket{\psi_{N+1}} \equiv \ket{\psi_1}$. Here, we see that the geometric for both curves is identical. This is one of the results of Samuel and Bhandari and it works perfectly fine in the general setting. The geodesic connecting $\rho_N$ to $\rho_1$ does not contribute to any geometric phase due to the very basic construction of a geodesic. 

\subsection{Continuous limit of Bargmann invariant}
\begin{figure}[H]
	\centering
	\includegraphics[width=11cm]{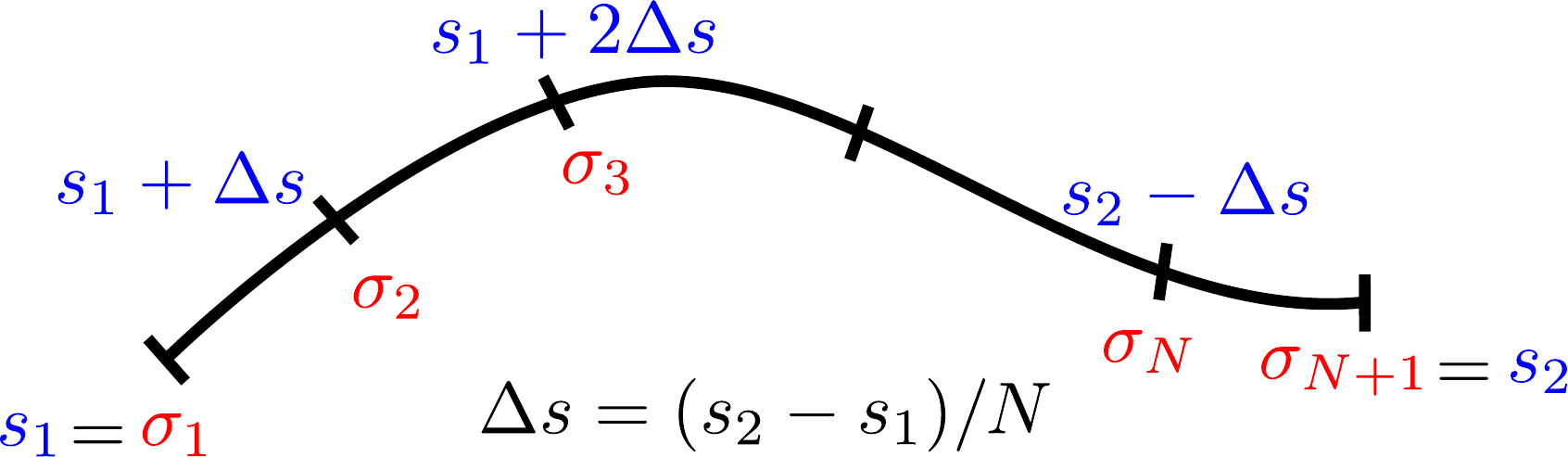}
	\caption{Continuous limit of Bargmann invariant}
	\label{fig:mukundabi}
\end{figure}

In this section, we will retrieve the main expression of the geometric phase Eq.~\eqref{eq:geometricphasefinal} staring from the Bargmann invariant. We take a smooth parameterized curve $\mathcal{C} \in \mathcal{B}$ (with projection $C \in \mathcal{R}$) with parameter $s \in [s_1, s_2]$ and divide it into $N$ bins of size $\Delta s$ as shown in Fig.\ref{fig:mukundabi}.
\noindent The corresponding points on $\mathcal{C}$ are given by
\begin{equation*}
	\ket{\psi(s_1)} \equiv \ket{\psi(\sigma_1)} = \ket{\psi_1}, \;\;\; \ket{\psi(\sigma_2)} = \ket{\psi_2}, \dots, \;\;\; \ket{\psi(\sigma_N)} = \ket{\psi_N}, \;\;\; \ket{\psi(s_2)} \equiv \ket{\psi(\sigma_{N+1})} = \ket{\psi_{N+1}}
\end{equation*}
and connect the successive pair of state vectors $\ket{\psi_j}$ and $\ket{\psi_{j+1}}$ by a geodesic arc such that the geometric phase is written in terms of the Bargmann invariant as
\begin{align*}
	\phi_g[C] =& \lim_{N \rightarrow \infty} \left\{-\arg\ip*{\psi_1}{\psi_2}\ip*{\psi_2}{\psi_3}\dots\ip*{\psi_{N}}{\psi_{N+1}}\ip*{\psi_{N+1}}{\psi_1} \right\} \nonumber \\
	=&\lim_{N \rightarrow \infty} \left\{ \arg\ip*{\psi(s_1)}{\psi(s_2)} - \arg \prod_{j = 1}^{N} \ip*{\psi(\sigma_j)}{\psi(\sigma_{j+1})} \right\} \nonumber \\
	=&\lim_{N \rightarrow \infty} \left\{ \arg\ip*{\psi(s_1)}{\psi(s_2)} - \arg \prod_{j = 1}^{N} \ip*{\psi(\sigma_j)}{\psi(\sigma_{j}) + \Delta s \dot{\psi}(\sigma_{j}) + \mathcal{O}(\Delta s^2)} \right\} \nonumber \\
	\approx&\lim_{N \rightarrow \infty} \left\{ \arg\ip*{\psi(s_1)}{\psi(s_2)} - \arg \prod_{j = 1}^{N} \left(1 + \Delta s \ip*{\psi(\sigma_j)}{\dot{\psi}(\sigma_{j})}\right) \right\} \nonumber \\
	\approx&\lim_{N \rightarrow \infty} \left\{ \arg\ip*{\psi(s_1)}{\psi(s_2)} - \arg \exp\left(\sum_{j = 1}^{N} \Delta s \ip*{\psi(\sigma_j)}{\dot{\psi}(\sigma_{j})}\right) \right\} \nonumber \\
	=& \arg\ip*{\psi(s_1)}{\psi(s_2)} - \arg \exp\left(\int_{s_1}^{s_2} ds \ip*{\psi(s)}{\dot{\psi}(s)}\right)
\end{align*}
which is the same as Eq.~\eqref{eq:geometricphasefinal}. The expression of geometric phase in terms of Bargmann invariant is instrumental in processes where we have measurements, which results in a discrete set of states ~\cite{Cho2019,Gefen2020,Facchi1999}. With this we end the discussion on the geometric phase for pure states. The geometric phase for mixed states has been experimentally verified in several systems such as photons~\cite{Chiao1986}, NMR~\cite{Suter1987}, neutrons~\cite{Dubbers1987}, sodium trimer~\cite{Meyer1998}, and in graphene~\cite{Zhang2005},  Mach-Zehnder interferometer~\cite{Loredo2009} and polarimetry~\cite{Wagh1995a,Wagh1995b}. There also exists a method where the geometric phase has been measured without interferometry~\cite{Alonso2018}.  

\section{Null phase curves}

We have seen that geodesics play an integral role in establishing a relation between the geometric phase and the Bargmann invariants because the system does not acquire any geometric phase when evolves along a geodesic. We connect two consecutive state vectors (nonorthogonal) by a geodesic. This notion can be extended to define a broader class of ray space curves with the property that the geometric phase vanishes for any connected stretch of any of these curves~\cite{Rabei1999}. These are known as "null phase curves", and these provide an alternative way to connect geometric phase with Bargmann invariants by replacing geodesics with null phase curves. Null phase curves (NPCs) are the larger class of curves and geodesics are the subset of them. We recall Eq.~\eqref{eq:BargmannIn}, 
\begin{align} 
	\phi_g[C] &= -\arg \ip*{\psi_1}{\psi_2}\ip*{\psi_2}{\psi_3}\dots\ip*{\psi_{N-1}}{\psi_N}\ip*{\psi_N}{\psi_1} \nonumber \\
	&= -\arg \Delta_N(\psi_1, \psi_2, \dots, \psi_{N})
\end{align}
where $C$ is the $N$-vertex polygon in the projective Hilbert space $\mathcal{R}$ connecting $\rho_1$ to $\rho_2$, $\rho_2$ to $\rho_3$, \dots, $\rho_N$ to $\rho_1$ by geodesics with $\rho_1$, $\rho_2$, \dots, $\rho_N$ being projections of $\ket{\psi_1}$, $\ket{\psi_2}$, \dots, $\ket{\psi_N}$, respectively in $\mathcal{R}$. Here, we note that the LHS of Eq.~\eqref{eq:BargmannIn} gives the geometric phase which depends on a discrete set of state vectors, whereas, to evaluate RHS we need an additional element, a curve that connects consecutive state vectors without contributing any extra geometric phase. Mathematically, a smooth parametrized curve $\mathcal{C} = \{\ket{\psi(s) \in \mathcal{B}\; |\; s \in [s_1, s_2]}\} \in \mathcal{B}$, with 
\begin{equation}
	\ip*{\psi(s)}{\psi(s')} \ne 0, \qquad \forall s,s' \in [s_1, s_2]
\end{equation}
and $ \ket{\psi(s)} $, $\rho(s)$ continuous once differentiable will be an NPC, if for any three mutually nonorthogonal consecutive vectors on it, the third-order Bargmann invariant is real and positive~\cite{Arvind2003} i.e.,
\begin{equation}
	\Delta_3(\psi(s), \psi(s'), \psi(s'')) = \Tr (\rho(s) \rho(s') \rho(s'')) > 0, \qquad s, s', s'' \in [s_1, s_2].
\end{equation}
We leave NPCs with this remark to maintain the continuity of the discussion on the geometric phase for mixed states. We will discuss NPCs in detail in Chapter~\ref{chapter:msr}.

\section{Geometric phase for mixed states}
So far, we have only considered pure quantum states, which are projectors of rank one and represent only a very limited class of quantum states. For example, the dynamics of a quantum system interacting with another system (an external environment) cannot be fully captured by a pure state. These are the so-called open quantum systems. Even if we start initially in a pure state of the form $\ket{\psi} = \ket{\phi_1} \otimes \ket{\phi_2} \in \mathcal{H}_{SE} = \mathcal{H}_S \otimes \mathcal{H}_E$, the system evolves to a state (entangled state) that cannot be written anymore as the tensor product of the states of individual systems. However, in such cases, a quantum state can be represented by the statistical average (or the ensemble) of the pure states. These statistical averages are called \emph{mixed} quantum states and are represented by \emph{density operators} (or \emph{density matrix}). Suppose that a quantum system is in either one of the states $\ket{\psi_i}$ with respective probabilities $p_i$ where $i = 1,2,3, \dots, N$. The density operator for this system is defined as
\begin{equation} \label{eq:densityoperator}
	\rho \equiv \sum_{i} p_i \dyad{\psi_i}.
\end{equation} 
with $\sum_{i} p_i = 1$. An important point to appreciate here is that statistical averages are different from quantum superposition. If we have a system that is in the pure states $\ket{\psi_1}$ and $\ket{\psi_2}$ with respective probabilities $p_1$ and $p_2$, then the superposed state will look like
\begin{equation}
	\ket{\Psi} = \sqrt{p_1} \ket{\psi_1} + e^{i \phi} \sqrt{p_2} \ket{\psi_2}
\end{equation}  
with an arbitrary and unknown phase $\phi \in \mathds{R}$ and the corresponding density matrix will be
\begin{equation}
	\rho = p_1 \dyad{\psi_1} + p_2 \dyad{\psi_2}
\end{equation}
which are not the same. A rank-one projector, representing a pure state, is a special case of a density operator. However, a general density operator can have a rank greater than 1. The density operator of a statistical average of a number of pure states is a convex sum of the density operators of the individual states. A density operator $\rho$ is represented by a positive operator with unit trace, i.e.
\begin{equation}
	\expval{\rho}{\phi} \ge 0 \;\;\forall \;\;\ket{\phi} \in \mathcal{H}, \;\;\;\;\;\; \Tr\rho = 1.
\end{equation}

\noindent For a given distribution of states $ \{p_i, \ket{\psi_i}\} $, we have a unique density operator given by Eq.~\eqref{eq:densityoperator}, however, the reverse is not true. The given density operator $\rho$ can be decomposed into infinite ways representing distinct ensembles. To illustrate this, let us take $ \rho $ and write its spectral decomposition as
\begin{equation}
	\rho = \sum_{i} \lambda_i \dyad{\psi_i}, \qquad \lambda_i \ge 0, \;\;\;\; \sum_{i} \lambda_i = 1
\end{equation}
where $\{\ket{\psi_i}\}$ forms an orthonormal set of basis. Now we define a new set of vectors $ \ket{\phi_k} $ using a unitary operator $W$ as
\begin{equation}
	\ket{\phi_k} = \sum_{i} W_{ki} \sqrt{\lambda_i} \ket{\psi_i}
\end{equation}
and we see
\begin{align*}
	\sum_{k} \dyad{\phi_k} =& \sum_k \left( \sum_{i} W_{ki} \sqrt{\lambda_i} \ket{\psi_i} \right) \left( \sum_{j} W^*_{jk} \sqrt{\lambda_j} \bra*{\psi_j} \right) \\
	=&  \sum_{ij} \sqrt{\lambda_i \lambda_j} \left(\sum_{k} W_{ki} W^*_{jk}\right) \dyad{\psi_i}{\psi_j} \\
	=& \sum_{ij} \delta_{ij}\sqrt{\lambda_i \lambda_j} \dyad{\psi_i}{\psi_j} \\
	=& \sum_{i} \lambda_i \dyad{\psi_i} = \rho
\end{align*}
which results in the same density operator. Therefore, we get a distinct decomposition by choosing a distinct unitary $W$, and therefore infinitely many decompositions exist for a given $\rho$.

\noindent In this section, we extend the definition of a geometric phase for \emph{mixed} quantum states. The fundamental criterion like gauge invariance and reparametrization invariance will remain the same. We also have a parallel-transport law for mixed states. Due to the complex structure of mixed states, there have been many attempts to settle the definition of geometric phase for the mixed state, and the literature is vast on the same. As a result, we have several possibilities to define the geometric phase that are not consistent, in general, with each other (we will cover some examples to illustrate this). We will try our best to cover critical results in the development of the field. We begin our discussion on the geometric phase for mixed states, starting with the interferometric approach by Sjoqvist et al. \cite{Sjoqvist}. It has been generalized to include non-degenerate states \cite{Singh} but still for unitary evolution. It was further generalized to provide a consistent definition for geometric phase for the mixed states that undergo a non-unitary evolution by a kinematic approach \cite{Tong}. There are number of articles where the geometric phase has been calculated in open quantum system settings~\cite{Bassi2006,Carollo2003,Laflamme2010,Yi2005,Yakovleva2019,Carollo2014} We will also discuss the first experimental measurement of the mixed state geometric phase using photon interferometry~\cite{Ericsson}. 

\noindent Apart from this, there is another approach to define the geometric phase for mixed state by Uhlmann ~\cite{Uhlmann1986,Uhlmann1991}. A key idea used by Uhlmann in his analysis is to lift the given density operator $\rho$ for a system, acting on the Hilbert space $\mathcal{H}$, to an extended Hilbert space $\mathcal{H^{\text{ext}}} = \mathcal{H} \otimes \mathcal{H'}$ where $\mathcal{H'}$ is another Hilbert space. This process is known as purification. Uhlmann was probably the first to address the issue of mixed state holonomy, but as a purely mathematical problem. We will discuss a work by Ericsson et al. \cite{Ericsson2003} that presents more information on this approach. There were several studies in which topological classification has been shown using the Uhlmann phase ~\cite{Viyuela2014,Viyuela2014b,Viyuela2015}. The definition by Uhlmann works for both unitary and non-unitary processes. This approach is not usually preferred because of its complex mathematics and lack of experimental observations. There are several theoretical and experimental studies that talk about the compatibility of two approaches ~\cite{Slater2002,Zanardi2006,Andersson2013,Zhu2011,Gong2018}. A differential geometric approach was also given by Chaturvedi et al.~\cite{Chaturvedi2004} for the mixed state geometric phase. 
\subsection{Mixed state geometric phase in interferometry}
For the pure states, we define a relative phase (Pancharatnam) between two non-orthogonal state vectors $ \ket{A} $ and $ \ket{B} $ as 
\begin{equation}
	\phi = \arg\ip{A}{B}.
\end{equation}
In the context of interferometry this phase difference can be measured by shifting the phase of $ \ket{A} $ by $\chi$ i.e. $ \ket{A} \rightarrow e^{i \chi} \ket{A} $ and making it interfere with $ \ket{B} $ which results in the following intensity
\begin{align} \label{eq:interference}
	\mathcal{I} = &\abs{e^{i \chi} \ket{A} + \ket{B}}^2 \nonumber \\
	=&2 + 2 \abs{\ip{A}{B}} \cos \left[\chi - \arg\ip{A}{B}\right].
\end{align} 
The intensity attains the maximum value precisely at the Pancharatnam phase $\phi = \arg\ip{A}{B}$. By following the same argument for a mixed state $\rho_0$ that is undergoing a unitary evolution such that $\rho_0 \rightarrow \rho(t) = U(t) \rho_0 U^{\dagger}(t)$. We now write a spectral decomposition for $\rho_0$ and $\rho(t)$ as
\begin{equation} \label{eq:spectraldecom}
	\rho_0 = \sum_{k}w_k\dyad{k}, \qquad \rho(t) = \sum_{k}w_k\dyad{k(t)}
\end{equation}
where $ \{\ket{k}\} $,  $ \{\ket{k(t)}\} $ form a set orthonormal basis and are related by $\ket{k(t)} = U(t)\ket{k}$.

\noindent Now, we introduce a phase shift in $\ket{k} \mapsto \text{e}^{i \chi}\ket{k}$ in a similar fashion to that we did before, and see the total interference profile,
\begin{align}
	\mathcal{I} &= \sum_{k} \mathcal{I}_k \nonumber \\
	&= \sum_{k} w_k \lvert \text{e}^{i \chi} \ket{k} +   \ket{k(t)} \rvert ^2 \nonumber\\
	&= \sum_{k} w_k \left( \bra{k}\ket{k} + \bra{k(t)}\ket{k(t)} + \text{e}^{-i \chi} \bra{k}\ket{k(t)} + \text{e}^{i \chi} \bra{k(t)}\ket{k}  \right) \nonumber\\
	&= 2 + 2 \sum_{k} w_k \lvert \bra{k}\ket{k(t)} \rvert\cos \left(\chi - \arg\bra{k}\ket{k(t)} \right)  \qquad \qquad   \because \;\sum_{k}w_k = 1
\end{align}
Here we note that the total intensity $\mathcal{I}$ is an incoherent average of the interference profiles $\mathcal{I}_k$ of individual pure states. Now, we will present the above expression in a form similar to Eq.~\eqref{eq:interference} which will provide physical insights into the picture. In order to do that, we define  
\begin{equation}
	\ip{k(t)}{k(t)} = \expval{U(t)}{k} \equiv v_k e^{i \varphi_k}
\end{equation}
such that $\varphi_k = \arg\expval{U(t)}{k} $ which is the relative phase between $\ket{k}$ and $\ket{k(t)}$ and $v_k = \abs{\expval{U(t)}{k}}$ is defined as \emph{visibility factor}. Using these notation, we get
\begin{equation}
	\begin{aligned}
		\mathcal{I} &= 2 + 2 \sum_{k} w_k v_k \cos \left(\chi - \varphi_k\right), \\
		&= 2 + 2 \left[ \cos\chi \left( \sum_{k} w_k v_k \cos\varphi_k \right) + \sin\chi \left( \sum_{k} w_k v_k \sin\varphi_k \right) \right],\\
		&= 2 + 2 \left[  \cos\chi \bigg \lvert \sum_{k} w_k v_k e^{i \varphi_k} \bigg \rvert \cos \left(\arg  \sum_{k} w_k v_k e^{i \varphi_k}  \right) \right. \\ 
		& \left. \qquad \qquad \qquad +  \sin\chi \bigg \lvert \sum_{k} w_k v_k e^{i \varphi_k} \bigg \rvert \sin\left( \arg  \sum_{k} w_k v_k e^{i \varphi_k} \right)	\right].
	\end{aligned}
\end{equation}
by defining
\begin{equation}
	\varphi \equiv \arg  \left(\sum_{k} w_k v_k \text{e}^{i \varphi_k}\right), \;\;\; \text{and} \;\;\;\; v \equiv \abs{\sum_{k} w_k v_k e^{i \varphi_k}}
\end{equation}
the expression for intensity reduces to 
\begin{equation}
	\mathcal{I} = 2 + 2 v \cos \left(\chi-\varphi\right)
\end{equation}
which is very similar to Eq.~\eqref{eq:interference}. On further inspection, one finds 
\begin{equation}
	\varphi = \arg \left( \sum_k w_k \expval{U(t)}{k} \right) = \arg \Tr\left[U(t) \rho_0 \right] 
\end{equation}
and
\begin{equation}
	v = \abs{\sum_k w_k \expval{U(t)}{k}} = \abs{\Tr\left[U(t) \rho_0 \right]}
\end{equation}
that is 
\begin{equation}
	\Tr \left(U(t) \rho_0 \right) = v \text{e}^{i \varphi}
\end{equation}
So, for a continuous evolution of $\rho_0$, given by $\rho(t) = U(t) \rho_0 U^{\dagger}(t)$, we see that the final state $\rho(t)$ acquires a phase with respect to the initial state $\rho_0$ when
\[ \Tr \left(U(t) \rho_0 \right) \ne 0.\] 

\subsection{Parallel transport}
To interpret the total phase acquired by the state $\varphi = \arg \Tr\left[U(t) \rho_0 \right] $ as the geometric phase, we need to ensure that the state is parallel transported. In extension of the notion of parallel transport for pure states, here we demand that at each time the $\rho(t)$ must be \emph{in phase} with the state $\rho(t + dt)$ at an infinitesimal time apart form $\rho(t)$. The state $\rho(t + dt)$ is related to $\rho(t)$ as
\begin{equation}
	\rho(t + dt) = U(t+dt) \rho_0 U^{\dagger}(t + dt) = U(t+dt) U^{\dagger}(t) \rho(t) U(t) U^{\dagger}(t + dt).
\end{equation} 
From this we can see the phase difference between $\rho(t + dt)$ and $\rho(t)$ as
\begin{equation} \label{eq:dphi}
	d\varphi = \arg \left[\Tr \left(U(t+dt)U^{\dagger}(t)\rho(t)\right)\right]
\end{equation}
and to get rid of this phase $d\varphi$, $ \Tr \left(U(t+dt)U^{\dagger}(t)\rho(t)\right) $ must be real and positive. This is one of the generalizations of Pancharatnam's connection for pure states. We can further write
\begin{equation} 
	\Tr \left[U(t+dt)U^{\dagger}(t)\rho(t)\right] = 1 + \Tr\left[\dot{U}(t)U^{\dagger}(t)\rho(t)\right]dt + \mathcal{O}(dt^2)
\end{equation}
and by using the normalization and hermiticity of $\rho(t)$, we can deduce 
\begin{equation}
	\Tr[\rho(t)\dot{U}(t)U^{\dagger}(t)] = \text{pure imaginary.}
\end{equation} 
From here, the parallel transport condition for mixed states evolving unitarily reads
\begin{equation} \label{eq:paralleltransportmixed}
	\Tr\left[\rho(t)\dot{U}(t)U^{\dagger}(t)\right] = 0.
\end{equation} 
The above condition together with a $\rho(t)$ determines the $N \times N$ unitary, up to $N$ phase factors, where $N$ is the $\dim[\rho(t)]$. These $N$ factors can be fixed by demanding a condition that is more strong than the condition in Eq.~\eqref{eq:paralleltransportmixed} which is stated as
\begin{equation}
	\expval{\dot{U}(t) U^{\dagger}(t)}{k(t)} = 0, \qquad k = 1,2,3,\dots,N.
\end{equation} 
where $ \{\ket{k(t)}\}$ are defined in Eq.~\eqref{eq:spectraldecom}. The above condition basically demands the parallel transport of individual eigenstates of $\rho(t)$ independently. These two conditions together are sufficient to find the parallel transport operator $U(t)$ for a given non-degenerate density matrix $\rho(t)$. 

\noindent The last component in this part is the dynamical phase. The dynamical phase is given by the time integral of the average Hamiltonian $H$ of the system, which reads
\begin{equation}
	\phi_{\text{dyn}} = - \dfrac{1}{\hbar} \int_{0}^{T} dt \Tr[\rho(t) H(t)].
\end{equation}
Using Eq.~\eqref{eq:dphi}, we can further write it as 
\begin{equation}
	\phi_{\text{dyn}} = \int_{0}^{T} d \varphi = -i \int_{0}^{T} dt \Tr\left[\rho_0U^{\dagger}(t)\dot{U}(t)\right]
\end{equation} 
which vanishes identically if the state $\rho_0$ is undergoing parallel transport. Therefore, for a state that traced a curve $C: t \in [0, T] \rightarrow \rho(t) = U(t) \rho_0 U^{\dagger}(t)$ where $U(t)$ satisfies the parallel-transport condition, the geometric phase is expressed as 
\begin{equation} \label{eq:sjoqovistgp}
	\phi_g[C] = \varphi = \arg \Tr\left[U(t) \rho_0 \right] = \arg \left( \sum_k w_k v_k e^{i \varphi_k} \right)
\end{equation}

\begin{figure}[H]
	\centering
	\includegraphics[width=10cm]{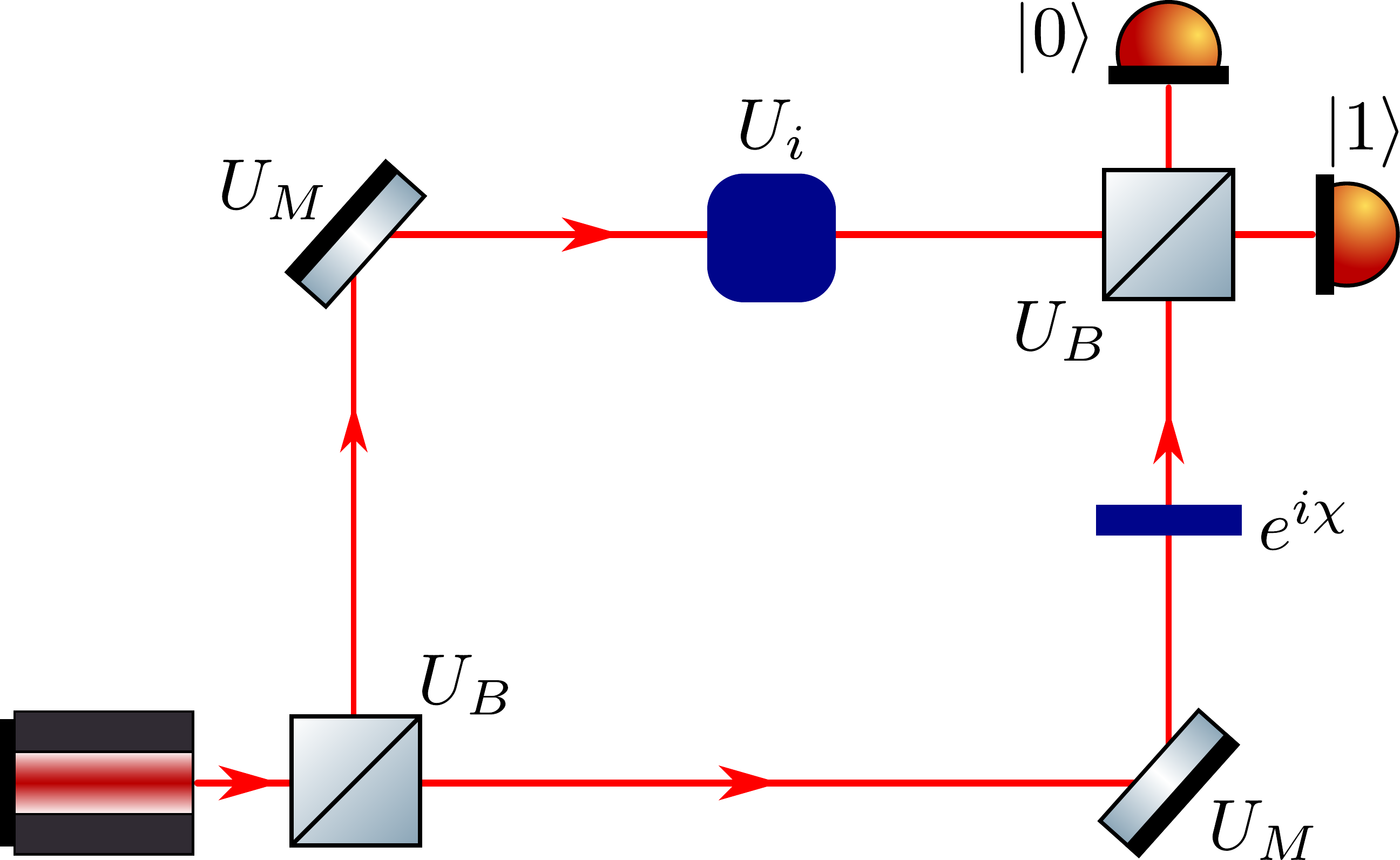}
	\caption{Sj\"oqvist interferometric setup.}
	\label{fig:sjoqvist2000}
\end{figure}

\noindent The geometric phase defined above satisfies the following properties:
\begin{enumerate}
	\item For pure states $\rho_0 = \dyad{\psi_0}$, the phase difference $\varphi$ reduces to Pancharatnam (or relative) phase between $\ket{\psi}$ and $U(t)\ket{\psi}$.
	\item For a pure state density operator $\rho(t) = \dyad{\psi (t)}$, the parallel transport condition Eq.~\eqref{eq:paralleltransportmixed} reduces to 
	\begin{equation}
		\ip*{\psi(t)}{\dot{\psi}(t)} = 0
	\end{equation}
	which was discussed in the context of pure state geometric phase.
	\begin{align}
		\Tr\left[\rho(t)\dot{U}(t)U^{\dagger}(t)\right] = \Tr\left[\dyad{\psi (t)}\dot{U}(t)U^{\dagger}(t)\right]
	\end{align}
\end{enumerate}
The Mach-Zehnder interferometer as shown in Fig.~\ref{fig:sjoqvist2000} was used by Sj\"oqvist \emph{et al} \cite{Sjoqvist} to arrive at the intensity interference pattern which reads
\begin{equation}
	\mathcal{I} = 2 + 2 \abs{\Tr\left[U_i \rho_0 \right]} \cos \left[ \chi - \arg\Tr\left(U_i \rho_0 \right) \right].
\end{equation}
for a mixed state undergoing a unitary evolution,
\begin{equation}
	\rho_{0} \rightarrow \rho(t) = U(t)\rho_{0}U^{\dagger}(t)
\end{equation}
with $\chi$ being a relative $U(1)$ phase as shown in Fig.~\ref{fig:sjoqvist2000}.
\subsection{Gauge invariance}
%
The expression of $\phi_g[C]$ given in Eq.~\eqref{eq:sjoqovistgp} defines the geometric phase for a mixed state evolving unitarily and the unitary is completely specified by the condition of parallel transport \eqref{eq:paralleltransportmixed}. However, if $U(t)$ does not satisfy the parallel transport condition, we have an additional contribution to the dynamical phase and $\phi_g[C] = \arg \Tr\left[U(t) \rho_0 \right]$ is no longer a purely geometrical quantity (or not gauge invariant). For example, consider a group element $G$ (which is $U(N)$) given by
\begin{equation}
	G = \underbrace{U(1) \times U(1) \times \dots \times U(1)}_{\text{N times}} \equiv \sum_{n = 1}^{N} e^{i \theta_n} \dyad{n}
\end{equation}
where \{$\theta_k$\} are arbitrary phases and $\{\ket{n}\}$ are the orthonormal bases of $\rho(0)$. Now, under such a transformation
\begin{equation} \label{eq:gaugetransSingh}
	U(t) \in U(N) \longrightarrow U'(t) = U(t)\sum_{n = 1}^{N} e^{i \theta_n(t)} \dyad{n}.
\end{equation}
leaves the evolution of the density matrix unaffected~\cite{Singh}
\begin{equation}
	\begin{aligned}
		\rho_0 \longrightarrow \rho'(t) = U'(t) \rho_0 U'^{\dagger}(t) = U(t) \rho_0 U^{\dagger}(t) = \rho(t).
	\end{aligned}
\end{equation}
where we make use of the spectral decomposition for $\rho(0)$. Therefore, depending on the choice of parameters $\theta_n$'s, we have an infinite number of transformations that correspond to the same evolution of $\rho(t)$. As we have observed in the case of pure state, we get a gauge invariant quantity by subtracting the dynamical phase from the total. However, under the transformation given by Eq.~\eqref{eq:gaugetransSingh}, the total phase transforms as 
\begin{equation}
	\phi_t \longrightarrow \phi_t' = \arg \{\Tr\left[\rho_0 U'(T)\right]\}= \arg \left[ \sum_{k}  w_k e^{i \theta_k (T)} \expval{U(T)}{k}\right],
\end{equation}
whereas dynamical phase transforms as
\begin{equation}
	\phi_{\text{dyn}} \longrightarrow \phi_{\text{dyn}}' = - i \int_{0}^{T} dt \; \Tr \left[\rho_0 U'^{\dagger}(t) \dot{U}'(t)\right] = - i \int_{0}^{T} dt \; \Tr \left[ \rho_0 U^{\dagger}(t) \dot{U}(t) \right] + \sum_{k} w_k \theta_n(T).
\end{equation}
It is very clear from the expressions of the total and dynamical phases that we cannot get rid of  the $\theta_k(t)$ dependence in the total phase just by subtracting the dynamical. In order to resolve this problem, a gauge invariant functional~\cite{Singh} can be defined that reads
\begin{equation} \label{eq:gp-singh}
	\phi_g[U] = \arg \left[ \sum_{k} w_k \expval{U(T)}{k} \text{exp} \left\{- \int_{0}^{\tau} dt \expval{U^{\dagger}(t) \dot{U}(t)}{k}\right\} \right].
\end{equation}
The above expression is gauge invariant, and we can very easily verify that
\begin{align*}
	\phi_g[U'] &= \arg \left[ \sum_{k} w_k \expval{U'(T)}{k} \text{exp} \left\{- \int_{0}^{\tau} dt \expval{U'^{\dagger}(t) \dot{U}'(t)}{k}\right\} \right] = \phi_g[U].
\end{align*}
In addition to that, for the unitary $U(t)$ which satisfies the parallel-transport condition, this expression reduces to one derived in~\cite{Sjoqvist}. Further for pure states (density matrix of rank 1), it reduces to the existing results in~\cite{Mukunda}.

\subsection{Explicit example of a 2-level (spin -1/2) system}
Consider a case of a 2-level (spin -1/2) system with density matrix
\begin{equation}
	\rho_0 = \dfrac{1}{2} \left(\mathds{1} + \vb{r} \vdot \boldsymbol{\sigma}\right)
\end{equation}
With the choice of $\vb{r} = \left(r \sin \theta, 0, r \cos \theta \right)$, we get
\begin{equation}
	\rho_0 = \dfrac{1}{2} \begin{pmatrix}
		1 + r \cos \theta & r \sin \theta \\
		r \sin \theta & 1 - r \cos \theta
	\end{pmatrix} .
\end{equation}
The above density matrix can be diagonalized as 
\begin{equation}
	\rho_0 = \dfrac{1}{2} \begin{pmatrix}
		1 + r  & 0 \\
		0 & 1 - r 
	\end{pmatrix} 
\end{equation}
with the following eigenvectors
\begin{equation}
	\ket{k+} = \begin{pmatrix}
		\cos (\theta/2) \\
		\sin (\theta/2)
	\end{pmatrix},  \; \; \;  \text{and}  \; \; \; \ket{k-} = \begin{pmatrix}
		\sin (\theta/2) \\
		-\cos (\theta/2)
	\end{pmatrix}.
\end{equation}
The geometric phase can be obtained using Eq.~\eqref{eq:gp-singh} as
\begin{equation} \label{eq:phig1}
	\begin{aligned}
		\phi_g[U] &= \arg \left[- \dfrac{1+r}{2} e^{i \pi \cos \theta} - \dfrac{1-r}{2} e^{-i \pi \cos \theta}\right], \\
		&= \arg \left[- \left( \cos(\pi\cos\theta) + i r \sin(\pi\cos\theta) \right) \right],\\
		&= \tan^{-1} [r \tan(\pi \cos \theta)].
	\end{aligned}
\end{equation}
which can be further simplified as
\begin{equation} \label{eq:phig2}
	\begin{aligned}
		\phi_g[U] &= \tan^{-1} [r \tan(\pi \cos \theta)], \\
		&= \tan^{-1} [- r \tan(\pi - \pi \cos \theta)], \\
		&= \tan^{-1} [-r \tan(\pi(1 -  \cos \theta))], \\
		&= - \tan^{-1} \left[-r \tan\dfrac{\Omega}{2}\right],
	\end{aligned}
\end{equation}
with $ \Omega = 2 \pi (1 - \cos \theta) $ being the solid angle subtended by the Bloch vector at the origin. This is the same expression obtained by Sj\"{o}qvist \emph{et al.}~\cite{Sjoqvist}, however, by imposing the condition of parallel transport, unlike here. The above result reduces to the usual expression for the geometric phase of the pure state, $ \phi_g[U] = -\Omega/2 $, in the limit $r \rightarrow 1$. Note, the expression of $\phi_g[U]$ in Eqs.~\eqref{eq:phig1},~\eqref{eq:phig2} is ill defined in the limit of $r \rightarrow 0$.


\section{Non-unitary evolution}
The definition of geometric phase was first generalized for the nonunitary evolution using operator sum representation or \emph{Kraus decomposition}~\cite{Faria2003,Ericsson2003}, but it was found that in this approach, the geometric phase depends explicitly on the choice of Kraus operators, which are not unique. Also, in~\cite{Carollo} the geometric phase of a system subjected to decoherence has been proposed through a quantum jump approach that discusses only a particular system. This ambiguity was resolved by Tong \emph{et al.}\cite{Tong} by giving a formalism to evaluate geometric phases in nonunitary evolution by taking a kinematic approach. In this section, we will elucidate the main results in~\cite{Tong}.

%
Let us start by consider a $ N $ dimensional quantum system $ S $ described by a Hilbert space $\mathcal{H}_S$ such that the initial state is written as
\begin{equation}
	\rho(0) = \sum_k \omega_k(0) \dyad{\phi_k(0)}.
\end{equation}
An evolution of this state can be written as
\begin{equation}
	\mathcal{P}: t \in [0, \tau] \rightarrow \rho(t) = \sum_k \omega_k(t) \dyad{\phi_k(t)}
\end{equation}
where $\omega_k(t) \ge 0$ and $ \{\ket{\phi_k (t)}\} $ are the eigenvalues and eigenvectors of $\rho$ at time $ t $. Now, we write purification for $\rho(t)$ by introducing an ancilla $ A $ such that we get a tensor product structure $\mathcal{H}_S \otimes \mathcal{H}_A$ and the lifted pure state in extended Hilbert space can be written as
\begin{equation}
	\ket{\Psi(t)} = \sum_k \sqrt{\omega_k(t)} \ket{\phi_k(t)} \otimes \ket{a_k} \in \mathcal{H}_s \otimes \mathcal{H}_A
\end{equation} 
with $ t \in [0, \tau] $. It is very straightforward to check that
\begin{equation}
	\rho(t) = \Tr_A \dyad{\Psi(t)}.
\end{equation}
Now, we can define the Pancharatnam relative phase between $ \ket{\Psi(\tau)} $ and $ \ket{\Psi(0)} $ as
\begin{equation}
	\alpha(\tau) = \arg \ip{\Psi(0)}{\Psi(\tau)} = \arg \left( \sum_k \sqrt{\omega_k(0) \omega_k(\tau)} \ip{\phi_k(0)}{\phi_k(\tau)}\right)
\end{equation}
Since $ \{ \ket{\phi_k(0)} \} $ and $ \{ \ket{\phi_k(t)} \} $ are orthonormal basis from the same Hilbert space $\mathcal{H}_S$, there must exist a unitary operator $ V (t) $ for $ t \in [0, \tau] $ such that
\begin{equation}
	\ket{\phi_k(t)} = V(t) \ket{\phi_k(t)}
\end{equation}
with $ V(0) = \mathds{1} $, $\mathds{1}$ being the identity operator in $\mathcal{H}_S$. We can also write $ V(t) $ explicitly as
\begin{equation}
	V(t) = \dyad{\phi_1(t)}{\phi_1(0)} + \dyad{\phi_2(t)}{\phi_2(0)} + \dots + \dyad{\phi_N(t)}{\phi_N(0)}.
\end{equation}
With this expression of $ V(t) $ we can rewrite $\alpha(\tau)$ as
\begin{equation}
	\alpha(\tau) = \arg \ip{\Psi(0)}{\Psi(\tau)} = \arg \left( \sum_k \sqrt{\omega_k(0) \omega_k(\tau)} \expval{V(\tau)}{\phi_k(0)} \right)
\end{equation}
This is the relative phase between the final and initial state, not the geometric phase. In order to interpret this as the geometric phase, we need to ensure that the evolution satisfies the parallel transport condition, i.e.
\begin{equation}
	\ip*{\Psi(t)}{\dot{\Psi}(t)} = 0.
\end{equation}
There is only one condition or constraint that is sufficient only for a pure state. However, in case of mixed state we require $ N $ conditions to fix the arbitrary phases of $ V(t) $ and to ensure that the geometric phase is independent of purification.  We note that there exists a set of unitaries $ \tilde{V}(t) $ for $ t \in [0, \tau] $ that realize the same evolution. These are nothing but the one with includes gauge transformation and written as
\begin{equation} \label{eq:V-tilde}
	\tilde{V}(t) = V(t) \sum_{k=1}^N e^{i \theta_k(t)} \dyad{\phi_k(0)}
\end{equation}
for real time-dependent parameters $\theta_k(t)$ such that $\theta_k(0) = 0$. We can particularly choose one unitary $ V^{\parallel}(t) $ from the whole set such that it satisfies the parallel transport condition
\begin{equation} \label{eq: V-parallel}
	\expval{V^{\parallel \dagger} \dot{V}^{\parallel}(t)}{\phi_k(0)} = 0, \;\;\;\;\; k = 1,2,\dots,N.
\end{equation}
in which case $\alpha(\tau)$ can be interpreted as the geometric phase for $\mathcal{P}$ of the mixed state. By substituting, $ V^{\parallel}(t) =  \tilde{V}(t) $ we have
\begin{equation*}
	V^{\parallel}(t) = V(t) \sum_{k} e^{i \theta_k(t)} \dyad{\phi_k(0)}
\end{equation*}
\begin{equation*}
	\dot{V}^{\parallel}(t) = \dot{V}(t) \sum_{k} e^{i \theta_k(t)} \dyad{\phi_k(0)} + i \dot{\theta}_k(t) V(t) \sum_{k} e^{i \theta_k(t)} \dyad{\phi_k(0)}
\end{equation*}
and by substituting in \eqref{eq: V-parallel}, we get
\begin{gather}
	V^{\parallel \dagger}(t) \dot{V}^{\parallel}(t)=  \sum_{l,m} e^{-i (\theta_l(t) - \theta_m(t))} \dyad{\phi_l(0)} V^{\dagger}(t) \dot{V}(t) \dyad{\phi_m(0)} + i  \dot{\theta}_l(t) \nonumber \\
	\expval{V^{\parallel \dagger}(t) \dot{V}^{\parallel}(t)}{\phi_k(0)} = \expval{V^{\dagger}(t) \dot{V}(t)}{\phi_k(0)} + i \dot{\theta}_k(t) = 0 
\end{gather}
which gives us
\begin{equation}
	\theta_k(t) = i \int_{0}^{t}ds\expval{V^{ \dagger}(s) \dot{V}(s)}{\phi_k(0)} 
\end{equation}
Finally, taking all the above expressions into account, we can write the geometric phase $\gamma[\mathcal{P}]$ for the path $\mathcal{P}$ as
\begin{align}
	\gamma[\mathcal{P}] &= \arg \left( \sum_k \sqrt{\omega_k(0) \omega_k(\tau)} \expval{V^{\parallel}(\tau)}{\phi_k(0)} \right) \nonumber \\
	&=\arg \left( \sum_k \sqrt{\omega_k(0) \omega_k(\tau)} \expval{V(\tau)}{\phi_k(0)} e^{- \int_{0}^{\tau}dt\expval{V^{\dagger}(t) \dot{V}(t)}{\phi_k(0)}} \right) \nonumber \\
	&=\arg \left( \sum_k \sqrt{\omega_k(0) \omega_k(\tau)} \ip{\phi_k(0)}{\phi_k(\tau)} e^{- \int_{0}^{\tau}dt\ip*{\phi_k(t)}{\dot{\phi}_k(t)}} \right) \label{eq:gammaP}
\end{align} 
There are certain basic requirements which an expression of geometric phase must satisfy. These are
\begin{enumerate}
	\item[(a)] it must be invariant under gauge transformation,
	\item[(b)] it should be reduced to the already exiting results in the limit of unitary evolution and pure state, and
	\item[(c)] it should be feasible to get it tested experimentally.
\end{enumerate}
The geometric phase in \eqref{eq:gammaP} is gauge invariant, i.e.
\begin{align}
	\gamma[\mathcal{P}] \vert_{\tilde{V}(t)} &= \arg \left( \sum_k \sqrt{\omega_k(0) \omega_k(\tau)} \expval{\tilde{V}(\tau)}{\phi_k(0)} e^{- \int_{0}^{\tau}dt\expval{\tilde{V}^{\dagger}(t) \dot{\tilde{V}}(t)}{\phi_k(0)}} \right) \nonumber \\
	&= \arg \left( \sum_k \sqrt{\omega_k(0) \omega_k(\tau)} \expval{V(\tau)}{\phi_k(0)} e^{i \theta_k (\tau)} e^{- \int_{0}^{\tau}dt\expval{V^{\dagger}(t) \dot{V}(t)}{\phi_k(0)}} e ^{-\int_{0}^{\tau} i \dot{\theta}_k(t)}\right) \nonumber \\
	&= \arg \left( \sum_k \sqrt{\omega_k(0) \omega_k(\tau)} \expval{V(\tau)}{\phi_k(0)} e^{- \int_{0}^{\tau}dt\expval{V^{\dagger}(t) \dot{V}(t)}{\phi_k(0)}} \right) \nonumber \\
	&= \gamma[\mathcal{P}] \vert_{V(t)} \nonumber
\end{align}
where we used $\theta_k(0) = 0$ and
\begin{align*}
	&\expval{\tilde{V}^{\dagger}(t) \dot{\tilde{V}}(t)}{\phi_k(0)} = \expval{V^{\dagger}(t) \dot{V}(t)}{\phi_k(0)} + i \dot{\theta}_k(t)
\end{align*}
Second, when the evolution is unitary, which corresponds to the case where $\omega_k$ are time independent and the operator $V(t)$ is identified with the time evolution operator of the system, the geometric phase Eq.~\eqref{eq:gammaP} reduces to the well-known results~\cite{Singh}. This phase is experimentally testable by lifting the given state $\rho(t)$ to a higher dimensional state $ \ket{\Psi(t)} $ using purification techniques~\cite{Tong}. We now illustrate the method described above by considering an explicit example.

Let us consider a two-level system with a free Hamiltonian $ H = (\eta/2) \sigma_z $, interacting with an environment represented by the Lindblad operator $\Gamma = (\sqrt{\Lambda/2}) \sigma_z$. It is a kind of pure dephasing environment (bosonic bath). We can solve the Lindblad equation like 
\begin{align*}
	\dfrac{d}{dt} \rho &= -i [H, \rho] + \dfrac{1}{2}\sum_k \left( 2 \Gamma_k \rho \Gamma_k^{\dagger} - \Gamma_k \Gamma_k^{\dagger} \rho - \rho \Gamma_k \Gamma_k^{\dagger}\right) \\
	&= -i \dfrac{\eta}{2} [\sigma_z, \rho] + \dfrac{\Lambda}{4} \left( 2 \sigma_z \rho \sigma_z - \rho - \rho \right) \\
	&= -i \dfrac{\eta}{2} [\sigma_z, \rho] + \dfrac{\Lambda}{2} \left( \sigma_z \rho \sigma_z - \rho\right) \\
	&= \begin{bmatrix}
		0 & (-i \eta - \Lambda) \rho_{01} \\
		(i \eta - \Lambda) \rho_{10} & 0
	\end{bmatrix}
\end{align*}
For an initial state given by
\begin{equation}
	\rho(0) = \dfrac{1}{2} \left( \mathds{1} + \vb{r} \vdot \boldsymbol{\sigma} \right)
\end{equation}
with $ \vb{r} = (\sin \theta_0, 0, \cos \theta_0) $, we can solve for $\rho(t)$ and will get
\begin{align*}
	\rho_{00}(t) &= \rho_{00}(0) = \dfrac{1}{2} (1 + \cos \theta_0)\\
	\rho_{01}(t) &= \rho_{00}(0) e^{-i \eta t - \Lambda t} = \dfrac{1}{2} \sin \theta_0 e^{-i \eta t - \Lambda t} \\
	\rho_{10}(t) &= \rho_{10}(0) e^{i \eta t - \Lambda t} = \dfrac{1}{2} \sin \theta_0 e^{i \eta t - \Lambda t} \\
	\rho_{11}(t) &= \rho_{11}(0) = \dfrac{1}{2} (1 - \cos \theta_0)\\
\end{align*}
which results in
\begin{equation}
	\rho(t) = \dfrac{1}{2} \begin{bmatrix}
		1 + \cos \theta_0 & \sin \theta_0 e^{-i \eta t - \Lambda t}\\
		\sin \theta_0 e^{i \eta t - \Lambda t} & 1 - \cos \theta_0
	\end{bmatrix}.
\end{equation}
It is characterized by the following eigenvalues 
\begin{equation*}
	\lambda_{1,2}(t) = \dfrac{1}{2} \left( 1 \pm \sqrt{\cos^2 \theta_0 + e^{-2 \Lambda t} \sin^2 \theta_0} \right)
\end{equation*}
and we can also write
\begin{equation}
	\rho(t) = \dfrac{1}{2} \left( \mathds{1} + \vb{r}(t) \vdot \boldsymbol{\sigma} \right)
\end{equation}
where
\begin{equation*}
	\vb{r} = (\sin \theta_0 \cos (\eta t) e^{-\Lambda t}, \sin \theta_0 \sin (\eta t) e^{-\Lambda t}, \cos \theta_0).
\end{equation*}
Therefore, we will have
\begin{equation}
	\tan \phi(t) = \tan (\eta t), \qquad \tan \theta_t = e^{-\Lambda t} \tan \theta_0,
\end{equation}
and the corresponding eigenstates are 
\begin{align}
	\ket{\phi_1(t)} &= \cos (\theta_t/2) \ket{0} + e^{i \eta t} \sin (\theta_t/2) \ket{1}, \nonumber \\ 
	\ket{\phi_2(t)} &= - e^{-i \eta t} \sin (\theta_t/2) \ket{0} +  \cos (\theta_t/2) \ket{1}
\end{align}
where $ \{\ket{0}, \ket{1}\} $ are the standard computational basis. Since $\lambda_2(0) = 0$, the only contribution comes from the $\ket{\phi_1(t)}$ in geometric phase. We have
\begin{align*}
	\ip*{\phi_1(t)}{\dot{\phi}_1(t)} =  i \eta \sin^2 (\theta_t/2) = \dfrac{i \eta}{2} \left[1 - \dfrac{1}{\sqrt{1 + e^{-2 \Lambda t} \tan^2 \theta_0}}\right] 
\end{align*}
By using all these relations in Eq.~\eqref{eq:gammaP} we get
\begin{equation}
	\gamma[\mathcal{P}] = -\pi + \dfrac{\eta}{4 \Lambda} \ln \left[ \dfrac{(\cos\theta_0 -1)(\cos \theta_0 + \sqrt{\cos^2 \theta_0 + e^{-4 \pi \Lambda / \eta} \sin^2 \theta_0})}{(\cos\theta_0 +1)(\cos \theta_0 - \sqrt{\cos^2 \theta_0 + e^{-4 \pi \Lambda / \eta} \sin^2 \theta_0})} \right]
\end{equation}
where we made use of a trigonometric identity $\tanh^{-1}x = \tfrac{1}{2} \ln\left(\tfrac{1 + x}{1 - x}\right)$ for $-1 \le x \le 1$. For small dephasing, i.e., for $\Lambda/\eta << 1$, we can use the Taylor expansion for the second term and obtain the geometric phase up to a first order as
\begin{equation} \label{eq:firstordergp}
	\gamma[\mathcal{P}] \approx -\pi(1 - \cos \theta_0) + \pi^2 \cos\theta_0 \sin^2\theta_0 \left(\dfrac{\Lambda}{\eta}\right).
\end{equation}
The effect of dephasing on the geometric phase has been analyzed in~\cite{Carollo} using the quantum jump approach. However, in quantum-jump approach, we effectively have a pure state at all the time. The robust nature of the geometric phase has been explicitly shown in~\cite{Carollo} against dephasing. The geometric phase calculated using mixed state, results in first order correction~Eq.\eqref{eq:firstordergp} due to dephasing which reduces to the results of~\cite{Carollo} for $\theta_0 = \pi/2$ which corresponds to the precession in the equatorial plane of the Bloch sphere.

\begin{figure}
	\centering
	\includegraphics[width=7cm]{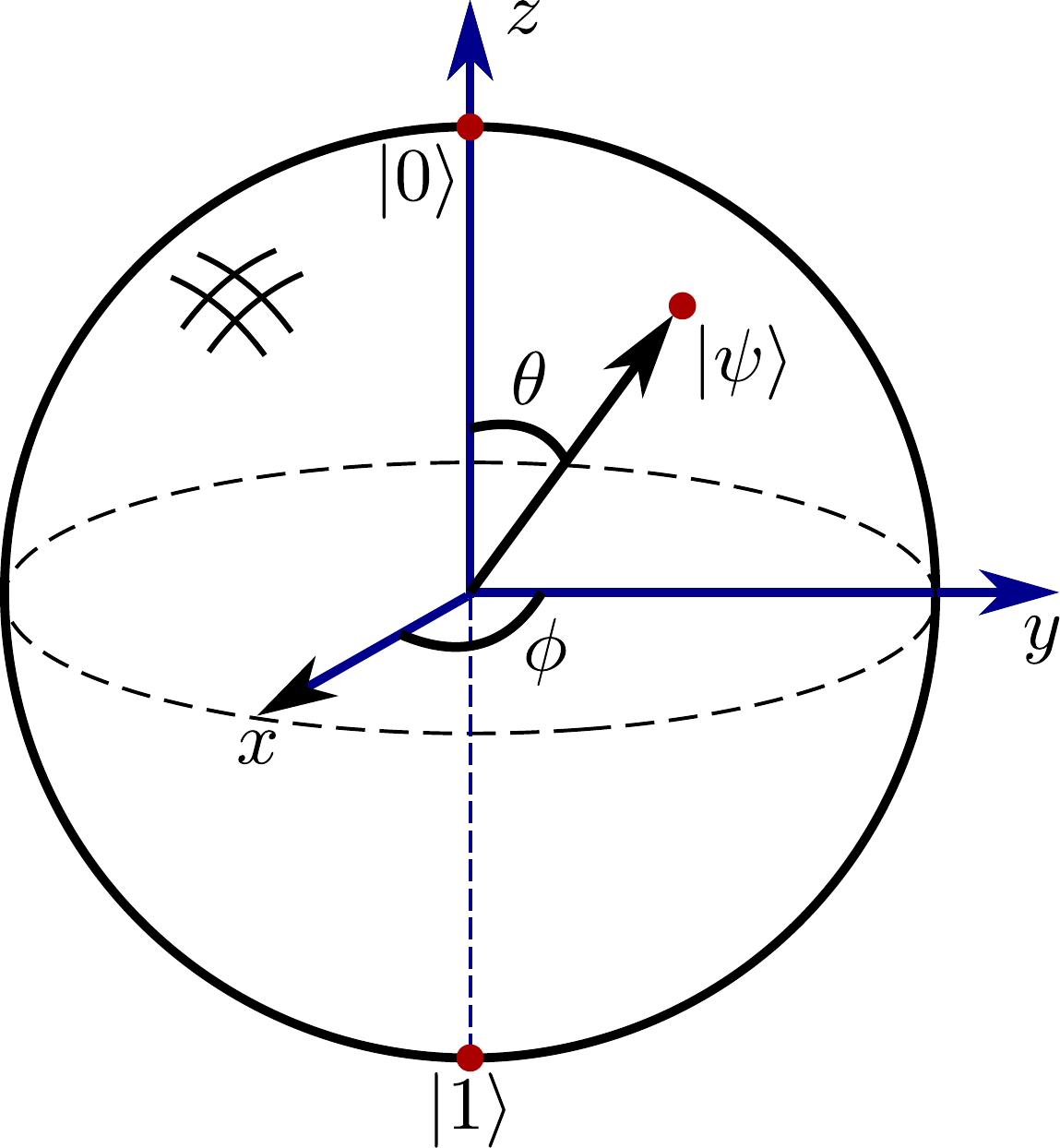}
	\caption{Bloch sphere representation of a qubit.}
	\label{fig:Blochsphere}
\end{figure}

\begin{tcolorbox}
	NOTE: A general state for a two-level system in computational basis is written as 
	\begin{equation*}
		\ket{\psi} = \cos (\theta/2) \ket{0} + e^{i \phi} \sin (\theta/2) \ket{1} = \begin{bmatrix}
			\cos (\theta/2) \\
			e^{i \phi} \sin (\theta/2)
		\end{bmatrix}
	\end{equation*}
	where $\theta$ and $\phi$ define a point on the unit three-dimensional sphere, known as "Bloch sphere" as shown in Fig.~\ref{fig:Blochsphere}. We can further write the state $ \ket{\psi} $ in the Bloch sphere representation as
	\begin{equation*}
		\rho = \dyad{\psi} = \dfrac{1}{2} (\mathds{1} + \vb{r} \vdot \sigma)
	\end{equation*}
	where $ \vb{r} \in \mathbb{R}^3$ such that $ \abs{r} \le 1$. This vector is known as the Bloch sphere and is given by
	\begin{equation*}
		\vb{r} = (r_x, r_y, r_z) = (\sin \theta \cos \phi, \sin \theta \sin \phi, \cos \phi).
	\end{equation*}
	with
	\begin{align*}
		\tan \phi &= \dfrac{r_y}{r_z}, \qquad \tan \theta = \dfrac{\sqrt{r_x^2 + r_y^2}}{r_z}.
	\end{align*}
\end{tcolorbox}
\section{Measurement of Mixed state Geometric Phase in Interferometry}
The first experimental verification of the mixed state geometric phase was given in~\cite{Ericsson2005} based on Single Photon Interferometry. We briefly discuss the experimental setup here. The composite Hilbert space of the system is given by the tensor product of Hilbert spaces corresponding to spatial ($\mathcal{H}_s$) and internal modes ($\mathcal{H}_i$) i.e. we have
\begin{equation}
	\mathcal{H} = \mathcal{H}_s \otimes \mathcal{H}_i
\end{equation}
Here, we start with 
\begin{align}
	\rho_{in} &=  \dyad{a}{a} \otimes \dyad{\psi_0}{\psi_0} = \dyad{a}{a} \otimes \rho_0  \nonumber \\
	&= \left[
	\begin{array}{c|c}
		\rho_0 & 0 \\
		\hline
		0 & 0
	\end{array}
	\right]
\end{align}
Then it gets split by BS1 whose action can be written as
\begin{align}
	U_{_{BS}} &= \dfrac{1}{\sqrt{2}} \begin{pmatrix}
		1 & i \\
		i & 1
	\end{pmatrix} \otimes \mathds{1} \nonumber \\
	& =  \dfrac{1}{\sqrt{2}} \left[
	\begin{array}{c|c}
		\mathds{1} & i \\
		\hline
		i & \mathds{1}
	\end{array}
	\right]
\end{align}
\begin{align}
	\rho_{in} \mapsto U_{_{BS}} \rho_{in} U_{_{BS}}^{\dagger} = \rho_1 &= \dfrac{1}{2} \left[
	\begin{array}{c|c}
		\mathds{1} & i \\
		\hline
		i & \mathds{1}
	\end{array}
	\right] \left[
	\begin{array}{c|c}
		\rho_0 & 0 \\
		\hline
		0 & 0
	\end{array}
	\right] \left[
	\begin{array}{c|c}
		\mathds{1} & -i \\
		\hline
		-i & \mathds{1}
	\end{array}
	\right] \nonumber \\
	&= \dfrac{1}{2} \left[
	\begin{array}{c|c}
		\mathds{1} & i \\
		\hline
		i & \mathds{1}
	\end{array}
	\right] \left[
	\begin{array}{c|c}
		\rho_0 & -i \rho_0 \\
		\hline
		0 & 0
	\end{array}
	\right]	\nonumber \\
	&= \dfrac{1}{2} \left[
	\begin{array}{c|c}
		\rho_0 & -i \rho_0 \\
		\hline
		i \rho_0 & \rho_0
	\end{array}
	\right]	
\end{align}
Then we have two half-wave plates (HWP) in mode $b$ and a phase shift, $\chi$ in mode $a$,
\begin{align}
	\rho_1 \mapsto U \rho_1 U^{\dagger} = \rho_2
\end{align}
where
\begin{align}
	U = \left\{ \dyad{a}{a} \otimes H + \dyad{b}{b} \otimes e^{i \chi} \right\}
\end{align}
with $H = HWP_{\theta_2} \times HWP_{\theta_1}$.
Therefore
\begin{align}
	\rho_2 &= \dfrac{1}{2} \left[
	\begin{array}{c|c}
		H & 0 \\
		\hline
		0 & e^{i \chi}
	\end{array}
	\right] \left[
	\begin{array}{c|c}
		\rho_0 & -i \rho_0 \\
		\hline
		i \rho_0 & \rho_0
	\end{array}
	\right] \left[
	\begin{array}{c|c}
		H^{\dagger} & 0 \\
		\hline
		0 & e^{-i \chi}
	\end{array}
	\right]
	\nonumber \\
	&= \dfrac{1}{2} \left[
	\begin{array}{c|c}
		H & 0 \\
		\hline
		0 & e^{i \chi}
	\end{array}
	\right] \left[
	\begin{array}{c|c}
		\rho_0 H^{\dagger} & -i \rho_0 e^{-i \chi} \\
		\hline
		i \rho_0 H^{\dagger} & \rho_0 e^{-i \chi}
	\end{array}
	\right] \nonumber \\
	&= \dfrac{1}{2} \left[
	\begin{array}{c|c}
		H \rho_0 H^{\dagger} & -i H \rho_0 e^{-i \chi} \\
		\hline
		i \rho_0 H^{\dagger} e^{i \chi} & \rho_0
	\end{array}
	\right]
\end{align}
Then we have two mirrors $M_1 \& M_2$ and their action would be
\begin{align}
	M &= \begin{pmatrix}
		0 & 1 \\
		1 & 0
	\end{pmatrix} \otimes \mathds{1} =  \left[
	\begin{array}{c|c}
		0 & \mathds{1} \\
		\hline
		\mathds{1} & 0
	\end{array}
	\right]
\end{align}
\begin{align}
	\rho_3 &\mapsto M \rho_3 M^{\dagger} \nonumber\\
	&= \dfrac{1}{2} \left[
	\begin{array}{c|c}
		0 & \mathds{1} \\
		\hline
		\mathds{1} & 0
	\end{array}
	\right]  
	\left[
	\begin{array}{c|c}
		H \rho_0 H^{\dagger} & -i H \rho_0 e^{-i \chi} \\
		\hline
		i \rho_0 H^{\dagger} e^{i \chi} & \rho_0
	\end{array}
	\right]
	\left[
	\begin{array}{c|c}
		0 & \mathds{1} \\
		\hline
		\mathds{1} & 0
	\end{array}
	\right] \nonumber \\
	&= \dfrac{1}{2} \left[
	\begin{array}{c|c}
		0 & \mathds{1} \\
		\hline
		\mathds{1} & 0
	\end{array}
	\right]  
	\left[
	\begin{array}{c|c}
		-i H \rho_0 e^{-i \chi} & H \rho_0 H^{\dagger} \\
		\hline
		\rho_0 & i \rho_0 H^{\dagger} e^{i \chi} 
	\end{array}
	\right] \nonumber \\
	&= \dfrac{1}{2} \left[
	\begin{array}{c|c}
		\rho_0 & i \rho_0 H^{\dagger} e^{i \chi}  \\
		\hline
		-i H \rho_0 e^{-i \chi}& H \rho_0 H^{\dagger}
	\end{array}
	\right]
\end{align}
And finally BS2
\begin{align}
	\rho_{op} &= \dfrac{1}{4} \left[
	\begin{array}{c|c}
		\mathds{1} & i \\
		\hline
		i & \mathds{1}
	\end{array}
	\right]  
	\left[
	\begin{array}{c|c}
		\rho_0 & i \rho_0 H^{\dagger} e^{i \chi}  \\
		\hline
		-i H \rho_0 e^{-i \chi}& H \rho_0 H^{\dagger}
	\end{array}
	\right]
	\left[
	\begin{array}{c|c}
		\mathds{1} & -i \\
		\hline
		-i & \mathds{1}
	\end{array}
	\right] \nonumber \\
	&= \dfrac{1}{4} \left[
	\begin{array}{c|c}
		\mathds{1} & i \\
		\hline
		i & \mathds{1}
	\end{array}
	\right]  
	\left[
	\begin{array}{c|c}
		\rho_0 + \rho_0 H^{\dagger} e^{i \chi} & -i \rho_0 + i\rho_0 H^{\dagger} e^{i \chi}  \\
		\hline
		-i H \rho_0 e^{-i \chi} -iH \rho_0 H^{\dagger} & -H \rho_0 e^{-i \chi} + H \rho_0 H^{\dagger}
	\end{array}
	\right] \nonumber \\
	&= \dfrac{1}{4} \; \text{diag} \left[\rho_0 + H \rho_0 H^{\dagger} + \rho_0 H^{\dagger} e^{i \chi} + H \rho_0 e^{-i \chi} , \right. \nonumber \\
	&\qquad \left. \rho_0 + H \rho_0 H^{\dagger} - \rho_0 H^{\dagger} e^{i \chi} - H \rho_0 e^{-i \chi} \right]
\end{align}
So, the intensity along $\dyad{a}{a}$ or D1 is
\begin{align}
	\mathscr I_a &= \dfrac{1}{4} \; \Tr \left[ \rho_0 + H \rho_0 H^{\dagger} + \rho_0 H^{\dagger} e^{i \chi} + H \rho_0 e^{-i \chi} \right] \nonumber \\ 
	&= \dfrac{1}{4} \left(2\Tr[\rho_0] + \Tr[\rho_0 H^{\dagger}] e^{i \chi} + \Tr[H \rho_0] e^{-i \chi} \right).
\end{align}
Using
\begin{equation}
	\Tr[\rho_0 H^{\dagger}] = \left( \Tr[H \rho_0] \right)^* = z^* = r e^{- i \phi}
\end{equation}
and assuming $ \Tr[\rho_0] = 1 $ we can write
\begin{align}
	\mathscr I_a &= \dfrac{1}{4} \left[ 2 + z^*e^{i \chi} + z e^{-i \chi}  \right] \nonumber \\
	&= \dfrac{1}{4} \left[ 2 + r e^{i (\chi - \phi)} + r e^{-i (\chi - \phi)} \right] \nonumber \\
	& \propto 1 + r \cos (\chi - \phi).
\end{align}
and similarly along $\dyad{b}{b}$ or D2, will be
\begin{equation}
	\mathscr I_b = \dfrac{1}{4} \; \Tr \left[ \rho_0 + H \rho_0 H^{\dagger} - \rho_0 H^{\dagger} e^{i \chi} - H \rho_0 e^{-i \chi}\right] \propto 1 - r \cos (\chi - \phi).
\end{equation}
where
\begin{equation}
	\phi = \arg \left(\Tr[H \rho_0]\right).
\end{equation}
\subsection{Experimental setup}
\begin{figure}[H]
	\centering
	\includegraphics[width=13cm]{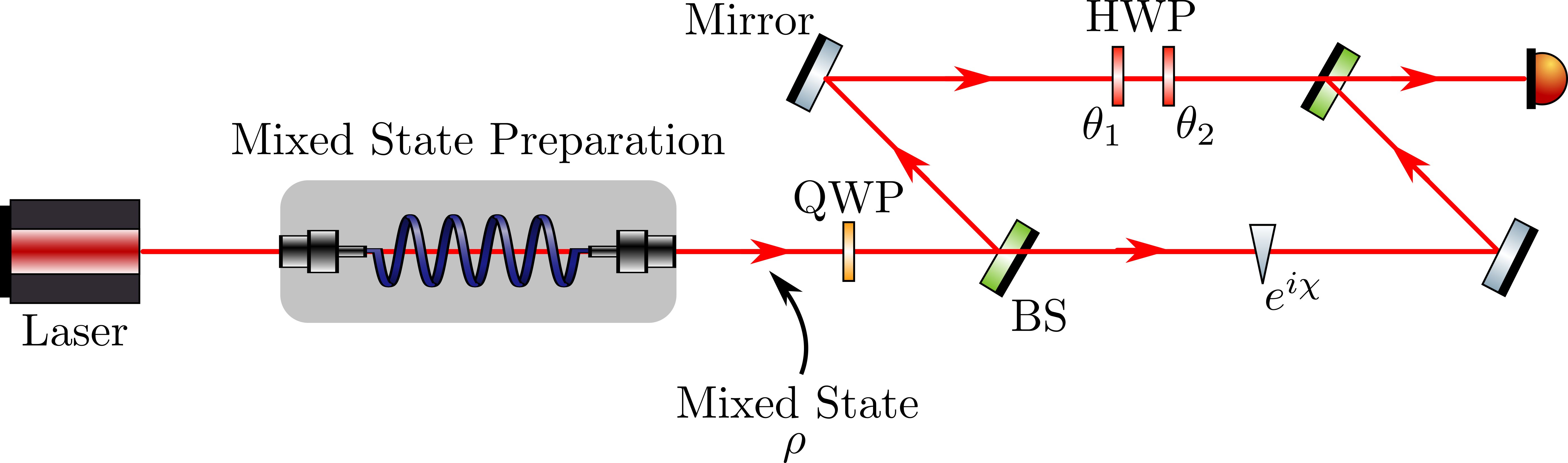}
	\caption{Schematic of the experimental setup in Ericsson et al~\cite{Ericsson2005}.}
	\label{fig:gpinterferometry}
\end{figure}
\begin{figure}[H]
	\centering
	\includegraphics[width=5cm]{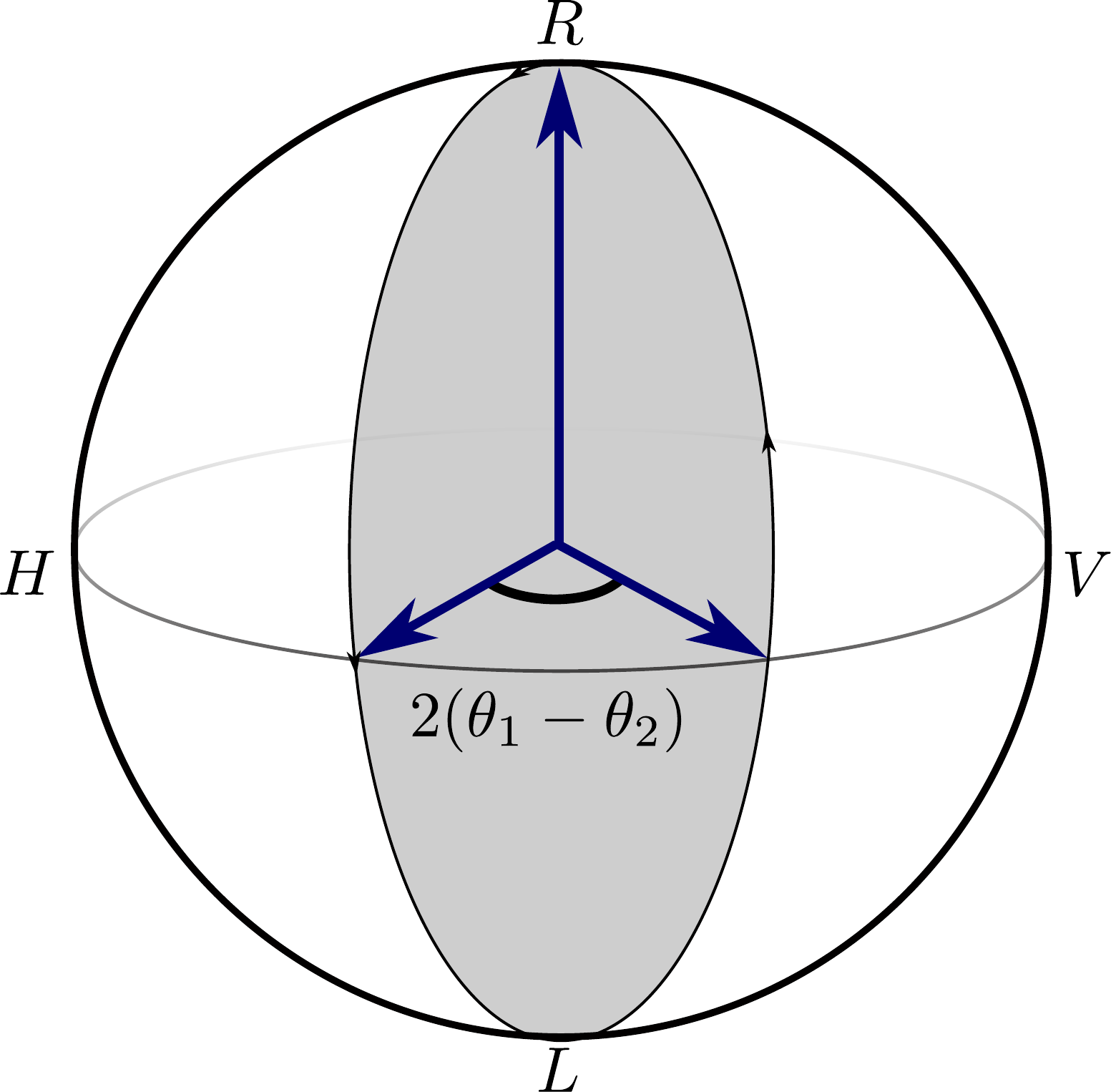}
	\caption{The solid angle $\Omega$, shown by shaded region, by one of the eigenvector of the density matrix $\rho$. The other eigenvector traces the same path, except in the clockwise direction.}
	\label{fig:gpinterferometry2}
\end{figure}

The system is prepared in a mixed state given by
\begin{equation}
	\rho = \cos^2\theta_p \dyad{H} + \sin^2\theta_p \dyad{V} = \begin{bmatrix}
		\cos^2\theta_p & 0 \\
		0 & \sin^2\theta_p
	\end{bmatrix}
\end{equation}
with purity, $ r = \abs{\cos 2 \theta_p} $. It is sent to a QWP to create a mixture of $ \ket{R} $ and $ \ket{L} $ polarization states (it is just a change of basis),
\begin{equation}
	\rho \mapsto \rho_0 = Q \rho Q^{-1} = \cos^2\theta_p \dyad{R} + \sin^2\theta_p \dyad{L}, 
\end{equation}
with
\begin{equation}
	Q = \begin{bmatrix}
		e^{i \pi/4} & 0 \\
		0 & e^{-i \pi/4}
	\end{bmatrix} = e^{i \pi/4}\begin{bmatrix}
		1 & 0 \\
		0 & -i
	\end{bmatrix}
\end{equation}
such that
\begin{equation}
	Q \ket{H} = Q \begin{bmatrix}
		1 \\
		0
	\end{bmatrix} = \ket{R} = \dfrac{e^{i \pi/4}}{\sqrt{2}}\begin{bmatrix}
		1 \\
		-i
	\end{bmatrix}
\end{equation}
and
\begin{equation}
	Q \ket{V} = Q \begin{bmatrix}
		0 \\
		1
	\end{bmatrix} = \ket{L} = \dfrac{e^{i \pi/4}}{\sqrt{2}}\begin{bmatrix}
		1 \\
		i
	\end{bmatrix}.
\end{equation}
Therefore, the $\rho_0$ in $ \{\ket{H},\ket{V}\} $ (it is necessary because $ Q $ is written in the same) basis can be written as
\begin{align}
	\rho_0 &= \dyad{H} + i \cos 2 \theta_p \dyad{H}{V} - i \cos 2 \theta_p \dyad{V}{H} + \dyad{V} \nonumber \\
	&= \dfrac{1}{2}\begin{bmatrix}
		1 & i \cos 2 \theta_p \\
		-i \cos 2 \theta_p & 1
	\end{bmatrix} = \dfrac{1}{2}\begin{bmatrix}
		1 & i r \\
		-i r & 1
	\end{bmatrix} = \dfrac{1}{2} \left( \mathds{1} - r \sigma_z \right).
\end{align}
In one mode, we have two HWP at angles $\theta_1$ and $\theta_2$ 
\begin{equation}
	H_{\theta} = R(\theta) H_0 R(\theta)^{-1} = R(\theta) i \sigma_z R(\theta)^{-1} = i \begin{bmatrix}
		\cos 2 \theta & \sin 2 \theta \\
		\sin 2 \theta & - \cos 2 \theta
	\end{bmatrix} 
\end{equation} 
with
\begin{equation}
	R(\theta) = \begin{bmatrix}
		\cos \theta & - \sin \theta \\
		\sin \theta & \cos \theta
	\end{bmatrix}
\end{equation}
which gives us
\begin{equation}
	U \equiv H_{\theta_2} H_{\theta_1} = - \begin{bmatrix}
		\cos 2 (\theta_1 - \theta_2) & \sin 2(\theta_1 - \theta_2) \\
		\sin 2 (\theta_1 - \theta_2) &  \cos 2(\theta_1 - \theta_2)
	\end{bmatrix} 
\end{equation}
In this case, the dynamical phase vanishes because of the fact that both components of $\rho$ are parallel-transported, as shown in Fig.~\ref{fig:gpinterferometry2}. So, the total phase has contribution only from the geometric phase which is given by
\begin{align}
	\gamma_g &= \arg \left(\Tr[H \rho_0]\right), \nonumber \\
	&= \arg \left( - \cos 2 (\theta_1 - \theta_2) + i r \sin 2 (\theta_1 - \theta_2) \right), \nonumber \\
	&= -\arctan \left( r \tan2 (\theta_1 - \theta_2)  \right).
\end{align}

\section{Uhlmann Holonomies}
\begin{figure}[H] 
	\centering
	\includegraphics[scale=1.6]{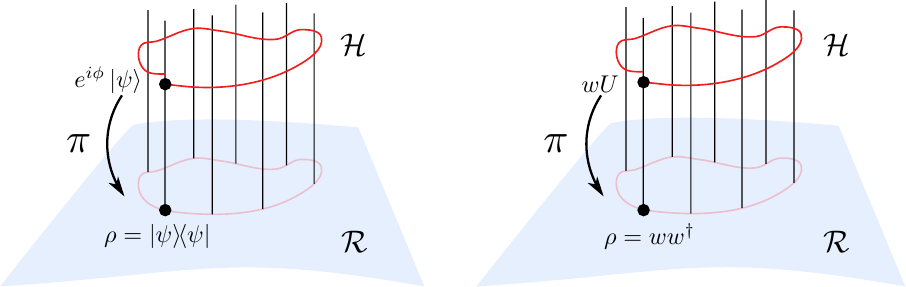}
	\caption{Comparison of Berry's phase for pure states and Uhlmann phase for mixed states.}
	\label{UhlmannPhase}
\end{figure}
The extension of geometric phase for pure states to mixed states is not trivial and straightforward. We have already discussed the interferometric approach~\cite{Sjoqvist} in previous sections which is more appealing from a physical perspective. However, Uhlmann was the first to address the issue of geometric phase for mixed states using a rigorous mathematical framework~\cite{Uhlmann1986,Uhlmann1989,Uhlmann1991, Uhlmann1995, Uhlmann2011} and provided a satisfactory solution. The two key elements in Uhlmann's approach were the purification of states in an extended Hilbert space and the parallel transport of states. The purification of mixed states in Uhlmann's sense can be understood in analogy with the case of pure state, which we developed in Sect.~\ref{sec:mathematical} and also shown in Fig.~\ref{UhlmannPhase}. The purification of states can be represented by an operator (Hilbert-Schmidt) in the extended space and is referred to as an amplitude. For a given path of the density operator, a corresponding path of the amplitude can be constructed, which projects down to the original path. There exists a certain special path of amplitudes which obeys the parallel transport condition and leads to a unique path in the extended Hilbert space. Such a choice of amplitude is a property only of the path of the states and defines a holonomy invariant. One advantage of Uhlmann's approach is that it does not distinguish between unitary or nonunitary evolution. The main aim of this part is to discuss the compatibility of the two approaches for the mixed state geometric phase. We start with a discussion on the construction of the amplitudes and the parallel transport law.

\subsection{Uhlmann's amplitudes}
As we have seen, in the case of pure states, we can define a projective Hilbert space $\mathcal{R}$ that consists of physical states $\rho = \dyad{\psi}$. By physical, we mean that the states $\ket{\psi}$ and $e^{i \phi }\ket{\psi}$ describe the same state of the system $\rho$. However, this concept cannot be trivially extended for the mixed states. The fundamental problem which arises while extending the idea of a geometric phase for the mixed state is, for a given set of real numbers $\{p_i\}$ and density matrices $ \{\rho_i\} $, the linear combination $ \sum_{i} p_i \rho_i $ is not a valid density matrix until and unless $ p_i  \le $ 0 $\forall \;i$ and $ \sum_{i} p_i = 1$. Our aim is to find such an analogous construction for a mixed state represented by a density matrix $\rho$. To achieve that, we consider a density matrix $\rho$ and define an \emph{amplitude} as a matrix $ w \in \mathcal{H}_w $ such that
\begin{equation}
	\rho = w w^{\dagger}
\end{equation}
These amplitudes form another Hilbert space $ \mathcal{H}_w $ (space of Hilbert-Schmidt operators) with an inner product or a Hilbert-Schmidt product defined as 
\begin{equation}
	(w_1, w_2) = \Tr(w_1^{\dagger} w_2)
\end{equation}
We can clearly see from the last expression that there is a $ U(n) $ gauge freedom to choose because $ w $ and $ w U $, $ U $ being a unitary operator corresponds to the same state. We note here that, as in the case of pure states, we had a gauge freedom of $U(1)$ in the states, here we have a $U(n)$-gauge freedom for the amplitudes as shown in Fig.~\ref{UhlmannPhase}.  Furthermore, the projection map, $\pi$, is defined as [see Fig.~\ref{UhlmannPhase}]
\begin{equation}
	\pi: \mathcal{H}_w \rightarrow \mathcal{R}, \; \; \pi(w) = w w^{\dagger} = \rho.
\end{equation}
In the case of pure states, the projective space $\mathcal{R}$ consists of projection operators, $ P(\psi) = \dyad{\psi} $ which corresponds to all states $e^{i \phi} \ket{\psi} \in \mathcal{H}$ with $\phi \in \mathds{R}$. We can extend this idea to mixed states by writing $ \rho $ as a linear combination of $ P(\psi_i) $'s
\begin{equation}
	\rho = \sum_{i} p_i P(\psi_i) = \sum_{i} p_i \dyad{\psi_i}
\end{equation}
The amplitude $ w $ is written as 
\begin{equation}
	w = \sqrt{\rho}  U = \sum_{i} \sqrt{p_i} \dyad{\psi_i} U,
\end{equation}
which is nothing but the polar decomposition of $w$ and is known as the 'square root' section of the Uhlmann fibre bundle~\cite{Viyuela2015}. It can be seen as 
\begin{equation}
	w = \sum_{i} \sqrt{p_i} \ket{\psi_i} \otimes \bra{\phi_i} U,
\end{equation}
which corresponds to an operator in extended Hilbert space $\mathcal{H}_w$ which is a tensor product of the systems Hilbert space $\mathcal{H}_S$ and an ancilla (of dimension at least equal to the dimension of the system) $ \mathcal{H}_A $ i.e. $ \mathcal{H}_w = \mathcal{H}_S \otimes \mathcal{H}_A$ ,$ \{\bra{\phi_i}\} \in \mathcal{H}_A $ is the set of orthonormal basis vectors. This process is known as \emph{purification}. The main difference between this and the standard one is that the density matrices are further purified by operators instead of vectors. This is how Uhlmann defined purification \cite{Uhlmann1986}. We can easily verify that
\begin{align}
	w w^{\dagger} &= \left( \sum_{i} \sqrt{p_i} \ket{\psi_i} \otimes \bra{\phi_i} U \right) \left( \sum_{j} \sqrt{p_j} \bra{\psi_j} \otimes U^{\dagger} \ket{\phi_j} \right) \nonumber \\
	&= \sum_{ij} \sqrt{p_i p_j} \dyad{\psi_i}{\psi_j} \otimes \mel{\phi_i}{U U^{\dagger}}{\phi_j} \nonumber \\
	&= \sum_{i} p_i \dyad{\psi_i} \qquad \qquad \ip*{\phi_i}{\phi_j} = \delta_{ij}.
\end{align}
But later, an isomorphism was defined~\cite{Viyuela2015} between the operator $ w \in \mathcal{H}_w $ and the vectors $ \ket{w} \in \mathcal{H}_S \otimes \mathcal{H}_A $ as
\begin{equation}
	w = \sum_{i} \sqrt{p_i} \ket{\psi_i} \otimes \bra{\psi_i} U \longleftrightarrow \ket{w} = \sum_{i} \sqrt{p_i} \ket{\psi_i} \otimes U^T \ket{\psi_i} 
\end{equation}
where $ U^T $ denotes the complex transposition of $ U $ with respect to the eigenbasis of $\rho$. Using
\begin{equation*}
	U^T \ket{\psi_i} \equiv \ket{\phi_i} \implies \bra{\psi_i} U = \bra{\phi_i}
\end{equation*}
again we can show that
\begin{align}
	\rho &= \Tr_A \dyad{w} \nonumber \\
	&= \Tr_A \left( \sum_{i} \sqrt{p_i} \ket{\psi_i} \otimes U^T \ket{\psi_i} \right) \left( \sum_{j} \sqrt{p_j} \bra{\psi_j} \otimes  \bra{\psi_j} U \right)\nonumber \\
	&= \Tr_A \left( \sum_{i} \sqrt{p_i} \ket{\psi_i} \otimes \ket{\phi_i} \right) \left( \sum_{j} \sqrt{p_j} \bra{\psi_j} \otimes  \bra{\phi_j} \right)\nonumber \\
	&= \sum_k \sum_i \sum_j \sqrt{p_i p_j} \dyad{\psi_i}{\psi_j} \otimes \ip*{\phi_k}{\phi_i} \ip*{\phi_j}{\phi_k}, \nonumber \\
	&= \sum_i p_i \dyad{\psi_i}.
\end{align}
where $\Tr_A$ represents the partial trace over the ancilla in $\mathcal{H}_S \otimes \mathcal{H}_A$. Therefore, any amplitude $ w $ can be seen as a pure state $ \ket{w} $ in an extended Hilbert space $ \mathcal{H}_S \otimes \mathcal{H}_A $ such that the partial trace over the ancilla gives us $\rho$. 
\subsection{Parallel amplitudes}
In order to understand the parallel transport condition by Uhlmann and the holonomy, we need the concept of parallel amplitudes. For two given pairs of density matrices, $\rho_1$ and $\rho_2$, there are two corresponding amplitudes $ w_1 $ and $ w_2 $, respectively. They are parallel if they minimize the Hilbert space distance in $\mathcal{H}_w$ which is given by
\begin{equation}
	\norm{w_1 - w_2}^2 = \min_{w_1', w_2'} \| w_1' - w_2' \|^2
\end{equation}
with $\rho_1 = w_1'w_1^{\dagger}$ and $\rho_2 = w_2'w_2^{\dagger}$.
\begin{align}
	\min_{w_1', w_2'} \| w_1' - w_2' \|^2 &= \min_{w_1', w_2'} (w_1' - w_2',w_1' - w_2'), \nonumber \\
	&= \min_{w_1', w_2'} \Tr[(w_1' - w_2')^{\dagger}(w_1' - w_2')], \nonumber \\
	&= \min_{w_1', w_2'} \Tr[w_1'^{\dagger}w_1' -w_1'^{\dagger}w_2' -w_2'^{\dagger}w_1' + w_2'^{\dagger}w_2'] , \nonumber \\
	&= \Tr\rho_1 + \Tr\rho_2 - \max_{w_1', w_2'} \Tr[w_1'^{\dagger}w_2' + w_2'^{\dagger}w_1'], \nonumber \\
	&= 2 - 2  \max_{w_1', w_2'} \Re\Tr[w_1'^{\dagger}w_2']
\end{align}
Since we know $ \text{Re}(x) \le \abs{x} $, therefore, the expression inside the parentheses can be maximized by choosing $ w_1'$ and $ w_2'$ in such a way that
\begin{equation}
	w_1'^{\dagger}w_2' = w_2'^{\dagger}w_1' > 0
\end{equation}
i.e. $w_1'^{\dagger}w_2'$ is self-adjoint and positive definite, which ensures that diagonal entries are real. According to the polar decomposition theorem, any operator $ A $ can be decomposed into a hermitian $ \abs{A} $ and a unitary factor $ U_A $ as
\begin{equation}
	A = \abs{A} U_A
\end{equation}
where $ \abs{A} = \sqrt{AA^{\dagger}}$. It is an extension of the conventional polar decomposition of complex numbers $ z = \abs{r} e^{i \arg(r)} $. For a given unitary matrix $ U $, we can write \cite{Nielsen,Jozsa1994}
\begin{align}
	\Re [\Tr(A U)] &\le \abs{\Tr(A U)}, \nonumber \\
	&= \abs{\Tr(\abs{A} U_A U)}, \nonumber \\
	&= \abs*{\Tr(\sqrt{\abs{A}} \sqrt{\abs{A}} U_A U)}, \nonumber \\
	&\le \sqrt{ (\Tr \abs{A}) [\Tr (U^{\dagger}U_A^{\dagger} \abs{A} U_A U)]}, \nonumber \\
	&= \Tr\abs{A}.
\end{align}
where we have used the Cauchy-Schwarz inequality;
\begin{equation*}
	\abs*{\Tr\left(M^{\dagger}N\right)}^2 \le \Tr(M^{\dagger}M) \Tr(N^{\dagger}N) 
\end{equation*}
with $M = \sqrt{\abs{A}}$ and $N = \sqrt{\abs{A}}U_A U$. We finally have $ \text{Re} [\Tr(A U)] \le \Tr\abs{A}$ and the equality holds when $ U = U_A^{\dagger} $ i.e.
\begin{equation}
	\max_U \text{Re} [\Tr(A U)] = \Tr\abs{A}
\end{equation}
Now, we write the polar decomposition for the amplitudes as follows
\begin{equation}
	w_1'= \sqrt{\rho_1} U_1, \;\; w_2'= \sqrt{\rho_2} U_2
\end{equation}
and
\begin{align}
	\max_{w_1', w_2'} \text{Re}[ \Tr (w_1'^{\dagger}w_2')] &= \max_{U_1, U_2} \text{Re}[ \Tr (U_1^{\dagger}\sqrt{\rho_1} \sqrt{\rho_2} U_2) ], \nonumber \\
	&= \max_{U} \text{Re}[ \Tr (\sqrt{\rho_1} \sqrt{\rho_2} U) ],  \; \; U = U_2 U_1^{\dagger}\nonumber \\
	&= \Tr \abs{\sqrt{\rho_1} \sqrt{\rho_2}}, \nonumber \\
	&= \Tr \sqrt{\sqrt{\rho_1}\rho_2\sqrt{\rho_1}}
\end{align}
where we have used the following polar decomposition$ \sqrt{\rho_1} \sqrt{\rho_2} = \abs{\sqrt{\rho_1} \sqrt{\rho_2}} U_{\sqrt{\rho_1} \sqrt{\rho_2}} $. This is exactly the Bures distance~\cite{Nielsen, Hubner1993} between two states $\rho_1$ and $\rho_2$. Therefore, the final condition on $ w_1 $ and $ w_2 $ to achieve parallelity can be written as
\begin{align}
	\| w_1 - w_2 \|^2 &= \min_{w_1', w_2'} \| w_1' - w_2' \|^2, \nonumber \\
	&= 2 - 2  \max_{w_1', w_2'} \text{Re}[ \Tr (w_1'^{\dagger}w_2')], \nonumber \\ 
	&= 2 - 2 \Tr \sqrt{\sqrt{\rho_1}\rho_2\sqrt{\rho_1}}.
\end{align}
As stated earlier, equality holds when
\begin{equation}
	U = U_2U_1^{\dagger} = U^{\dagger}_{\sqrt{\rho_1} \sqrt{\rho_2}}.
\end{equation}
This unitary is defined only when $\rho_1$ and $\rho_2$ are full-rank operators. In that scenario
\begin{equation}
	U_{\sqrt{\rho_1} \sqrt{\rho_2}} = \abs{\sqrt{\rho_1} \sqrt{\rho_2}}^{-1} \sqrt{\rho_1} \sqrt{\rho_2}
\end{equation}
and
\begin{gather*}
	U_{\sqrt{\rho_1} \sqrt{\rho_2}} U^{\dagger}_{\sqrt{\rho_1} \sqrt{\rho_2}} = \mathds{1} \\
	(\abs{\sqrt{\rho_1} \sqrt{\rho_2}}^{-1} \sqrt{\rho_1} \sqrt{\rho_2}) U^{\dagger}_{\sqrt{\rho_1} \sqrt{\rho_2}} = \mathds{1} \\
	\implies	U^{\dagger}_{\sqrt{\rho_1} \sqrt{\rho_2}} = \sqrt{\rho_2^{-1}} \sqrt{\rho_1^{-1}} \abs{\sqrt{\rho_1} \sqrt{\rho_2}}
\end{gather*}
which gives us
\begin{equation}
	U^{\dagger}_{\sqrt{\rho_1} \sqrt{\rho_2}} = U_2U_1^{\dagger} = \sqrt{\rho_2^{-1}} \sqrt{\rho_1^{-1}} \sqrt{\sqrt{\rho_1} \rho_2 \sqrt{\rho_1}}
\end{equation}
Therefore, the freedom that we have to choose $U$ is used to define a parallel transport condition for given two density states $\rho_1$ and $\rho_2$ with amplitudes $w_1$ and $w_2$, respectively. 


\subsection{Uhlmann's phase for a two-level system}
In~\cite{Ericsson2003}, a two-level system undergoing a unitary precession has been studied. It was shown explicitly that the two approaches for the mixed state geometric phase yield two different results in general and converge only in certain limit. We go through the results step by step. We start by considering a path
\begin{equation*}
	\zeta: t \in [0, \tau] \rightarrow \rho_t
\end{equation*}
which can be lifted or purified with $ w_t \in \mathcal{H}$ where $ \mathcal{H} $ is Hilbert space of Hilbert-Schimdt operators with scalar product
\begin{equation}
	\ip*{w_t}{w_{t'}} = \Tr(w_t^{\dagger}w_{t'}) \; \; \text{such that} \; \; \rho_t = w_t w_t^{\dagger}. 
\end{equation}
\textit{NOTE: $ w_t $ is a matrix not vector and $ w_t = \sqrt{\rho_t} x_t $ is a purification of $ \rho_t $ for an arbitrary unitary $ x_t $.} Uhlmann phase associated with $\zeta$ is defined as
\begin{equation}\label{eq:Uhlmann}
	\Phi_g = \arg \lim_{N \rightarrow \infty} (\ip*{w_0}{w_{\tau/N}}\ip*{w_{\tau/N}}{w_{2\tau/N}} \dots \ip*{w_{(N-1)\tau/N}}{w_\tau} \ip*{w_{\tau}}{w_0})
\end{equation}
A parallel transport condition on $ w_t $ is imposed by demanding that $ w^{\dagger}_{t + dt}w_t $ must be Hermitian and positive $\forall$ $ t $. We can further see that
\begin{align*}
	w^{\dagger}_{t + dt}w_t &= (\mathds{1} + \dot{w}_t dt + \order{dt^2})^{\dagger} w_t,\\
	&= (\mathds{1} + \dot{w}_t^{\dagger} dt + \order{dt^2}) w_t,\\ 
	&= w_t + \dot{w}_t^{\dagger} w_t dt + \order{dt^2}.
\end{align*}
and
\begin{align*}
	w^{\dagger}_tw_{t + dt} &= w_t + w_t^{\dagger} \dot{w}_t dt + \order{dt^2}
\end{align*}
which reduces to
\begin{equation} \label{eq: Parallel-Transport}
	w_t^{\dagger} \dot{w}_t = \dot{w}_t^{\dagger} w_t
\end{equation}
In \eqref{eq:Uhlmann} the intermediate terms should be real and positive. It will ensure that the argument for these terms will be zero.
\begin{equation}
	\ip*{w_{t + dt}}{w_t} = \Tr(w^{\dagger}_{t + dt}w_t) > 0
\end{equation}
For such a parallel lift, \eqref{eq:Uhlmann} reduces to
\begin{equation}\label{eq:Parallel-Uhlmann}
	\Phi_g = \arg\ip*{w_{\tau}}{w_0} = \arg\ip*{w_0}{w_{\tau}}
\end{equation}
Now, let us consider a unitary evolution
\begin{equation}
	\rho_0 = \sum_{k} \lambda_k \dyad{k} \rightarrow \rho_t = u_t \rho_0 u_t^{\dagger}
\end{equation}
\begin{align}
	w_0 = \sqrt{\rho_0} \rightarrow w_t &= \sqrt{\rho_t}x_t = u_t\sqrt{\rho_0} u_t^{\dagger}x_t = u_t\sqrt{\rho_0} v_t \\
	&= \sum_{k} \sqrt{\lambda_k} u_t \dyad{k} v_t
\end{align}
where we used $ x_0 = \mathds{1} $ and $ v_t = u_t^{\dagger}x_t $. Using this and \eqref{eq:Parallel-Uhlmann} we get
\begin{align}
	\Phi_g &= \arg\ip*{w_0}{w_{\tau}} \nonumber\\
	&= \arg \left(\Tr \sum_k \sqrt{\lambda_k}\dyad{k} \sum_{l} \sqrt{\lambda_l} u_{\tau} \dyad{l} v_{\tau} \right) \nonumber\\
	&= \arg \left( \sum_{k,l} \sqrt{\lambda_k \lambda_l} \dyad{k} u_{\tau} \dyad{l} v_{\tau} \right) \nonumber\\
	&= \arg \sum_{k,l} \sqrt{\lambda_k \lambda_l} \bra{k}u_{\tau} \ket{l}\bra{l}v_{\tau}\ket{k}
\end{align}
We can establish a one-to-one connection between Uhlmann's purification and the conventional purification by considering a pure state $ \ket{\Psi_0} $ belonging to the combined Hilbert space of the system and ancilla, i.e., $ \ket{\Psi_0} \in \mathcal{H}_s \otimes \mathcal{H}_a $. The evolution of $ \ket{\Psi_0} $ will be governed by a bilocal operator $ u_t \otimes y_t $ (where $ y_t = v_t^{T} $) that is,
\begin{equation}
	w_t \leftrightarrow \ket{\Psi_t} = \sum_k \sqrt{\lambda_k} u_t \ket{k} \otimes y_t \ket{k},
\end{equation} 
such that
\begin{equation}
	\Phi_g = \arg \ip*{\Psi_0}{\Psi_{\tau}}
\end{equation}
Here, if we consider a unilocal operation of type $ u_t \otimes \mathds{1} $ instead of a bilocal operator, then
\begin{equation}
	\ket{\Psi_t} = \sum_k \sqrt{\lambda_k} u_t \ket{k} \otimes \ket{k}.
\end{equation}
Consequently, the phase difference between the initial and final states will be
\begin{equation}
	\arg \ip*{\Psi_0}{\Psi_{\tau}} = \arg \sum_{k} \expval{u_{\tau}}{k} = \arg[\Tr (\rho_0 u_{\tau})].
\end{equation}
In a situation where $ u_t $ satisfies parallel transport, it reduces to the geometric phase for mixed states as defined in \cite{Sjoqvist}
\begin{equation}
	\Phi_g = \arg \sum_{k} \lambda_k \nu_k e^{i \beta_k},
\end{equation}
where $ \expval{u_{\tau}}{k} = \nu_k e^{i \beta_k} $ and $\beta_k$ is the pure state geometric phase for $ \ket{k} $. Let us go back to the Uhlmann's geometric phase where we have a bilocal operator and $ u_t $ and $ v_t $ are related via the parallel transport condition given by \eqref{eq: Parallel-Transport}. We can solve this further to have an explicit expression 
\begin{gather*}
	w_t = u_t \sqrt{\rho_0} v_t, \\ 
	w_t^{\dagger} = v_t^{\dagger} \sqrt{\rho_0} u_t^{\dagger}, \\
	\dot{w}_t = \dot{u}_t \sqrt{\rho_0} v_t + u_t \sqrt{\rho_0} \dot{v}_t, \\
	\dot{w}_t^{\dagger} = \dot{v}_t^{\dagger} \sqrt{\rho_0} u_t^{\dagger} + v_t^{\dagger} \sqrt{\rho_0} \dot{u}_t^{\dagger}, \\
\end{gather*}
\begin{align*}
	w_t^{\dagger} \dot{w}_t & = v_t^{\dagger} \sqrt{\rho_0} u_t^{\dagger} (\dot{u}_t \sqrt{\rho_0} v_t + u_t \sqrt{\rho_0} \dot{v}_t) \\
	&= v_t^{\dagger} \sqrt{\rho_0} u_t^{\dagger} \dot{u}_t \sqrt{\rho_0} v_t + v_t^{\dagger} \rho_0 \dot{v}_t
\end{align*}
\begin{align*}
	\dot{w}_t^{\dagger} u_t &= (\dot{v}_t^{\dagger} \sqrt{\rho_0} u_t^{\dagger} + v_t^{\dagger} \sqrt{\rho_0} \dot{u}_t^{\dagger} ) u_t \sqrt{\rho_0} v_t \\
	&= \dot{v}_t^{\dagger} \rho_0 v_t + v_t^{\dagger} \sqrt{\rho_0} \dot{u}_t^{\dagger} u_t \sqrt{\rho_0} v_t
\end{align*}
Also, let us write
\begin{equation}
	u_t = \exp(-i H t), v_t = \exp(i \tilde{H} t)
\end{equation}
\begin{gather*}
	\dot{u}_t = -i H u_t , \; \; \dot{v}_t = i \tilde{H} v_t, \\
	\dot{u}_t^{\dagger} = i u_t^{\dagger} H , \; \; \dot{v}_t^{\dagger} = -i v_t^{\dagger} \tilde{H} 
\end{gather*}
Using the above relations and \eqref{eq: Parallel-Transport} we get
\begin{gather*}
	v_t^{\dagger} \rho_0 \dot{v}_t + v_t^{\dagger} \sqrt{\rho_0} u_t^{\dagger} \dot{u}_t \sqrt{\rho_0} v_t= \dot{v}_t^{\dagger} \rho_0 v_t + v_t^{\dagger} \sqrt{\rho_0} \dot{u}_t^{\dagger} u_t \sqrt{\rho_0} v_t \\
	v_t^{\dagger} \rho_0 \dot{v}_t -  \dot{v}_t^{\dagger} \rho_0 v_t = v_t^{\dagger} \sqrt{\rho_0} (\dot{u}_t^{\dagger} u_t - u_t^{\dagger} \dot{u}_t) \sqrt{\rho_0} v_t \\
	v_t^{\dagger} \rho_0 (i \tilde{H} v_t) -  (-i v_t^{\dagger} \tilde{H}) \rho_0 v_t = v_t^{\dagger} \sqrt{\rho_0} ((i u_t^{\dagger} H) u_t - u_t^{\dagger} (-i H u_t)) \sqrt{\rho_0} v_t, \\
	i (v_t^{\dagger} \rho_0 \tilde{H} v_t +  v_t^{\dagger} \tilde{H} \rho_0 v_t)	= 2 i v_t^{\dagger} \sqrt{\rho_0} H \sqrt{\rho_0} v_t, \\
	i v_t^{\dagger} (\rho_0 \tilde{H} + \tilde{H} \rho_0 ) v_t = 2 i v_t^{\dagger} \sqrt{\rho_0} H \sqrt{\rho_0} v_t
\end{gather*}
which gives us a relation between $ H $ and $ \tilde{H} $
\begin{equation}
	\rho_0 \tilde{H} + \tilde{H} \rho_0 = \sqrt{\rho_0} H \sqrt{\rho_0}.
\end{equation}
Now, using the spectral decomposition for $\rho_0 = \sum_k \lambda_k \dyad{k}$ we can write
\begin{gather*}
	\mathds{1} (\rho_0 \tilde{H} + \tilde{H} \rho_0) \mathds{1} = \sum_{k,l} 2 \sqrt{\lambda_k \lambda_l} \dyad{k}{l} \bra{k}H\ket{l} \\
	\sum_k \dyad{k} (\rho_0 \tilde{H} + \tilde{H} \rho_0) \sum_{k} \dyad{l} = \sum_{k,l} 2 \sqrt{\lambda_k \lambda_l} \dyad{k}{l} \bra{k}H\ket{l} \\
	\sum_{k,l} (\lambda_k + \lambda_l) \bra{k} \tilde{H} \ket{l} \dyad{k}{l} = \sum_{k,l} 2 \sqrt{\lambda_k \lambda_l} \dyad{k}{l} \bra{k}H\ket{l}
\end{gather*}
implies
\begin{equation}\label{eq: H-tilde}
	\tilde{H} = \sum_{k,l} \dfrac{2 \sqrt{\lambda_k \lambda_l}}{\lambda_k + \lambda_l} \dyad{k}{l} \bra{k}H\ket{l}.
\end{equation}
For example, consider a two-level system with the Hamiltonian given by
\begin{equation}
	H = \dfrac{1}{2} \hat{\vb{n}} \vdot \boldsymbol{\sigma} = \dfrac{1}{2} (n_x \sigma_x + n_z \sigma_z),\; \; n_x^2 + n_z^2 = 1.
\end{equation}
and initially in the state
\begin{equation}
	\rho_0 = \dfrac{1}{2}(\mathds{1} + r \sigma_z) = \dfrac{1+r}{2} \dyad{0} + \dfrac{1-r}{2} \dyad{1}
\end{equation}
By substituting $ H $ into \eqref{eq: H-tilde} we get
\begin{equation*}
	\tilde{H} = \dfrac{1}{2} (\sqrt{1 - r^2}n_x \sigma_x + n_z \sigma_z) = \dfrac{1}{2} \sqrt{1 - r^2n_x^2}(\tilde{n}_x \sigma_x + \tilde{n}_z \sigma_z) = \dfrac{1}{2} \sqrt{1 - r^2n_x^2} (\tilde{\vb{n}} \vdot \boldsymbol{\sigma})
\end{equation*}
where
\begin{equation}
	\tilde{\vb{n}} = \left(\dfrac{\sqrt{1 - r^2}n_x}{\sqrt{1 - r^2n_x^2}}, 0,  \dfrac{n_z}{\sqrt{1 - r^2n_x^2}}\right)
\end{equation}
Thus
\begin{align}
	\Phi_g &= \arg \sum_{k,l} \sqrt{\lambda_k \lambda_l} \bra{k}u_{\tau} \ket{l}\bra{l}v_{\tau}\ket{k}, \nonumber \\
	&= \arg \sum_{k,l} \sqrt{\lambda_k \lambda_l} \bra{k}e^{- i \tau H} \ket{l}\bra{l}e^{i \tau \tilde{H}}\ket{k}, \nonumber \\
	&= \arg \sum_{k,l} \sqrt{\lambda_k \lambda_l} \bra{k}e^{- i \tau (\hat{\vb{n}} \vdot \boldsymbol{\sigma})/2} \ket{l}\bra{l}e^{i \tilde{\tau} (\tilde{\vb{n}} \vdot \boldsymbol{\sigma})/2}\ket{k}, \nonumber
\end{align}
Using
\begin{gather*}
	\expval{e^{- i \tau (\hat{\vb{n}} \vdot \boldsymbol{\sigma})/2}}{0} = \cos \dfrac{\tau}{2} - i n_z \sin\dfrac{\tau}{2}, \\
	\bra{0}e^{- i \tau (\hat{\vb{n}} \vdot \boldsymbol{\sigma})/2}\ket{1} = -i n_x \sin \dfrac{\tau}{2}, \\
	\bra{1}e^{- i \tau (\hat{\vb{n}} \vdot \boldsymbol{\sigma})/2}\ket{0} = -i n_x \sin \dfrac{\tau}{2}, \\
	\expval{e^{- i \tau (\hat{\vb{n}} \vdot \boldsymbol{\sigma})/2}}{1} = \cos \dfrac{\tau}{2} + i n_z \sin\dfrac{\tau}{2}	
\end{gather*}
and similarly for $ e^{- i \tilde{\tau} (\tilde{\vb{n}} \vdot \boldsymbol{\sigma})/2} $ and 
\begin{equation*}
	\tilde{\tau} = \sqrt{1 - r^2n_x^2} \tau
\end{equation*}
we get the Uhlmann phase as
\begin{equation}
	\Phi_g = \tan^{-1} \left[ \dfrac{r \left( \tilde{n}_z \tan \dfrac{\tilde{\tau}}{2} - n_z \tan\dfrac{\tau}{2} \right)}{1 + (n_z \tilde{n}_z + \sqrt{1 - r^2}n_x \tilde{n}_x)\tan\dfrac{\tau}{2}\tan\dfrac{\tilde{\tau}}{2} }\right]
\end{equation}
First, for the cyclic case, that is, for $\tau = 2 \pi$, we have 
\begin{equation}
	\Phi_g = \tan^{-1} \left[ r \tilde{n}_z \tan \pi \sqrt{1 - r^2n_x^2} \right] = \tan^{-1} \left[ \dfrac{r n_z }{\sqrt{1 - r^2n_x^2}} \tan \pi \sqrt{1 - r^2n_x^2} \right]
\end{equation}
Now, if we consider the state to be pure, that is, $r = 1$, we have $ \tilde{\vb{n}} = (0,0,1)$ and $\tilde{\tau} = \sqrt{1 - n_x^2} = n_z$ which yields
\begin{align}
	\Phi_g = -  \tan^{-1} \left[ n_z \tan\dfrac{\tau}{2} \right] + n_z \dfrac{\tau}{2}.
\end{align}
This is exactly equal to the minus half of the geodesically closed solid angle subtended by the open path on the Bloch sphere. Compared to the results we get using the interferometric approach ~\cite{Sjoqvist}, where the geometric phase for a mixed state $\rho$ with $r \ne 0$, we have $\phi_g = -\tan^{-1}\left[r \tan(\Omega/2)\right]$ where $\Omega$ is the geodesically closed solid angle on the Bloch sphere. The two approaches converge only in the case of pure state or in a trivial case when both the system and the ancilla are not evolving. This convergence of two approaches has been experimentally verified using NMR in ~\cite{Zhu2011}.

\subsection{A comment on the importance of the two approaches, interferometric and Uhlmann's approach, for mixed geometric phases in the topological characterization}
Topological characterization [don't worry, we will discuss it in detail in upcoming chapters] using Uhlmann's approach to mixed state geometric phase was proposed~\cite{Viyuela2014,Arovas2014,Andersson2016,Bruno2017}. It was also pointed out that the topological properties cannot survive above a certain critical temperature using Uhlmann construction~\cite{Viyuela2014}, in contrast to the interferometric approach where topological properties survive for non-zero temperature and cease to survive only in the limit of infinite temperature. Furthermore, a modified Chern character was also proposed, whose integral gives the thermal Uhlmann Chern number~\cite{Chein2018}. In Ref.~\cite{Viyuela2018}, the measurement of Uhlmann’s phase has been demonstrated using superconducting qubits. Apart from these, there is something called "ensemble geometric phase" which is used to characterize topology of the system in mixed states~\cite{Diehl2015,Diehl2018,Fleischhauer2021}.

\section{Weak Measurement Approach to Measure GP}
The geometric phase can be associated with the complex-valued weak value that arises in certain experiments. In such experiments, we make the system interact with an ancilla in a limited amount in order to preserve the quantum nature of the system, such as coherence. For a system, prepared in a pre-selected state $ \ket{\psi_i} $ and subject to a post-selection state $ \ket{\psi_f} $, the weak value of an observable $A$ is given by~\cite{Vaidman1988,Aharonov1991,Boyd2014}
\begin{equation}
	A_w = \dfrac{\mel{\psi_f}{A}{\psi_i}}{\ip{\psi_f}{\psi_i}}.
\end{equation}
In this section, we will establish a connection between the geometric phase and the complex valued $A_w$. Let us start by considering a quantum system and a quantum measurement device prepared initially in the product state~\cite{Sjoqvist2006}
\begin{equation}
	\rho_0 = \rho_{qs} \otimes \rho_{md} = 
	\dyad{a} \otimes \dyad{M_0}
\end{equation}
The system and the measurement device are made to interact by an interaction Hamiltonian given by~\cite{Neumann1955,Jozsa2007} 
\begin{equation}
	H(t) = g(t) \mathcal{P}^b \otimes Q
\end{equation}
where $ \mathcal{P}^b = \dyad{b} $ is a one-dimensional projector acting on the quantum system, $ g(t) $ is the coupling parameter which controls the interaction, and $Q$ is the position operator for the measurement device or pointer. We take the initial state of the measurement device, in the position representation, as Gaussian i.e.,
\begin{equation}
	M_0(q) = \bra{q}\ket{M_0} \sim e^{-q^2/2 \sigma^2}
\end{equation}
where $\ket{q}$ is the eigenstate of the position operator $Q$ of the measurement device or the pointer. The unitary evolution is 
\begin{equation}
	\begin{aligned}
		U(t) &= e^{- \tfrac{i}{\hbar} \int g(t) \mathcal{P}^b \otimes Q} \\
		&= e^{- i \kappa \mathcal{P}^b \otimes Q} \\
	\end{aligned}
\end{equation}
with
\begin{equation}
	\kappa = \dfrac{1}{\hbar} \int g(t) dt
\end{equation}
After the interaction, the state of the combined system is as follows
\begin{equation}
	\rho(t) = U(t)\rho_0U^{\dagger}(t) = e^{- i \kappa \mathcal{P}^b \otimes Q} \dyad{a} \otimes \dyad{M_0} e^{ i \kappa \mathcal{P}^b \otimes Q}
\end{equation}
and conditioned on the post-selection of the state $ \dyad{c} $, we get
\begin{equation}
	\begin{aligned}
		\rho_{ps} &= \dyad{c} \otimes \bra{c} U(t)\rho_0U^{\dagger}(t) \ket{c} \\
		&= \dyad{c} \otimes \bra{c}e^{- i \kappa \mathcal{P}^b \otimes Q} \ket{a}\ket{M_0}\bra{M_0} \bra{a} e^{ i \kappa \mathcal{P}^b \otimes Q} \ket{c}
	\end{aligned}
\end{equation}
In the limit $\kappa$ $\ll$ 1,
\begin{equation}
	\begin{aligned}
		&\bra{c}e^{- i \kappa \mathcal{P}^b \otimes Q} \ket{a} \\
		&\qquad =\bra{c} \mathds{1} - i \kappa \mathcal{P}^b \otimes Q \ket{a} \\
		&\qquad =\bra{c} \mathds{1} - i \kappa \dyad{b} \otimes Q \ket{a} \\
		&\qquad =\bra{c}\ket{a} - i \kappa \bra{c}\ket{b}\bra{b}\ket{a} Q \\
		&\qquad =\bra{c}\ket{a} \left( \mathds{1} - i \kappa \dfrac{\bra{c}\ket{b}\bra{b}\ket{a}}{\bra{c}\ket{a}} Q \right) \\
		&\qquad =\bra{c}\ket{a} \left( \mathds{1} - i \kappa \mathcal{P}^b_w (a,c) Q \right) \\
		&\qquad =\bra{c}\ket{a} e^{- i \kappa \mathcal{P}^b_w (a,c) Q}
	\end{aligned}
\end{equation}
where
\begin{equation}
	\mathcal{P}^b_w (a,c) = \dfrac{\bra{c}\ket{b}\bra{b}\ket{a}}{\bra{c}\ket{a}} 
\end{equation}
is the weak value of the operator $\mathcal{P}^b$ with pre-selected state $\ket{a}$ and post-selected state $\ket{c}$. Thus,
\begin{equation}
	\rho_{ps} = \ket{c}\bra{c} \otimes \abs{\bra{c}\ket{a}}^2 e^{- i \kappa \mathcal{P}^b_w (a,c) Q}\ket{M_0} \bra{M_0} e^{ i \kappa \mathcal{P}^b_w (c,a) Q}
\end{equation}
The state of the pointer after the post-selection is
\begin{equation}
	\begin{aligned}
		\ket{M} &= e^{-i \kappa z Q}\ket{M_0} \sim e^{-i \kappa z Q} \left(\sum_{i} e^{q_i^2/2 \sigma^2}\ket{q_i}\right) \\
		&= \sum_{i} e^{q_i^2/2 \sigma^2}  e^{-i \kappa z q_i} \ket{q_i}
	\end{aligned}
\end{equation}
which further results in
\begin{equation}
	M(q) = \bra{q}\ket{M}  \sim e^{q^2/2 \sigma^2}  e^{-i \kappa z q}
\end{equation}
Writing $z = a + ib$ and by completing the square, we will get
\begin{equation}
	M(q) \sim \underbrace{\text{exp} \left( - \dfrac{(q - \kappa \sigma^2 b)^2}{2 \sigma^2} \right)}_{\text{shifted Gaussian}} \times \underbrace{\text{exp} \left( -i \kappa a q \right)}_{\text{shift in momentum}}
\end{equation}
Therefore, as a result of post-selection, the weak measurement results in a shift in position of the pointer 
\begin{equation}
	\delta q = \kappa \sigma^2 b = \kappa \sigma^2 \Im \mathcal{P}^b_w (a,c)
\end{equation}
and the momentum of the pointer.
\begin{equation}
	\delta p = - \hbar \kappa a =  - \hbar \kappa \Re \mathcal{P}^b_w (a,c)
\end{equation}
We have 
\begin{align}
	\mathcal{P}^b_w (a,c) &= \dfrac{\bra{c}\ket{b}\bra{b}\ket{a}}{\bra{c}\ket{a}}  \nonumber \\
	&= \dfrac{\ip{a}{c}\bra{c}\ket{b}\bra{b}\ket{a}}{\abs{\ip{c}{a}}^2}  \nonumber \\
	\implies \arg \mathcal{P}^b_w (a,c) &= \arg \ip{a}{c}\bra{c}\ket{b}\bra{b}\ket{a} = \arg \Delta_3(a,b,c)
\end{align}
where $\Delta_3$ is the third-order Bargmann invariant (as discussed earlier), the argument of which results in the geometric phase between the three mutually nonorthogonal states $ \{\ket{a}, \ket{b}, \ket{c}\} $. Therefore, we have
\begin{equation}
	\begin{aligned}
		\gamma_g = \arg \Delta_3(a,b,c) &= \tan^{-1} \left(\dfrac{\text{Im} \mathcal{P}^b_w (a,c)}{\text{Re} \mathcal{P}^b_w (a,c)}\right)\\
		&= \tan^{-1} \left(\dfrac{\delta q / \kappa \sigma^2}{- \delta p / \hbar \kappa}\right)\\
		& = - \tan^{-1} \left(\dfrac{\hbar \delta q}{\sigma^2 \delta p}\right)
	\end{aligned}
\end{equation}
Therefore, by measuring the shift in position and the momentum of the pointer after turning off the interaction, we can find the geometric phase. The above method can be extended for the case where we have $N$ number of mutually non-orthogonal states \{$\ket{\psi_0}$, $\ket{\psi_1}$, $\ket{\psi_2}$, \dots, $\ket{\psi_{N-1}} $\}. Recently, weak measurement sequence has been shown to lead to the accumulation of the geometric phase, depending on the strength of the measurement~\cite{Cho2019}.

\subsection{Weak value measurements with qubits}
Let's consider a spin-1/2 system pre-selected and post-selected states represented by Bloch vectors $\hat{\vb{n}}$ and $\hat{\vb{m}}$ respectively and is given by 
\begin{equation}
	\dyad{\psi_i} = \dfrac{1}{2} \left(\mathds{1} + \hat{\vb{n}} \vdot \boldsymbol{\sigma} \right)
\end{equation}
and 
\begin{equation}
	\dyad{\psi_f} = \dfrac{1}{2} \left(\mathds{1} + \hat{\vb{n}} \vdot \boldsymbol{\sigma} \right)
\end{equation}
We now consider the observable $Z$ of the form
\begin{equation}
	Z = \dfrac{1}{2} \left(\mathds{1} + \hat{\vb{k}} \vdot \boldsymbol{\sigma} \right)
\end{equation}
and calculate the weak value corresponding to $Z$ as
\begin{equation} \label{eq:zw}
	\begin{aligned}
		Z_w (\psi_i, \psi_f) &= \dfrac{\bra{\psi_f} Z \ket{\psi_i}}{\bra{\psi_f}\ket{\psi_i}} \\
		&= \dfrac{\bra{\psi_f} Z \ket{\psi_i}\bra{\psi_i}\ket{\psi_f}}{\abs{\bra{\psi_f}\ket{\psi_i}}^2} \\
		&= \dfrac{\Tr \left[ \ket{\psi_f}\bra{\psi_f} Z \ket{\psi_i}\bra{\psi_i} \right]}{\Tr \left[ \ket{\psi_f}\bra{\psi_f} \ket{\psi_i}\bra{\psi_i} \right]} \\
		&= \dfrac{1}{2} \dfrac{\Tr \left[ \left(\mathds{1} + \hat{\vb{n}} \vdot \boldsymbol{\sigma} \right) \left(\mathds{1} + \hat{\vb{n}} \vdot \boldsymbol{\sigma} \right) (\mathds{1} + \hat{\vb{k}} \vdot \boldsymbol{\sigma} ) \right]}{\Tr \left[ \left(\mathds{1} + \hat{\vb{n}} \vdot \boldsymbol{\sigma} \right) \left(\mathds{1} + \hat{\vb{n}} \vdot \boldsymbol{\sigma} \right) \right]} \\
		&=\dfrac{1}{2} \left(\dfrac{1 + \hat{\vb{n}}\vdot\hat{\vb{k}} + \hat{\vb{k}}\vdot\hat{\vb{m}} + \hat{\vb{m}}\vdot\hat{\vb{n}} + i \hat{\vb{k}}\vdot(\hat{\vb{n}} \times \hat{\vb{m}})}{1 + \hat{\vb{m}} \vdot\hat{\vb{n}}}\right)
	\end{aligned}
\end{equation}
Further, if we choose $Z = \dyad{\uparrow}$ i.e. $\hat{\vb{k}} = (0, 0, 1)$ then
\begin{equation}
	Z_w (\psi_i, \psi_f) = \dfrac{1}{2} \left(\dfrac{1 + n_z + m_z + \hat{\vb{m}}\vdot\hat{\vb{n}} + i (\hat{\vb{n}} \times \hat{\vb{m}})_z}{1 + \hat{\vb{m}}\vdot\hat{\vb{n}}}\right)
\end{equation}
Since $Z_w (\psi_i, \psi_f)$ is a complex quantity, we can write a polar decomposition for it as
\begin{equation}
	Z_w = \dfrac{\sqrt{(1 + n_z + m_z + \hat{\vb{m}}\vdot\hat{\vb{n}})^2 +  [(\hat{\vb{n}} \times \hat{\vb{m}})_z]^2}}{2(1 + \hat{\vb{m}}\vdot\hat{\vb{n}})} e^{-i \phi}
\end{equation}
where
\begin{equation}
	\phi = \tan^{-1} \left(\dfrac{(\hat{\vb{n}} \times \hat{\vb{m}})_z}{1 + n_z + m_z + \hat{\vb{m}}\vdot\hat{\vb{n}}}\right)
\end{equation}
Using the expression for the solid angle $\Omega$, subtended by given three vectors $\vb{R}_1, \vb{R}_2,$ and $\vb{R}_3$ on the center~\cite{Strackee1983} as
\begin{equation}
	\tan \dfrac{\Omega}{2} = \dfrac{\vb{R}_1\vdot (\vb{R}_2 \times \vb{R}_3)}{R_1R_2R_3 + (\vb{R}_1\vdot\vb{R}_2)R_3 + (\vb{R}_1\vdot\vb{R}_3)R_2 + (\vb{R}_2\vdot\vb{R}_3)R_1}
\end{equation}
we conclude from Eq.~\eqref{eq:zw} directly, that the phase $\phi$ is nothing but the half of the solid angle subtended by the Bloch vectors $\hat{\vb{m}}$, $\hat{\vb{n}}$ and $\hat{\vb{k}}$ at the origin i.e.,
\begin{equation}
	\phi = \dfrac{\Omega}{2}.
\end{equation}
In general, the weak value of the operator of the form $\hat{\vb{q}} \vdot \boldsymbol{\sigma}$ can be evaluated using Pauli's algebra as
\begin{equation}
	\begin{aligned}
		\langle \hat{\vb{q}} \vdot \boldsymbol{\sigma} \rangle_w &= \dfrac{\bra{\psi_f} \hat{\vb{q}} \vdot \boldsymbol{\sigma} \ket{\psi_i}}{\bra{\psi_f}\ket{\psi_i}} \\
		&= \dfrac{\Tr \left( \ket{\psi_f}\bra{\psi_f} \hat{\vb{q}} \vdot \boldsymbol{\sigma} \ket{\psi_i}\bra{\psi_i} \right)}{\Tr \left( \ket{\psi_f}\bra{\psi_f} \ket{\psi_i}\bra{\psi_i} \right)} \\
		&= \dfrac{\hat{\vb{q}}\vdot\hat{\vb{n}} + \hat{\vb{q}}\vdot\hat{\vb{m}} + i \hat{\vb{q}}\vdot(\hat{\vb{n}} \times \hat{\vb{m}})}{1 + \hat{\vb{n}}\vdot\hat{\vb{m}}}.
	\end{aligned}
\end{equation}
Next, we will take qubit as the measurement device or pointer and see how we can measure the real and imaginary parts of the weak value, which are required to calculate the geometric phase.

\section{Qubit as A Measurement Device}
Suppose we consider a qubit as a measurement device in an initial state $\ket{\phi_i}$. We again make the system and the pointer interact and after the post-selection, total state of the system and the measurement qubit will be
\begin{equation}
	\rho_{ps} = \dyad{\psi_f} \otimes \bra{\psi_f}e^{- i \kappa \mathcal{P}^b \otimes Q} \ket{\psi_i}\dyad{\phi_i}\bra{\phi_i} \bra{\psi_i}e^{ i \kappa \mathcal{P}^b \otimes Q} \ket{\psi_f}
\end{equation}
Therefore
\begin{equation}
	\begin{aligned}
		\ket{\phi_f} &= \bra{\psi_f}e^{- i \kappa \mathcal{P}^b \otimes Q} \ket{\psi_i}\ket{\phi_i} \\
		& \sim \bra{\psi_f} \ket{\psi_i} ( \mathds{1} - i \kappa \langle \mathcal{P}^b \rangle_w  Q) \ket{\phi_i}
	\end{aligned}
\end{equation}
NOw, we consider any general operation for a qubit which reads
\begin{equation}
	Q = \hat{\vb{n}} \vdot \boldsymbol{\sigma}
\end{equation}
where $ \hat{\vb{n}} $ is a unit vector and let us say $ \langle \mathcal{P}^b \rangle_w = z = x + iy$. After the interaction is switched off and the system is subject to post-selection, we are free to measure any operator of the form $\hat{\vb{q}} \vdot \boldsymbol{\sigma}$, the expectation value of which is given by~\cite{Molmer2009}
\begin{equation}
	\expval{\hat{\vb{q}} \vdot \boldsymbol{\sigma}}{\phi_f} = \dfrac{\bra{\phi_f} \hat{\vb{q}} \vdot \boldsymbol{\sigma} \ket{\phi_f}}{\bra{\phi_f}\ket{\phi_f}}
\end{equation}
Numerator
\begin{equation}
	\begin{aligned}
		&\bra{\phi_f} \hat{\vb{q}} \vdot \boldsymbol{\sigma} \ket{\phi_f} \\
		&= \abs{\bra{\psi_f}\ket{\psi_i}}^2 \bra{\phi_i} ( \mathds{1} + i \kappa z^* (\hat{\vb{q}} \vdot \boldsymbol{\sigma}) \hat{\vb{q}} \vdot \boldsymbol{\sigma} ( \mathds{1} - i \kappa z (\hat{\vb{n}} \vdot \boldsymbol{\sigma}) \ket{\phi_i}\\
		&\simeq \abs{\bra{\psi_f}\ket{\psi_i}}^2 \left( \bra{\phi_i}\hat{\vb{q}} \vdot \boldsymbol{\sigma}\ket{\phi_i} + i \kappa x \bra{\phi_i} [\hat{\vb{n}} \vdot \boldsymbol{\sigma}, \hat{\vb{q}} \vdot \boldsymbol{\sigma}]\ket{\phi_i} + \kappa y \bra{\phi_i} \{\hat{\vb{n}} \vdot \boldsymbol{\sigma}, \hat{\vb{q}} \vdot \boldsymbol{\sigma}\}_+\ket{\phi_i}\right) + \order{g^2}  
	\end{aligned}
\end{equation}
Denominator
\begin{equation}
	\begin{aligned}
		\bra{\phi_f}\ket{\phi_f} &= \abs{\bra{\psi_f}\ket{\psi_i}}^2 \bra{\phi_i} ( \mathds{1} + i \kappa z^* (\hat{\vb{n}} \vdot \boldsymbol{\sigma}) ( \mathds{1} - i \kappa z (\hat{\vb{n}} \vdot \boldsymbol{\sigma}) \ket{\phi_i}\\
		&= \abs{\bra{\psi_f}\ket{\psi_i}}^2 \left( \mathds{1} + 2gy\bra{\phi_i}\hat{\vb{n}} \vdot \boldsymbol{\sigma}\ket{\phi_i} + 2 g^2 \abs{z}^2 \right)\\
		&\simeq \abs{\bra{\psi_f}\ket{\psi_i}}^2 \left( \mathds{1} + 2gy\bra{\phi_i}\hat{\vb{n}} \vdot \boldsymbol{\sigma}\ket{\phi_i}\right) + \order{g^2}
	\end{aligned}
\end{equation}
Thus,
\begin{equation}
	\begin{aligned}
		\dfrac{\bra{\phi_f} \hat{\vb{q}} \vdot \boldsymbol{\sigma} \ket{\phi_f}}{\bra{\phi_f}\ket{\phi_f}} &= \dfrac{\bra{\phi_i}\hat{\vb{q}} \vdot \boldsymbol{\sigma}\ket{\phi_i} + i \kappa x \bra{\phi_i} [\hat{\vb{n}} \vdot \boldsymbol{\sigma}, \hat{\vb{q}} \vdot \boldsymbol{\sigma}]\ket{\phi_i} + \kappa y \bra{\phi_i} \{\hat{\vb{n}} \vdot \boldsymbol{\sigma}, \hat{\vb{q}} \vdot \boldsymbol{\sigma}\}_+\ket{\phi_i}}{\mathds{1} + 2gy\bra{\phi_i}\hat{\vb{n}} \vdot \boldsymbol{\sigma}\ket{\phi_i}}
	\end{aligned}
\end{equation}
By expanding the denominator in power series up to first order in $ g $, we get
\begin{equation} \label{eq:expectation}
	\begin{aligned}
		\dfrac{\bra{\phi_f} \hat{\vb{q}} \vdot \boldsymbol{\sigma} \ket{\phi_f}}{\bra{\phi_f}\ket{\phi_f}} &= \bra{\phi_i}\hat{\vb{q}} \vdot \boldsymbol{\sigma}\ket{\phi_i} + i \kappa x \bra{\phi_i} [\hat{\vb{n}} \vdot \boldsymbol{\sigma}, \hat{\vb{q}} \vdot \boldsymbol{\sigma}]\ket{\phi_i} + \kappa y \bra{\phi_i} \{\hat{\vb{n}} \vdot \boldsymbol{\sigma}, \hat{\vb{q}} \vdot \boldsymbol{\sigma}\}_+\ket{\phi_i} \\
		&\qquad -2 \kappa y\bra{\phi_i}\hat{\vb{n}} \vdot \boldsymbol{\sigma}\ket{\phi_i}\bra{\phi_i}\hat{\vb{q}} \vdot \boldsymbol{\sigma}\ket{\phi_i}
	\end{aligned}
\end{equation}
We further take
\begin{equation}
	\dyad{\phi_i} = \dfrac{1}{2} (\mathds{1} + \hat{\vb{m}} \vdot \boldsymbol{\sigma})
\end{equation}
and using the relation~\cite{Sakurai1985}
\begin{equation}
	(\hat{\vb{a}} \vdot \boldsymbol{\sigma})(\hat{\vb{b}} \vdot \boldsymbol{\sigma}) = (\vb{a} \vdot \vb{b}) \mathds{1} + i (\vb{a}\times\vb{b}) \vdot \boldsymbol{\sigma}.
\end{equation}
we further simply the expression for the expectation value. The first term would be 
\begin{equation}
	\begin{aligned}
		\bra{\phi_i}\hat{\vb{q}} \vdot \boldsymbol{\sigma}\ket{\phi_i} &= \Tr (\hat{\vb{q}} \vdot \boldsymbol{\sigma}\ket{\phi_i}\bra{\psi_i}) \\
		&= \dfrac{1}{2} \Tr (\hat{\vb{q}} \vdot \boldsymbol{\sigma}(\mathds{1} + \hat{\vb{m}} \vdot \boldsymbol{\sigma})) \\
		&= \dfrac{1}{2} \Tr (\hat{\vb{q}} \vdot \boldsymbol{\sigma} + (\hat{\vb{q}} \vdot \boldsymbol{\sigma})(\hat{\vb{m}} \vdot \boldsymbol{\sigma})) \\
		&= \hat{\vb{q}} \vdot \hat{\vb{m}}.
	\end{aligned}
\end{equation}
Second term
\begin{equation}
	\begin{aligned}
		&i \kappa x \bra{\phi_i} [\hat{\vb{n}} \vdot \boldsymbol{\sigma}, \hat{\vb{q}} \vdot \boldsymbol{\sigma}]\ket{\phi_i} \\
		&=\dfrac{i \kappa x}{2} \Tr ([\hat{\vb{n}} \vdot \boldsymbol{\sigma}, \hat{\vb{q}} \vdot \boldsymbol{\sigma}]\ket{\phi_i}\bra{\phi_i}) \\
		&=\dfrac{i \kappa x}{2} \Tr [((\hat{\vb{n}} \vdot \boldsymbol{\sigma})(\hat{\vb{q}} \vdot \boldsymbol{\sigma}) - (\hat{\vb{q}} \vdot \boldsymbol{\sigma})(\hat{\vb{n}} \vdot \boldsymbol{\sigma}))(\mathds{1} + \hat{\vb{m}} \vdot \boldsymbol{\sigma})] \\
		&=\dfrac{i \kappa x}{2} \Tr( (\hat{\vb{n}} \vdot \boldsymbol{\sigma})(\hat{\vb{q}} \vdot \boldsymbol{\sigma}) - (\hat{\vb{q}} \vdot \boldsymbol{\sigma})(\hat{\vb{n}} \vdot \boldsymbol{\sigma}) + (\hat{\vb{n}} \vdot \boldsymbol{\sigma})(\hat{\vb{q}} \vdot \boldsymbol{\sigma})(\hat{\vb{m}} \vdot \boldsymbol{\sigma}) - (\hat{\vb{q}} \vdot \boldsymbol{\sigma})(\hat{\vb{n}} \vdot \boldsymbol{\sigma})(\hat{\vb{m}} \vdot \boldsymbol{\sigma}) ) \\
		&=\dfrac{i \kappa x}{2} \left( i (\hat{\vb{n}} \times \hat{\vb{q}}) \vdot \hat{\vb{m}} - i (\hat{\vb{q}} \times \hat{\vb{n}}) \vdot \hat{\vb{m}} \right) \\
		&=2\kappa x(\hat{\vb{q}} \times \hat{\vb{n}}) \vdot \hat{\vb{m}}
	\end{aligned}
\end{equation}
Third term
\begin{equation}
	\begin{aligned}
		&\kappa y \bra{\phi_i} \{\hat{\vb{n}} \vdot \boldsymbol{\sigma}, \hat{\vb{q}} \vdot \boldsymbol{\sigma}\}_+\ket{\phi_i} \\
		&=\dfrac{\kappa y}{2} \Tr (\{\hat{\vb{n}} \vdot \sigma, \hat{\vb{q}} \vdot \boldsymbol{\sigma}\}_+\ket{\phi_i}\bra{\phi_i}) \\
		&=\dfrac{\kappa y}{2} \Tr [((\hat{\vb{n}} \vdot \boldsymbol{\sigma})(\hat{\vb{q}} \vdot \boldsymbol{\sigma}) + (\hat{\vb{q}} \vdot \boldsymbol{\sigma})(\hat{\vb{n}} \vdot \boldsymbol{\sigma}))(\mathds{1} + \hat{\vb{m}} \vdot \boldsymbol{\sigma})] \\
		&=\dfrac{\kappa y}{2} \Tr( (\hat{\vb{n}} \vdot \boldsymbol{\sigma})(\hat{\vb{q}} \vdot \boldsymbol{\sigma}) + (\hat{\vb{q}} \vdot \boldsymbol{\sigma})(\hat{\vb{n}} \vdot \boldsymbol{\sigma}) + (\hat{\vb{n}} \vdot \boldsymbol{\sigma})(\hat{\vb{q}} \vdot \boldsymbol{\sigma})(\hat{\vb{m}} \vdot \boldsymbol{\sigma}) + (\hat{\vb{q}} \vdot \boldsymbol{\sigma})(\hat{\vb{n}} \vdot \boldsymbol{\sigma})(\hat{\vb{m}} \vdot \boldsymbol{\sigma}) ) \\
		&=2\kappa y \hat{\vb{q}} \vdot \hat{\vb{n}} 
	\end{aligned}
\end{equation}
Fourth term
\begin{equation}
	\begin{aligned}
		&-2 \kappa y\bra{\phi_i}\hat{\vb{n}} \vdot \boldsymbol{\sigma}\ket{\phi_i}\bra{\phi_i}\hat{\vb{q}} \vdot \boldsymbol{\sigma}\ket{\phi_i}\\
		&=-2 \kappa y \Tr (\hat{\vb{n}} \vdot \boldsymbol{\sigma}\ket{\phi_i}\bra{\phi_i}\hat{\vb{q}} \vdot \boldsymbol{\sigma}\ket{\phi_i}\bra{\phi_i}) \\
		&=-\dfrac{\kappa y}{2}\Tr (\hat{\vb{n}} \vdot \boldsymbol{\sigma}(\mathds{1} + \hat{\vb{m}} \vdot \boldsymbol{\sigma})\hat{\vb{q}} \vdot \boldsymbol{\sigma}(\mathds{1} + \hat{\vb{m}} \vdot \boldsymbol{\sigma})) \\
		&=-\dfrac{\kappa y}{2}\Tr \left( (\hat{\vb{n}} \vdot \boldsymbol{\sigma})(\hat{\vb{q}} \vdot \boldsymbol{\sigma}) + (\hat{\vb{n}} \vdot \boldsymbol{\sigma})(\hat{\vb{q}} \vdot \boldsymbol{\sigma})(\hat{\vb{m}} \vdot \boldsymbol{\sigma}) + (\hat{\vb{n}} \vdot \boldsymbol{\sigma})(\hat{\vb{m}} \vdot \boldsymbol{\sigma})(\hat{\vb{q}} \vdot \boldsymbol{\sigma}) +  (\hat{\vb{n}} \vdot \boldsymbol{\sigma})(\hat{\vb{m}} \vdot \boldsymbol{\sigma})(\hat{\vb{q}} \vdot \boldsymbol{\sigma})(\hat{\vb{m}} \vdot \boldsymbol{\sigma})\right)\\
		&=- \kappa y \left(\hat{\vb{n}} \vdot \hat{\vb{q}} + i(\hat{\vb{n}}\times\hat{\vb{q}})\vdot\hat{\vb{m}} + i(\hat{\vb{n}}\times\hat{\vb{m}})\vdot\hat{\vb{q}} + (\hat{\vb{n}}\vdot\hat{\vb{m}})(\hat{\vb{q}}\vdot\hat{\vb{m}})\right)\\
		&=- \kappa y \left(\hat{\vb{q}}\vdot\hat{\vb{n}} + (\hat{\vb{n}}\vdot\hat{\vb{m}})(\hat{\vb{q}}\vdot\hat{\vb{m}})\right)
	\end{aligned}
\end{equation}
Therefore, Eq.~\eqref{eq:expectation} becomes
\begin{equation} \label{eq:finalexpectation}
	\dfrac{\bra{\phi_f} \hat{\vb{q}} \vdot \boldsymbol{\sigma} \ket{\phi_f}}{\bra{\phi_f}\ket{\phi_f}}  = \hat{\vb{q}}\vdot\hat{\vb{m}} + 2\kappa \left[(\hat{\vb{q}} \times \hat{\vb{n}})\vdot\hat{\vb{m}}\right]x + 2\kappa \left[ \hat{\vb{q}}\vdot\hat{\vb{n}} - (\hat{\vb{n}}\vdot\hat{\vb{m}})(\hat{\vb{q}}\vdot\hat{\vb{m}}) \right]y. 
\end{equation}
where 
\begin{equation}
	x = \Re\expval{\mathcal{P}_B}_w, \qquad y = \Im \expval{\mathcal{P}_B}_w
\end{equation}
and $\hat{\vb{m}}$ is the Bloch corresponding to the initial state of the pointer. We observe that the expectation value given in Eq.~\eqref{eq:finalexpectation} depends on the initial state of the pointer. By choosing the appropriate initial state of the meter qubit, we can obtain the real and imaginary parts of the weak value $A_w$ in terms of the expectation value of an observable. For example, if we choose $\hat{\vb{m}} \perp \hat{\vb{n}}$ and $\hat{\vb{q}} = \hat{\vb{n}} \times \hat{\vb{m}}$, then
\begin{equation}
	\dfrac{\bra{\phi_f} \hat{\vb{q}} \vdot \boldsymbol{\sigma} \ket{\phi_f}}{\bra{\phi_f}\ket{\phi_f}}  = 2 \kappa x
\end{equation}
and if $\hat{\vb{m}} \perp \hat{\vb{n}}$ and $\hat{\vb{q}} = \hat{\vb{n}}$
\begin{equation}
	\dfrac{\bra{\phi_f} \hat{\vb{q}} \vdot \boldsymbol{\sigma} \ket{\phi_f}}{\bra{\phi_f}\ket{\phi_f}}  = 2 \kappa y.
\end{equation}

\noindent To conclude this chapter, we have reviewed the concept of geometric phases for pure states starting from Berry's derivation in the context of the adiabatic theorem and its subsequent generalizations. We have moved on to geometric phases for mixed states, where we discussed two approaches: the interferometric approach and Uhlmann's approach. The interferometric approach looks convenient from the experimental perspective. We will use it in later chapters to study the geometric response of a rotating two-level atom inside an electromagnetic cavity. Uhlmann's approach to the geometric phase is mathematically rigorous; however, it is used to study topological phase transitions in condensed matter systems~\cite{Galindo2021} and to study the topological indicators at finite-temperature~\cite{Chien2021}. We have also discussed some experimental studies to measure the geometric phase.

\chapter{Quantum Walks} \label{chap:qw}
Quantum walks are the quantum analogue of classical random walks \cite{Aharonov1993,Ambainis2001,Kempe2003,VenegasAndraca2012,Nayak2000} where a quantum walker propagates on a lattice and the direction of propagation is conditioned over the state of its coin. Due to the quantum nature of the walker and the coin, the position state of the walker is a superposition of multiple lattice sites. This provides a quadratically fast spread of the walker across the lattice compared to its classical counterpart \cite{Ambainis2001}. As
opposed to classical random walks, quantum walks are governed by quantum superpositions of amplitudes
rather than classical probability distributions. There are two kinds of quantum walks, continuous and discrete. In this thesis, we will consider only discrete-time quantum walks.

Quantum walks, continuous-time as well as discrete-time, are important in various fields including universal quantum computation \cite{Childs2009,Childs2013,Lovett2010}, quantum search algorithms \cite{Ambainis2003,Childs2004,Shenvi2003,Agliari2010,Xue2022}, quantum simulations \cite{Nicola2014}, quantum state transfer \cite{Wojcik2011} and simulation of physical systems \cite{Schreiber2012, Sansoni2012, Peruzzo2010}. Quantum walks have also been used in other branches of science, such as biology, to study energy transfer in photosynthesis \cite{Mohseni2008,Lambert2013}. They have also been shown to be a promising candidate to simulate the decoherence \cite{Romanelli2005,Romanelli2005b,Kendon2007} and for the implementation of generalized measurements, positive operator valued measures (POVM) \cite{Wojcik2013, Guo2015}. The discrete-time quantum walks have been realized on a variety of systems such as NMR~\cite{Laflamme2005}, trapped ions~\cite{Schmitz2009,Tobias2012,Roos2010}, in linear optical systems like linear cavity~\cite{Knight2003a}, optical rings~\cite{Knight2003b,Soriano2006,Schreiber2010,Schreiber2011}, interferometry~\cite{Pandey2011,Jeong2013}, optical lattices~\cite{Preiss2015,Esposito2022}, optical networks~\cite{Zhao2002,Jeong2004,Do2005,Broome2010}, classical light~\cite{Francisco2006}, using superconducting qubits~\cite{Flurin2017}, cavity QED~\cite{Sanders2003}, quantum optics~\cite{Zou2006,Zhang2007,Zhang2010,Su2019}, trapped ions~\cite{Schmitz2009,Zahringer2010}, neutral atom trap~\cite{Karski2009}, superconducting processors~\cite{Pan2019,Pan2021}, integrated photonics~\cite{Sansoni2012,Cardano2015}, BECs~\cite{Dadras2018, Yan2020}.

In this chapter, we will discuss various protocols of discrete-time quantum walk (DTQW) in one (1D) and two dimensions (2D). We discuss the generalization of 1D DTQW to 1D split-step quantum walk (SSQW), and we further show the decomposition of 1D SSQW and 2D DTQW into 1D DTQW.  

\section{1D Discrete Time Quantum Walk (DTQW)}
\begin{figure}[H]
	\centering
	\includegraphics[width=9cm]{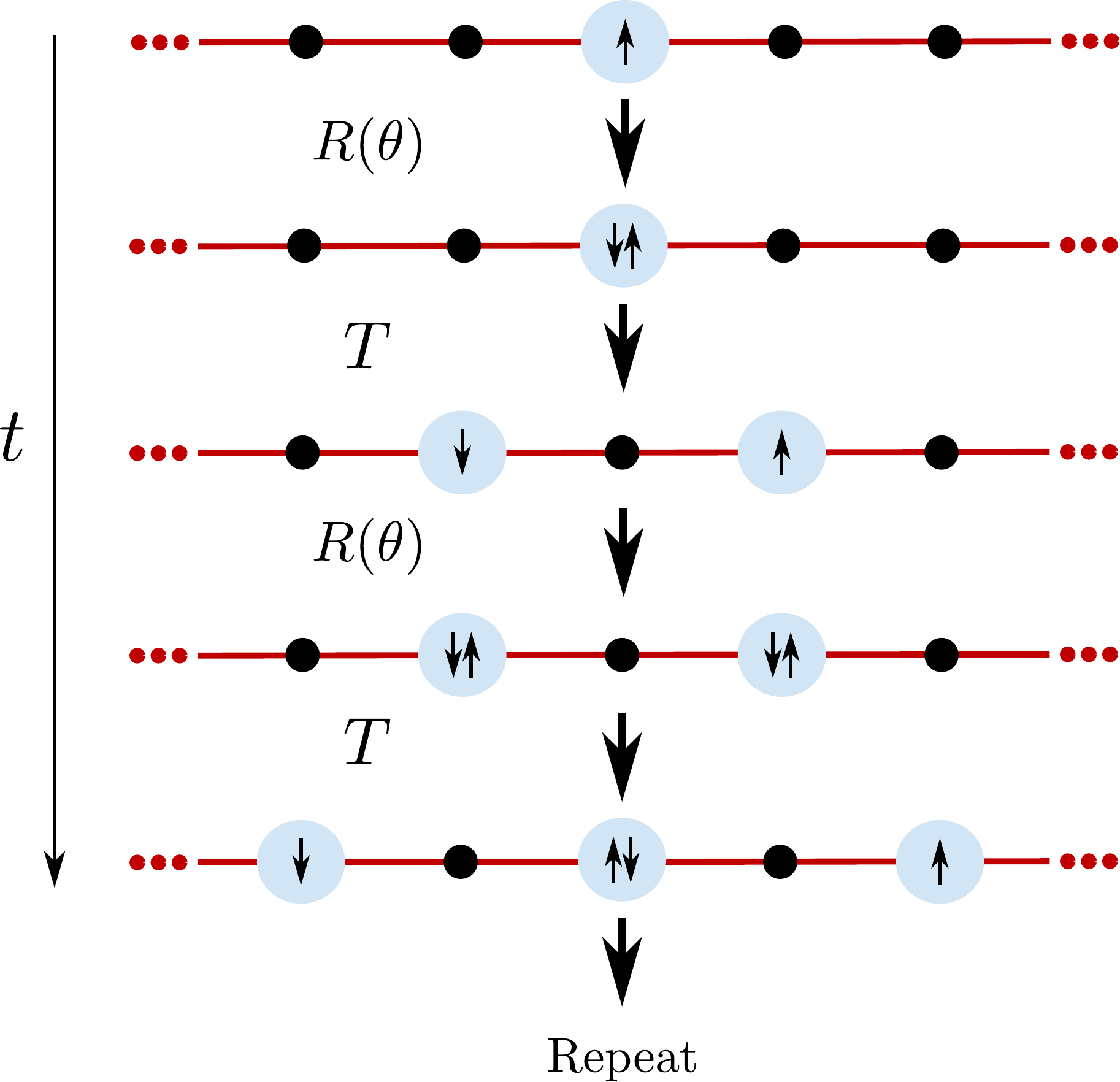}
	\label{fig:1DDTQW}
	\caption{Schematic of the protocol of two steps of 1D DTQW. The system consists of 1D lattice where site index runs from $ -n $ to $ +n $ and spin degrees of freedom. We can see the interference at $n = 0$ after the second step.}
\end{figure}
\begin{figure}[H]
	\centering
	\includegraphics[width=10cm]{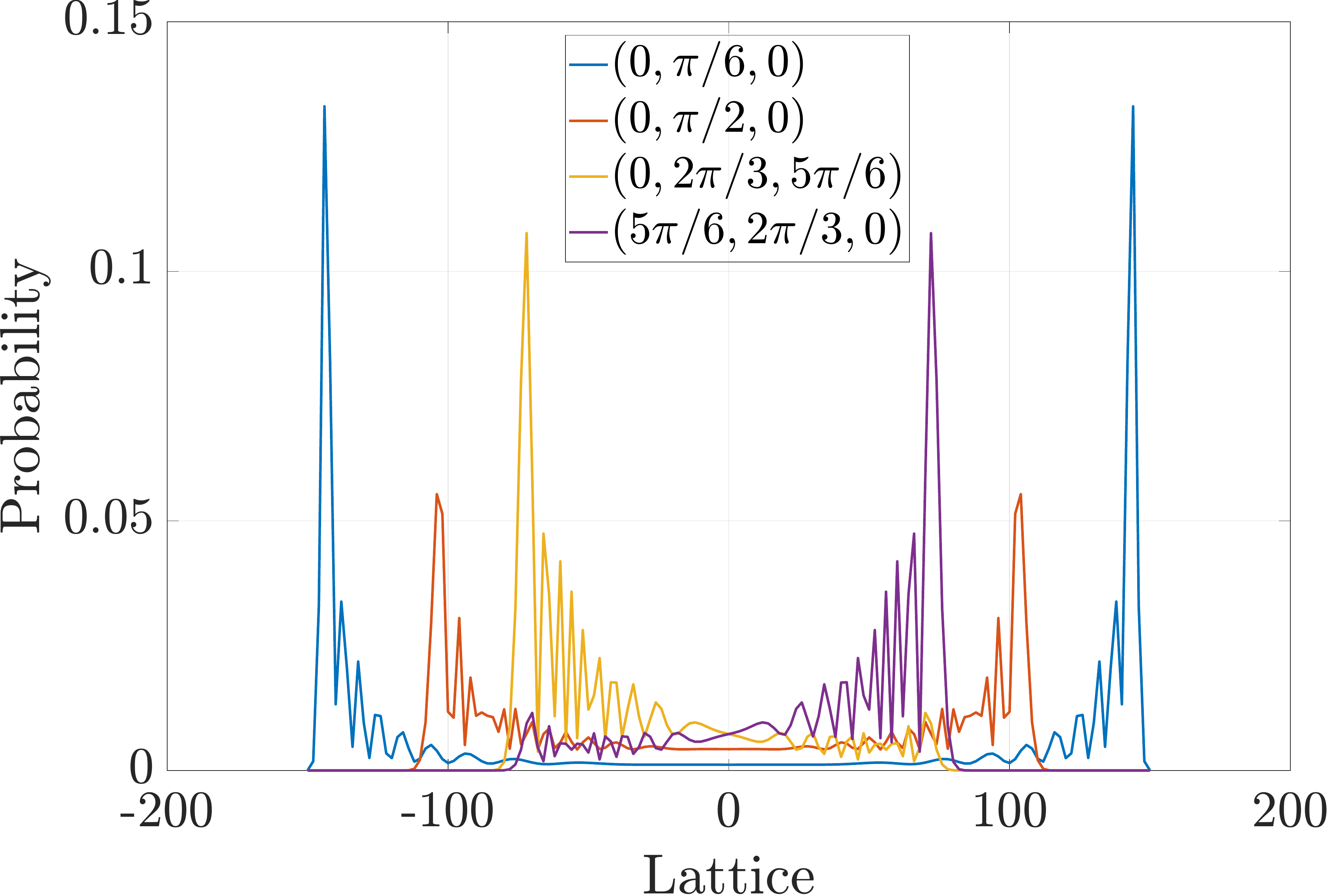}
	\caption{The probability distribution of walker after 150 time steps for different values of $(\xi, \theta, \zeta)$. }
	\label{fig:fullprobdist}
\end{figure}

DTQW protocol is defined for a particle hopping over a 1D lattice with an internal degree of freedom, spin, which is equivalent to the coin in a classical random walk. The particle has two orthogonal spin states which are referred to as spin `up' and `down'. DTQW consists of two operations
\begin{enumerate}
	\item a coin toss or spin flip operation
	\begin{equation}
		R(\xi, \theta, \zeta, \eta) = \sum_n R_n(\xi, \theta, \zeta, \eta) \otimes \dyad{n}.
	\end{equation} 
	where $R_n(\xi, \theta, \zeta, \eta)$ is a general U(2) operator given by~\cite{Chandrashekar2008}
	\begin{equation} \label{eq:CoinOperator}
		R_n(\xi, \theta, \zeta, \eta) = e^{i \eta_n }\begin{pmatrix}
			e^{-i \xi_n/2} \cos \theta_n/2 & -e^{-i \zeta_n/2} \sin \theta_n/2 \\
			e^{i \zeta_n/2} \sin \theta_n/2 & e^{i \xi_n/2} \cos \theta_n/2
		\end{pmatrix}
	\end{equation}  
	and the factor $n$ in the subscript represents the dependence on lattice sites. 
	
	\item a spin dependent translation operator $T$ which makes the particle move to the right (left) by one lattice site when the spin is up (down) i.e.
	\begin{gather*}
		T (\ket{\uparrow} \otimes \ket{n}) = \ket{\uparrow} \otimes \ket{n+1}, \\
		T (\ket{\downarrow} \otimes \ket{n}) = \ket{\downarrow} \otimes \ket{n-1},
	\end{gather*}
	\begin{equation} \label{eq:1D-Translation}
		T = \sum_n \dyad{\uparrow} \otimes \ket{n+1} \bra{n}  + \dyad{\downarrow} \otimes \ket{n-1} \bra{n}.
	\end{equation}
\end{enumerate}
In position basis $\{\ket{n}\} \in \mathcal{H}_{\text{pos}}$ and spin basis $\{\ket{\uparrow}, \ket{\downarrow} \} \in \mathcal{H}_{\text{spin}}$, the unitary operator which governs the time evolution of the walker for a unit step time reads
\begin{align} 
	U(\xi, \theta, \zeta) &= T R(\xi, \theta, \zeta) \nonumber \\
	&= \sum_n \dyad{\uparrow}R_n(\xi, \theta, \zeta) \otimes \ket{n+1} \bra{n}  + \dyad{\downarrow}R_n(\xi, \theta, \zeta) \otimes \ket{n-1} \bra{n} \label{eq:1D-Unitary}.
\end{align}
For the time being, we consider the case of homogeneous system where the spin flip operator does not depend on the lattice site and $\eta = 0$ which yields
\begin{equation} 
	R(\xi, \theta, \zeta) = \sum_n R_n(\xi, \theta, \zeta) \otimes \dyad{n} \mapsto R(\xi, \theta, \zeta) \otimes \mathds{1}
\end{equation}  
and consequently
\begin{equation} \label{eq:1D-UnitaryHomo}
	U(\xi, \theta, \zeta) = \sum_n \dyad{\uparrow}R(\xi, \theta, \zeta) \otimes \ket{n+1} \bra{n}  + \dyad{\downarrow}R(\xi, \theta, \zeta) \otimes \ket{n-1} \bra{n} .
\end{equation}
where $\mathds{1}$ denotes the identity operation on the lattice. The total Hilbert space $\mathcal{H}$ is the tensor product $ \mathcal{H}_{\text{pos}} $ and $ \mathcal{H}_{\text{spin}} $. Given an initial state $ \ket{\psi(0)} $, the wave function after $ t $ time steps is written as
\begin{equation}
	\ket{\psi(t)} = U^t \ket{\psi(0)} = \sum_n \left(\psi_{n, \uparrow} (t) \ket{n} \otimes \ket{\uparrow} + \psi_{n, \downarrow} (t) \ket{n} \otimes \ket{\downarrow}\right)
\end{equation}
where $\psi_{n, \uparrow} (t)$ and $ \psi_{n, \downarrow}(t) $ are the normalized probability amplitudes such that $\sum_n ( \abs*{\psi_{n, \uparrow}(t)}^2 + \abs*{\psi_{n, \downarrow}(t)}^2) = 1.$ The probability of finding the walker at the $ n $th site after $t$ time steps is given by
\begin{align}
	P(n,t) &= \abs{\bra{n} \otimes \bra{\uparrow}\ket{\psi(t)}}^2 + \abs{\bra{n} \otimes \bra{\downarrow}\ket{\psi(t)}}^2 \nonumber \\ 			
	&= \abs{\psi_{n, \uparrow} (t)}^2 + \abs{\psi_{n, \downarrow} (t)}^2.
\end{align}
In Fig.~\eqref{fig:fullprobdist}, we plotted the probability distribution of the walker after 150 time steps for different $(\xi, \theta, \zeta)$. The walker was initially localized at the origin, and the coin is taken to be in a symmetric state such that
\begin{equation}\label{eq:initialcond}
	\ket{\psi(0)} = \ket{0} \otimes \left(\dfrac{\ket{\uparrow} +i \ket{\downarrow}}{\sqrt{2}}\right).
\end{equation} 
\noindent We observe that for $\xi = \zeta = 0$, the probability distribution is symmetric. However, the distribution is dominant along one direction for non-zero values of $\xi$ and $\zeta$ despite the fact that the initial coin state is symmetric. 

\subsection{Unitary equivalence of quantum walks}
In this part, we show that the quantum walk, governed by the time evolution operator in Eq.~\eqref{eq:1D-Unitary}, can be reduced to a single-parameter family of quantum walks~\cite{Goyal2015}. We note that the two time evolution operator of quantum walk, given by
\begin{align}
	U = T (R \otimes \mathds{1}), \qquad U' = T' (R' \otimes \mathds{1})
\end{align}
are unitary equivalent, if
\begin{equation} \label{eq:unitary-equivalent}
	U' = V U V^{\dagger}
\end{equation}
where $V$ is unitary. The state of the quantum walker after $t$ time steps under the evolution of U is written as 
\begin{align}
	\ket{\psi(t)} = U^t \ket{\psi(0)} = (V^{\dagger} U' V)^t \ket{\psi(0)} = V^{\dagger} (U')^t V \ket{\psi(0)}. 
\end{align}
Therefore, the dynamics of the state $\ket{\psi(0)}$ under the action of $U$ is the same as the dynamics of the initial state $V\ket{\psi(0)}$ under the action of $U'$. Note that the factor $V^{\dagger}$ is just a unitary and leaves the underlying physics unaffected. That is, the statistics of any observable $O$, with the initial state of the system $\ket{\psi(0)}$, governed by $U$, will be equivalent to the statistics of the observable $V O V^{\dagger}$ in the system with the initial state $V\ket{\psi(0)}$ and under the action of $U'$. Now, by demanding the invariance of the entanglement between the coin and the lattice and the translational invariance under the action of $V$, we can choose it to be of the form
\begin{equation}
	V = X \otimes \mathds{1}
\end{equation}
i.e., $V$ acts only on the coin degree of freedom. The operator $R$ can be taken to any $SU(2)$ of the form 
\begin{equation}
	R = \exp(i \dfrac{\phi}{2} \vb{r} \vdot \boldsymbol{\sigma})
\end{equation}
and $X$ to be of the form 
\begin{equation}
	X = \dyad*{\uparrow}{\uparrow'} + \dyad*{\downarrow}{\downarrow'}
\end{equation}
such that
\begin{equation} \label{eq:blochrot}
	\vb{r} \vdot \boldsymbol{\sigma} \ket*{\uparrow'} = \ket*{\uparrow'}, \;\;\;  \vb{r} \vdot \boldsymbol{\sigma} \ket*{\downarrow'} = -\ket*{\downarrow'}.
\end{equation}
Under the action of this transformation and utilizing Eq.~\eqref{eq:blochrot} we get 
\begin{align}
	R \to R' = VRV^{\dagger} &= X \exp(i \dfrac{\phi}{2} \vb{r} \vdot \boldsymbol{\sigma}) X^{\dagger} \nonumber \\
	&= X \left( \exp{i \phi/2} \dyad*{\uparrow'}{\uparrow} + \exp{-i \phi/2} \dyad*{\downarrow'}{\downarrow} \right) X^{\dagger} \nonumber \\
	&= \exp{i \phi/2} \dyad{\uparrow} + \exp{-i \phi/2}  \dyad{\downarrow} =  \exp(i \phi \sigma_z/2).
\end{align}
Therefore, the time evolution operator $U$ transforms as
\begin{equation}
	U \to U' = VUV^{\dagger} = \exp(i \phi \sigma_z/2) \dyad*{\uparrow'} \otimes T + \exp(i \phi \sigma_z/2) \dyad*{\downarrow'} \otimes T^{\dagger}.
\end{equation}
Now, we consider the transformation of the form $V = W \otimes X$ which preserves the coin basis but does not need to be translationally invariant. The propagation operator in Euler parameterization is written as~\cite{Goyal2015}
\begin{equation}
	U(\eta, \theta, \xi) = e^{i \tfrac{\eta}{2} \sigma_z} e^{i \tfrac{\theta}{2} \sigma_y} e^{i \tfrac{\xi}{2}} \dyad{\uparrow} \otimes T + e^{i \tfrac{\eta}{2} \sigma_z} e^{i \tfrac{\theta}{2} \sigma_y} e^{-i \tfrac{\xi}{2}} \dyad{\downarrow} \otimes T^{\dagger}
\end{equation} 
where we utilized the following relations
\begin{equation}
	e^{i \tfrac{\xi}{2} \sigma_z} \ket{\uparrow} =  e^{i \tfrac{\xi}{2}} \ket{\uparrow}, \;\;\; e^{i \tfrac{\xi}{2} \sigma_z} \ket{\downarrow} =  e^{-i \tfrac{\xi}{2}} \ket{\downarrow}.
\end{equation}
Next, we choose $X$ to be of the form $X = e^{-i \tfrac{\eta}{2} \sigma_z}$ such that
\begin{equation*}
	U(\eta, \theta, \xi) \to (X \otimes \mathds{1}) U(\eta, \theta, \xi) (X^{\dagger} \otimes \mathds{1}) = e^{i \tfrac{\theta}{2} \sigma_y} e^{i \tfrac{\xi + \eta}{2}} \dyad{\uparrow} \otimes T + e^{i \tfrac{\theta}{2} \sigma_y} e^{-i \tfrac{\xi + \eta}{2}} \dyad{\downarrow} \otimes T^{\dagger}.
\end{equation*}
Finally, we introduce a phase shift $\Phi$ in the quasi-momentum, which results in breaking of the translation invariance and reads
\begin{equation}
	E_{\Phi} R E_{\Phi}^{\dagger} = e^{i \Phi} R,\;\;\;\;\; E_{\Phi} = \sum_n e^{i \Phi n} \dyad{n}.
\end{equation}
We introduced this phase factor to choose $W = E_{-(\eta + \xi)/2}$ so that, together with $X = e^{-i \tfrac{\eta}{2} \sigma_z}$, it yields
\begin{equation}
	(\mathds{1} \otimes W) U(\eta, \theta, \xi) (\mathds{1} \otimes W^{\dagger}) = e^{i \tfrac{\theta}{2} \sigma_y} \dyad{\uparrow} \otimes T + e^{i \tfrac{\theta}{2} \sigma_y} \dyad{\downarrow} \otimes T^{\dagger}.
\end{equation}
Therefore, the three-parameter quantum walk governed by $U(\eta, \theta, \xi)$ is unitary equivalent to a one-parameter quantum walk with $U(\theta)$. We can also rewrite the above equation as
\begin{equation} \label{eq:unitaryequivalence}
	U(\theta) = \sum_n R(\theta) \dyad{\uparrow}  \otimes \ket{n+1} \bra{n}  + R(\theta) \dyad{\downarrow} \otimes \ket{n-1} \bra{n}
\end{equation}
where $R(\theta) = e^{-i \theta \sigma_y/2}$ ($-i$ and $+i$ are equivalent and just a matter of convention) with $\theta \in [-2 \pi, 2 \pi]$ being a real parameter and $\sigma_y$ the Pauli matrix along the $y$-axis. We will continue our discussion with the time evolution operator $U(\theta)$. It has been shown that the angle $\theta$ plays a role in characterizing phase and group velocity~\cite{Kempf2009}.

\begin{figure}
	\centering
	\subfigure[]{
		\includegraphics[width=7cm]{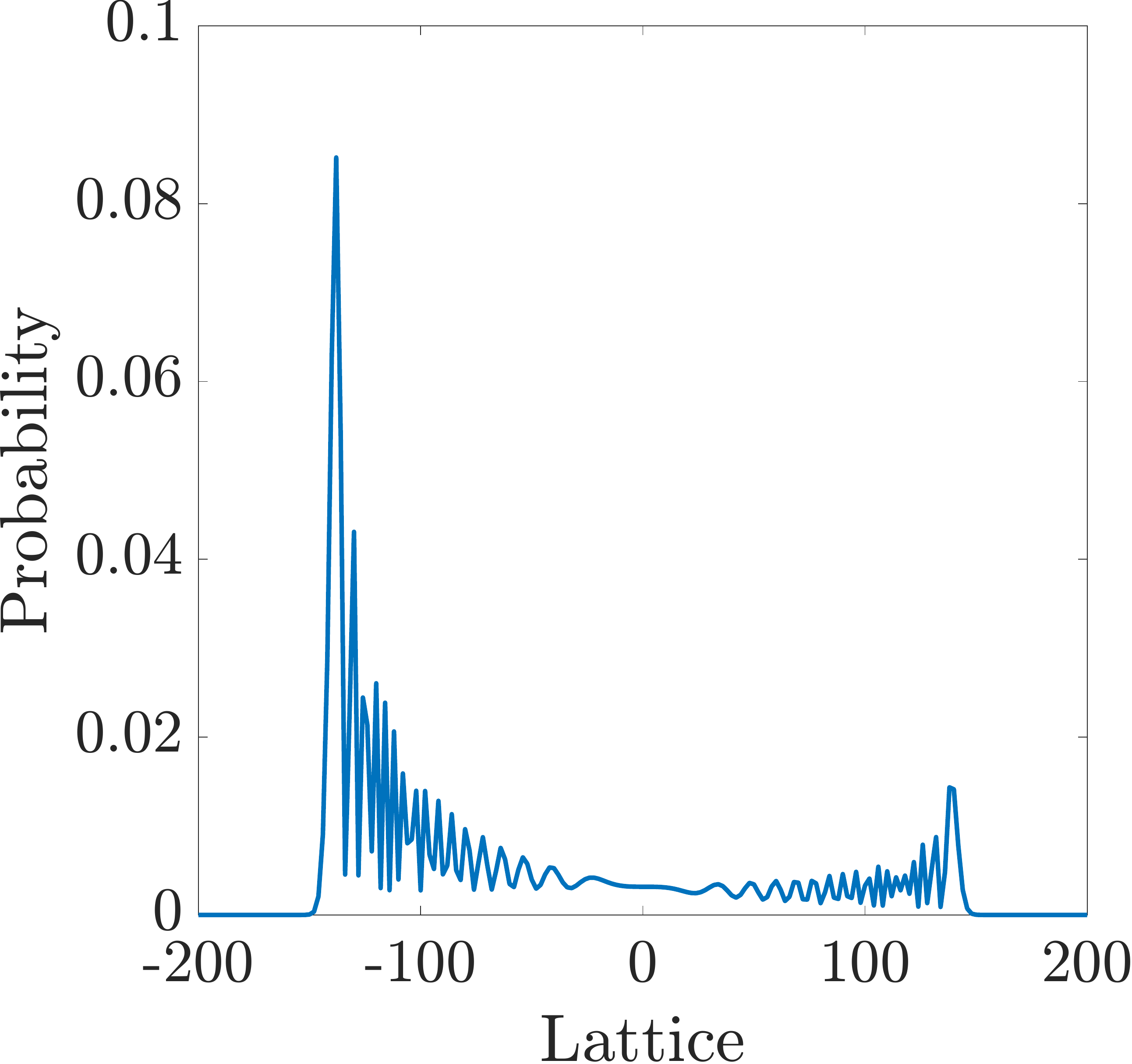}
		\label{fig:1DEvolutiondown}}
	\subfigure[]{
		\includegraphics[width=7cm]{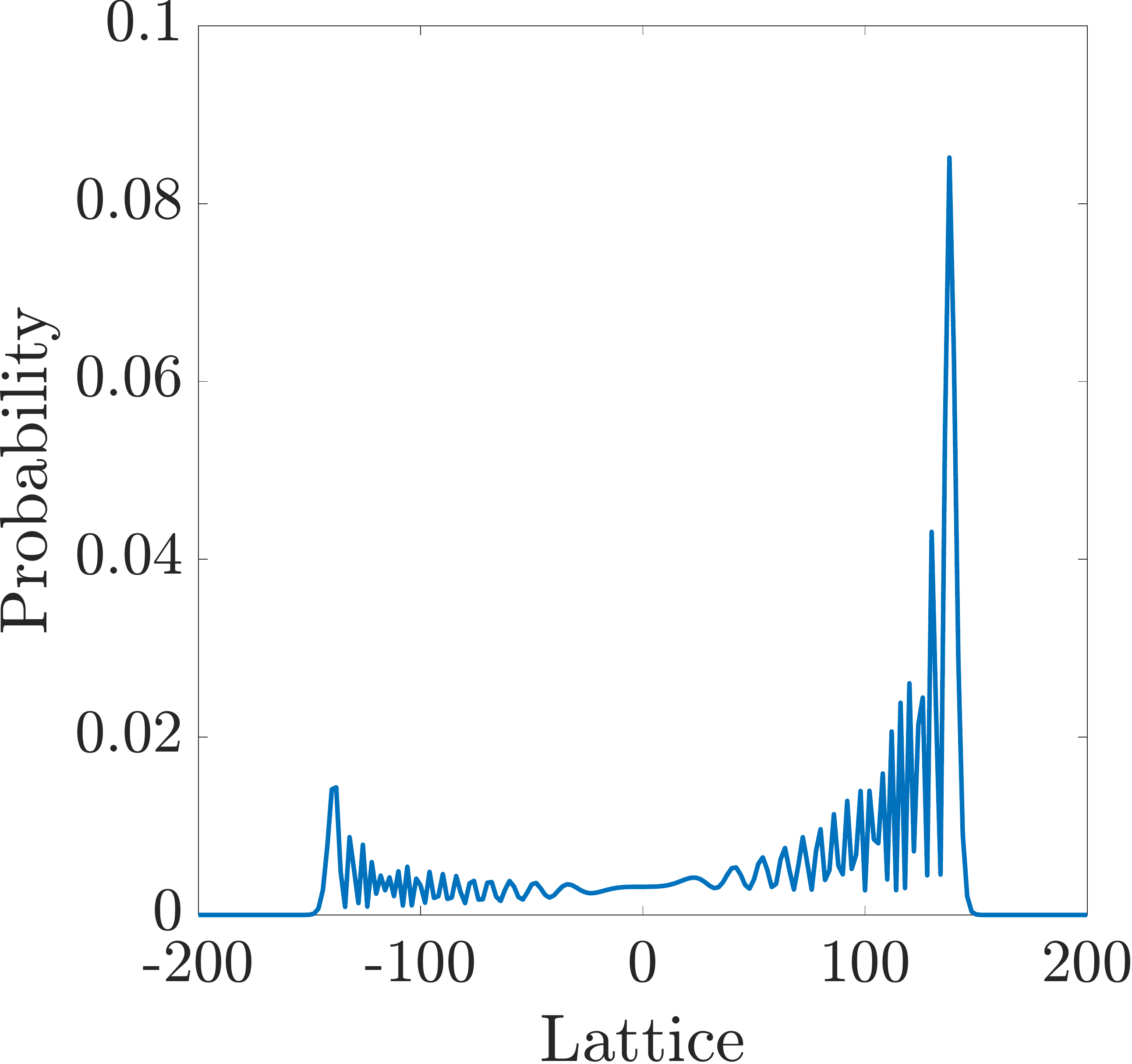}
		\label{fig:1DEvolutionup}}
	\subfigure[]{
		\includegraphics[width=7cm]{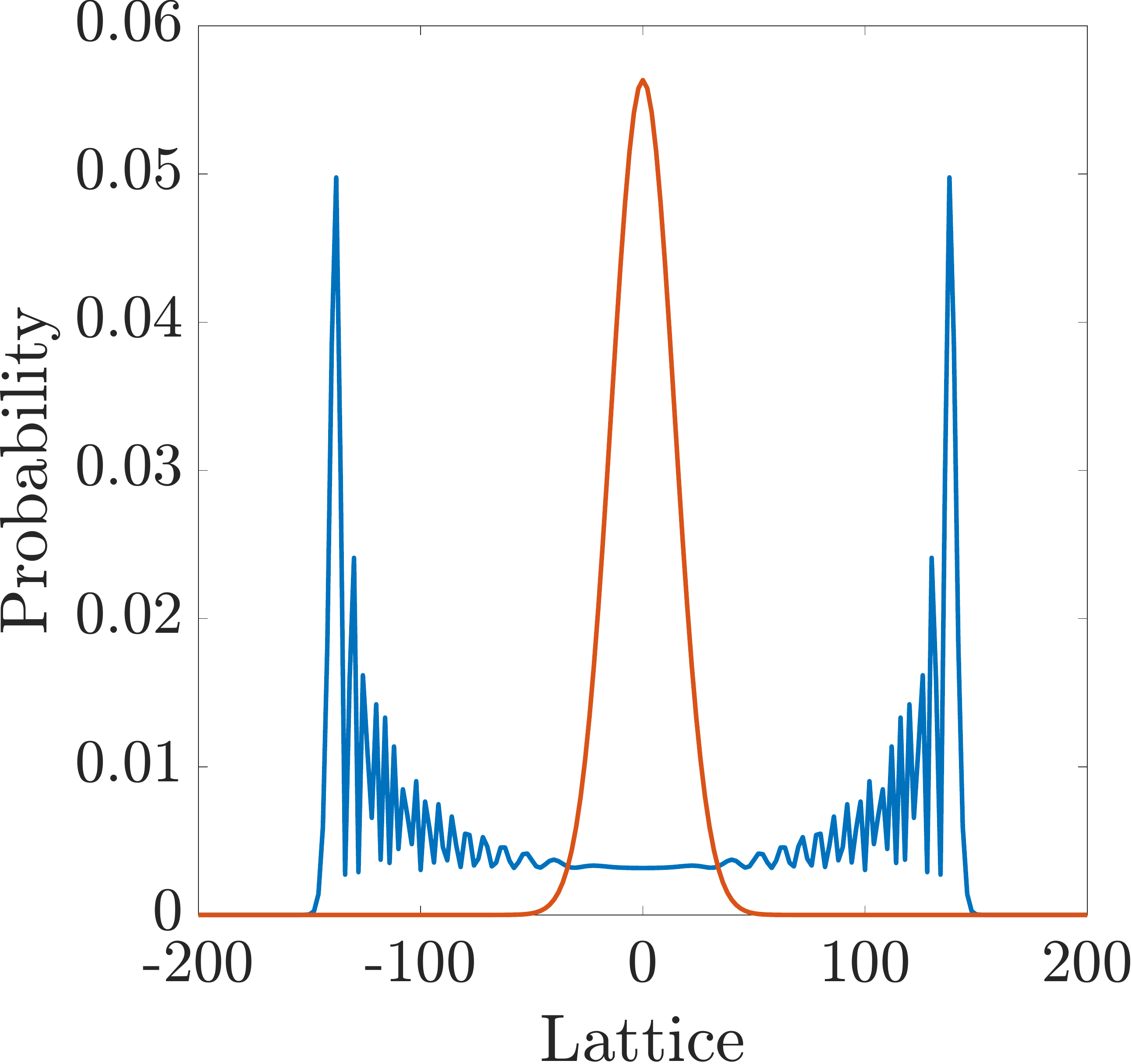}
		\label{fig:dtqw}}	
	\subfigure[]{
		\includegraphics[width=6.6cm]{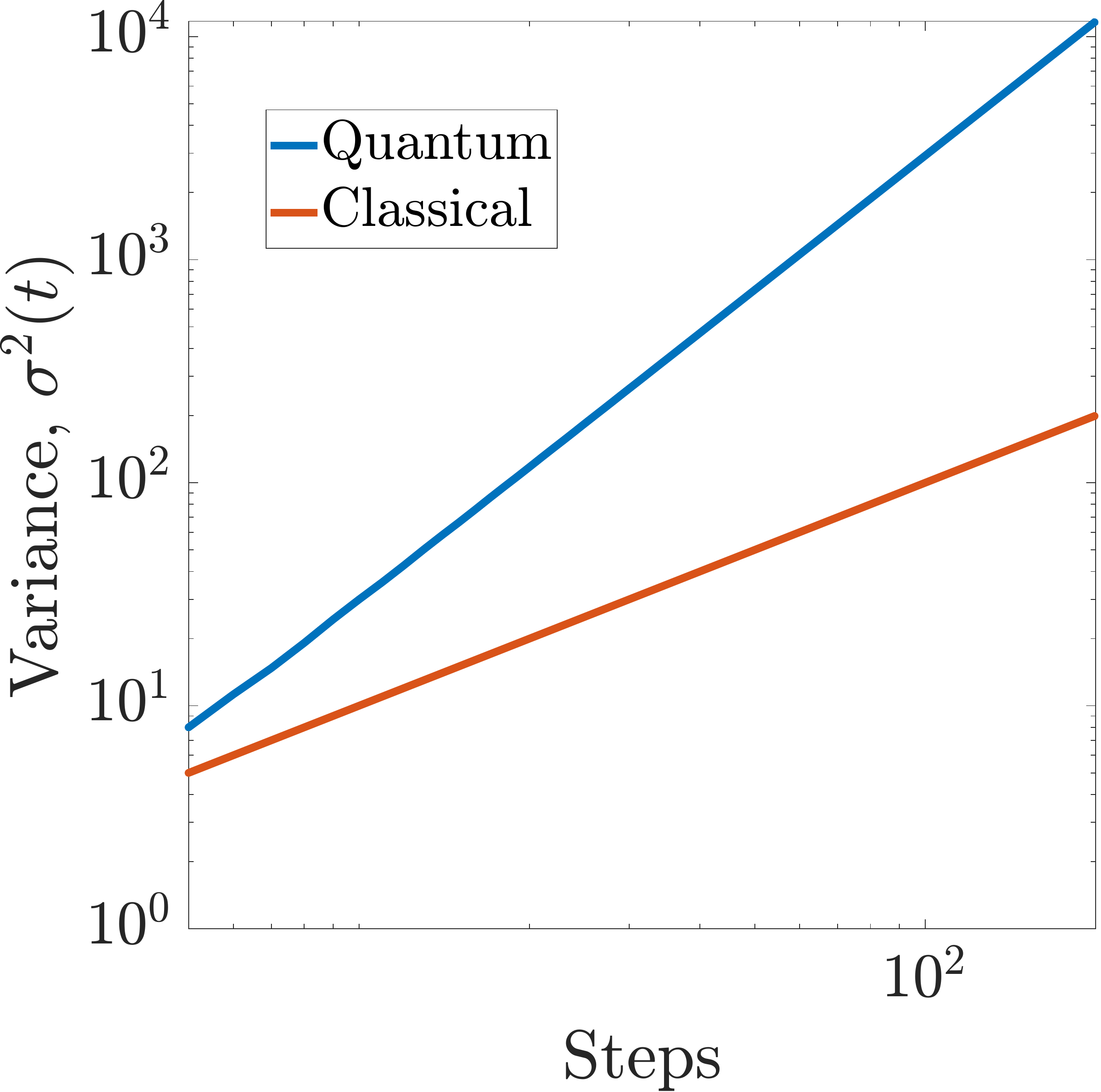}
		\label{fig:variance}}
	\caption{Probability distribution of a walker after 200 time steps for different initial states. \subref{fig:1DEvolutiondown} $ \ket{\psi(0)} = \ket{0} \otimes \ket{\downarrow} $, \subref{fig:1DEvolutionup} $ \ket{\psi(0)} = \ket{0} \otimes \ket{\uparrow} $, \subref{fig:dtqw} $ \ket{\psi(0)} = \ket{0} \otimes (\ket{\uparrow} + i \ket{\downarrow})/\sqrt{2} $ in contrast to classical walk (red). Only the points with non-zero probability are plotted. \subref{fig:variance} The variance, $\sigma^2(t)$ for quantum and classical walks with time steps $ t $. The axes are chosen on logarithmic scale.} 
	\label{fig:ProbDis}
\end{figure}

\noindent We now plot the probability distribution of the walker dynamics, which is governed by Eq.~\eqref{eq:unitaryequivalence} in Fig.~\ref{fig:1DEvolutiondown},~\ref{fig:1DEvolutionup},~\ref{fig:dtqw} for 200 time steps and for different initial states. In Fig.~\ref{fig:dtqw} we have compared the probability distribution for the classical and quantum walk. We can see a contrasting behavior between the two. In quantum walk, we let the probability amplitude evolve with time, which leads to constructive and destructive interference. It interferes constructively at the extremes and destructively at the origin. However, classical lack coherence, and thus we do not see any interference effects and get a normal distribution. The variance for DTQW for the same initial state as in Fig.~\ref{fig:dtqw} is proportional to the square of the step number $ t $, i.e. $\sigma^2 \propto t^2 $, which is quadratically faster than that of the classical random walk. In Fig.~\ref{fig:variance} we plotted the variance for the classical and quantum walk on the logarithmic scale. On the logarithmic scale, the variance for DTQW and the classical random walk will be a straight line; however, the slope in DTQW is twice ($\log(t^2)$) as compared to the classical one ($\log(t)$). Due to this, the DTQW and classical walk processes are referred to as ballistic and diffusive processes, respectively.
\subsection{Effective Hamiltonian}
\begin{figure}[H]
	\centering
	\subfigure[]{
		\includegraphics[width=5.1cm]{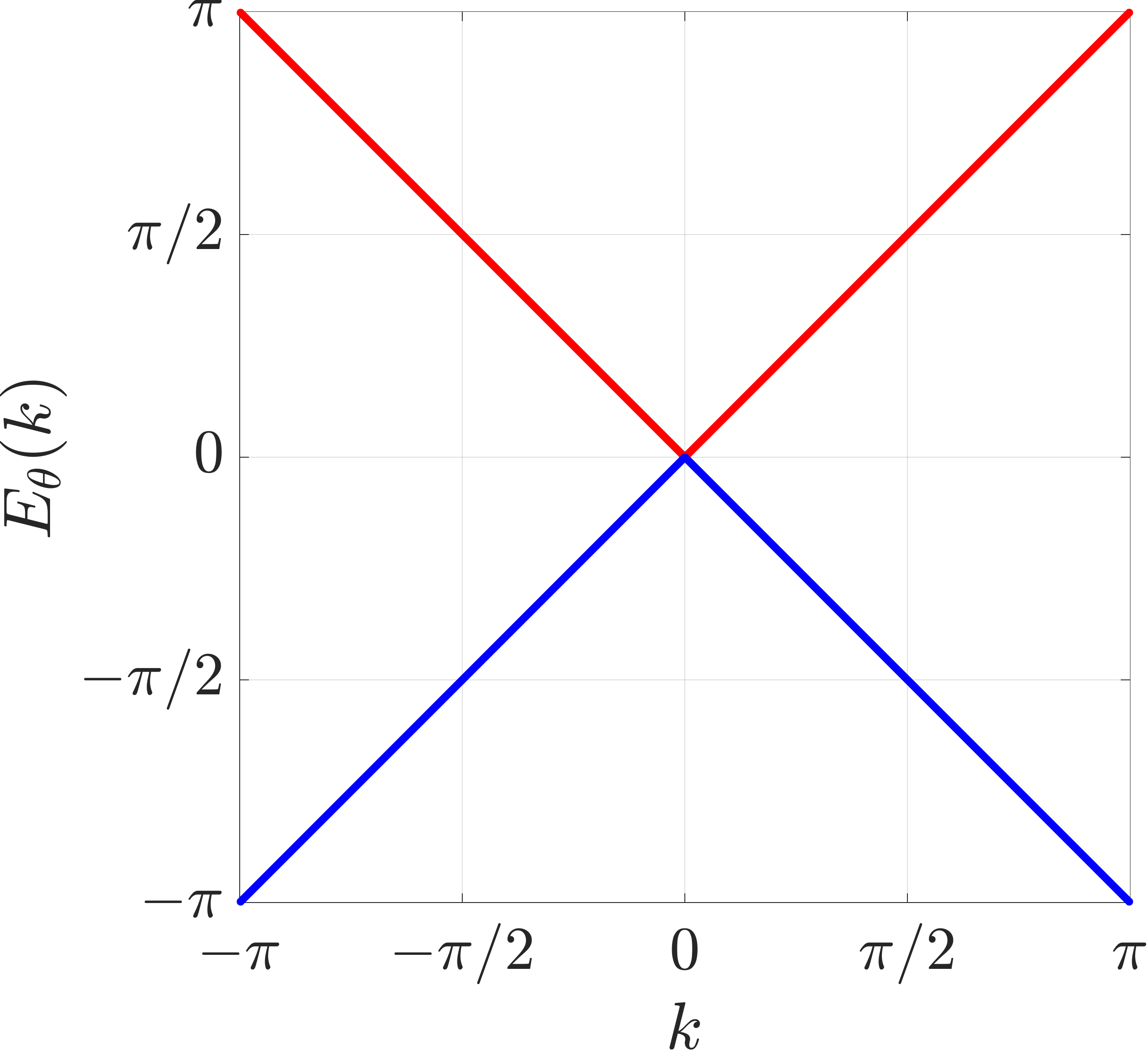}
		\label{fig:1DEnergy}} \hspace{-1mm}%
	\subfigure[]{
		\includegraphics[width=5.1cm]{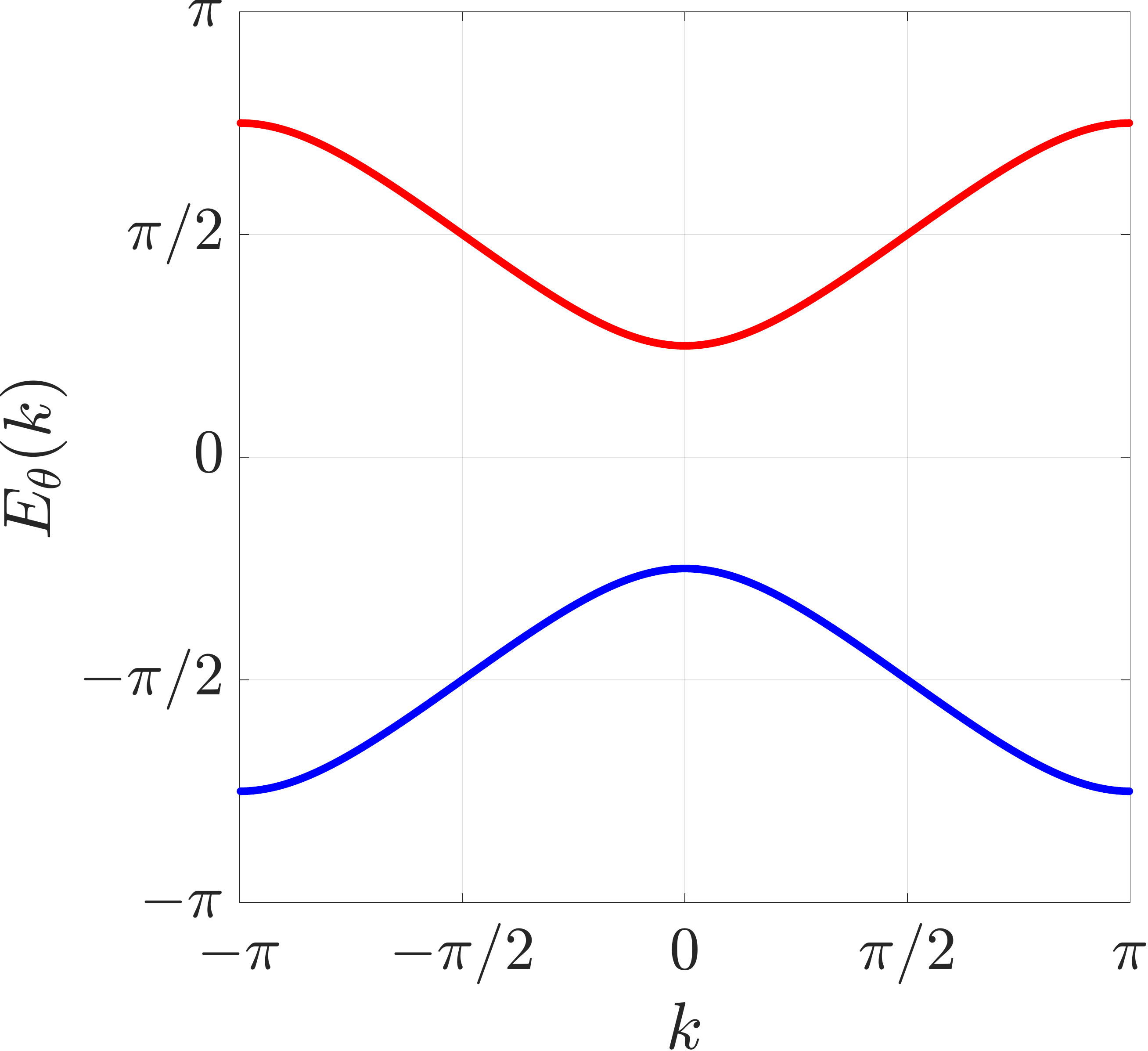}
		\label{fig:1DEnergy2}} \hspace{-1mm}%
	\subfigure[]{
		\includegraphics[width=5.1cm]{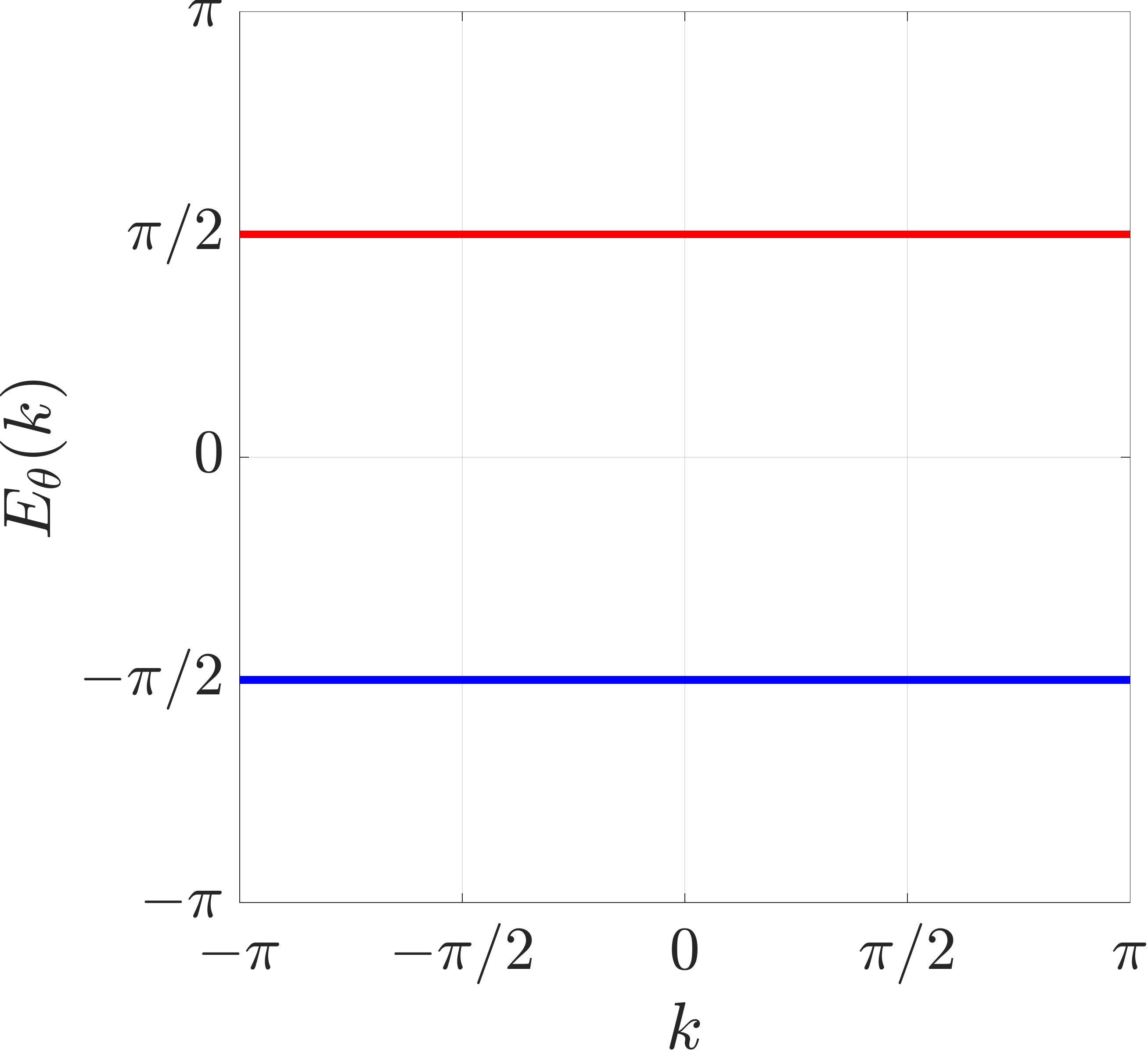}
		\label{fig:1DEnergy3}}
	\caption{The quasi-energy spectrum of the effective Hamiltonian $ H(\theta) $ for \subref{fig:1DEnergy} $\theta = 0$, \subref{fig:1DEnergy2} $\theta = \pi/2$ and \subref{fig:1DEnergy3} $\theta = \pi$.}
	\label{fig:1DSpectrum}
\end{figure}

The unitary evolution of the quantum walk governed by Eq.~\eqref{eq:unitaryequivalence} can also be generated by an underlying time independent Hamiltonian $H(\theta)$ ~\cite{Kitagawa2010} over a unit time such that
\begin{equation}
	U(\theta) = e^{-i H(\theta)}
\end{equation}
For simplicity, we have assumed $\hbar = 1$ and the periodic boundary condition with $N$ number of lattice sites. Since the unitary operator $U(\theta)$ and the Hamiltonian are translationally invariant, these will be reduced to block diagonal form in (quasi) momentum or Fourier basis $\{\ket{k}\}$ which are defined as
\begin{equation} \label{eq:FourierMode}
	\ket{k} = \dfrac{1}{\sqrt{N}} \sum_{n}  e^{-i 2\pi k n/N} \ket{n},
\end{equation}
with $ k $ being the quasi-momentum that can take discrete values between $-\pi$ and $\pi$ in integer multiples of $ 2 \pi/N $.
Hence, in the momentum basis $\ket{k} \otimes \ket{\sigma} = \dfrac{1}{\sqrt{N}} \sum_{n}  e^{-i k n} \ket{n} \otimes \ket{\sigma}$, translational operator in \eqref{eq:1D-Translation} transforms as 
\begin{equation}
	T(k) = \sum_{k} e^{i k \sigma_z} \otimes \dyad{k}{k}
\end{equation}
however $ R(\theta) $ remains unchanged. Using these, we can write the evolution operator $ U(\theta) $ quasi-momentum basis as
\begin{equation} \label{eq: 1DUnitary k}
	\begin{aligned}
		U(\theta) &= T R(\theta) \\ 
		&= \sum_{k} e^{i k \sigma_z} R(\theta) \otimes \ket{k}\bra{k}, \\ 
	\end{aligned}
\end{equation}
We can find the effective Hamiltonian by comparing \eqref{eq: 1DUnitary k} with the most general unitary for a two-level system given by
\begin{equation}
	U = \sum_{k} e^{-i \vb{n}(k) \vdot \boldsymbol{\sigma} } \otimes \dyad{k}
\end{equation} 
which represents the most general Hamiltonian for the two-level system in block-diagonal form. Resultantly, the effective Hamiltonian $H(\theta)$ in the quasi-momentum space reads~\cite{Kitagawa2010}
\begin{equation} \label{eq:1D-Hamil}
	H(\theta) = \sum_{k} [E_{\theta}(k)\, \vb{n}_{\theta}(k)\vdot \boldsymbol{\sigma}] \otimes \dyad{k}, 
\end{equation}
where the quasienergy $E_{\theta}(k)$ and the unit Bloch vector $\vb{n}_{\theta}(k)$ read
\begin{equation}\label{eq:1D-Energy}
	E_{\theta}(k) = \cos^{-1} \left[\cos (\theta/2) \cos k\right]
\end{equation} 
and
\begin{equation}\label{eq:1D-BlochVector}
	\vb{n}_{\theta}(k) = \dfrac{1}{\sin E_{\theta}(k)} (\sin(\theta/2) \sin k, \sin(\theta/2) \cos k, -\cos(\theta/2) \sin k).
\end{equation}
We have plotted the quasi-energy spectrum in Fig. \eqref{fig:1DSpectrum} for $\theta = 0$, $\theta = \pi/2$ and $\theta = \pi$. There are two bands due to the two internal states, and there exists a gap between them. We see that the gap closes for $\theta = 0$ at $ E = 0 $ and at $ E = \pi$.
Note that the quasienergy has a periodicity of $2\pi$. For $\theta = 0, 2\pi$, the quasienergy and the Bloch vector are ill defined, and the spectrum becomes gapless.  

\section{1D Split-Step Quantum Walk (SSQW)}
\begin{figure}[H]
	\centering
	\subfigure[]{
		\includegraphics[width=5cm]{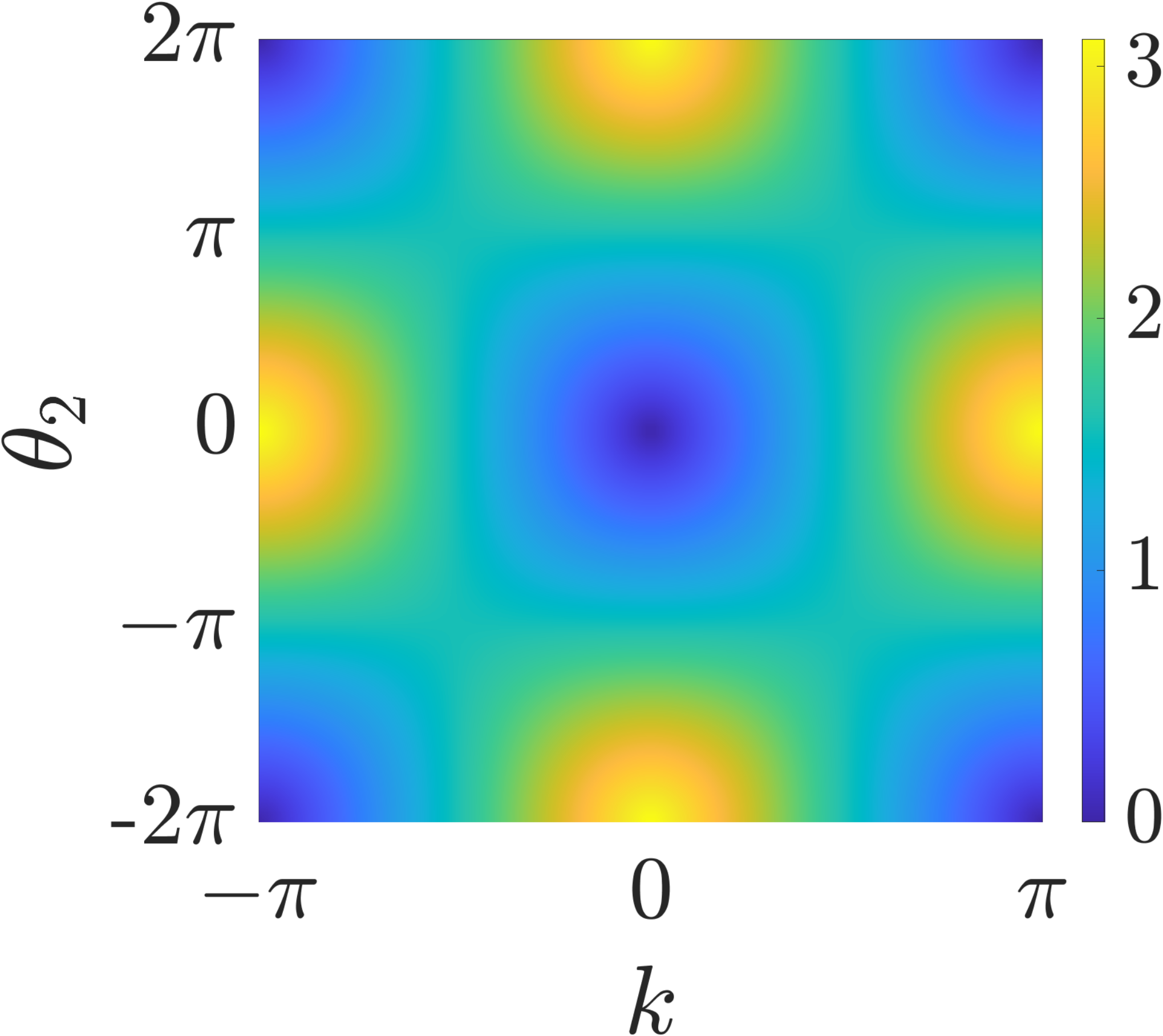}
		\label{fig:1DSSQWDis0}}
	\subfigure[]{
		\includegraphics[width=5cm]{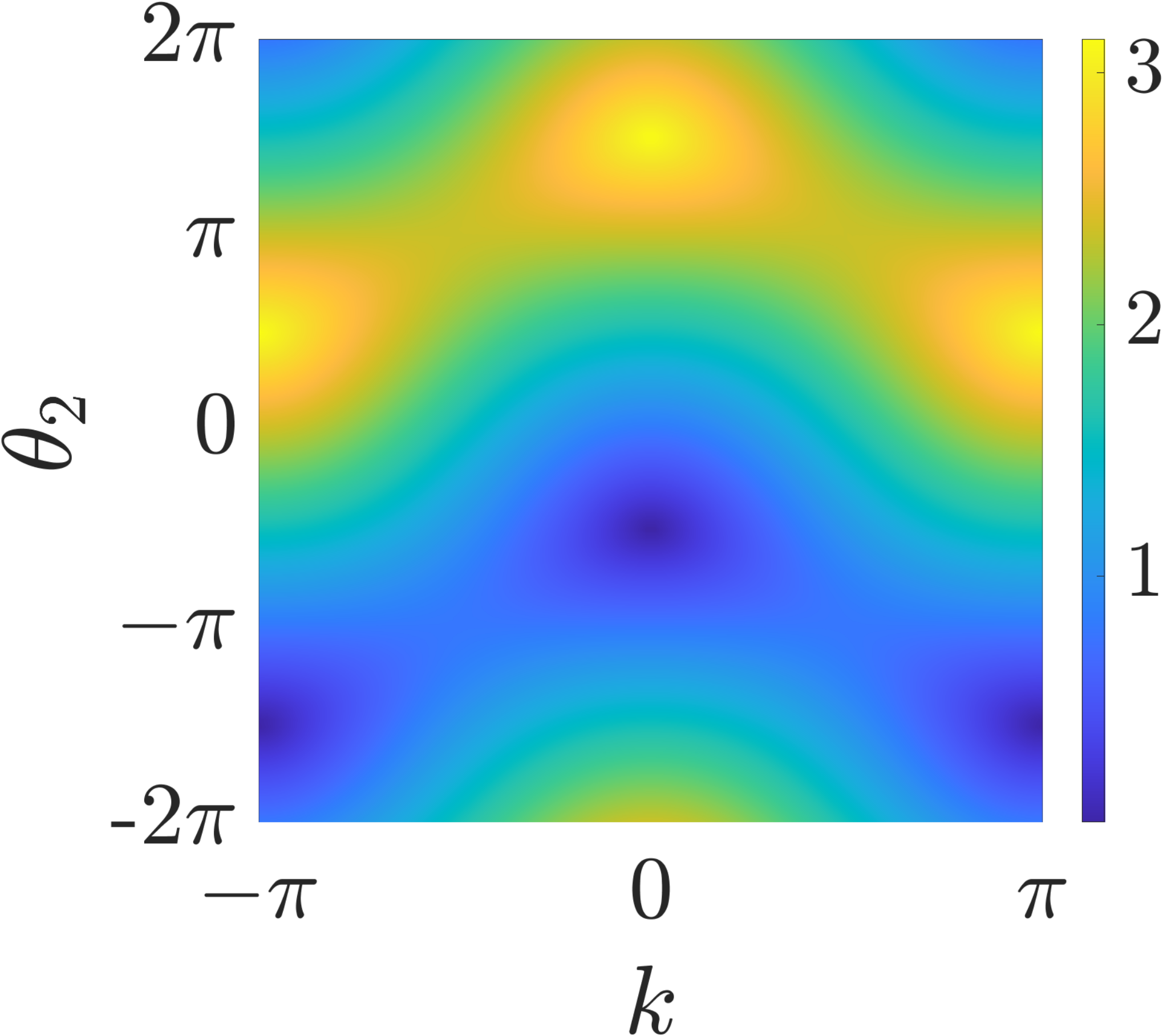}
		\label{fig:1DSSQWDis1}}
	\subfigure[]{
		\includegraphics[width=5cm]{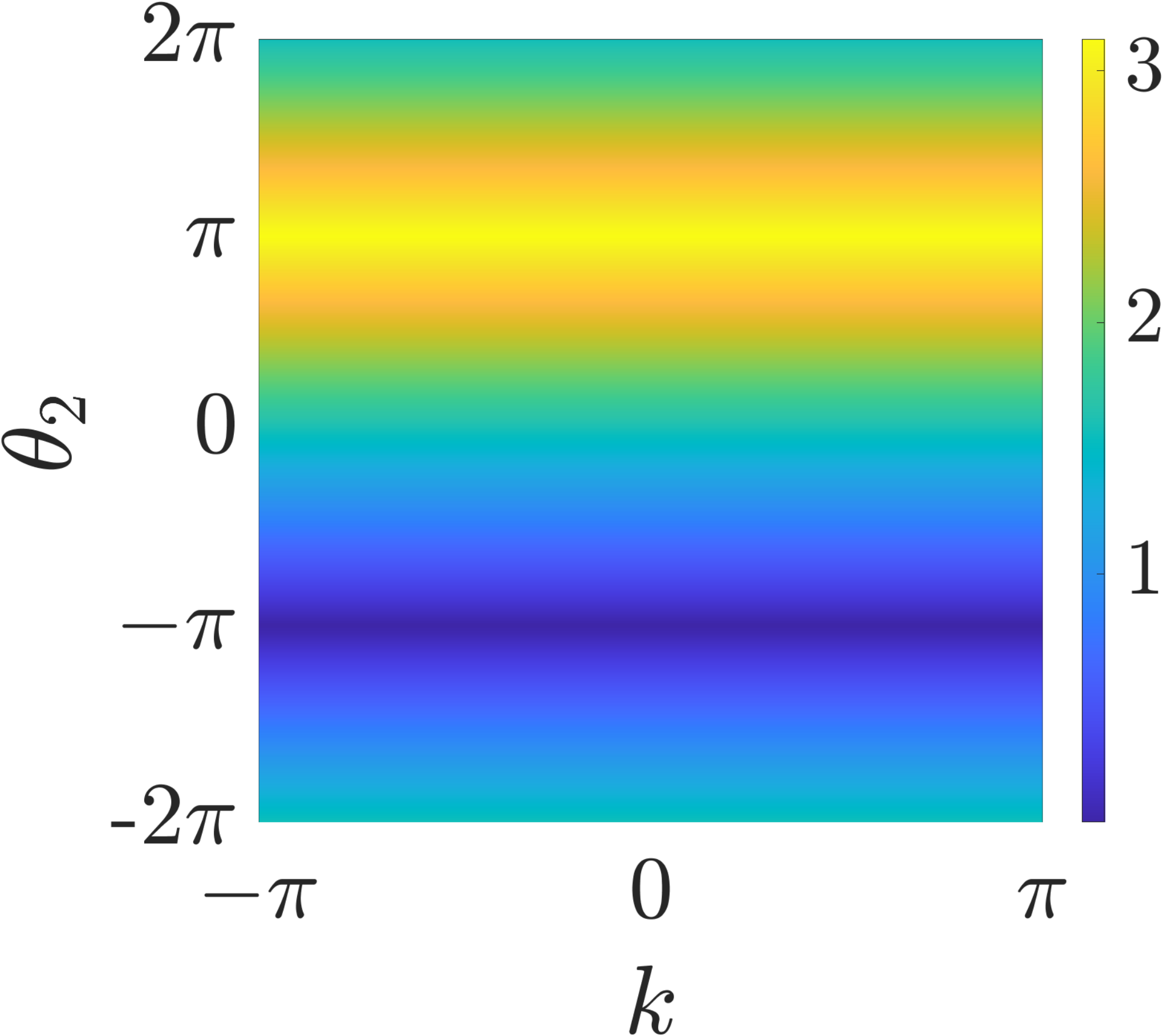}
		\label{fig:1DSSQWDis2}}
	\caption{The dispersion relation of the 1D SSQW given in Eq.~\eqref{eq:SSQW-Energy} as function of rotation angle $\theta_2$ and the quasi-momenum $k$ for \subref{fig:1DSSQWDis0} $\theta_1=0$, \subref{fig:1DSSQWDis0} $\theta_1=\pi/2$, and \subref{fig:1DSSQWDis0} $\theta_1=\pi$.} 
	\label{fig:1DSSQWDispersion}
\end{figure}

\begin{figure}[H]
	\centering
	\subfigure[]{
		\includegraphics[height=6cm]{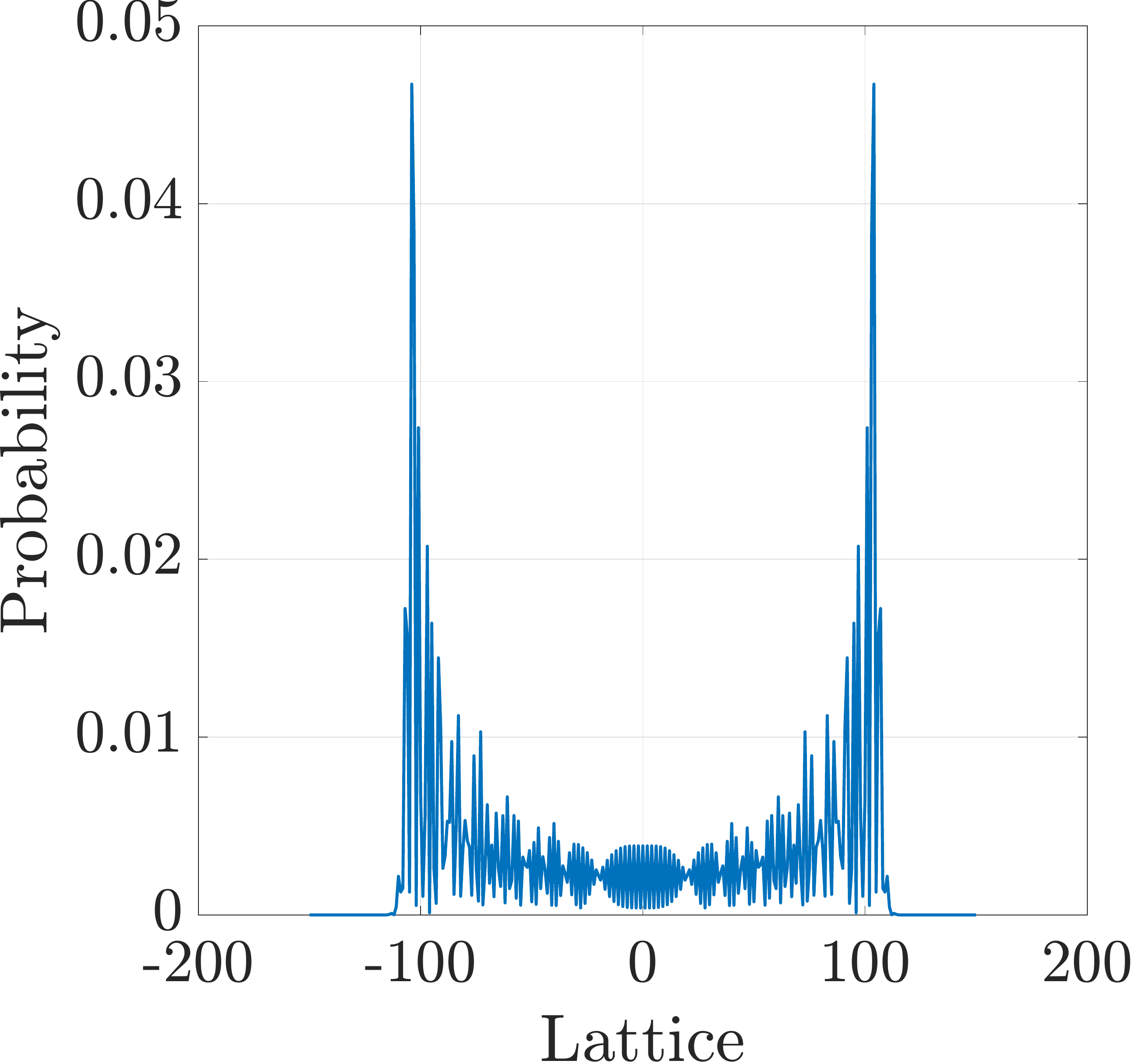}
		\label{fig:1DSSQWPD1}}
	\subfigure[]{
		\includegraphics[height=6cm]{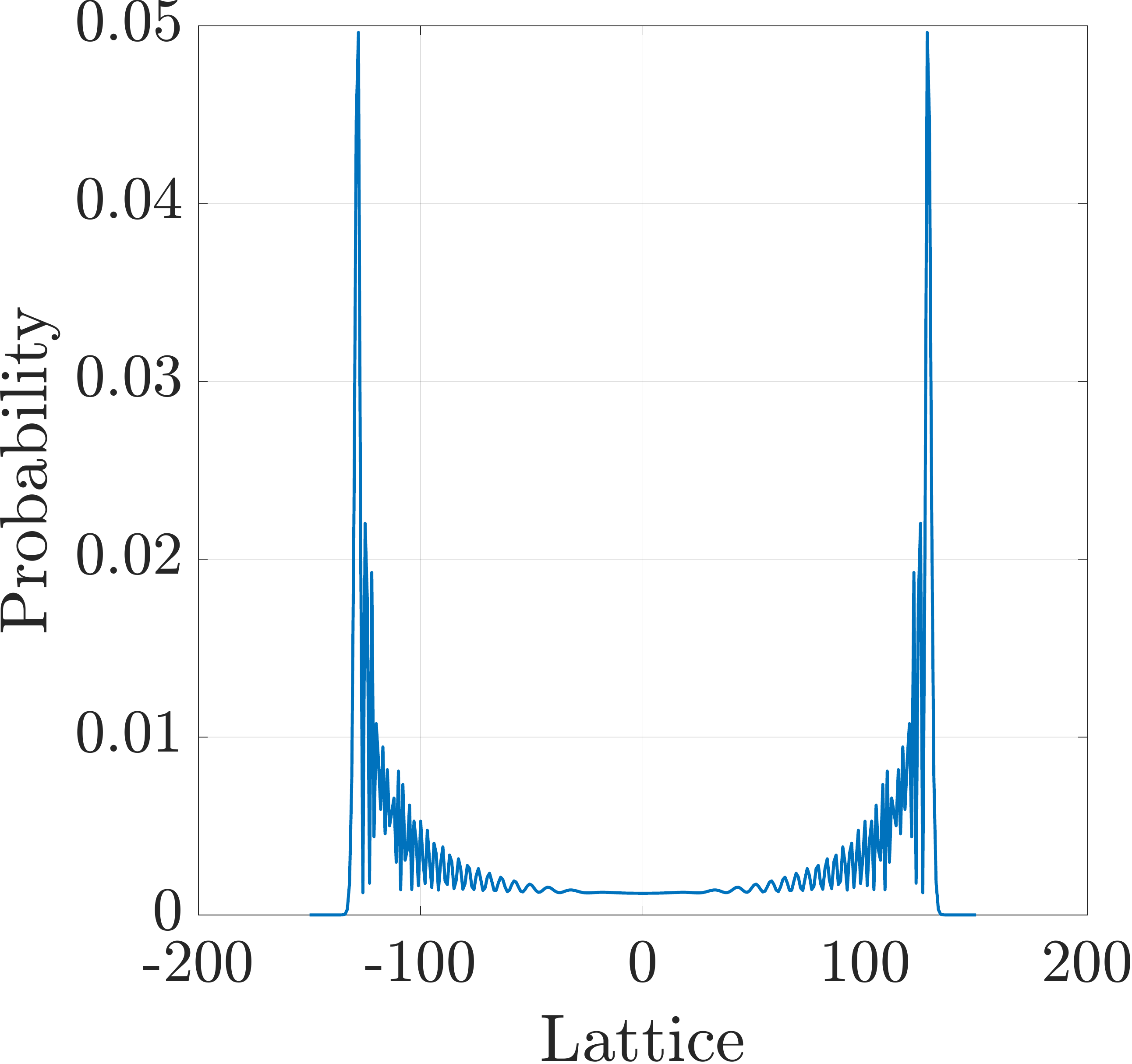}
		\label{fig:1DSSQWPD2}}
	\subfigure[]{
		\includegraphics[height=6cm]{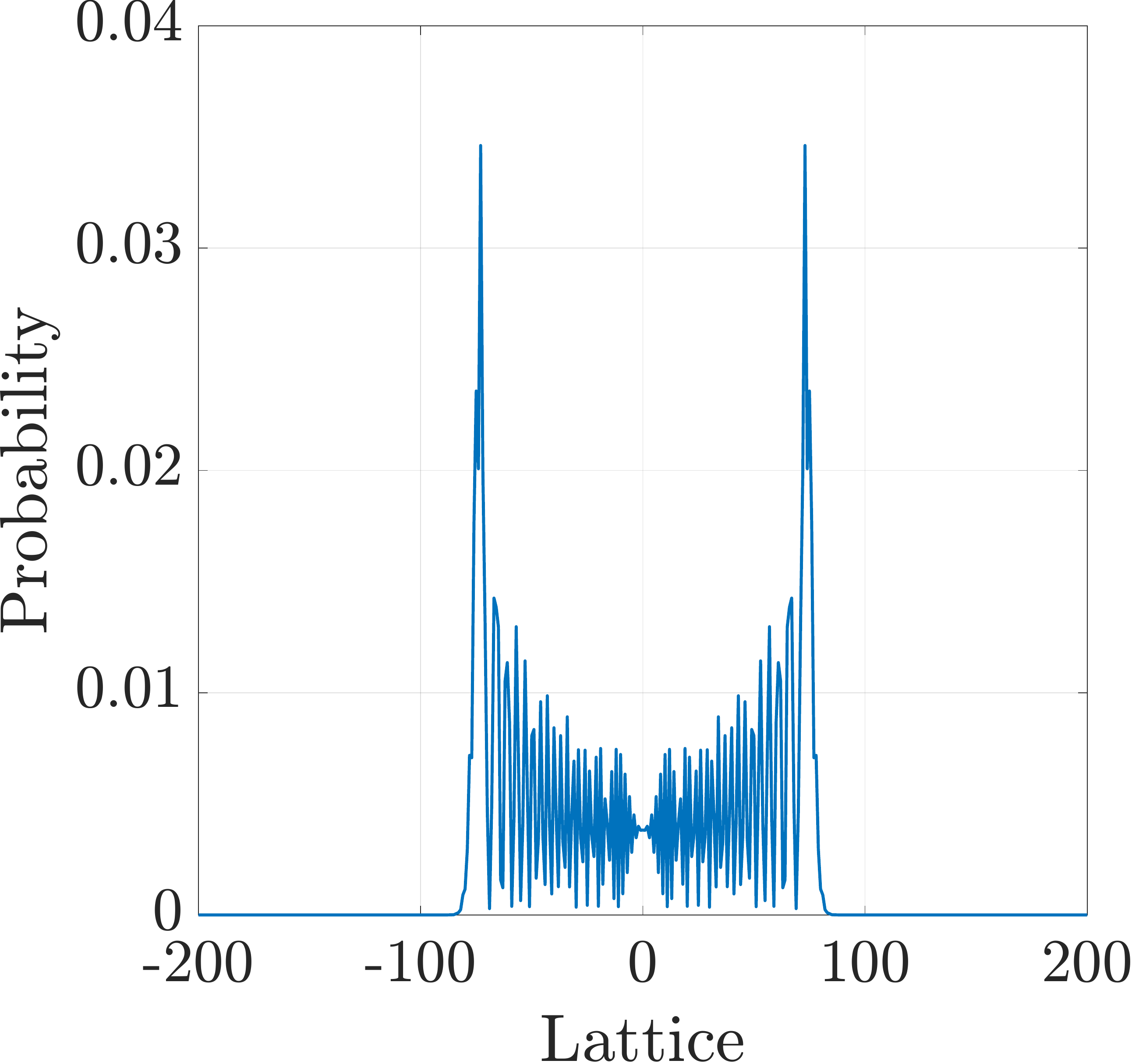}
		\label{fig:1DSSQWPD3}}
	\subfigure[]{
		\includegraphics[height=6cm]{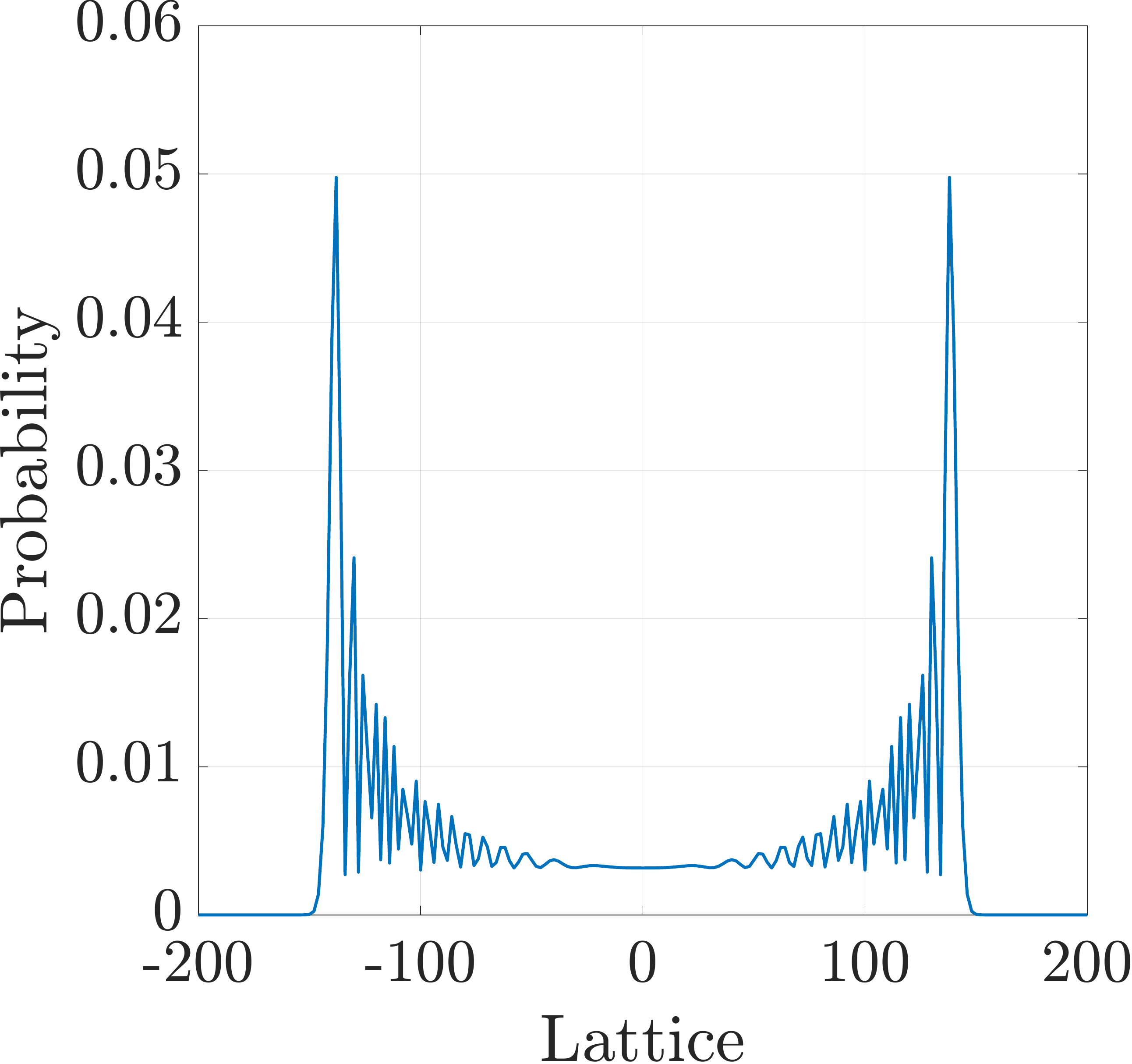}
		\label{fig:1DSSQWPD4}}
	\caption{Probability distribution of a walker after 200 time steps for
		\subref{fig:1DSSQWPD1} $(\theta_1, \theta_2) = (\pi/2, \pi/3)$, 
		\subref{fig:1DSSQWPD2} $(\theta_1, \theta_2) = (\pi/3, \pi/3)$, 
		\subref{fig:1DSSQWPD3} $(\theta_1, \theta_2) = (-\pi/2, 3\pi/4)$, 
		\subref{fig:1DSSQWPD3} $(\theta_1, \theta_2) = (\pi/2, 0)$. All the plots are symmetric due to the choice of symmetric initial state $ \ket{\psi(0)} = \ket{0} \otimes (\ket{\uparrow} + i \ket{\downarrow})/\sqrt{2} $. In \subref{fig:1DSSQWPD4} we recovered the probability distribution of 1D DTQW by putting $\theta_2 = 0$. } 
	\label{fig:1DSSQWPD}
\end{figure}

A more enriched class of 1D DTQW is SSQW, which involves splitting the conditional shift operator $T$ into left-shift ($ T_{\downarrow} $) and right-shift ($ T_{\uparrow} $) operators, separated by an additional coin toss $R(\theta_2)$~\cite{Kitagawa2010}. The resultant time evolution operator for split-step quantum walks (in one dimension) reads
\begin{equation} \label{eq:SSQW-Unitary}
	U_{_{\text{SS}}}(\theta_1, \theta_2) = T_{\downarrow} R(\theta_2) T_{\uparrow} R(\theta_1),
\end{equation}
where
\begin{align*}
	T_{\downarrow} &= \sum_n \ket{\uparrow} \bra{\uparrow} \otimes \mathds{1}_{2j+1}  + \ket{\downarrow} \bra{\downarrow} \otimes \ket{n-1} \bra{n}, \\
	T_{\uparrow} &= \sum_n \ket{\uparrow} \bra{\uparrow} \otimes \ket{n+1} \bra{n}  + \ket{\downarrow} \bra{\downarrow} \otimes \mathds{1}_{2j+1}.
\end{align*}
We plot the probability distribution of the walker after 200 steps in Fig.~\ref{fig:1DSSQWPD}. It is evident from Fig.~\ref{fig:1DSSQWPD4} that for $\theta_2 = 0$, SSQW reduces to conventional DTQW which was explained in the previous section. As before, we can go to the Fourier basis using Eq. \eqref{eq:FourierMode} which yields
\begin{align*}
	T_{\downarrow}(k) &= \sum_{k} e^{ik (\sigma_z - \mathds{1})/2} \otimes \dyad{k}{k}, \\
	T_{\uparrow}(k) &= \sum_{k} e^{ik (\sigma_z + \mathds{1})/2} \otimes \dyad{k}{k}.
\end{align*}
Consequently, in this case, the effective Hamiltonian $H_{\text{SS}}(\theta_1, \theta_2)$ can be written down in quasi-momentum space as
\begin{equation} \label{eq:SSQW-Hamil}
	H_{_{\text{SS}}}(\theta_1, \theta_2) = \sum_{k} [E_{\theta_1, \theta_2}(k) \vb{n}_{\theta_1, \theta_2}(k)\vdot \boldsymbol{\sigma}] \otimes \dyad{k}.
\end{equation}
The quasi-energy [see Fig.~\ref{fig:1DSSQWDispersion}] and the components of the Bloch vector are given by 
\begin{align} \label{eq:SSQW-Energy}
	\cos E_{\theta_1, \theta_2}(k) &= \cos(\theta_1/2) \cos(\theta_2/2) \cos k -\sin(\theta_1/2) \sin(\theta_2/2),
\end{align}
and $\vb{n}_{\theta_1, \theta_2}(k) = n_x(k) \vb{\hat{i}} + n_y(k) \vb{\hat{j}} + n_z(k) \vb{\hat{k}}$ with
\begin{align} \label{eq:SSQW-BlochVector}
	n_x(k) &= \dfrac{\sin(\theta_1/2)\cos(\theta_2/2)\sin k}{\sin E_{\theta_1, \theta_2}(k)}, \nonumber \\
	n_y(k) &= \dfrac{\cos(\theta_1/2) \sin(\theta_2/2) +   \sin(\theta_1/2) \cos(\theta_2/2) \cos k }{\sin E_{\theta_1, \theta_2}(k)},\nonumber \\
	n_z(k) &= \dfrac{-\cos(\theta_1/2)\cos(\theta_2/2) \sin k}{\sin E_{\theta_1, \theta_2}(k)}.
\end{align}
1D SSQW possess a rich topological structure depending on the symmetries in the system. We will discuss symmetries and topological characterization in quantum walks in detail in upcoming chapters.

\section{2D DTQW}
\begin{figure}[H]
	\centering
	\includegraphics[width=11cm]{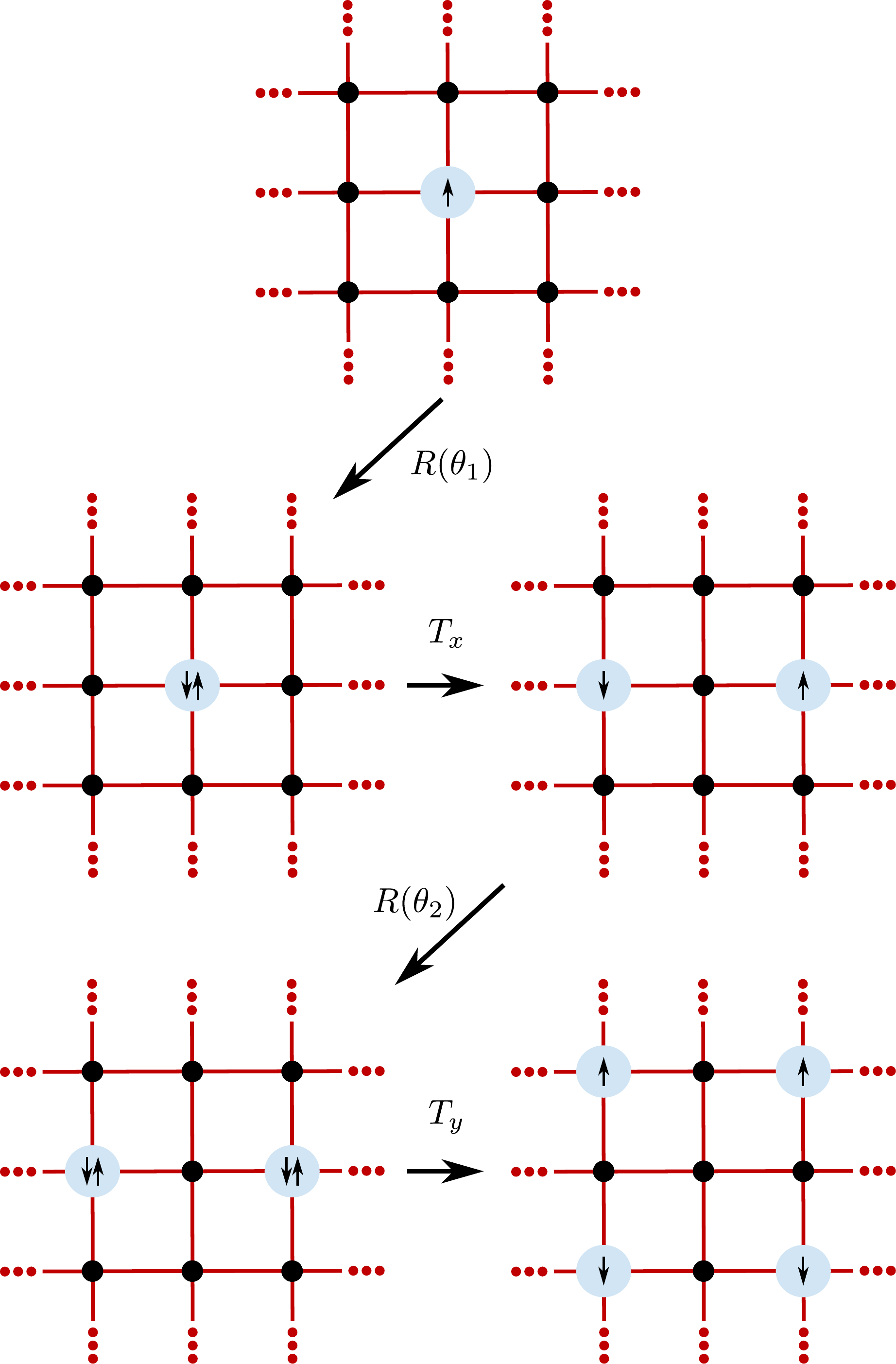}
	\caption{Schematic of the protocol of 2D DTQW. The system consists of 2D lattice and site index runs from $ -n $ to $ +n $ in both the directions.}
	\label{fig:2DDTQW}
\end{figure}
\begin{figure}
	\centering
	\includegraphics[width=9cm]{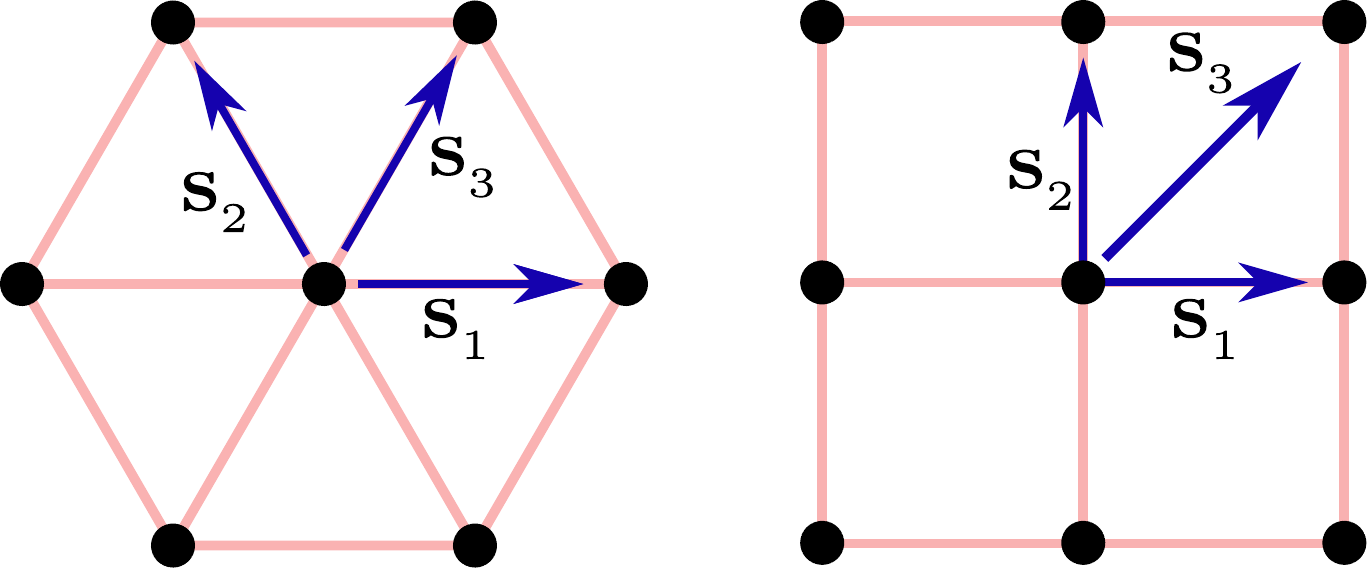}
	\caption{(Color online) 2D DTQW with nontrivial topology on a triangular lattice and its equivalent square lattice.}
	\label{fig:tri}
\end{figure}
\begin{figure}
	\centering
	\subfigure[]{
		\includegraphics[width=5.1cm]{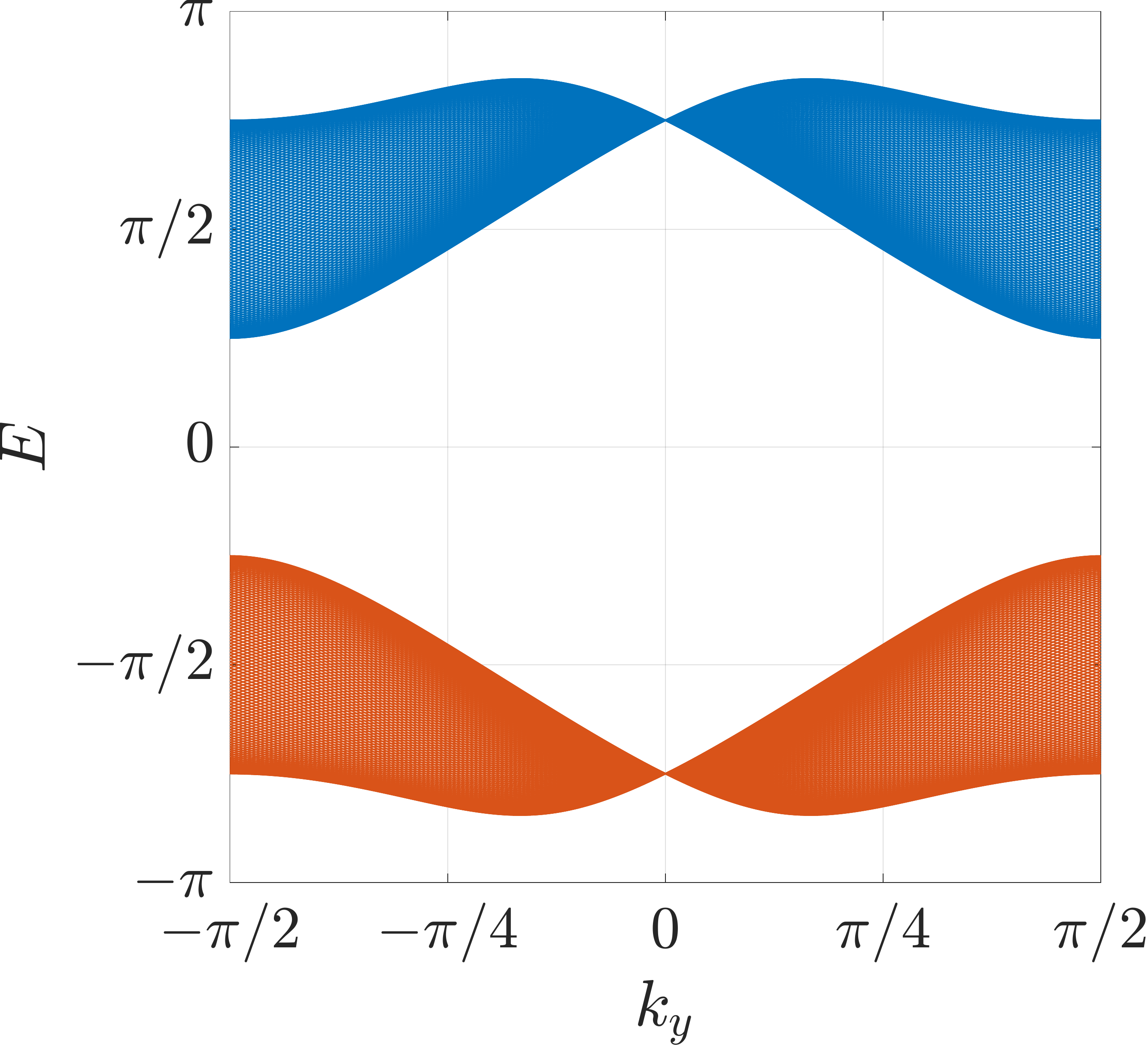}
		\label{fig:2DEnergy}} \hspace{-1mm}%
	\subfigure[]{
		\includegraphics[width=5.1cm]{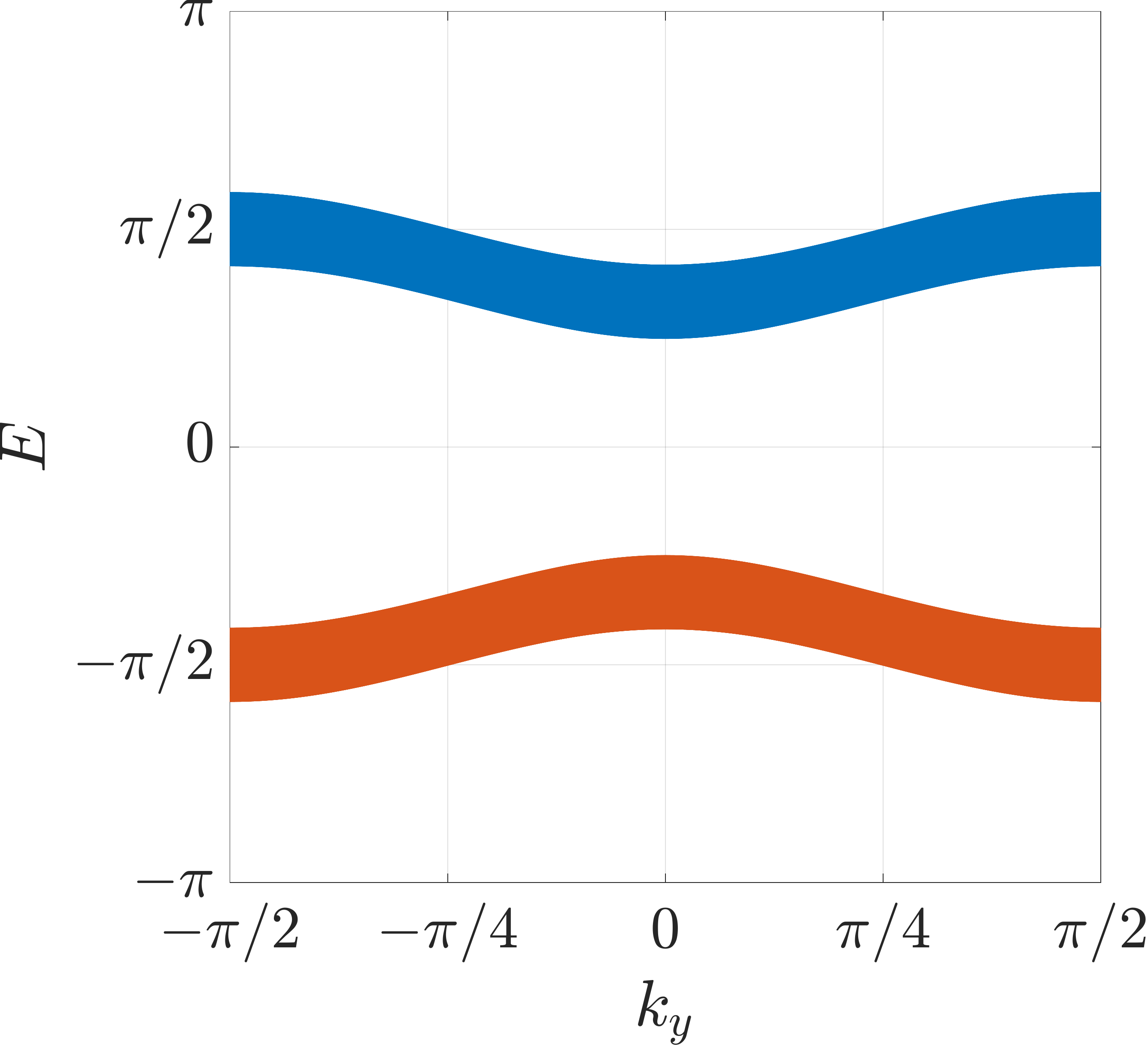}
		\label{fig:2DEnergy2}} \hspace{-1mm}%
	\subfigure[]{
		\includegraphics[width=5.1cm]{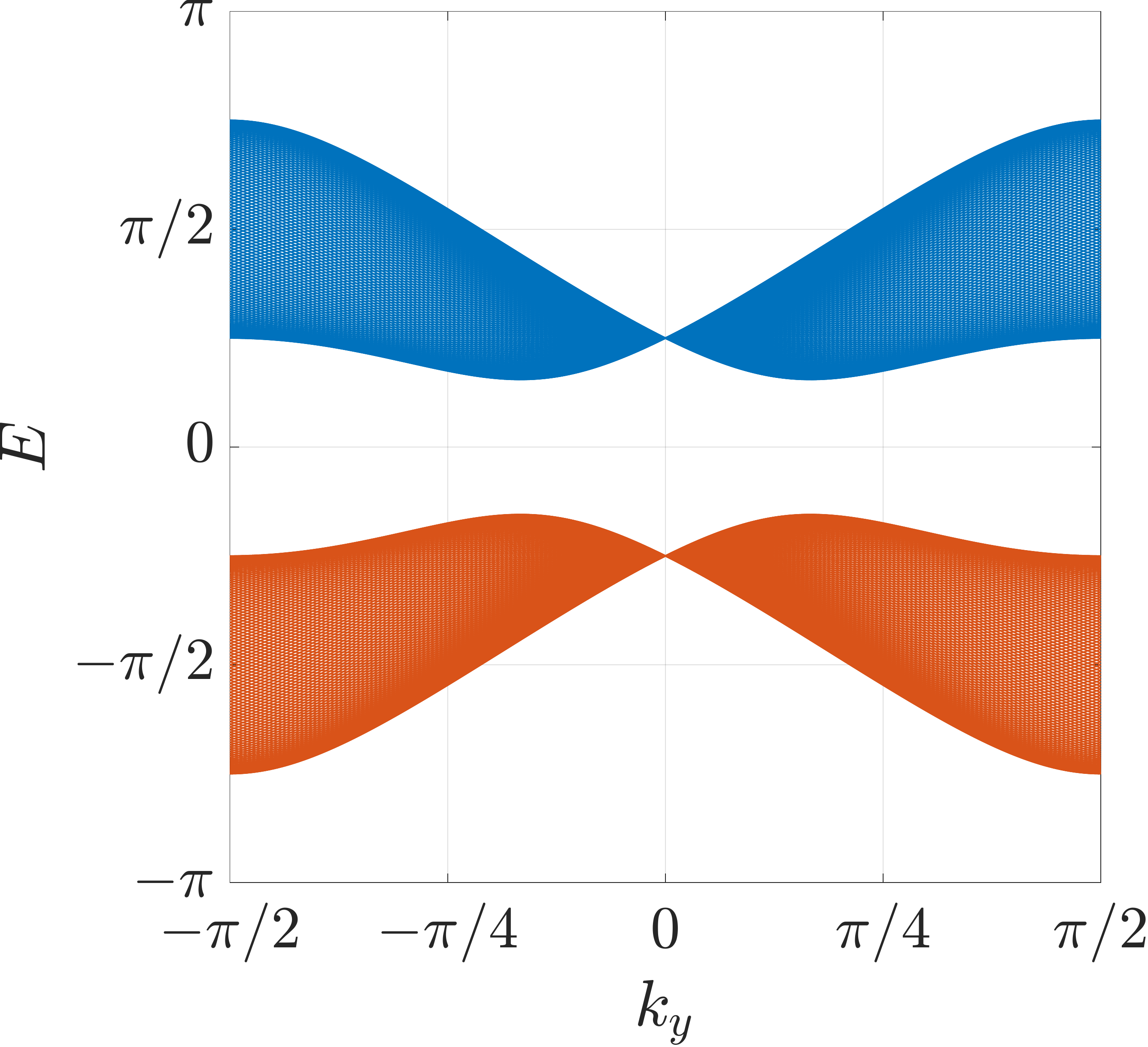}
		\label{fig:2DEnergy3}}
	\caption{The quasi-energy band structure of the effective Hamiltonian $ H_{2D}(\theta_1, \theta_2) $ for \subref{fig:2DEnergy} $\theta_1 = \pi/2 = \theta_2$, \subref{fig:2DEnergy2} $\theta_1 = 7\pi/6 = \theta_2$ and \subref{fig:2DEnergy3} $\theta_1 = 3\pi/2 = \theta_2$.}
	\label{fig:2DSpectrum}
\end{figure}
\begin{figure}
	\centering
	\subfigure[]{
		\includegraphics[width=6cm]{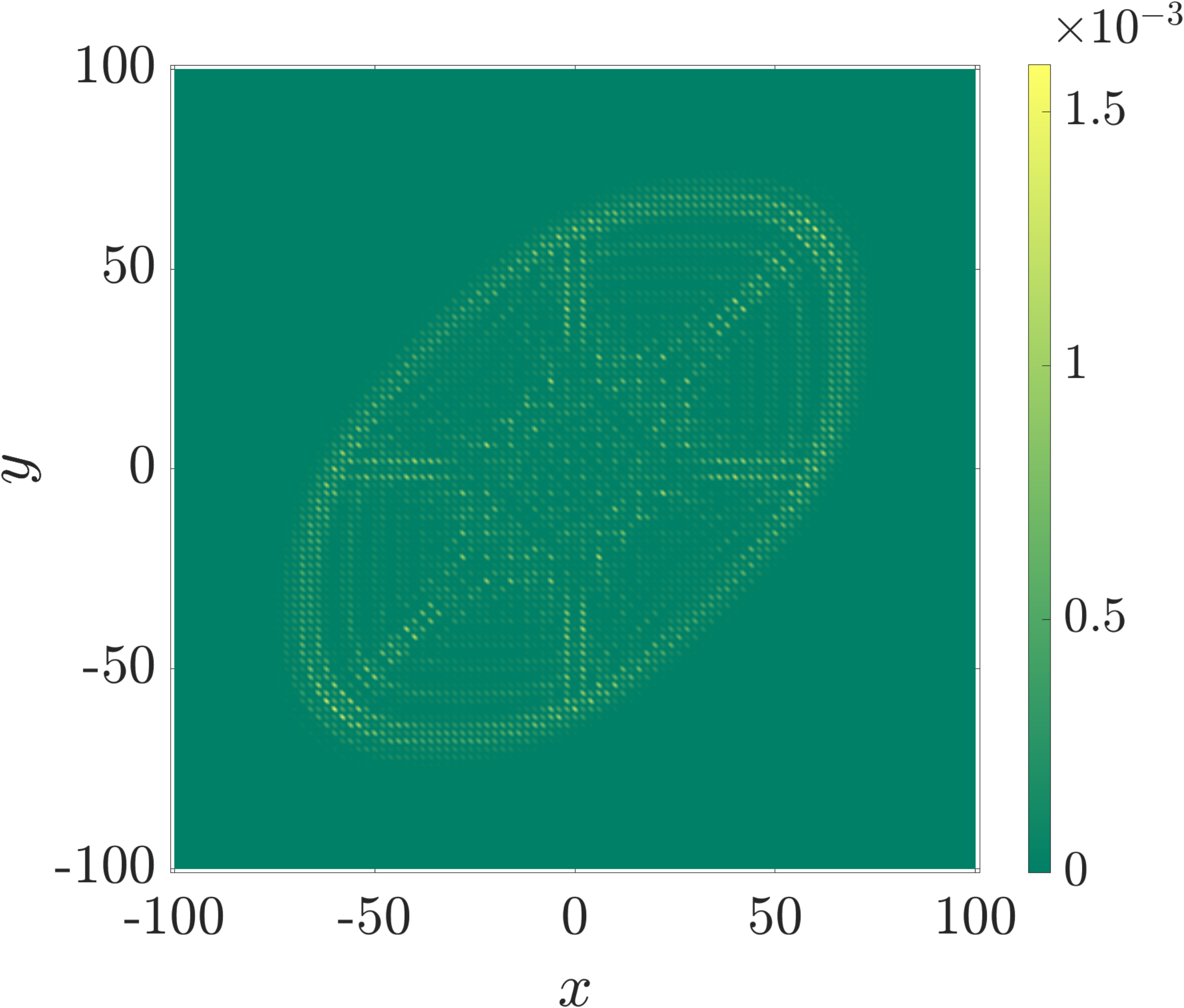}
		\label{fig:2DDTQWPD1}}
	\subfigure[]{
		\includegraphics[width=6cm]{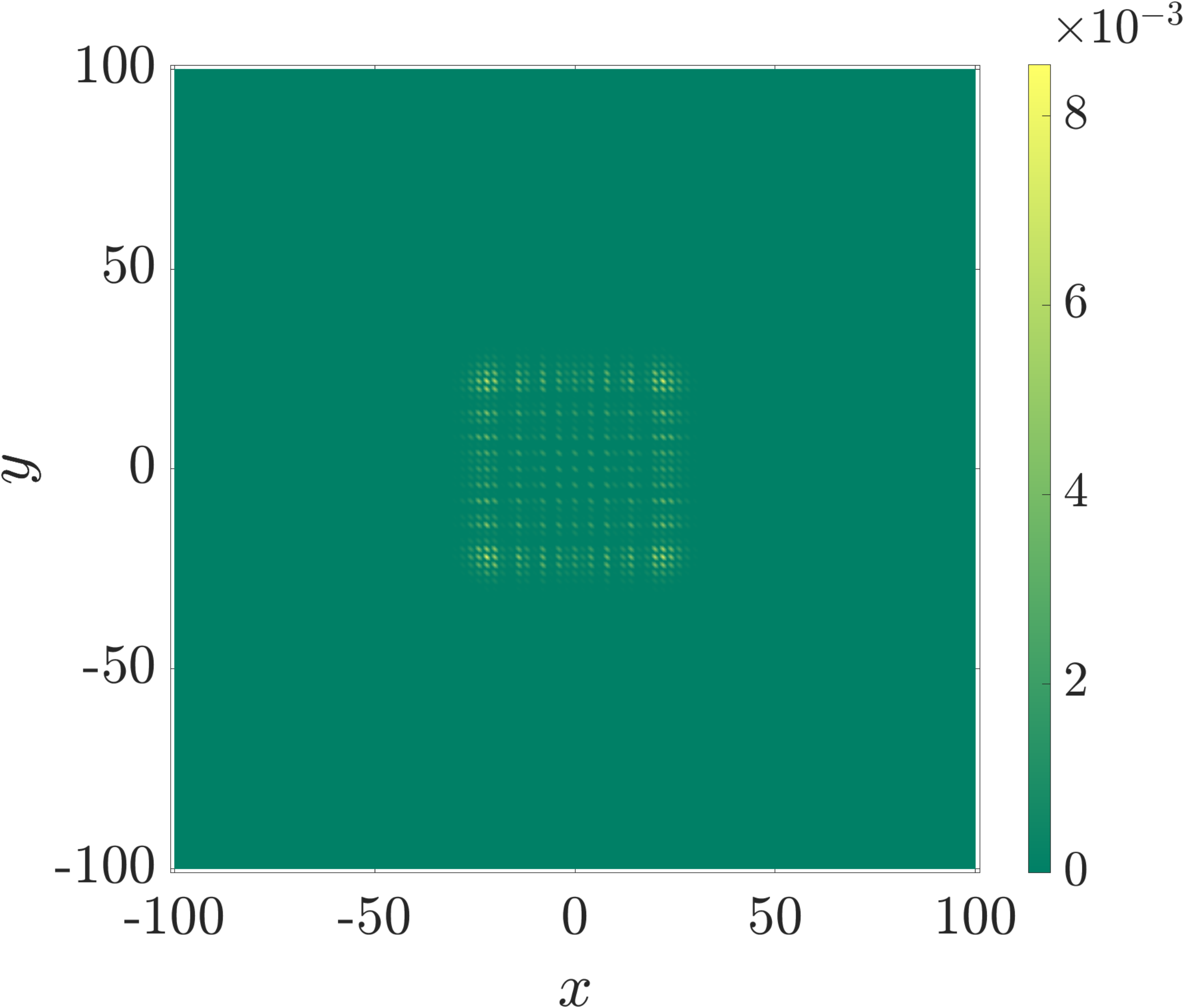}
		\label{fig:2DDTQWPD2}}
	\subfigure[]{
		\includegraphics[width=6cm]{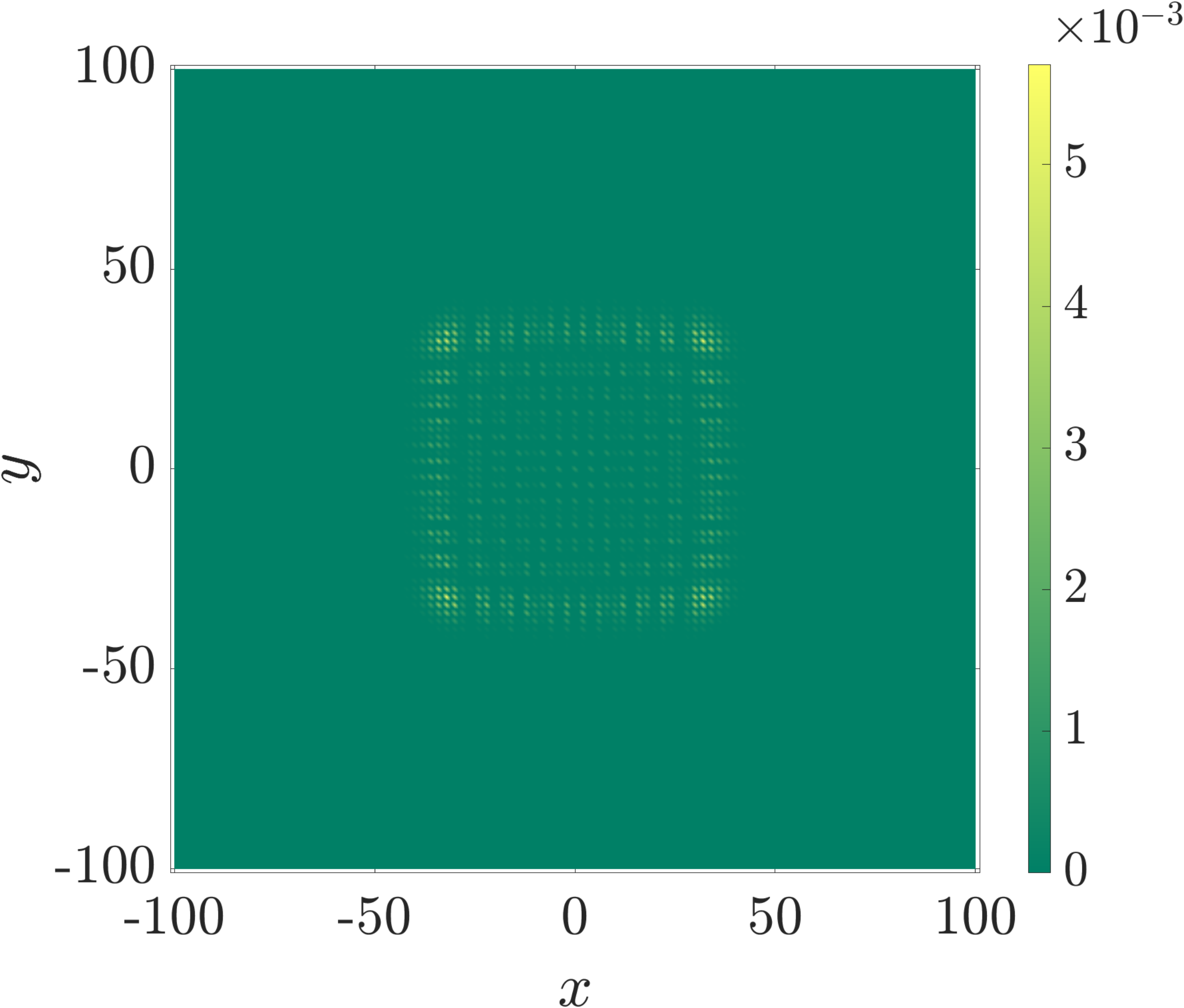}
		\label{fig:2DDTQWPD3}}
	\subfigure[]{
		\includegraphics[width=6cm]{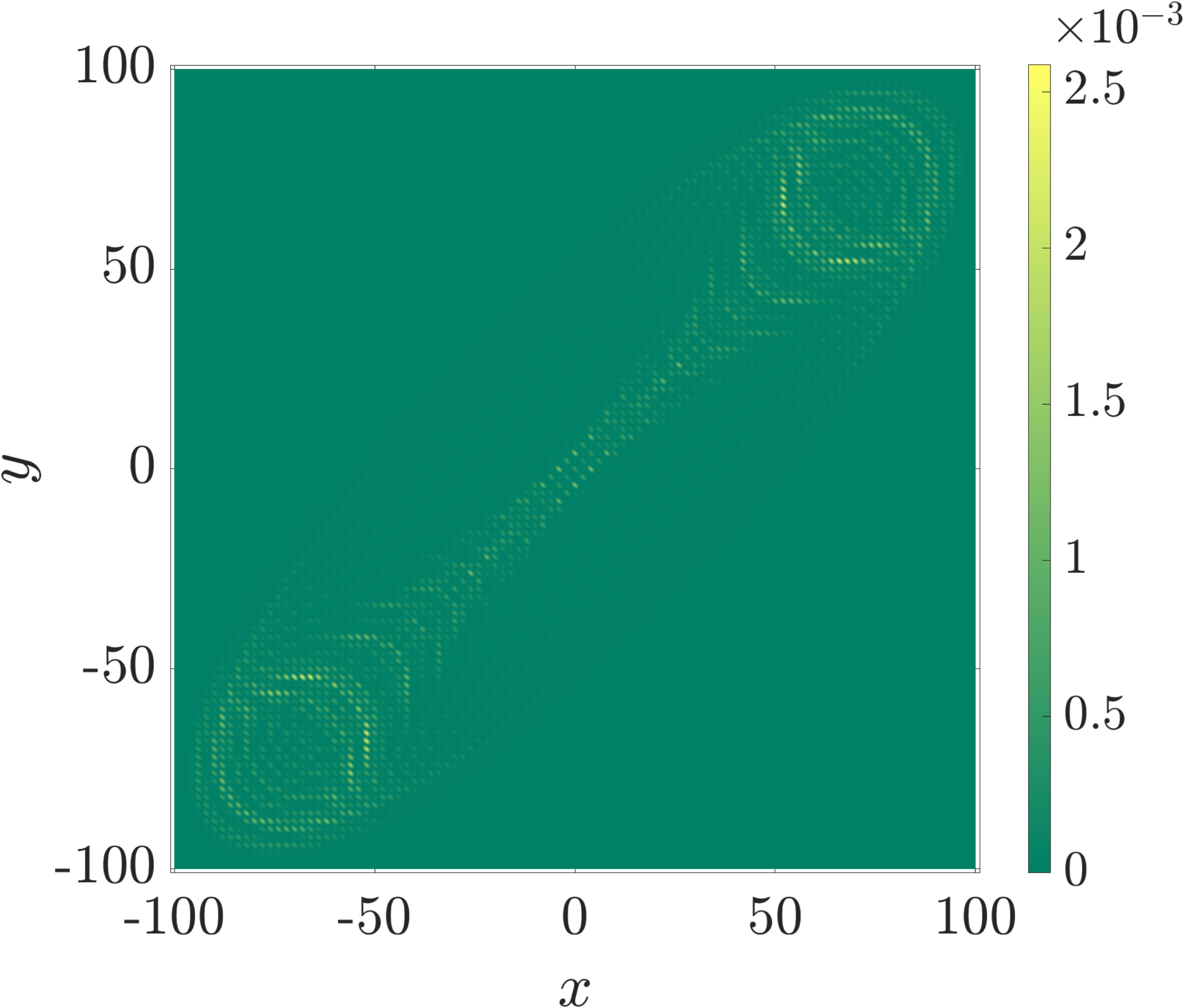}
		\label{fig:2DDTQWPD4}}
	\caption{The probability distribution of the 2D DTQW given in Eq.~\eqref{Eq:Qwalk2D} after 100 number of time steps for different values of $\theta_1$ and $\theta_2$. In \subref{fig:2DDTQWPD1} $\theta_1=\pi/2 = \theta_2$, \subref{fig:2DDTQWPD2} $\theta_1=7\pi/6 = \theta_2$, \subref{fig:2DDTQWPD3} $\theta_1= 3\pi/4 = \theta_2$ and \subref{fig:2DDTQWPD4} $\theta_1= -\pi/5, \theta_2 = 3\pi/7$. The system size is taken to be $201 \times 201$.} 
	\label{fig:2DDTQWPD}
\end{figure}

We can extend the idea of a quantum walk to higher dimensions as well. There are several ways of defining a 2D DTQW in a lattice. First, we start with a simple one as shown in Fig.~\ref{fig:2DDTQW}. It consists of the following operations 
\begin{enumerate}
	\item Rotations of the spin around $ y $ with an angle $\theta_1$ which is written as $ R(\theta_1) = e^{-i \sigma_y \theta_1/2} $,
	\item Translation operator $T_x$ in $ x $ direction which shifts the up spin to the right and spin down to the left. It is written as 
	\begin{equation}
		T_x = \sum_{x,y} \ket{\uparrow} \bra{\uparrow} \otimes \ket{x+1,y} \bra{x,y}  + \ket{\downarrow} \bra{\downarrow} \otimes \ket{x-1,y} \bra{x,y},
	\end{equation}
	\item Rotations of the spin around $ y $ with an angle $\theta_2$ which is written as $ R(\theta_2) = e^{-i \sigma_y \theta_2/2} $,
	\item Translation operator $T_y$ in $ y $ direction which shifts the up and down spin to up and down vertically respectively. It is written as 
	\begin{equation}
		T_y = \sum_{x,y} \ket{\uparrow} \bra{\uparrow} \otimes \ket{x,y+1} \bra{x,y}  + \ket{\downarrow} \bra{\downarrow} \otimes \ket{x,y-1} \bra{x,y},
	\end{equation}
\end{enumerate}
such that the time evolution operator for one step reads
\begin{equation} \label{eq:2D-Unitary}
	U_{_{2D}}(\theta_1, \theta_2) = T_y R(\theta_2) T_x R(\theta_1),
\end{equation}
Note that there are other ways too to define higher-dimensional quantum which consists of more internal degrees of freedom (for example, with a 4 dimensional coin). We restrict ourselves to the one where we use only two internal degrees of freedom. Now, we define an extended version of 2D DTQW on a triangular lattice, which consists of three spin-dependent translations separated by coin-flip operations. In that case, the unitary operator which governs the time evolution is written as \cite{Kitagawa2012}
\begin{equation} \label{eq:U2D}
	\tilde U_{_{2D}}(\theta_1, \theta_2) = T_{xy} R(\theta_1) T_y R(\theta_2) T_x R(\theta_1),
\end{equation}
where $ T_i (i=x,y,xy) $ are the translations along the $ \vb{s}_i  $ directions with $ T_{xy} = T_xT_y $, as shown in the Fig.\,\ref{fig:tri}. We can further derive another two-dimensional quantum walk that is unitarily equivalent to $\tilde U_{_{2D}}(\theta_1, \theta_2)$ as $\tilde U_{_{2D}} \to U_{_{2D}} = T_x^\dagger \tilde U_{_{2D}} T_x$. The resulting time-evolution unitary operator can be written as
\begin{equation} \label{Eq:Qwalk2D}
	U_{_{2D}}(\theta_1, \theta_2) = T_y R(\theta_1) T_y R(\theta_2) T_x R(\theta_1) T_x.
\end{equation}
The underlying Hamiltonian for this quantum walk (in quasi-momentum space) reads
\begin{equation} \label{eq:Hamil2D}
	H_{_{2D}}(\theta_1, \theta_2) = \sum_{k_x, k_y} E(k_x,k_y) \hat{\vb{n}}(k_x,k_y) \vdot \boldsymbol{\sigma} \otimes \dyad{k_x, k_y},
\end{equation}
\noindent where the expression of quasi-energy [see Fig.~\ref{fig:2DSpectrum}] reads
\begin{align}
	\cos E(k_x, k_y) &=  \cos \theta_1 \cos(\theta_2/2) \cos^2(k_x + k_y) -\cos(\theta_2/2) \sin^2(k_x + k_y) \nonumber\\
	&\;  - \sin \theta_1\sin(\theta_2/2)\cos(k_x + k_y)\cos(k_x - k_y)  ,
\end{align}
and the Bloch vector reads \cite{Kitagawa2010}
\begin{equation*}
	\hat{\vb{n}}(k_x,k_y) = \dfrac{n_x(k_x,k_y) \hat{\vb{i}} + n_y(k_x,k_y) \hat{\vb{j}} + n_z(k_x,k_y) \hat{\vb{k}}}{\sin E(k_x, k_y)},
\end{equation*}
with
\begin{align}
	n_x(k_x,k_y) = &- \sin \theta_1\cos(\theta_2/2) \cos(k_x + k_y) \sin (k_x - k_y) \nonumber \\
	&- \cos^2 \theta_1 \sin(\theta_2/2) \sin 2(k_x - k_y), \nonumber \\
	n_y(k_x,k_y) = &\sin \theta_1 \cos(\theta_2/2) \cos(k_x + k_y) \cos(k_x - k_y) \nonumber \\
	+ &\cos \theta_1  \cos^2(k_x - k_y) \sin(\theta_2/2) \nonumber \\ 
	- &\sin^2(k_x - k_y) \sin(\theta_2/2),\nonumber \\
	n_z(k_x,k_y) = &-\cos^2(\theta_1/2) \cos(\theta_2/2) \sin 2(k_x + k_y) \nonumber \\
	&+ \sin \theta_1\sin(\theta_2/2)\sin(k_x + k_y)\cos(k_x - k_y) \nonumber.
\end{align}
In this case, each step of the quantum walk moves the walker from the even (odd) site to the even (odd) site, so the lattice constant of the effective Hamiltonian is 2. Thus, the first Brillouin zone is defined as $ -\pi/2 \le k_x \le \pi/2$ and $ -\pi/2 \le k_y \le \pi/2$. We plot the probability distribution of a particle initially localized at the origin $(0,0)$ in Fig.~\ref{fig:2DDTQWPD} for different settings of $\theta_1$ and $\theta_2$. The coin starts in the symmetric state and the system size is taken to be $201 \times 201$. We get a symmetric probability distribution as a consequence of the initial symmetric state of the coin. 

\subsection{2D DTQW with 4D Coin}
For our purpose, to study the persistence of topological order in quantum walks, we will consider 2D DTQW with 2D coin only. However, there are protocols in the literature \cite{Mackay2002,Tregenna2003,Donangelo2006} where one can have a 4D coin. For the sake of completeness, we briefly discuss a simple protocol of 2D DTQW using a 4D coin. Ror such quantum walks, the time evolution operator for one step is given by
\begin{equation}
	U = T (C \otimes \mathds{1})
\end{equation}
with
\begin{align}
	T = &\sum_{x,y} \dyad{\uparrow} \otimes \dyad{x+1,y+1}{x,y}  + \dyad{\uparrow}{\downarrow} \otimes \dyad{x+1,y-1}{x,y} \nonumber \\
	&\;\; + \dyad{\downarrow}{\uparrow} \otimes \dyad{x-1,y+1}{x,y} + \dyad{\downarrow} \otimes \dyad{x-1,y-1}{x,y}
\end{align}
and the $C$ operator can be one of the following: \\
1. Hadamard coin: This is one of the most general extensions of the 1D Hadamard transformation given as
\begin{align} \label{eq:HadamardCoin}
	C &= H_2 = H \otimes H = \dfrac{1}{\sqrt{2}} \begin{bmatrix}
		1 & 1 \\
		1 & -1
	\end{bmatrix} \otimes \dfrac{1}{\sqrt{2}} \begin{bmatrix}
		1 & 1 \\
		1 & -1
	\end{bmatrix} = \dfrac{1}{2} \begin{bmatrix}
		1 & 1 & 1 & 1 \\
		1 & -1 & 1 & -1 \\
		1 & 1 & -1 & -1 \\
		1 & -1 & -1 & 1 
	\end{bmatrix}
\end{align}
and in $d$-dimensional quantum walks, it takes a very simple separable form, which reads $H_d = H \otimes H \otimes \dots \otimes H$. 

\noindent 2. Fourier coin: Another generalization is the $2^d$-dimensional discrete Fourier transform $F_d$. Given the basis states $\{\ket{u}, i = 0,1,2, \dots, 2^d - 1\}$, the action of $F_d$ on the bases states reads \cite{Mackay2002} 
\begin{equation}
	F_d \ket{i} =  \dfrac{1}{\sqrt{2^d}}\sum_{v = 0}^{2^d - 1} e^{2 \pi i u v /2^d} \ket{v}.
\end{equation}
For $d=1$, the discrete Fourier transform $F_1$ is just Hadamard. For $d=2$, it reads 
\begin{align}\label{eq:FourierCoin}
	C &= F_2 =\dfrac{1}{2} \begin{bmatrix}
		1 & 1 & 1 & 1 \\
		1 & i & -1 & -i \\
		1 & -1 & 1 & -1 \\
		1 & -i & -1 & i 
	\end{bmatrix}
\end{align}

\begin{figure}[H]
	\centering
	\subfigure[]{
		\includegraphics[width=7.1cm]{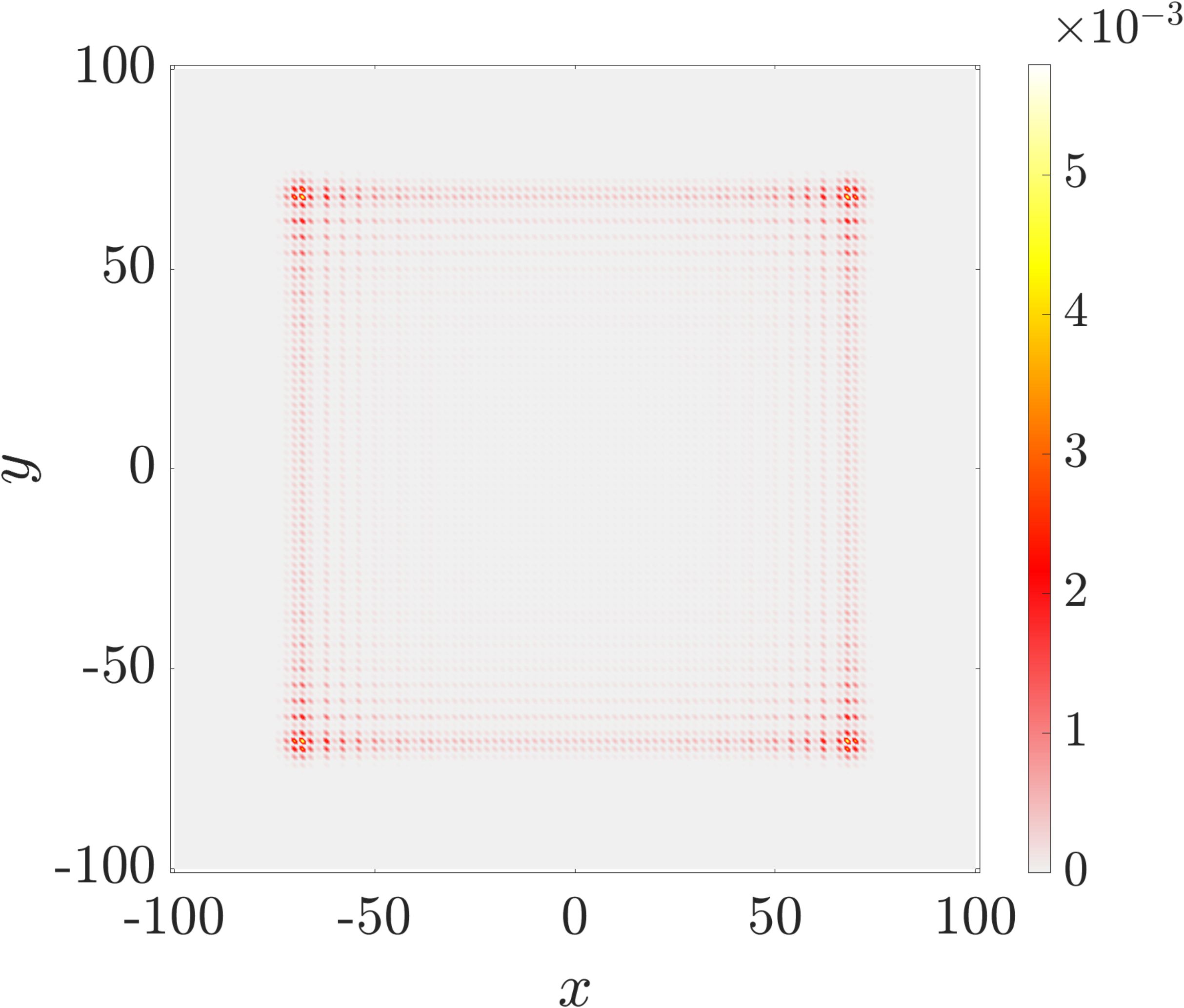}
		\label{fig:2DTQWH1}}
	\subfigure[]{
		\includegraphics[width=7.1cm]{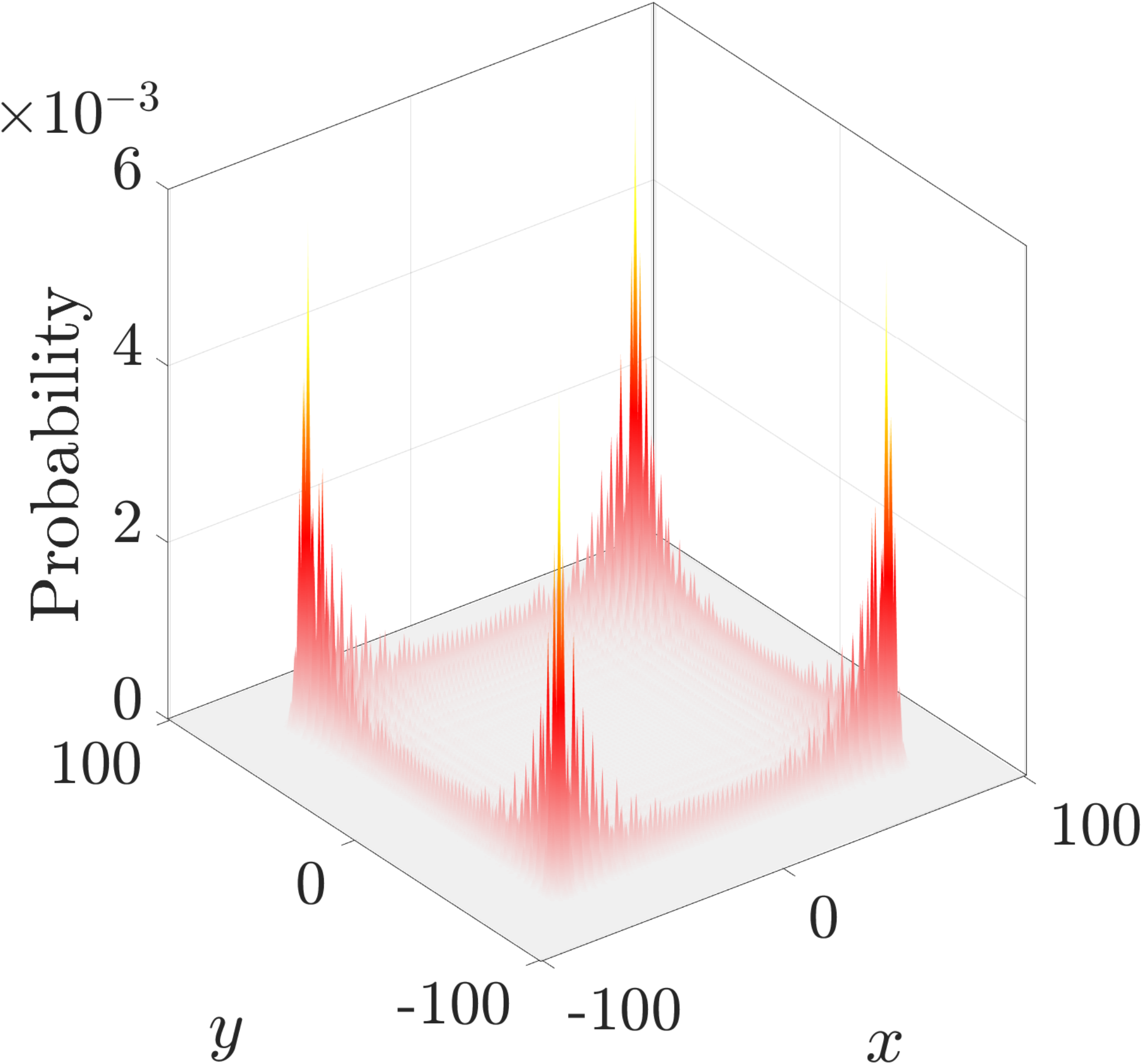}
		\label{fig:2DTQWH2}}
	\subfigure[]{
		\includegraphics[width=7.1cm]{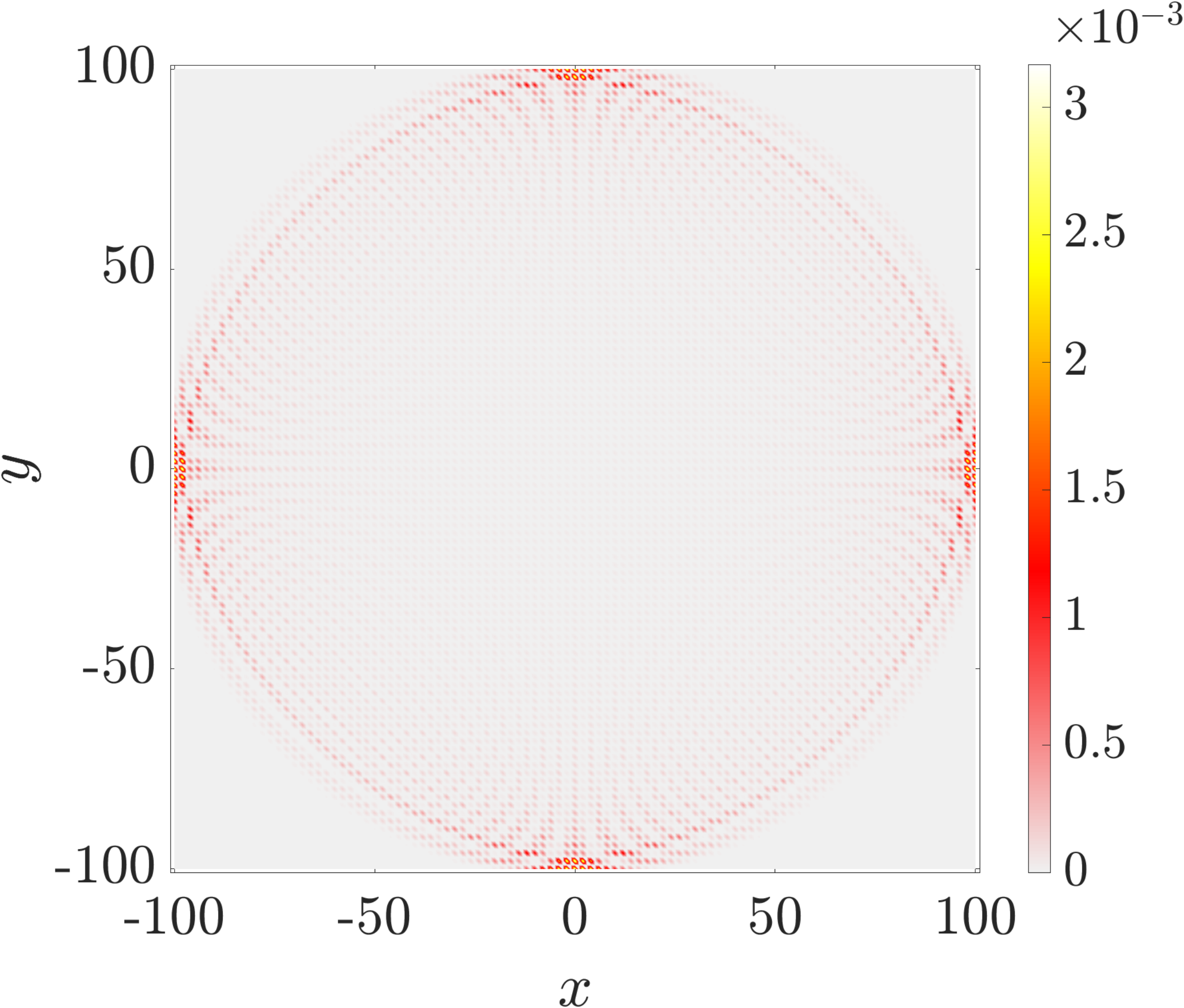}
		\label{fig:2DTQWG1}}
	\subfigure[]{
		\includegraphics[width=7.1cm]{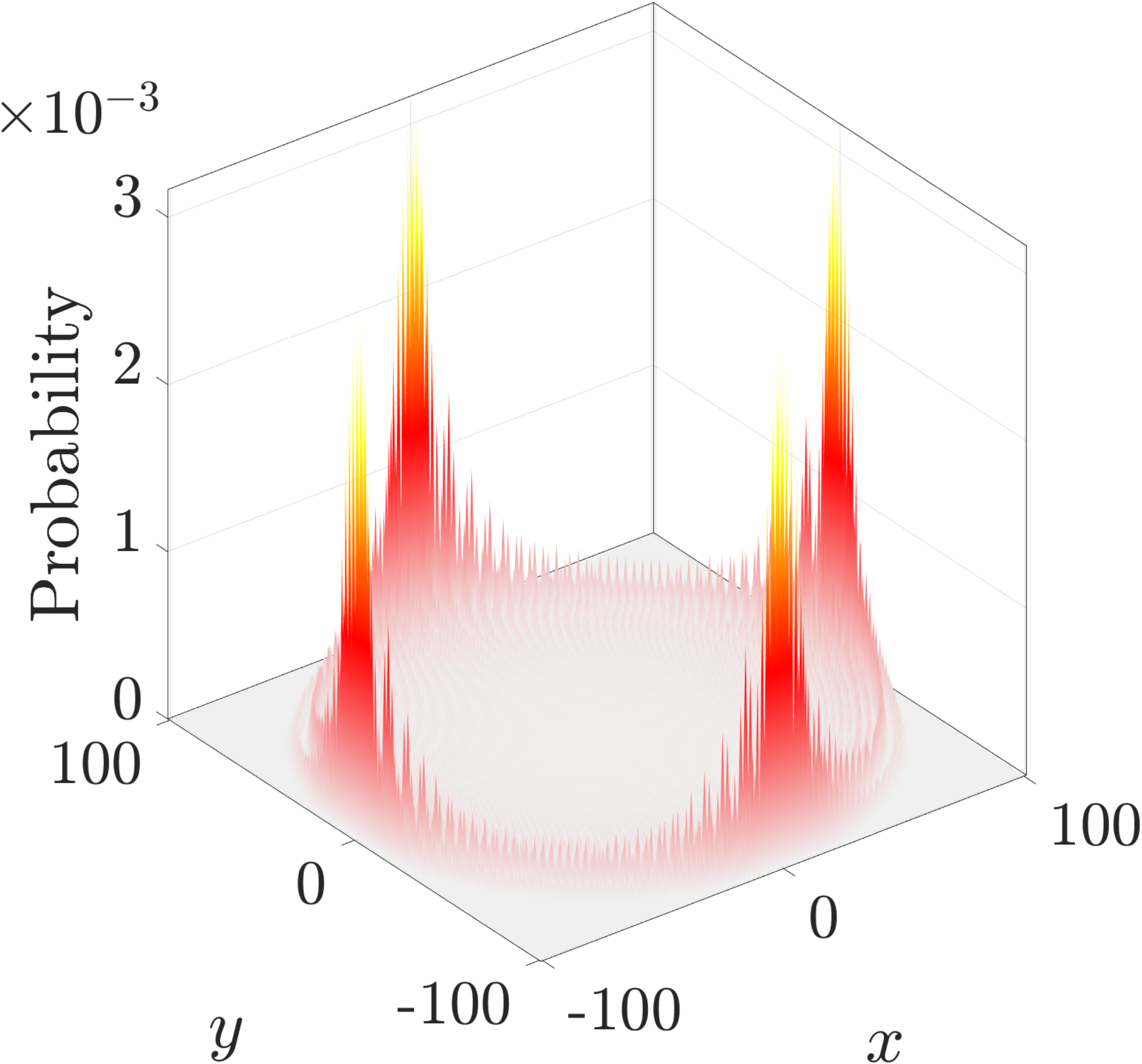}
		\label{fig:2DTQWG2}}
	\subfigure[]{
		\includegraphics[width=7.1cm]{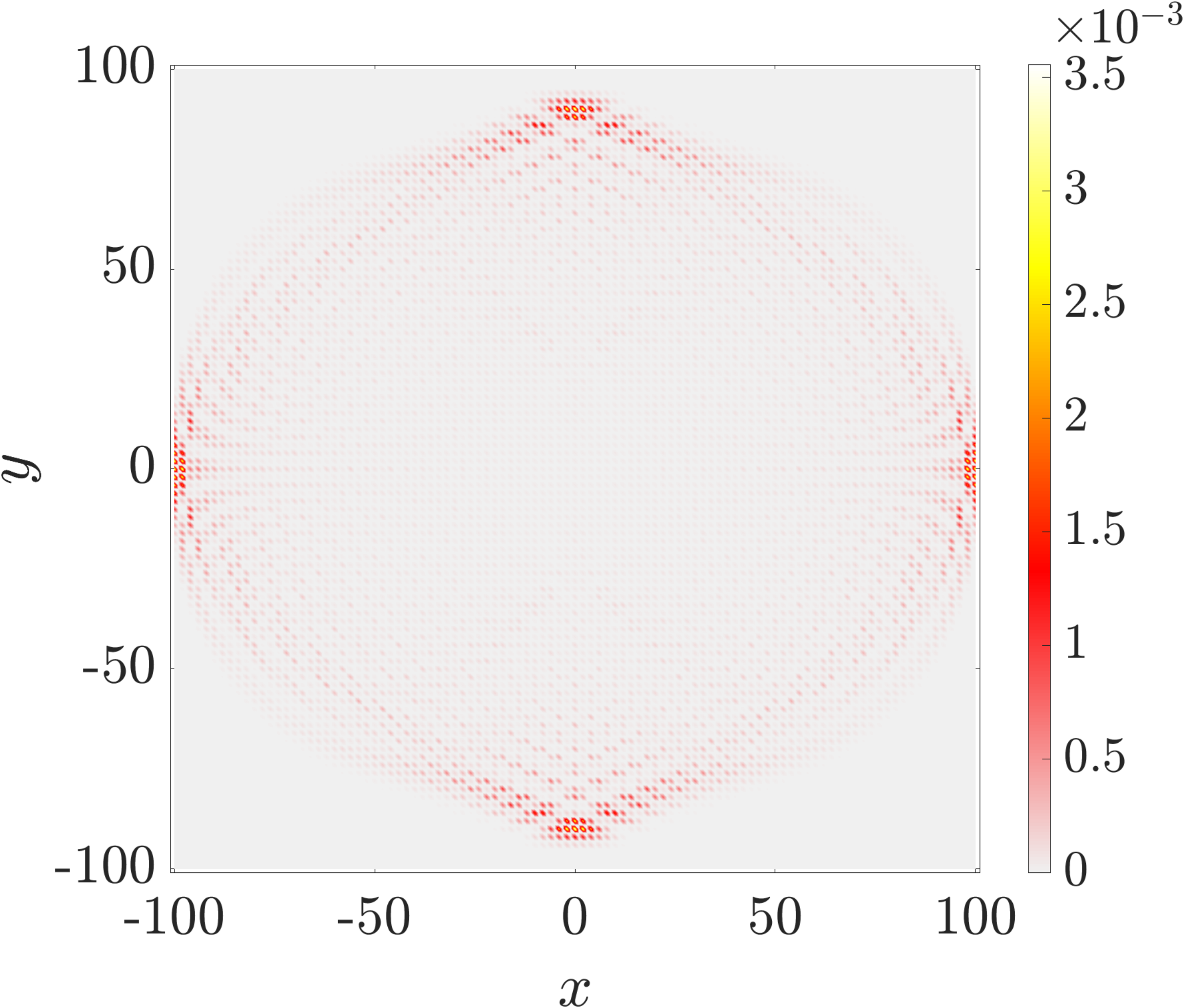}
		\label{fig:2DTQWF1}}
	\subfigure[]{
		\includegraphics[width=7.1cm]{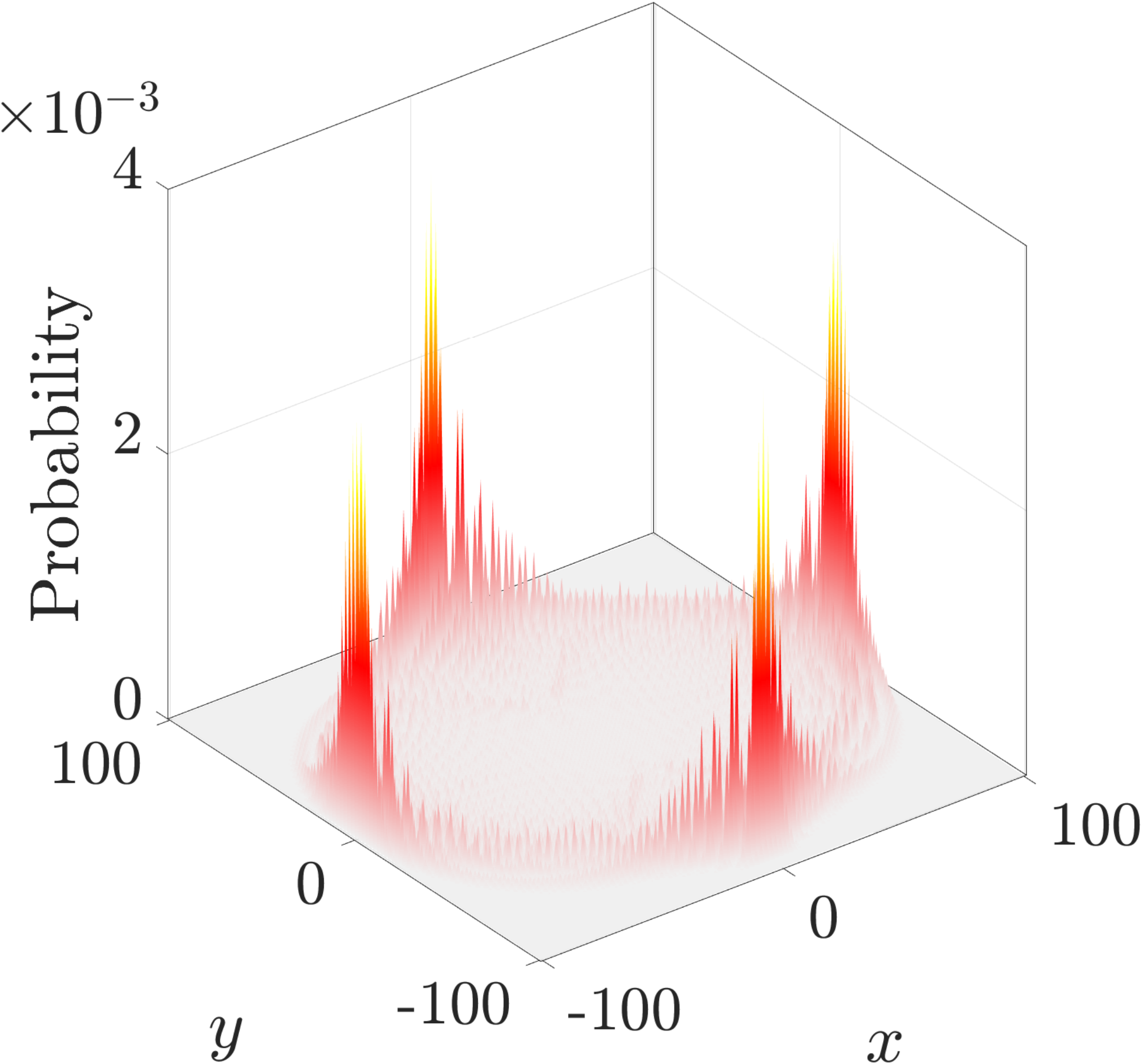}
		\label{fig:2DTQWF2}}
	\caption{Probability distribution for \subref{fig:2DTQWH1} Hadamard, \subref{fig:2DTQWG1} Grover, and \subref{fig:2DTQWF1} Fourier coin. The lattice size is taken to be $201 \times 201$ for all the plots.}
	\label{fig:Prob_2DDTQW}
\end{figure}

\noindent 3. Grover coin: Lastly, we talk about the Grover operator, which is given for $d$-dimensional systems as \cite{Mackay2002}
\begin{equation}
	G_d \ket{u} = \dfrac{1}{\sqrt{2^d}} \left( -2 \ket{u} + \sum_{v = 0}^{2^d - 1}\ket{v} \right)
\end{equation}
and in $d=2$, it reads
\begin{align} \label{eq:GroverCoin}
	C &= G_2 = \dfrac{1}{2} \begin{bmatrix}
		-1 & 1 & 1 & 1 \\
		1 & -1 & 1 & 1 \\
		1 & 1 & -1 & 1 \\
		1 & 1 & 1 & -1 
	\end{bmatrix}
\end{align}	
With the initial state of the composite system of the 2D lattice and the coin given by 
\begin{equation}
	\ket{\Psi(0)} = \sum_{x,y} \sum_{j, k} \Psi_{j,k}(x,y,0) \ket{j,k}\otimes\ket{x,y}
\end{equation}
the wave function after $t$ time step is written as
\begin{equation}
	\ket{\Psi(t)} = U^t \ket{\Psi(0)} = \sum_{x,y} \sum_{j, k} \Psi_{j,k}(x,y,t) \ket{j,k}\otimes\ket{x,y}
\end{equation}
and the probability of finding the walker at site $(x,y)$ after $t$ steps is
\begin{equation}
	P(x,y,t) = \sum_{j, k}\abs{\Psi_{j,k}(x,y,t)}^2.
\end{equation}
We have plotted the probability distribution in Fig.~\ref{fig:Prob_2DDTQW} for various coins. In Figs.~\ref{fig:2DTQWH1},~\ref{fig:2DTQWH2} we have chosen the Hadamard coin, $H_2$~\eqref{eq:HadamardCoin} with the initial state given by
\begin{equation}
	\ket{\Psi(0)} = \dfrac{1}{2} \left( \ket{0} - i \ket{1}\right)\left( \ket{0} - i \ket{1}\right) \ket{0,0}.
\end{equation}
Since we have considered the symmetric state for the coin, it produces a symmetric probability distribution after $t = 100$ time steps. It is the same as the one we get in 1D DTQW in both the $x$ as well as $y$ directions because the coin operator does not create any entanglement between two directions. Second, we plot the probability distribution with Grover coin $G_2$~\eqref{eq:GroverCoin} in Figs.~\ref{fig:2DTQWG1},~\ref{fig:2DTQWG2} with the initial state 
\begin{equation}
	\ket{\Psi(0)} = \dfrac{1}{2} \left( \ket{0} - \ket{1}\right)\left( \ket{0} - \ket{1}\right) \ket{0,0}.
\end{equation}
In this case, the probability distribution is highly symmetric about the origin and has 4-fold symmetry. Lastly, we use the Fourier coin $F_2$~\eqref{eq:FourierCoin} and use
\begin{equation}
	\ket{\Psi(0)} = \dfrac{1}{2} \left( \ket{0,0} + \dfrac{1 - i}{\sqrt{2}} \ket{0,1} + \ket{1,0} - \dfrac{1 - i}{\sqrt{2}} \ket{1,1}\right) \ket{0,0}
\end{equation} 
this as the initial state. The probability distribution is plotted in Figs.~\ref{fig:2DTQWF1}, ~\ref{fig:2DTQWF2}. In this case, if we observe carefully, we see a 2-fold symmetry.

\subsection{Decomposition of SSQW}
\begin{figure}[H]
	\centering
	\subfigure[]{
		\includegraphics[width=12cm]{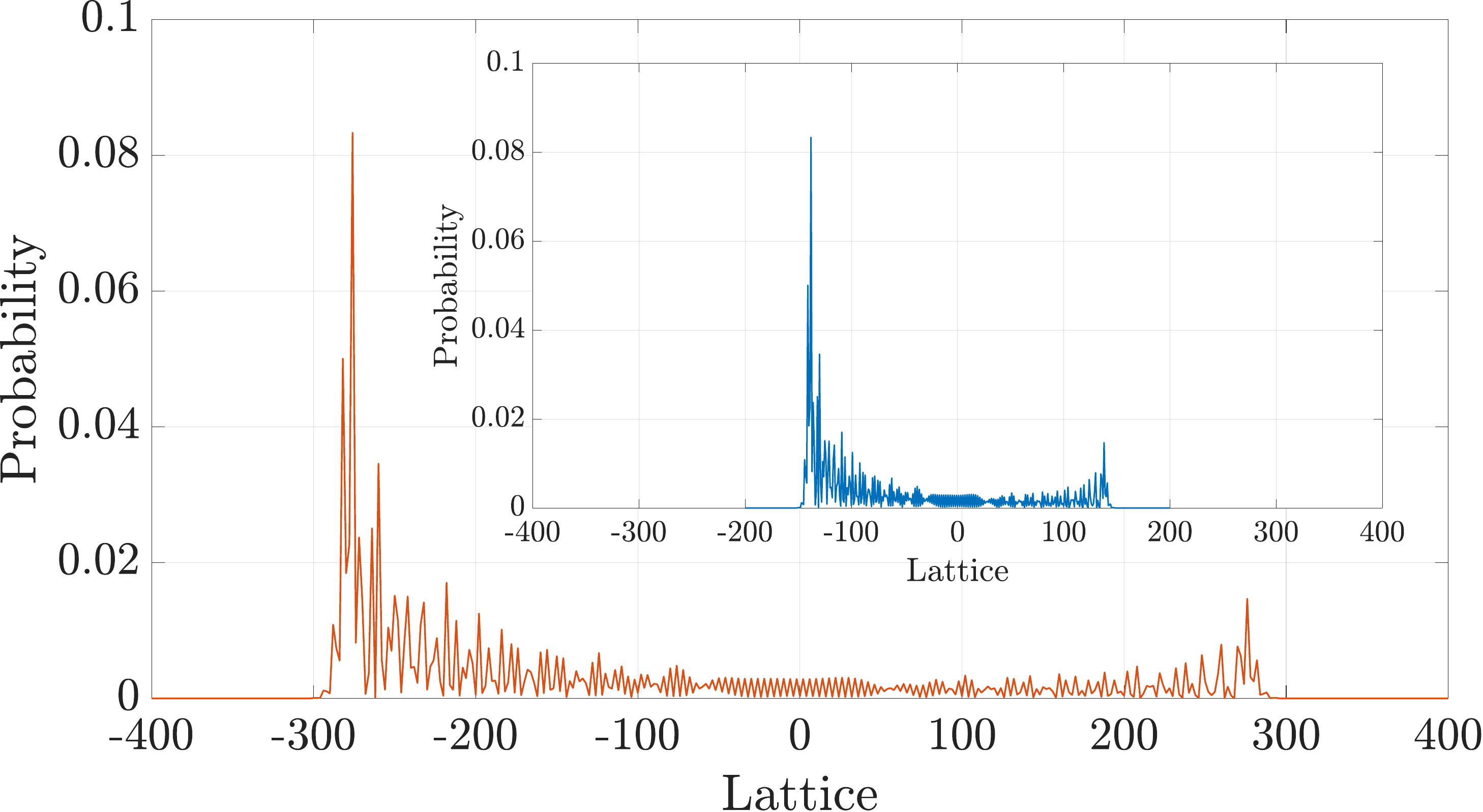}
		\label{fig:1DSSQW1}}
	\caption{Probability distribution of the walker initially localized at the origin and with the symmetric state of the coin for $\theta_1 = \pi/4, \theta_2 = \pi/7$ after 200 steps. We took the system size to be $ N = 400 $. with unitary given by \eqref{eq:SSQW-Unitary} and in the inset with unitary given by \eqref{eq:U3}.
		\label{fig:1DSSQWEquivalence}}
\end{figure}

In this section, we present a hierarchical structure to decompose a 1D SSQW and 2D DTQW into 1D DTQWs~\cite{Zhang2017}. We start with the time evolution operator,  $U_{_{\text{SS}}}(\theta_1, \theta_2)$ of a 1D SSQW given in Eq.~\eqref{eq:SSQW-Unitary} and replace $ T_{\downarrow}$ and $ T_{\uparrow} $ with $ T_{\downarrow}^2$ and $ T_{\uparrow}^2 $ respectively such that
\begin{equation}\label{eq:U2}
	U_{_{\text{SS}}}(\theta_1, \theta_2) = T_{\downarrow}^2R(\theta_2)T_{\uparrow}^2R(\theta_1).
\end{equation}
where
\begin{equation}
	T_{\uparrow}^2 = \sum_{n} \ket{\uparrow} \bra{\uparrow} \otimes \ket{n+2} \bra{n}  + \ket{\downarrow} \bra{\downarrow} \otimes \mathds{1}_{2j+1},
\end{equation}
\begin{equation}
	T_{\downarrow}^2 = \sum_{n} \ket{\uparrow} \bra{\uparrow} \otimes \mathds{1}_{2j+1}   + \ket{\downarrow} \bra{\downarrow} \otimes \dyad{n-2}{n}.
\end{equation}
Qualitatively, there is no difference between~\eqref{eq:SSQW-Unitary} and~\eqref{eq:U2}. In the former, the walker moves on the neighboring sites, whereas in the latter one, the walker skips one lattice site in each jump. So, in the latter case, the probability of finding the particle at odd sites is always zero irrespective of the number of steps. We can also choose lattice with even or odd numbered sites, which would result in an identical probability distribution (as shown in Fig.~\ref{fig:1DSSQWEquivalence}). Now, using the fact that SSQW is translation invariant, we can write
\begin{equation}\label{eq:U3}
	U_{_{\text{SS}}}(\theta_1, \theta_2) = T_+ T_{\downarrow}^2R(\theta_2)T_{\uparrow}^2R(\theta_1) T_-
\end{equation}
with
\begin{equation}
	T_+ = \sum_{n} \mathds{1} \otimes \dyad{n+1}{n} = T_-^{\dagger}.
\end{equation}
Since $ T_{\pm} $ acts only on the lattice part, we can rewrite it as
\begin{equation}
	U_{_{\text{SS}}}(\theta_1, \theta_2) = T_+ T_{\downarrow}^2R(\theta_2)T_{\uparrow}^2 T_-R(\theta_1) = T_+ T_{\downarrow}^2R(\theta_2)T_-T_{\uparrow}^2 R(\theta_1)
\end{equation}
and we can easily show 
\begin{equation}
	T_+ T_{\downarrow}^2 = T = T_-T_{\uparrow}^2.
\end{equation}
where
\begin{equation}
	T = \sum_n \dyad{\uparrow} \otimes \ket{n+1} \bra{n}  + \dyad{\downarrow} \otimes \ket{n-1} \bra{n}.
\end{equation}
is the translation operator from 1D DTQW. Hence,
\begin{equation} \label{eq:U4}
	U_{_{\text{SS}}}(\theta_1, \theta_2) = (T R(\theta_2)) (T R(\theta_1)) = U(\theta_2)U(\theta_{1}).
\end{equation}
So, even though 1D SSQW seems complicated when it comes to implementation, it is not much different from ordinary 1D DTQW. From \eqref{eq:U4} it is clear that the 1D SSQW is equivalent to a quantum walker performing the ordinary 1D DTQW, with alternate coin operations at each step. Therefore, each step of a 1D SSQW can be decomposed into two steps of the 1D DTQW with alternating spin-flip operations. Furthermore, this decomposition connects the Hamiltonian of the 1D DTQW to the Hamiltonian for the 1D SSQW. We plotted the probability distribution of the walker after 200 steps in Fig. \eqref{fig:1DSSQWEquivalence}. We observe that both plots are nearly identical and only differ in their spread on the $x$-axis. 
\noindent We extend this structure to decompose the 2D DTQW given in Eq.~\eqref{Eq:Qwalk2D} into two 1D SSQW performed in two independent degrees of freedom (for example, directions) with the same coin~\cite{Zhang2017}. We can rewrite the time evolution operator in $U_{_{2D}}(\theta_1, \theta_2)$ given in Eq.~\eqref{Eq:Qwalk2D} as
\begin{equation}
	U_{_{2D}}(\theta_1, \theta_2) = U^y_{_{\text{SS}}}(\theta_1, 0) U^x_{_{\text{SS}}}(\theta_1, \theta_2),
\end{equation}
where $ U^i_{_{\text{SS}}} (i = x, y) $ is the time evolution operator of the 1D SSQW \eqref{eq:SSQW-Unitary} in the $ i $th direction. 

\subsection{Electric Quantum Walk}
\begin{figure}[H]
	\centering
	\subfigure[]{
		\includegraphics[width=7cm]{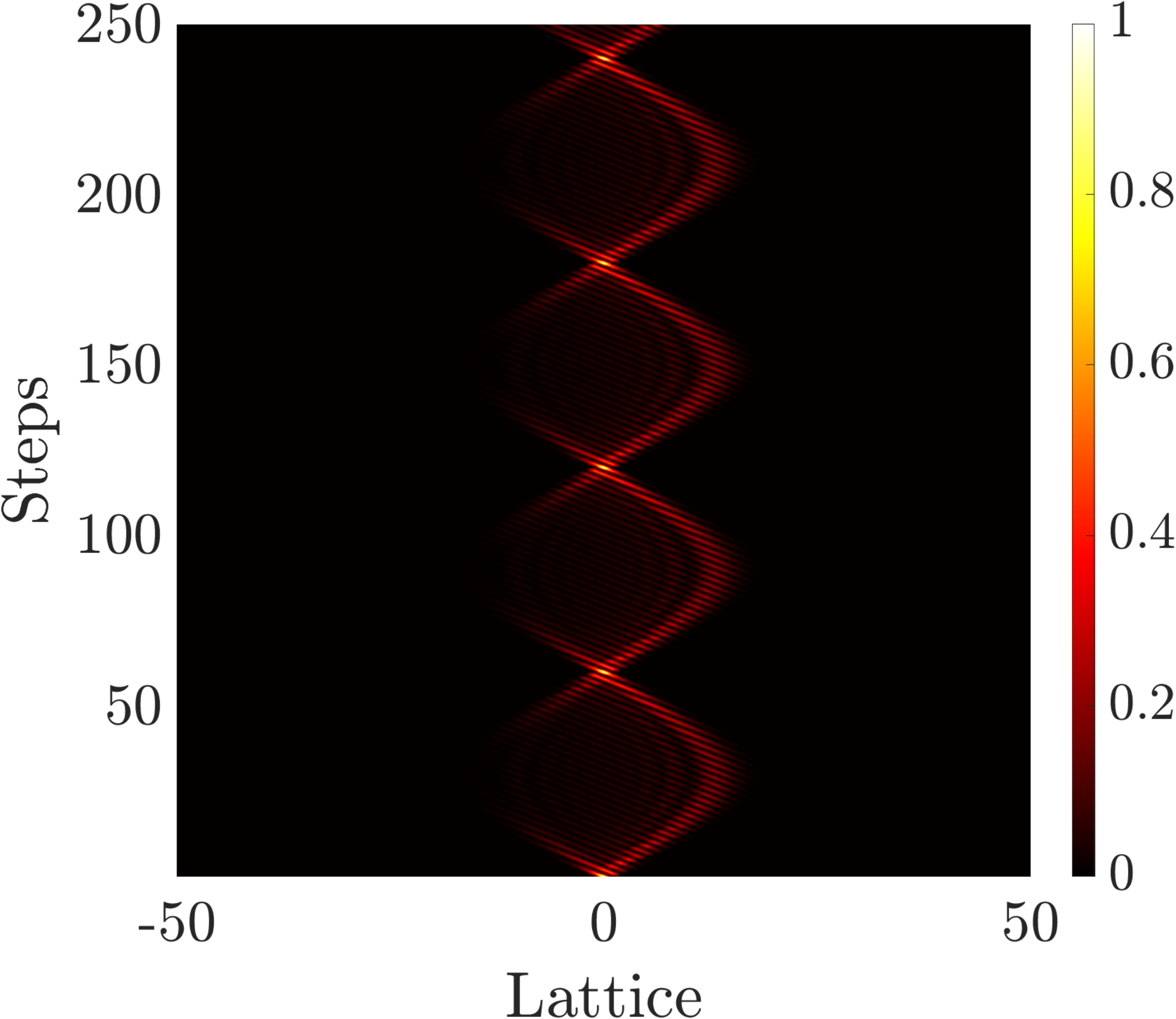}
		\label{fig:1DEQW1}}
	\subfigure[]{
		\includegraphics[width=7cm]{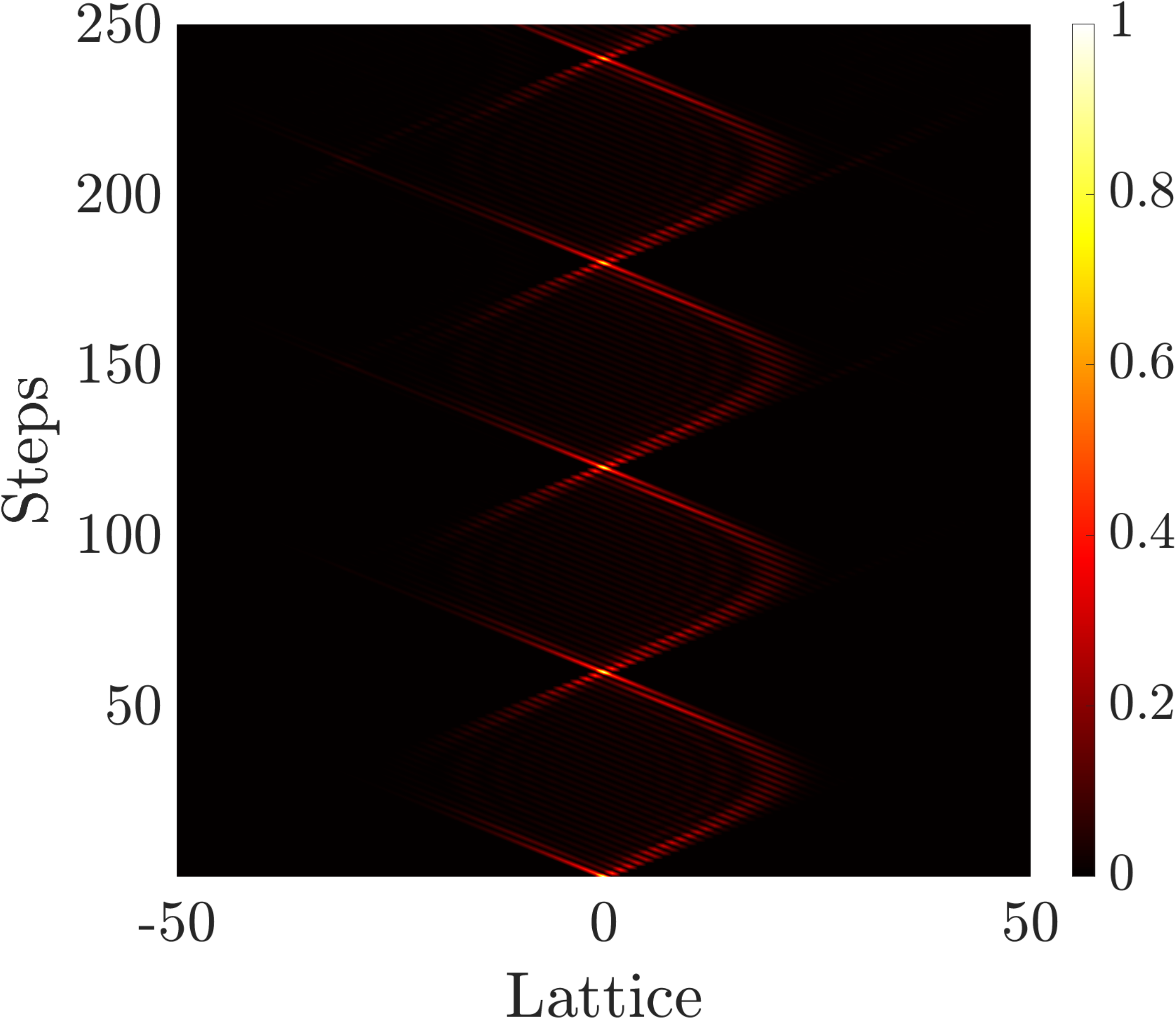}
		\label{fig:1DEQW2}}
	\caption{The evolution of probability density on the lattice with number of time steps with lattice size, $N = 101$. \subref{fig:1DEQW1} $\theta_2 = \pi/2$, and \subref{fig:1DEQW2} $\theta_2 = \pi/4$. Here we chose $\phi = 2 \pi/60$.}
	\label{fig:1DElectricQW}
\end{figure}
\begin{figure}[H]
	\centering
	\subfigure[]{\includegraphics[width=7.5cm]{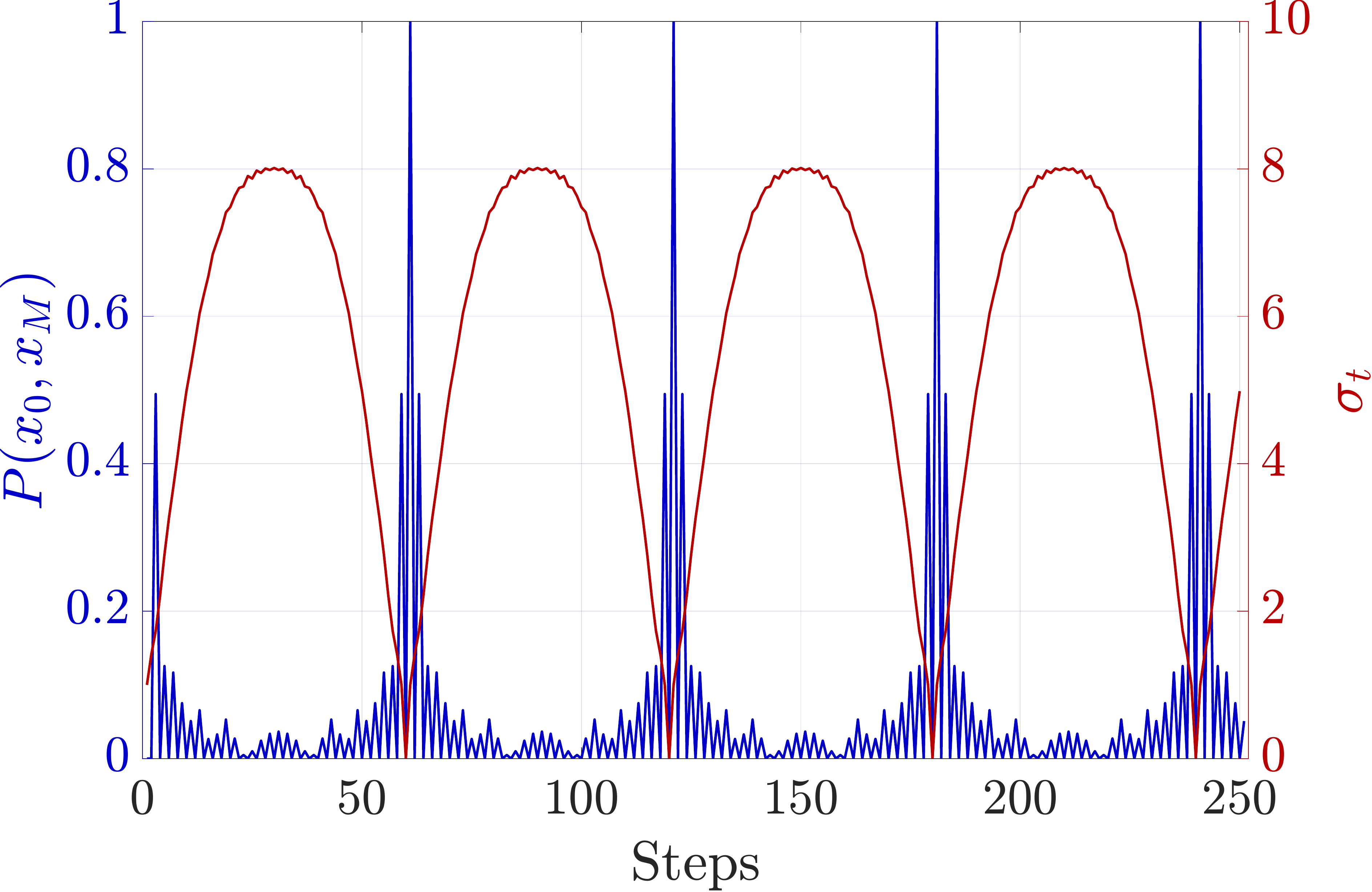}
			\label{fig:EQWVa}}
	\subfigure[]{\includegraphics[width=7.5cm]{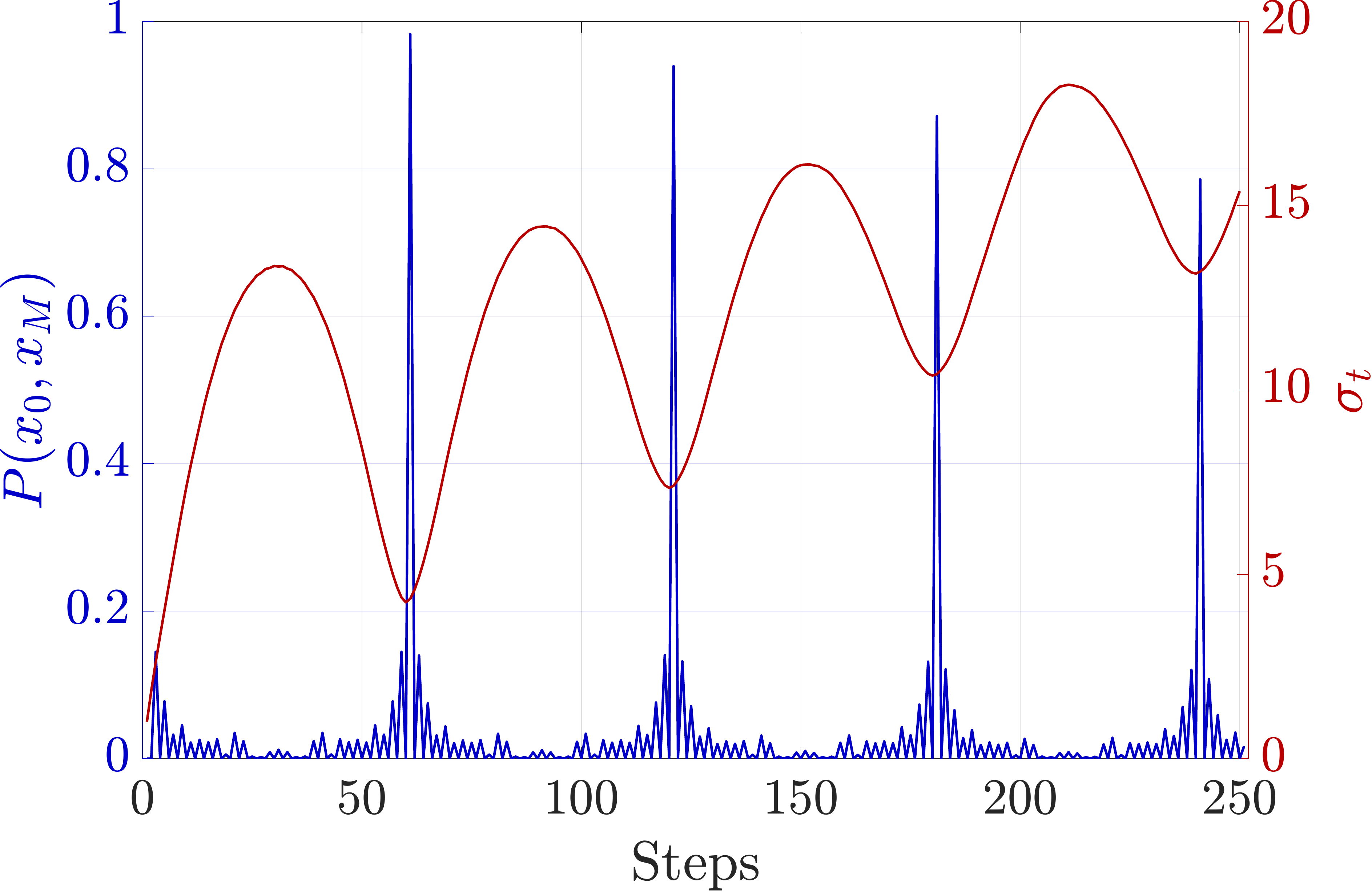} 
			\label{fig:EQWVb}}
	\caption{The plot of the returning probability $P(x_0,x_M)$ after $M$ number of steps and the variance as a function of time steps for \subref{fig:EQWVa} $\theta_2 = \pi/2$, \subref{fig:EQWVb} $\theta_2 = \pi/4$.}
	\label{fig:EQWV}
\end{figure}

We have introduced a phase factor, given by  
\begin{equation}
	E_{\phi} = \sum_n e^{i \phi n \mathds{1}}  \otimes \dyad{n}
\end{equation}
during the discussion on unitary equivalence of quantum walks. This extra phase results in a modified time evolution for 1D DTQW which reads
\begin{equation}
	U = E_{\phi} T R(\theta).
\end{equation}
This modified quantum walk is referred to as ``Electric Quantum Walk"~\cite{Wojcik2004,Romanelli2005,Soriano2006,Alberti2013,Cedzich2013,Eugenio2016}. We plot the variation of probability distribution with the number of steps in Fig. \eqref{fig:1DElectricQW} for the initial state given by
\begin{equation}
	\ket{\psi(t=0)} = \dfrac{1}{\sqrt{2}} \left( \ket{0} + i \ket{1} \right) \otimes \ket{0}.
\end{equation}
and observe the Bloch oscillations~\cite{Pablo2020} which are the results of quantum particle superimposed with electric field. This is one of the main feature of these quantum walks. We also plot the probability of the return of the walker to the initial state and the variance as a function of time steps in Fig.~\ref{fig:EQWV}.

\noindent To summarize, in this chapter, we have discussed general protocols for 1D and 2D DTQW and other related concepts. In Chapter~\ref{chap:sym} and~\ref{ch:pers}, we will describe the topological phases exhibit by quantum walks and discuss the persistence of topological behaviour in such systems with more general setting. 
\chapter{Symmetries and Topology}\label{chap:sym}
One of the central aims in condensed matter physics is to classify different states (or phases) of matter and study the behavior of a system in the vicinity of a phase transition. A state or phase of matter is defined as a particular order in which the constituent particles are organized in the system, leading to a distinct set of physical properties. This is referred as the ‘\emph{\textbf{principle of emergence}}’, which states that the arrangement of the constituent particles, not the constituent particles themselves, determines the physical properties of the material~\cite{Wen2015,Stanescu2016}. Different settings of the constituent particles lead to different phases of matter, which result in distinct properties. An immediate question one can ask here is that "Is there a way to tell whether the two phases of a system are same or equivalent?" The answer to the question which has been posed is that, the phases or states of a system are identical if they are connected to each other without any \emph{phase transition}. And by connecting the two states, we mean that we start with one state of the system and start deforming it by changing the local parameters such as temperature, pressure, external fields, etc. until we reach the desired phase. A \emph{phase transition} in the system is indicated by a discontinuity in one of the local quantities, known as \emph{order parameter}. 

The first step to classify the distinct phases of matter was taken by one of the great physicists of all time, Lev Landau, in the first half of the twentieth century by introducing the theory of phase transitions~\cite{Landau1965,Landau1958,Ginzburg2009,Sachdev2011}. These phase transitions, known as ``Landau-like phase transitions ``, are characterized by symmetry breaking. Landau proposed that different phases correspond to the existence of contrasting symmetries in a system, and a phase transition is simply a transition that changes the symmetry of the system. In a phase transition, a higher symmetry group is broken into a lower symmetry subgroup or vice versa. When a fluid which has rotational and translational symmetries turns into a crystal, where these symmetries are broken, it is an example of a phase transition. Another example of the same kind of phase transition is when the temperature of a magnet is raised past a critical value, it goes from the ferromagnetic state to the paramagnetic state. The ferromagnetic state has a nonzero magnetization (because the spins are aligned in a particular direction), which breaks the rotation symmetry, which is reinstated in the paramagnetic state. In a phase transition in the ferromagnet, the net magnetization is the order parameter, whereas the shear modulus is the order parameter in the liquid-solid phase transition.	

Landau's theory of phase transition was successful in classifying 230 different kinds of crystal that can exist in three dimensions. It managed to make people believe that the quest to understand the different phases of matter has come to an end. However, our understanding of phases took a sharp turn with the discovery of integer~\cite{Klitzing1980} and fractional quantum Hall effect~\cite{Stormer1982} and the high-temperature superconductivity~\cite{Bednorz1986}. In these systems, for example, in the quantum Hall effect, the system supports different states with same symmetries. The discovery of these phases led physicists to take a break from Landau's theory of phase transitions and resort to topological concepts to understand these phases~\cite{Kosterlitz1973,TKNN1982,Thouless1983,Haldane1983,Haldane1988}. The order in such systems is beyond the laws of symmetries, and a new term was given to them, "topological order"~\cite{Wen2015}.    

The topological phases of matter have started to gain popularity very quickly due to their unique characteristics. The ground wavefunction of the system in such phases cannot be described by localized orbitals and exhibits entanglement (short-range or long-range)~\cite{Wen2010}. Further, the topological phases are defined globally, for example, to calculate a topological invariant, we need to integrate over the whole parameter space in order to get the information of the phase. This global behavior makes these phases robust against local noise, such as defects and impurities. In addition to this, the topological phases have a quantized response against the control parameter (like the external field), which only depends on fundamental constants such as $h, c, e$~\cite{Klitzing1980,Wu2016}. For example, the Hall conductivity in a two-dimensional electron gas is given by an integer multiple of $e^2/h$~\cite{Klitzing1980,TKNN1982}. Topological properties are found in many systems, such as electrons in insulators~\cite{HasanKane2010,QiZhang2011}, metals~\cite{Ashvin2018}, photonics~\cite{Ozawa2019}, strongly interacting systems~\cite{Wen2017} and in amorphous matter~\cite{Grushin2020}. 

As we discussed earlier, in ferromagnets the local magnetization is the order parameter which accounts for the spontaneous symmetry breaking and hence the phase transition. However, topological phases, being global in nature, cannot be described by local order parameters. The topological phases in a material can be described by \emph{topological invariants}, which takes different quantized values in different phases, depending on the dimensions of the system. Topological invariants are deeply connected with a theorem from differential geometry known as \emph{Gauss-Bonnet theorem}~\cite{Nakahara1990,HasanKane2010}
\begin{equation} \label{eq:gaussbonnet}
	\dfrac{1}{2 \pi} \int_S K dA = \chi.
\end{equation}
It states that the integral of the Gaussian curvature $ K $ on a surface of the manifold divided by $2 \pi$ is always an integer $\chi$ (also known as the Euler characteristic). The integer $\chi$ is related to the genus as $\chi = 1 - g$ which represents the number of holes in the surface. For example, the surface of a sphere has $ g = 0 $ and the surface of a doughnut has $ g = 1 $, making them two topologically different objects. We encounter integrals, as one appearing in the Gauss-Bonnet theorem, while dealing with topological phases (we will see this in preceding sections) and $\chi$ is directly related to the topological invariants. Note that the small perturbations affect the curvature of the surface only locally and the integration over the new surface yields the same value on the right hand side of the Eq.~\eqref{eq:gaussbonnet} and hence the same topological invariants. This is the fundamental idea behind the robustness of topological phases.

In this chapter, we will discuss unitary and anti-unitary symmetries and how these symmetries define topological classes. We then discuss the SSH model model to understand the concepts of topological phases, topological invariants, and bulk-boundary correspondence.

\section{Classification of topological phases}
We start by looking for the following information about the system: a) the statistics of the constituents particles: bosons, fermions, and even anyons (particles that acquire an arbitrary phase $e^{i \phi}$ on exchange, rather than just $\pm1$~\cite{Sankar2008}), b) whether the system under consideration is gapped or gapless, and c) the entanglement (short-range or long-range) in the system. Once we have this information, the topological phases are characterized by the dimensionality and symmetries of the systems~\cite{Schnyder2008,Schnyder2009,Ryu2010,Kitaev2009,Ludwig2015}. Here, by classifying, we mean finding the topological invariants of the system. These topological invariants can be a Chern number, a winding number, or some other mathematical index~\cite{Budich2013,SchnyderRyu2016,Stanescu2016}.

In topological materials, topological insulators are of particular interest~\cite{Ando2013}. An ordinary insulator has a bulk with a band gap and lacks states that can support electrical conductivity. On the other hand, topological insulators are materials that also have a gapped bulk but are characterized by some topological invariant. This bulk invariant allows and predicts a number of low-energy eigenstates, which are robust against perturbations, at the edges, which is known as \emph{bulk-boundary correspondence}~\cite{LudwigRyu2010,Kane2010,Asboth2016}. In some materials, these topological phases and, hence, low-energy eigenstates are protected against perturbations due to discrete symmetries of the system~\cite{Mourik2012,Heeger2001}.

\section{Symmetries}
Symmetry\cite{Sarkar2018,Ryu2016,Stan2016,LudwigRyu2010,Ryu2010} is an operation that leaves the properties of the system unaffected. The concept of symmetry has always been useful in understanding physical systems. Although systems without any symmetries can also exhibit topological phases, understanding the symmetries allows us to classify the topological phases and the topological invariants that define their characteristics. In this section, we will go through the unitary and non-unitary symmetries and how they are useful in the context of classifying topological phases of matter. 

\subsection{Unitary symmetries}
\begin{figure}
	\centering
	\includegraphics[width=12cm]{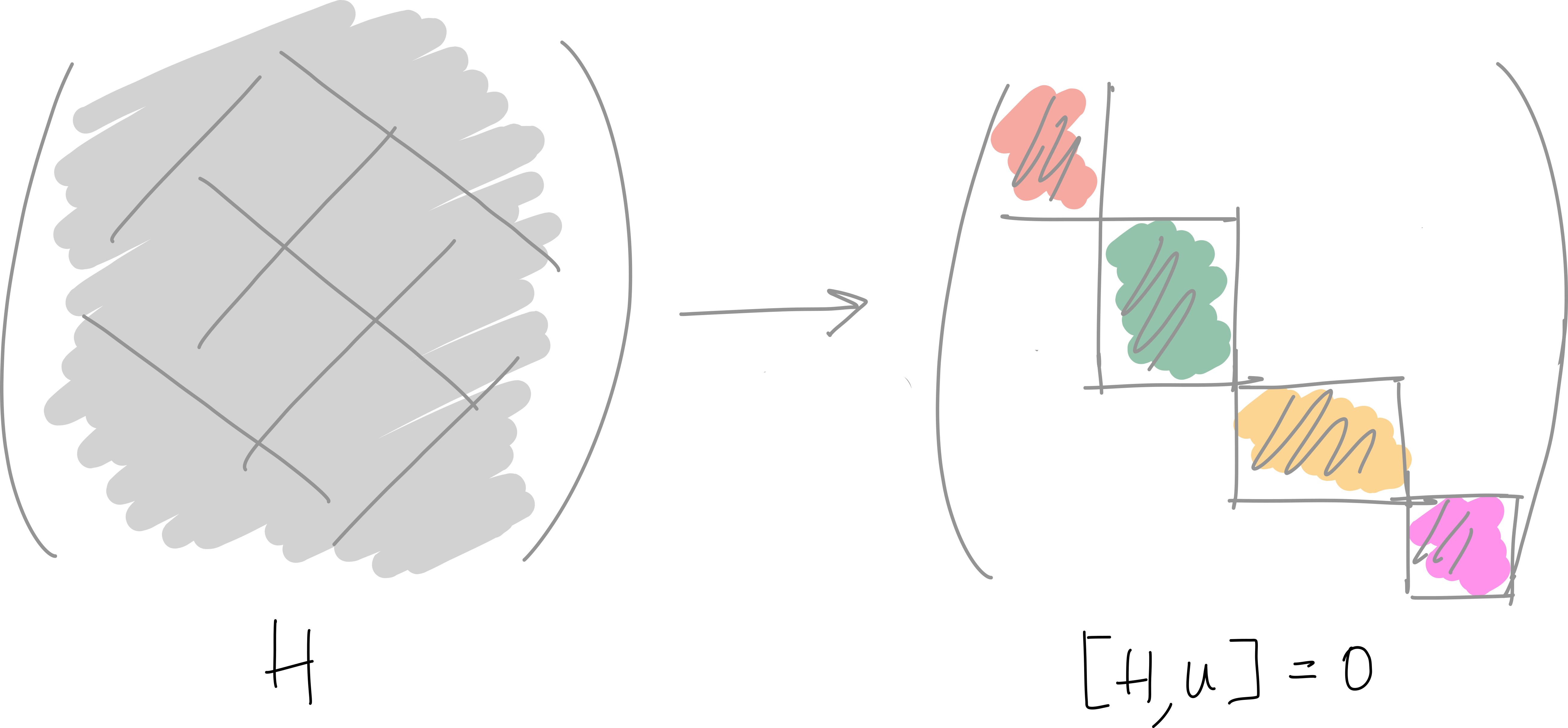}
	\caption{Block diagonalization of $\hat{H}$ as the consequence of a unitary symmetry.}
	\label{fig:blockdiagonal}
\end{figure}

Before beginning the discussion on symmetries, let us first write a general (non-superconducting) Hamiltonian written in the second quantization~\cite{Fetter2003} as
\begin{equation}
	\hat{H} = \sum_{ij} \hat{b}_i^{\dagger} H_{ij} \hat{b}_j = \hat{b}^{\dagger} H \hat{b}											
\end{equation}
where $H$ is the first quantized Hamiltonian, and $ \{\hat{b}_i^{\dagger}\} $ and $ \{\hat{b}_j\} $ are the creation and annihilation operators, respectively. Note that all statistics of the particle and operator properties of the Hamiltonian are contained in the creation and destruction operators $ \hat{b}^{\dagger} $ and $ \hat{b} $. In the subsequent section, we will understand this structure with the SSH model. A system exhibits unitary symmetry if the first-quantized Hamiltonian $H$ of the system commutes with $U$ such that
\begin{equation}
	U H U^{\dagger} = H.
\end{equation} 
In terms of second quantization, this symmetry corresponds to a linear operator $\mathcal{U}$ that acts on fermionic operators as~\cite{Ludwig2015}
\begin{equation}
	\hat{b}_i \rightarrow \hat{b}'_i = \sum_j U_{ij} \hat{b}_j = \mathcal{U} \hat{b}_i \mathcal{U}^{-1}, \qquad \hat{b}^{\dagger}_i \rightarrow (\hat{b}^{\dagger}_i)' = \sum_j \hat{b}^{\dagger}_j U^*_{ji}= \mathcal{U} \hat{b}^{\dagger}_i \mathcal{U}^{-1}.
\end{equation}
and commutes with the Hamiltonian as
\begin{equation}
	\mathcal{U} \hat{H} \mathcal{U}^{-1} = \hat{H}.
\end{equation}
From here on-wards we will consider first-quantized Hamiltonian $H$ only. Furthermore, in such scenarios, we can find a basis in which $H$ and $U$ both take a block diagonal form and each block is referred to as an irreducible representation of the symmetry~\cite{Stan2016} as shown in Fig.~\ref{fig:blockdiagonal}. Note that the individual blocks do not need to be of the same dimension.
An example of unitary symmetries is \emph{Translational symmetry}, which we will use in studying the SSH model in the next section. Another example is \emph{Inversion/Parity symmetry}, $\mathcal{P}$ under the action of which the position and the momentum transform as
\begin{equation}
	\mathcal{P} \hat{x} \mathcal{P}^{-1} = -\hat{x}, \qquad \mathcal{P} \hat{p} \mathcal{P}^{-1} = -\hat{p}, \qquad \mathcal{P} \comm{\hat{x}}{\hat{p}} \mathcal{P}^{-1} = \mathcal{P} i \hbar \mathcal{P}^{-1} = i \hbar.
\end{equation}  
The third identity explicitly shows that the inversion symmetry is a unitary symmetry. Let us consider a simple example to understand the inversion symmetry: a molecule having two atoms with an electron hopping between them with amplitude $t$. The Hamiltonian for this system is written as
\begin{equation}
	H = \begin{pmatrix}
		0 & t \\
		t & 0
	\end{pmatrix} = t \sigma_x.
\end{equation}
The eigenvalues and the eigenvectors are
\begin{equation}
	E_{\pm} = \pm t, \qquad \ket{\psi_{\pm}} = \dfrac{1}{\sqrt{2}} \begin{pmatrix}
		\pm 1 \\
		1
	\end{pmatrix}.
\end{equation}
In this case, the inversion interchanges two atoms and hence $\mathcal{P} = \sigma_x$ such that $\comm{H}{\mathcal{P}} = 0$. Furthermore, we can construct a matrix $U_D$ using $ \ket{\psi_{\pm}} $ as its column, which will take $H$ to a diagonal basis as
\begin{equation}
	H = U_D H' U_D^{\dagger} \implies H' = U_D^{\dagger} H U_D =  \dfrac{1}{2} \begin{pmatrix}
		1 & 1 \\
		-1 & 1
	\end{pmatrix} \begin{pmatrix}
		0 & t \\
		t & 0
	\end{pmatrix} \begin{pmatrix}
		1 & -1 \\
		1 & 1
	\end{pmatrix} = t \sigma_z.
\end{equation}
\subsection{Non-unitary/Anti-unitary symmetries}
As we have seen in the previous section, the symmetry operation $U$ commutes with the Hamiltonian $H$ (i.e. $U H U^{\dagger} = H$) allows us to write $H$ in block diagonalized form in the irreducible representation of the symmetry group. However, such block diagonalization is not possible with anti-unitary symmetries. Hence, it is assumed that all of the unitary symmetries that commute with the Hamiltonian can be eliminated by writing the Hamiltonian in block form and by focusing attention on a block of the Hamiltonian. So, we are left with only the anti-unitary symmetries that need to be further analyzed for each block. The \emph{time-reversal} and \emph{particle-hole} symmetries are the anti-unitary symmetries that commutes with the Hamiltonian $H$. We can combine these two to get a unitary symmetry, called the \emph{chiral or sublattice} symmetry, but that anticommutes with $H$. These possibilities have been shown to be exhaustive~\cite{LudwigRyu2010} and lead to 10 different classes of topological phases~\cite{Schnyder2008,Schnyder2009,Kitaev2009,Ludwig2015}.
\subsubsection{Time-reversal symmetry (TRS)}
Time reversal symmetry~\cite{Sachs1987,Sato2012} is one of the fundamental symmetry of a system. This symmetry leaves the Hamiltonian of the system unchanged under the time reversal transformation of the form $\TR: t \rightarrow t' = -t$. Mathematically, time-reversal invariance is written as
\begin{equation}
	\TR H \TR^{-1}  = H.
\end{equation}
It implies that the time-reversal operator commutes with the Hamiltonian and $H$ and $\TR$ share the same set of eigenstates. If $\ket{\psi}$ is an eigenstate of $H$ with energy $E$, then $\TR\ket{\psi}$ will also be an eigenstate of $H$ with the same energy. We will come back to this discussion later in the section. Furthermore, the effect of the time-reversal operator on the position, momentum, and spin operators, respectively, is as follows
\begin{align}
	\TR \hat{x} \TR^{-1} = \hat{x} \nonumber, \\
	\TR \hat{p} \TR^{-1} = -\hat{p} \nonumber, \\
	\TR \hat{\sigma} \TR^{-1} = -\hat{\sigma}.
\end{align}
Consequently, we have
\begin{align}
	\TR \left[\hat{x}, \hat{p}\right] \TR^{-1} &=  \TR \hat{x} \hat{p} \TR^{-1} - \TR \hat{p} \hat{x} \TR^{-1} \nonumber \\
	&= \TR \hat{x} \TR^{-1} \TR \hat{p} \TR^{-1} - \TR \hat{p} \TR^{-1} \TR \hat{x} \TR^{-1} \nonumber \\
	&= -\left[\hat{x}, \hat{p}\right]
\end{align}
which implies
\begin{equation}
	\TR i \TR^{-1} = -i.
\end{equation}
From all of the above properties, we can conclude that $\TR$ must include the operator $ \K $ that takes any complex number $ z $ into its complex conjugate 
\begin{equation}
	\K z \K^{-1} = z^*, \qquad \K^{-1} = \K.
\end{equation}
and $\TR$ is an \emph{anti-unitary} operator that can be written, in general, as the product of a unitary operator and the complex conjugation operator 
\begin{equation}
	\TR = U \K
\end{equation}
with $ U^{-1} = U^{\dagger} $. As far as $\K$ is concerned, we have 
\begin{equation}
	\K \ket{\psi} = \ket{\psi^*} \implies \K^2 = \mathds{1}.
\end{equation}
For a time-reversal operator $\TR$, we have 
\begin{align}
	\TR (\alpha \ket{\psi} + \beta \ket{\phi}) &= U \K (\alpha \ket{\psi} + \beta \ket{\phi}) \nonumber \\ 
	&= U (\alpha^* \K \ket{\psi} + \beta^* \K \ket{\phi}) \nonumber \\ 
	&= \alpha^* U \K \ket{\psi} + \beta^* U \K \ket{\phi} \nonumber \\
	&= \alpha^* \TR \ket{\psi} + \beta^* \TR \ket{\phi} \nonumber.
\end{align} 
These are the properties of an anti-linear operator. Apart from these $\TR$ has one more property
\begin{align} \label{eq:antiunit}
	\ip{\TR \psi}{\TR \phi} &= \ip{U \K \psi}{U \K \phi}  \nonumber \\
	&= \ip{U \psi^*}{U \phi^*} = \ip*{\psi^*}{U^{\dagger} U\phi^*} = \ip{\psi^*}{\phi^*} \nonumber \\
	&= \ip{\psi}{\phi}^* = \ip{\phi}{\psi}.
\end{align}
An operator which satisfies Eq.~\eqref{eq:antiunit} is called the \emph{anti-unitary} operator~\cite{Wigner1960}. Now, if we apply the time-reversal operator twice, it should leave the system unchanged, i.e.
\begin{align}
	\TR^2 = U \K U \K = U U^* \K \K = U U^*
\end{align}
\emph{NOTE: From this we can show that $U U^*$ is equivalent (related by a unitary transformation) to its complex conjugate
	\begin{align}
		U U^* = U U^* U U ^{-1} = U ( U U^*)^*  U ^{-1}.
\end{align}}Since $\TR^2$ leaves the system unchanged, i.e.
\begin{align}
	U U^* &= U (U^{-1})^T = e^{i \phi} \nonumber \\
	U (U^{-1})^T U^T &= U ( U U^{-1})^T = U = e^{i \phi} U^T \nonumber \\
	\implies U &= e^{i \phi} (e^{i \phi} U^T)^T = e^{2 i \phi} U\nonumber
\end{align}
which implies $e^{i \phi} = \pm 1$ and consequently 
\begin{equation}
	\TR^2 = U U^* = e^{i \phi} = \pm \mathds{1}.
\end{equation}	
Using the action of the complex conjugation operator $ \K $ on the position eigenvector which is $\K \ket{\psi} = \ket{\psi}$, we obtain the action of the complex conjugation operator $ \K $ on the position operator as 
\begin{equation}
	\K \hat{x} \K^{-1} = \hat{x}
\end{equation}
and the action on the momentum operator can be obtained as
\begin{align}
	\K \hat{p} \K^{-1} \ket{x} &= \K \hat{p} \ket{x} = \K \left(- i \hbar \dfrac{\partial}{\partial x}\right) \ket{x} = i \hbar \dfrac{\partial}{\partial x} \ket{x} = -p \ket{x}.
\end{align}
which gives us
\begin{equation}
	\K \hat{p} \K^{-1} = -\hat{p}.
\end{equation}
Now, for a spinless particle, we have only $\hat{x}$ and $\hat{p}$ and the action of $\TR$ on $\hat{x}$ and $\hat{p}$ is given by
\begin{align}
	\TR \hat{x} \TR^{-1} = \hat{x}, \qquad \TR \hat{p} \TR^{-1} = -\hat{p}
\end{align}
which implies $U = \mathds{1}$ and $ \TR = \K $. Consequently, we have
\begin{equation}
	\TR^2 = \K^2 = + \mathds{1}.
\end{equation}
On the other hand, if we consider a particle with a spin (say spin-1/2). Then, there is one more job that $\TR$ has to do. The time reversal operation flips the spin, i.e.
\begin{equation}
	\TR \boldsymbol{\sigma} \TR^{-1} = - \boldsymbol{\sigma}.
\end{equation}
where the components of $ \boldsymbol{\sigma} $ are given by 
\begin{equation}
	\sigma_x = \begin{pmatrix}
		0 & 1\\
		1 & 0
	\end{pmatrix}, \;\;\; \sigma_y = \begin{pmatrix}
		0 & -i\\
		i & 0
	\end{pmatrix}, \;\;\; \sigma_z = \begin{pmatrix}
		1 & 0\\
		0 & -1
	\end{pmatrix}
\end{equation}
\begin{equation}
	\implies \K \sigma_x \K^{-1} = \sigma_x, \;\;\; \K \sigma_y \K^{-1} = -\sigma_y, \;\;\; \K \sigma_z \K^{-1} = \sigma_z
\end{equation}
Furthermore, using $ \TR \boldsymbol{\sigma} \TR^{-1} = - \boldsymbol{\sigma} $ and $ \TR = U \K $ we can derive the following
\begin{equation}
	U \sigma_x U^{-1} = -\sigma_x, \;\;\; U \sigma_y U^{-1} = \sigma_y, \;\;\; U \sigma_z U^{-1} = -\sigma_z
\end{equation}
which implies $U = c_0 \sigma_y $, where $c_0$ is some constant. We can choose $c_0 = 1$, using the unitarity of $U$, which results in
\begin{equation}
	\TR = \sigma_y \K
\end{equation}
such that $ \TR^2 =  \sigma_y \K \sigma_y \K = \sigma_y \sigma_y^* \K^2 = - \mathds{1} $. In conclusion, for a quantum system with \textbf{half-integer} spin, the operator $ \TR $ squares to negative of the identity, i.e. $\TR^2$ = $-\mathds{1}$. Similarly, one can show that for a quantum system without spin or \textbf{integer} spin, the time reversal operator squares to $\TR^2$ = $\K^2$ = $\mathds{1}$.

\noindent A quantum particle moving in a periodic potential exhibits translation symmetry. In that case, the eigenvalues of the position and momentum are the same, and the Hamiltonian will be block-diagonalized in momentum basis and written in second-quantization as
\begin{equation}
	\hat{H} = \sum_{\vb{k}} b^{\dagger}_{\vb{k}} H (\vb{k}) b_{\vb{k}}
\end{equation} 
The time-reversal invariance for $H(\vb{k})$ for quasi-momentum $\vb{k}$ is expressed as
\begin{equation}
	\TR H (\vb{k}) \TR^{-1} = U \K H(\vb{k}) \K^{-1} U^{-1} = U H^*(\vb{k}) U^{-1} = H(-\vb{k}).
\end{equation}
In continuation, we observe that if $\ket{\psi(\vb{k})}$ is an eigenstate of $H(\vb{k})$ with energy $E(\vb{k})$ then $\TR \ket{\psi(\vb{k})}$ is an eigenstate of $H(-\vb{k})$ with energy $E(-\vb{k})$ which is the same as $E(\vb{k})$. Further, we can show 
\begin{equation*}
	\ip{\TR \psi(\vb{k})}{\psi(\vb{k})} = \ip{\TR \psi(\vb{k})}{\TR^2 \psi(\vb{k})} = - \ip{\TR \psi(\vb{k})}{\psi(\vb{k})}
\end{equation*}
\begin{equation}
	\implies \ip{\TR \psi(\vb{k})}{\psi(\vb{k})} = 0
\end{equation}
where we used the fact that $ \TR $ preserves the norm (or probability) and $ \TR^2 = - \mathds{1} $. Thus, $ \ket{\psi(\vb{k})} $ and $ \TR \ket{\psi(\vb{k})} $ are two different states with the same energy $E(\vb{k})$. This is called \emph{Kramer's degeneracy}. Note that only the system with time-reversal symmetry and $\TR^2 = - \mathds{1}$ exhibits this kind of degeneracy. 

\subsection{Particle-hole or Charge conjugation symmetry (PHS)}
The next symmetry that plays an integral role in the classification of single particle Hamiltonians is the particle-hole ($\mathcal{C}$). In the context of high-energy physics, charge conjugation refers to a transformation of a particle into its antiparticle. However, in a condensed matter system, this symmetry turns creating particles into creating holes~\cite{SchnyderRyu2016}. The invariance of the Hamiltonian under charge conjugation results the following condition 
\begin{equation}
	\mathcal{C} H \mathcal{C}^{-1} = -H \implies U_{\mathcal{C}} H U_{\mathcal{C}}^{-1} = -H
\end{equation}
Therefore, the particle-hole operator $\mathcal{C} = U_{\mathcal{C}} \K$ is an anti-unitary operator that anti-commutes with the Hamiltonian. Like in the case of time-reversal, we can show that the action of the particle-hole operator on the single-particle Hamiltonian, in momentum space, reads
\begin{equation}
	\mathcal{C} H(\vb{k}) \mathcal{C}^{-1} = -H(\vb{-k}) \implies U_{\mathcal{C}} H^*(\vb{k}) U_{\mathcal{C}}^{-1} = -H(\vb{-k})
\end{equation} 
The operator $\mathcal{C}$ also squares to $\pm 1$. In this case, given $\ket{\psi(\vb{k})}$ is an eigenstate of $H(\vb{k})$ with energy $E_+(\vb{k})$, it has a correspondent $\mathcal{C} \ket{\psi(\vb{k})}$, with energy $E_-(-\vb{k}) = -E_+(\vb{k})$. The PHS is responsible for a symmetric energy band structure about $k = 0$ as shown in Fig.~\ref{fig:symmetry}.  

\subsection{Chiral or Sub-lattice symmetry (CS)} \label{subsec:chiralsymm}
\begin{figure}[H]
	\centering
	\includegraphics[width=8cm]{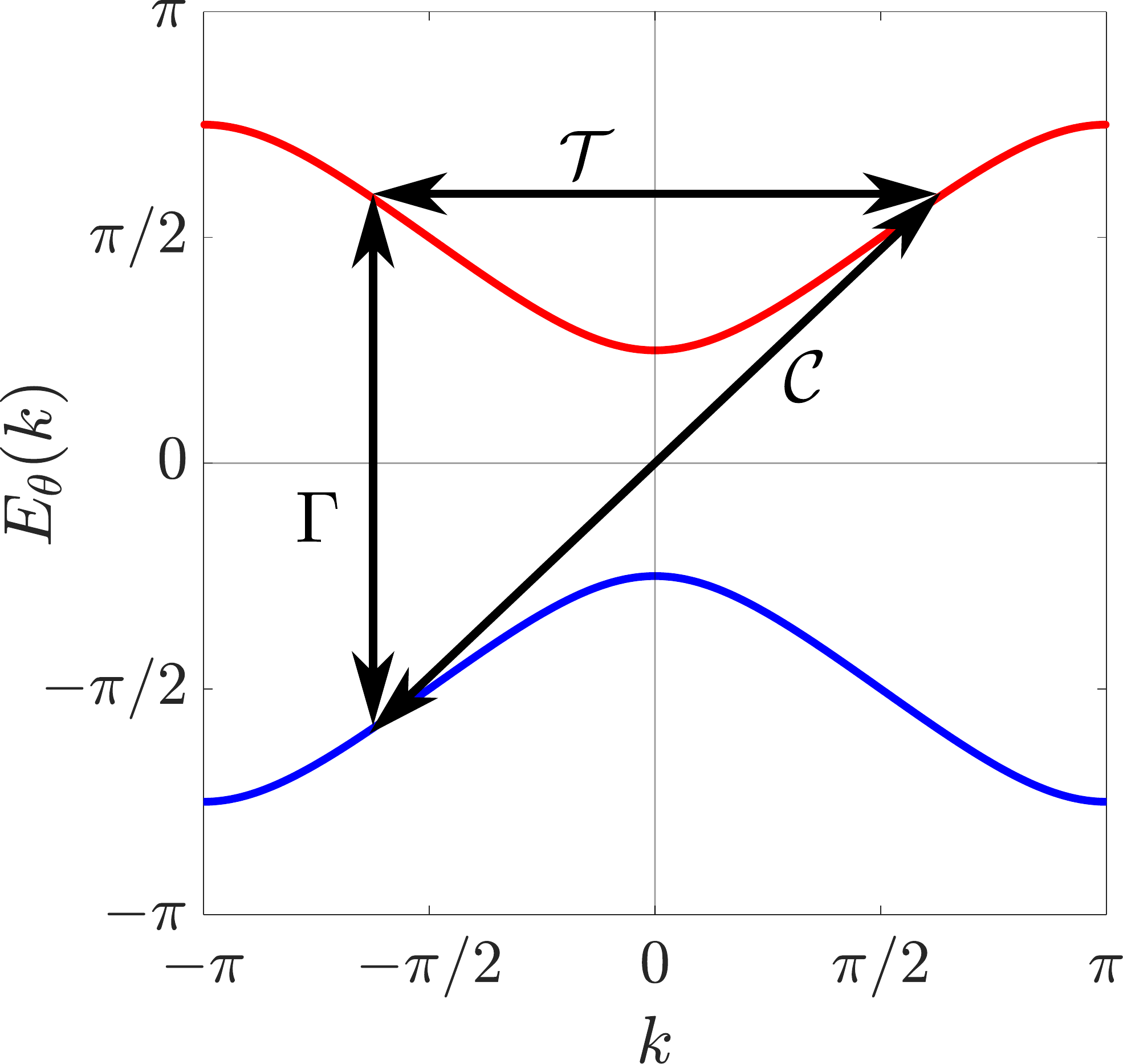}
	\caption{Dispersion relation of a system that exhibits all three, time-reversal, charge-conjugation and chiral symmetry. The arrows represent the relation between the states $ \ket{\psi(\vb{k})} $ and $ \TR \ket{\psi(\vb{k})} $, $ \mathcal{C} \ket{\psi(\vb{k})} $, $ \Gamma\ket{\psi(\vb{k})} $ which is discussed in the text.}
	\label{fig:symmetry}
\end{figure}

Having talked about TRS and PHS, we can construct another symmetry by combining the previous two symmetries (time-reversal and charge-conjugation) as $\Gamma = \TR \vdot \mathcal{C}$. (Note that we can also take $ \Gamma = \mathcal{C} \vdot \mathcal{T} $, which only corresponds to the change of the basis.) This new symmetry $\Gamma$ is a unitary symmetry that does not commute but anti-commutes with the Hamiltonian~\cite{LudwigRyu2010} and hence $ \TR \vdot \mathcal{C} $ does not correspond to an 'ordinary' Hamiltonian symmetry. This is the reason why we consider chiral symmetry $\Gamma$ as an additional essential ingredient for the classification of the blocks of the irreducible representation, other than time-reversal $\mathcal{T}$ and charge-conjugation (particle–hole) symmetries $\mathcal{C}$. Mathematically, it satisfies the following relation
\begin{equation}
	\Gamma H \Gamma^{-1} = - H.
\end{equation} 
It is interesting to note here that this unitary symmetry does not represent a conventional symmetry in quantum mechanics because it anti-commutes with the Hamiltonian, whereas conventionally we talk about symmetries which commute with the Hamiltonian. In the momentum space, it acts as
\begin{equation}
	\Gamma H(\vb{k}) \Gamma^{-1} = -\Gamma(\vb{k}).
\end{equation}
Since $\Gamma$ is unitary, it always squares to $ +\mathds{1} $. The spectrum of a system exhibiting chiral symmetry has eigenstates $ \ket{\psi(\vb{k})}$ and $ \Gamma\ket{\psi(\vb{k})} $ as shown in Fig.~\ref{fig:symmetry}. This is the last ingredient in order to talk about topological classes based on the existence of the symmetries in the system.

\section{Tenfold Way by AZ}
Having all the symmetries that a system can possess at our disposal, we can talk about symmtery classes. Depending on the type of the symmetry our system possess among TRS, PHS and CS we have symmetry classification. The system can have TRS in three possible ways: i) absence of time-reversal ($\TR = 0$), ii) presence of time-reversal with $\TR^2 = +1$, iii) presence of time-reversal with $\TR^2 = -1$. Similarly, we have such distinct possibilities with PHS and CS. Consequently, we have precisely ten symmetry classes given in Table~\eqref{table:AZClass} which are due to Altland and Zirnbauer~\cite{Zirnbauer1996,Altland1997,Heinzner2005}. which have completed the earlier classification due to Wigner and Dyson~\cite{Dyson1962}.
\begin{table}[ht]
	\centering 
	\begin{tabular}{c c c c c} 
		\hline\hline
		& \textbf{Cartan label} & \textbf{TRS} &\textbf{ PHS} & \textbf{CS} \\ [0.5ex] 
		\hline \hline
		\textbf{Standard Classes} & A & 0 & 0 & 0 \\ 
		& AI & $+1$ & 0 & 0 \\
		& AII & $-1$ & 0 & 0 \\ [2ex]
		\textbf{Chiral Classes} & AIII & 0 & 0 & 1 \\ 
		& BDI & $+1$ & $+1$ & 1 \\
		& CII & $-1$ & $-1$ & 1 \\ [2ex]
		\textbf{BdG classes} & D & 0 & $+1$ & 0 \\
		& C & 0 & $-1$ & 0 \\
		& DII & $-1$ & $+1$ & 1 \\
		& CI & $+1$ & $-1$ & 1 \\ [1ex] 
		\hline 
	\end{tabular}
	\caption{10-fold way.}
	\label{table:AZClass} 
\end{table}

\section{1D Tight-Binding Model}
\begin{figure}[H]
	\centering
	\includegraphics[width=12cm]{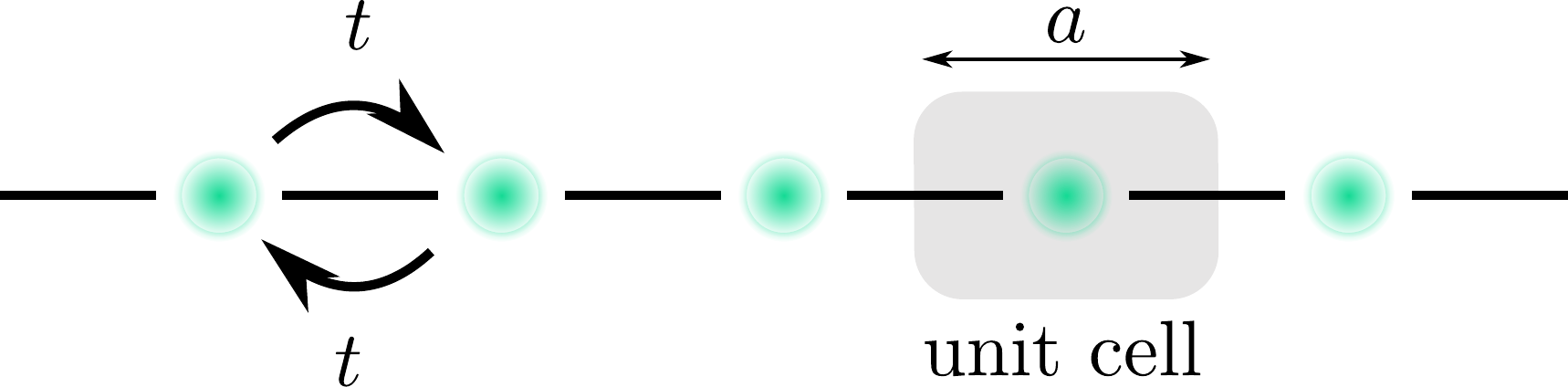}
	\caption{1D lattice with lattice spacing $a$ and the hopping amplitude between nearest neighbors $t$.}
	\label{fig:TB}
\end{figure}
\begin{figure}[H]
	\centering
	\includegraphics[width=8cm]{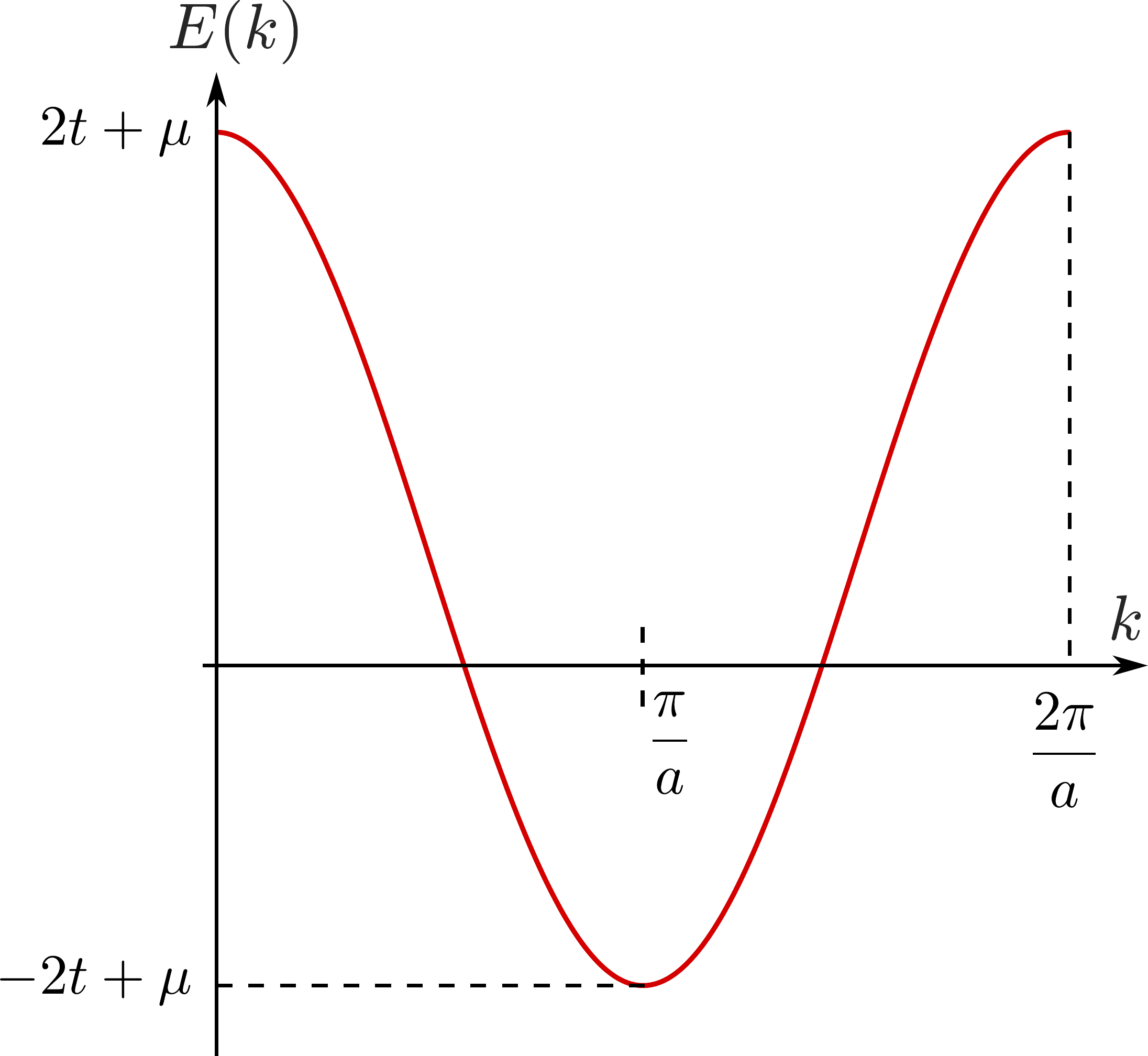}
	\caption{The plot of the dispersion relation derived in Eq.~\eqref{eq:DispersionTB}.}
	\label{fig:DispersionTB}
\end{figure}

In this section, consider an infinite 1D chain having $N$ sites with only nearest-neighbor interaction denoted by $t$ as shown in Fig.~\ref{fig:TB}. We will illustrate how unitary symmetries result in block diagonalization with this very simple model and extend it to the SSH model in the next section. For electrons (spinless) hopping on this 1D chain with amplitude $t$, we can write the Hamiltonian in second-quantized basis as
\begin{equation} \label{eq:1DTightBinding}
	\hat{H} = \sum_{i}^{N} t \hat{b}_{i+1}^{\dagger} \hat{b}_{i} + t \hat{b}_i^{\dagger} \hat{b}_{i+1} + \mu \hat{n}_i
\end{equation}
where $ \hat{b}_i^{\dagger} $ and $ \hat{b}_i $ are the fermionic (because of electrons) creation and annihilation operators, respectively, and $ \hat{n}_i \equiv \hat{b}_i^{\dagger} \hat{b}_i $ is the number or density operator. The creation and annihilation operators satisfy the commutation relations
\begin{equation}
	\acomm*{\hat{b}_i}{\hat{b}_j^{\dagger}} = \delta_{ij}, \qquad \acomm*{\hat{b}_i}{\hat{b}_j} = 0 = \acomm*{\hat{b}^{\dagger}_i}{\hat{b}^{\dagger}_j}.
\end{equation} 
These operators act on a state written in occupation number representation that looks like $ \ket{n_1, n_2, \dots, n_N} $ where $n_i$ is the occupation number of the $i$th site~\cite{Bruus2004, Fetter2003,Altland2010}. For fermions, these occupation numbers can only take a value of either zero or one, in accordance with Pauli's exclusion principle. For spinless electrons, we have
\begin{gather}
	\hat{b}^\dagger_i \ket{0,0,\dots,0} = \ket{0,\dots,1_i,\dots,0}, \;\;\; \hat{b}_i \ket{0} = 0 \;\;\; \hat{b}_i \ket{0,0,\dots,1_i, \dots, 0} = 0 \nonumber \\
	\hat{b}^\dagger_i \hat{b}_i \ket{n_1, n_2, \dots,n_i,\dots n_N} = \hat{n}_i \ket{n_1, n_2, \dots,n_i,\dots n_N} = n_i \ket{n_1, n_2, \dots,n_i,\dots n_N}, \;\;\; n_i \in [0,1].
\end{gather}
The last term in Eq.~\eqref{eq:1DTightBinding} corresponds to the chemical potential $\mu$. Now, we make use of the translational symmetry of the chain and write
\begin{equation} \label{eq:momtransformation}
	\hat{b}_i^{\dagger} = \dfrac{1}{\sqrt{N}} \sum_k e^{i k x_i} \hat{b}_k^{\dagger}, \;\;\; \hat{b}_i = \dfrac{1}{\sqrt{N}} \sum_k e^{-i k x_i} \hat{b}_k
\end{equation}
where $x_i = i a$ and $k = 2 \pi n/ N$ are the allowed momenta. Using the above relations, we can have
\begin{align} \label{eq:momentumoperator}
	\sum_{i} \hat{b}_{i+n}^{\dagger} \hat{b}_i &= \sum_{i} \left( \dfrac{1}{\sqrt{N}} \sum_k e^{i k x_{i+n}} \hat{b}_k^{\dagger} \right)  \left( \dfrac{1}{\sqrt{N}} \sum_{k'} e^{-i k' x_i} \hat{b}_{k'} \right) \nonumber \\
	&= \dfrac{1}{N} \sum_{k, k'} \sum_{i} e^{i k (x_i + na)} e^{-i k' x_i} \hat{b}_k^{\dagger} \hat{b}_{k'} \nonumber \\
	&= \sum_{k, k'} e^{i k na} \hat{b}_k^{\dagger} \hat{b}_{k'}\left( \dfrac{1}{N} \sum_{i} e^{i(k - k')x_i} \right) \nonumber \\ 
	&= \sum_{k, k'} e^{i k na} \hat{b}_k^{\dagger} \hat{b}_{k'} \delta_{k k'} \nonumber \\ 
	&= \sum_k e^{i k na} \hat{b}_k^{\dagger} \hat{b}_{k}.
\end{align} 
Therefore, utilizing Eq.~\eqref{eq:momentumoperator}, the Hamitonian in Eq.~\eqref{eq:1DTightBinding} can be written in momentum space as
\begin{equation}
	\hat{H} = \sum_k \left[t e^{i k a} + t e^{-i k a} + \mu \right] \hat{b}_k^{\dagger} \hat{b}_{k} = \sum_{k} E(k) \hat{b}_k^{\dagger} \hat{b}_{k}
\end{equation}
where
\begin{equation} \label{eq:DispersionTB}
	E(k) = 2t \cos(ka) + \mu
\end{equation}
which we plotted in Fig.~\ref{fig:DispersionTB}. In this example, it was straightforward to find a dispersion relation because we had only one degree of freedom. We will now extend the same analysis to a model in which we have two degrees of freedom in the next section. 
\section{Su-Schrieffer-Heeger (SSH) model}
\begin{figure}[H]
	\centering
	\includegraphics[width=14cm]{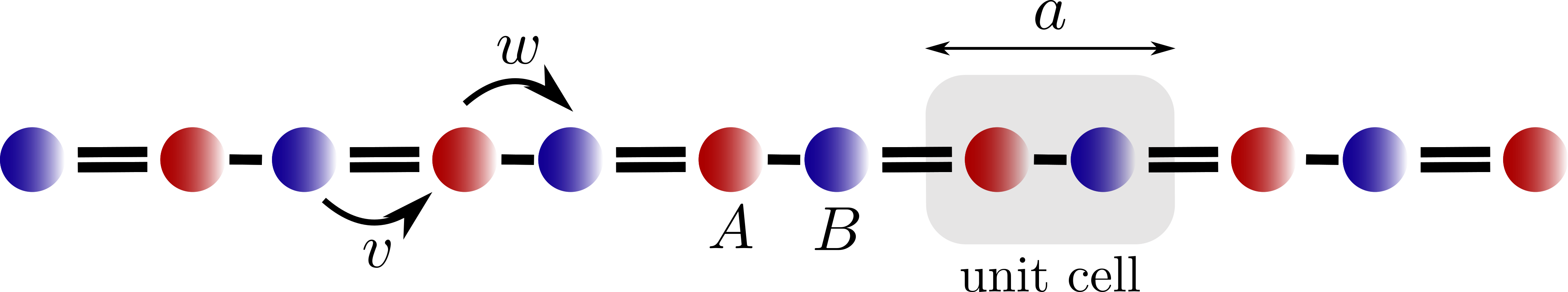}
	\caption{The geometry of SSH with lattice spacing $ a $. One unit cell consists of two sites $ A $ and $ B $, which is shown with shaded part. A chain consists of $ N $ unit cells.}
	\label{fig:SSH}
\end{figure}

\begin{figure}[H]
	\centering
	\subfigure[]{
		\includegraphics[height=3.55cm]{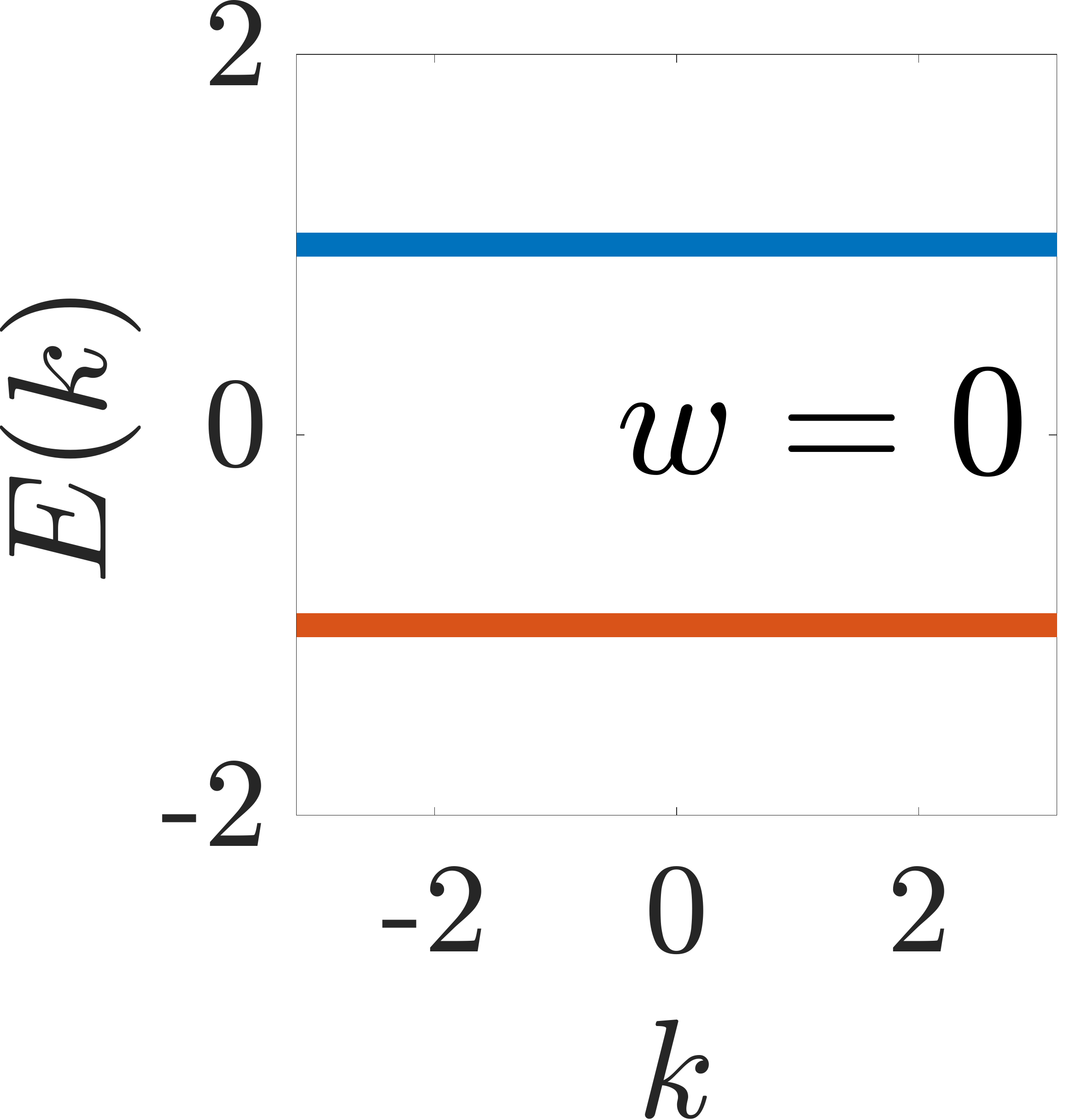}}
	\subfigure[]{
		\includegraphics[height=3.55cm]{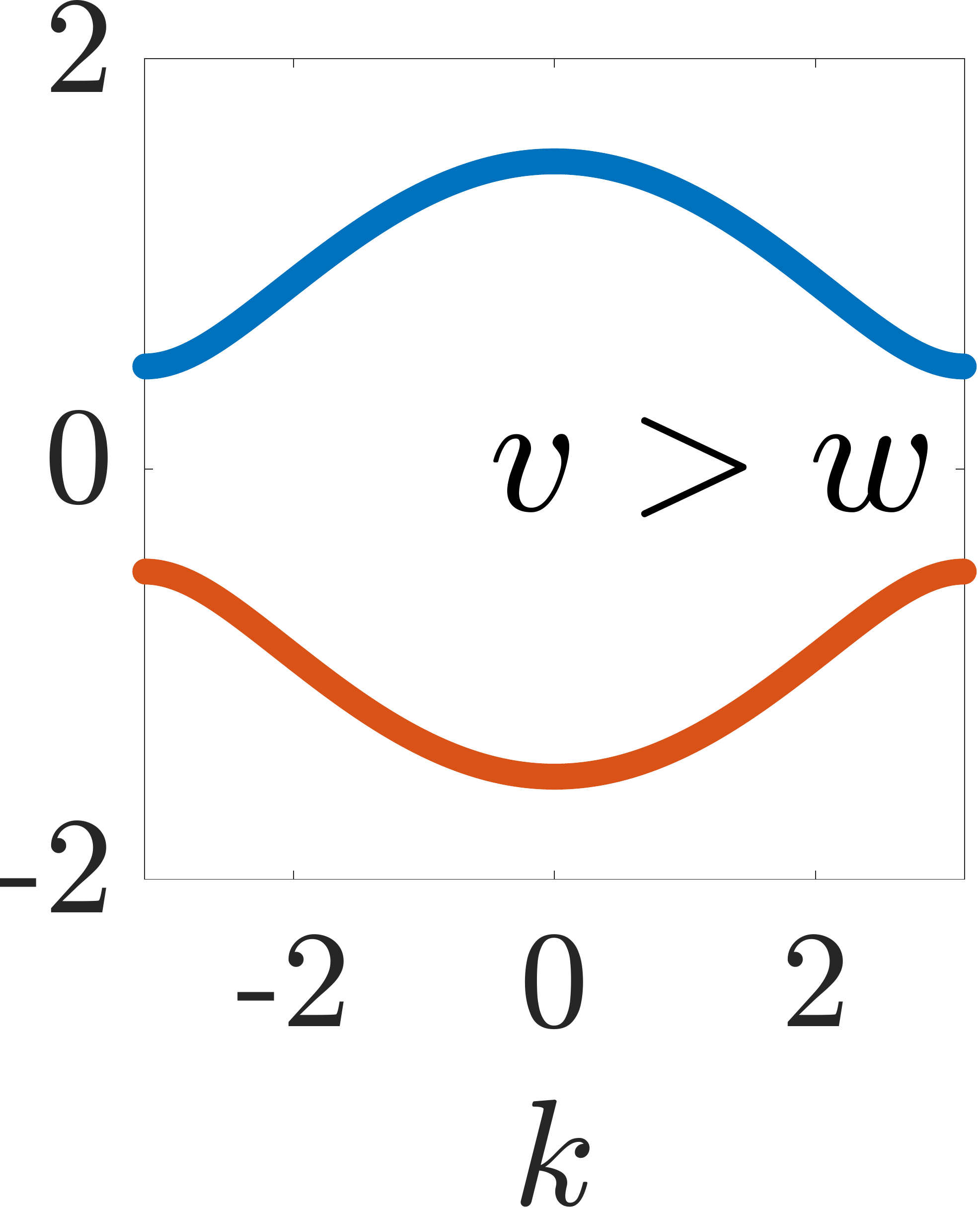}}
	\subfigure[]{
		\includegraphics[height=3.55cm]{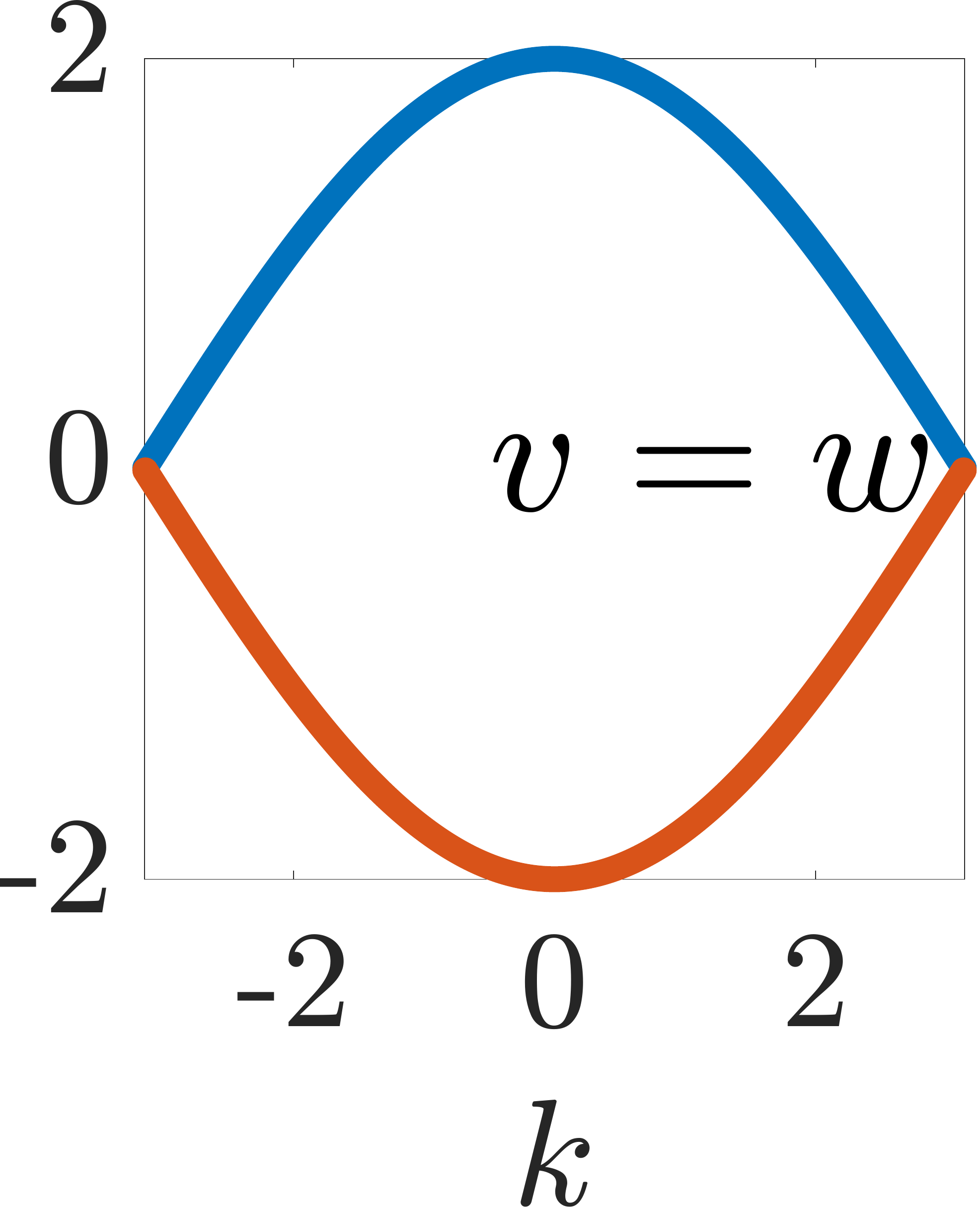}}
	\subfigure[]{
		\includegraphics[height=3.55cm]{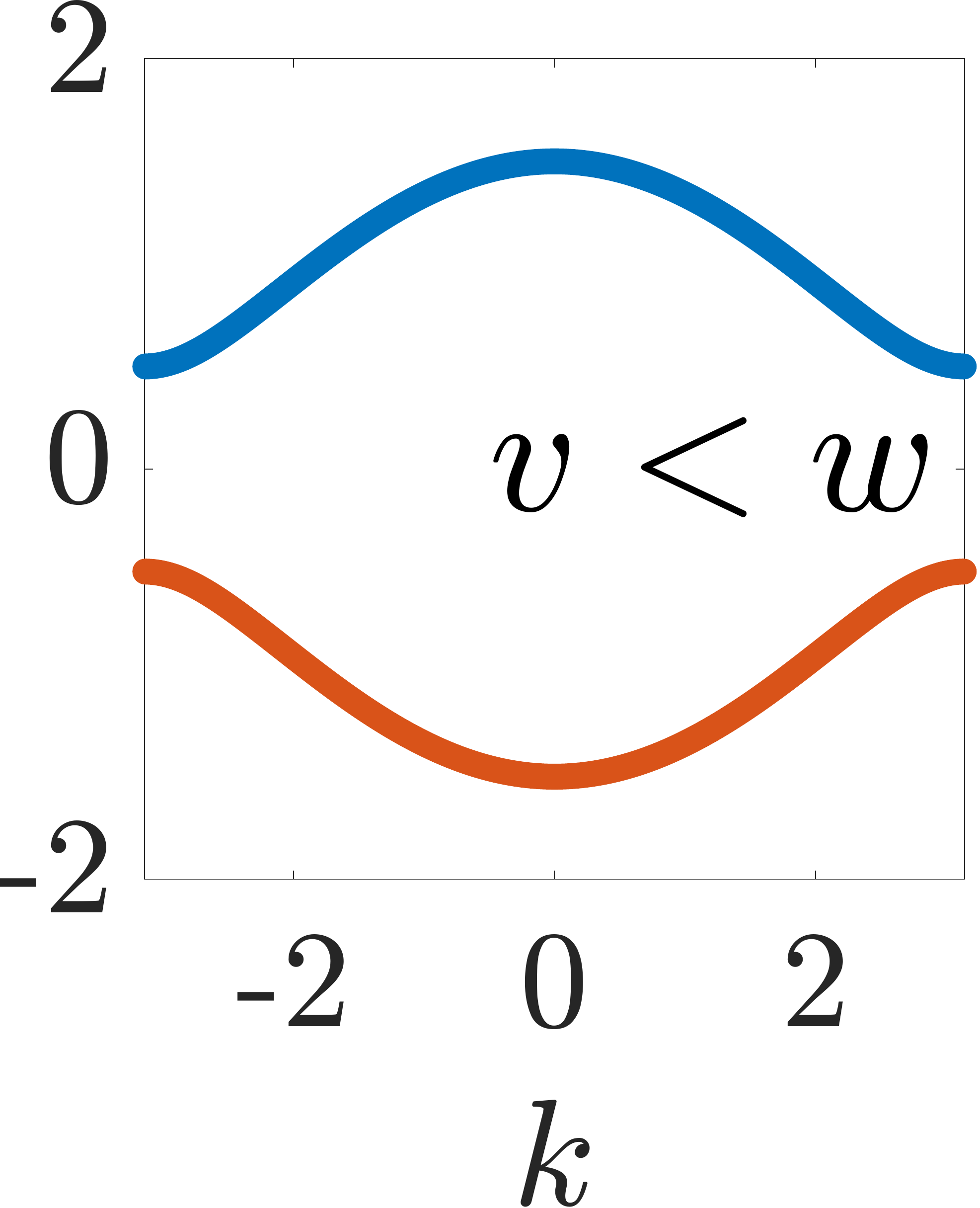}}
	\subfigure[]{
		\includegraphics[height=3.55cm]{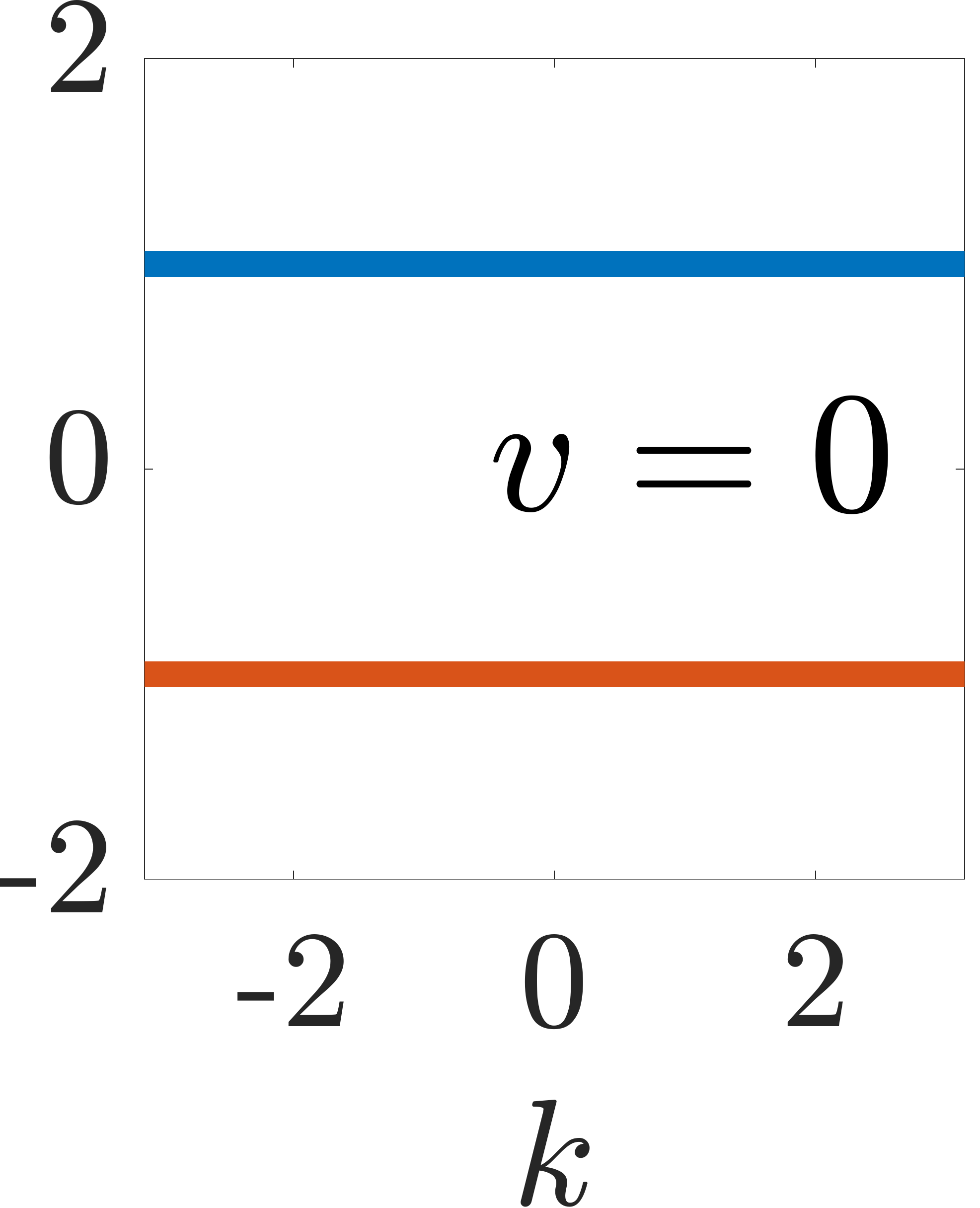}}
	
	\subfigure{
		\includegraphics[width=3.01cm]{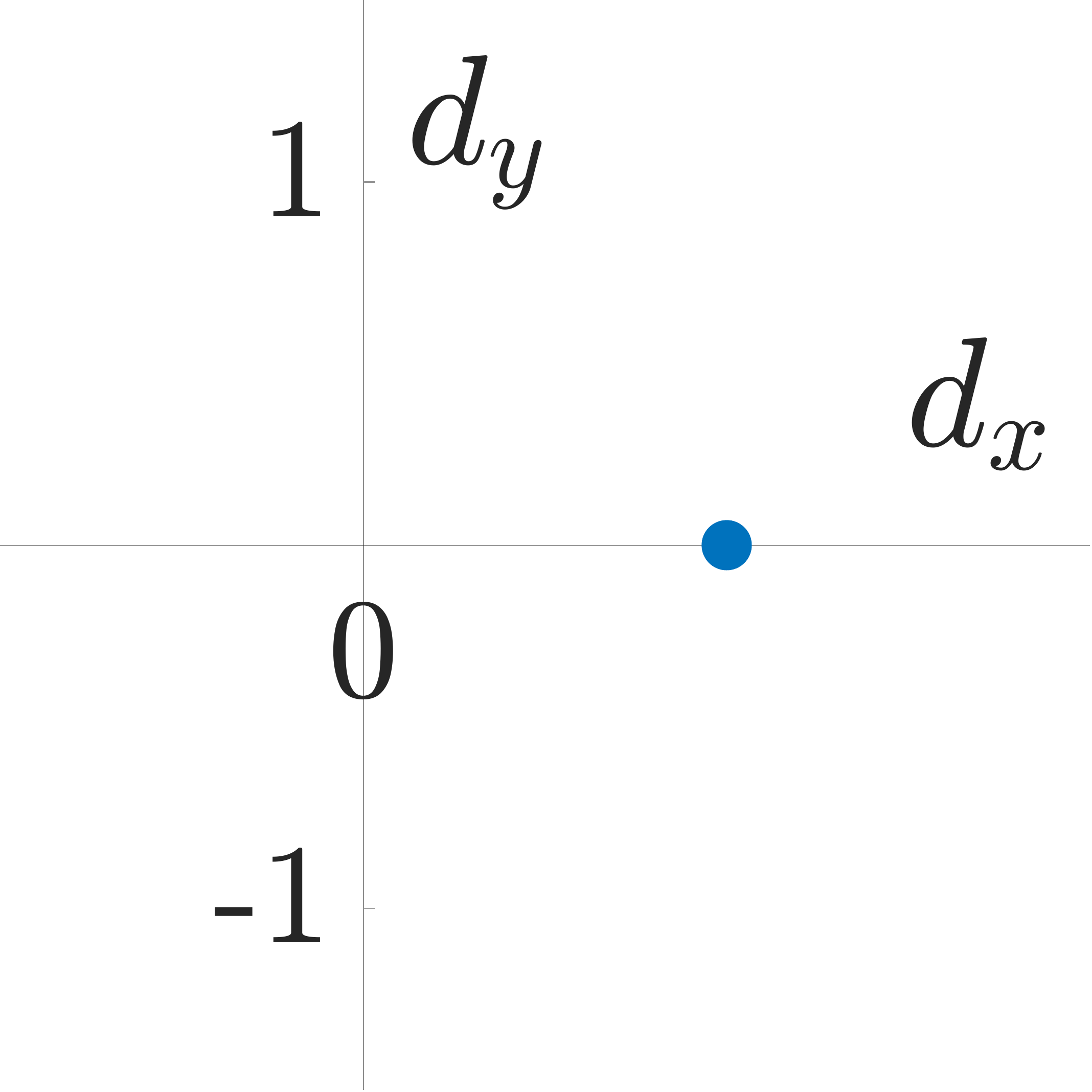}}
	\subfigure{
		\includegraphics[width=3.01cm]{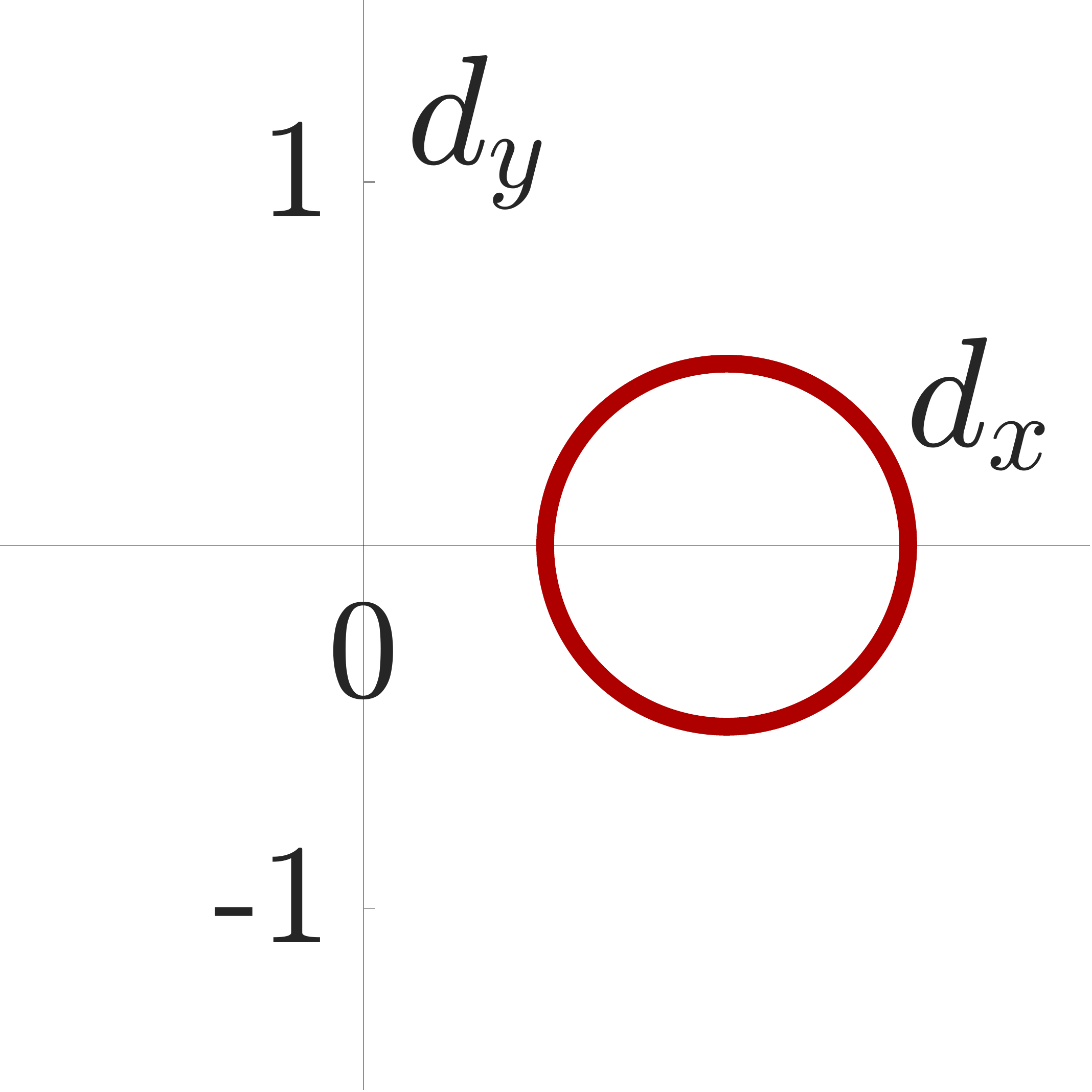}}
	\subfigure{
		\includegraphics[width=3.01cm]{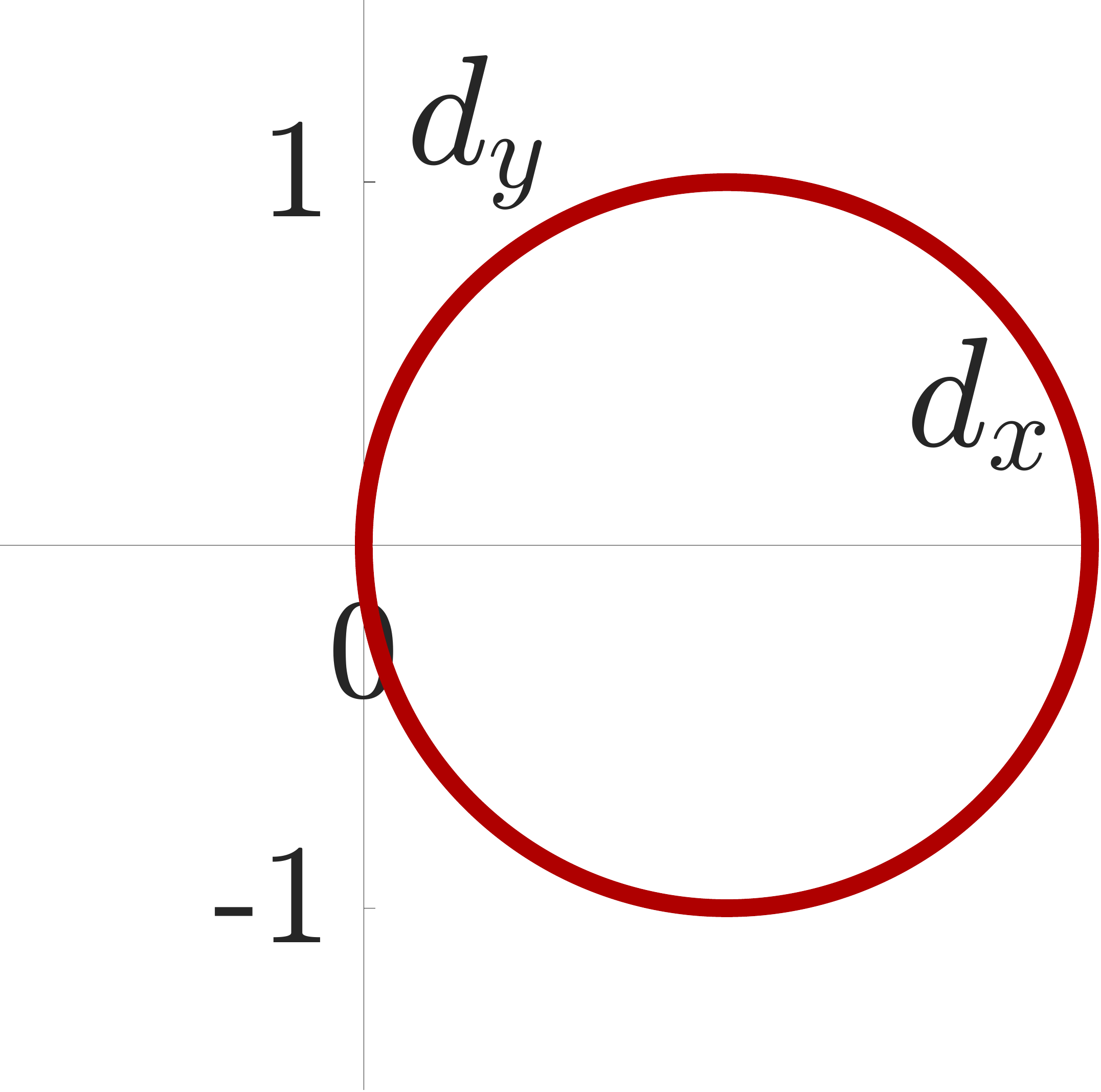}}
	\subfigure{
		\includegraphics[width=3.01cm]{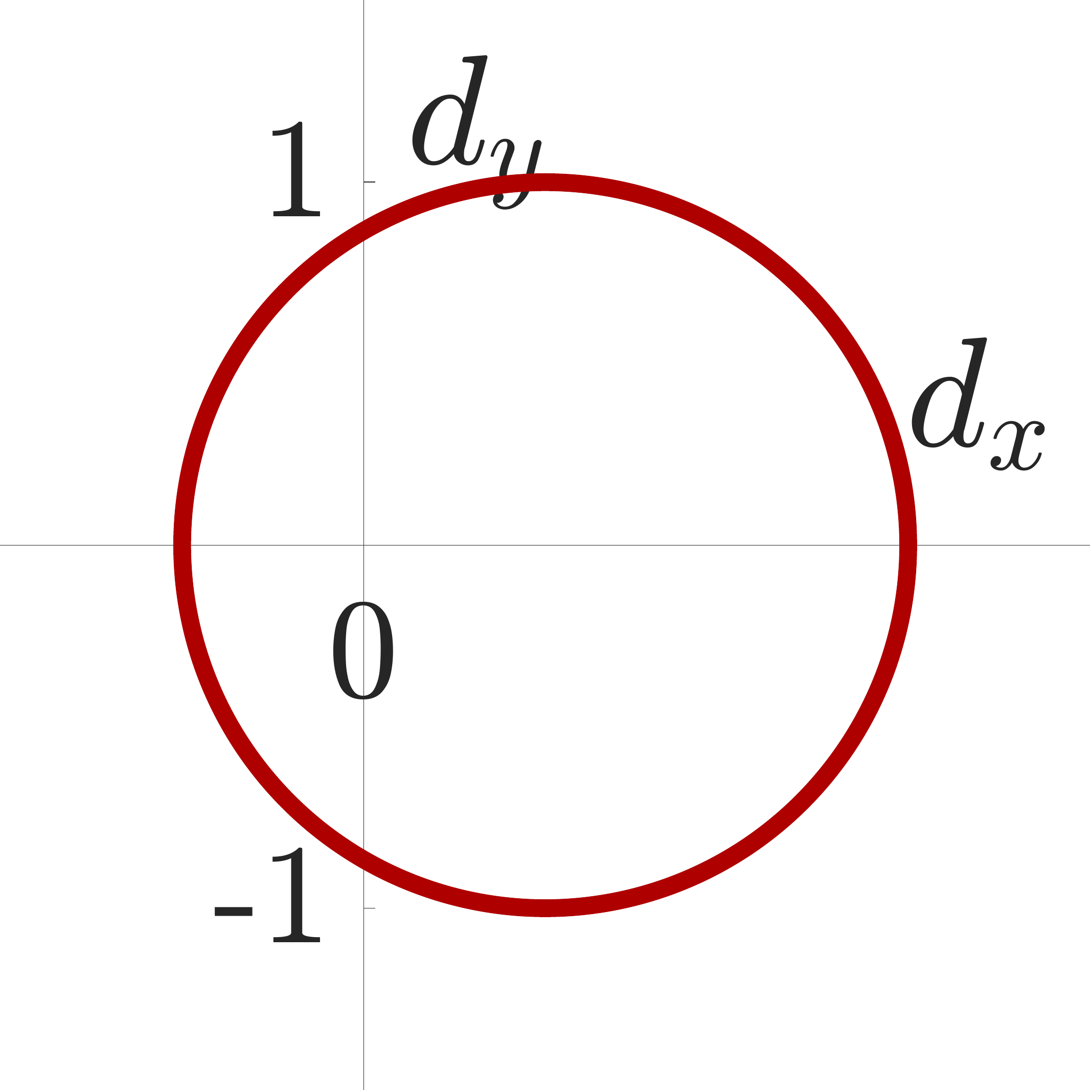}}
	\subfigure{
		\includegraphics[width=3.01cm]{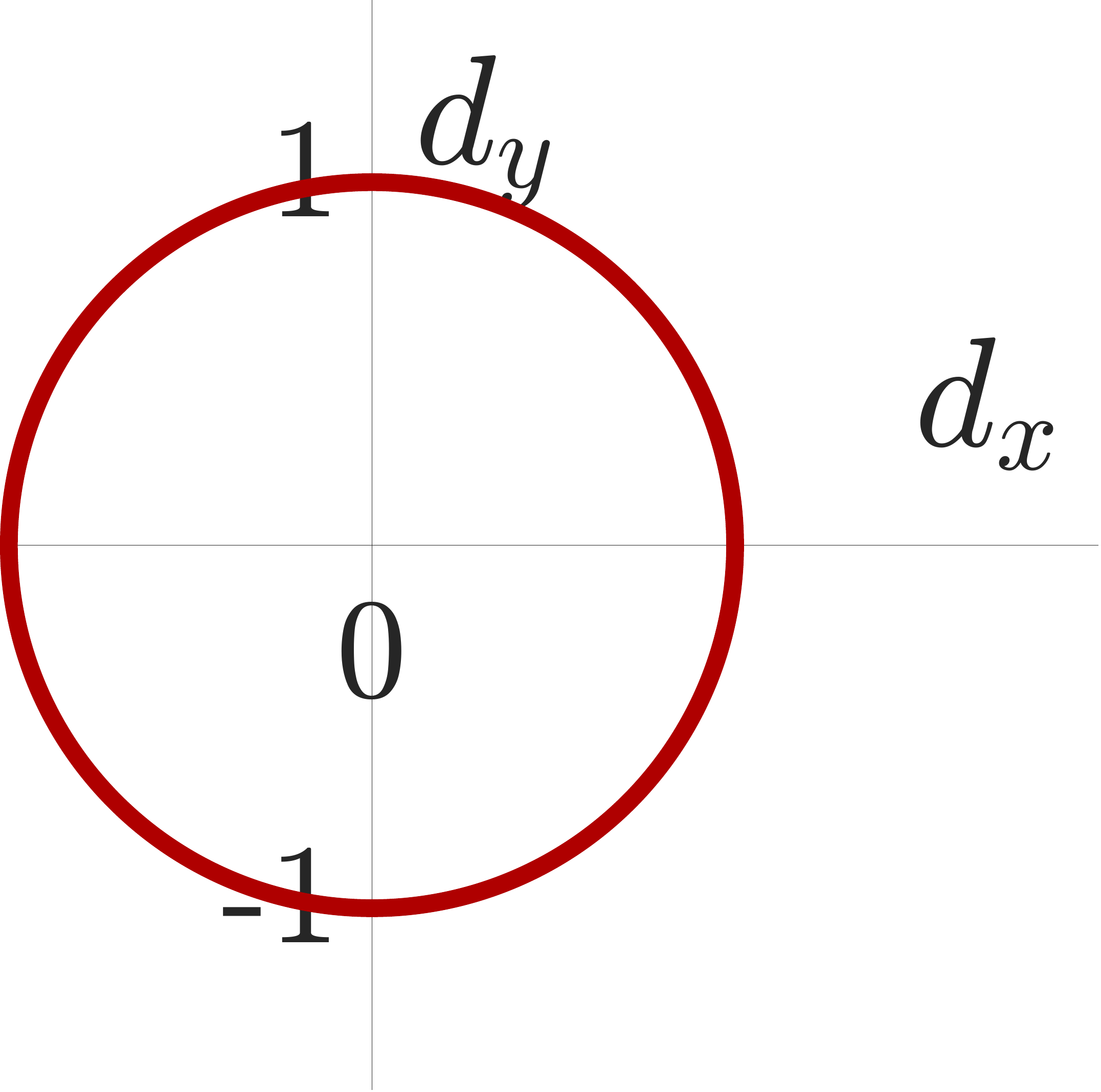}}
	\caption{Dispersion relation of the SSH model $ E_{\pm}(k) $ given in Eq.~\eqref{eq:DispersionSSH} for different settings of $v$ and $w$. (a) $v = 1, w = 0$, (b) $v = 1, w = 0.5$, (c) $v = 1, w = 1$, (d) $v = 0.5, w = 1$, (e) $v = 0, w = 1$. In the bottom row, the endpoints of the vectors $ \vb{d}(k)$ [given in Eq.~\eqref{eq:SSHBlochVector}] in the $d_x$-$d_y$ plane as $k$ takes values form 0 to $2\pi$ in first Brillouin zone.}
	\label{fig:SSHDispersion}
\end{figure}

The SSH was introduced to model polyactylene, which is a 1D carbon chain~\cite{SSH1979,SSH1980}. The model consists of a one-dimensional lattice (chain) in which each lattice site consists of a unit cell that also hosts two sites $A$ and $B$. The electrons jump on this one-dimensional chain with amplitudes $v$ and $w$ as shown in Fig.~\ref{fig:SSH}. Using this model, which has been implemented experimentally using ultracold atoms~\cite{Leder2016,Nakajima2016,Lohse2016,XiDai2013,Atala2013}, we illustrate the concepts of the single-particle Hamiltonian, symmetries of the Hamiltonian (unitary and anti-unitary), topological invariance, and bulk-boundary correspondence.
\noindent We consider here non-interacting particles, hence the dynamics of the system is governed by a single particle Hamiltonian which is written as~\cite{Franz,Asboth2016}
\begin{equation}
	\hat{H} = \sum_{n = 1}^{N} \left[ v \dyad{n, B}{n, A} + h.c \right] + \sum_{n = 1}^{N-1} \left[w \dyad{n+1,A}{n, B} + h.c\right]
\end{equation}
where $ h.c. $ is the Hermitian conjugate. For the case where we have $ N = 3 $ lattice sites, the Hamiltonian takes the form
\begin{equation}
	H = \begin{pmatrix}
		0 & v & 0 & 0 & 0 & 0 \\
		v & 0 & w & 0 & 0 & 0 \\
		0 & w & 0 & v & 0 & 0 \\
		0 & 0 & v & 0 & w & 0 \\
		0 & 0 & 0 & w & 0 & v \\
		0 & 0 & 0 & 0 & v & 0 \\
	\end{pmatrix}
\end{equation}
We separate the two degrees of freedom in external (unit cell) and internal (orbital/site index) as $\ket{n, \sigma} = \ket{n} \otimes \ket{\sigma} $ which belongs to a composite Hilbert space $\mathcal{H} = \mathcal{H}_{\text{external}} \otimes \mathcal{H}_{\text{internal}}$ where $n = 1,2, \dots, N$ and $\sigma = A, B$. Using this tensor product basis and choosing the computational basis $ \{\ket{0}, \ket{1}\} $ for $ \{A, B\} $, we can write Hamiltonian as
\begin{align}
	\hat{H} = &\sum_{n = 1}^{N} v \left[\dyad{n} \otimes (\dyad{1}{0} + \dyad{0}{1})  \right] + \sum_{n = 1}^{N-1} w \left[ \dyad{n+1}{n} \otimes \dyad{0}{1}  + \dyad{n}{n+1} \otimes \dyad{1}{0} \right] \nonumber \\
	= & v \sum_{n = 1}^{N} \left[\dyad{n} \otimes \sigma_x \right] + w \sum_{n = 1}^{N-1} \left[ \dyad{n+1}{n} \otimes \sigma_+  +  \dyad{n}{n+1} \otimes \sigma_- \right]
\end{align}
where $\sigma_x$ is the Pauli $x$ matrix and $\sigma_+ = \sigma_-^{\dagger}$ is the step-up operator in computational basis. First,we will concentrate on the bulk part of the chain by connecting the edges, leading to periodic boundary conditions. This corresponds to a closed ring with the slight different Hamiltonian which reads
\begin{equation}
	\hat{H}_{\text{bulk}} = \sum_{n = 1}^{N} \left[ v \dyad{n, B}{n, A} + w \dyad{(n\mod N)+1,A}{n, B} \right] + h.c
\end{equation}
By further exploiting the translational invariance of the system, we can go to quasi-momentum space by defining
\begin{equation}
	\ket{n, \sigma} = \dfrac{1}{\sqrt{N}} \sum_{k} e^{ikx_n} \ket{k, \sigma}; \sigma = A, B.
\end{equation}
where $ x_n = an $, and let us take $ a = 1 $ for the sake of simplicity. The bulk Hamiltonian in momentum basis reads
\begin{align}
	\hat{H}_{\text{bulk}} &= \dfrac{1}{N} \sum_{n = 1}^{N} \left[ v  \sum_{k,l} e^{ikn}e^{-iln} \dyad{k, A}{l, B} + w \sum_{s,t} e^{is(n+1)}e^{-itn} \dyad{s, B}{t, A} + h.c \right]	\nonumber \\
	&= \sum_{k} \left[ v \dyad{k, A}{k, B} + w e^{ik} \dyad{k, B}{k, A} + v \dyad{k, B}{k, A} + w e^{-ik} \dyad{k, A}{k, B}\right] \nonumber \\
	&= \sum_{k} \left[ \left(v + w e^{-ik} \right) \dyad{k, A}{k, B} + \left(v + w e^{ik} \right) \dyad{k, B}{k, A} \right] \nonumber.
\end{align}
where we used 
\begin{equation}
	\dfrac{1}{N} \sum_{n = 1}^{N} e^{i(k-l)n} = \delta_{kl}.
\end{equation}
We can further write it as
\begin{equation}
	\hat{H}_{\text{bulk}} = \sum_{k} \dyad{k} \otimes \begin{bmatrix}
		0 & v + w e^{ik} \\
		v + w e^{-ik} & 0
	\end{bmatrix} = \bigoplus_k H(k)
\end{equation}
where
\begin{equation} \label{eq:SSHHamil}
	H(k) = \vb{d}(k) \vdot \boldsymbol{\sigma}
\end{equation}
with the vector $\vb{d}(k)$ given by
\begin{equation} \label{eq:SSHBlochVector}
	\vb{d}(k) = (d_x, d_y, d_z) = (v+w\cos k, w \sin k, 0)
\end{equation}
and the dispersion relation;
\begin{equation} \label{eq:DispersionSSH}
	E_{\pm}(k) = \pm \sqrt{d_x^2 + d_y^2} = \pm \sqrt{v^2 + w^2 + 2vw\cos k} .
\end{equation}
We now plot the dispersion relation and the vector $\hat{\vb{d}}(k)$ for different settings of $v$ and $w$. We can write the Hamiltonian in the second-quantization basis also as
\begin{equation}
	\hat{H} = \sum_{n} \left[v \hat{b}^{\dagger}_{B,n} \hat{b}_{A,n} + w \hat{b}^{\dagger}_{A,n+1} \hat{b}_{B,n} + h.c.\right]
\end{equation} 
and again going to momentum basis using 
\begin{equation}
	\hat{b}_{i, \alpha} = \dfrac{1}{\sqrt{N}} \sum_k e^{i k x_i} \hat{b}_{k, \alpha}, \qquad \alpha = A,B
\end{equation} 
we get
\begin{equation}
	\hat{H} = \sum_{k,\alpha\beta} \hat{b}^{\dagger}_{k, \alpha} H_{\alpha \beta}(k)\hat{b}_{\beta, k}
\end{equation}
with $ H_{\alpha \beta}(k) $ same as $H(k)$ from Eq.~\eqref{eq:SSHHamil}. We plot the dispersion relation for different settings of the amplitudes $v$ and $w$ in Fig.~\ref{fig:SSHDispersion}. 
\noindent We observe that, as long as the hopping amplitudes are staggered, $v \ne w$, the two bands are separated by a band gap $\sim \abs{v - w}$ as shown in Fig.~\ref{fig:SSHDispersion}a,b,d,e. When the two amplitudes are the same, we see the closing of the band gap [see Fig.~\ref{fig:SSHDispersion}c] and the chain behaved as a conductor. We naturally come across this type of staggering in systems like polyacetylene, which is the result of Peierls' instability~\cite{Peierls1991,Thorne1996,Solyom2010}. 

\begin{figure}
	\centering
	\includegraphics[width=9.0cm]{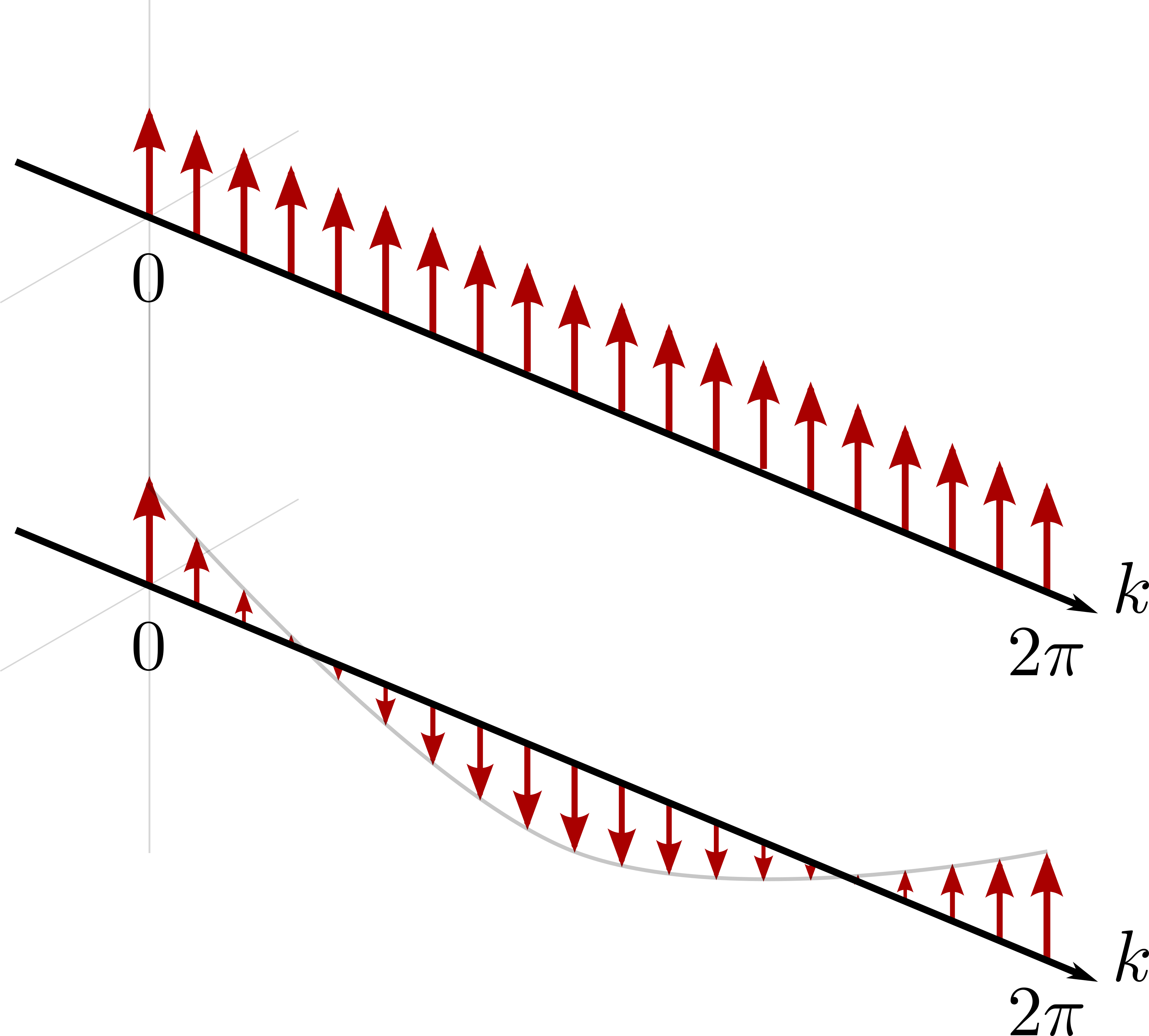}
	\caption{A schematic to illustrate the winding of the wavefunction $ \ket{\psi(k)} $, as $k$ goes around in the Brillouin zone, for the trivial (top) and the topological phase (bottom). Note, that the two phases shown above cannot are not smoothly connected due to the presence of a twist in the topologically non-trivial phase (bottom).}
	\label{fig:kwinding}
\end{figure}

\noindent So far, we have studied the dispersion relation to study the characteristics of the SSH model. Looking at Fig.~\ref{fig:SSHDispersion}, we see that the dispersion looks exactly the same in the two regimes, namely $v > w$ and $v < w$ and does not provide any interesting information in more detail. However, we can still further investigate the system using information about eigenstates, encoded in the vector $ \vb{d}(k) $. As $k$ takes values from $0$ to $2 \pi$ in the Brillouin zone, the tip of the vector $ \vb{d}(k) $ traces out a closed curve due to periodic boundary conditions and
\begin{gather}
	d_x = v + w\cos k, \;\;\; d_y = w \sin k \implies (d_x - v)^2 + d_y^2 = w^2
\end{gather}  
i.e. it traces out a circle of radius $w$, centered at $ (v,0) $. Note that in general, it does not need to be a circle. Here, we can define the topology of the loop by an integer called the bulk winding number $W$. The winding number refers to the number of times the loop winds around the origin of the $d_x$-$d_y$ plane. Therefore, it is evident from the bottom row of Fig.~\ref{fig:SSHDispersion}, that the two insulating phases (gapped phases) in the two regions, namely $v > w$ and $v < w$ are different. For example, the winding number, $W = 0$ for $v > w$ [Fig.~\ref{fig:SSHDispersion}a,b] and $W = 1$ for $v < w$ [Fig.~\ref{fig:SSHDispersion}d,e], whereas, for $v = w$, the circle just touched the origin and therefore the winding number is not defined for this parameter setting. As plotted in Fig.~\ref{fig:kwinding}, we can imagine a two-dimensional vector corresponding to each value of $k$ and see that the rotation of it as $k$ goes from 0 to 2$\pi$. We will refer to the insulating phase with $W = 0$ as \emph{trivial} and with $W = 1$ as \emph{topological}. In the next section, we will come out of the bulk and explore the edges.

\subsection{Edge states}
Till now, we have looked at the dispersion of the Hamiltonian under periodic boundary conditions. We also define the bulk winding number, which decides whether we have trivial ($ W = 0 $) or non-trivial topology ($ W = 1 $) in the system depending on the hopping amplitudes. In most practical situations, we have finite-size lattices, where we cannot use translational invariance and study bulk characteristics, as we did in the last section. In such cases, we need to consider open chains and solve the Hamiltonian for the dispersion in the real space. Here, an interesting question to ask is "What are the physical consequences of the non-zero winding number in the bulk we see on the edges or boundaries?" 

\begin{figure}[H]
	\centering
	\includegraphics[width=12cm]{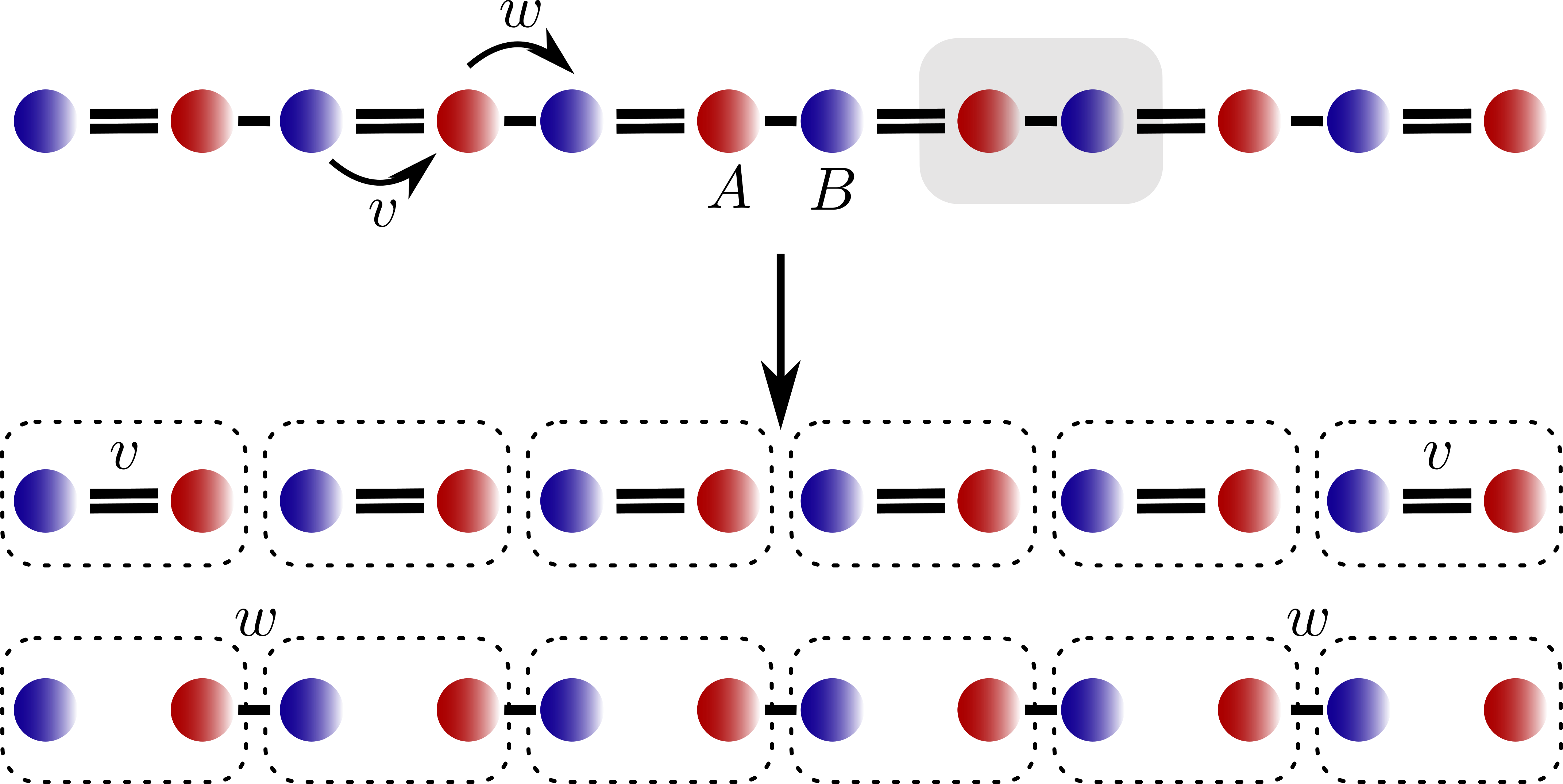}
	\caption{The two extremes cases of open chain of SSH in the dimerized limit.}
	\label{fig:EdgeStatesSSH}
\end{figure}

\noindent We start by looking at the two extreme cases, known as the dimerized limit~\cite{Asboth2016}: a) when we allow only intracell hopping and intercell hopping to vanish, i.e. $v \ne 0$ (set $1$ for simplicity), $w = 0$, b) when we have the reverse situation, i.e. $w \ne 0$ (set $1$ for simplicity), $v = 0$. In both cases, the SSH chain breaks down into dimers, as shown in Fig.~\ref{fig:EdgeStatesSSH}.
\noindent Since, we do not have periodicity anymore, we cannot find a dispersion relation as the function of momentum $k$, but we can still calculate it numerically by diagonalizing the Hamiltonian written in Eq.~\eqref{eq:SSHHamil}. The bulk has flat bands in the dimerized limits. In the \emph{trivial} case of $v = 1, w = 0$, we have $H(k) = \sigma_x$ and the energy eigenstates would simply be a superposition of the $ A $ and $ B $ sites for each dimer, i.e. 
\begin{equation}
	H ( \ket{n,A} \pm \ket{n,B}) = \pm ( \ket{n,A} \pm \ket{n,B}).
\end{equation}
Similarly, in the \emph{non-trivial} or \emph{topological} case of $v = 0, w = 1$, we have $H(k) = \sigma_x \cos k + \sigma_y \sin k$ and the energy eigenstates would be a superposition of the $ A $ and $ B $ sites of neighboring sites, i.e. 
\begin{equation}
	H ( \ket{n,A} \pm \ket{n+1,B}) = \pm ( \ket{n,A} \pm \ket{n+1,B}).
\end{equation}
However, in addition to these states, we have two additional states which come from the two single sites at the end of the chain [as shown at the bottom in Fig.~\ref{fig:EdgeStatesSSH}]. Each of these sights hosts a single eigenstate with zero energy because of the absence of onsite potentials in SSH. 

\begin{figure}[H]
	\centering
	\subfigure[]{
		\includegraphics[height=5.0cm]{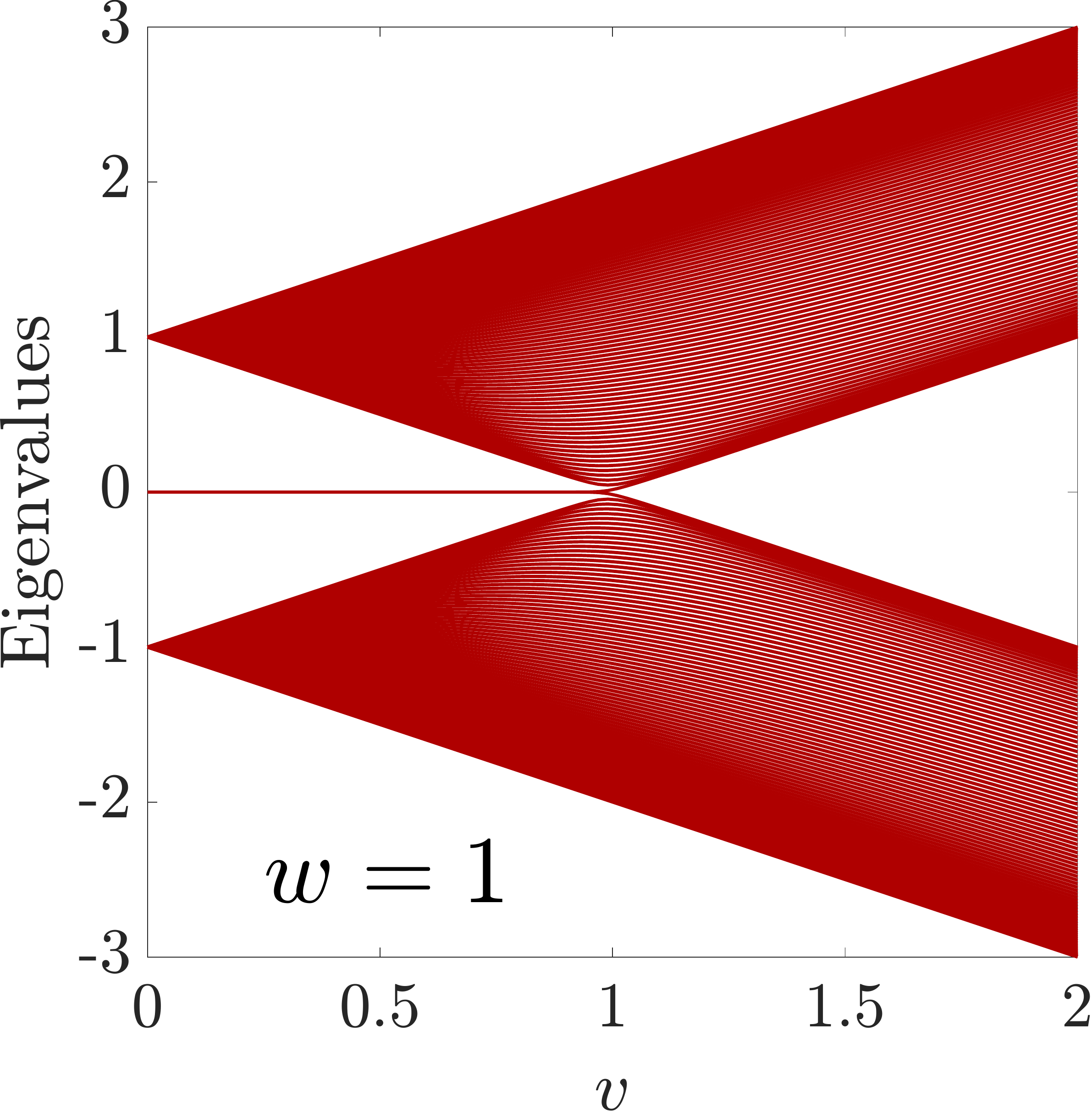}
		\label{fig:ev1}}
	\subfigure[]{
		\includegraphics[height=5.0cm]{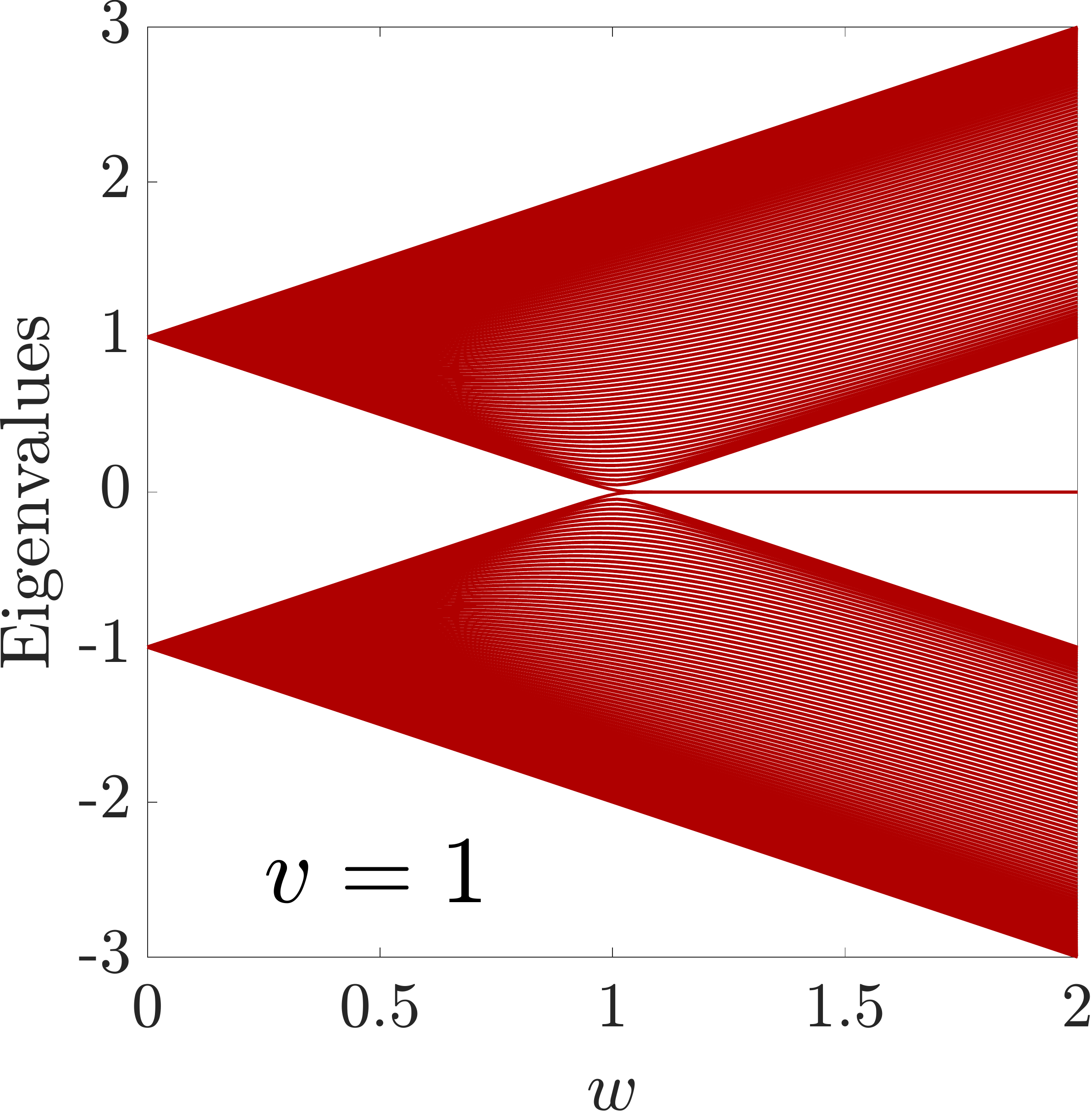}
		\label{fig:ev2}}
	\subfigure[]{
		\includegraphics[height=5.0cm]{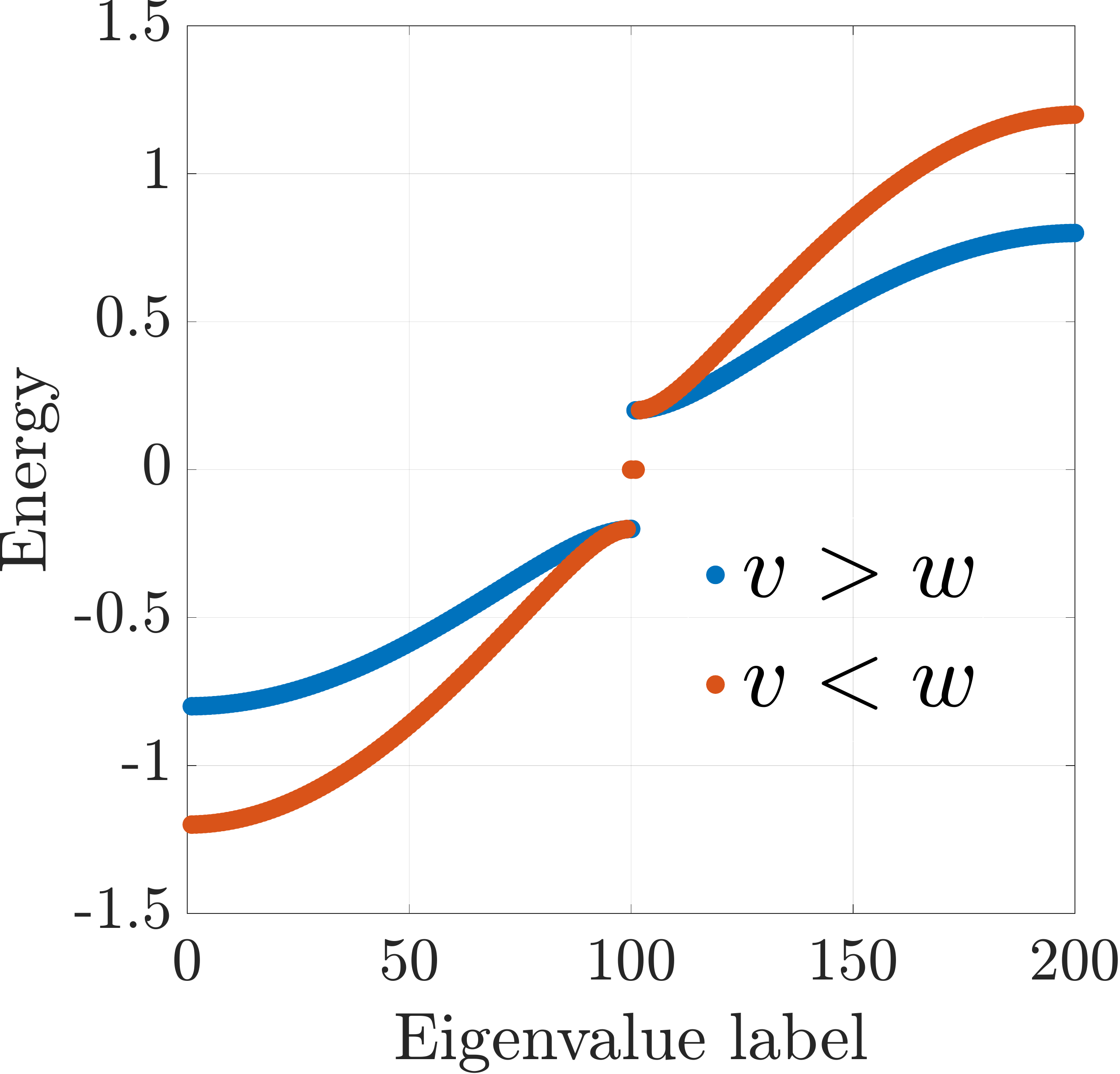}
		\label{fig:ev3}}
	\caption{(Energy spectrum of the finite chain with $N = 100$ unit cells and for \subref{fig:ev1} fixed $ w = 1$ and varying $ v $, \subref{fig:ev2} fixed $ v = 1$ and varying $ w $. \subref{fig:ev1} shows that $v > w$ and $v < w$ corresponds to trivial and topological phases respectively.}
	\label{fig:eigenvaluesSSH}
\end{figure}

\begin{figure}[H]
	\centering
	\subfigure[]{
		\includegraphics[width=12cm]{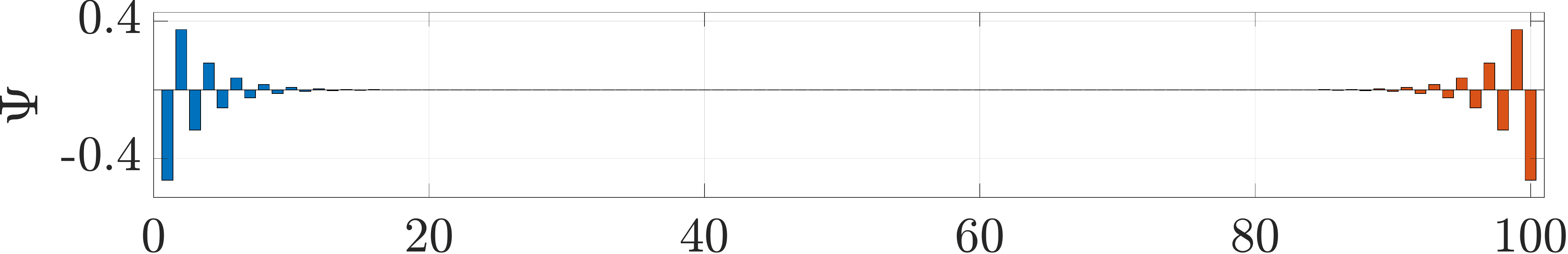}
		\label{fig:SSHWaveFunction1}}
	\subfigure[]{
		\includegraphics[width=12cm]{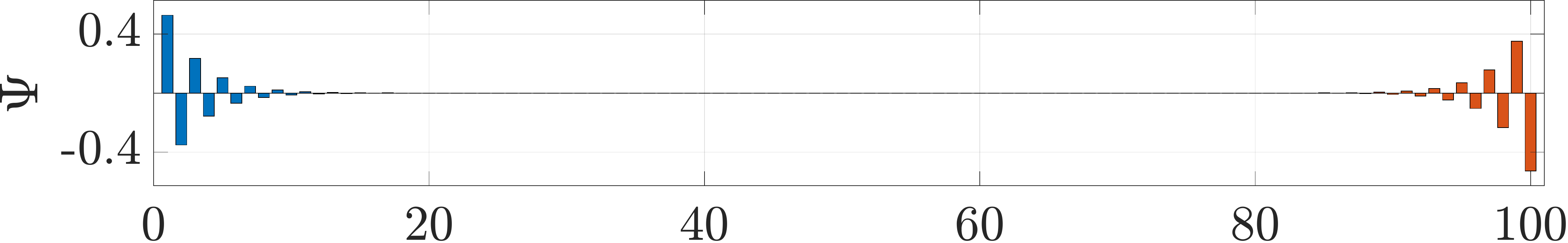}
		\label{fig:SSHWaveFunction2}}
	\subfigure[]{
		\includegraphics[width=12cm]{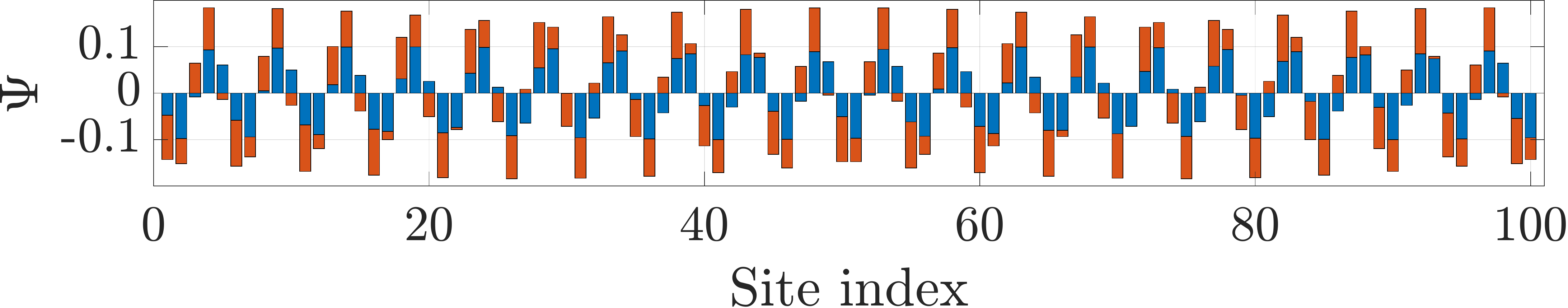}
		\label{fig:SSHWaveFunction3}}
	\caption{\subref{fig:SSHWaveFunction1}, \subref{fig:SSHWaveFunction2} The plot of the wavefunction corresponding to zero-energy edge states. The blue and orange bars represent the components on the sublattice $A$ and $B$ respectively. \subref{fig:SSHWaveFunction3} shown any bulk state for the reference which is delocalized all over the chain.} 
	\label{fig:SSHWaveFunction}
\end{figure}

\noindent In Fig.~\ref{fig:ev1},~\ref{fig:ev2}, we have plotted the spectrum of eigenvalues of the Hamiltonian corresponding to the open chain, having $N = 100$ unit cells, by varying the intra-cell hopping amplitude $v$ and fixed $w = 1 $ and vice versa. For the \emph{topological} case, we see the two states with zero energy, \emph{edge states}, in the band gap at $v = 0$ in Fig.~\ref{fig:ev1} and in \emph{trivial} case, we do not observe any zero mode states for $w = 0$ and $v = 1$ as shown in Fig.~\ref{fig:ev2}. Interestingly, we see that the system exhibits energy eigenstates that are very close to zero when we move away from the dimerized limit by switching on the intracell hopping parameter $v$. If these are really edge states, then the wavefunctions of these states have to be localized exponentially at both edges. When we plot the wavefunction, corresponding to these states in Figs.~\ref{fig:SSHWaveFunction1} and ~\ref{fig:SSHWaveFunction2}, we observe that the zero energy states are exponentially localized at the two edges of the chain. For reference, we have also plotted the wavefunction corresponding to a non-zero state in Fig.~\ref{fig:SSHWaveFunction3} and we see that this state is delocalized throughout the chain. We note another interesting thing in the plots of the zero-energy eigenstates, the left (right) edge state has a nonvanishing contribution only on sublattice $A$ ($B$). These edge states exist as long as $v < w$, which makes this regime different from the regime $v > w$, where we do not see any edge states. Lastly, in Fig.~\ref{fig:ev1}, we plot the spectrum of eigenvalues for the trivial $v > w$ and topological $v < w$ region.

\subsection{Symmetries of SSH model} 
The Hamiltonian of the SSH model exhibits sublattice or chiral symmetry (discussed in Sec.~\eqref{subsec:chiralsymm}) and hence belongs to AIII class. The chiral symmetry is represented by the operator 
\begin{equation}
	\Gamma = \sigma_z.
\end{equation}
The existence of chiral symmetry confines the vector $ \vb{d}(k) $ to the $d_x$-$d_y$ plane
\begin{equation}
	\sigma_z H(k) \sigma_z = -H(k)  \implies d_z = 0.
\end{equation}
Furthermore, with the given chiral symmetry operator, we can define the projectors that act on the sublattices $A$ and $B$ as
\begin{align}
	\mathcal{P}_A &= \dfrac{1}{2} \left( \mathds{1} + \Gamma \right) = \mathds{1}_{\text{lattice}} \otimes \dyad{A} \nonumber \\
	\mathcal{P}_B &= \dfrac{1}{2} \left( \mathds{1} - \Gamma \right) = \mathds{1}_{\text{lattice}} \otimes \dyad{B}
\end{align}
and can rewrite the Hamiltonian as
\begin{equation}
	H = \mathcal{P}_A H \mathcal{P}_B + \mathcal{P}_B H \mathcal{P}_A.
\end{equation}
\subsection{Winding number}
Formally, for a given Hamiltonian $H(k)$, the winding number $W_m$ for the $ m $th band (we have two bands in the SSH model)  is defined as
\begin{equation} \label{eq: Winding-Number}
	W_m = \dfrac{1}{\pi} \int_{-\pi}^{\pi} i \mel*{\Gamma \psi_m(k)}{\dfrac{\partial}{\partial k}}{\psi_m(k)},
\end{equation}
where $\Gamma$ is the chiral symmetric operator and $\ket{\psi_m(k)} $ is the eigenstate that belongs to the $ m $th band of $H(k)$ for the parameter value $ k $.  
Note that the expression of the winding number is very familiar with the Gauss-Bonnet theorem. In SSH, we have the Hamiltonian in momentum space which reads
\begin{equation}
	H(k) = \begin{bmatrix}
		0 & v + w e^{ik} \\
		v + w e^{-ik} & 0
	\end{bmatrix} \equiv \begin{bmatrix}
		0 & r e^{i \theta (k)} \\
		r e^{-i \theta (k)} & 0
	\end{bmatrix}
\end{equation}
with eigenvalues $\lambda_{\pm} = \pm r$ and eigenvectors
\begin{equation}
	\ket{\psi_+(k)} = \dfrac{1}{\sqrt{2}}\begin{bmatrix}
		e^{i \theta(k)} \\
		1
	\end{bmatrix}, \; \ket{\psi_-(k)} = \dfrac{1}{\sqrt{2}} \begin{bmatrix}
		-1 \\
		e^{-i \theta(k)}
	\end{bmatrix}
\end{equation}
such that
\begin{equation}
	\det \left(\dfrac{1}{\sqrt{2}} \begin{bmatrix}
		e^{i \theta (k)} & -1 \\
		1 & e^{-i \theta(k)}
	\end{bmatrix}\right) = +1.
\end{equation}
and now we calculate the Zak phase~\cite{Zak1989} for the ground state $ \ket{\psi(k)} $ of the Hamiltonian as
\begin{align}
	\gamma &= i \oint_k \ip*{\Gamma \psi(k)}{\partial_k \psi(k)} dk =  \dfrac{1}{2} \oint_k \dfrac{\partial}{\partial k} \theta(k) dk.
\end{align}
Hence
\begin{equation}
	\gamma =  \dfrac{1}{2} \oint_k \dfrac{\partial \theta(k)}{\partial k} dk = \dfrac{\pi}{2} \left( 1 -\dfrac{1}{\text{sign}(v-w)}\right).
\end{equation}
Therefore
\begin{equation}
	\gamma= 
	\begin{cases}
		0,& \text{if } v > w\\
		\pi,& \text{if } v < w\\
		\text{undefined}, & \text{if } v = w
	\end{cases} \implies W_{\pm} = \begin{cases}
		0,& \text{if } v > w\\
		1,& \text{if } v < w\\
		\text{undefined}, & \text{if } v = w
	\end{cases}
\end{equation}
There exist other ways to calculate the winding number also. We define a unit vector $ \hat{\vb{d}}(k) $ as $ \hat{\vb{d}}(k) = \vb{d}(k)/\abs{\vb{d}(k)}$ and the winding number is written as~\cite{Asboth2016}
\begin{equation}
	W = \dfrac{1}{2 \pi} \int_{-\pi}^{\pi} \left( \hat{\vb{d}}(k) \times \dfrac{d}{dk} \hat{\vb{d}}(k) \right)_z dk.
\end{equation}

\subsection{Bulk-edge correspondence}
We have shown the existence of edge states in the dimer limit $w = 1, v = 0$. These appear at the edges of the open chain and are robust even after turning on the intracell hopping amplitude $v$ and remain the same in number. The total number of zero-energy states is finite, and they are restricted to a single sublattice site (either $A$ or $B$) as shown in Fig.~\ref{fig:SSHWaveFunction1},~\ref{fig:SSHWaveFunction2}. If we denote the edge states on the sublattice $A$ and $B$ by $N_A$ and $N_B$, respectively, then 
\begin{equation}
	N_A - N_B = \text{invariant}
\end{equation}
i.e., the net number of edge states on the sublattice $A$ ($B$) on the left (right) edge is a topological invariant~\cite{Asboth2016,Bernevig2013}. So, we have two topological invariants, one from the bulk part of the system, and one from the information on the edges. We have shown that both topological invariants are $0$ in the trivial case $v >w$ and both are $1$ in the topological case $v < w$. This is referred to as \emph{bulk-edge correspondence}. The bulk-boundary correspondence is a physical concept that relates the topological properties of the bulk with the number of gapless edge modes. The bulk-edge correspondence for higher-dimensional systems has also been studied~\cite{Kane2007,Ortiz2011}.

\section{Topological characterization in 2D systems}
\begin{figure}[H]
	\centering
	\includegraphics[width=10cm]{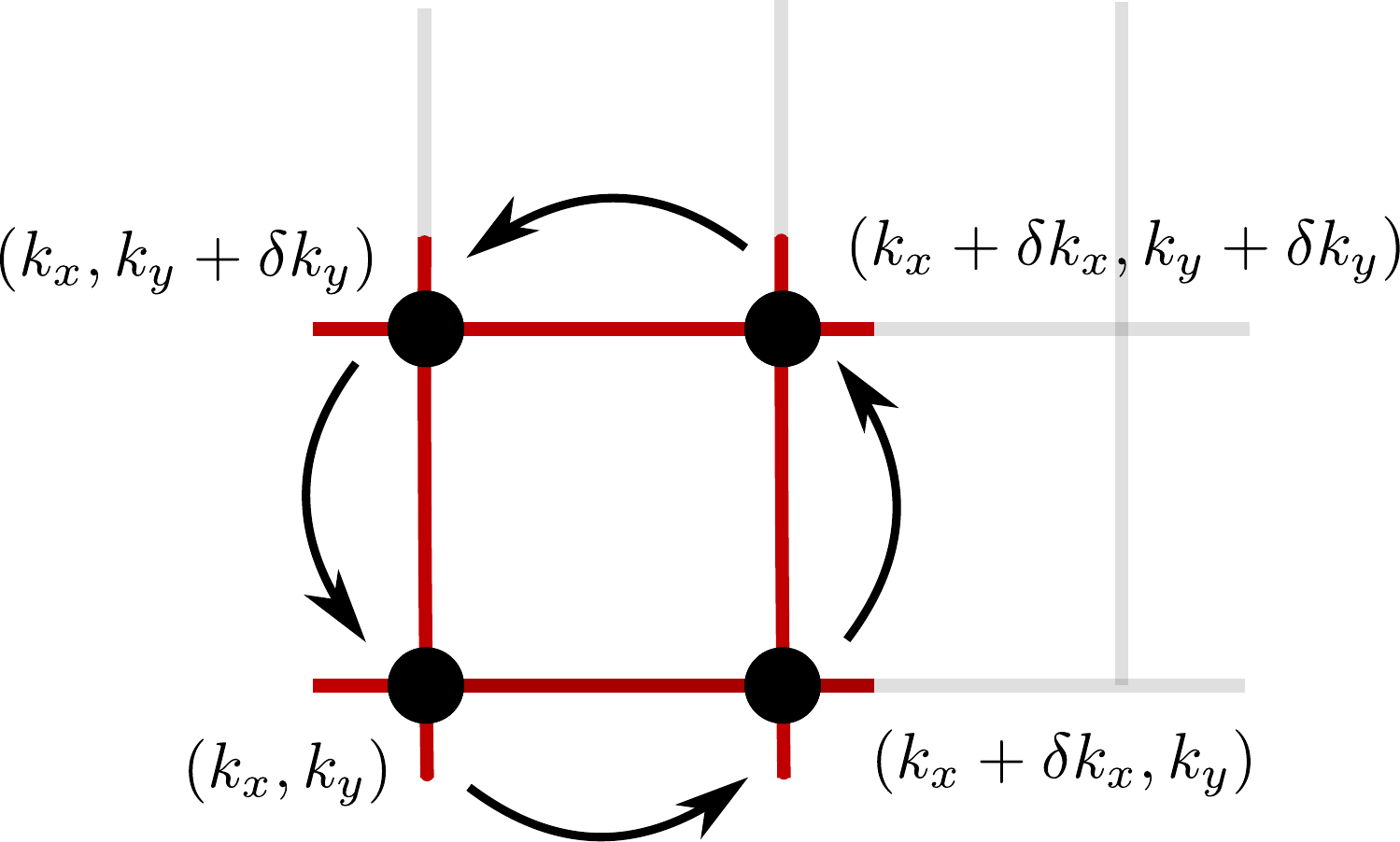}
	\caption{The calculation of the Berry curvature using the $U(1)$-link on a discretized two-dimensional Brillouin zone. The two momenta $k_x$, $k_y$ goes from $0$ to $2\pi/q_x$ and $0$ to $2\pi/q_y$ in a discrete step of $2 \pi/q_xN_x$ and $2 \pi/q_yN_y$ respectively. $N_x$, $N_y$ being the total number of points in respective directions.}
	\label{fig:Fukui}
\end{figure}

In 2D systems, for example in the integer quantum Hall effect, quantized Hall conductances are related to the Chern number~\cite{TKNN1982,Kohmoto1985,Berry1984,Barry1983}. The Chern number is defined for two-dimensional quantum systems with periodic parameters, for example two-dimensional momentum $ \vb{k} = (k_x, k_y) $ and is given by the integral
\begin{equation} \label{eq:Chern-Number}
	C_n = \dfrac{1}{2 \pi} \oint_S \mathcal{F}_{12}(k) d^2 k,
\end{equation}
where the Berry connection $A_{\mu}(k); \mu = 1,2$ (for example $\mu = x, y$) and the field strength $ \mathcal{F}_{12} $ are defined as~\cite{TKNN1982,Berry1984,Barry1983}
\begin{equation}
	A_{\mu}(k) = -i \ip*{\psi_n(k)}{\partial_{\mu}\psi_n(k)}, \;\;\; \mathcal{F}_{12}(k) = \partial_{1} A_2(k) - \partial_{2} A_2(k)
\end{equation}
where $ \ket{\psi_n(k)} $ is the normalized wavefunction of the $n$th band such that
\begin{equation}
	H(k) \ket{\psi_n(k)} = E_n(k) \ket{\psi_n(k)}.
\end{equation}
Since in a 2D system, the Hamiltonian is periodic in both directions, i.e. 
\begin{equation}
	H(k_1,k_2) = H(k_1 + 2\pi/q_1, k_2) = H(k_1, k_2 +  2\pi/q_2)
\end{equation}
where $q_1, q_2$ are some integers (lattice constants), the Brillouin zone can be imagined as a two-dimensional torus $T^2$.  Now, we discuss an efficient method to calculate the Chern number for a discretized Brillouin zone using a $ U(1) $ link variable~\cite{Fukui2005}. The $U(1)$ link variable is defined as
\begin{equation}
	U^{\hat{\vb{e}}}_n(\vb{k}) \equiv \dfrac{\ip*{\psi_n(\vb{k})}{\psi_n(\vb{k} + \hat{\vb{e}})}}{\abs{\ip*{\psi_n(\vb{k})}{\psi_n(\vb{k} + \hat{\vb{e}})}}}
\end{equation} 
where $\hat{\vb{e}}$ is the unit vector in the direction of either $k_x$ or $k_y$ and such that
\begin{equation}
	(k_x, k_y) + \hat{\vb{e}}_x = (k_x + \delta k_x, k_y), \;\;\; (k_x, k_y) + \hat{\vb{e}}_y = (k_x, k_y + \delta k_y)
\end{equation} 
and as shown in Fig.~\ref{fig:Fukui}. Using the link variable, we can now define the field strength which would give us the Berry curvature as
\begin{equation}
	\mathcal{F}_n (\vb{k}) = \dfrac{1}{i} \ln U^{\hat{\vb{e}}_x}_n(k_x, k_y) U^{\hat{\vb{e}}_y}_n(k_x+\delta k_x, k_y) U^{-\hat{\vb{e}}_x}_n(k_x+\delta k_x, k_y+\delta k_y) U^{-\hat{\vb{e}}_y}_n(k_x, k_y+\delta k_y).
\end{equation}
\noindent The summation of all $\mathcal{F}_n (\vb{k})$ over the Brillouin zone will result in
\begin{equation}
	\sum_{BZ} \mathcal{F}_n (\vb{k}) = 2 \pi C_n
\end{equation}
where $C_n$ is the Chern number of the $n$th band. We use these methods to calculate the topological invariants in 1D and 2D quantum walks in the preceding chapters.

\section{Topological Phases in Quantum Walk}
\begin{figure}[H]
	\centering
	\subfigure[]{
		\includegraphics[width=5cm]{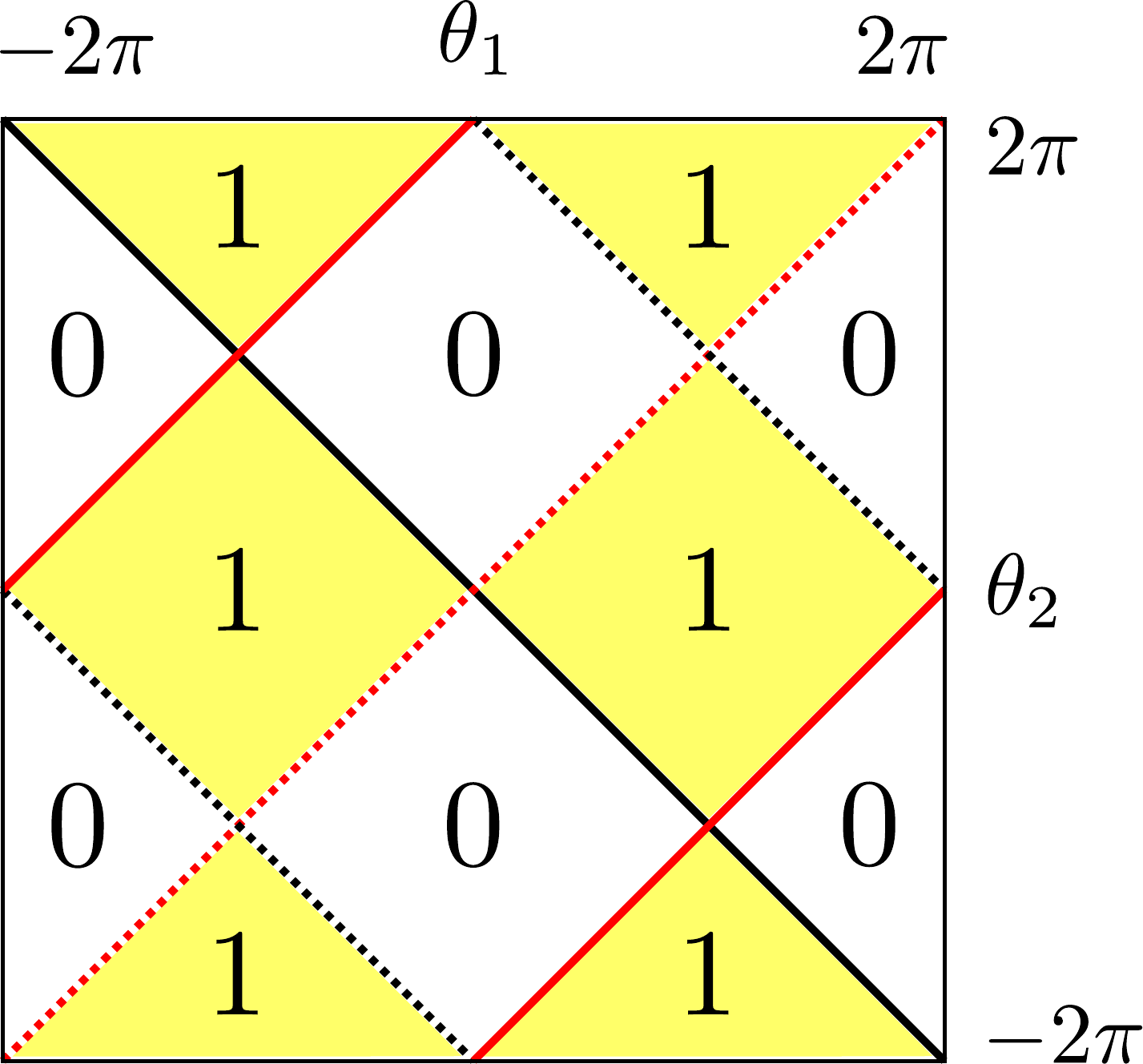}
		\label{fig:PSQW}}
	\subfigure[]{
		\includegraphics[width=5cm]{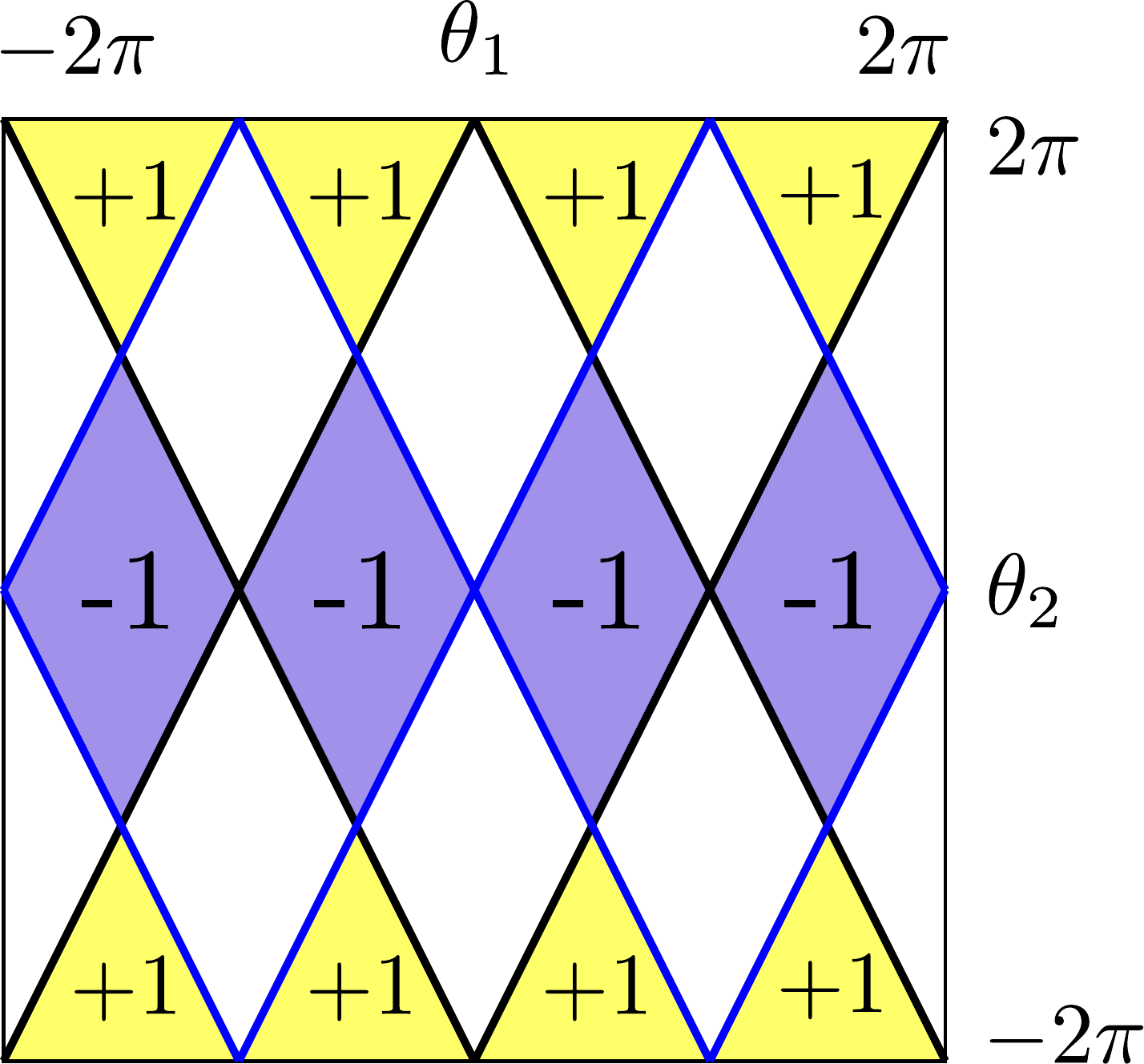}
		\label{fig:PS2DQW}}
	\caption{(Color online) \subref{fig:PSQW} Different topological phases realized in 1D SSQW as a function of $\theta_1$ and $\theta_2$. We observe two topological phases here corresponding to $W = 0$ and $1$. Here, black and red lines represent closing of energy band at $k = 0$ and $k = \pi$, respectively, and solid and dotted lines demonstrate the closing at $E = 0$ and $E = \pi$, respectively. \subref{fig:PS2DQW} Topological phases which exist in 2D DTQW for different values of $\theta_1$ and $\theta_2$. Here, blue and black lines show the closing of energy gap at $ E = 0 $ and $E = \pi$, respectively. The yellow, violet and white regions correspond to $ C = +1, -1 $ and $ 0 $, respectively.}
	\label{fig:PS}
\end{figure}

Quantum walk Hamiltonian possesses a rich topological structure. For example, 1D SSQW with an effective Hamiltonian $H_{\text{SS}}(\theta_1, \theta_2)$ with parameters $\theta_1$ and $\theta_2$ [discussed in Chapter~\ref{chap:qw} in Eq.~\ref{eq:SSQW-Hamil}] exhibits two different topological phases characterized by the winding number $W = 0$ and $W = 1$, as shown in Fig.~\ref{fig:PSQW}~\cite{Kitagawa2010, KitagawaNature}. The 1D SSQW possess TRS and PHS with the choice of 
\begin{equation}
	\mathcal{T} = \sigma_x \K, \;\;\;  \mathcal{C} = \sigma_z \K
\end{equation} 
respectively. The presence of both TRS and PHS results in the existence of chiral symmetry, which is represented by the operator
\begin{equation}
	\Gamma = \mathcal{C} \vdot \mathcal{T} = i \sigma_y.
\end{equation} 
Interestingly, the presence of two out of three (PHS, TRS, CS) symmetries in the system ensures the existence of the remaining~\cite{Kitaev2009,Schnyder2008}, which is also evident from the Table~\eqref{table:AZClass}. 

\noindent In order to see the existence chiral symmtery, we redefine the time evolution operator for SSQW (written in quasi-momentum space) by performing a unitary transformation which reads~\cite{Asboth2013} as
\begin{equation} \label{eq:SSQW-Unitary_TS}
	\tilde{U}(\theta_1, \theta_2, k) \rightarrow \tilde{U}^{'}(\theta_1, \theta_2, k) = R(\theta_1/2) T_{\downarrow}(k) R(\theta_2) T_{\uparrow}(k) R(\theta_1/2),
\end{equation}
where by tilde we meant the part which acts only on the coin space. The $ \tilde{U}^{'}(\theta_1, \theta_2, k) $ is related to $\tilde{U}(\theta_1, \theta_2, k)$ as $\tilde{U}^{'}(k) = R(\theta_1/2) \tilde{U}(\theta_1, \theta_2, k) R^{-1}(\theta_1/2)$.  We shifted the origin of time so that the time-evolution operator has symmetric order of individual operators in the time direction. This is referred as time-symmetric representation~\cite{Asboth2013}. The motivation behind this transformation is to show the existence of CS in 1D SSQW explicitly. In addition to that, there exists another CS time-frame in which the redefined time-evolution reads
\begin{equation}
	\tilde{U}^{''}(k) = R(\theta_2/2) T_{\downarrow}(k) R(\theta_1) T_{\uparrow}(k) R(\theta_2/2).
\end{equation}
Using these two evolution operators, together, oen can identify the two topological invariants, which are needed to completely characterize the chiral symmetric quantum walks. 

\noindent The effective Hamiltonian corresponding to 2D DTQW exhibits only PHS with
\begin{equation}
	\mathcal{C} = \K
\end{equation}In Fig.~\ref{fig:PS2DQW}, we plot the  topological phases with Chern number $ C = 0, \pm 1 $ exhibited by the Hamiltonian $H_{_{2D}}(\theta_1, \theta_2)$ for 2D DTQW~\cite{Kitagawa2010}. 

\section{Real-space representation of Winding number}
In an extended SSH model where losses are added at every second sublattice site ($A$ or $B$), it has been shown that the average distance traveled by a particle, which was initialized at a site without losses, is quantized and is directly related to the winding number~\cite{Asboth2017}. This correspondence holds for all 1D systems that exhibit chiral symmetry. This addition of the losses leads to non-Hermiticity (we will discuss it in detail in the next chapter) in systems that have already been studied in the SSH model~\cite{Rudner2009,Rudner2015}. The idea to have a real space representation of winding number that serves as a topological invariant for one-dimensional systems~\cite{Prodan2014}. The idea was further generalized to the systems in the presence of disorders that break translation symmetry~\cite{Lee2021}. These approaches are instrumental in experimental studies of the topological phases in various systems  In this section, we will elucidate the connection between the average displacement and the bulk winding number. We will end the discussion by considering 1D SSQW.

\begin{figure}[H]
	\centering
	\includegraphics[width=7cm]{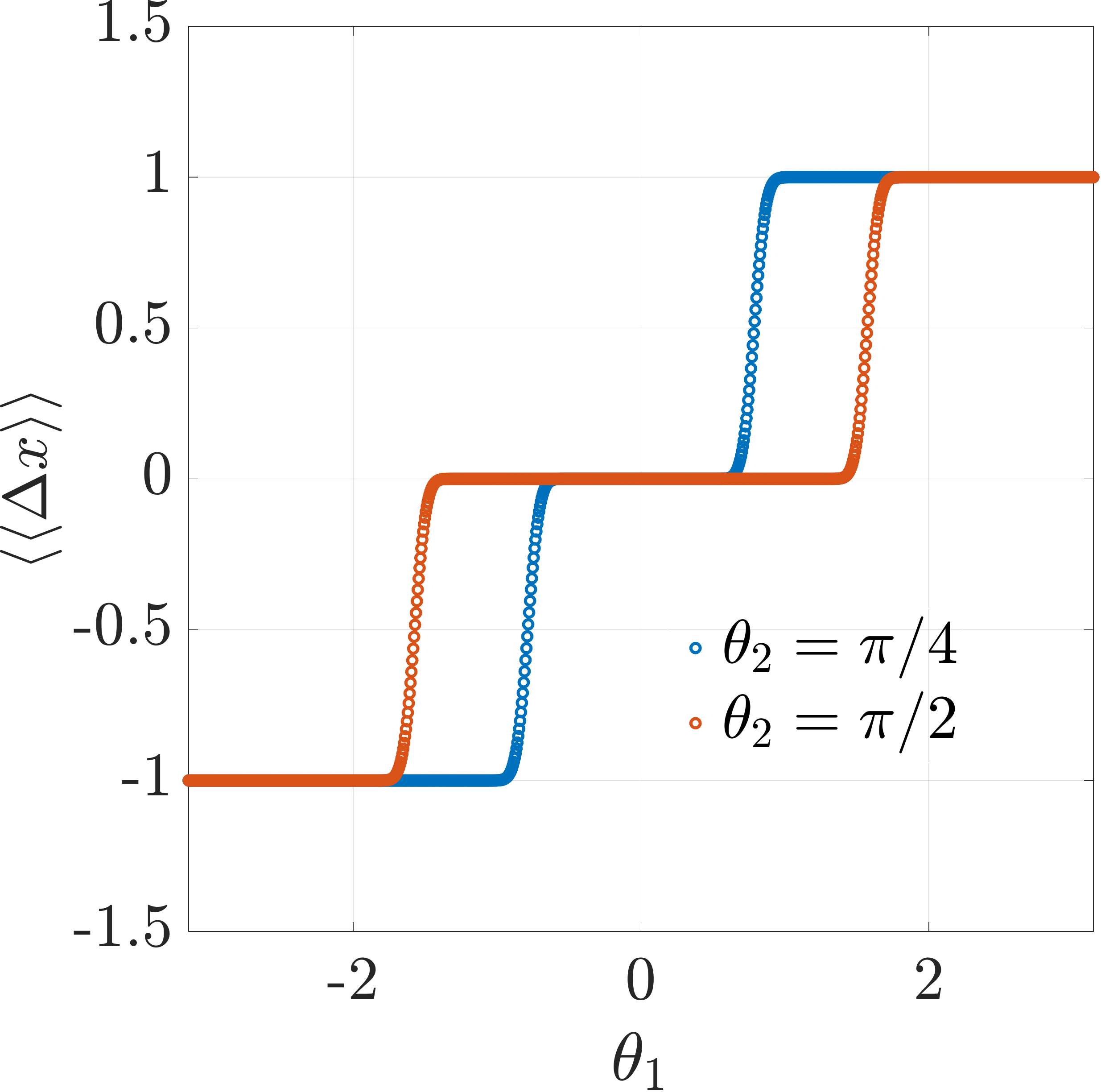}
	\caption{Plot of the average displacement of the walker, initially localized at $ \ket{-}\otimes\ket{0} $, as a function of $\theta_1$ after $M = 200 $ steps for two different values of $\theta_2$. The number of sites $L$ are taken to be 51.}
	\label{fig:Asboth}
\end{figure}

\noindent We take the SSH model except that each lattice site can host $2N$ internal sites or sublattice sites. The basis in this case would be $\{\ket{x, \sigma}\}$ where $ n = 1,2, \dots, L $ lattice site, $ \sigma = 1,2, \dots, 2N$ internal state. We group internal sites into two sublattices $ A $ \& $ B $ and 
\begin{equation}
	\hat{P}_A = \sum_{n \in \mathbb{Z}} \sum_{a=1}^{N} \dyad{n,a} = \mathds{1} \otimes \sum_{a=1}^{N} \dyad{a}
\end{equation}
with
\begin{equation*}
	\hat{P}_B = \mathds{1} - \hat{P}_A
\end{equation*}
and the corresponding chiral symmetry operator is given as
\begin{equation*}
	\hat{\Gamma} = \hat{P}_A - \hat{P}_B \implies \hat{\Gamma} \hat{H} \hat{\Gamma}^{-1} = - \hat{H}. 
\end{equation*}
As a consequence of chiral symmetry, the eigenstates come in pair $ \{\ket{n}, \hat{\Gamma} \ket{n}\} $ with energy $ \{-\varepsilon_n,\varepsilon_n\} $ respectively. We can correspondingly define the projectors for the lower and upper spectrum as
\begin{equation*}
	\hat{Q}_- = \sum_{n-} \dyad{n-}, \; \hat{Q}_+ = \hat{\Gamma} \hat{Q}_- \hat{\Gamma}
\end{equation*}
with
\begin{equation}
	\hat{Q} = \hat{Q}_+ -\hat{Q}_-
\end{equation}
The presence of chiral symmetry allows the system to have non-trivial topology which is characterized by the winding number in real space representation given by~\cite{Prodan2014,Lee2021}
\begin{align}
	W &= -\dfrac{1}{L} \tr \{ \hat{P}_B \hat{Q} \hat{P}_A [\hat{X}, \hat{P}_A \hat{Q} \hat{P}_B] \} \nonumber
\end{align}
where $ \hat{X} \equiv \sum_x \sum_{c = 1}^{2N} x \dyad{x,c}$ is the position operator. Now to detect the winding number, we localize the particle at sublattice $A$ and introduce weak partial measurement after each time step to measure the position of the particle at sublattice $B$ only (which is why partial measurement). Mathematically, such a weak measurement can be modeled with a non-unitary operator $M_1$ given by  
\begin{equation}
	\hat{M}_1 = \hat{P}_A + \sqrt{1 - p_M} \hat{P}_B.
\end{equation}
where $p_M$ is the measurement strength such that $0 \le p_M \le 1$. Here, the second operator is given by $\hat{M}_2 = \sqrt{p_M} \hat{P}_B$ such that 
\begin{equation}
	M_1^{\dagger} M_1 + M_2^{\dagger} M_2 = \mathds{1}.
\end{equation}
When this measurement yields a positive result, i.e., when the particle is detected at the site $B$ with conditional probability $p_M$, we stop the evolution. However, if the measurement yields a negative result when the particle is not detected, we will have the time evolution followed by the next measurement. The wave function after the $ j $ steps reads as follows
\begin{equation*}
	\ket{\psi (t = jT)} = \hat{U} [\hat{M}_1 \hat{X}]^{j-1} \ket{\psi(0)}.
\end{equation*}
The probability of finding the particle at $ \ket{y,b} $ after $ j $ steps, initially localized at $ \ket{x,a} $ is given by
\begin{equation}
	s_{(x,a) \rightarrow (y,b)}(j) = p_M \abs{\ip*{y,b}{\psi(t = jT)}}^2
\end{equation} 
Basically, we have $\hat{M}_2$ at the $ j $th step, due to which we get a factor $ p_M $ and it projects the state in the $B$ sublattice, which is responsible for the inner product with $ \ket{y,b} $ in the expression above. The average displacement is then defined as
\begin{equation}
	\expval{\Delta x}_{(x,a)} \equiv \sum_{j = \mathbb{N}} \sum_{y=1}^{L} (y-x) \sum_{b = 2N+1}^{2N} s_{(x,a) \rightarrow (y,b)}(j)
\end{equation}
and to get a more general expression valid for any arbitrary $ N $ we sum over all the initial states $\ket{x,a}$ and write
\begin{equation}
	\expval{\expval{\Delta x}} \equiv \dfrac{\sum_{x,a} \expval{\Delta x}_{(x,a)}}{NL} = \dfrac{1}{NL} \sum_{x,a}\sum_{j = \mathbb{N}} \sum_{y=1}^{L} (y-x) \sum_{b = 2N+1}^{2N} s_{(x,a) \rightarrow (y,b)}(j)
\end{equation}
It can be solved further to write it in a compact form as~\cite{Asboth2017}
\begin{equation}
	\expval{\expval{\Delta x}} = -\dfrac{2}{NL} \tr(\hat{X} \hat{\Gamma} \hat{Q}_-)
\end{equation}
with
\begin{equation*}
	\hat{\Gamma} = \hat{P}_A - \hat{P}_B = \sum_{x = 1}^{L} \left[ \sum_a \dyad{x,a} - \sum_b \dyad{x,b}\right].
\end{equation*}
and after simplifying it further, we arrive at the very interesting connection between the average displacement and the topological invariant, winding number in this case, which reads
\begin{equation}
	\expval{\expval{\Delta  x}} = \dfrac{W}{N}.
\end{equation}
As an example, we consider the SSQW quantum walk where $N=1$ as described in the previous chapter. The time evolution is governed by the following unitary operator 
\begin{equation}
	U(\theta_1, \theta_2) = R(\theta_1/2) T_{\downarrow} R(\theta_2) T_{\uparrow} R(\theta_1/2)
\end{equation}
The chiral symmetry operator is given by 
\begin{equation}
	\Gamma = \sigma_x \otimes \mathds{1}
\end{equation} 
which decides the sublattices corresponding to the internal states 
\begin{equation}
	\ket{\pm} = \dfrac{\ket{\uparrow} \pm \ket{\downarrow}}{\sqrt{2}}.
\end{equation}
Initially the particle is localized at $ \ket{\psi_0} = \ket{-} \otimes \ket{0}$ and we remove the particle if it is detected at $ \ket{+} $. This is realized by the operator 
\begin{equation}
	\hat{M} = \dyad{-} \otimes \mathds{1}.
\end{equation}
We plot the average displacement of the walker after $M = 200$ number of steps in Fig.~\ref{fig:Asboth} for two different settings of $\theta_2$. We can clearly observe the quantized behavior of the average displacement, which exactly coincides with the winding number. This approach has already been used to measure the winding number in experimental settings~\cite{Xue2017,Xue2018,Guo2018,Maffei2018}.

\section{Gauss-Bonnet theorem}
As we discussed in the beginning of this chapter, the Gauss-Bonnet theorem reads
\begin{equation}
	\int_S \kappa dA = 4 \pi (1-g)
\end{equation}
where 
\begin{align*}
	\kappa  &= \dfrac{1}{r_1 r_2} = \text{Gaussian curvature}, \\
	g &= \text{genus (\# of holes)} .
\end{align*}
\begin{figure}[H]
	\centering
	\includegraphics[width=7cm]{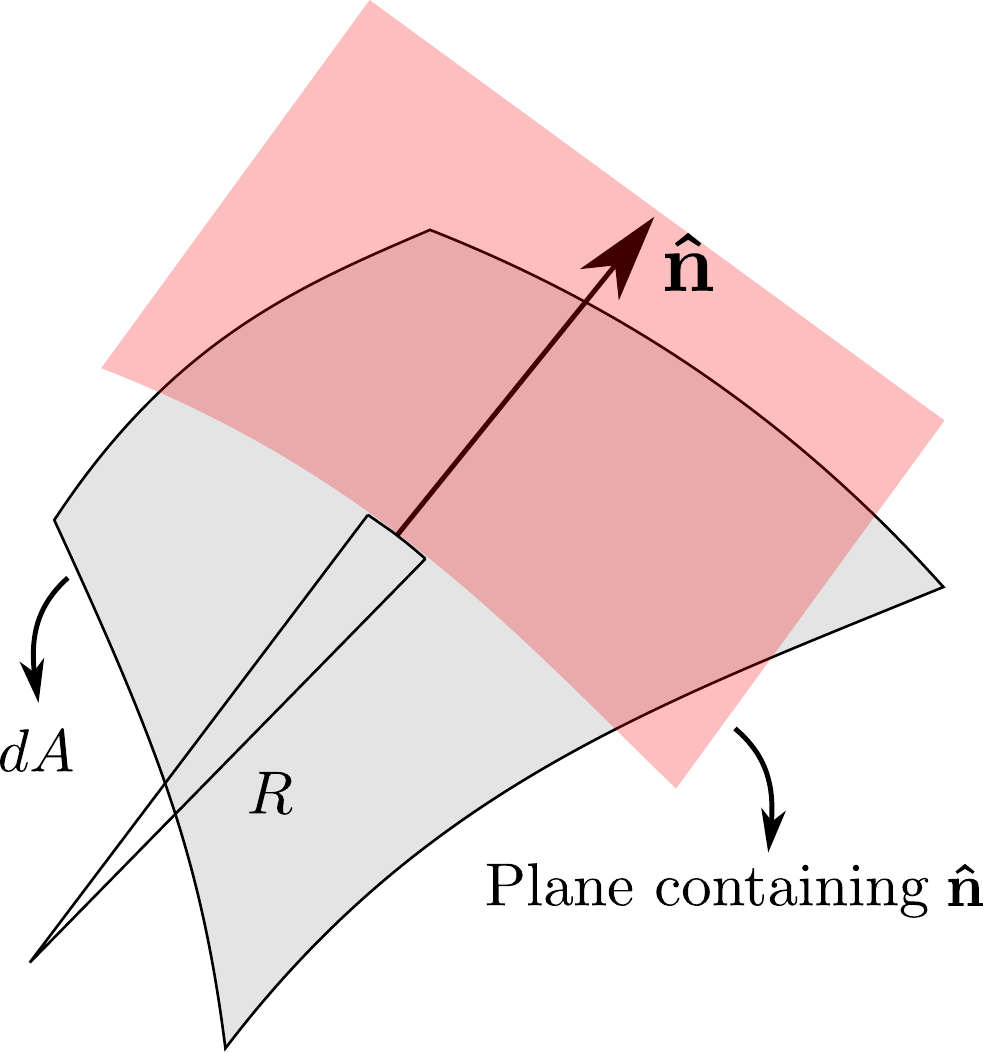}
	\caption{Gaussian curvature.}
	\label{fig:gausscur}
\end{figure}
\noindent If we consider a surface element $ dA $ and a unit vector $\hat{n}$. The intersection of a plane containing the unit vector and $ dA $ gives a 1D curve that can be approximated by a circle of radius $ R $ (as shown in Fig. \ref{fig:gausscur}). $ r_1 $ and $r_2$ are the principal curvatures and are defined as 
\begin{equation}
	r_1 = \min (R), r_2 = \max (R)
\end{equation}
This particular form of the Gauss-Bonnett theorem is applicable for "closed" surfaces. We can also write a more generic form of it, which can be used for open surfaces, too.
\begin{figure}[H]
	\centering
	\includegraphics[width=13cm]{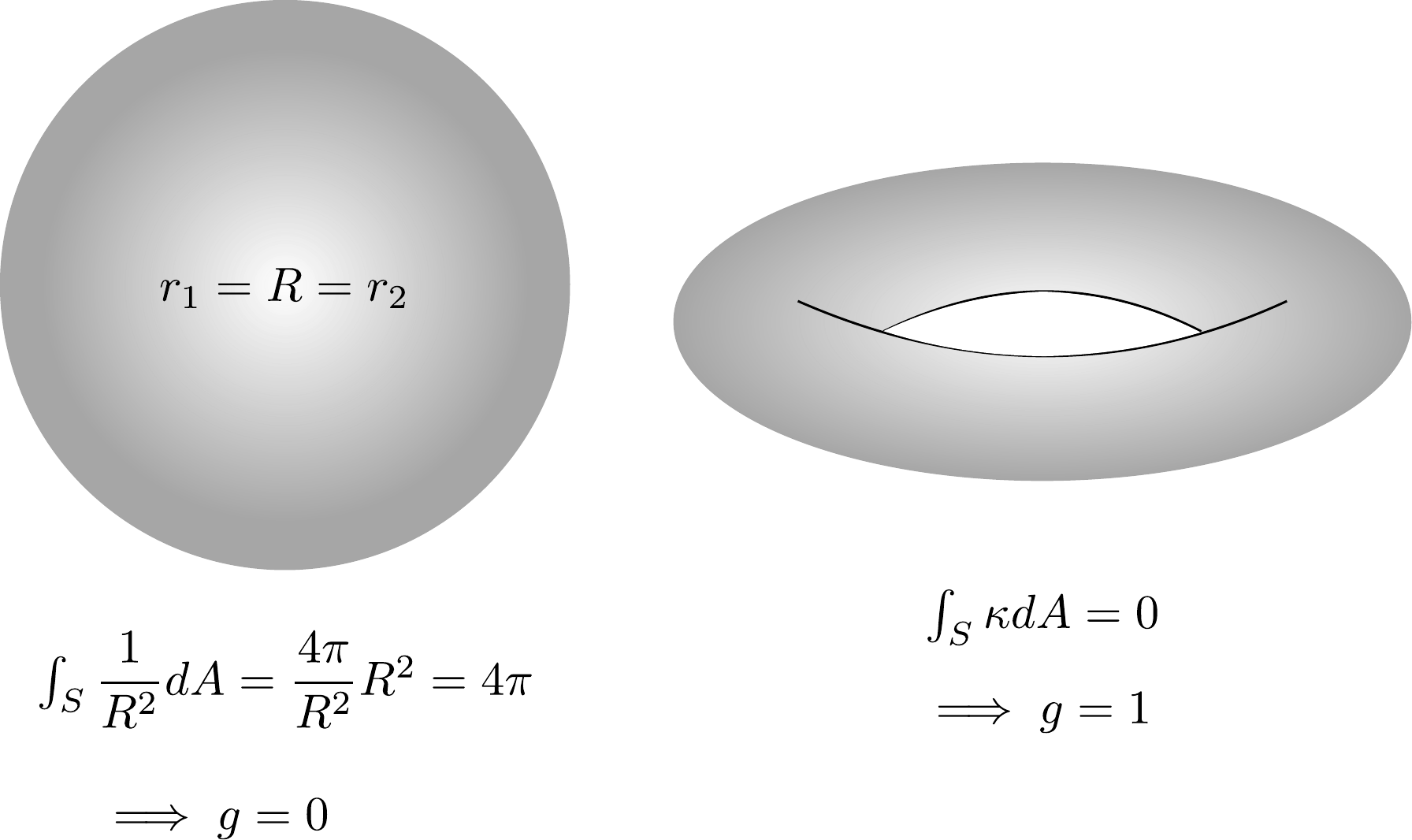}
	\caption{Explaing Gaussian curvature.}
	\label{fig:genus}
\end{figure}
\noindent In case of torus ($ T^2 $), we have positive and negative that will add up to zero. The integral is invariant under any smooth transformations or disorder in the system.

\chapter{Non-Hermitian Systems}\label{chap:nonher}
In conventional quantum mechanics, in order to have a real spectrum (eigenvalues) and unitary evolution (probability conservation), we demand that the physical observables of a closed quantum system should be represented by hermitian operators~\cite{Sakurai1985}. However, in practical situations, our system is coupled to an environment, or at least with a measuring device, which leads to non-unitary dynamics and non-Hermitian observables, especially Hamiltonian. In such systems, we resort to the master equation of formalism~\cite{Breuer2007,Lindblad1976}. In 1998, Bender \emph{et al.} proposed a class of Hamiltonians that exhibit real spectrum despite the fact that they are non-Hermitian~\cite{Bender1998,Bender1999}. This novel class of Hamiltonians has an addition property which is that they respect the combined time-reversal ($\mathcal{T}$) and parity ($\mathcal{P}$) symmetry and are called $\mathcal{PT}$ symmetric Hamiltonian. The parity $\mathcal{P}$ is a unitary operator that reverses the position as ${\vb{x}} \to-{\vb{x}}$. On the other hand, the time reversal operator $\mathcal{T}$ is an anti-unitary operator that reverses the arrow of time, i.e., $t \rightarrow -t$ (we discussed it in detail in Chapter~\ref{chap:sym}). Therefore, a Hamiltonian $H$ is said to be $\mathcal{PT}$-symmetric if
\begin{equation} \label{eq:PT-Symmetry}
	(\mathcal{PT}) H (\mathcal{PT})^{-1} = H.
\end{equation}
Basically, when $\mathcal{PT}$ operator commutes with the Hamiltonian $H$. Due to the anti-linear nature of the $\mathcal{PT}$ operator, even if the Hamiltonian $H$ commutes with the operator $\mathcal{PT}$, they need not necessarily share the same set of eigenvectors. Note that the Hamiltonian can satisfy Eq.~\eqref{eq:PT-Symmetry} even when it does not possess these symmetries individually. In addition to the existence of a anti-unitary operator $\mathcal{PT}$, we demand that the $\mathcal{PT}$ operator and the non-Hermitian Hamiltonian must have the same set of eigenvectors. Together, these two conditions are sufficient to ensure the real spectrum of the Hamiltonian~\cite{Bender2002,Bender2002b,Bender2010,Bender2019,Moiseyev2011}. Furthermore, the notion of pseudo-hermitcity was introduced to show that all the Hamiltonian with a real spectrum are pseudo-hermitian~\cite{Mostafazadeh2002,Mostafazadeh2002b,Mostafazadeh2002c,Mostafazadeh2003}. A given Hamiltonian is pseudo-hermitian with respect to a metric $\eta$ if~\cite{Mostafazadeh2002}
\begin{equation}
	H^{\dagger} = \eta H \eta^{-1}.
\end{equation}
It has also been argued that $\mathcal{PT}$-symmetry is a necessary and sufficient condition for unitary time evolution~\cite{Mannheim2013}. Since then, there have been numerous studies, theoretical and experimental both, where $\mathcal{PT}$ symmetric systems have been realized~\cite{Kottos2011,Bittner2012,Fleury2015,Vladimir2013,Min2016,Regensburger2012,Samuel2013,Wimmer2015,El-Ganainy2018,Feng2014}. Recently, non-Hermitian systems have been used in quantum information masking~\cite{Lv2022}. Motivated by these studies, the notion of topological phases has been extended to non-Hermitian systems~\cite{GongUeda2018,Kunst2018,Kunst2019,WangYao2018,Kawabata2019,Wang2021,Ueda2020,Kawabata2019b,Kawabata2022} and a bulk-boundary correspondence has also been proposed with the presence of a non-Hermitian skin effect~\cite{Kunst2018,Torres2018,Xiong2018,Lee2016,Sato2020}  and exceptional points~\cite{Ozdemir2019,Flore2021}. The exceptional points are sort of degeneracies of a non-Hermitian system, where the real parts and the corresponding imaginary parts of certain eigenvalues coincide and the eigenvectors coalesce~\cite{Heiss2012}. The most fundamental model, the SSH model (which we discussed in the last chapter in detail), has been extended to include non-hermiticity~\cite{Chen2014,Lieu2018,Zhong2018,Li2020}. The topological features of the periodically driven non-Hermitian SSH model was also introduced~\cite{Vyas2021}.  Further, a geometrical interpretation of topological invariant is provided for non-hermitian systems~\cite{Chen2018}. In this chapter we will discuss how non-hermiticity results in symmetry ramification, then we discuss non-Hermitian 1D SSQW and conclude with a discussion on exceptional points.

\section{Symmetry ramification and unification in non-Hermitian systems}
In the last chapter we discussed that particle-hole symmetry for the Hermitian systems is represented by an anti-unitary operator, $\mathcal{C}$ which satisfies  
\begin{equation}\label{eq:PHS1}
	\mathcal{C} H \mathcal{C}^{-1} = -H \implies U_{\mathcal{C}} H^* U_{\mathcal{C}}^{-1} = -H
\end{equation}
where we have used the fact that any anti-unitary operator can be written as $\mathcal{C} = U_{\mathcal{C}} \mathcal{K}$. Here, $U_{\mathcal{C}}$ is a unitary matrix and $\mathcal{K}$ is the complex conjugation operator. Since $H$ is hermitian, we have
\begin{equation*}
	H = H^{\dagger} \implies H^T = H^*
\end{equation*}
and the relation in Eq.~\eqref{eq:PHS1} is equivalent to
\begin{equation}\label{eq:PHS2}
	\mathcal{C} H^T \mathcal{C}^{-1} = -H.
\end{equation}
However, Eqs.~\eqref{eq:PHS1} and \eqref{eq:PHS2} are not equivalent for a non-Hermitian system. We see similar kind of symmetry ramification in other symmetries as well. For example, chiral symmetry is represented by a unitary operator $\Gamma$ and for Hermitian systems it is defined as 
\begin{equation}\label{eq:CS1}
	\Gamma H \Gamma^{-1} = -H.
\end{equation}
For Hermitian system, Eq.~\eqref{eq:CS1} is equivalent to
\begin{equation}\label{eq:CS2}
	\Gamma H \Gamma^{-1} = -H^{\dagger}
\end{equation}
however, Eqs.~\eqref{eq:CS1} and \eqref{eq:CS2} are not same for non-Hermitian systems. Due to the distinction between the complex conjugation and the transposition in non-Hermitian systems, we see such ramifications and unification in symmetries which decides the topological classes as discussed in the last Chapter~\ref{chap:sym}. As a consequence, 10-fold classification is no longer valid and we will have 38-fold classification~\cite{Kawabata2019}. For our purpose, we are only interested on Chiral symmetry and we will show our system exhibits CS and satisfies Eq.~\eqref{eq:CS2} even in the non-Hermitian regime.


\section{Non-unitary/Non-hermitian Quantum Walk}
Generally, quantum walk dynamics is given by a unitary time evolution operator. However, limitations in physical implementation and the environmental effects can cause losses which can cause the dynamics to deviate from unitary nature. 
In general, one can extend 1D SSQW to a non-unitary quantum walk by introducing a scaling (or gain/loss) operator, $G_i$ and a phase operator, $\Phi_i$~\cite{Mochizuki2016}, with tunable parameters in the dynamics. The resulting time evolution operator for a non-unitary quantum walk can be written as
\begin{equation} \label{eq:NUSSQWUni}
	U = T_{\downarrow} G_2 \Phi_2 R(\theta_2)T_{\uparrow} G_1 \Phi_1 R(\theta_1)
\end{equation}
with
\begin{align}
	G_i &= \sum_{n} \tilde{G}_{i,n} \otimes \dyad{n}{n} \; ;  \; \;  \tilde{G}_{i,n} = \begin{pmatrix}
		g_{i, \uparrow}(n) & 0 \\
		0  &  g_{i, \downarrow}(n)
	\end{pmatrix} \\
	\Phi_i &= \sum_{n} \tilde{\Phi}_{i,n} \otimes \dyad{n}{n} \; ;  \;  \; \tilde{\Phi}_{i,n} = \begin{pmatrix}
		\phi_{i, \uparrow}(n) & 0 \\
		0  &  \phi_{i, \downarrow}(n)
	\end{pmatrix}.	
\end{align}

\subsection{Homogeneous system}
\begin{figure}
	\centering
	\subfigure[]{
		\includegraphics[height=3.85cm]{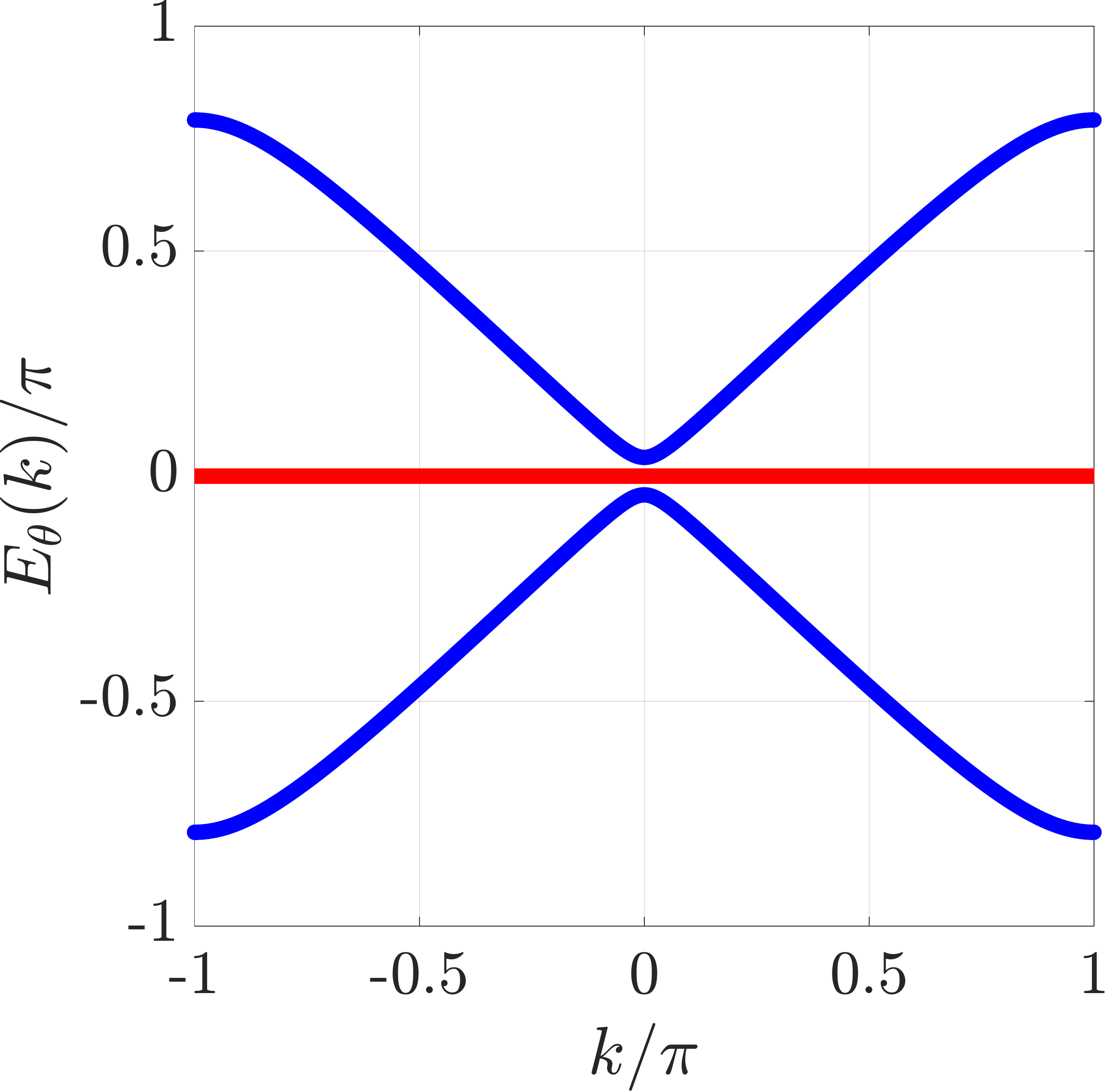}
		\label{fig:1DNUQWEnergy1}}
	\subfigure[]{
		\includegraphics[height=3.85cm]{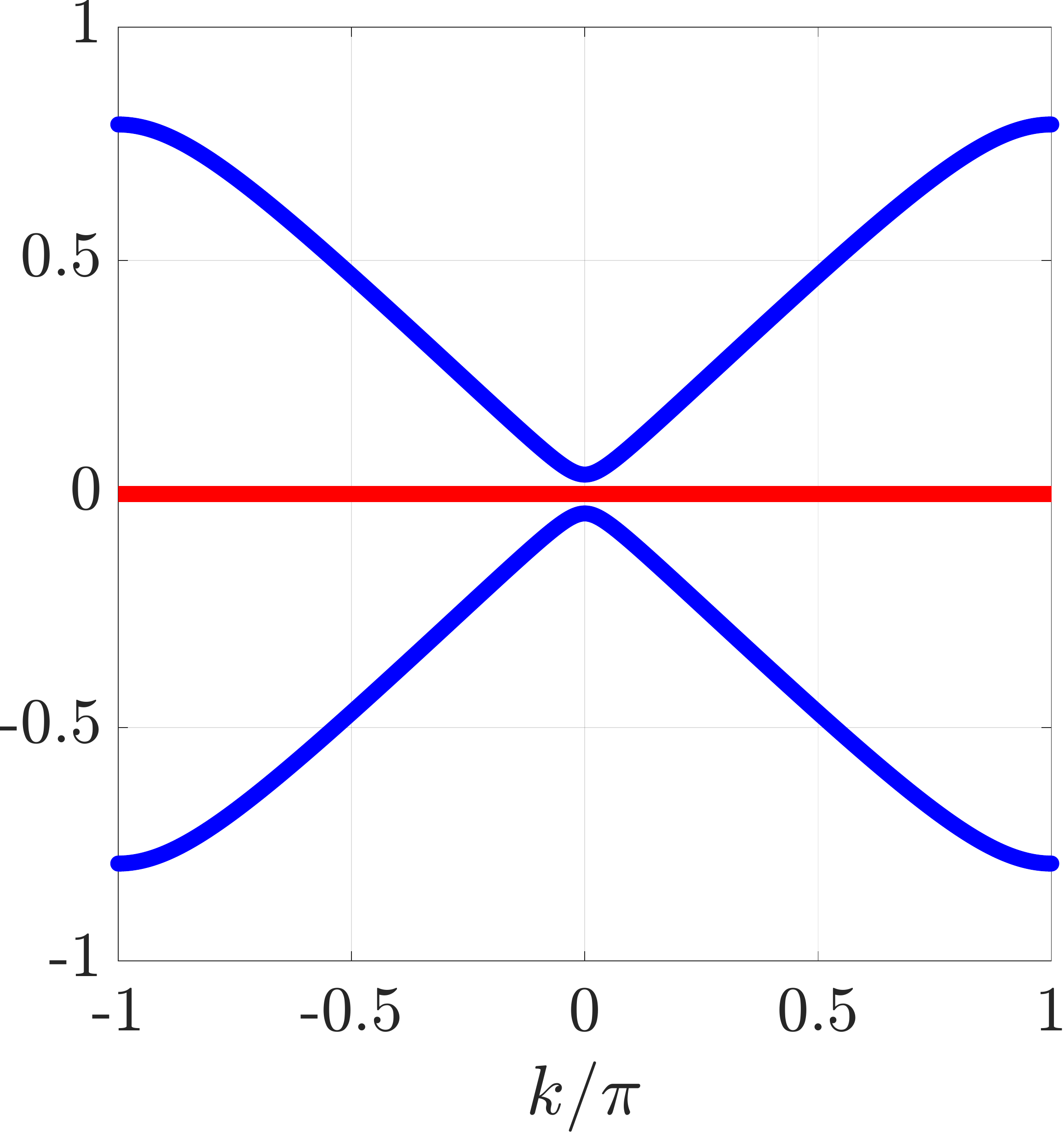}
		\label{fig:1DNUQWEnergy2}}
	\subfigure[]{
		\includegraphics[height=3.85cm]{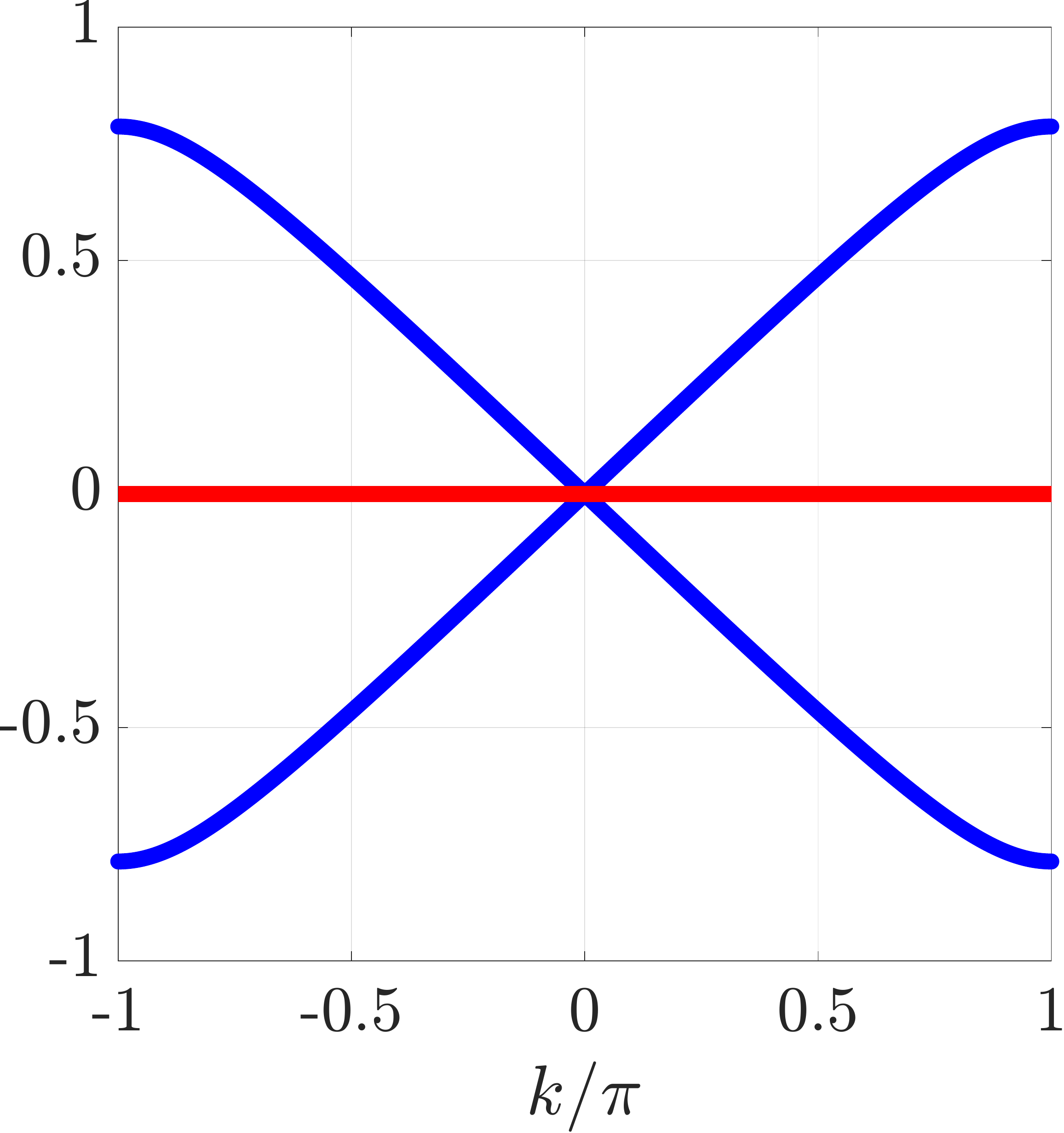}
		\label{fig:1DNUQWEnergy3}}
	\subfigure[]{
		\includegraphics[height=3.85cm]{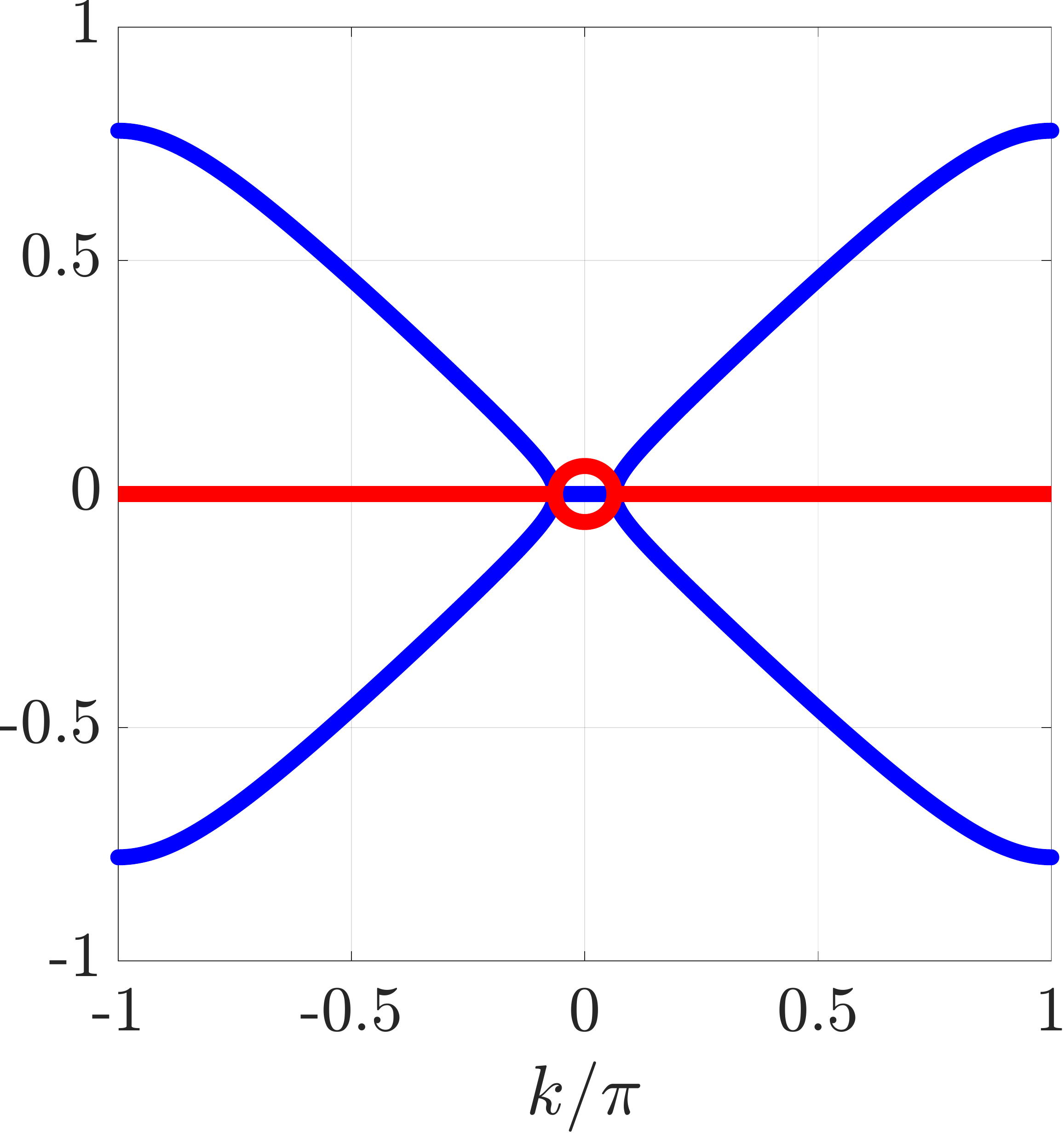}
		\label{fig:1DNUQWEnergy4}}
	
	\subfigure{
		\includegraphics[height=3.85cm]{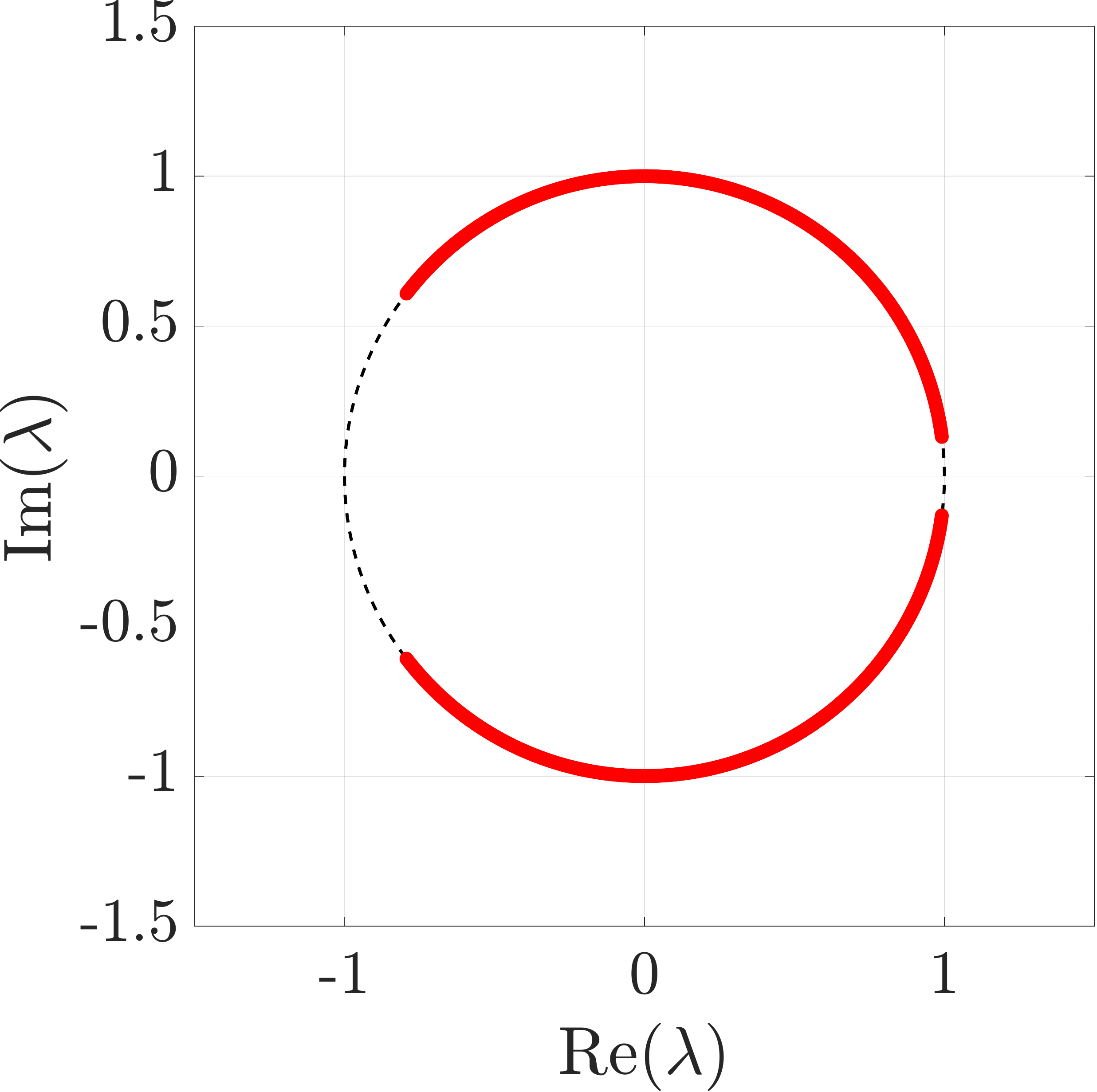}
		\label{fig:1DNUQWEigen1}}
	\subfigure{
		\includegraphics[height=3.85cm]{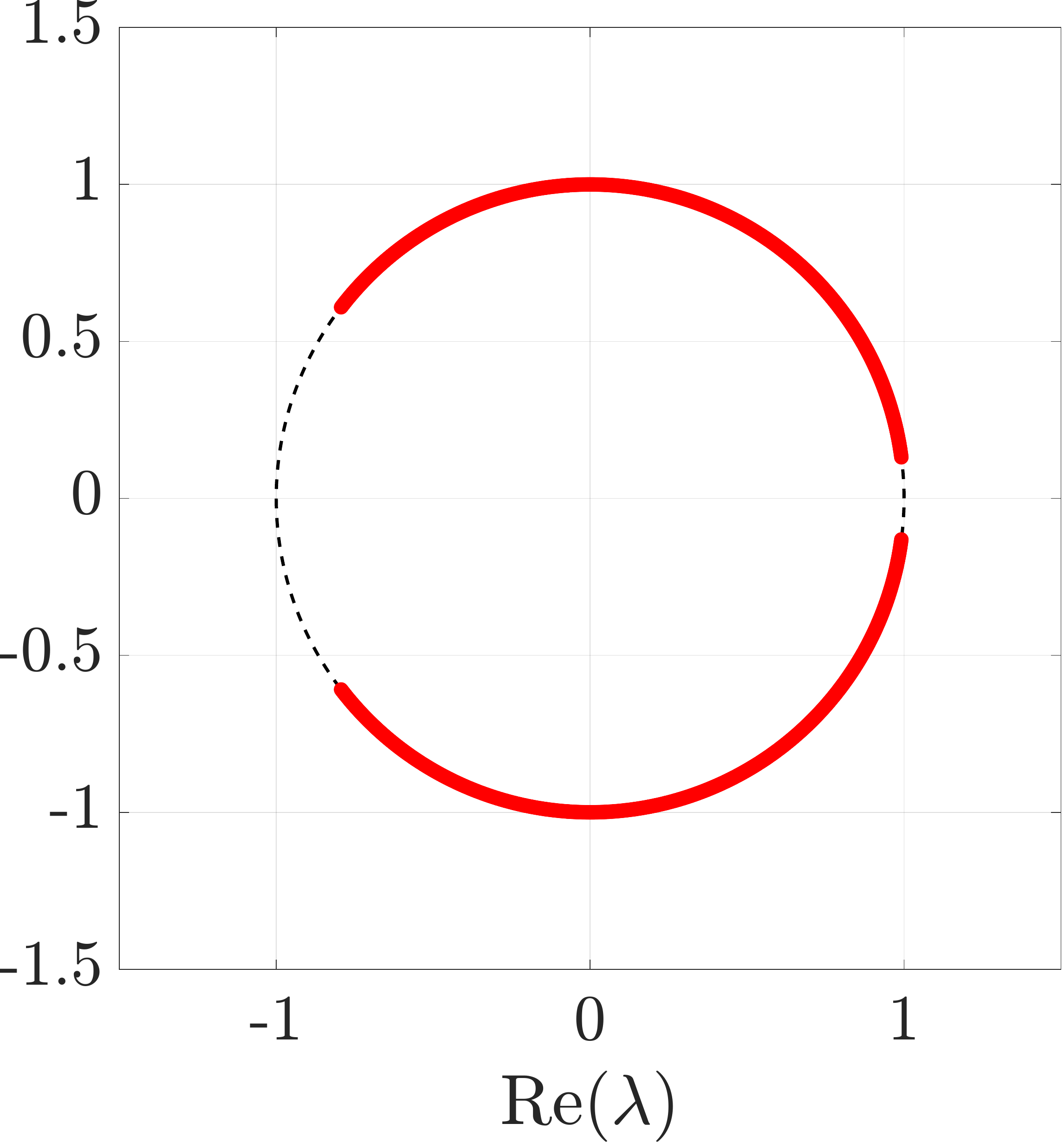}
		\label{fig:1DNUQWEigen2}}
	\subfigure{
		\includegraphics[height=3.85cm]{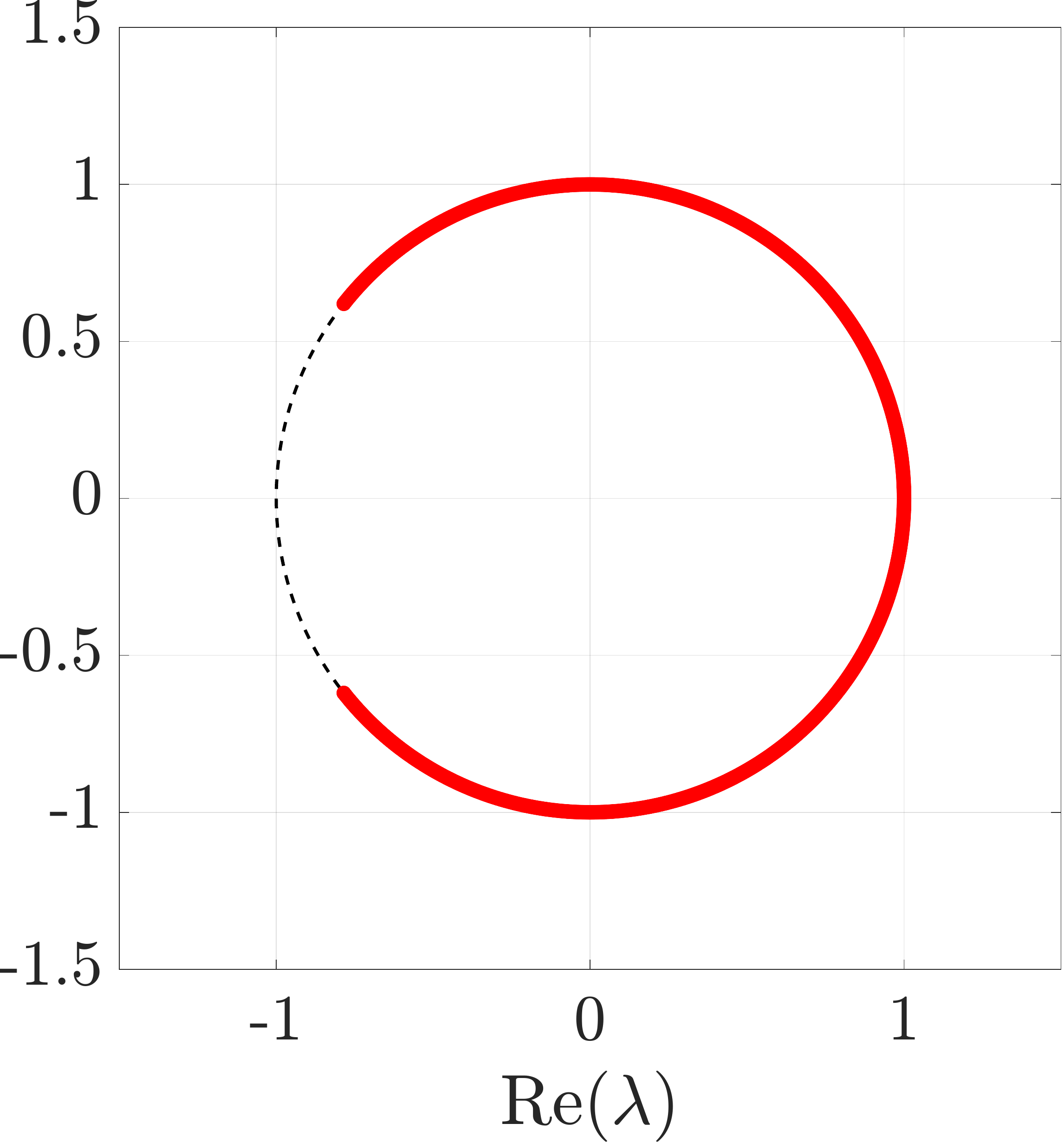}
		\label{fig:1DNUQWEigen3}}
	\subfigure{
		\includegraphics[height=3.85cm]{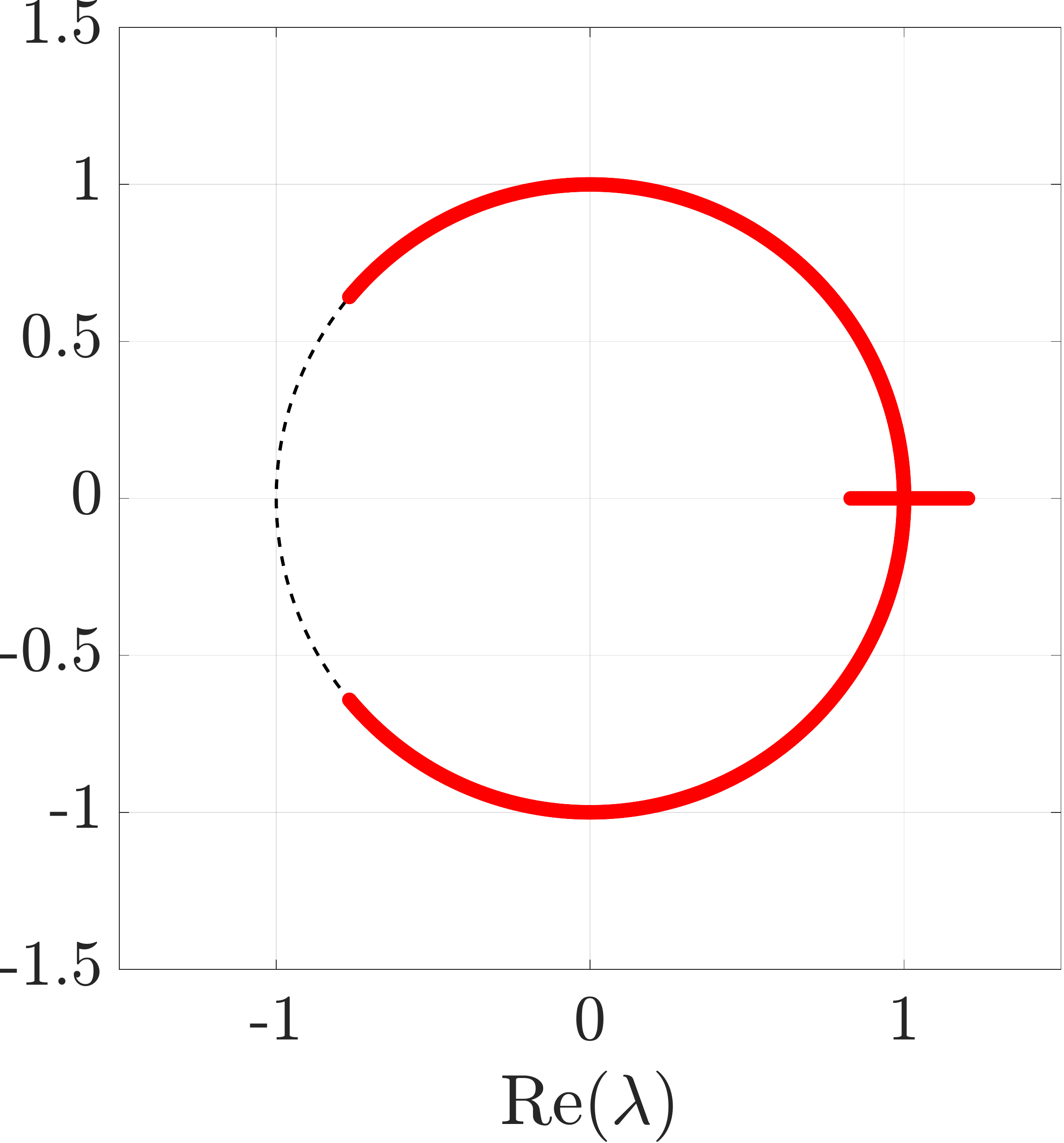}
		\label{fig:1DNUQWEigen4}}
	\caption{The quasienergy for $\theta_1 = \pi/4, \theta_2 = -\pi/6,$ and $\phi = 0$. In top row, the blue (red) corresponds to real (imaginary) part of the quasienergy, whereas in the bottom row we have eigenvalues of the unitary operator Eq.~\eqref{eq:NonUnitary-SSQW} on a unit circle. \subref{fig:1DNUQWEnergy1} $e^{\gamma} = 1$, \subref{fig:1DNUQWEnergy2} $e^{\gamma} = 1.1$, \subref{fig:1DNUQWEnergy3} $e^{\gamma} = e^{\gamma_c}$, \subref{fig:1DNUQWEnergy4} $e^{\gamma} = 1.4$.}
	\label{fig:1DNUQWEnergy}
\end{figure}

\begin{figure}
	\centering
	\subfigure[]{
		\includegraphics[height=3.75cm]{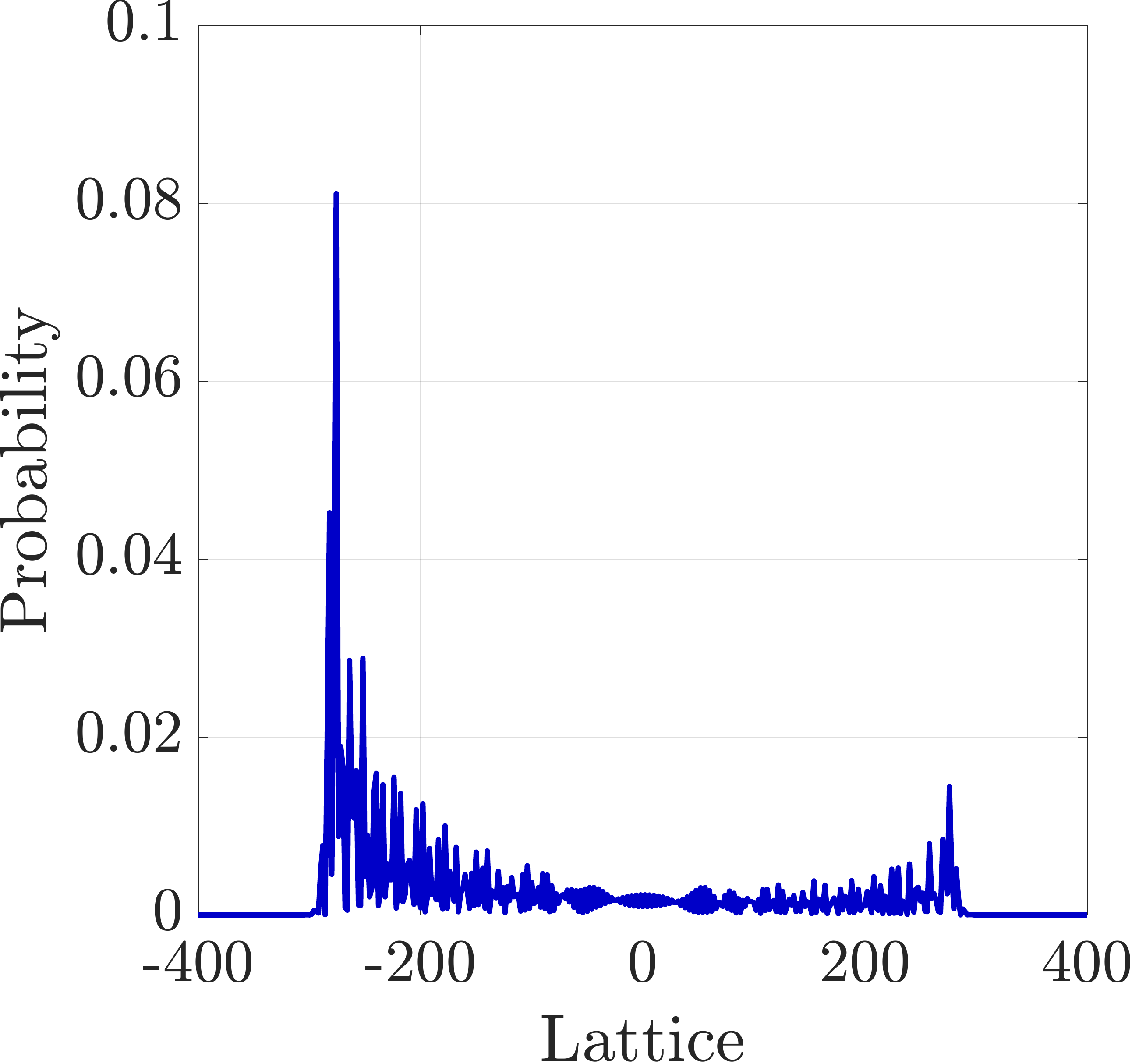}
		\label{fig:1DNUQWE1}}
	\subfigure[]{
		\includegraphics[height=3.75cm]{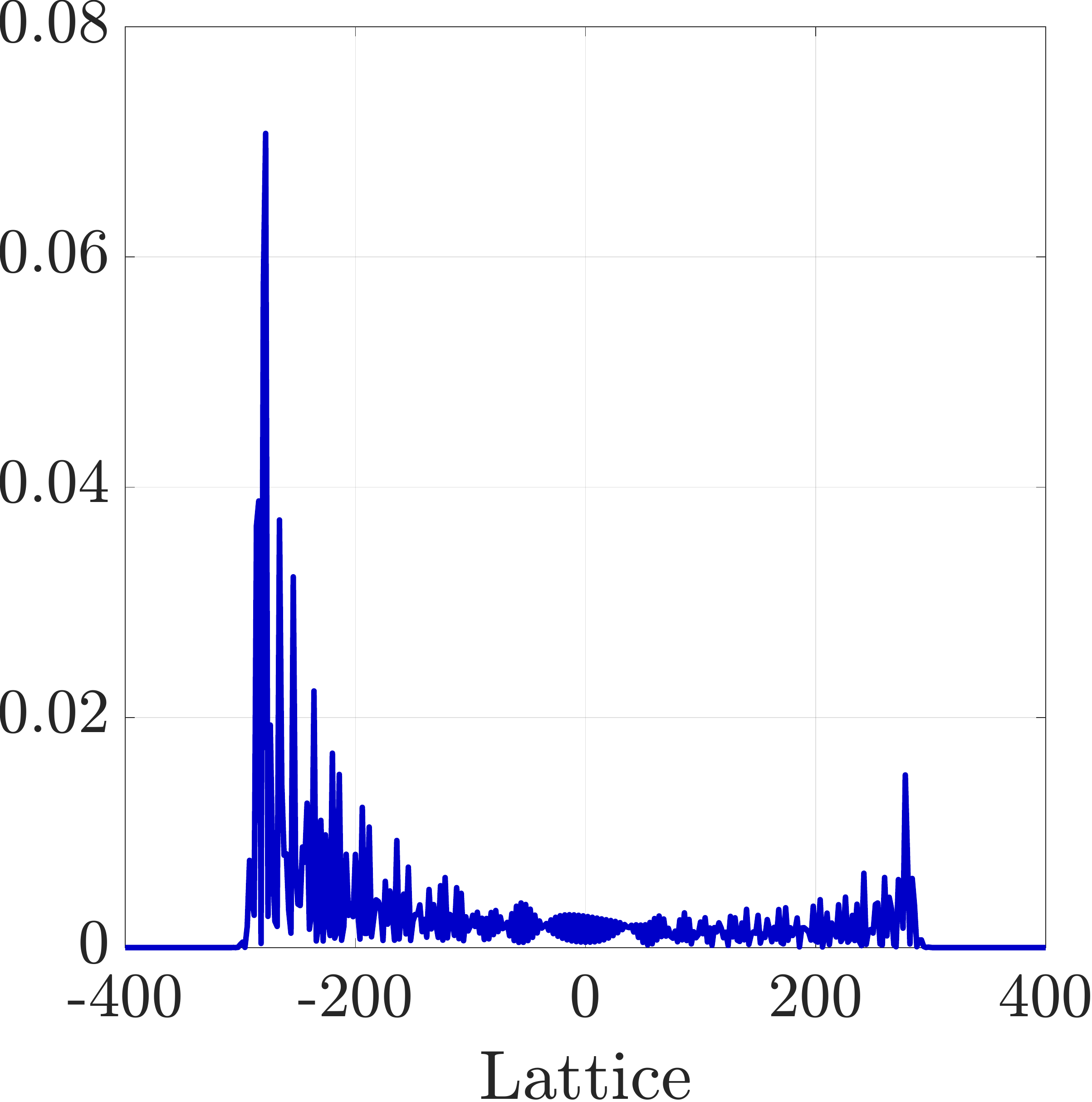}
		\label{fig:1DNUQWE2}}
	\subfigure[]{
		\includegraphics[height=3.75cm]{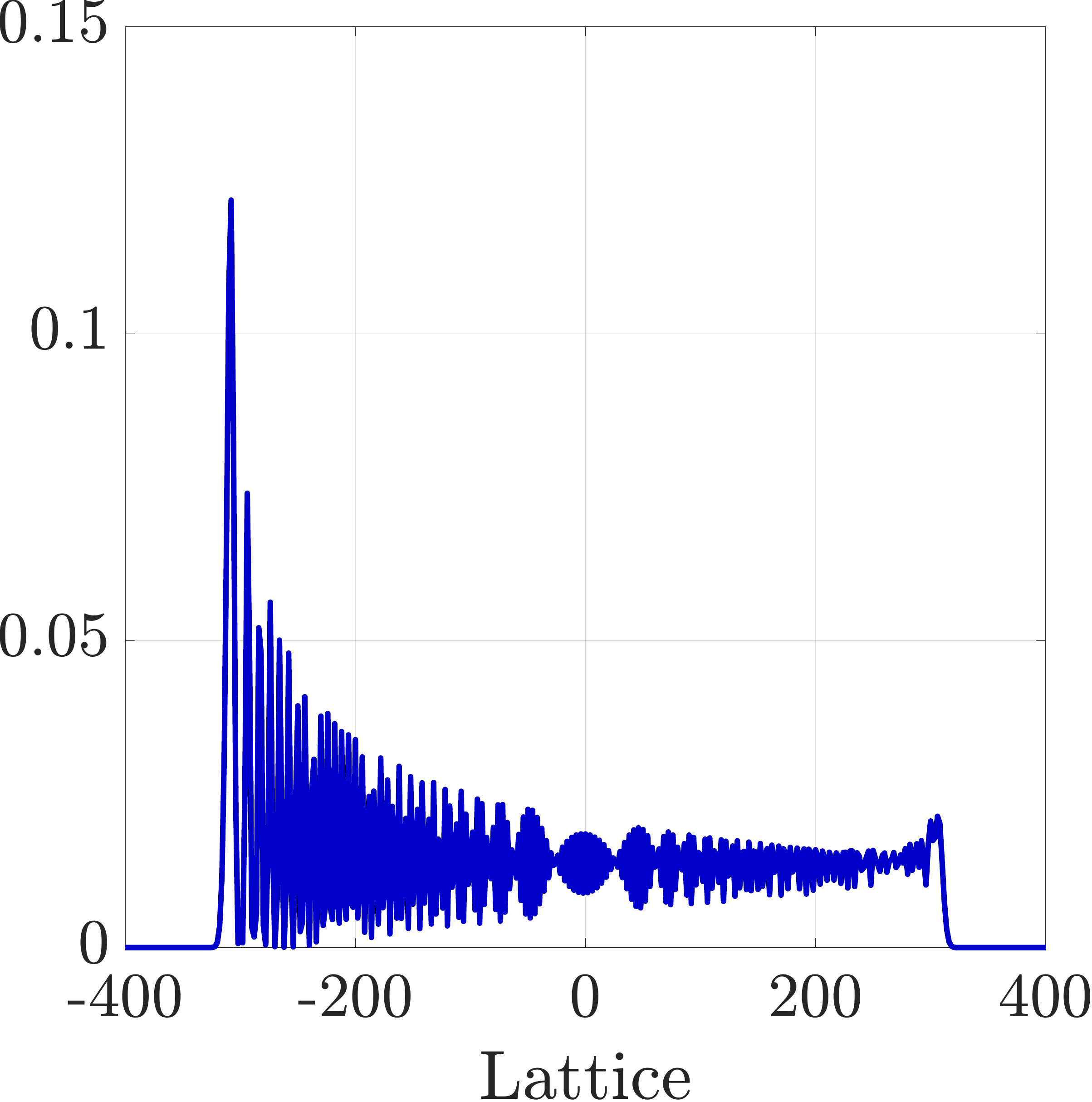}
		\label{fig:1DNUQWE3}}
	\subfigure[]{
		\includegraphics[height=3.80cm]{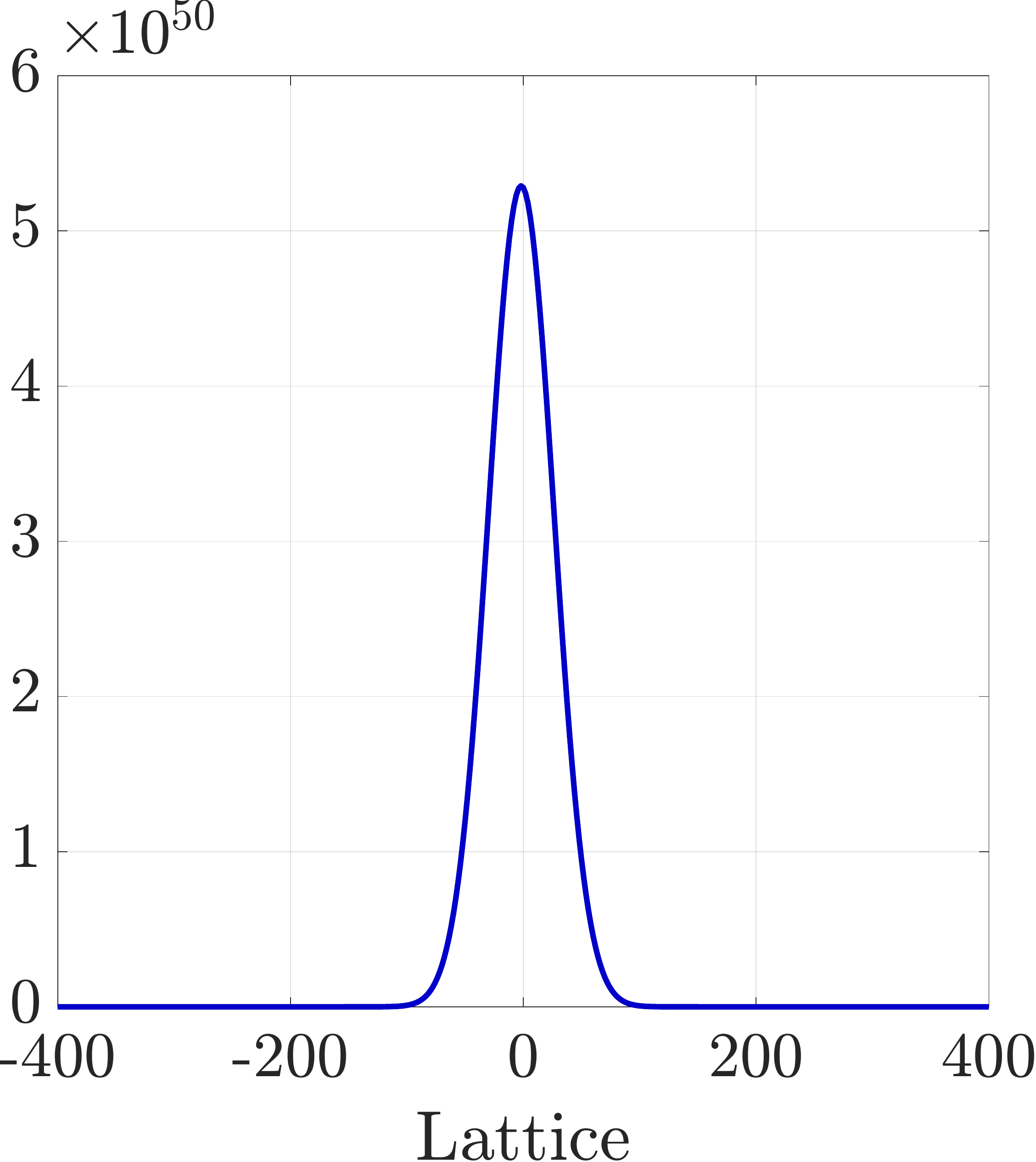}
		\label{fig:1DNUQWE4}}
	
	\subfigure{
		\includegraphics[height=3.75cm]{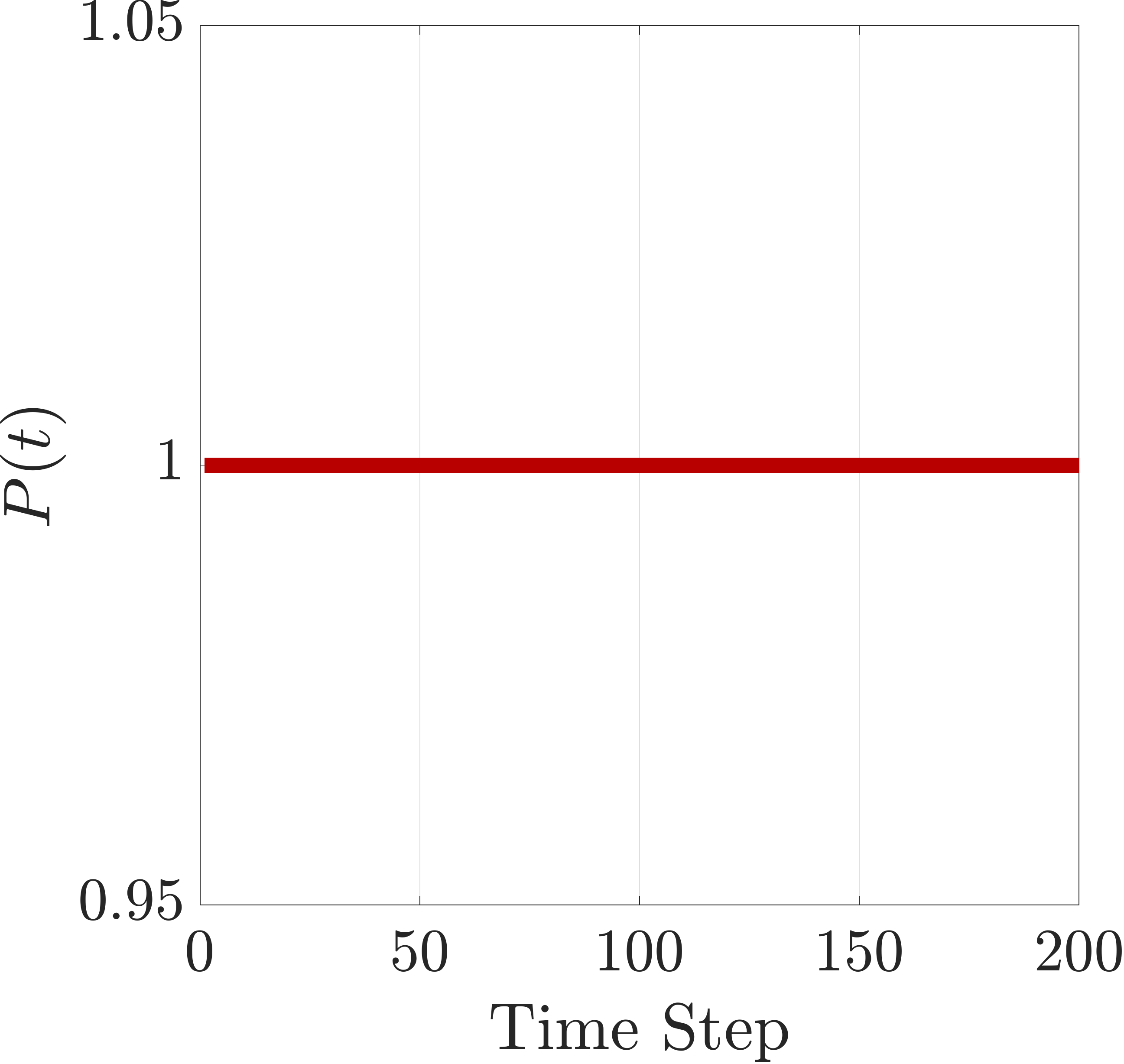}
		\label{fig:1DNUQWN1}}
	\subfigure{
		\includegraphics[height=3.75cm]{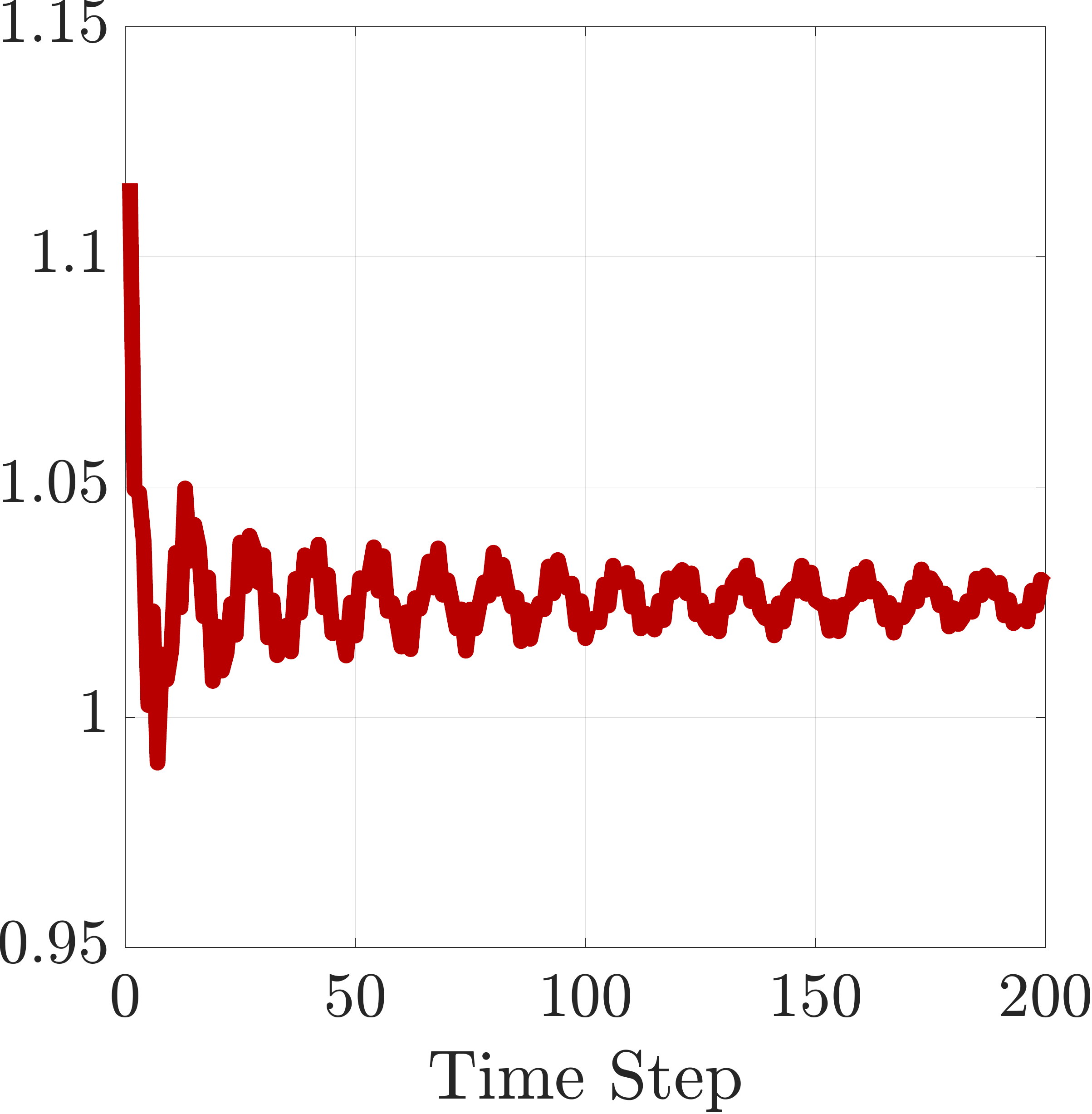}
		\label{fig:1DNUQWN2}}
	\subfigure{
		\includegraphics[height=3.75cm]{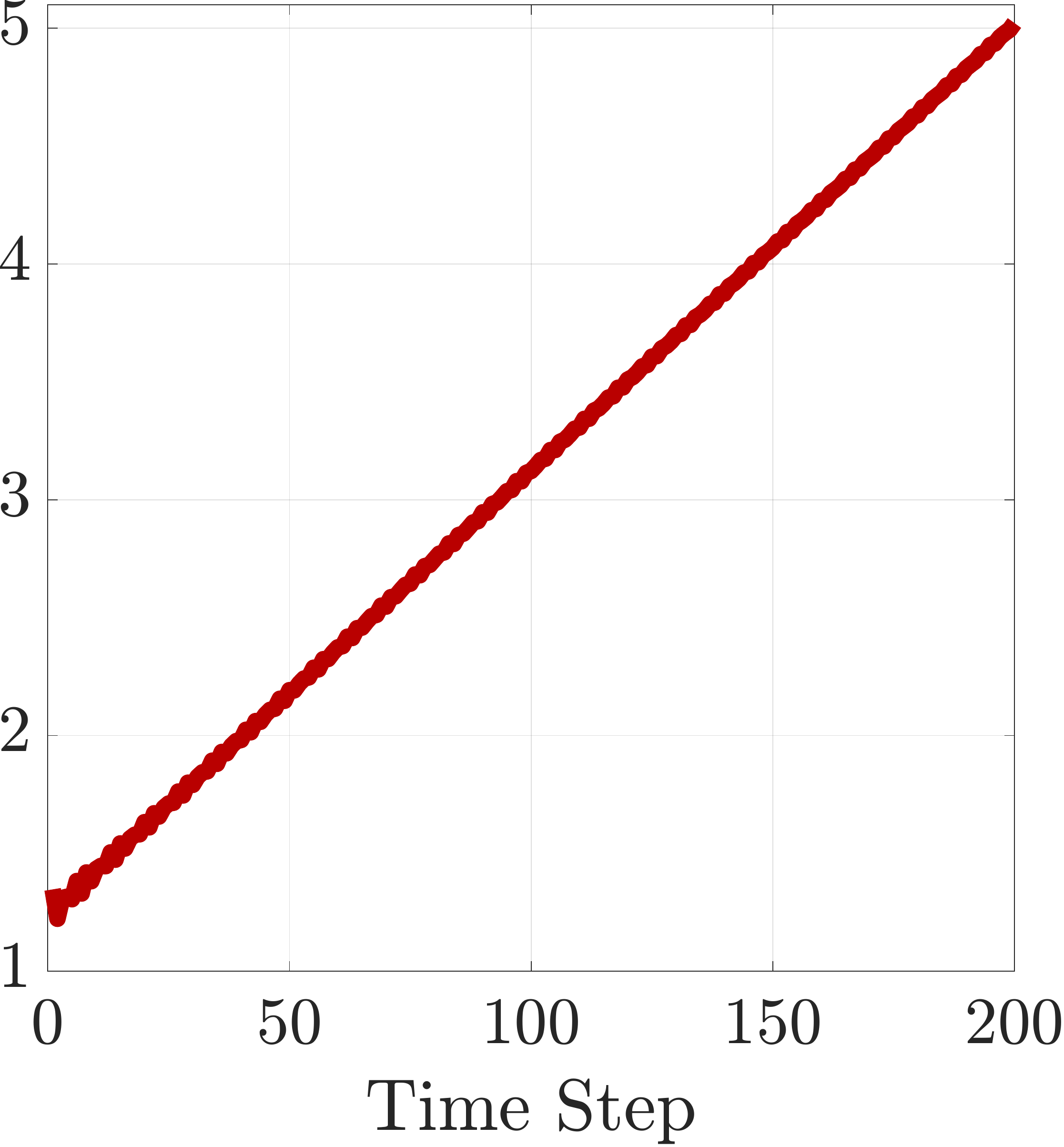}
		\label{fig:1DNUQWN3}}
	\subfigure{
		\includegraphics[height=3.75cm]{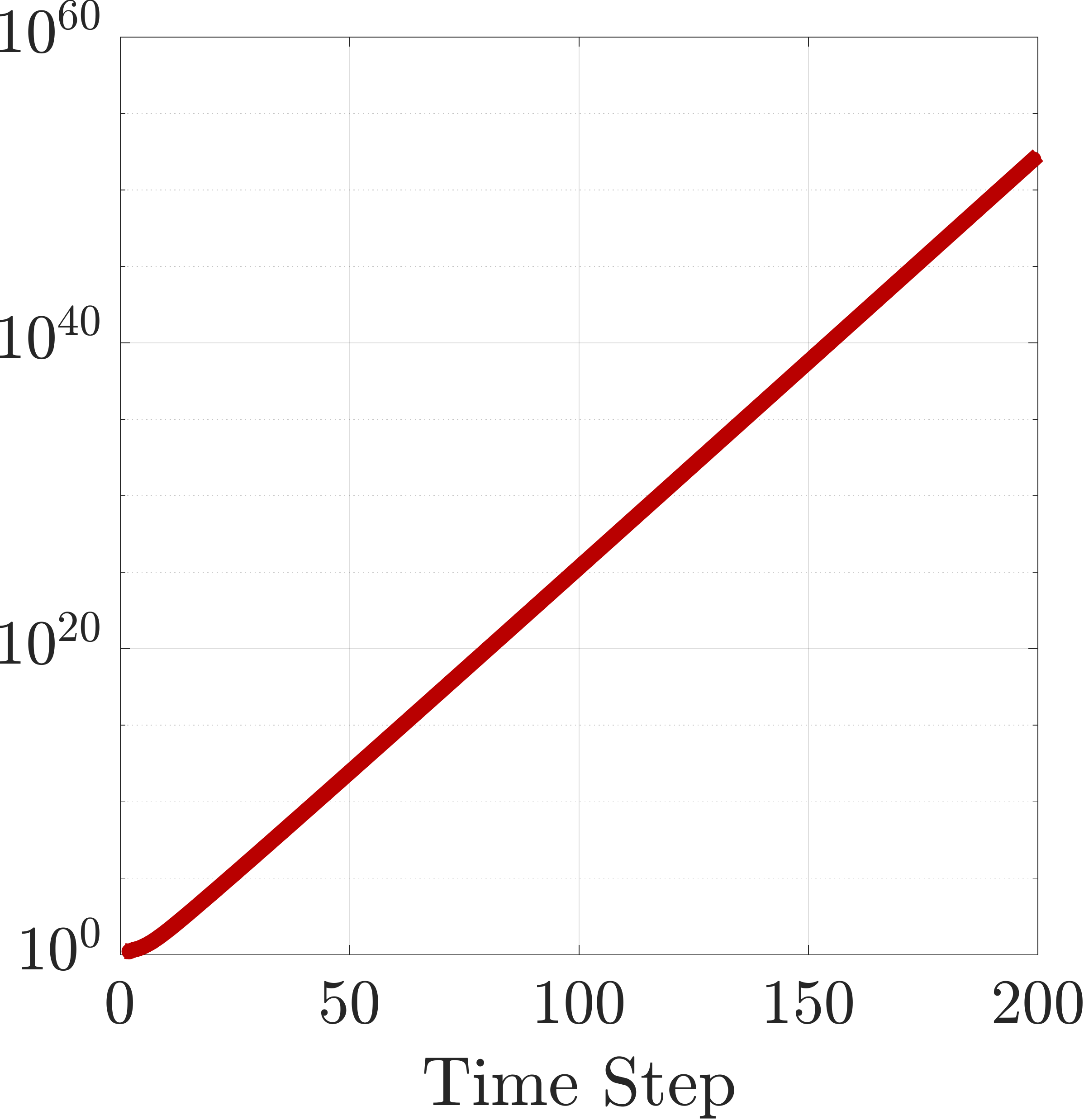}
		\label{fig:1DNUQWN4}}
	\caption{Probability distribution of the walker initially localized at the origin and with the symmetric state of the coin for $\theta_1 = \pi/4, \theta_2 = -\pi/6$ after 200 steps in top row. In bottom row we plot the time evolution of the norm for different settings of $\gamma$. \subref{fig:1DNUQWE1} $e^{\gamma} = 1$, \subref{fig:1DNUQWE3}, $e^{\gamma} = 1.1$ $e^{\gamma} = e^{\gamma_c}$, \subref{fig:1DNUQWE4} $e^{\gamma} = 1.4$. The system size is taken to be $ N = 400 $.}
	\label{fig:1DNUQW}
\end{figure}

For the homogeneous system, all the operators are $n$ independent i.e. $\{ \tilde{R}(\theta_{i,n}), \tilde{G}_{i,n}, \tilde{\Phi}_{i,n}\}$ become $\{ \tilde{R}(\theta_{i}), \tilde{G}_{i}, \tilde{\Phi}_{i} \}$ and are given by
\begin{align}
	\tilde{G}_2 &= \tilde{G}_1^{-1} = \tilde{G} = \begin{pmatrix}
		e^{\gamma} & 0 \\
		0 & e^{-\gamma}
	\end{pmatrix} = e^{\gamma \sigma_z} \\
	\tilde{\Phi}_2 &= \tilde{\Phi}_1 = \tilde{\Phi} = \begin{pmatrix}
		e^{i \phi} & 0 \\
		0 & e^{-i\phi}
	\end{pmatrix} = e^{i \phi \sigma_z}
\end{align}
where $\sigma_z$ is the Pauli $z$ matrix in computational basis. The time evolution operator in Eq.~\eqref{eq:NUSSQWUni} reads 
\begin{equation} \label{eq:NUSSQWUni1}
	U = T_{\downarrow} G \Phi R(\theta_2)T_{\uparrow} G^{-1} \Phi R(\theta_1)
\end{equation}
With the initial state $\ket{\psi(0)}$, the state after $t$ time step will be~\cite{NielsenChuang2010}
\begin{equation}
	\ket{\psi(t)} = U^t \ket{\psi(0)} = \sum_{n, s = \uparrow, \downarrow} C_{n,s}(t) \ket{n} \otimes \ket{s}
\end{equation}
where $C_{n,s}(t)$'s are the normalized probability amplitudes. We can now write the eigenvalue equation to define the quasienergy as
\begin{equation}
	U \ket{\psi_{\lambda}} = \lambda \ket{\psi_{\lambda}} ; \lambda = e^{i \varepsilon}.
\end{equation}
For a unitary quantum walk $\abs{\lambda} = 1$ and $\epsilon$ is real. 
For the case $\phi = 0$, we have
\begin{equation} \label{eq: Unitary NonUSSQW}
	U^{^{\text{NU}}}_{_{\text{SS}}} = T_{\downarrow} G_2 R(\theta_2) T_{\uparrow} G_1 R(\theta_1),
\end{equation}    
with
\begin{align}
	G_2 &= G_1^{-1} = G_{\gamma} = \begin{pmatrix}
		e^{\gamma} & 0 \\
		0 & e^{-\gamma}
	\end{pmatrix} \otimes \mathds{1}. 
\end{align} 
The above choice of operators is motivated by the experimental setup used in~\cite{Regensburger2012}. The factor $\gamma$ is known as the loss and gain factor as the operator $G$ results in increasing (decreasing) the amplitude of spin-up (down). The time evolution operator for the non-unitary 1D SSQW becomes 
\begin{equation} \label{eq:SSQW-TimeEvolution}
	U^{^{\text{NU}}}_{_{\text{SS}}} = T_{\downarrow} G_{\gamma} R(\theta_2) T_{\uparrow} G_{\gamma}^{-1} R(\theta_1).
\end{equation}
This particular choice of the scaling operator leaves the translational symmetry of the quantum walk intact. Hence, the dynamical operator can be block diagonalized in the momentum basis as
\begin{equation} \label{Uk}
	U^{^{\text{NU}}}_{_{\text{SS}}} = \sum_{k} \tilde{U}^{^{\text{NU}}}_{_{\text{SS}}}(k) \otimes \dyad{k}, 
\end{equation}
where
\begin{equation} \label{eq:NonUnitary-SSQW}
	\tilde{U}^{^{\text{NU}}}_{_{\text{SS}}}(k) = T_{\downarrow}(k) G_{\gamma} R(\theta_2) T_{\uparrow}(k) G_{\gamma}^{-1} R(\theta_1),
\end{equation}      
with $ T_{\downarrow}(k) = e^{i k (\sigma_z - \mathds{1})/2} $, $ T_{\uparrow}(k) = e^{i k (\sigma_z + \mathds{1})/2} $ and it acts only on the coin part. The corresponding generator or an effective Hamiltonian $H_{_{\text{NU}}}(\theta_1, \theta_2,  \gamma)$ reads
\begin{equation} \label{eq:Hamil-SSQW}
	H_{_{\text{NU}}}(\theta_1, \theta_2, \gamma) = \bigoplus_k  E(k)\, \hat{\vb{n}}(k) \vdot \vb{\sigma},
\end{equation}
with quasi-energy
\begin{equation} \label{eq:Energy-SSQWL}
	\cos E(k) = \cos (\theta_1/2) \cos (\theta_2/2) \cos k - \sin (\theta_1/2) \sin (\theta_2/2) \cosh 2 \gamma,
\end{equation}
and $\vb{\hat{n}} = n_x(k) \vb{\hat{i}} + n_y(k) \vb{\hat{j}} + n_z(k) \vb{\hat{k}}$ with
\begin{align}
	n_x(k) &= \dfrac{\sin(\theta_1/2) \cos(\theta_2/2) \sin k  - i \cos(\theta_1/2) \sin(\theta_2/2) \sinh 2\gamma}{\sin E(k)}, \nonumber \\
	n_y(k) &= \dfrac{\sin(\theta_1/2) \cos(\theta_2/2) \cos k + \cos(\theta_1/2) \sin(\theta_2/2) \cosh 2\gamma}{\sin E(k)}, \nonumber \\
	n_z(k) &= \dfrac{- \cos(\theta_1/2) \cos(\theta_2/2) \sin k - i \sin(\theta_1/2) \sin(\theta_2/2) \sinh 2\gamma}{\sin E(k)}.
\end{align}
Note that for $\gamma \ne 0$, $ G $ and $ U^{^{\text{NU}}}_{_{\text{SS}}} $ are no longer unitary operators and the norm of the state in evolution may not be preserved. Consequently, $H_{_{\text{NU}}}(\theta_1, \theta_2, \gamma)$ is not Hermitian, but we still have a real spectrum up to a certain critical value of $\gamma = \gamma_c$. Given the fact that the energy band closes at $k = 0, E = 0$ and from Eq.~\eqref{eq:Energy-SSQWL} we have an expression for $\gamma_c$ which reads
\begin{equation} \label{Eq:Delta-c}
	\gamma_c = \dfrac{1}{2}\cosh^{-1}\left[\dfrac{\cos (\theta_1/2) \cos (\theta_2/2)  - 1}{\sin (\theta_1/2) \sin (\theta_2/2)}\right].
\end{equation}
The argument of $\cosh^{-1}$ in the above equation is positive (or negative) when $\theta_1$ and $\theta_2$ have the opposite (or same) sign. The negative argument results in a complex value of $\gamma_c$. 
Therefore, $ \gamma $ can be complex in general and it would be convenient to use the form given by $\gamma = a + i b$.
We observe that the negative argument of $\cosh^{-1}$ results in  a critical value of $b$, $\pi/2$. It amounts to shifting $k \to k+\pi/2$, leaving all the results unaffected.  Therefore, we restrict ourselves to the case when $\gamma_c$ is real, i.e., opposite signs for $\theta_1$ and $\theta_2$ and refer $\gamma$ as the scaling factor. The $\gamma_c $ is the point where the exact $\mathcal{PT}$-symmetry of the system breaks spontaneously (also known as the exceptional point~\cite{Ozdemir2019}), and we will have complex energies for $\gamma > \gamma_c$. We plotted quasi-energy Eq.~\eqref{eq:Energy-SSQWL} and the eigenvalues of the unitary operator Eq.~\eqref{eq:NonUnitary-SSQW} as a function of $k$ and for several values of $\gamma$ in Fig.~\eqref{fig:1DNUQWEnergy}. We can clearly see that the spectrum of quasi-energy becomes imaginary for values of $\gamma > \gamma_c$ and the eigenvalues of the unitary operator do not constrain to the circle of unit radius.
\noindent We plot the time evolution of sum of probability distribution $P(t)$ (norm)
\begin{equation}
	P(t) = \sum_{n} \abs{C_n(t)}^2
\end{equation}
and the probability distribution
\begin{equation}
	\abs{C_n(t)}^2 = \abs{C_{n, \uparrow}(t)}^2 + \abs{C_{n, \downarrow}(t)}^2
\end{equation} 
in Fig.~\eqref{fig:1DNUQW} for different values of the scaling parameter $\gamma$. Till the critical point $\gamma_c$ we see that the time evolution is not very different from that of unitary quantum walk and we observe that the norm of the wavefunction oscillates around as shown in Fig.~\ref{fig:1DNUQWE2}. However, at the critical or exceptional point $\gamma_c$, the situation changes very drastically and we see linear growth and beyond the point $\gamma_c$ we observe exponential growth as shown in Fig.~\ref{fig:1DNUQWE3} and ~\ref{fig:1DNUQWE4}, respectively. Also, the time evolution beyond the exceptional point can be approximated by the Gaussian distribution centered around the initial position [Fig.~\ref{fig:1DNUQWE4}].

\section{Physics of degenerate and exceptional points}
Here, we illustrate the difference between degenerate points and exceptional points by considering an explicit example. Let us take a non-symmetric matrix~\cite{Francisco2021}
\begin{equation}
	H(\beta) = \begin{pmatrix}
		0 & 1 & 1\\
		1 & 0 & 1\\
		\beta & 1 & 0
	\end{pmatrix}
\end{equation}
which has the following eigenvalues
\begin{equation}
	E_1 = -1, E_{\pm} = \dfrac{1}{2} \left(1 \pm \sqrt{5 + 4 \beta}\right).
\end{equation}
Here we observe that $E_1$ and $E_-$ become degenerate for $\beta = 1$ and the corresponding orthonormal vectors will be
\begin{gather}
	E_- = -1, \;\;\;\ket{E_-} = \dfrac{1}{\sqrt{2}}\begin{pmatrix}
		-1 \\
		0 \\
		1
	\end{pmatrix}, E_1 = -1, \;\;\;\ket{E_1} = \dfrac{1}{\sqrt{2}}\begin{pmatrix}
		-1 \\
		1 \\
		0
	\end{pmatrix}, \nonumber\\
	E_+ = 2, \;\;\;\ket{E_+} = \dfrac{1}{\sqrt{3}}\begin{pmatrix}
		1 \\
		1 \\
		1
	\end{pmatrix}.
\end{gather} 
The symmetric matrix $H(1)$ exhibits 3 linearly independent eigenvectors and can be diagonalized by a matrix $S$ formed by the three eigenvectors such that
\begin{equation}
	S^{-1} H(1) S = \begin{pmatrix}
		2 & 0 & 0\\
		0 & -1 & 0\\
		0 & 0 & -1
	\end{pmatrix}, \qquad S = \{ \ket{E_+}, \ket{E_-}, \ket{E_{1}} \}.
\end{equation}
This degeneracy is what we generally come across in quantum mechanics. 

\begin{figure}
	\centering
	\subfigure[]{
		\includegraphics[width=6cm]{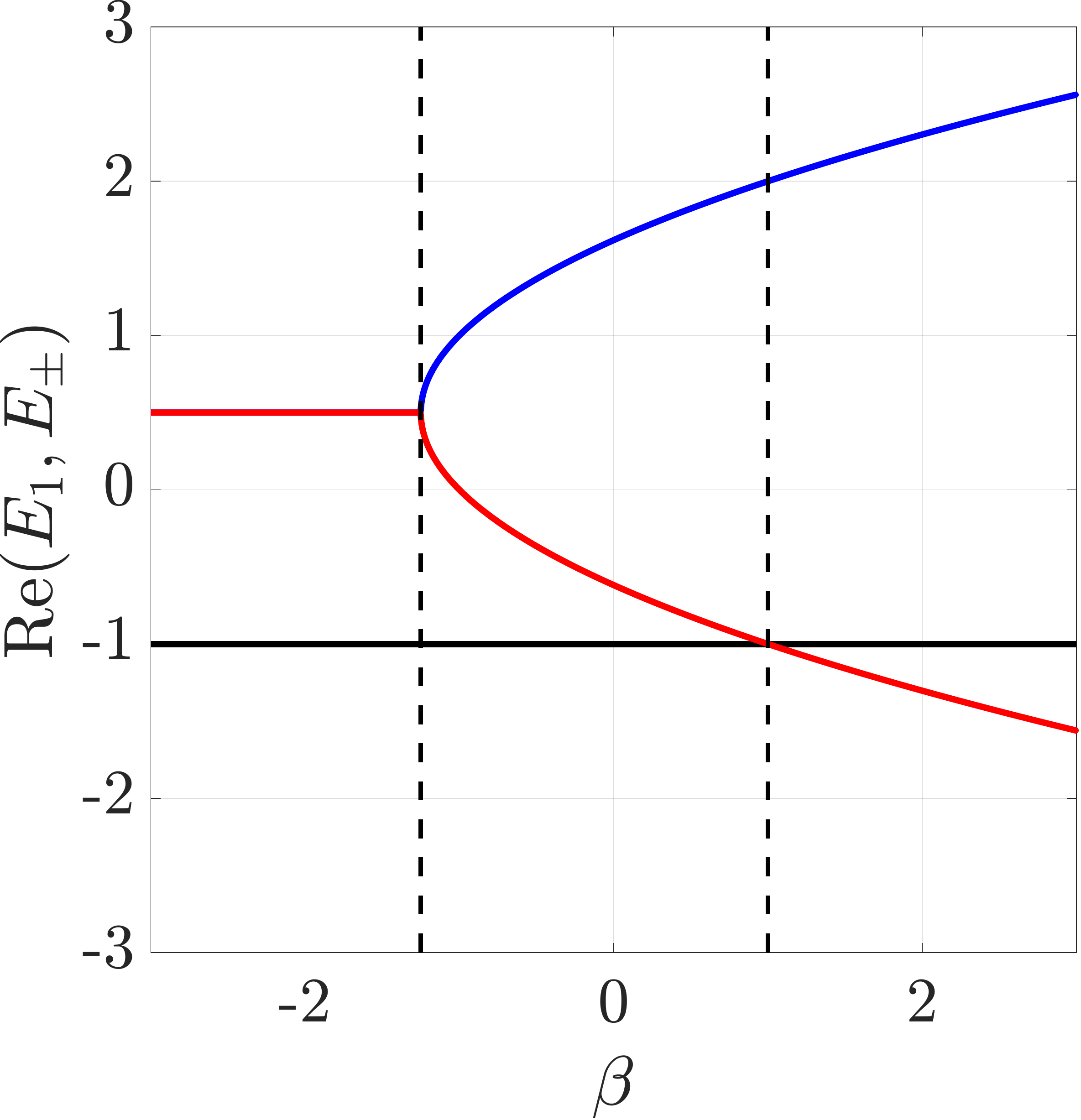}
		\label{fig:hbetareal}}
	\subfigure[]{
		\includegraphics[width=6cm]{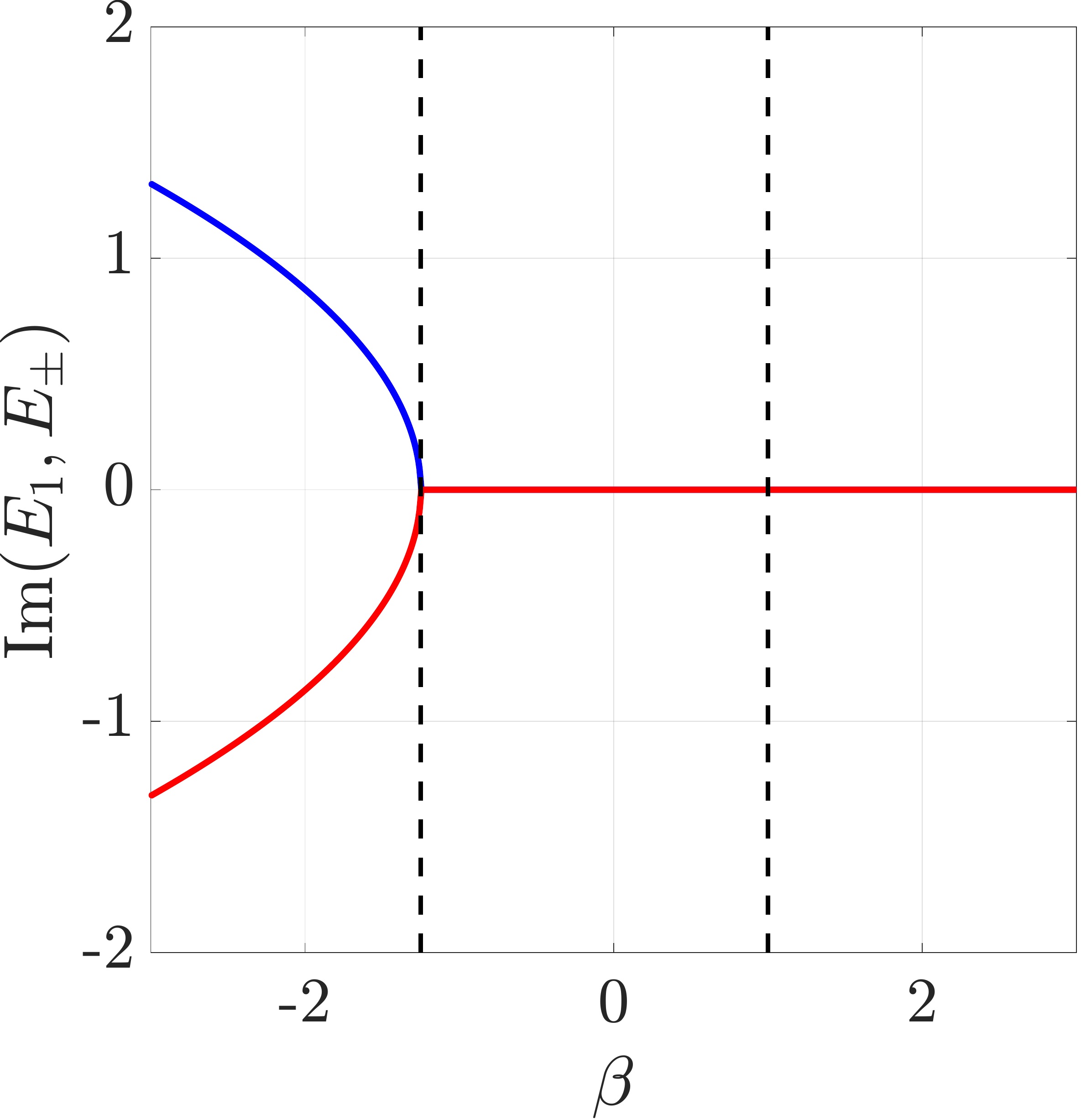}
		\label{fig:hbetaimag}}
	\caption{Real~\subref{fig:hbetareal} and imaginary~\subref{fig:hbetaimag} part of the eigenvalues are plotted with the parameter $\beta$. The blue and red corresponds to $E_+$ and $E_-$ respectively.}
	\label{fig:hbeta}
\end{figure}

However, the eigenvalues $E_+$ and $E_-$ coalesce at $\beta = -5/4$. The real and imaginary parts of the eigenvalues are plotted in Fig.~\eqref{fig:hbeta}. The eigenvalues become complex for $\beta < -5/4$ as shown in Fig.~\eqref{fig:hbeta}. The Hamiltonian $H(-5/4)$ has the eigensystem given by 
\begin{gather}
	E_- = E_+ = 1/2, \;\;\;\ket{E_{\pm}} = \dfrac{1}{3}\begin{pmatrix}
		-2 \\
		-2 \\
		1
	\end{pmatrix}, \nonumber \\ E_1 = -1, \;\;\;\ket{E_1} = \dfrac{1}{\sqrt{2}}\begin{pmatrix}
		0 \\
		-1 \\
		1
	\end{pmatrix}.
\end{gather} 
We note here that $H(-5/4)$ exhibit only two linearly independent vectors and $H(-5/4)$ is no longer diagonalizable because $S^{-1}$ does not exist. Therefore, coalescence is different from degeneracy because at the exceptional point there is only one linearly independent eigenvector. It is sometimes referred to as non-Hermitian degeneracy as opposed to Hermitian degeneracy.
\chapter{Persistence of Topological Phases in Non-Hermitian Quantum Walk} \label{ch:pers}
In this work we introduce non-Hermiticity and establish the \textit{persistent} nature of topological phases in non-Hermitian quantum walks~\cite{Mittal2021}. We show that the topological nature of the underlying Hamiltonian does not change in the lossy environment within certain limits.  We show that in 1D SSQW the topological phase persists as long the system possess exact $\mathcal{PT}$-symmetry. In the case of a 2D quantum walk, such correspondence between the topological order and the $\mathcal{PT}$-symmetry is missing. In this systems, an interesting observation is the loss-induced topological phase transition, which is absent in the 1D case.
\section{Symmetries of the Hamiltonian}
The system under consideration is quantum walk which is performed on position space with the aid of coin states. Since, the time evolution operator $U(k)$  is block diagonal in the momentum space i.e. $U(k) = \sum_{k} \tilde{U}(k) \otimes \dyad{k}$ and $U(k) = e^{-i H(k)}$, we write the condition for $\tilde{U}(k)$ in order to have $\mathcal{PT}$ symmetry as
\begin{equation}
	(\mathcal{\tilde{\mathcal{P}}\tilde{\mathcal{T}}}) \tilde{U}(k) (\mathcal{\tilde{\mathcal{P}}\tilde{\mathcal{T}}})^{-1} = \tilde{U}^{-1}(k),
\end{equation} 
where the operators $\tilde{\mathcal{P}}$, $\tilde{\mathcal{T}}$ act only on the coin Hilbert space.

In the case of non-Hermitian 1D SSQW, $\tilde{U}(k)$ given in Eq. \eqref{eq:NonUnitary-SSQW} satisfies the above mentioned conditions with the choice of $\tilde{\mathcal{P}} = \sigma_y$ and $\tilde{\mathcal{T}} = \sigma_x \mathcal{K}$ such that the combined operator becomes $\mathcal{\tilde{\mathcal{P}}\tilde{\mathcal{T}}} = i \sigma_z \mathcal{K}$ and we have
\begin{equation}
	\sigma_z \tilde{U}^*\sigma_z^{-1} = \tilde{U}^{-1}(k),
\end{equation}
where $\mathcal{K}$ is the complex conjugation operator. Therefore, the 1D SSQW is $\mathcal{PT}$-symmetric for all the values of $\delta$ (and $\gamma$). However, at the exceptional point~\cite{Ozdemir2019} $\gamma_c$, the eigenstates and eigenvectors become degenerate. Beyond this point, the eigenvectors of the Hamiltonian and the $\mathcal{PT}$ operator are not the same~\cite{Mochizuki2016}; hence the system no longer possesses exact-$\mathcal{PT}$-symmetry, which results in a complex spectrum, as shown in the previous section.

Next we discuss the particle hole symmetry (PHS) represented by an antiunitary operator $\Xi$, and the chiral symmetry (CS) represented by a unitary operator $\Gamma$ for the time evolution operator and they read~\cite{Sato2019}
\begin{align} 
	\Xi U(k) \Xi^{-1} &=  U(-k) \label{eq: PHS-symmetry-U}, \\
	\Gamma U(k) \Gamma^{-1} &=  U^{\dagger}(k) \label{eq: chiral-symmetry-U}.
\end{align}
We redefine the time evolution operator given in Eq.~\eqref{eq:NonUnitary-SSQW} by performing a unitary transformation which reads~\cite{Asboth2013}
\begin{equation} \label{eq:NonUnitary-SSQW-TS}
	\tilde{U}^{'}(k) = R(\theta_1/2) T_{\downarrow}(k) G_{\delta} R(\theta_2) T_{\uparrow}(k) G_{\delta}^{-1} R(\theta_1/2),
\end{equation}
which is related to $\tilde{U}^{^{\text{NU}}}_{_{\text{SS}}}(k)$ as $\tilde{U}^{'}(k) = R(\theta_1/2) \tilde{U}^{^{\text{NU}}}_{_{\text{SS}}}(k) R^{-1}(\theta_1/2)$. This is done to make the evolution operator symmetric in time and known as time-symmetric representation. 
The motivation behind this transformation is  to show the existence of CS and PHS in non-Hermitian 1D SSQW explicitly. We can clearly see that $\tilde{U}^{'}(k)$ satisfies Eq.~\eqref{eq: PHS-symmetry-U} and Eq.~\eqref{eq: chiral-symmetry-U} with the choice of $\Gamma = \sigma_x $ and $\Xi = \mathcal{K}$. Hence, with the existence of these symmetries, $\tilde{U}^{^{\text{NU}}}_{_{\text{SS}}}(k)$ belongs to a symmetry class (BDI$^{\dagger}$~\cite{Sato2019}) which supports $\mathds{Z}$ topological invariant. Note, we can also define another time-symmetric that reads~\cite{Xue2017}
\begin{equation}
	\tilde{U}^{''}(k) = R(\theta_2/2) T_{\downarrow}(k) G_{\delta} R(\theta_1) T_{\uparrow}(k) G_{\delta}^{-1} R(\theta_2/2).
\end{equation}
For the current purpose, we consider only the $ \tilde{U}^{'}(k) $ to define the topological invariants and see the persistence of with the introduction of gain and loss. 

\section{Results}\label{Sec:Results}
In this section, we study the behavior of the topological phases in 1D SSQW and 2D DTQW by introducing a nonzero scaling factor $\gamma$ which, essentially, makes the system non-Hermitian. In 1D SSQW, we find that the topological phases are unaffected even when the system is non-Hermitian (i.e., $\gamma \ne 0$), as far as the system possesses a real spectrum following the exact $\mathcal{PT}$-symmetry. However, the topological nature of the system vanishes  as we cross the exceptional point $\gamma_c$, which means the quantity  $W$ becomes a non-integer number which decays asymptotically to zero for $\gamma >\gamma_c$.
We observe the persistence of the Chern number $C$ in 2D DTQW as well until the scaling factor $\gamma$ reaches a critical value. However, unlike the 1D case, we cannot associate exact $\mathcal{PT}$-symmetry breaking with the point where the topological phase transition happens due to the absence of the $\mathcal{PT}$-symmetry in 2D DTQW. Since the $\mathcal{PT}$-symmetry is absent in 2D DTQW even in the unitary region, we can not associate the persistence of the topological phase with this particular symmetry. Furthermore, we observe a loss-induced topological phase transition in 2D DTQW.

\subsection{Topological phases in 1D non-unitary quantum walk}
We start our analysis with non-unitary 1D SSQW, with the associated non-Hermitian Hamiltonian $H_{_{\text{NU}}}(\theta_1, \theta_2, \gamma)$ being given in \eqref{eq:Hamil-SSQW}. Since the Hamiltonian is traceless for all values of $\gamma$, the corresponding eigenvalues will always be of the form of $\pm E(k)$. For each momentum $k$, we compute the energy eigenstates $\ket{\psi_\pm(k)}$ corresponding to energies $\pm E(k)$ and, we call the set $\{\ket{\psi_-(k)}\}$ and $\{\ket{\psi_+(k)}\}$ as the lower and upper energy bands, respectively. Using the expression for the winding number $W$ from \eqref{eq: Winding-Number}, we calculate the winding numbers for the lower and upper bands and we name them as $W_-$ and $W_+$, respectively. 

\begin{figure*}
	\centering
	\subfigure[]{
		\includegraphics[width=3.6cm]{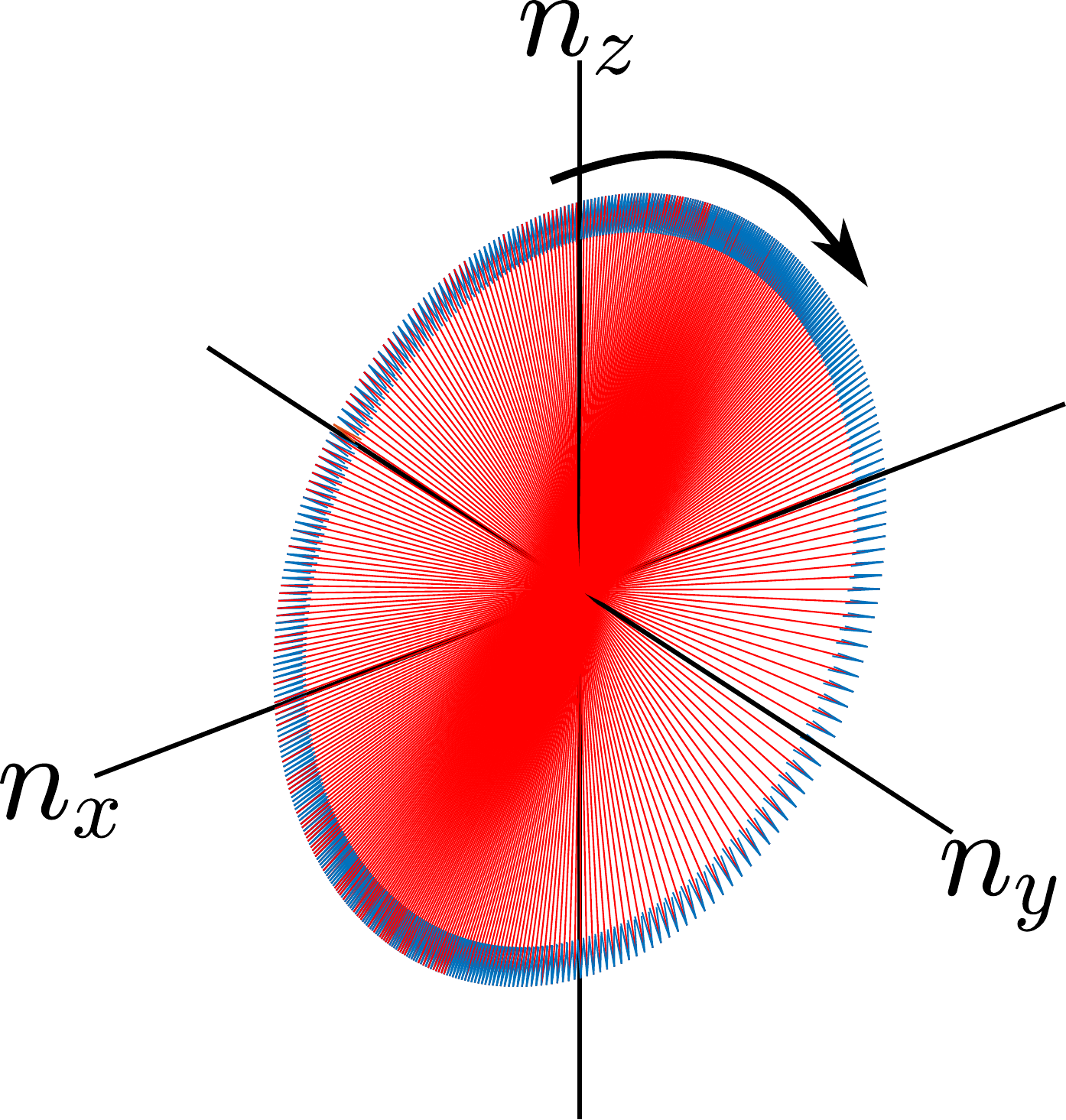}
		\label{fig:NVEC1}}
	\subfigure[]{
		\includegraphics[width=3.6cm]{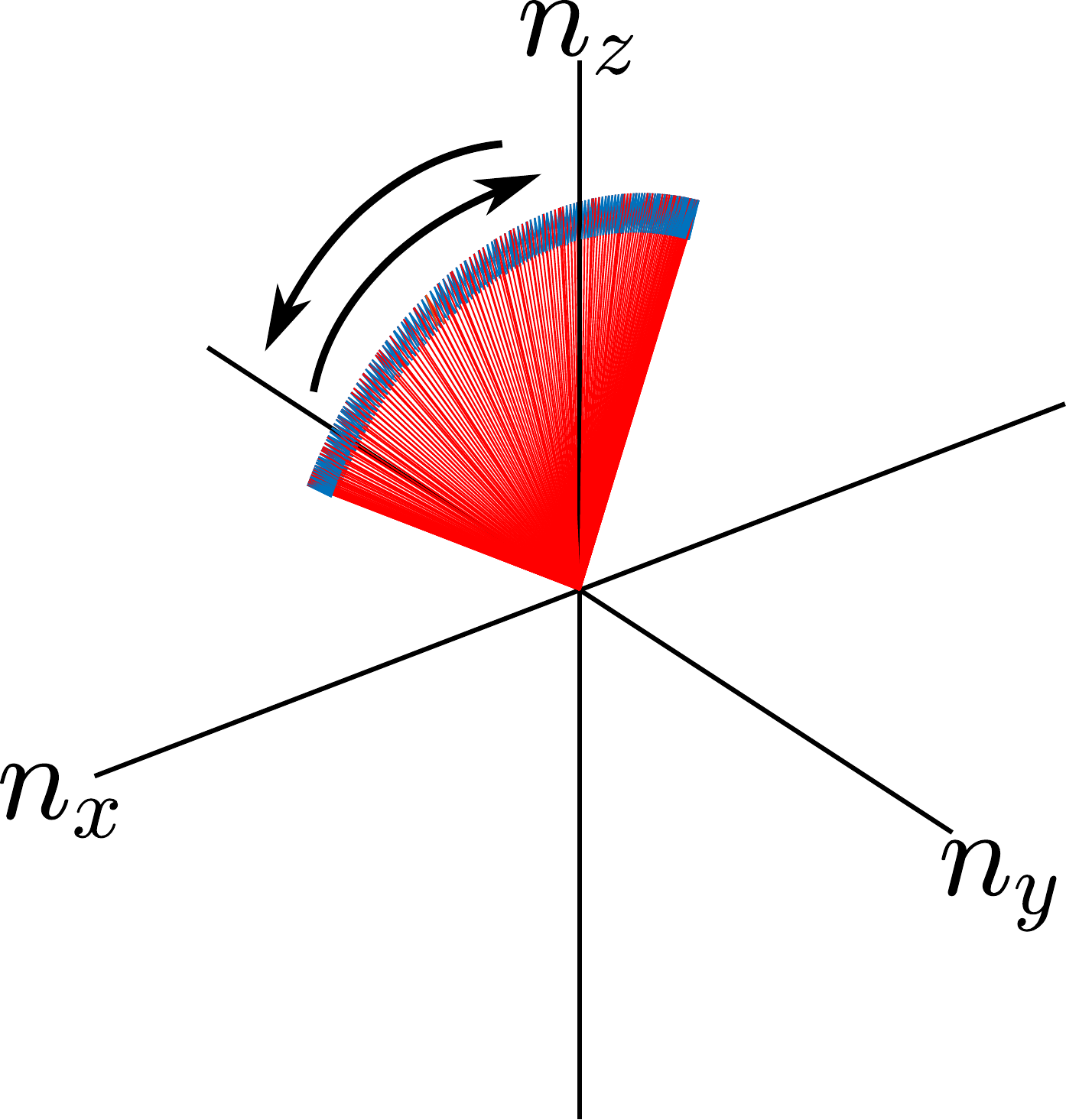}
		\label{fig:NVEC2}} 
	\subfigure[]{
		\includegraphics[width=3.6cm]{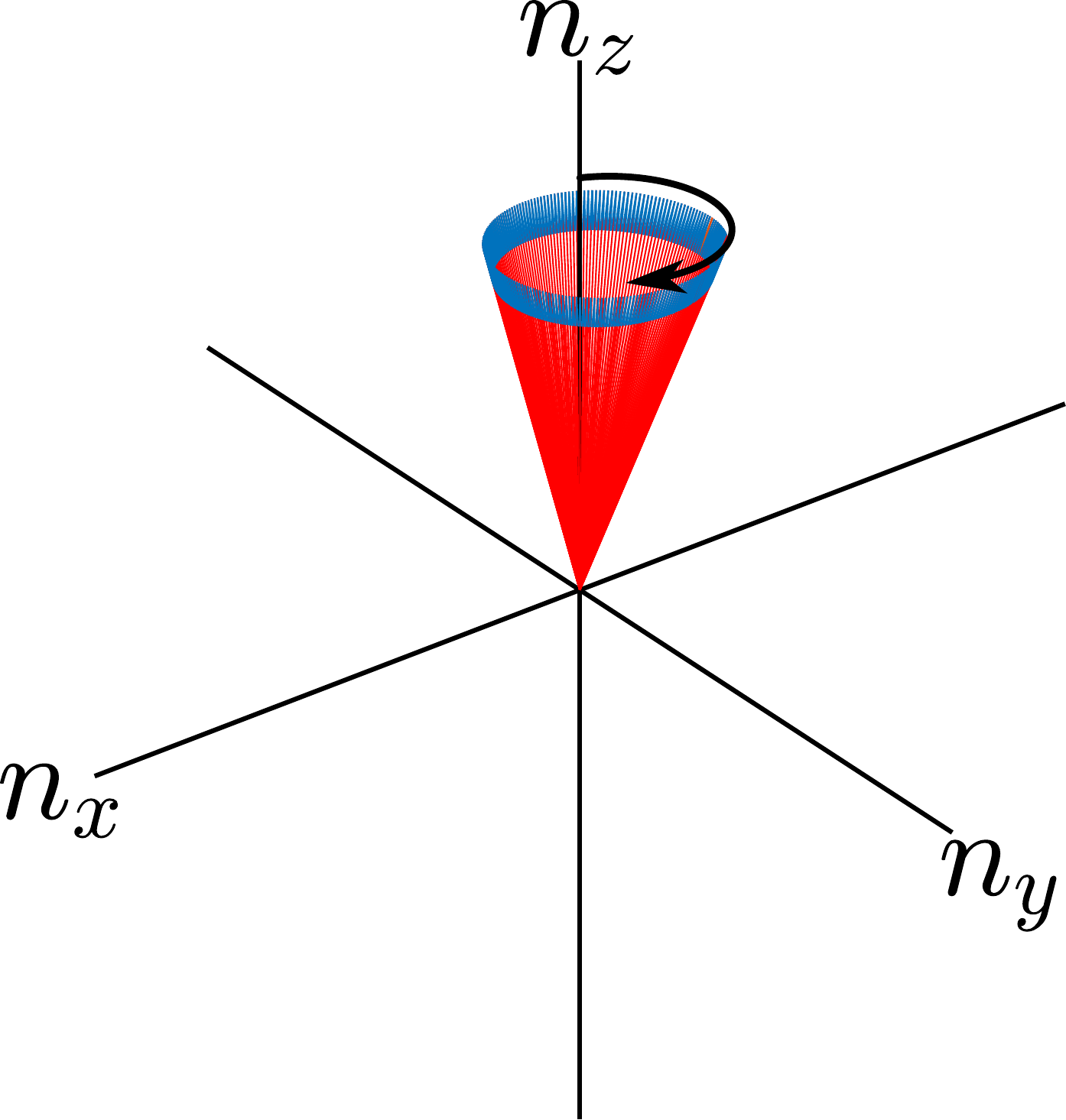}
		\label{fig:NVEC3}}
	\subfigure[]{
		\includegraphics[width=3.6cm]{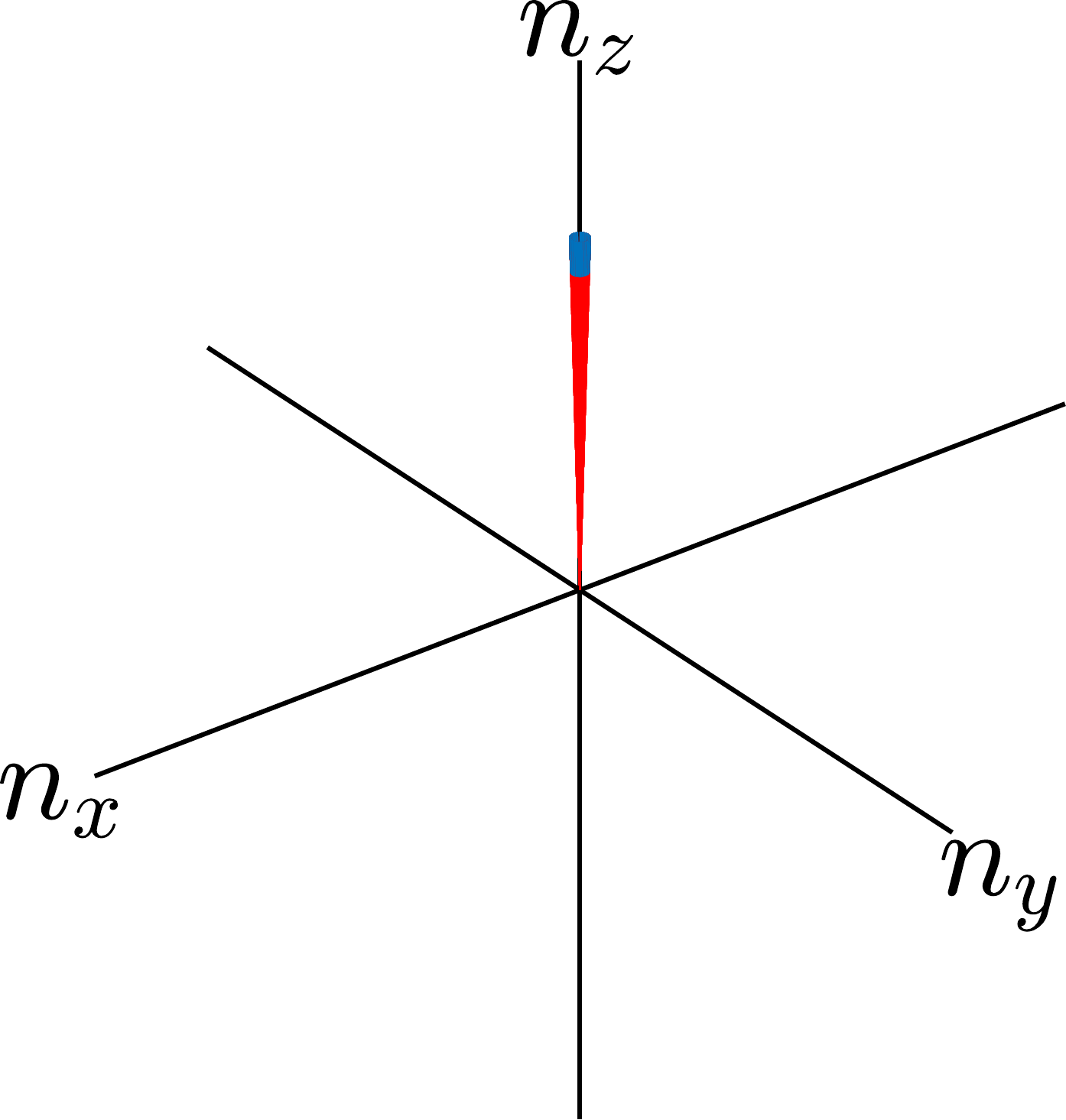}
		\label{fig:NVEC4}}
	\caption{(Color online) Winding of the Bloch vector around the origin with the lattice size, $N = 201~ \subref{fig:NVEC1}~\theta_1 = -3\pi/8,~\theta_2 = \pi/8,~\gamma=0.25~\subref{fig:NVEC2}~\theta_1 = -3\pi/8,~\theta_2 = 5\pi/8,~\gamma=0.25~\subref{fig:NVEC3}~\theta_1 = -3\pi/8,~\theta_2 = \pi/8, \gamma = 1.8~\subref{fig:NVEC4}~\theta_1 = -3\pi/8,~\theta_2 = \pi/8, \gamma = 3.0$.}
	\label{fig:NVEC}
\end{figure*}

\begin{figure*}
	\centering
	\subfigure[]{
		\includegraphics[width=4.6cm]{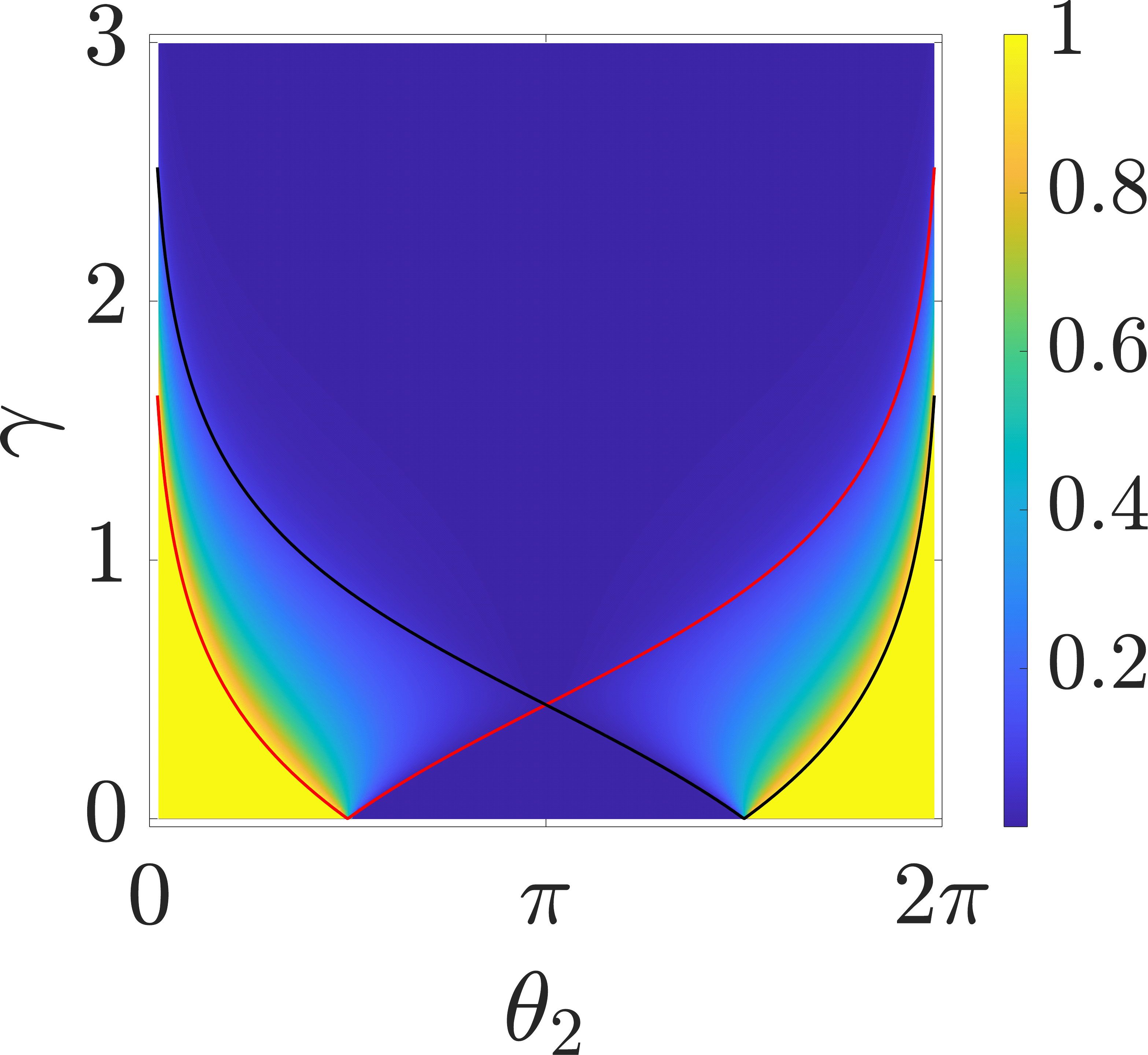}
		\label{fig:SSQWL1}}
	\subfigure[]{
		\includegraphics[width=4.6cm]{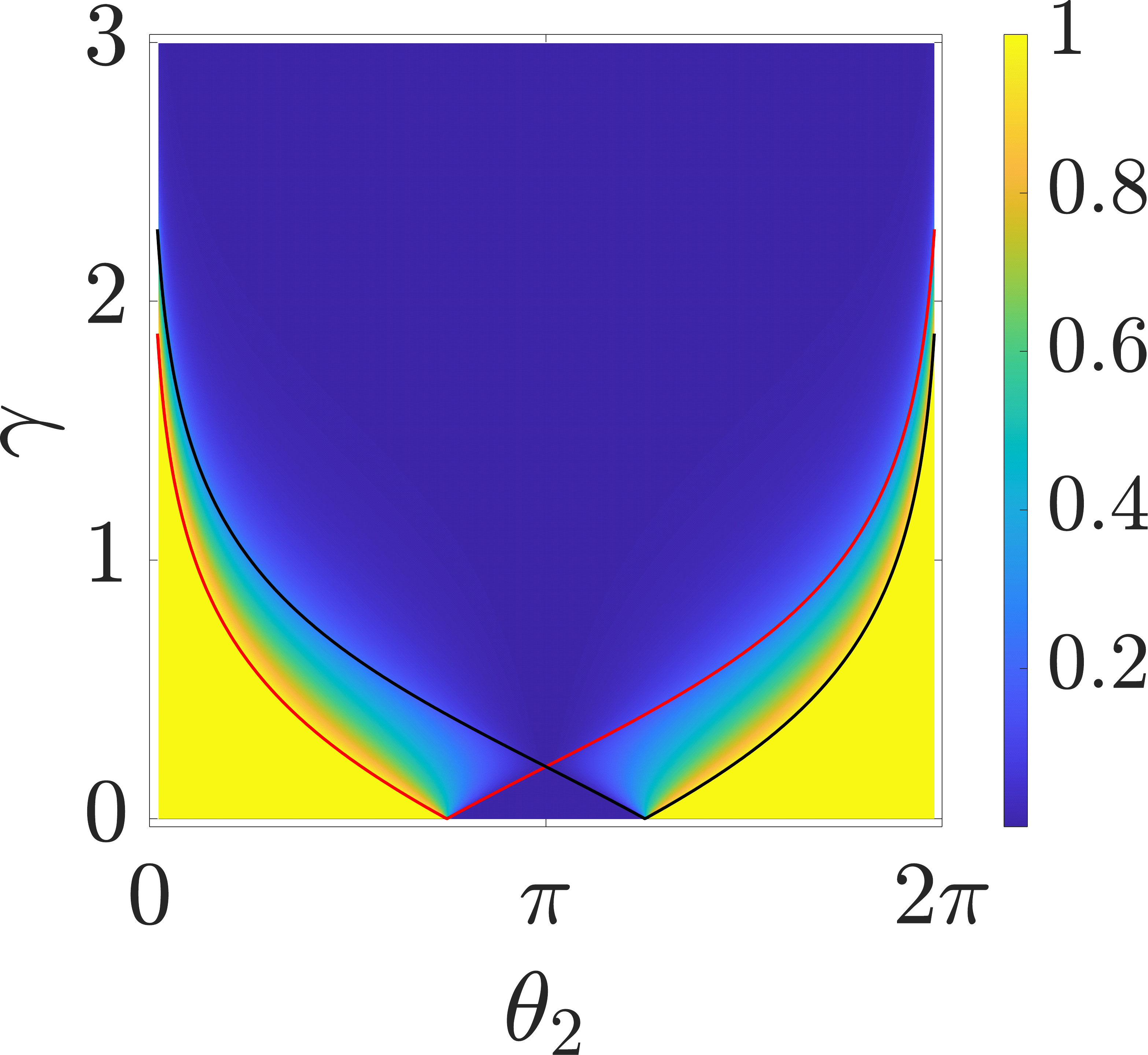}
		\label{fig:SSQWL2}}
	\subfigure[]{
		\includegraphics[width=4.6cm]{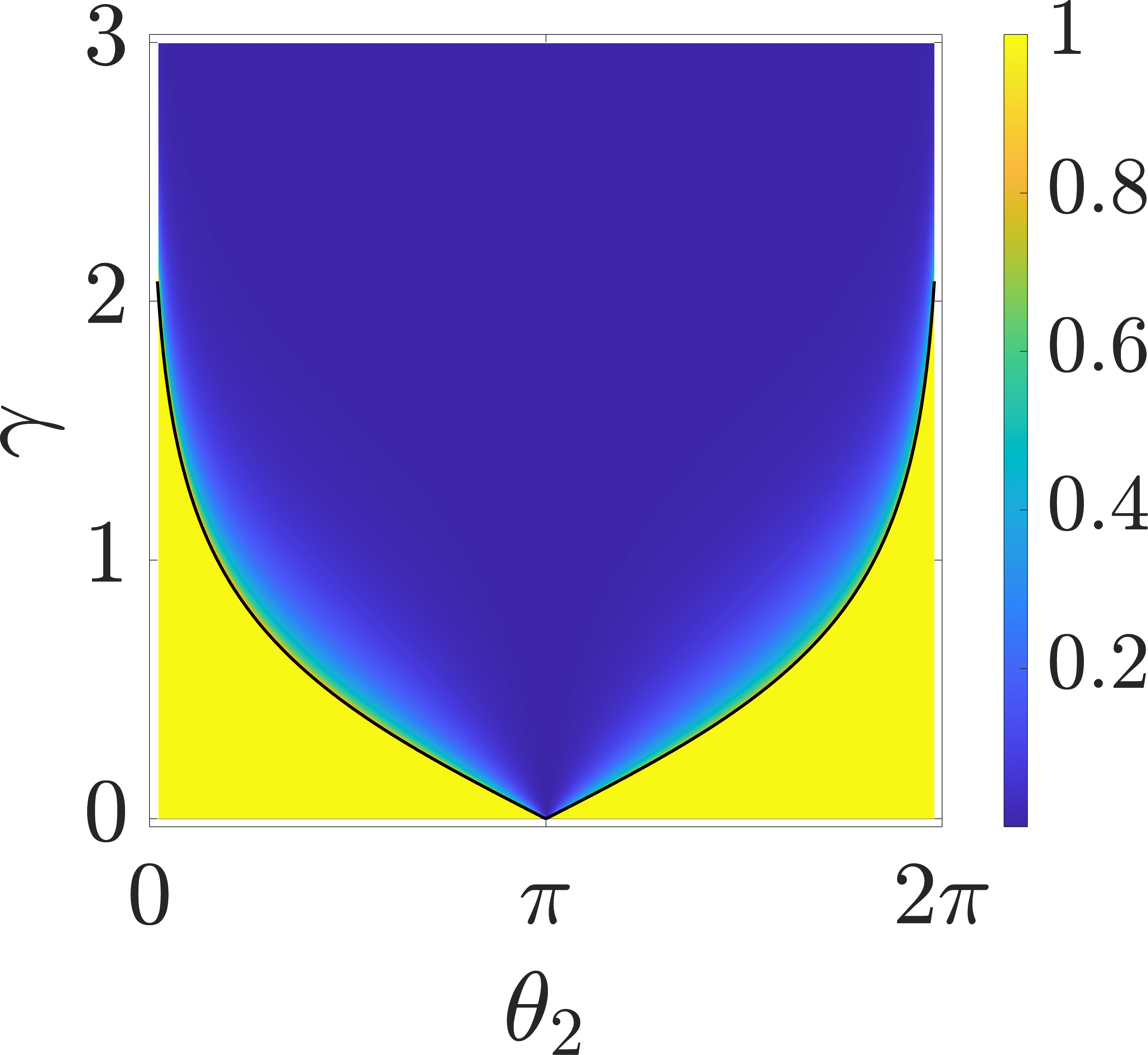}
		\label{fig:SSQWL3}}
	\caption{(Color online) Plot for $W_-$  for lower energy band  as a function of $\gamma$ and $\theta_2$, and \subref{fig:SSQWL1} $\theta_1 = -\pi/2$ \subref{fig:SSQWL2} $\theta_1 = -3 \pi/4$ \subref{fig:SSQWL3} $\theta_1 = -\pi$. The system size is taken to be $N = 201$. The red and black lines in all of the panels represent $\gamma_c$ for $(k,E)=(0,0)$ and $(k,E)=(\pi,0)$, respectively.}
	\label{fig:SSQWL}
\end{figure*}

Since the eigenstates and eigenvalues depend on $\gamma$, $\theta_1$ and $\theta_2$,  the winding numbers are also expected to depend upon these parameters. In Fig.\,\ref{fig:SSQWL}, we plot the  winding number for the lower band $W_-$ as a function of $\gamma$ and $\theta_2$ for different values of $\theta_1$. In all figures, we notice that for $\gamma=0$, the winding number can take two different values, zero and one, depending on the choice of $\theta_1$ and $\theta_2$. Focusing on the case of $W_-=1$ for a vanishing $\gamma$, we observe that for a given $(\theta_1,\theta_2)$ if we increase the value of $\gamma$, the winding number remains unaffected until we approach the critical value of $\gamma$, i.e., $\gamma_c$ \eqref{Eq:Delta-c}. Crossing the $\gamma_c$ causes a phase transition and the value of $W$ starts decreasing and approaches zero asymptotically.  Whereas, if initially the winding number $W_-= 0$, it remains zero until we approach $\gamma_c$, and then it starts to increase momentarily approaching some maximum value and then deteriorates to zero asymptotically.

By definition, the winding number is an integer quantity. In other words, the geometric phase acquired by the eigenstates of the Hamiltonian in the $k$-space is quantized and is a multiple of $\pi$, which is possible only when all the states in an energy band lie in a plane on the Bloch sphere. The winding number must always be an integer for all the Hermitian Hamiltonians. However, beyond the exceptional points, $W$ acquires non-integer values, hence it does not qualify as the winding number. This non-integer value of $W$ can be explained by observing the behaviour of the eigenstates of the non-Hermitian Hamiltonian. In Fig.\,\ref{fig:NVEC}, we plot the Bloch vectors corresponding to the eigenstates $\ket{\psi_-(k)}$ of the Hamiltonian $H_{_{\text{NU}}}(\theta_1,\theta_2,\gamma)$ on the Bloch sphere. In Fig.~\ref{fig:NVEC1}, the Bloch Vector moves in the clockwise direction and comes back to the same point, winding around the origin once resulting in $W=1$. However, in Fig. \ref{fig:NVEC2}, it first goes clockwise and reverses its direction, and; therefore, $W=0$. Figs.~\ref{fig:NVEC1} and \ref{fig:NVEC2} are for $\gamma \le \gamma_c$ whereas Figs.~\ref{fig:NVEC3} and \ref{fig:NVEC4} are for $\gamma > \gamma_c$. The animation of Bloch vectors can be found in the Supplementary Material online. We can clearly see that in the exact $\mathcal{PT}$-symmetric region, the eigenstates lie in a plane and results in an integer value of $W$, whereas in the exact $\mathcal{PT}$-symmetry broken region the eigenvectors trace a path which lies outside the plane. Hence geometric phase is not a multiple of $\pi$ resulting in a non-integer value of $W$. 

In summary, we have shown that the topological phase in 1D SSQW remains invariant as long as the energy eigenvalues are real, even though the Hamiltonian is not Hermitian, i.e., the topological order persists as long as the Hamiltonian respects exact $\mathcal{PT}$ symmetry. Next, we extend our study to the case of 2D DTQW.
\begin{figure*}
	\centering
	\subfigure[]{
		\includegraphics[width=4.5cm]{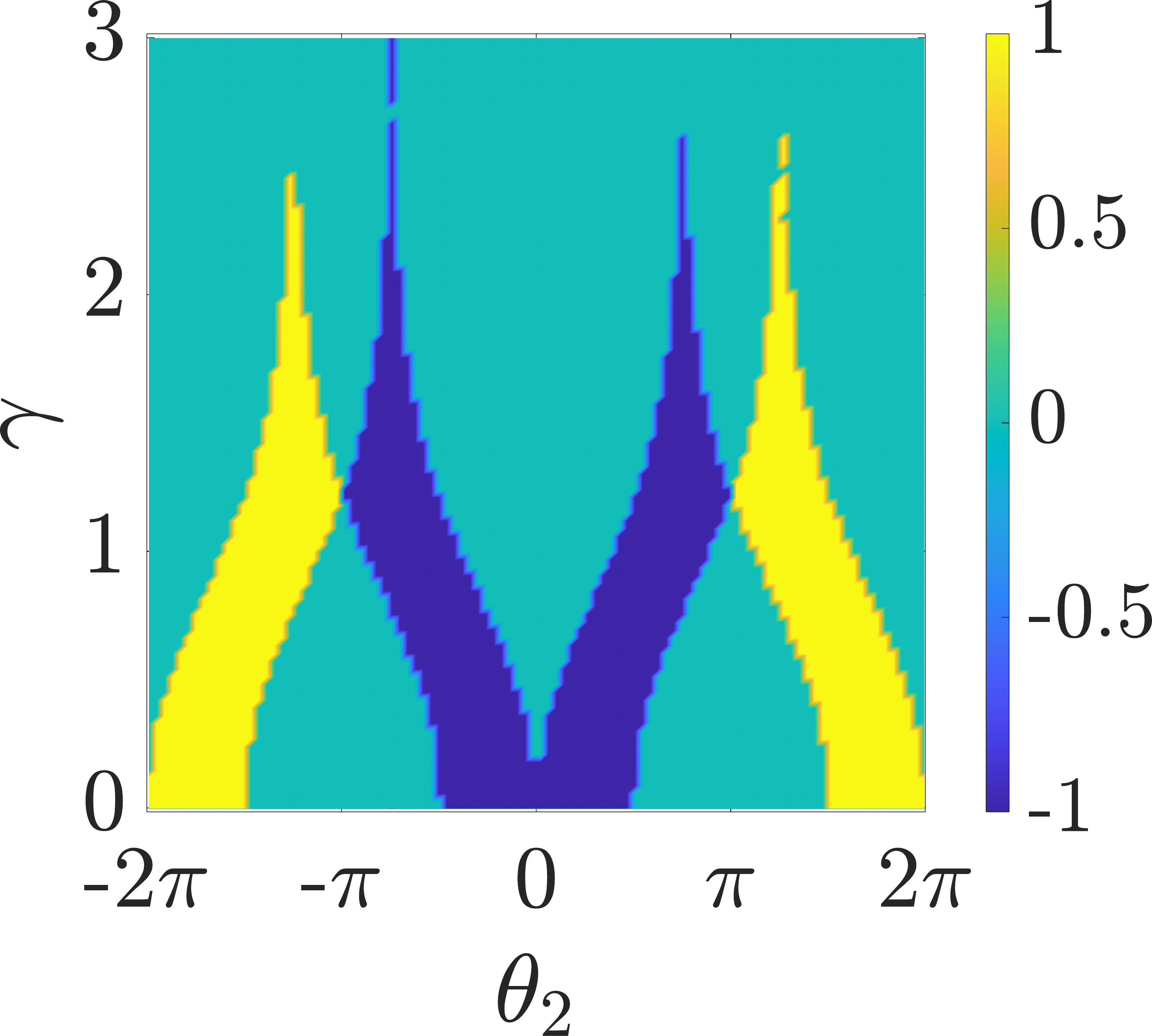}
		\label{fig:DTQWpiby4a}}
	\subfigure[]{
		\includegraphics[width= 4.5cm]{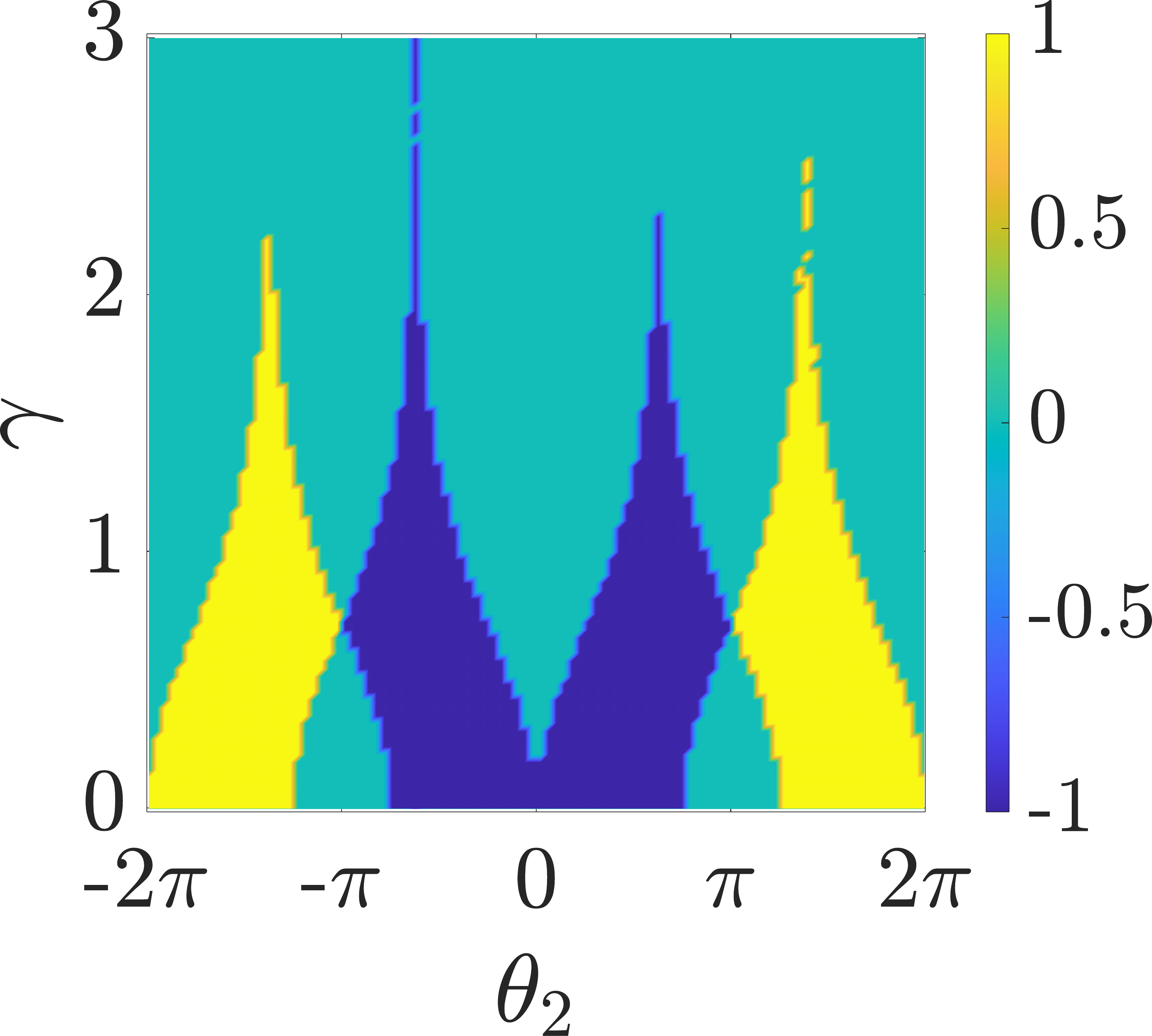}
		\label{fig:DTQW3piby8a}}
	\subfigure[]{
		\includegraphics[width= 4.5cm]{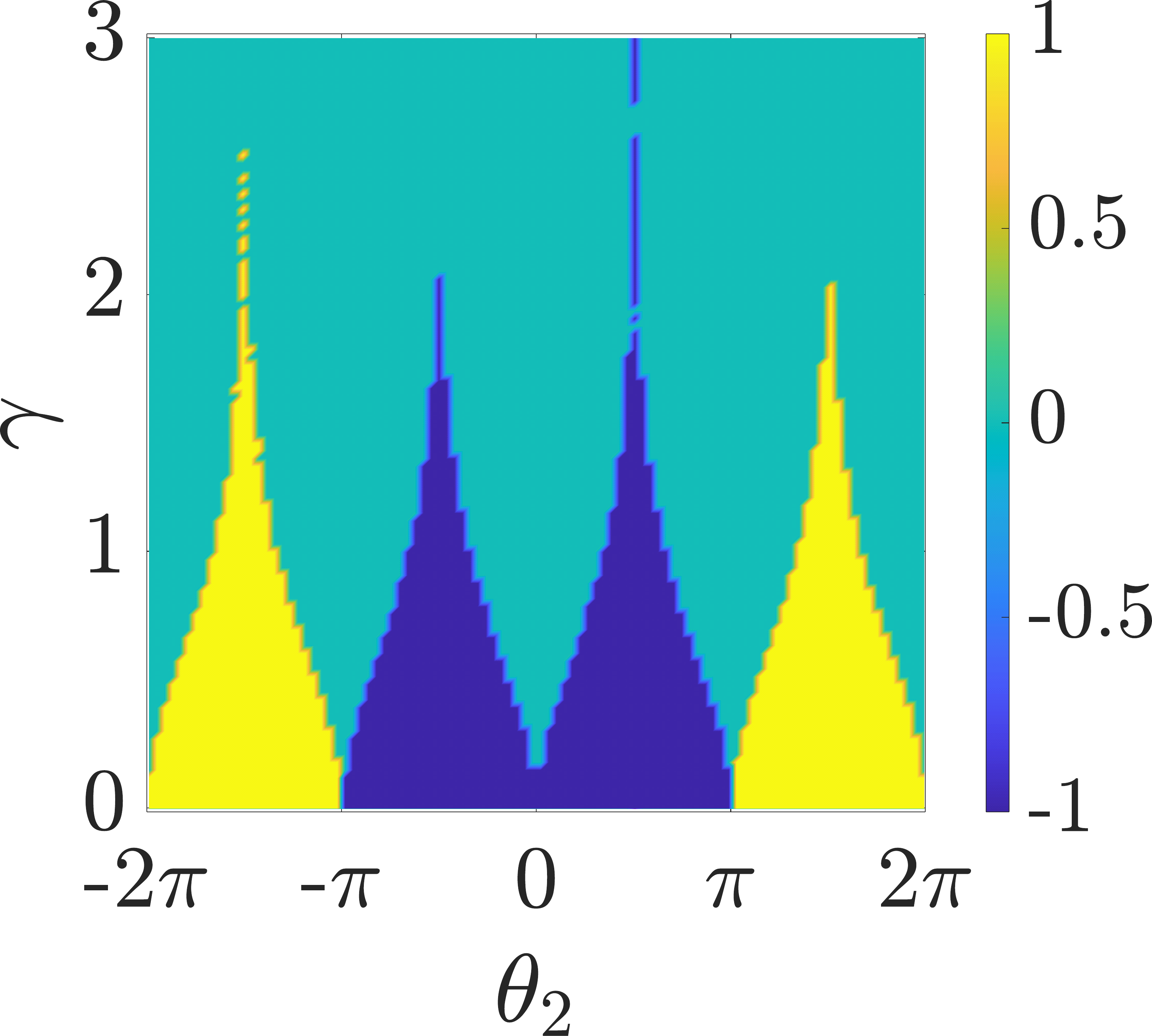}
		\label{fig:DTQW3piby2a}}
	\subfigure[]{
		\includegraphics[width=4.5cm]{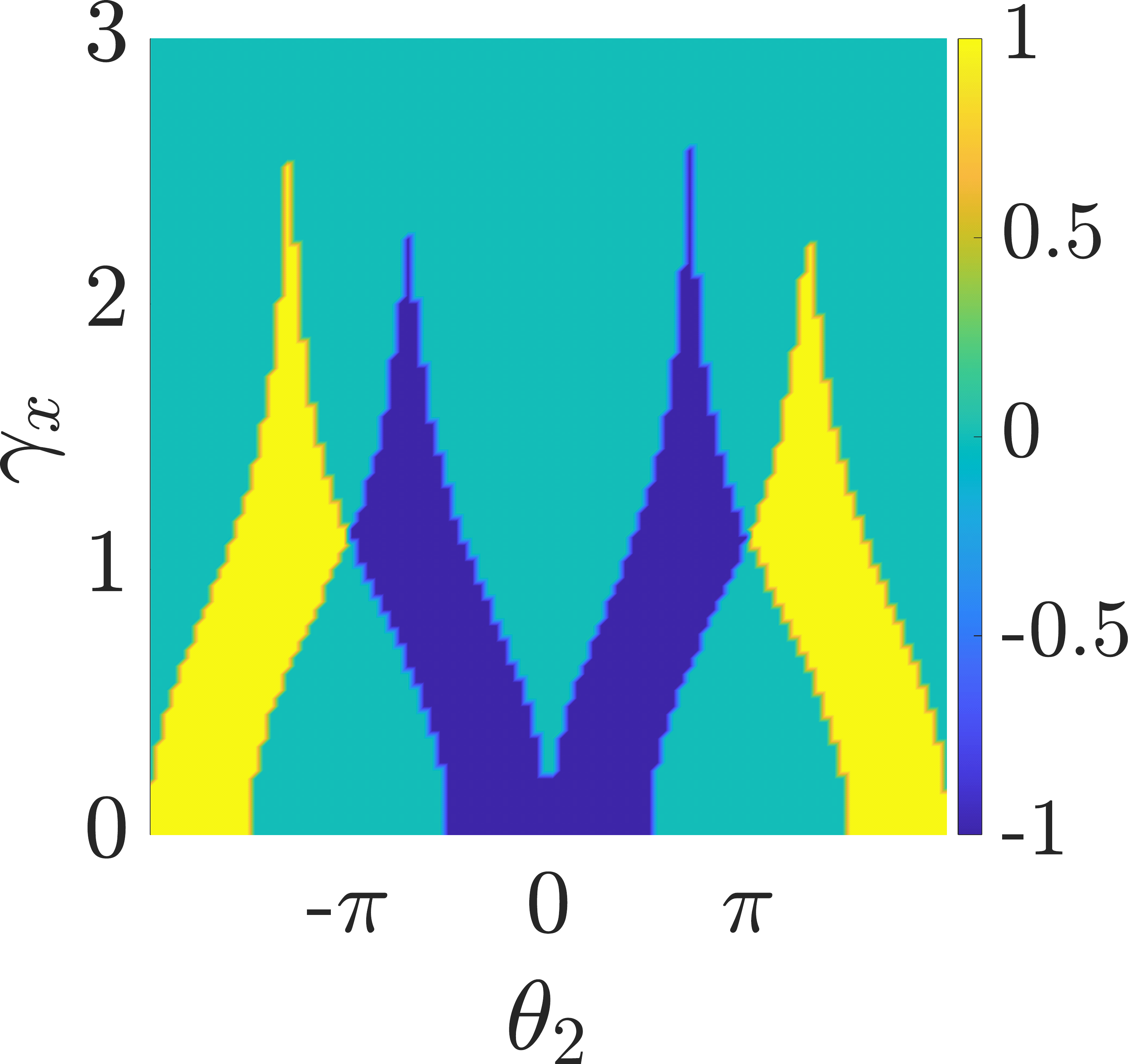}
		\label{fig:DTQWpiby4b}}
	\subfigure[]{
		\includegraphics[width= 4.5cm]{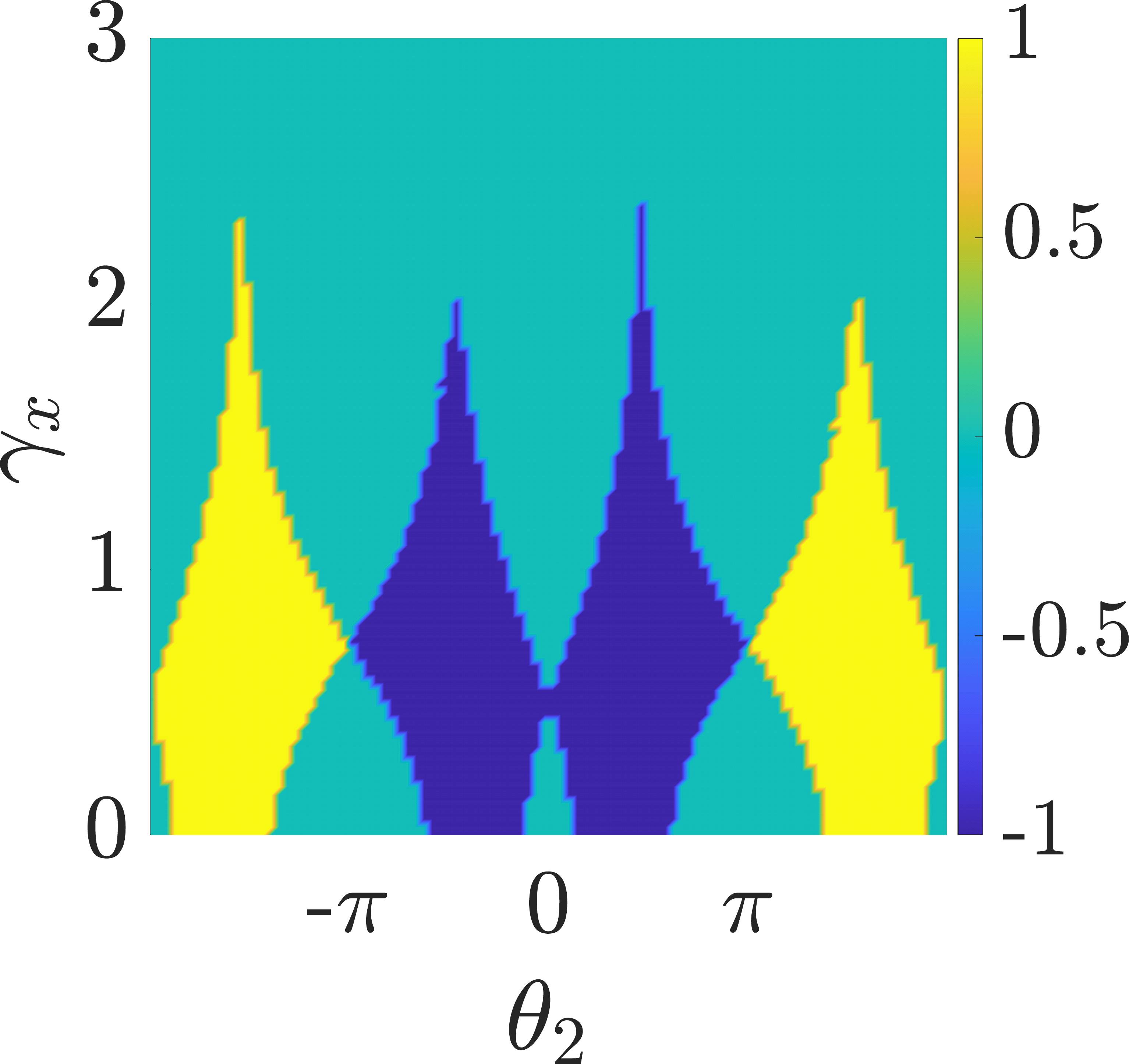}
		\label{fig:DTQW3piby8b}}
	\subfigure[]{
		\includegraphics[width= 4.5cm]{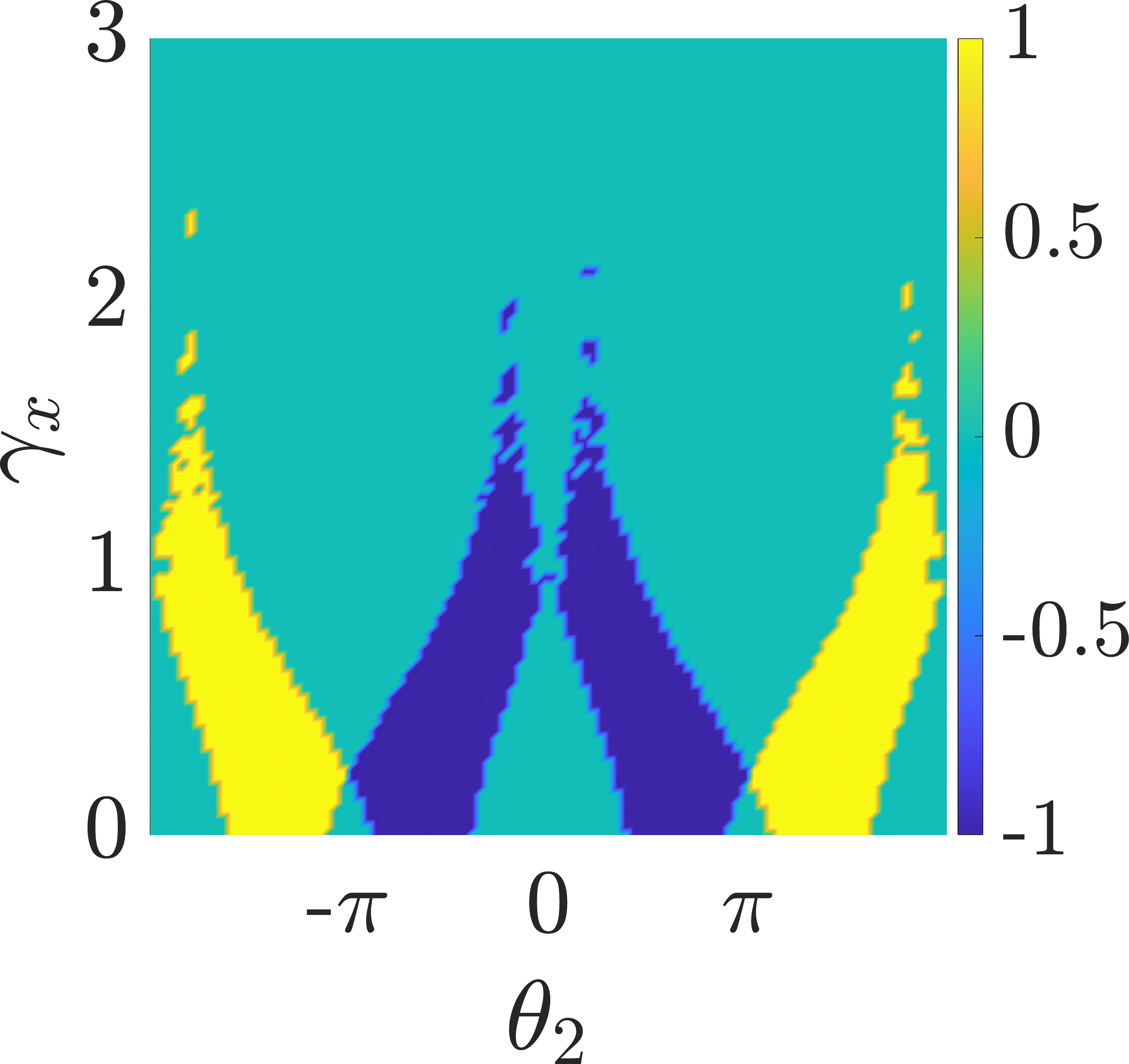}
		\label{fig:DTQW3piby2b}}
	\caption{(Color online) Effect of $\gamma_x$ on Chern number is plotted with varying $\theta_2$ for $\gamma_y = 0$ \subref{fig:DTQWpiby4a} $\theta_1 = \pi/4$ \subref{fig:DTQW3piby8a} $\theta_1 = 3\pi/8$ \subref{fig:DTQW3piby2a} $\theta_1 = 3\pi/2$. In the bottom row, \subref{fig:DTQWpiby4b} $\gamma_y = 0.1$, \subref{fig:DTQW3piby8b} $\gamma_y = 0.5$, \subref{fig:DTQW3piby2b} $\gamma_y = 1.0$, respectively. The lattice size is taken to be 201 $\times$ 201.}
	\label{fig:2DDTQWG}
\end{figure*}

\subsection{Topological phases in 2D non-unitary quantum walk}
Since 2D DTQW can be decomposed as a product of two 1D SSQW, we can easily extend 2D DTQW to non-unitary limits by introducing the scaling operator  $G$ along the $x$- as well as the $y$-axis. The time evolution operator can be written as
\begin{equation} \label{eq:Nonunitary-2DDTQW}
	U^{^{\text{NU}}}_{_{2D}}(\theta_1, \theta_2,\gamma_x, \gamma_y) = G_{\gamma_y} T_y R(\theta_1) G_{\gamma_y}^{-1} T_y R(\theta_2) G_{\gamma_x} T_x R(\theta_1) G_{\gamma_x}^{-1}T_x.
\end{equation}
The corresponding non-Hermitian Hamiltonian of this system reads
\begin{align}
	H^{^{\text{NU}}}_{_{2D}}(\theta_1, \theta_2,\gamma_x, \gamma_y) = \bigoplus_{k_x,k_y} E(k_x,k_y,\gamma_x, \gamma_y) {\bf n}(k_x,k_y,\gamma_x, \gamma_y)\cdot {\bf \sigma},
\end{align}
where
\begin{align}
	&\cos E(k_x, k_y,\gamma_x, \gamma_y) \nonumber \\
	=  & \cos \theta_1 \cos (\theta_2/2) \cos (k_x + k_y - i \gamma_x + i \gamma_y) \cos (k_x + k_y + i \gamma_x - i \gamma_y) \nonumber \\
	&- \cos (\theta_2/2) \sin(k_x + k_y - i \gamma_x + i \gamma_y)\sin(k_x + k_y +i \gamma_x - i \gamma_y) \nonumber \\
	&- \sin \theta_1\sin (\theta_2/2) \cos(k_x - k_y - i \gamma_x - i \gamma_y) \cos(k_x + k_y + i \gamma_x - i \gamma_y),
\end{align}
and
\begin{equation}
	\hat{\vb{n}}(k_x,k_y,\gamma_x, \gamma_y) = \dfrac{n_x(k_x,k_y,\gamma_x, \gamma_y) \hat{\vb{i}} + n_y(k_x,k_y,\gamma_x, \gamma_y) \hat{\vb{j}} + n_z(k_x,k_y,\gamma_x, \gamma_y) \hat{\vb{k}}}{\sin E(k_x, k_y,\gamma_x, \gamma_y)},
\end{equation}
with 
\begin{align}
	n_x(k_x, k_y,\gamma_x, \gamma_y) =& - \sin \theta_1\cos(\theta_2/2) \cos(k_x + k_y -i \gamma_x + i \gamma_y) \sin (k_x - k_y +i \gamma_x + i \gamma_y) \nonumber \\
	&- \cos \theta_1 \sin(\theta_2/2) \cos (k_x - k_y - i \gamma_x - i \gamma_y) \sin (k_x - k_y + i \gamma_x + i \gamma_y) \nonumber \\
	&- \sin(\theta_2/2)  \sin (k_x - k_y - i \gamma_x - i \gamma_y) \cos (k_x - k_y + i \gamma_x + i \gamma_y), \nonumber \\ 
	n_y(k_x, k_y,\gamma_x, \gamma_y)=& \sin \theta_1\cos(\theta_2/2) \cos(k_x + k_y - i \gamma_x + i \gamma_y) \cos (k_x - k_y + i \gamma_x + i \gamma_y) \nonumber \\
	&+ \cos \theta_1 \sin(\theta_2/2) \cos (k_x - k_y - i \gamma_x - i \gamma_y) \cos (k_x - k_y + i \gamma_x + i \gamma_y ) \nonumber \\
	&- \sin(\theta_2/2)  \sin (k_x - k_y - i \gamma_x - i \gamma_y) \sin (k_x - k_y + i \gamma_x + i \gamma_y), \nonumber \\
	n_z(k_x, k_y,\gamma_x, \gamma_y)=& -\cos \theta_1 \cos(\theta_2/2) \cos (k_x + k_y - i \gamma_x + i \gamma_y )\sin (k_x + k_y + i \gamma_x - i \gamma_y) \nonumber \\
	&-\cos(\theta_2/2) \sin (k_x + k_y - i \gamma_x + i \gamma_y) \cos (k_x + k_y + i \gamma_x - i \gamma_y) \nonumber \\
	&+ \sin \theta_1 \sin(\theta_2/2) \cos(k_x - k_y-i \gamma_x - i \gamma_y) \sin(k_x + k_y+ i \gamma_x - i \gamma_y) \nonumber.
\end{align}

The 2D DTQW is different from the 1D SSQW as the former case does not support $\mathcal{PT}$-symmetry even in the unitary region. The energy eigenvalues become complex even for very small values of the scaling factor. 
If we take $\gamma_x << 1$ and $\gamma_y = 0$, the expression for the energy reads
\begin{equation} \label{eq: No PT-Symmetry}
	\cos E(\gamma_x) = \cos E(\gamma_x = 0) + i \gamma_x \sin \theta_1 \sin (\theta_2/2) \sin (2 k_y),
\end{equation}
which makes the quasi-energy complex for infinitesimal scaling parameter $\gamma_x$. 

For 2D DTQW we will have $\vb{k} = (k_x,k_y)$ and the time evolution operator in Eq. \eqref{Eq:Qwalk2D} in momentum space must satisfy $	\Xi U(\vb{k}) \Xi^{-1} = U(-\vb{k})$ in order to possess PHS \cite{Kitagawa2010, Mochizuki2016}, which is satisfied by choosing $\Xi = \mathcal{K}$ for all the values of scaling factor $G_{\gamma_x}$ and $G_{\gamma_y}$. Their exist PHS in 2D DTQW and the topological phases in the class with $\mathds{Z}$ topological invariant can be realized~\cite{Schnyder2008,Kitaev2009,Kitagawa2010}.

Similar to the case of 1D SSQW, in 2D quantum walks also the energy eigenvalues appear in pairs $\pm E(k_x, k_y,\gamma_x, \gamma_y)$ resulting in two energy bands. Introducing loss and gain (scaling factor $\gamma$) in $x$ and $y$-direction results in complex pairs of energy eigenvalues. We can choose the lower energy state by looking at the sign of the real part of the energy eigenstate and calculate the Chern number. 

We use \eqref{eq:Chern-Number} to calculate the Chern number for the lower energy band and plot it against  $\gamma_x$ and $\theta_2$ for some fixed values of $\theta_1$ and $\gamma_y$ (Fig.\,\ref{fig:2DDTQWG}). Despite the absence of a real spectrum, we see the persistence of the topological phase as we turn on the scaling . In other words, the system remains in the same topological phase as we introduce loss and gain factors. 
In 2D DTQW  we observe another interesting feature, namely, for some particular values of $\theta_1$ and $\theta_2$, the Chern number can change abruptly from one integer value to another as $\gamma_x$ increases, resulting in a topological phase transition. This is a loss-induced topological phase transition. Furthermore, unlike the 1D SSQW, the Chern number in 2D DTQW changes abruptly and for sufficiently large values of  $\gamma_x$ and $\gamma_y$ the Chern number for all the parameters becomes zero.

\subsection{Bulk-boundary correspondence}
\begin{figure*}
	\centering
	\subfigure[]{
		\centering
		\includegraphics[width=8.5cm]{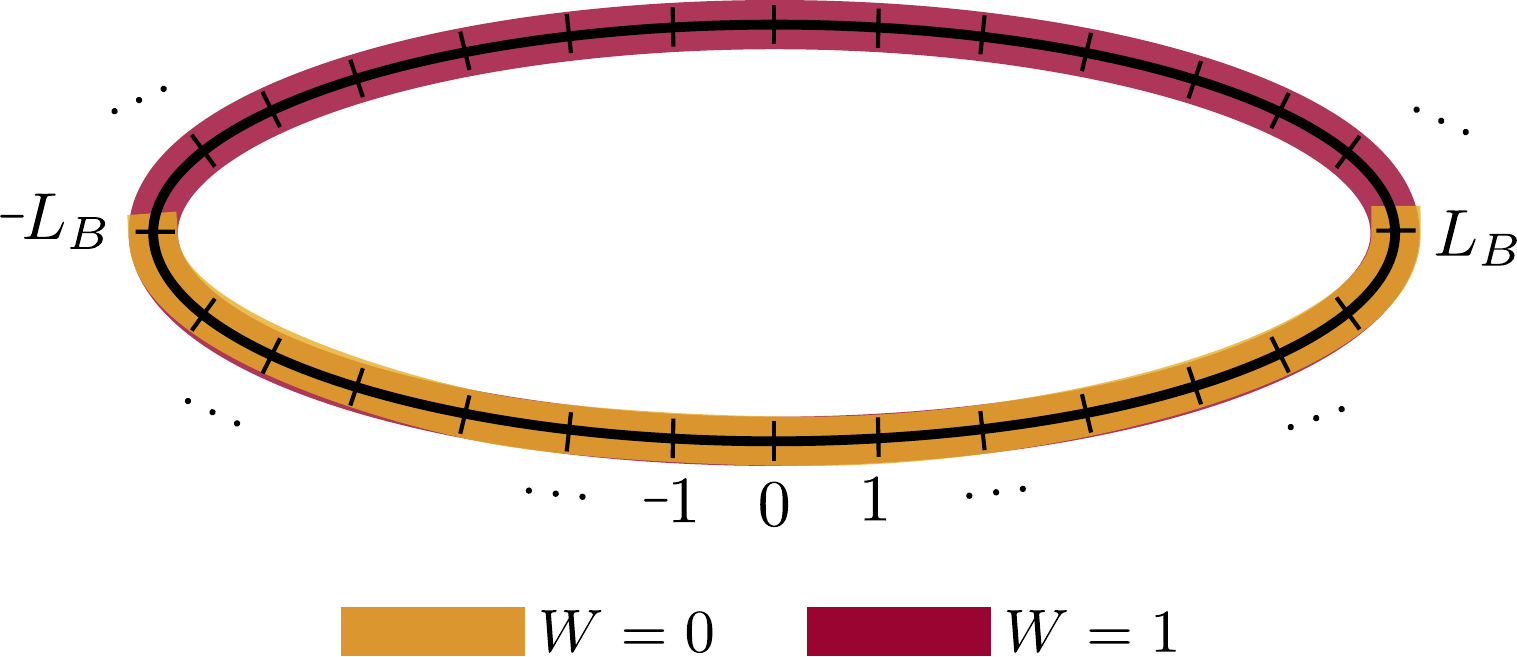}
		\label{fig:BulkEdge1D}}
	\hspace{2cm}
	\subfigure[]{
		\centering
		\includegraphics[width=7cm]{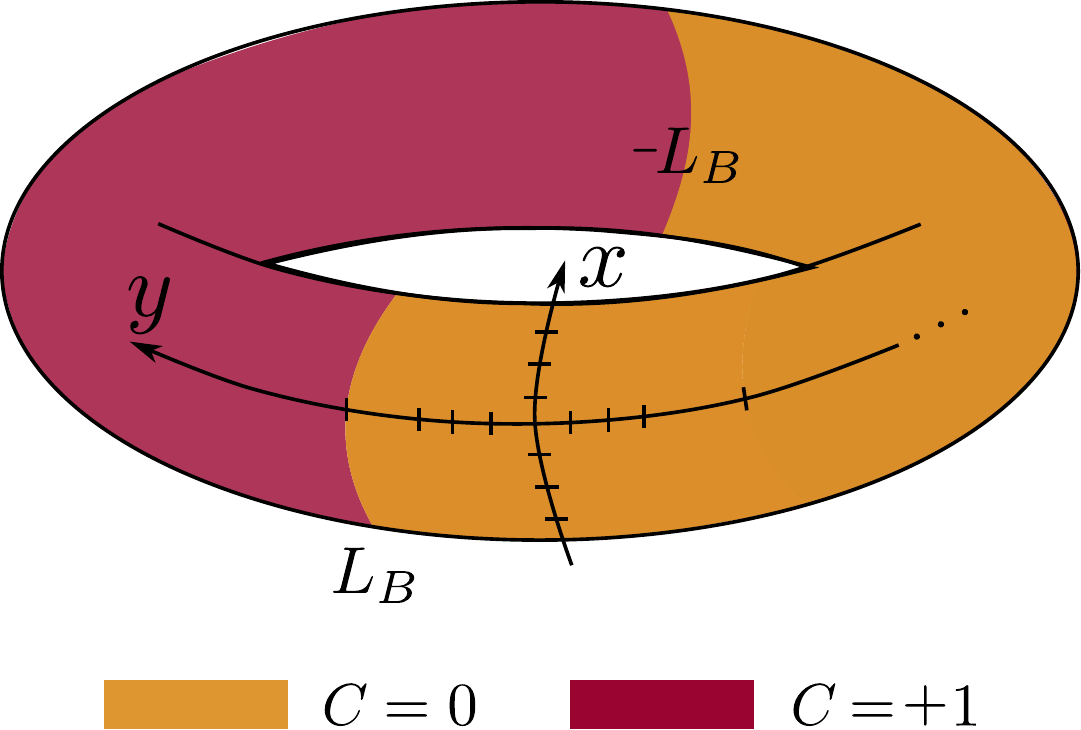}
		\label{fig:BulkEdge2D}}
	\caption{(Color online) The bulk boundary correspondence is studied by dividing the lattice in two parts which are characterized by distinct topological phases locally. \subref{fig:BulkEdge1D} 1D lattice is divided in two equal parts with two boundaries at $\pm L_B$ with $L_B = 50$ and system size $N = 201$. \subref{fig:BulkEdge2D} A two dimensional periodic lattice is divided into two equal parts where the partition is made in the $y$-direction while retaining the periodicity in the $x$-direction. The boundary on the $y$-axis is chosen at $\pm L_B$ with a lattice size 201 $\times$ 201.}
	\label{fig:BulkEdge}
\end{figure*}

In the case of infinite lattice or with periodic boundary condition, we characterize our system with topological invariants such as Winding number and Chern number, however, when we have finite lattice with open boundary conditions, we observe topologically protected states on the boundary \cite{Asboth2016, Kane2013}. In the bulk of topological insulators, the system behaves like an ordinary insulator but on the edges, we find conducting edges states. This is referred to as bulk-boundary correspondence. In this section, we study the edge states in the 1D SSQW and the 2D DTQW systems to ensure the persistence of the topological states and hence topological order.

The 1D SSQW is generally performed on an infinite lattice or a closed chain. In order to create a boundary in this system, we still consider the quantum walk on a closed chain, but divide the lattice into two regions with different  rotation angles $(\theta_1, \theta_2)$, thus making the lattice inhomogeneous. The parameters for the two parts are chosen such that the two parts locally have different topological phases, as shown in Fig.~\ref{fig:BulkEdge1D}. At the boundary of these two phases, we should see edge states which establish the topological nature of 1D SSQW.

\begin{figure*}
	\centering
	\subfigure[]{
		\includegraphics[width=5.5cm]{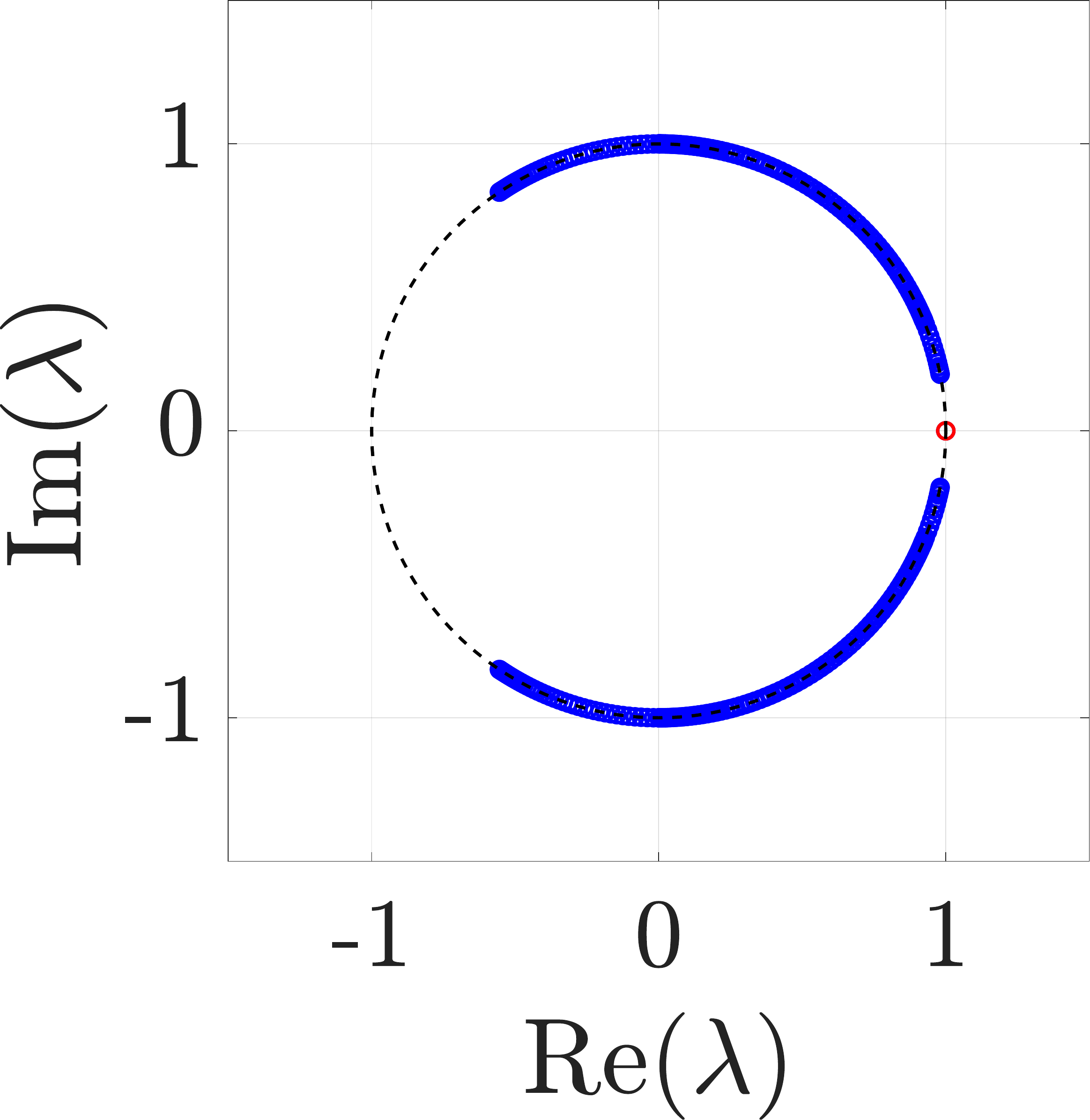}
		\label{fig:1D1a}}
	\subfigure[]{
		\includegraphics[width=5.5cm]{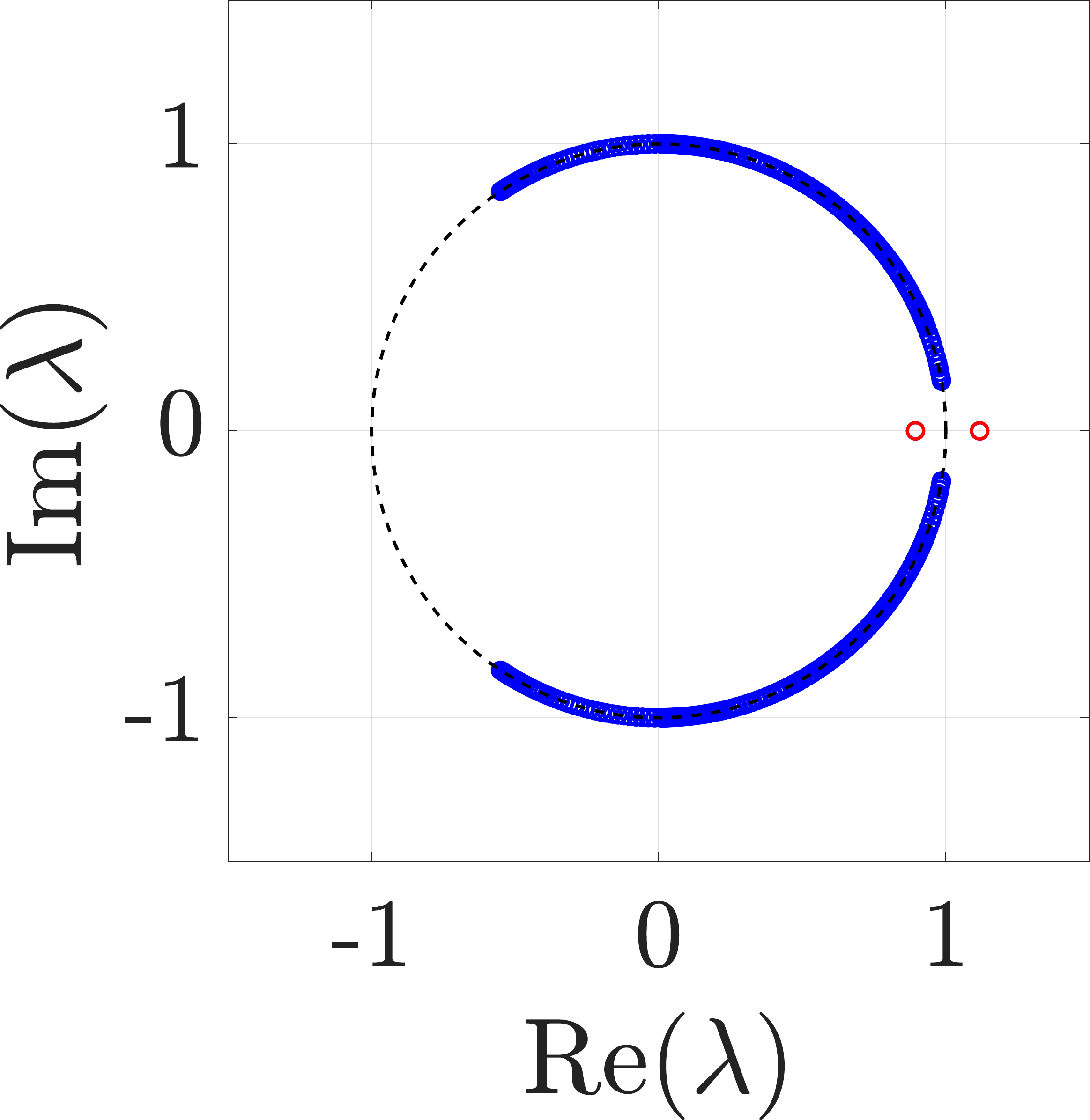}
		\label{fig:1D2a}}
	
	\subfigure[]{
		\includegraphics[width=5.5cm]{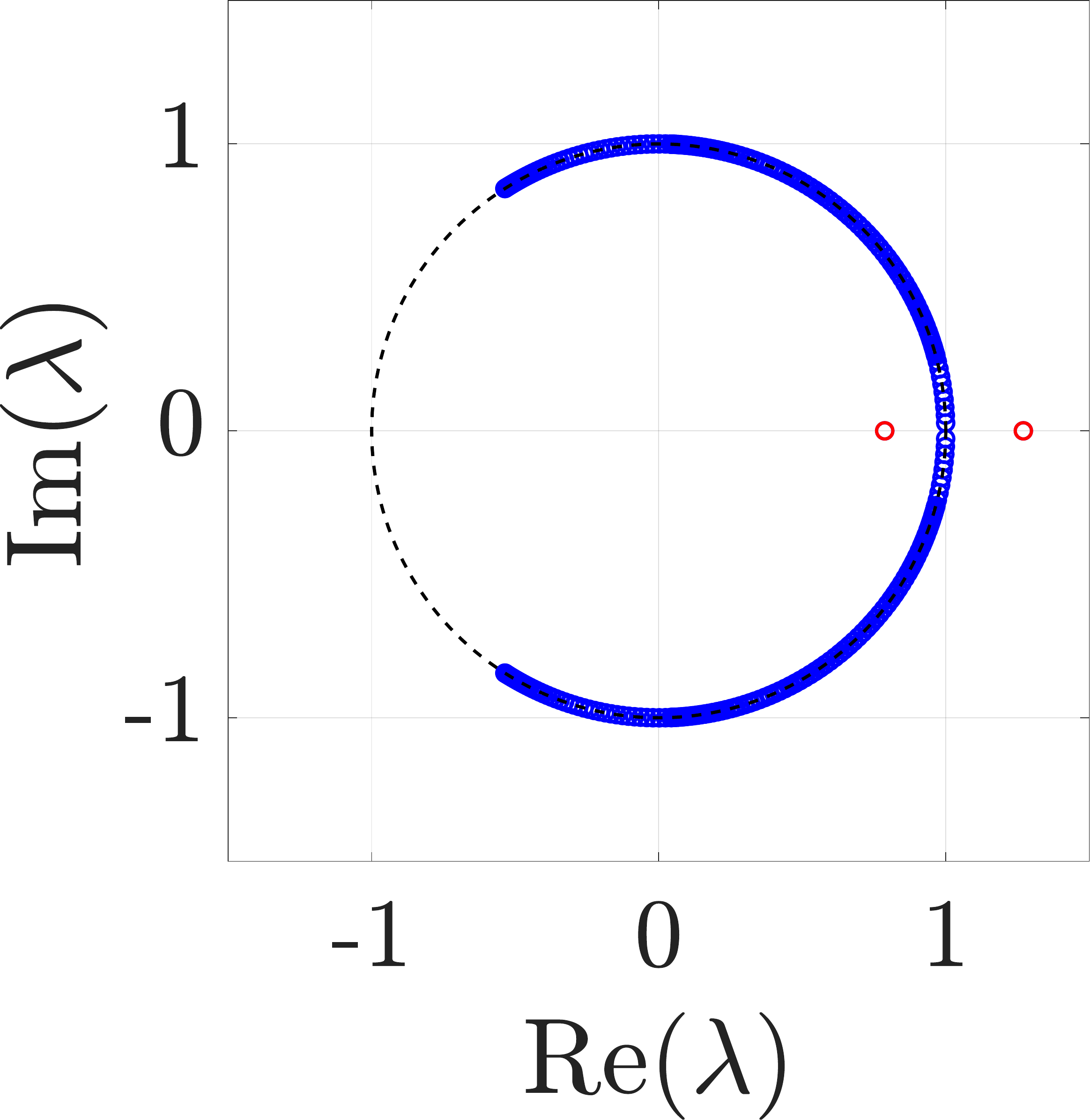}
		\label{fig:1D3a}}
	\subfigure[]{
		\includegraphics[width=5.5cm]{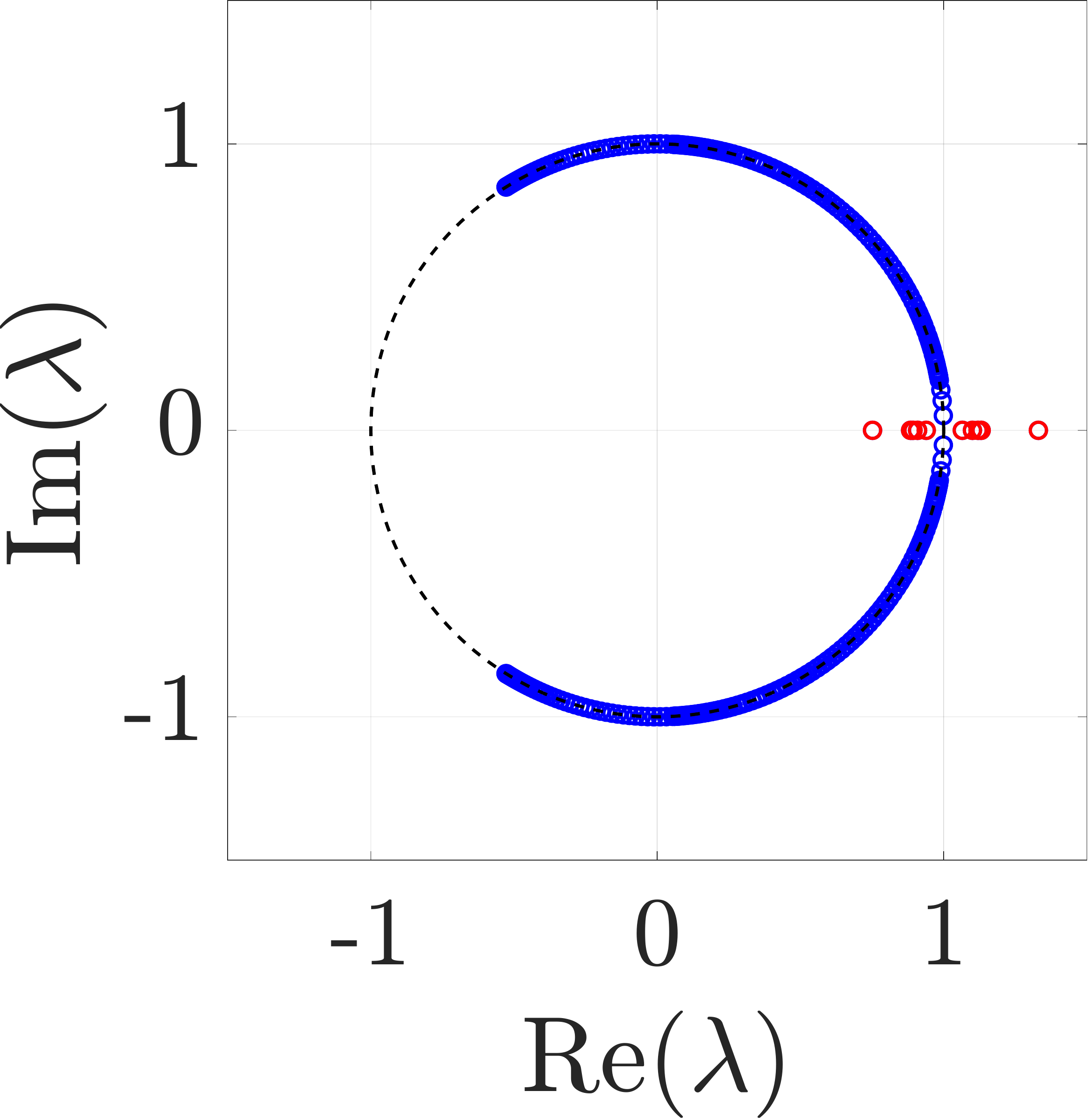}
		\label{fig:1D4a}}
	\caption{(Color online) The eigenvalues, $\lambda$ of the time evolution operator in Eq. \eqref{eq:SSQW-TimeEvolution} are plotted for $(\theta_1^1, \theta_2^1) = (-3 \pi/8, \pi/4)$ and $(\theta_1^0, \theta_2^0) = (-3 \pi/8, 5\pi/8)$ and different values of $\gamma$.  In \subref{fig:1D1a} $\gamma = 0$, \subref{fig:1D2a} $\gamma = 0.2$, \subref{fig:1D3a} $\gamma = $ min $(\gamma^1, \gamma^2) = 0.2110$, and \subref{fig:1D4a} $\gamma = 0.25$.}
	\label{fig:1DBE}
\end{figure*}

In Fig.~\ref{fig:1DBE}, we plot the complex eigenvalues $\lambda$ of the non-unitary evolution operator $U$ given by \eqref{eq:SSQW-TimeEvolution}. Here the close chain of length $201$ is divided into two parts of length $L = 100$ and $L = 101$ lattice sites. The boundaries are denoted by points $L_B = \pm50$. We have chosen $(\theta_1^1, \theta_2^1) = (-3 \pi/8, \pi/4)$ for $n > \abs{L_B}$ and $(\theta_1^0, \theta_2^0) = (-3 \pi/8, 5\pi/8)$ for  $n \le \abs{L_B}$ corresponding to winding numbers $W = 1$ and $W = 0$, respectively.

In Fig.~\ref{fig:1D1a}, we observe that two of the eigenvalues of the operator $U$ lying on the real axis, signifying the states with energy $0$ or $\pi$ for $\gamma = 0$. These states were absent in the homogeneous case; therefore, they are the edge states.  As we introduce scaling factor i.e. $\gamma \ne 0$,  the same behaviour persists until we reach the critical value of $\gamma$.  Since we have two sets of $\theta_1,~\theta_2$ which correspond to two different energy landscapes, we will have different exceptional points. Using Eq. \eqref{Eq:Delta-c}, these exceptional points come out to be $\gamma_c^1 = 0.2110$ and $\gamma_c^0 = 0.2832$ for the given choice of rotation parameters $\theta$'s. We find that the edge states persist till the point given by min($\gamma_c^0, \gamma_c^1$) after which we will have a complex spectrum for the Hamiltonian and we get many states with pure real $\lambda$ which have a contribution from broken exact $\mathcal{PT}$-symmetry.

In the case of non-Hermitian 2D DTQW, it is more difficult to establish the bulk-edge correspondence. This is mainly due to the fact that 2D DTQW does not support $\mathcal{PT}$-symmetry. The spectrum becomes complex as soon as we introduce the scaling which is evident from Eq. \eqref{eq: No PT-Symmetry}. In order to see the persistence of edge states, we only plot the real part of the eigenvalues of the Hamiltonian. In the case of 2D DTQW, we introduce the boundary by considering position-dependent coin operator only along the $y$-axis while keeping the $x$-direction periodic, as shown in Fig.~\ref{fig:BulkEdge2D}. For one part of the lattice, we choose $(\theta_1^{+1}, \theta_2^{+1}) = (7 \pi/6, 7 \pi/6)$ and for the other, we choose $(\theta_1^0, \theta_2^0) = (3 \pi/2, 2\pi/2)$; hence, the Chern numbers are  $C = +1$ and $0$ for the two parts.

\begin{figure*}
	\centering
	\subfigure[]{
		\includegraphics[width=5.5cm]{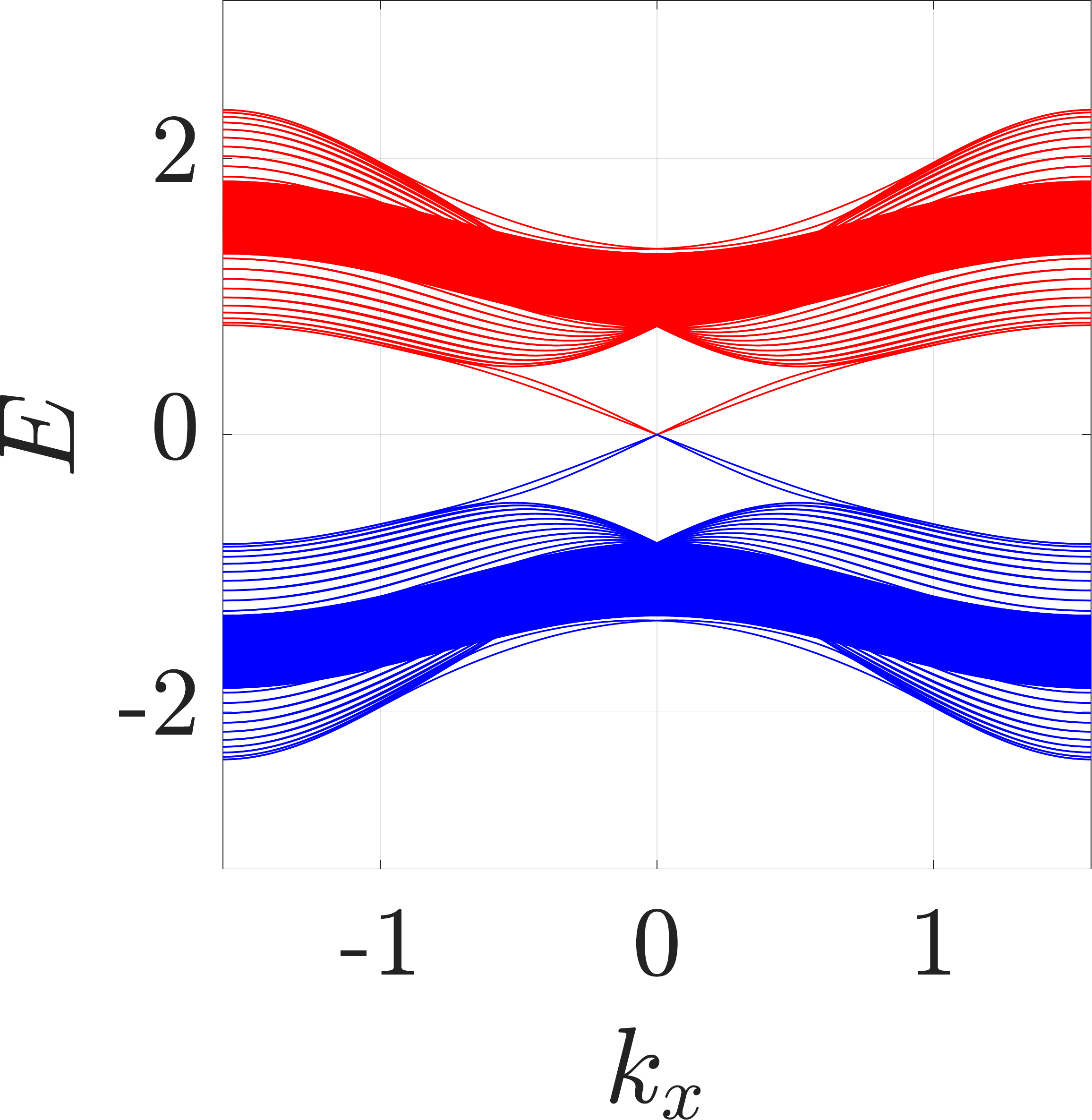}
		\label{fig:2D2a}}
	\subfigure[]{
		\includegraphics[width=5.5cm]{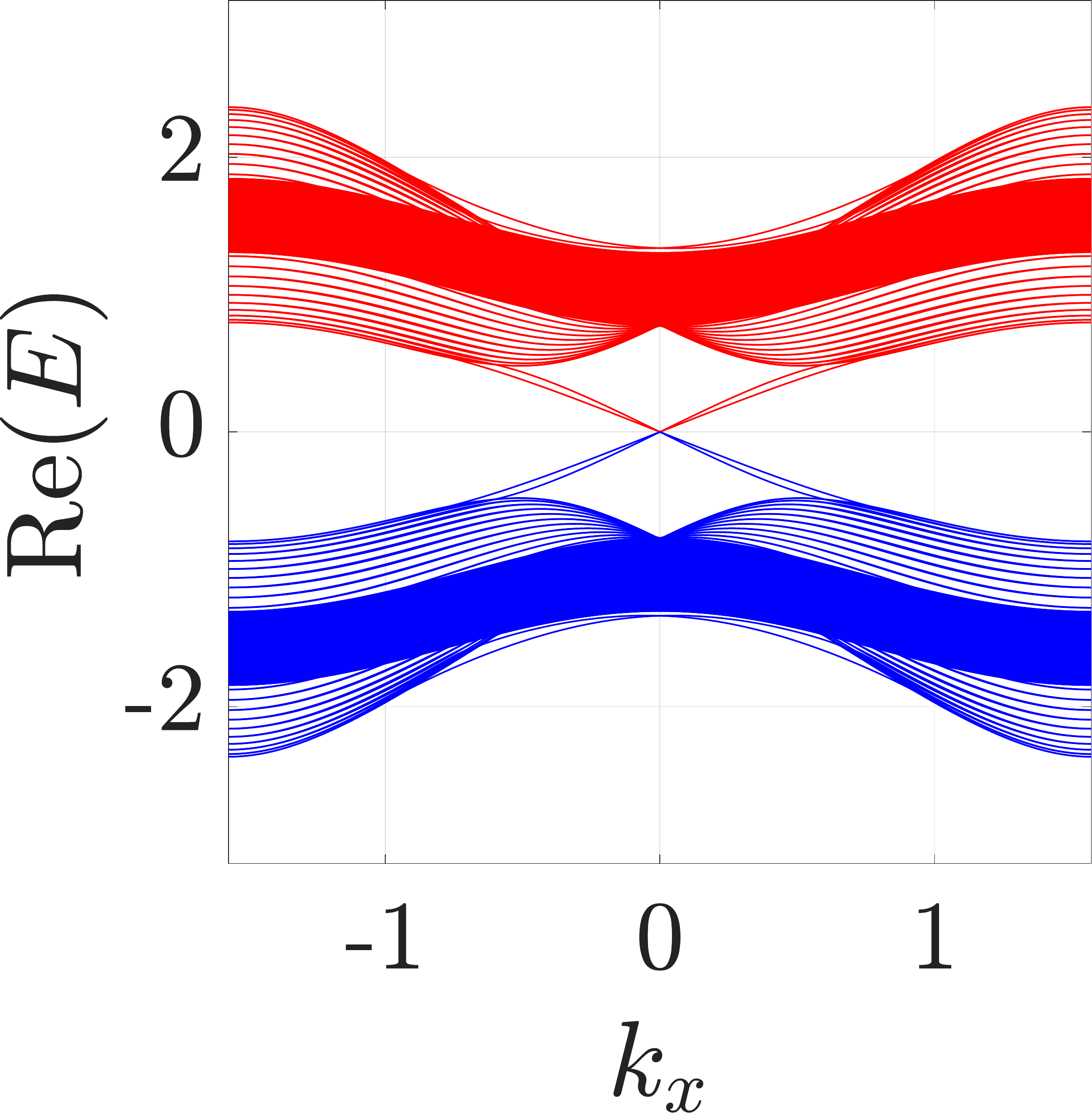}
		\label{fig:2D3a}}
	
	\subfigure[]{
		\includegraphics[width=5.5cm]{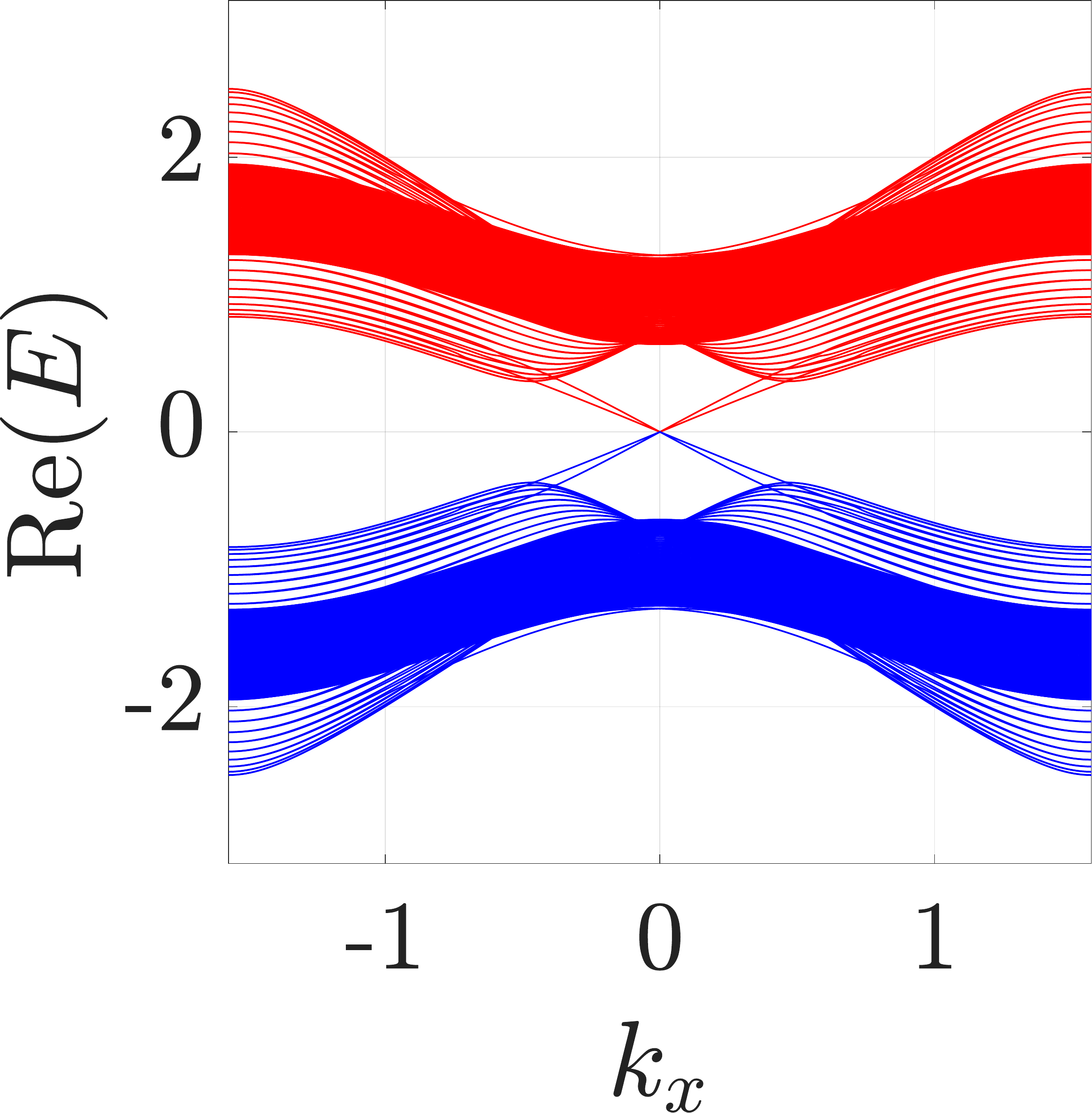}
		\label{fig:2D4a}}
	\subfigure[]{
		\includegraphics[width=5.5cm]{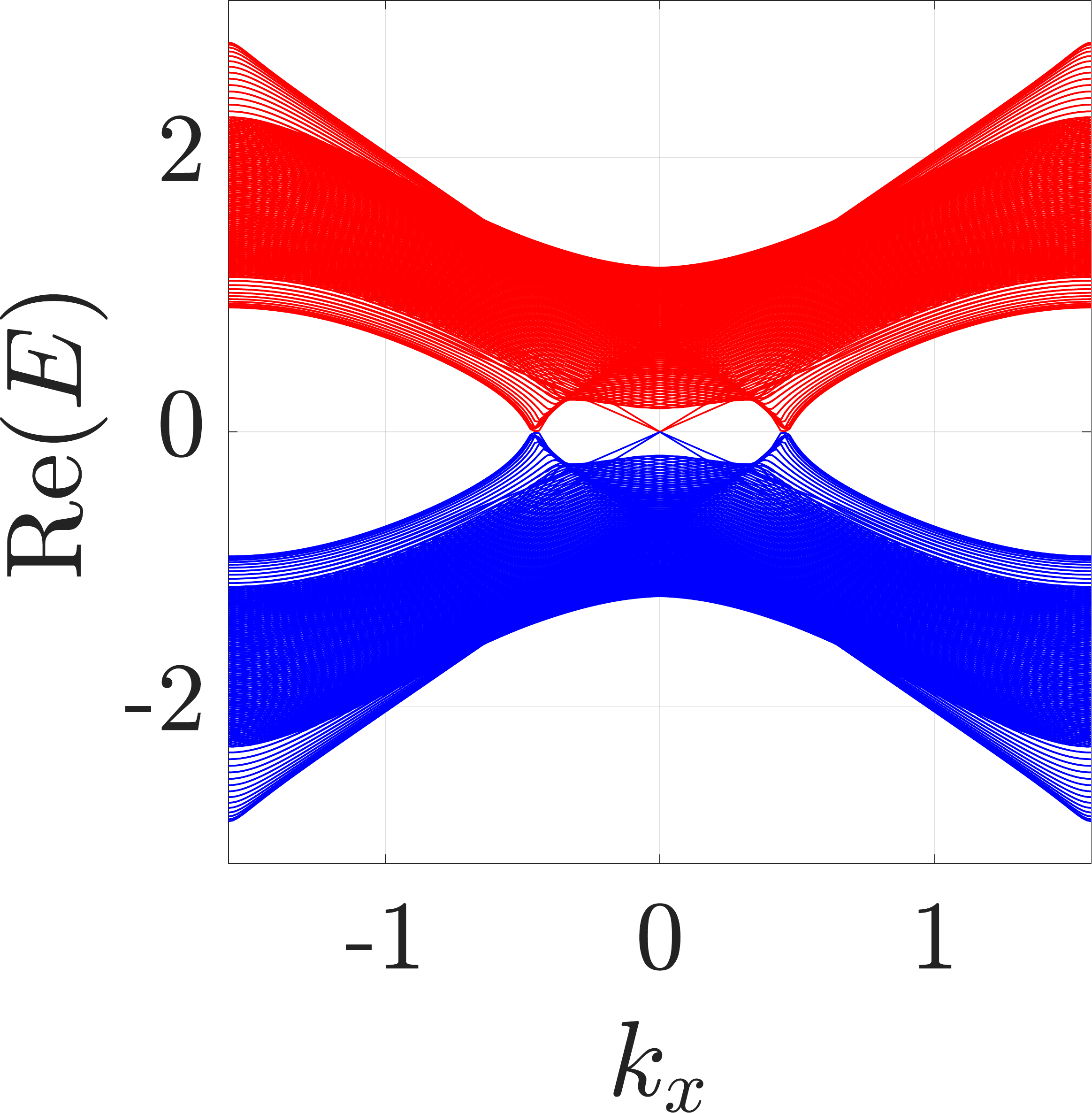}
		\label{fig:2D5a}}
	\caption{(Color online) Energy bands for the 2D DTQW are plotted for inhomogeneous lattice with lattice size $201 \times 201$. We have chosen  $(\theta_1^0, \theta_2^0) = (3 \pi/2, 2\pi/2)$ and $(\theta_1^{+1}, \theta_2^{+1}) = (7 \pi/6, 7\pi/6)$ which correspond to $C = 0$ and $C = +1$, respectively, for the two parts of the lattice. The scaling parameters  are chosen to be $\gamma_x =  \gamma_y = 0,~ 0.2,~0.3$ for \subref{fig:2D2a}, \subref{fig:2D3a} and \subref{fig:2D4a}, respectively. In all these figures we can see the edge states appearing on the boundaries of the two parts of the lattice. For larger values of the scaling parameter, i.e., $\gamma_x=\gamma_y = 0.47$ in Fig.~\subref{fig:2D5a} we see a large number of states between the two bands, which is due to the losses.}
	\label{fig:2DBE}
\end{figure*}

In Fig.~\ref{fig:2DBE}, we plot the real part of the spectrum as a function of the quasi-momentum in the $x$-direction. From these plots, we can see the persistence of the edge states even after introducing the scaling factor $\gamma$. For the large value of the scaling factors, we see the gap vanishes which is predominately due to the losses. Thus, it becomes very difficult to study the bulk-edge correspondence. 

To conclude, we have studied the effect of a balanced gain and loss environment on the topological properties of discrete-time quantum walks. Specifically, we have studied the 1D SSQW and 2D DTQW and observed the persistence of topological phases  against losses in these systems. The loss is incorporated using the non-Hermitian Hamiltonian approach, where we include a scaling parameter $\gamma$ which characterizes the non-Hermiticity. 
We find a strong correspondence between the spontaneous exact $\mathcal{PT}$-symmetry breaking and the loss of topological order in 1D SSQW, i.e, the system retains its topological order for any value of $\gamma$, as long as the system respects the exact $\mathcal{PT}$-symmetry. Due to the absence of $\mathcal{PT}$-symmetry in 2D DTQW, we do not observe such correspondence in these systems. 
However, we observe loss-induced topological phase transition where we see that increasing the scaling parameter $\gamma$ may transfer the system from one non-trivial topological phase to another. We studied the bulk-boundary correspondence in 1D and 2D DTQW and observe the robustness of edge states against the losses. Our results confirm the robustness of topological properties of DTQWs and the role of losses in a topological phase transition.

\chapter{Bloch decomposition of geodesics and null phase curves (NPCs)} \label{chapter:msr}
The shortest path between two (non-antipodal or non-orthogonal) points on a given surface is a minimal geodesic\footnote{In the rest of the chapter, when we refer to geodesic, we shall mean the minimal geodesic.}~\cite{Kimmel1995,Do2016}. In the state space of a quantum system, the geodesics are also the curves along which the acquired geometric phase is zero~\cite{Samuel1988,Mukunda1993} as discussed in detail in Chapter~\ref{chap:gp}. Hence, they play a crucial role in the theory of geometric phases. Further, they are also used in designing optimal quantum circuits, which turned out to be equivalent to finding the shortest path between two points in a certain curved geometry~\cite{Nielsen2006,Nielsen2008}. Geodesics can be generalized to a larger class of curves, known as null phase curves (NPCs)~\cite{Rabei1999}. An NPC is defined as a curve between two pure quantum states on the state space along which the acquired geometric phase is zero~\cite{Arvind2003,Chaturvedi2013}. Unlike geodesics, the NPCs need not be the shortest path between the two states. The role of the geometric phase in characterizing topological phases of matter~\cite{Zak1989}, in precision measurements~\cite{Martinez2011,HuYu2012,AryaMittal2022}, and in robust quantum information processing~\cite{Jones2000,Vedral2003} highlights the importance of understanding NPCs and geodesics. In this chapter, we present a geometric decomposition of the geodesics and NPCs for the higher-dimensional quantum systems to a set of curves on the Bloch sphere. This decomposition reveals the hidden symmetries of higher-dimensional geodesics and NPCs and may facilitate a deeper understanding of the state space structure for such systems.

To find the geometric decomposition of the geodesics and NPCs, we use the Majorana star (MS) representation, which enables the representation of a state of an $n$-level quantum system by a symmetric combination of  $(n-1)$ states of a two-level systems~\cite{Majorana1932, BlochRabi1945}. This representation has found applications in quantum information~\cite{Solano2009,Bastin2010}, quantum entanglement~\cite{Messina2010, Makela2010, Markham2011, Devi2012, Fu2016, Aulbach2011}, geometric phases~\cite{Hannay1998, Fu2014, Fu2016, Akhilesh2020} and topological phases of matter~\cite{Chen2015, Gong2020}. Recently, the bulk topology and the bulk-boundary correspondence have been studied in the non-hermitian tight-binding model using the MS representation~\cite{Zhao2021}.

Since in the MS representation, a state $\ket{\Psi}$ of an $n$-level quantum system can be mapped to a symmetric state of $n-1$ number of two-level systems, it can be represented by a set of $n-1$ points on the Bloch sphere. Hence, a curve on the state space of $n$-level system can be mapped to $n-1$ curves on the Bloch sphere. The $n-1$ points corresponding to the state $\ket{\Psi}$ are often called MSs, and the collection of the points is referred to as a constellation.

In this chapter, we decompose the geodesics and NPCs of higher-dimensional quantum systems into curves on the Bloch sphere using MS representation~\cite{AryaMittal2022}. The key findings of this paper are the following: (i) geodesics of the $n$-level quantum system decompose to $n-1$ circular segments on the Bloch sphere when the end states are represented by $(n-1)$-fold degenerate MSs. When $n$ is odd, the $n-1$ curves occur in pairs that are reflection of each other about the great circle on the Bloch sphere connecting the end states. For $n$ even, one curve is along the great circle connecting the end states, and the remaining $n-2$ curves occur in pairs reflective about the same great circle. (ii) for odd $n$, a class of NPCs can be constructed using $(n-1)/2$ pairs of curves which are reflective about a great circle, whereas, for even $n$,  $(n-2)/2$ pairs of curves are reflective, and the remaining curve can be chosen along the great circle connecting the two states. 

Our treatment provides a deeper understanding and inherent symmetries of geodesics in higher-dimensional state space. For example, geodesics in three-dimensional state space, where the end states are chosen such that each one of them is represented by degenerate MSs, decomposes into two curves. We found that these two curves together form a unique circle on the Bloch sphere, where the end states are the diagonally opposite points on the circle. The radius of the circle depends solely on the inner product between the end states. Therefore, we can generate a geodesic between two states in three-dimensional state space by constructing a circle on the Bloch sphere between the corresponding end points. Since any pair of three-dimensional states can be mapped to states represented by a degenerate MSs using a unitary transformation, we can construct a geodesic between any arbitrary states. 

Using our geometric decomposition,  we construct a prominent class of NPCs for $(n>2)$-dimensional state space~\cite{Mittal2022}. These NPCs can be constructed by choosing curves in pairs such that the curves within a pair are reflections of each other. If the total number of curves is odd, then one curve can be chosen along the great circle connecting the end states on the Bloch sphere. Since there exists an infinite number of such pairs between any two end states, we can construct infinitely many NPCs. A special subset of these NPCs is where the curves are reflections of themselves, i.e., all the curves are along the great circle connecting the end points. This subset can be of experimental importance while designing quantum circuits.

\section{MS representation} \label{subsec:MSR}
Symmetric subspace of $n-1$ number of two-level quantum systems is $n$-dimensional which is isomorphic to $n$-level quantum system. Hence, the state of an $n$-level system can be geometrically represented as a collection of $n-1$ number of points on the Bloch sphere, which is known as MS representation~\cite{Majorana1932,BlochRabi1945}. In this section, we briefly outline the details of MS representation.

Consider a general $n$-level state $\ket{\Psi}$ written as
\begin{equation} \label{eq:spin-j state}
	\ket{\Psi}=\sum_{r=0}^{n-1}c_{r}\ket{r},
\end{equation}
where $c_{r}$ are the expansion coefficients such that $\sum_r|c_r|^2= 1$ and $\{\ket{r}\}$ is the computational basis. The same state $\ket{\Psi}$ can also be written as a symmetric superposition of $n-1$ number of two-level systems as
\begin{equation} 
	\ket*{\tilde{\Psi}} = \mathcal{N} \sum_{P}\big[ \ket{\psi_1} \otimes \ket{\psi_2}\otimes \dots \otimes \ket{\psi_{n-1}} \big]. \label{Eq:Sym}
\end{equation}
Here $\sum_{P}$ corresponds to the sum over all $(n-1)!$ permutations of the qubits and $\mathcal{N}$ is the normalization factor. From here onwards, we denote the state in the MS representation with a `tilde' sign on the state, i.e., a state $\ket{\Psi}$ in the normal representation will read $\ket*{\tilde\Psi}$ in MS representation.

The state $\ket{\psi_k} = \alpha_k \ket{0} +\beta_k\ket{1}$ represents a state of a two-level system. In order to arrive at the MS representation, $\ket{\psi_k}$ is expressed in dual-rail representation~\cite{Milburn2007} i.e. $\ket{\psi_k} \equiv (\alpha_ka_1^\dagger + \beta_ka_2^\dagger)\ket{0,0}$ where $a^{\dagger}_1$, $a^{\dagger}_2$ are the bosonic creation operators for two independent modes and $\ket{0,0}$ is the two-mode vacuum state. The symmetrized state of $n-1$ two-level systems in this representation can simply be written as $\prod_{k=1}^{n-1}(\alpha_k a^{\dagger}_1+\beta_ka^{\dagger}_2)\ket{0,0}$ due to the  indistinguishable nature of $n-1$  bosons. Hence,  
\begin{align} \label{eq:dual-ray-qubit}
	\ket*{\tilde{\Psi}} &\equiv \prod_{k=1}^{n-1}(\alpha_k a^{\dagger}_1+\beta_ka^{\dagger}_2)\ket{0,0},\\
	&\equiv\sum_{r=0}^{n-1}c_r\dfrac{(a_1^\dagger)^{n-1-r}(a^{\dagger}_2)^{r}}{\sqrt{r! (n-1-r)!}}\ket{0,0}.
\end{align}
Comparing Eqs.~\eqref{eq:spin-j state} and~\eqref{eq:dual-ray-qubit}, we get
\begin{equation}
	\ket{r} =\dfrac{(a^{\dagger}_1)^{n-1-r}(a^{\dagger}_2)^{r}}{\sqrt{r! (n-1-r)!}}\ket{0,0}
\end{equation}
and coefficients $c_r$ are functions of $\alpha_k$'s and $\beta_k$'s. Now the task is to evaluate the $\alpha_k$ and $\beta_k$ from given  $c_r$'s. This can be achieved by constructing a polynomial of the form:
\begin{equation}
	\prod_{k=1}^{n-1}(\alpha_k x - \beta_k) \equiv \sum_{r=0}^{n-1} f_{r} x^{n-1-r} = 0 \label{eq:Majorana-polynomial}
\end{equation}
where
\begin{equation}\label{eq:Majorana-coefficients}
	f_{r} = (-1)^{r} \dfrac{c_{r}}{\sqrt{r! (n-1-r)!}}.
\end{equation}
Hence, solving the polynomial equation~\eqref{eq:Majorana-polynomial} yields the $\alpha_k$'s and $\beta_k$'s. For example, the vector space corresponding to two-qubit system is 4 dimensional, however, it's symmetric subspace is 3 dimensional. Whenever, we say symmetric, it is symmetric under particle exchange. The 4D space is spanned by $\{\ket{\uparrow \uparrow}, \ket{\uparrow \downarrow}, \ket{\downarrow \uparrow}, \ket{\downarrow \downarrow}\}$ and the corresponding symmetric subspace will be spanned by
\begin{align*}
	\ket{1,+1} &= \ket{\uparrow \uparrow} \\  \noalign{\vskip5pt}
	\ket{1,0} &= \dfrac{1}{\sqrt{2}} \left( \ket{\uparrow \downarrow} + \ket{\downarrow \uparrow} \right) \\ \noalign{\vskip5pt}
	\ket{1,-1} &= \ket{\downarrow \downarrow}
\end{align*}
and, anti-symmetric subspace
\begin{align*}
	\ket{0,0} &= \dfrac{1}{\sqrt{2}} \left( \ket{\uparrow \downarrow} - \ket{\downarrow \uparrow} \right).
\end{align*}
It is exactly same as addition of angular momentum. When we add the spin angular momentum of two spin-half particles we write
\begin{equation}
	2 \otimes 2 = 3 \oplus 1
\end{equation}
where $3$ and $1$ are the dimensions of the symmetric and anti-symmetric subspace respectively.
\subsection{Bargmann invariant (BI) and geometric phase }           
In this subsection, we define the BI and its relation with the geometric phase. Given three non-orthogonal states $\{\ket{\Psi_1}, \ket{\Psi_2}, \ket{\Psi_3}\}$ from the Hilbert space $\mathcal{H}$, i.e., $\ip{\Psi_i}{\Psi_j}\ne 0; \forall ~i\ne j$, the BI of third order is defined as
\begin{equation} \label{third-order-BI}
	\Delta_3(\Psi_1,\Psi_2,\Psi_3)= \ip*{\Psi_1}{\Psi_2}\ip*{\Psi_2}{\Psi_3}\ip*{\Psi_3}{\Psi_1}.
\end{equation}     
The BI is invariant under unitary transformation $\ket{\Psi_i} \to U\ket{\Psi_i}$. It plays a crucial role in the theory of geometric phase. Consider a closed curve $\mathcal{C}$ constructed by connecting the three non-orthogonal states $\{\ket{\Psi_i}\}$ by geodesics, then the geometric phases $\Phi_{\text{g}}$ associated with this closed curve is  given by~\cite{Mukunda1993}
\begin{equation}
	\label{BI-geom}
	\Phi_{\text{g}}[\mathcal{C}]=-\arg\Delta_3(\Psi_1,\Psi_2,\Psi_3).
\end{equation}
It is straightforward to generalize Eq.~\eqref{third-order-BI} to define $n$th order BI as
\begin{align} \label{eq:nth-order-BI}
	\Delta_n(\Psi_1,\dots,\Psi_n) = \ip*{\Psi_1}{\Psi_2}\ip*{\Psi_2}{\Psi_3} \dots\ip*{\Psi_n}{\Psi_1}.
\end{align}
Further, any higher order BIs can be reduced to third order BIs~\cite{Mukunda1993} and correspondingly the geometric phase for a closed curve constructed by connecting $n$ number of states via geodesics can be expressed as the sum of geometric phases for $n-2$ third order BIs.

\subsection{Geodesic curves} 
Geodesic is the path of shortest distance between two points on a surface. In this subsection, we introduce a differential equation for the geodesic in the state space of a quantum system. 
To define geodesic curves, we need a continuously parametrized smooth curve $\mathcal{C}$ in the Hilbert space  $\mathcal{H}$ given by
\begin{equation}
	\label{g24}
	\mathcal{C}=\{\ket{\Psi(s)}\in \mathcal{H}\;|\;s_1\le s\le s_2\},
\end{equation}
where $s$ is the real parameter varies over $[s_1,s_2]$. 
The quantity called `length' associated with $\mathcal{C}$ is defined as~\cite{Mukunda1993}
\begin{equation} \label{length}
	\mathcal{L}  = \int_{s_1}^{s_2}ds \ip*{u_{\perp}(s)}^{1/2},
\end{equation}
where
\begin{equation*}
	\ket{u_{\perp}(s)} = \ket{u(s)}-\ip*{\Psi(s)}{u(s)}\ket{\Psi(s)},
\end{equation*}
and $\ket{u(s)}$ is the tangent to $\ket{\Psi(s)}$ which reads
\begin{equation*}
	\ket{u(s)} = \frac{d}{ds} \ket{\Psi(s)} \equiv \ket*{\dot{\Psi}(s)}.
\end{equation*}
By requiring $\delta{\cal{L}}=0$, we obtain a differential equation obeyed by $\mathcal{C}$ to be a geodesic:
\begin{equation} \label{eq:diffgeodesic}
	\left(\frac{d}{ds}-\ip*{\Psi(s)}{u(s)}\right)\frac{\ket{u_{\perp}(s)}}{\norm{u_{\perp}(s)}}=f(s)\ket{\Psi(s)},
\end{equation} 
where $f(s)$ is a real function of $s$ which is yet to be determined.

\noindent The geodesics are invariant under the $U(1)$ transformation of the form $\ket{\Psi(s)} \to e^{i\alpha(s)} \ket{\Psi(s)}$. By exploiting this freedom and the freedom of reparameterization, we can make $\ip{\Psi(s)}{u(s)} = 0$ and $ \norm{u} = $ constant, respectively, which yields~\cite{Mukunda1993,Arvind2003}
\begin{equation}
	\dfrac{d^2}{ds^2} \ket{\Psi(s)} = f(s)\ket{\Psi(s)}.
\end{equation}
It is important to note that such reparametrisations, referred to as \emph{affine} reparametrisations, are unique only up to linear transformations i.e.,
\begin{equation}
	s \rightarrow s' = as + b
\end{equation}  
where $a$ and $b$ are constants. Further, the above equation can be shown equivalent to a differential equation of the form~\cite{Mukunda1993}:
\begin{align} \label{reduced-geodesic-eq}
	\dfrac{d^2}{ds^2} \ket{\Psi(s)} = -\ip*{\dot{\Psi}(s)}{\dot{\Psi}(s)}\ket{\Psi(s)}.
\end{align}
A formal solution of Eq.~\eqref{reduced-geodesic-eq}, for the given two end states $\ket{\Psi(s_1)}$ and $\ket{\Psi(s_2)}$ with 
\begin{equation}
	\ip*{\Psi(s_1)}{\Psi(s_2)} \equiv \xi \;\; \text{real and positive},
\end{equation}
is given by
\begin{equation}
	\label{geodesic-curve}
	\ket{\Psi(s) } = \cos(s)\ket{\Psi(s_1)}+\frac{\ket{\Psi(s_2)}-\xi\ket{\Psi(s_1)}}{(1-\xi^2)^{1/2}}\sin(s).
\end{equation}
Hence, using Eq.~\eqref{geodesic-curve} we can generate the geodesic between any two points in the state space.

\subsection{NPCs} 
NPCs are the curves between two points on the quantum state space along which the acquired geometric phase is zero~\cite{Rabei1999,Arvind2003,Chaturvedi2013}. Mathematically, these curves can be defined as follows: consider a differentiable curve $\{\ket{\Psi(s)}\}$ for the real parameter $s \in (s_1,s_2)$ such that $\ip{\Psi(s)}{\Psi(s')}\ne 0 $ for all $s,s'\in(s_1,s_2)$. 
The curve $\{\ket{\Psi(s)}\}$ is an  NPC if for any three points  on the  curve, the BI is real and positive, i.e.,
\begin{equation} \label{null-phase-def-1}
	\Delta_3 \left(\Psi(s), \Psi(s'), \Psi(s'')\right) > 0 , \; s,s',s''\in [s_1,s_2].
\end{equation}
From the above definition it is clear that if the curve $\{\ket{\Psi(s)}\}$ is an NPC, then $\{e^{i \beta} \ket{\Psi(s)}\}$ will also be an NPC. Exploiting this condition, we can always choose a curve in the $\mathcal{H}$ such that
\begin{equation}
	\label{NPC in unit sphere}
	\ip{\Psi(s)}{\Psi(s')} > 0
\end{equation}
for any $s,s'\in(s_1,s_2)$~\cite{Arvind2003}. Hence, there exist infinitely many NPCs between any two points in the state space.

\section{Bloch sphere decomposition of geodesics} \label{Sec:Geodesic}
The geodesic between any two states of a two-level quantum system is the segment of the great circle connecting these states on the Bloch sphere. This is the consequence of the spherical geometry  of the state space of a two-level system. However, the geodesics in three or higher dimensional state spaces are notoriously difficult to understand, even though the expression to calculate these geodesics is given in Eq.~\eqref{geodesic-curve}. 
In this section, we present the Bloch sphere decomposition of higher dimensional geodesics using MS representation. This Bloch sphere decomposition reveals intrinsic symmetries of geodesics which may help to understand the geometric structure of higher dimensional  state space. We start with geodesics in three-dimensional state space and extend these results to higher dimensions.

\subsection{Geodesics in three-dimensional state space} \label{sec:geodesics3}
Consider two states $\{\ket{\Psi_1}, \ket{\Psi_2} \}$ in the three-dimensional state space. For simplicity, we   these choose states  of the following form
\begin{equation} \label{eq:endstates}
	\ket{\Psi_1} = \begin{pmatrix}
		1\\
		0\\
		0
	\end{pmatrix},\,\,\, \ket{\Psi_2} = \begin{pmatrix}
		\alpha^2 \\
		\sqrt{2} \alpha \beta \\
		\beta^2
	\end{pmatrix},
\end{equation}
such that each one of them individually are represented by degenerate MSs. Here $\alpha$ and $\beta$ are real, with $ \ip*{\Psi_1}{\Psi_2} = \alpha^2 \equiv \cos\theta$ ; $ 0 \le \theta < \pi/2$ and $\alpha^2 + \beta^2 = 1$.

In the MS representation, the states considered in Eq.~\eqref{eq:endstates} takes the form
\begin{eqnarray}
	\ket*{\tilde{\Psi}_1} &=& \ket{0} \otimes \ket{0},\nonumber\\
	\ket*{\tilde{\Psi}_2} &=& \ket{\phi} \otimes \ket{\phi}, \label{Eq:End-states}
\end{eqnarray} where $ \ket{0} = \begin{pmatrix} 1 & 0\end{pmatrix}^T$ and $ \ket{\phi} = \begin{pmatrix} \alpha & \beta \end{pmatrix}^T$. 

From Eq.\eqref{geodesic-curve} we can find a geodesic $ \{\ket{\Psi(s)} \;|\; 0 \le s \le \theta \} $ connecting $\ket{\Psi_1}$ and $\ket{\Psi_2}$  which reads
\begin{align} \label{eq:geodesic}
	\ket{\Psi(s)} &= \begin{pmatrix}
		\cos (s) \\
		a \sin (s) \\
		b \sin (s)
	\end{pmatrix} \equiv \dfrac{1}{\mathcal{N}(s)}\left[\ket{\psi_+(s)}\ket{\psi_-(s)}+\ket{\psi_-(s)}\ket{\psi_+(s)}\right].
\end{align}
Here $a=\sqrt{2} \alpha \beta/\sin \theta $, $b= \beta^2/\sin \theta$ and $\mathcal{N}(s)$ is the normalisation constant. The states $\ket{\psi_\pm}$ in the MS representation of $\ket{\Psi(s)}$ are given by 
\begin{equation} 
	\label{Majorana-decomposition}
	\ket{\psi_{\pm}(s)} = \dfrac{1}{\sqrt{1 + \abs*{x_{\pm}(s)}^2}} \begin{pmatrix}
		1 \\
		x_{\pm}(s)
	\end{pmatrix},
\end{equation}
where  
\begin{equation}
	\label{solutions}
	x_{\pm}(s) = \dfrac{a\sin(s)\pm i \sqrt{b\sin(2s)-a^2\sin^2(s)}}{\sqrt{2}\cos(s)}.
\end{equation}
Here $x_{\pm}(s)$ are the solutions of the Majorana polynomial Eq.~\eqref{eq:Majorana-polynomial} corresponding to the $\ket{\Psi(s)}$ in  Eq.~\eqref{eq:geodesic}.
So, the geodesic curve $\{\ket{\Psi(s)}\}$ decomposes into two curves $ \{\ket{\psi_+(s)} \;|\; 0 \le s \le \theta \} $ and $ \{\ket{\psi_-(s)} \;|\; 0 \le s \le \theta \} $ belonging to  two-dimensional space. Note that the curves $ \{\ket{\psi_{\pm}(s)} \;|\; 0 \le s \le \theta \} $ themselves do not satisfy the differential equation for a geodesic~\eqref{eq:diffgeodesic}. Therefore, they are not geodesics in two-dimensional space.

The Bloch vectors corresponding to  $ \{\ket{\psi_{\pm}(s)\}}  $  are denoted as $\{\vb{n}_{\pm}(s)\}$ and obtained by
\begin{equation}
	\label{components on the Bloch sphere}
	\vb{n}_{\pm}(s) = \ev{\boldsymbol{\sigma}}{\psi_{\pm}(s)},
\end{equation}
where $\boldsymbol{\sigma} = \{\sigma_x, \sigma_y, \sigma_z\}$ is the vector of Pauli matrices~\cite{Nielsen2010}. 
The components of the curves $\{\vb{n}_{\pm}(s)\}$ along $x,y,z$ read
\begin{eqnarray}
	\label{reflection-bloch-sphere}
	\left(\vb{n}_{+}\right)_x&=&\left(\vb{n}_{-}\right)_x=\dfrac{\sqrt{2}a\sin(s)}{\cos(s)+b\sin(s)}, \nonumber\\
	\left(\vb{n}_{+}\right)_y&=&-\left(\vb{n}_{-}\right)_y=\dfrac{\sqrt{b\sin(2s)-a^2\sin^2(s)}}{\cos(s)+b\sin(s)}, \nonumber\\
	\left(\vb{n}_{+}\right)_z&=&\left(\vb{n}_{-}\right)_z=\dfrac{\cos(s)-b\sin(s)}{\cos(s)+b\sin(s)}.
\end{eqnarray} 
\begin{figure}
	\centering
	\subfigure[]{
		\includegraphics[width=5cm]{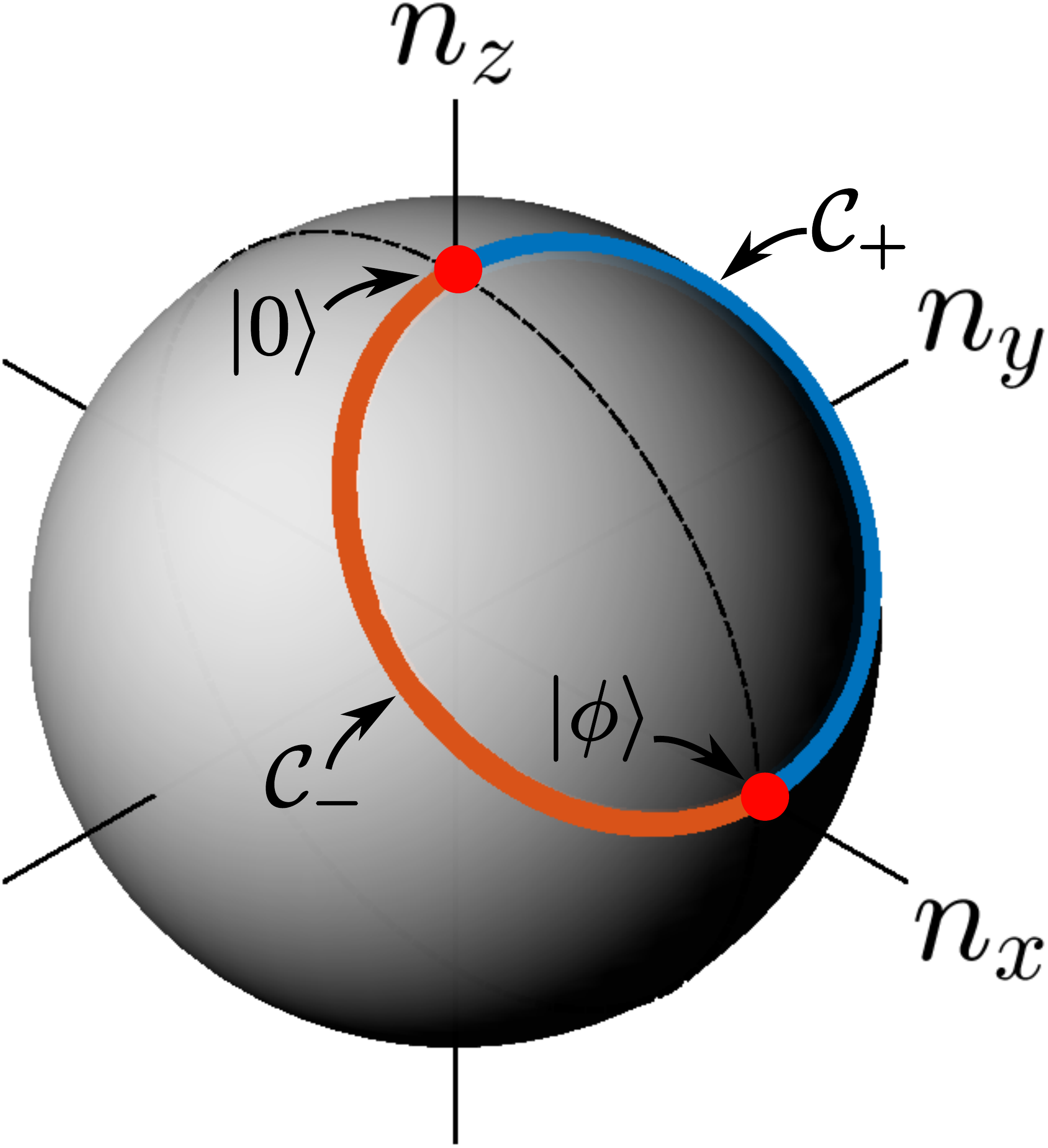}
		\label{fig:geod1}}
	\subfigure[]{
		\includegraphics[width=5cm]{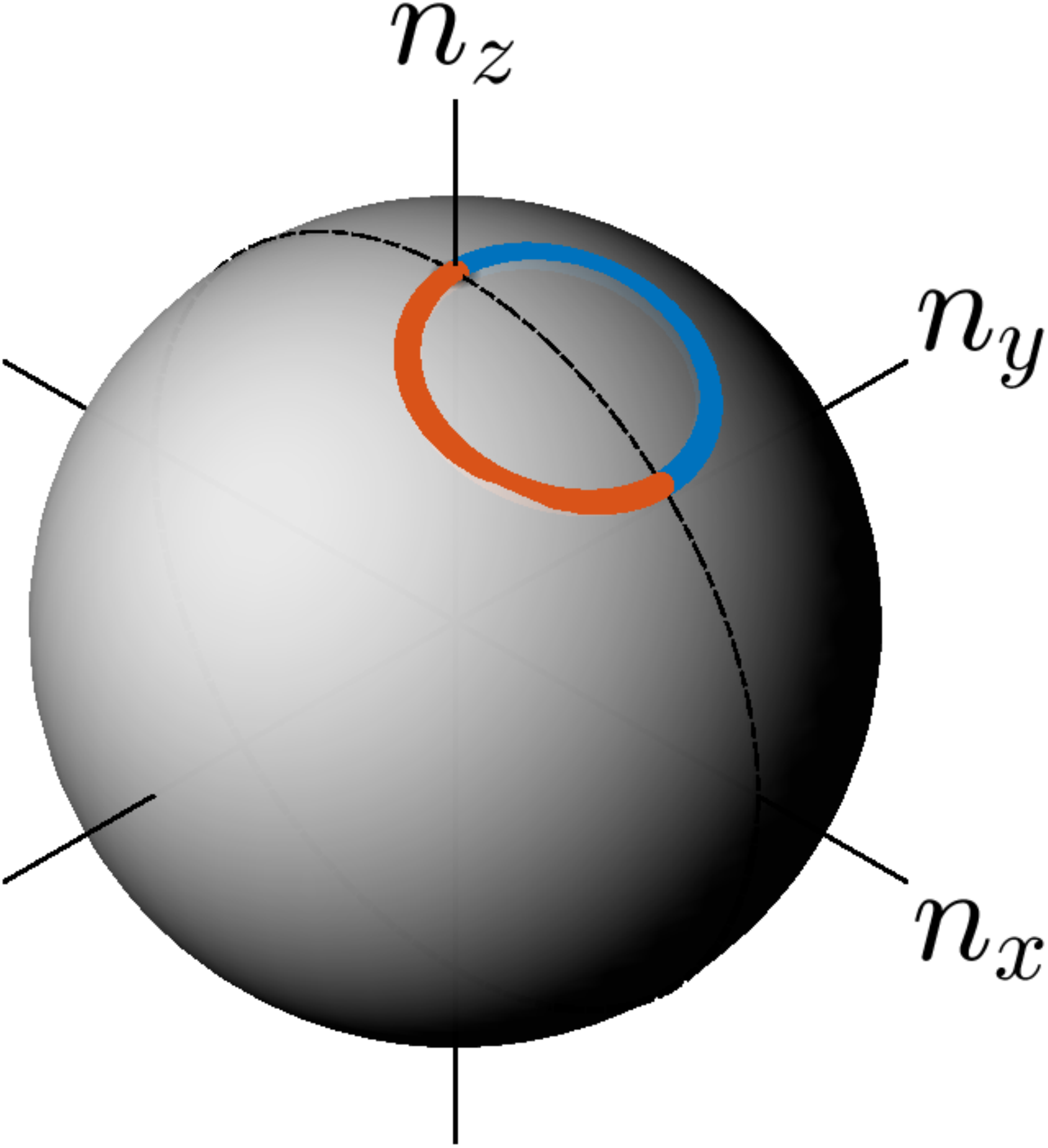}
		\label{fig:geod2}}
	\caption{(Color online) Here we plot the geometric decomposition of a geodesic between two states given in Eq.~\eqref{Eq:End-states} where we have chosen \subref{fig:geod1} $\theta = \pi/3$, \subref{fig:geod2} $\theta = \pi/5$. The blue and orange curve correspond to $\{\vb{n}_+\}$ and $\{\vb{n}_-\}$ respectively, given in Eq. \eqref{reflection-bloch-sphere}. }
	\label{fig:geodesic}
\end{figure}
Since,  the end states $\ket{0}$ and $\ket{\phi}$ in Eq.~\eqref{Eq:End-states} lie on the $xz$-plane on the Bloch sphere and the solutions $\{x_\pm(s)\}$ \eqref{solutions} form a complex conjugate pair, the components of the curves $\{\vb{n}_{\pm}(s)\}$ differ only along $ y $-axis by a negative sign. Therefore, the two curves are reflective about the $xz$-plane. We will call these curves as `dual' of each other.

On a careful observation,  we can see that the components of the pair of dual curves  along $x,y,z$ axes satisfy an equation of the form
\begin{equation}
	\label{eq-of-circle}
	\bigg((\vb{n}_{\pm})_x-\alpha\beta\bigg)^2+\left(\vb{n}_{\pm}\right)_y^2+\bigg((\vb{n}_{\pm})_z-\alpha^2\bigg)^2=\beta^2,
\end{equation}
which is an equation of a circle with center at $(\alpha\beta,0,\alpha^2)$, the midpoint of the line joining the end states on the  Bloch sphere. Hence, the two curves are segments of a circle, and share the same center and the same radius $\beta$. Therefore, the geodesic curve connecting two states in the three-dimensional state space can be identified by a complete circle of radius $\beta =\sqrt{1-\ip{\Psi_1}{\Psi_2}}$ on the Bloch sphere constructed by joining two semi-circular arcs $\{\vb{n}_\pm(s)\}$.
The  pair of dual curves  $\mathcal{C}_{\pm} \equiv \{ \vb{n}_{\pm}(s)\}$ traced by the two states in Eq.~\eqref{Majorana-decomposition} are shown in Fig.~\ref{fig:geodesic} for different values of $\theta$.

So far we have considered the geodesics between the end states which are represented by degenerate MSs given in Eq.~\eqref{eq:endstates}. Interestingly, this formalism can be applied to the geodesics between two  arbitrary  three-dimensional states.

We note that any two states $\{\ket{\Psi_1},\ket{\Psi_2}\}$ in three-dimensional Hilbert space can be transformed to degenerate MSs  by a common unitary transformation $U$. We can show it explicitly by considering two arbitrary states $ \{\ket{\Psi_1}, \ket{\Psi_2}\} $ in three-dimensional state space with the inner product given by $\ip*{\Psi_1}{\Psi_2} = \cos \theta$, i.e., real and positive. We can write $ \ket{\Psi_2} $ as
\begin{align}
	\ket{\Psi_2} = \cos\theta \ket{\Psi_1} + \gamma \ket*{\bar{\Psi}_1}
\end{align}
where $\gamma = \sin\theta e^{i \phi}$, $\phi \in \mathbb{R}$ and $\ket*{\bar{\Psi}_1}$ is orthogonal to $\ket{\Psi_1}$. Now, we take unitary of the form 
\begin{align}
	U =\dyad{0}{\Psi_1} + e^{-i \phi} \dyad*{1}{\bar{\Psi}_1} + \dyad*{2}{{\tilde{\Psi}}_1}
\end{align} 
where $\{\ket{0}, \ket{1}, \ket{2}\}$ and $\{ \ket{\Psi_1}, \ket*{\bar{\Psi}_1}, \ket*{{\tilde{\Psi}}_1}\}$ form the orthonormal bases. On application of this unitary on the two states $ \{\ket{\Psi_1}, \ket{\Psi_2}\} $ results in the states of the form
\begin{align} \label{eq:general-states}
	U\ket{\Psi_1} \equiv \ket{\Psi'_1}= \begin{pmatrix}
		1 \\
		0 \\
		0
	\end{pmatrix}, \;\;\; U\ket{\Psi_2} \equiv \ket{\Psi'_2}= \begin{pmatrix}
		\cos\theta \\
		\sin\theta \\
		0
	\end{pmatrix}.
\end{align} 
One can see that $\ket{\Psi'_2}$ cannot be represented by a degenerate MSs. For that one need to apply one more unitary transformation in order to bring $\ket{\Psi'_2}$ to a state represented by degenerate MSs. An appropriate unitary for that is given by
\begin{align}
	U = \begin{pmatrix}
		1 & 0 & 0 \\
		0 & a & b \\
		0 & -b & a 
	\end{pmatrix}
\end{align}
where $a=\sqrt{2} \alpha \beta/\sin \theta $, $b= \beta^2/\sin \theta$, $\alpha^2 = \cos \theta$ and $\alpha^2 + \beta^2 = 1$. After applying the following unitary, we will get
\begin{align}
	\ket*{\Psi_2} = \begin{pmatrix}
		\alpha^2 \\
		\sqrt{2} \alpha \beta \\
		\beta^2
	\end{pmatrix}
\end{align}
which is written as 
\begin{align}
	\ket*{\tilde\Psi_2} = \begin{pmatrix}
		\alpha \\
		\beta
	\end{pmatrix} \otimes\begin{pmatrix}
		\alpha \\
		\beta
	\end{pmatrix}.
\end{align}
Therefore, any two arbitrary states with real inner product can be brought to the degenerate MSs states and we can study the structure of the geodesics between these states by first mapping them to degenerate MSs and constructing the geodesic using the semicircular curves $\mathcal{C}_\pm$. Applying $U^\dagger$ on these curves will result the actual curves corresponding to the geodesic between $\{\ket{\Psi_1},\ket{\Psi_2}\}$.

To summarize  the results obtained in this section: (i) A geodesic connecting the  three-dimensional states $\ket{\Psi_1}$ and $\ket{\Psi_2}$, which are represented by degenerate MSs on the Bloch sphere,  decomposes into two unique curves on the Bloch sphere.
(ii) These two curves are reflective about the great circle connecting the two end points on the Bloch sphere,  and constitute a circle of radius  $\sqrt{1 -\ip*{\Psi_1}{\Psi_2}}$.
(iii) One can obtain the geodesic connecting any two arbitrary states in three-dimensional space by first converting the two end states to degenerate MSs states by using a unitary transformation and then constructing the unique circle between the end states on the Bloch sphere.

\subsection{Geodesics in higher-dimensional state space}
We extend our analysis to study the structure of geodesics in higher-dimensional state space. Let us start by considering non-orthogonal end states $ \{\ket{\Psi_1}, \ket{\Psi_2}\} $ in an $n$-dimensional state space which map to $(n-1)$-fold degenerate MSs individually on the Bloch sphere. In the MS representation, the end states can be written as
\begin{align}\label{Eq:end-state}
	\ket*{\tilde{\Psi}_1} &= \ket{0}_0 \otimes \ket{0}_1 \otimes \dots \otimes \ket{0}_{n-2}, \nonumber \\
	\ket*{\tilde{\Psi}_2} &= \ket{\phi}_0 \otimes \ket{\phi}_1\otimes \dots \otimes \ket{\phi}_{n-2},
\end{align}
where $ \ket{0} = \begin{pmatrix} 1 & 0\end{pmatrix}^T$ and $ \ket{\phi} = \begin{pmatrix} \alpha & \beta \end{pmatrix}^T$. Here $\alpha$ and $\beta$ are real, with $ \ip*{\Psi_1}{\Psi_2} = \alpha^{n-1} \equiv \cos\theta$ ; $ 0 \le \theta < \pi/2$ and $\alpha^2 + \beta^2 = 1$.
From Eq.~\eqref{geodesic-curve}, the geodesic curve  $\{\ket{\Psi(s)};0\le s\le \theta\}$ connecting the end states in the MS representation turns out to be
\begin{align} \label{eq:PsiTilde}
	\ket*{\tilde\Psi(s)} &= \ket{0}^{\otimes n - 1} + A^{(n-1)}(s) \ket{\phi}^{\otimes n-1},\\
	&\equiv \mathcal{N} \sum_{P} [\ket{\chi_1(s)} \otimes \ket{\chi_2(s)} \otimes \dots \otimes \ket{\chi_{n-1}(s)}]
\end{align}
where $\sum_{P}$ corresponds to the sum over all symmetric permutations of the $n-1$ number of states $ \ket{\chi_k(s)} $  of two-level systems.  $\mathcal{N}$ is the normalization constant, and 
\begin{align}
	A^{(n-1)}(s) = \dfrac{\sin s}{\cos s (1 - \alpha^{n-1})^{1/2} - \alpha^{n-1} \sin s}.
\end{align}
Since the end states considered in Eq.~\eqref{Eq:end-state} are real, the Majorana polynomial [Eq.~\eqref{eq:Majorana-polynomial}] is also real. Therefore,  the roots occur as complex conjugate pairs along with a real root depending on the dimension of the state space. By solving the Majorana polynomial Eq.~\eqref{eq:Majorana-polynomial}, one can find the curves $\{\ket{\chi_k(s)}\}$, $i = 1, \dots , n-1$ as
\begin{align}\label{eq:ansatz}
	\ket{\chi_k(s)} = \ket{0} + \Delta\, \omega_k A(s)\ket{\phi}.
\end{align}
Here, $\omega_k$'s are the $(n-1)$th roots of unity given by $\omega_k = e^{2 \pi i k/n-1}$ with $k = 0,1,\dots, n-2$ and $\Delta = \left(\prod_{k=0}^{n-2} \omega_k\right)^{1/(n-1)}$. This shows that a geodesic curve in an $n$-dimensional state space decomposes to $n-1$ curves on the Bloch sphere.

The Bloch vector corresponding to the state $\{\ket{\chi_k(s)}\}$ can be written as 
$\{\vb{n}_{k}(s) = \ev{\boldsymbol{\sigma}}{\chi_{k}(s)}\}$.  Next we show that the curve  $\{\ket{\chi_k(s)}\}$ traced by the states $\ket{\chi_k(s)}$ for all the values of $k$ constitute circular segments on the Bloch sphere.

Consider three Bloch vectors $ \vb{p_1},~ \vb{p_2},~ \vb{p_3}$  corresponding to three states on the curve $\{\ket{\chi_k(s)}\}$. The unit vector $\vb{m}$ orthogonal to the plane  containing these three Bloch vectors  can be written as 
\begin{equation} \label{normal-vector}
	\vb{m}=\dfrac{(\vb{p}_2-\vb{p}_1) \times (\vb{p}_3-\vb{p}_1)}{\norm{(\vb{p}_2-\vb{p}_1)\times (\vb{p}_3-\vb{p}_1)}}.
\end{equation}
The intersection of the plane containing the three vectors $ \vb{p_1},~ \vb{p_2},~ \vb{p_3}$ and the Bloch sphere constitute a circle. The three Bloch vectors $\vb{p}_1,~\vb{p}_2,~\vb{p}_3$ also lie on the circle whereas the unit vector $\vb{m}$ passes through the center of this circle. Therefore, the projection of any of these three Bloch vectors will be the same on the vector $\vb{m}$.

\begin{figure}
	\centering
	\subfigure[]{
		\includegraphics[width=4.5cm]{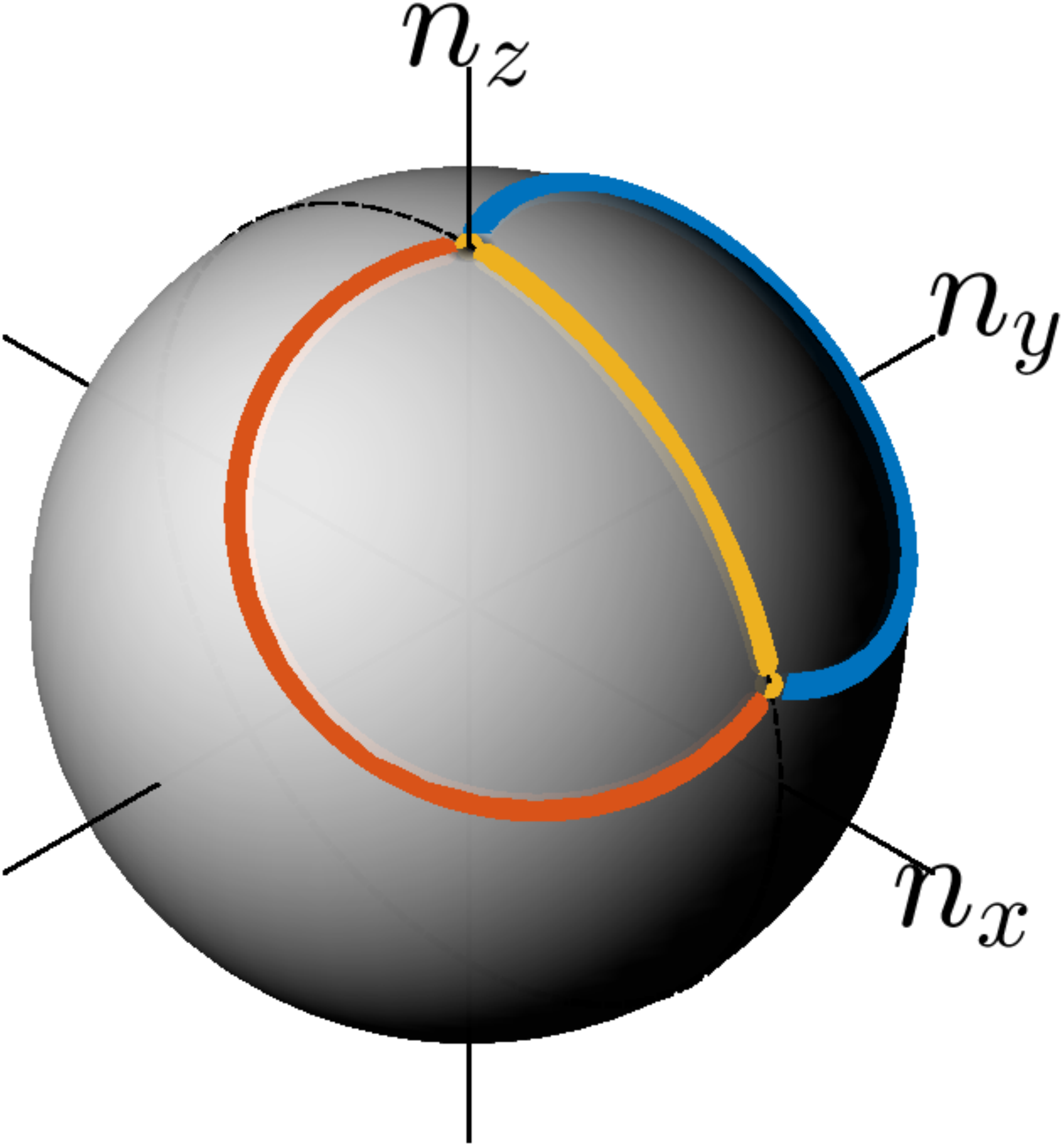}
		\label{fig:geo4D}}
	\subfigure[]{
		\includegraphics[width=4.5cm]{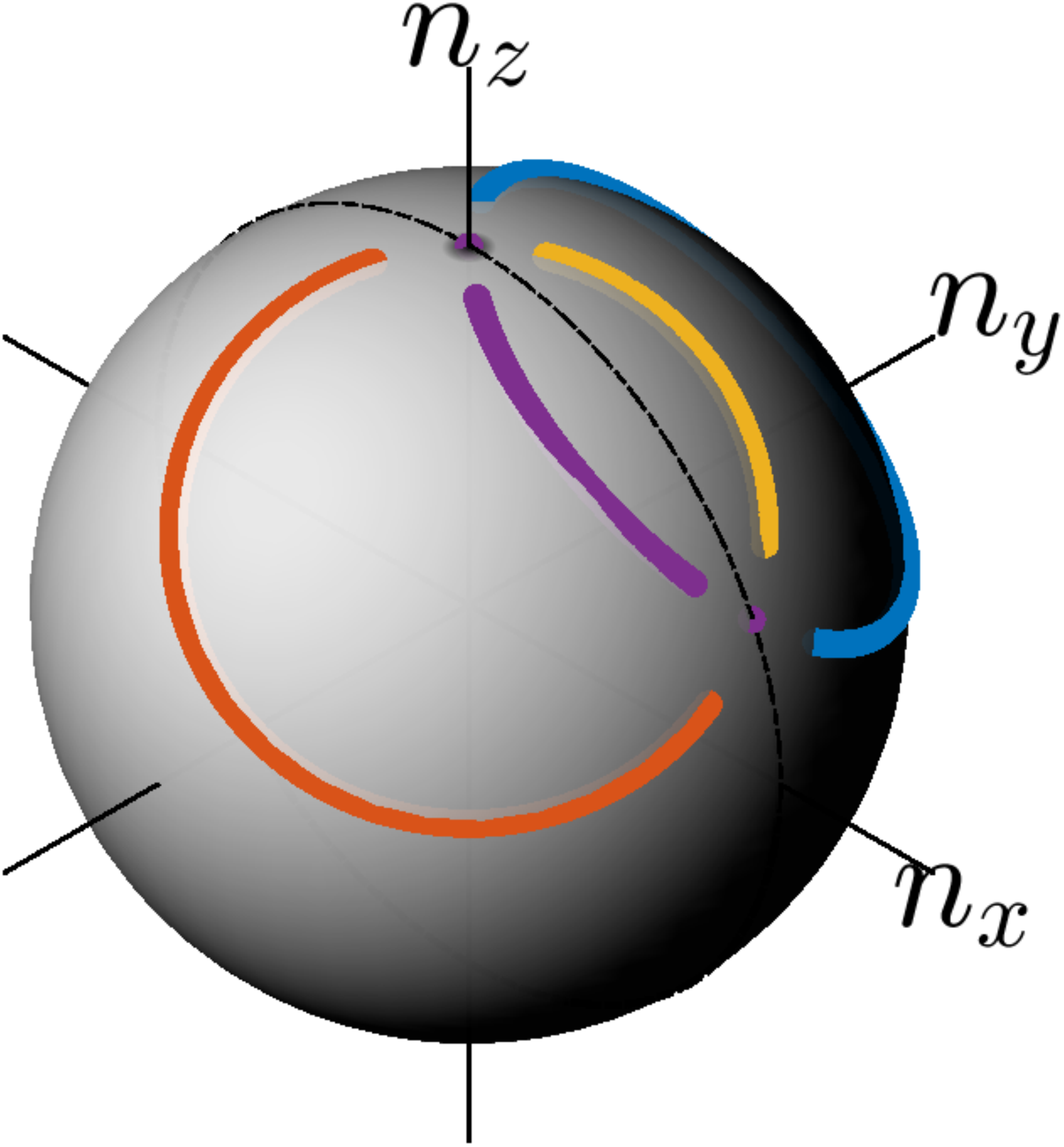}
		\label{fig:geo5D}}
	\subfigure[]{
		\includegraphics[width=4.5cm]{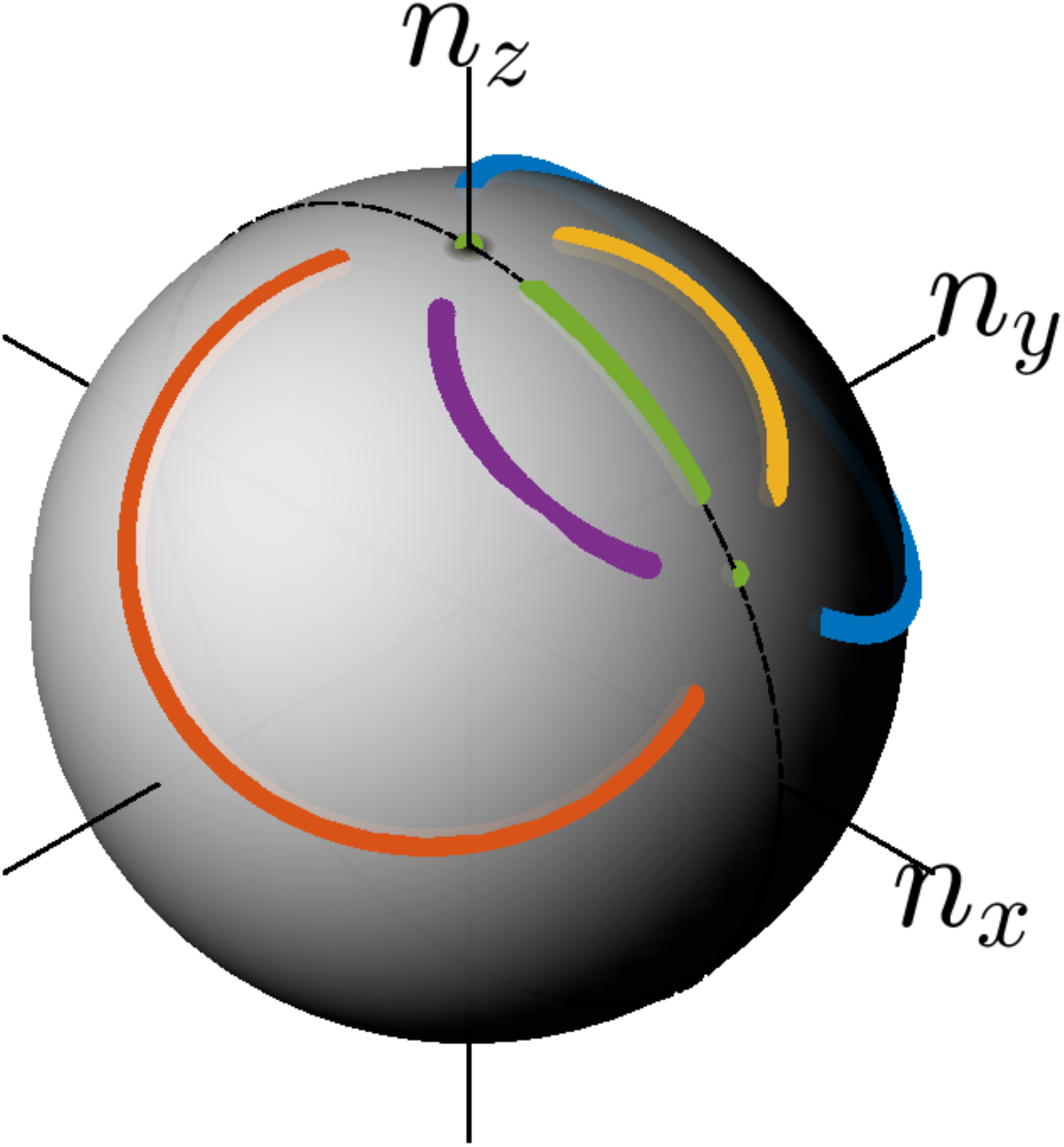}
		\label{fig:geo6D}}
	\subfigure[]{
		\includegraphics[width=4.5cm]{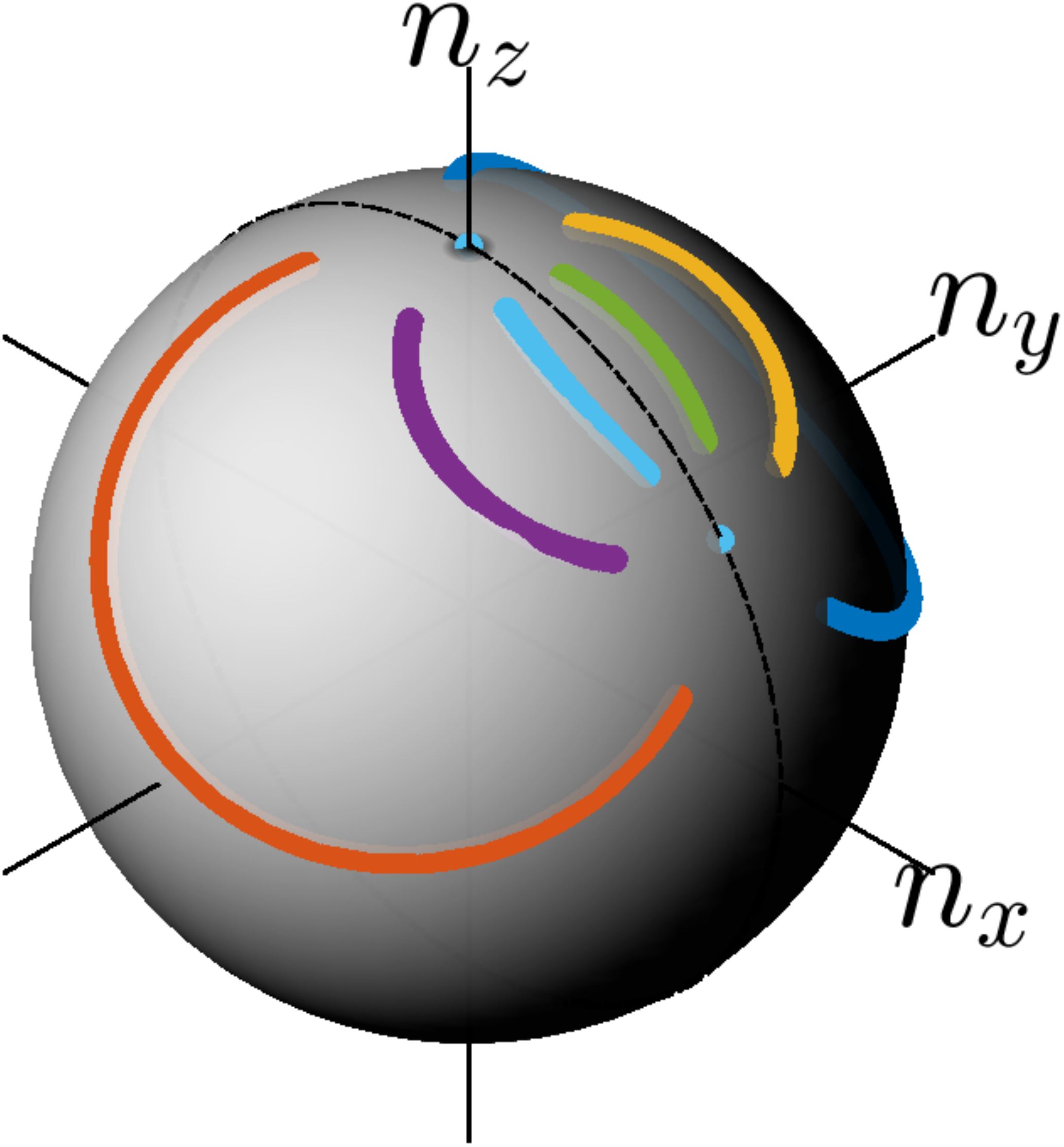}
		\label{fig:geo7D}}
	\subfigure[]{
		\includegraphics[width=4.5cm]{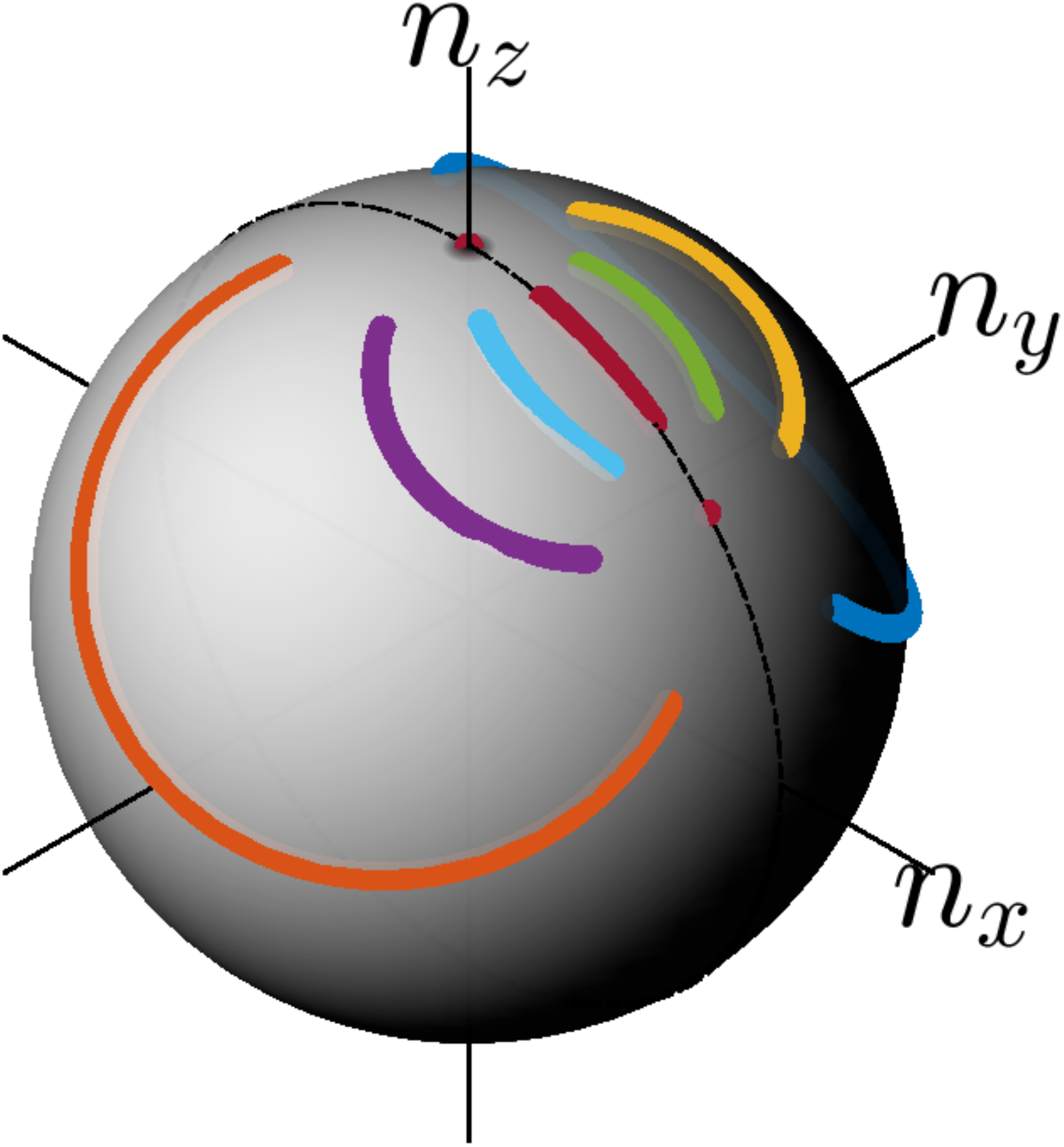}
		\label{fig:geo8D}}
	\caption{(Color online) Geometric decomposition of  geodesics in \subref{fig:geo4D} 4- \subref{fig:geo5D} 5- \subref{fig:geo6D} 6-  \subref{fig:geo7D} 7- \subref{fig:geo8D} 8-dimensional state space for $\theta = \pi/3$. We can see that we obtain $n-1$ number of  curves for $n$-dimensional system which is consistent with our decomposition.}
	\label{fig:highergeo}
\end{figure}

We find that the $\vb{m}$ is the same for any choice of the three Bloch vectors on the curve $\{\ket{\chi_k(s)}\}$. Moreover, the projection of all the Bloch vectors $\vb{p}(s)$ on the curve $\{\ket{\chi_k(s)}\}$ with $\vb{m}$ remains constant. This indicates that the curve traced by $ \ket{\chi_k(s)} $ is circular segment for all the values of $k$. The radius corresponding to the $k$-th circular segment is 
\begin{align}
	\label{eq:radius-geodesic}
	R_k = \dfrac{2 \beta}{\sqrt{4 \beta^2 - \alpha^2 (\Delta^* \omega_k^* - \Delta \omega_k)^2}}.
\end{align}
From the above expression, we see that the radii depends only on the inner product between the end states and not explicitly on the states. Furthermore, the circular segments are unique for the geodesic between a given set of end states. Therefore, once the end states are uniquely identified on the Block sphere, one can construct the desired geodesic by using Eq.~\eqref{eq:radius-geodesic} alone.  In Fig.~\ref{fig:highergeo}, we have plotted the structure of geodesic curve on the Bloch sphere for $n=4$- and $5$-dimensional state space. 

There are certain intrinsic symmetries in the MS representation of a higher dimensional geodesic. These symmetries reflect differently for even and odd dimensions. For example, if the dimension $n$ is odd, we get even number of curves which appear in dual pairs, i.e., the pair of curves which are reflections of each other. The two curves $\{\ket{\chi_{i}(s)}\},\{\ket{\chi_{j}(s)}\}$, which are  dual of each other, satisfy the condition $i + j = 0 \mod n-1 $, where $i,j = 0,1,2,\dots, n-2$ and $i\ne j$.  However, in the case of even $n$, one curve occurs along the great circle connecting the end states and the remaining $n-2$ curves occur in dual pairs. For even $n$, the two dual curves $\{\ket{\chi_{i}(s)}\},\{\ket{\chi_{j}(s)}\}$ satisfy the condition $i + j = n-2 \mod n-1$. From Eq.~\eqref{normal-vector} and Eq.~\eqref{eq:radius-geodesic}, it is evident that the dual curves give the same value of radius with different centers.

Now, we have a way of constructing a unique geodesic between any two states $ \{\ket{\Psi_1}, \ket{\Psi_2}\} $ in the $n$-dimensional state space. Given $\ket{\Psi_1},~\ket{\Psi_2}$ we first map them to $(n-1)$-fold degenerate MSs states $\ket*{\tilde\Psi_1},~\ket*{\tilde\Psi_2}$ in the MS representation using a unitary transformation $U$. Thereafter, we construct circular segments on the Bloch sphere which are defined by Eqs.~\eqref{normal-vector} and~\eqref{eq:radius-geodesic} yielding the  geodesic in $n$-dimensional state space, connecting the states $ \ket*{\tilde{\Psi}_1} $ and $ \ket*{\tilde{\Psi}_1} $. Finally, we apply the $U^\dagger$  to get the desired geodesic between the original states.


\section{Bloch Sphere decomposition of NPCs}\label{Sec:NPC}
So far we have seen that the geometric decomposition of the geodesic has revealed an interesting underlying symmetry. This not only have given us better understanding of the geodesics but also have provided us geometric ways of constructing one.  In this section, we investigate the geometric structure of NPCs in higher-dimensional state space. We mostly deal with the three-dimensional case but the analysis can easily be extended to $n$-dimensional state space. Here we propose a way to construct NPCs by choosing suitable set of curves on the Bloch sphere. 

Once again, we consider the  end states 
$		\ket*{\tilde\Psi_1} = \ket{0} \otimes \ket{0},$ and 
$\ket*{\tilde\Psi_2} = \ket{\phi} \otimes \ket{\phi}$ as defined earlier. In order to construct an NPC between $\ket{\Psi_1}$ and $\ket{\Psi_2}$ we propose the following:

{\bf Proposition :} An arbitrary curve $\mathcal{C}$ connecting the states $\{\ket*{\tilde\Psi_1},\ket*{\tilde\Psi_2}\}$, and its dual curve $\mathcal{C}^*$  together form an NPC in the three-dimensional state space.

{\it Proof}:
The most general curve  $\mathcal{C} \equiv \{\ket{\psi(s)}|s_1\le s \le s_2\}$ connecting the states $\{\ket*{\tilde\Psi_1},\ket*{\tilde\Psi_2}\}$, is given by  
\begin{equation}
	\label{NPC-curve1}
	\ket{\psi(s)} = \begin{pmatrix}
		\cos(\eta(s)/2)\\
		e^{i\Gamma(s)} \sin(\eta(s)/2)\\
	\end{pmatrix},
\end{equation}
where $\eta(s)$ and $\Gamma(s)$ are arbitrary real functions of $s$. The functions $\eta(s)$ and $\Gamma(s)$ satisfy the relations $\eta(s_1)=0,\eta(s_2)=2\cos^{-1}(\alpha)$ and $\Gamma(s_1)=\Gamma(s_2)=0$. The dual curve $\mathcal{C}^* \equiv \{\ket{\psi(s)}|s_1\le s \le s_2\}$ can then be defined as
\begin{equation}
	\label{NPC-curve2}
	\ket{\psi'(s)}=\begin{pmatrix}
		\cos(\eta(s)/2)\\
		e^{-i\Gamma(s)} \sin(\eta(s)/2)\\
	\end{pmatrix}
\end{equation}
Using the curves $\mathcal{C}$ and $\mathcal{C}^*$ the states on the curve in the three-dimensional state space can be written as
\begin{equation}
	\label{Majorana-representation-NPC}
	\ket*{\tilde\Psi(s)}=\dfrac{1}{\mathcal{N}(s)}\big[\ket{\psi(s)}\ket{\psi'(s)}+\ket{\psi'(s)}\ket{\psi(s)}\big],
\end{equation}
where $\mathcal{N}(s)$ is the normalization constant. Now, we will find the third order BI between any three states on the curve given in Eq.~\eqref{Majorana-representation-NPC}. The BI of third order $\Delta_3$ for any given three mutually non-orthogonal states $ \{\ket{\Psi_1}, \ket{\Psi_2}, \ket{\Psi_3}\} $ is written as
\begin{align}
	\Delta_3(\Psi_1, \Psi_2, \Psi_3) = \ip*{\Psi_1}{\Psi_2}\ip*{\Psi_2}{\Psi_3}\ip*{\Psi_3}{\Psi_1}.
\end{align}
In the MS representation, these states are written as 
\begin{align}
	\ket{\Psi_1}&=\frac{1}{\mathcal{N}_1}\left[\ket{\psi_1}\ket{\psi'_1}+\ket{\psi'_1}\ket{\psi_1} \right]\nonumber\\
	\ket{\Psi_2}&=\frac{1}{\mathcal{N}_2}\left[\ket{\psi_2}\ket{\psi'_2}+\ket{\psi'_2}\ket{\psi_2} \right]\nonumber\\
	\ket{\Psi_3}&=\frac{1}{\mathcal{N}_3}\left[\ket{\psi_3}\ket{\psi'_3}+\ket{\psi'_3}\ket{\psi_3} \right]
\end{align}
where $\mathcal{N}_i$ are the normalization constants. Using this representation, we can expand $ \Delta_3(\Psi_1, \Psi_2, \Psi_3) $ as follows
\begin{align*}
	&\ip*{\Psi_1}{\Psi_2}\ip*{\Psi_2}{\Psi_3}\ip*{\Psi_3}{\Psi_1} \\
	&= \ip*{\psi_1}{\psi_2}\ip*{\psi_2}{\psi_3}\ip*{\psi_3}{\psi_1}  \ip{\psi'_1}{\psi'_2} \ip{\psi'_2}{\psi'_3} \ip{\psi'_3}{\psi'_1} \\
	& + \ip*{\psi_1}{\psi_2}\ip*{\psi_2}{\psi'_3}\ip{\psi'_3}{\psi'_1} \ip{\psi'_1}{\psi'_2} \ip{\psi'_2}{\psi_3} \ip*{\psi_3}{\psi_1}   \\
	&+ \ip*{\psi_1}{\psi'_2}\ip{\psi'_2}{\psi'_3}\ip{\psi'_3}{\psi'_1} \ip{\psi'_1}{\psi_2}\ip*{\psi_2}{\psi_3} \ip*{\psi_3}{\psi_1}  \\ 
	&+ \ip*{\psi_1}{\psi'_2}\ip{\psi'_2}{\psi_3}\ip*{\psi_3}{\psi_1} \ip{\psi'_1}{\psi_2}\ip*{\psi_2}{\psi'_3}  \ip{\psi'_3}{\psi'_1} \\
	&+ \ip*{\psi_1}{\psi_2}\ip*{\psi_2}{\psi_3}\ip*{\psi_3}{\psi'_1} \ip{\psi'_1}{\psi'_2} \ip{\psi'_2}{\psi'_3}  \ip{\psi'_3}{\psi_1} \\
	& + \ip*{\psi_1}{\psi_2}\ip*{\psi_2}{\psi'_3}\ip{\psi'_3}{\psi_1} \ip{\psi'_1}{\psi'_2} \ip{\psi'_2}{\psi_3} \ip*{\psi_3}{\psi'_1}   \\
	&+ \ip*{\psi_1}{\psi'_2}\ip{\psi'_2}{\psi'_3}\ip{\psi'_3}{\psi_1} \ip{\psi'_1}{\psi_2}\ip*{\psi_2}{\psi_3} \ip*{\psi_3}{\psi'_1}  \\ 
	&+ \ip*{\psi_1}{\psi'_2}\ip{\psi'_2}{\psi_3}\ip*{\psi_3}{\psi'_1} \ip{\psi'_1}{\psi_2}\ip*{\psi_2}{\psi'_3}  \ip{\psi'_3}{\psi_1} 
\end{align*}
Now we choose,
\begin{align}
	\ket{\psi(s)} = \begin{pmatrix}
		\cos(\eta(s)/2)\\
		e^{i\Gamma(s)} \sin(\eta(s)/2)\\
	\end{pmatrix},  
\end{align}
and 
\begin{align}
	\ket{\psi'(s)} = \begin{pmatrix}
		\cos(\eta(s)/2)\\
		e^{-i\Gamma(s)} \sin(\eta(s)/2)\\
	\end{pmatrix}.
\end{align}
It is very clear from the above choice, that a pairs of inner products $\ip*{\psi_i}{\psi_j}$ and $\ip*{\psi'_i}{\psi'_j}$ or $\ip*{\psi_i}{\psi'_j}$ and $\ip*{\psi'_i}{\psi_j}$ or $\ip*{\psi_i}{\psi'_j}$ and $\ip*{\psi'_i}{\psi_j}$ are complex conjugate of each other. The $\Delta_3(\Psi_1, \Psi_2, \Psi_3)$ has eight terms where each term contains three such pairs. Hence, it is very straightforward to show that $\Delta_3(\Psi_1, \Psi_2, \Psi_3)$ is real and positive for the above choice of $ \ket{\psi(s)} $ and $ \ket{\psi'(s)} $. Hence the curve $\{\ket*{\tilde\Psi(s)}\}$ is an NPC. For the given inner product between the end states, by choosing  $\eta(s)$ and $\Gamma(s)$, one can generate NPCs.
Since, one can construct infinitely many curves $\mathcal{C}$ connecting the states $\{\ket*{\tilde\Psi_1},\ket*{\tilde\Psi_2}\}$, we can construct infinitely many NPCs between these two states. 

There is a particularly interesting subclass of  NPC which can be useful. An arc of a great circle passing through the initial and final states is dual to itself and referred as ``self-dual''.  This kind of curve on the Bloch sphere is given by (up to a unitary transformation)
\begin{equation}
	\label{self}
	\ket{\psi(s)}=\begin{pmatrix}
		\cos(\eta(s)/2)\\
		\sin(\eta(s)/2)\\
	\end{pmatrix}; \,\,s_1\le s \le s_2
\end{equation}
with $\eta(s_1)=0, \; \eta(s_2)=2\cos^{-1}(\alpha)$. A curve $\{\ket*{\tilde\Psi(s)}=\ket{\psi(s)}\otimes \ket{\psi(s)}\}$ is also an NPC, which we shall call a self dual curve.

\begin{figure}
	\centering
	\subfigure[]{
		\includegraphics[width=5cm]{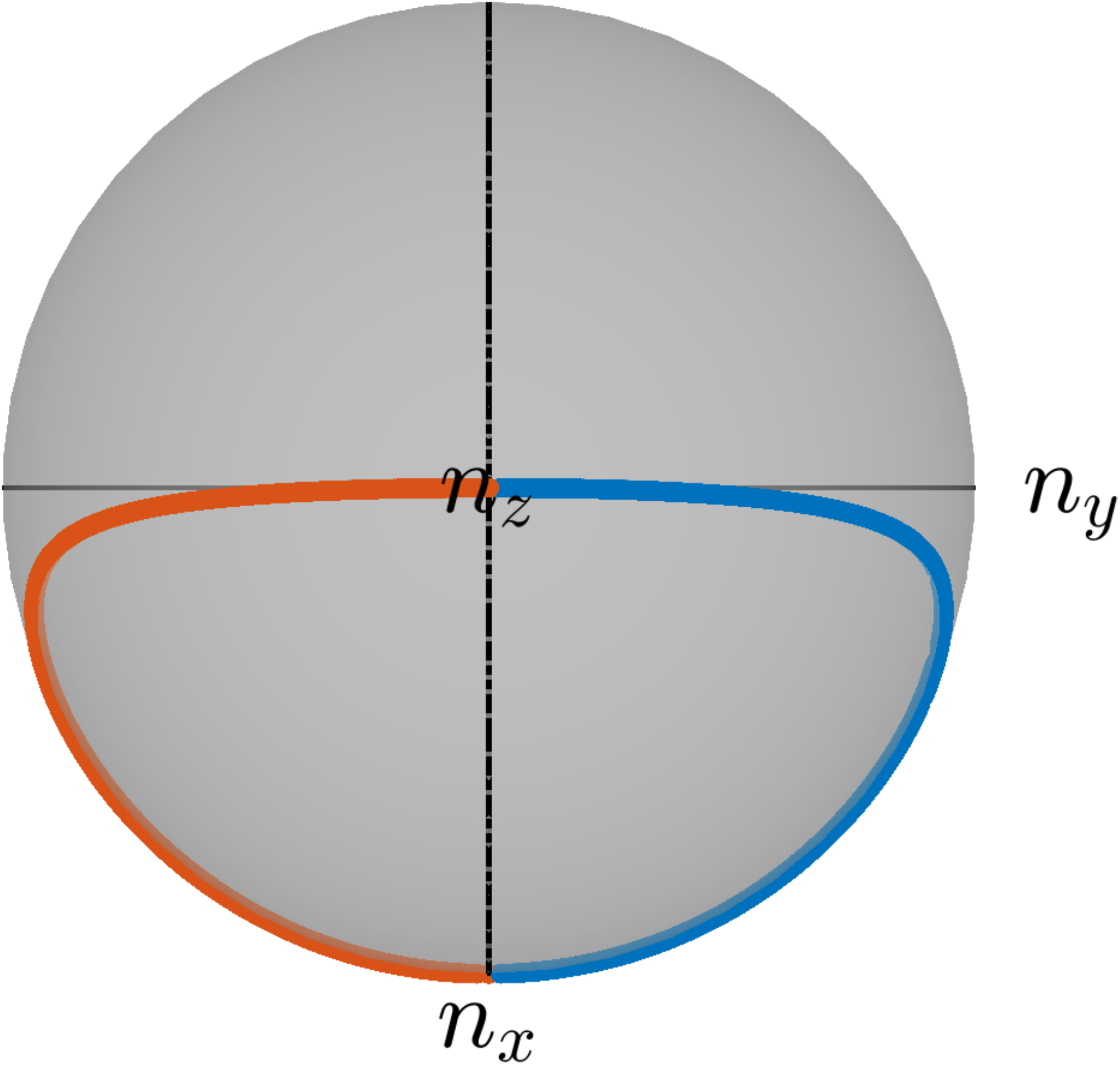}
		\label{fig:npc1}}
	\subfigure[]{
		\includegraphics[width=5cm]{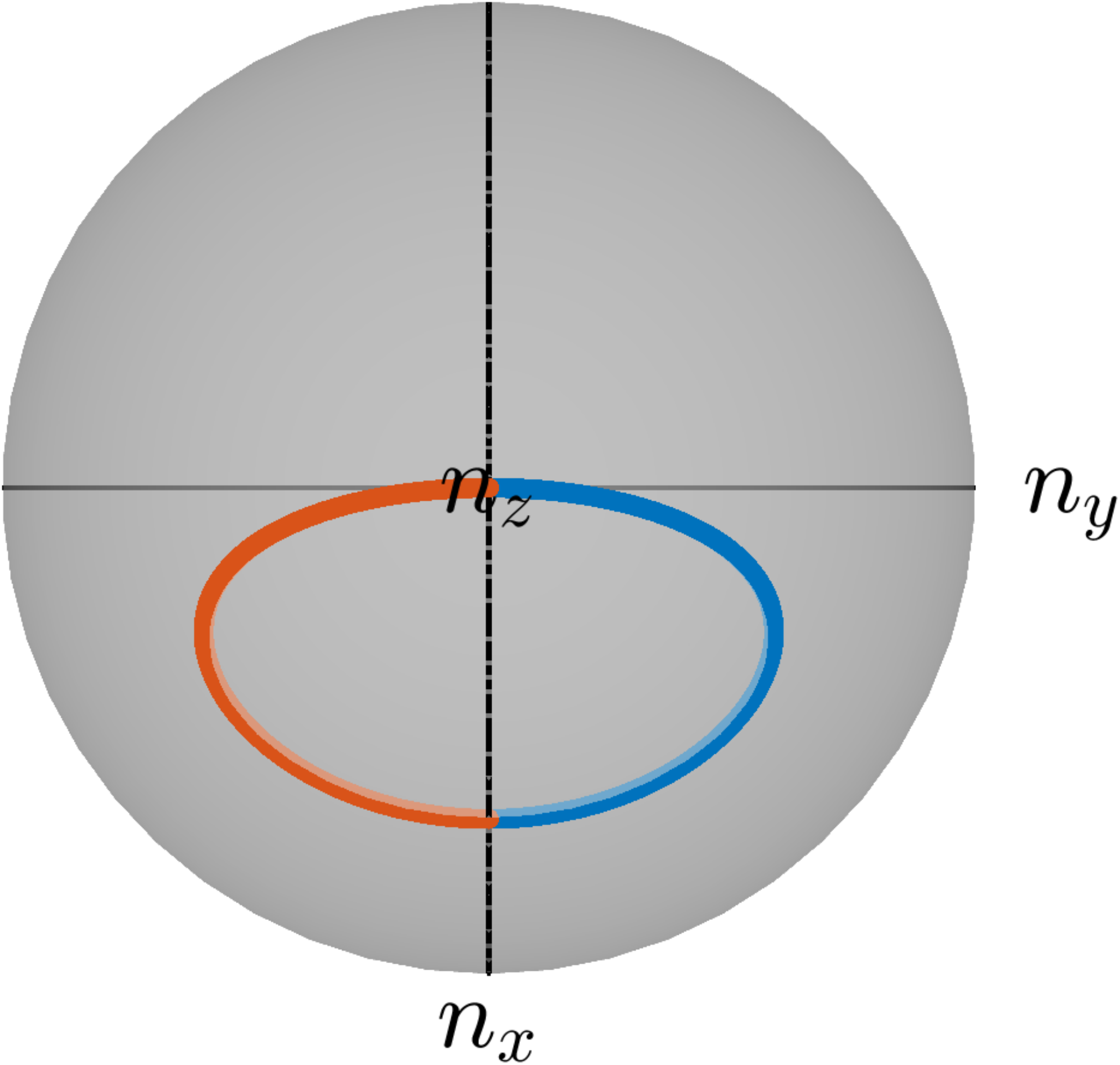}
		\label{fig:npc2}}
	
	\subfigure[]{
		\includegraphics[width=5cm]{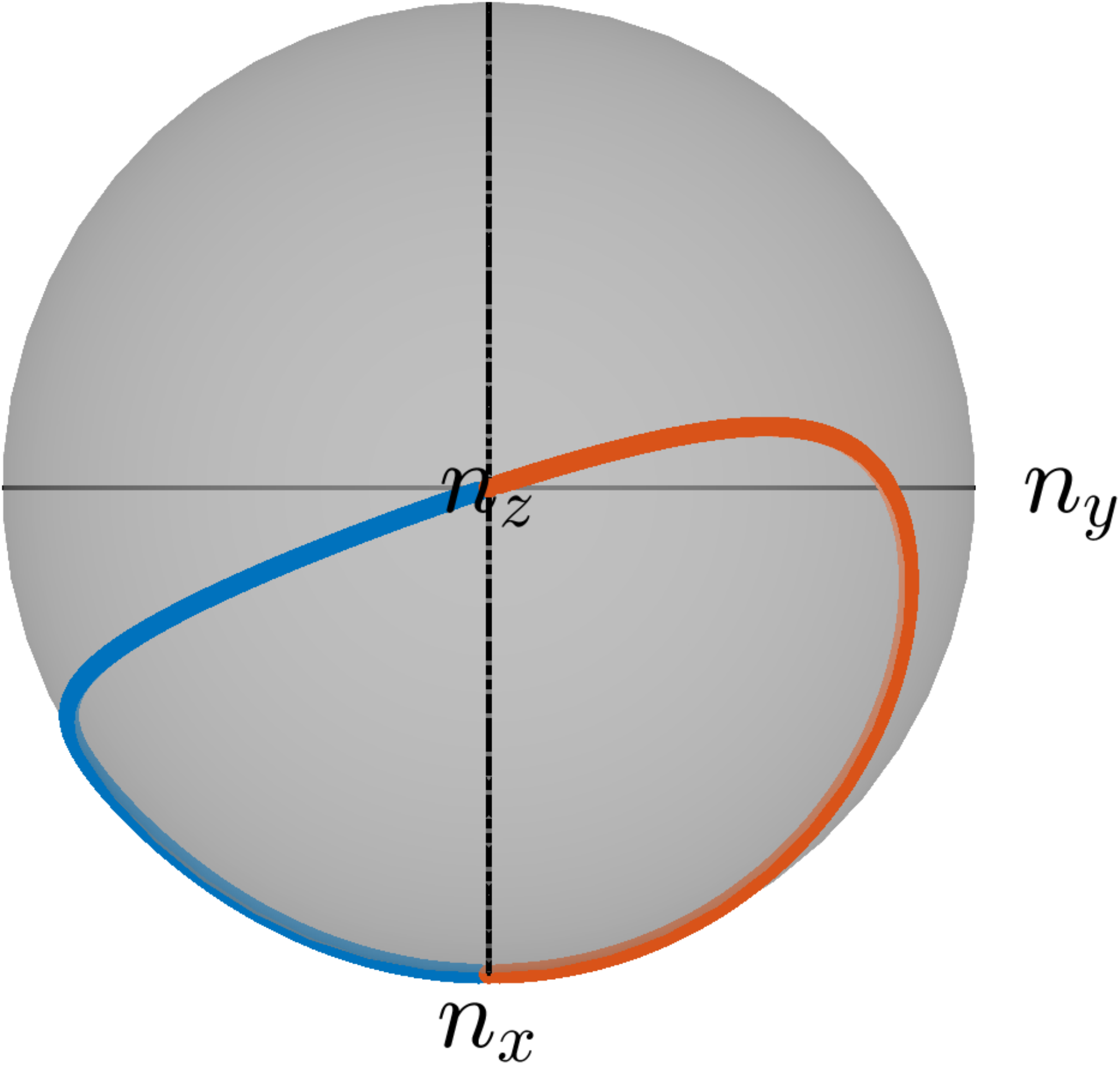}
		\label{fig:npc1b}}
	\subfigure[]{
		\includegraphics[width=5cm]{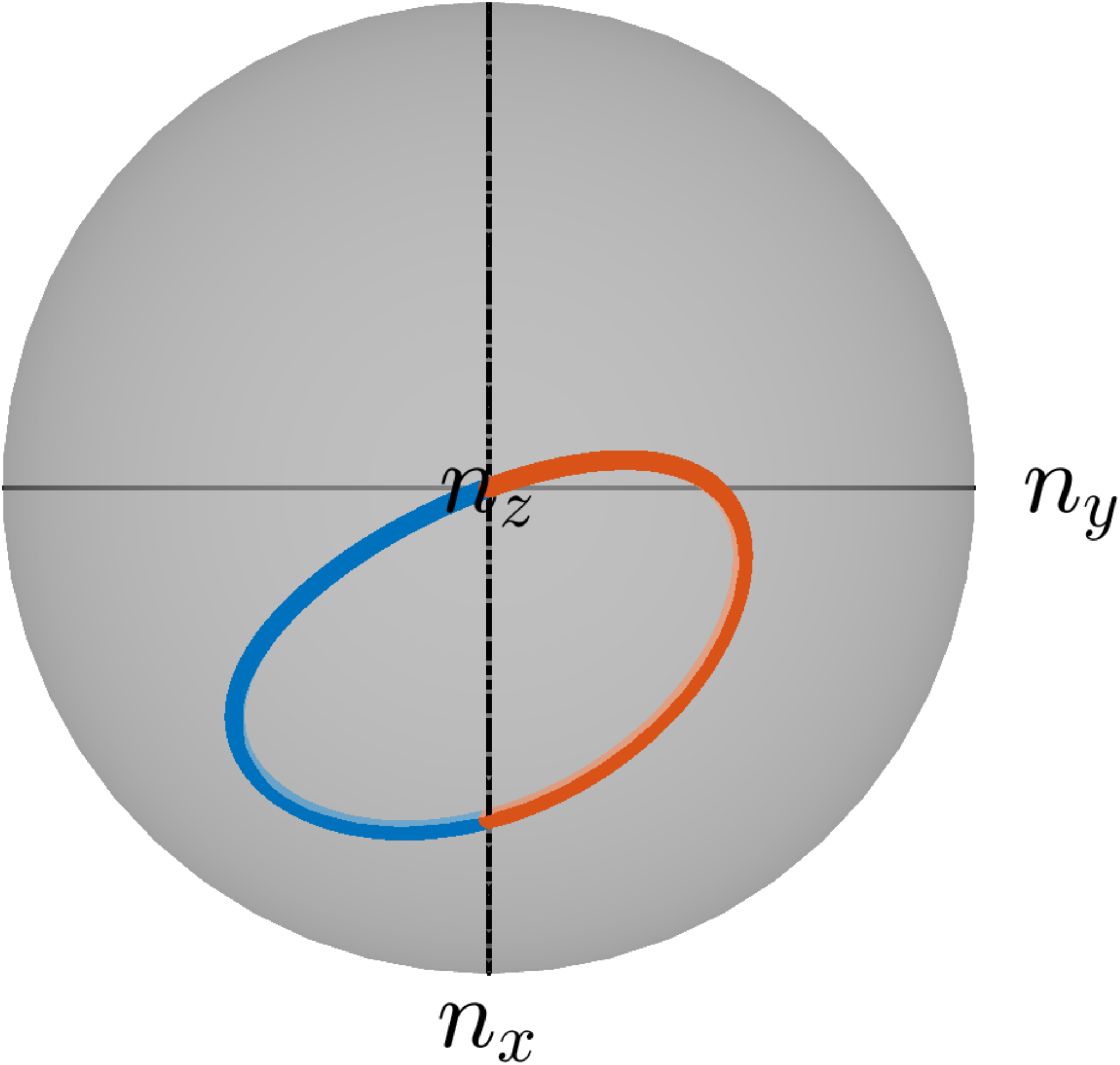}
		\label{fig:npc2b}}
	\caption{(Color online) Here we plot the Bloch geometric decomposition of NPCs in three-dimensional state space between the two states given in Eq.~\eqref{Eq:39}. In \subref{fig:npc1} $\theta = \pi/3$, \subref{fig:npc2} $\theta = \pi/6$ we plot the NPCs which are constructed by considering dual pairs of curves. Whereas, in \subref{fig:npc1b} and \subref{fig:npc2b} we plot the NPCs which are constructed by geometric decomposition of the curve in Eq.~\eqref{eq:nullphasecurve-II}. We have chosen $\chi = \pi/3$ and the same values of $\theta$ as in \subref{fig:npc1} and \subref{fig:npc2}.}
	\label{fig:npc}
\end{figure}

Although, we have presented the construction of NPCs for the end states which can be represented by degenerate MSs, same technique can be used to construct NPCs for arbitrary end states. As an example, let us construct an NPC connecting the two non-orthogonal states $\ket{\Psi_1} = (1 \; 0 \; 0)^T$ and $\ket{\Psi_2} = (\cos\theta \; \sin\theta \;\; 0)^T$ in three-dimensional state space. Since the state $\ket{\Psi_2}$ cannot be represented by degenerate MSs, we first apply a common unitary transformation $U$ to bring this to the state represented by degenerate MSs. The appropriate unitary transformation $U$ is of the form
\begin{align}
	U = \begin{pmatrix}
		1 & 0 & 0 \\
		0 & a & b \\
		0 & -b & a 
	\end{pmatrix},
\end{align}
where $a=\sqrt{2} \alpha \beta/\sin \theta $, $b= \beta^2/\sin \theta$, $\alpha^2 = \cos \theta$ and $\alpha^2 + \beta^2 = 1$. The states after applying $U$ reads

\begin{align}\label{Eq:39}
	\ket*{\Psi'_1} = \begin{pmatrix}
		1 \\
		0 \\
		0
	\end{pmatrix}, \;\;	\ket*{\Psi'_2} =  \begin{pmatrix}
		\cos \theta \\
		a \sin \theta \\
		-b\sin \theta
	\end{pmatrix}
\end{align}
which can further  be written as 
\begin{align*}
	\ket*{\tilde{\Psi}_1} &= \ket{0} \otimes \ket{0} \\
	\ket*{\tilde{\Psi}_2} &= \ket{\phi} \otimes \ket{\phi} \\
\end{align*}
where $\ket{0}$ and $ \ket{\phi} $ are the same as defined earlier. We now take a pair of dual curves given in Eq.~\eqref{NPC-curve1} and Eq.~\eqref{NPC-curve2} with appropriate boundary conditions on $\eta(s)$ and $\Gamma(s)$. Consequently, the state in three-dimensional state is written as 
\begin{align*}
	\ket*{\Psi'} = \mathcal{N} \begin{pmatrix}
		2 \cos^2(\eta(s)/2) \\
		\sqrt{2} \cos(\eta(s)/2) \sin(\eta(s)/2) \cos(\Gamma(s)) \\
		2 \sin^2(\eta(s)/2)
	\end{pmatrix}
\end{align*}  
Now we apply $U^{\dagger}$ on $ \ket*{\Psi'} $ to get the state $\ket{\Psi}$ back. With the appropriate choice of $\eta$ and $\Gamma$  we can bring it to the form which is already given in~\cite{Arvind2003} as an example of an NPC. If we choose the functions $\eta(s)$ and $\Gamma(s)$ with $0\le s \le \theta$ as
\begin{align}
	\eta(s)&=\cos^{-1}\Big[\frac{A(s)-C(s)}{A(s)+C(s)}\Big], \nonumber \\
	\Gamma(s)&=\tan^{-1}\Big[\frac{\sqrt{4 A(s) C(s) - B^2(s)}}{B(s)}\Big],
\end{align}
where
\begin{align}
	A(s) &= g(s) \cos s, \nonumber\\
	B(s) &= b\left(1 - g(s)^2\right)^{1/2}-a g(s) \sin s \nonumber\\
	C(s) &= 	a  \left(1 - g(s)^2\right)^{1/2} + b g(s) \sin s
\end{align}
with $0\le g(s)\le 1$, and $g(0)=g(\theta)=1$. This particular choice of the functions, results in an NPC given by
\begin{align} \label{eq:nullphasecurve-I}
	\ket{\Psi(s)} = \begin{pmatrix}
		g(s) \cos s  \\
		g(s) \sin s  \\
		\left(1 - g(s)^2\right)^{1/2}
	\end{pmatrix};\,\,\,0\le s \le \theta.
\end{align}
that is exactly the same derived in~\cite{Arvind2003} between the states $\ket{\Psi_1} = (1 \; 0 \; 0)^T$ and $\ket{\Psi_2} = (\cos\theta \; \sin\theta \;\; 0)^T$. In Figs.~\ref{fig:npc1},~\ref{fig:npc2} we plot the geometric construction of NPCs on the Bloch sphere where we have chosen  $g(s)=\cos [s(s-\theta)]$.

We have provided  methods to generate NPCs in three-dimensional state space using pairs of dual curves on the Bloch sphere. Our method of constructing NPCs can be easily extended to $n$-dimensional state space. For example, for odd $n$, we can construct $(n-1)/2$ number of pairs of dual curves to get an NPC. For the case when $n$ is even, we take $(n-2)/2$ number of pairs of dual curves and one curve  along the great circle connecting the end states, to construct the NPC. 

{\it Example of NPCs which do not come under this category:}     
Interestingly, there exist NPCs connecting the considered end states $\ket{\Psi_1}$ and $\ket{\Psi_2}$ which cannot be constructed either by a pair of dual curves or by self-dual curves. These NPCs are obtained from Eq.~\eqref{eq:nullphasecurve-I} by applying a unitary transformation which reads
\begin{align}
	V = \begin{pmatrix}
		1 & 0 & 0 \\
		0 & 1 & 0  \\
		0 & 0 & e^{i \chi} 
	\end{pmatrix}.
\end{align}
The BI is invariant under such unitary transformations and results in another NPC. The new NPC reads
\begin{equation} \label{eq:nullphasecurve-II}
	\ket{\Psi(s)} \mapsto V \ket{\Psi(s)} = \begin{pmatrix}
		g(s) \cos s  \\
		g(s)\sin s  \\
		e^{i\chi} \left(1 - g(s)^2)\right)^{1/2} 
	\end{pmatrix},
\end{equation}
where the parameter $0\le s \le \theta$, $\chi \in \mathbb{R}$ is a constant phase factor, $0\le g(s)\le 1$ and $g(0)=g(\theta)=1$. After a common unitary transformation, these NPCs decompose to a pair of curves which are not reflective about the great circle connecting the degenerate MSs on the Bloch sphere as shown in Fig.~\ref{fig:npc1b}, \ref{fig:npc2b}. Therefore, this kind of NPCs cannot be constructed by choosing $\eta(s)$ and $\Gamma(s)$.

In conclusion, we have developed a Bloch sphere geometric decomposition for geodesics and NPCs in higher-dimensional state space.  We have shown that geodesics connecting the two $(n-1)$-fold degenerate MS states in $n$-dimensional state space decompose to $n-1$ number of circular segments on the Bloch sphere. The circular segments, which are uniquely determined by the radius and the end states, can be completely characterized by the MSs corresponding to the end states and the inner product between them. Therefore, once the MSs for the end states is known, we can construct the geodesic between any given states in $n$-dimensional state space. 
We have also shown that NPCs can be constructed between any given end states by constructing arbitrary pairs of dual curves between the end states. 
A particularly interesting class of NPCs is the one where all the curves are dual of themselves, i.e., self dual curves. The geometric decomposition presented in this paper reveals intrinsic symmetries in the geodesics and NPCs and improve our understanding of the geometric structure of the higher-dimensional state space.

\chapter{Geometric Phase and non-inertial Effect}
The experimental detection of the non-inertial effects in quantum field theory, e.g., the Unruh effect~\cite{fulling1973,davies1975,unruh1976,crispino2008}, the Hawking radiation~\cite{HawkingG1974, Hawking1975}, has remained elusive till date because of the requirement of extreme conditions. For example, in the Unruh effect, to perceive a thermal effect of $1$~K, typically an acceleration of the order of $10^{21}$~{m/s$^2$} is required~\cite{crispino2008}. Numerous attempts have been made to relax extreme conditions using sophisticated techniques and precise measurements, but only to a limited success~\cite{Rogers1988,Tajima1999,fuentes2010,scully2003,Lochan2021,Vanzella2001,Barshay1978,Barshay1980,Kharzeev2006,bell1983,bell1987,unruh1998}.

It has been further advocated that quantum features such as geometric phase (GP) may be of much usage in bringing down the scales involved. Moreover, due to its accumulative and sensitive nature, the GP can be helpful in capturing weak effects such as the non-inertial effects in quantum field theory.

It has been proposed~\cite{Martinez2011} that using the GP the Unruh effect can be detected for accelerations as low as  $10^{17}$~{m/s$^2$}, which, though a significant reduction in the acceleration typically required, is still extremely challenging to achieve. In~\cite{HuYu2014}, the GP acquired by a circularly rotating detector in free space was studied, but the detector's non-inertial response remains too feeble to be detectable at physically realizable accelerations. 

Interestingly, the usage of an electromagnetic cavity has been argued to further relax the acceleration requirement by a few orders~\cite{scully2003,lochan2020,Lochan2021}.
For example, by studying the spontaneous emission rate of an atom circularly rotating inside an electromagnetic cavity, it has been shown that the non-inertial effects can be detected at accelerations as low as $10^{12}$~{m/s$^2$}~\cite{lochan2020}. 

In this work, we study the GP acquired by a circularly rotating two-level atom inside an electromagnetic cavity~\cite{AryaMittal2022}. Since the GP is sensitive to transition rates and we already know that the transition rates become significantly modified inside the cavity~\cite{Purcell1995}, we expect the non-inertial component of the GP to be correspondingly modified. The accumulative nature of the GP~\cite{Berry1984,Tong2004} facilitates the detection of weak effects such as the non-inertial modifications to the field correlators, whereas using the electromagnetic cavity the atom's non-inertial response can be isolated from, or strengthened relative to, the inertial response~\cite{lochan2020}.

By studying the GP response of the atom inside an electromagnetic cavity, we show that the acceleration-induced modifications to the field correlators can be detected at much lower accelerations and with much relaxed parameters for the experimental setup. Specifically, we show that the atom acquires an experimentally detectable GP at accelerations of the order of $10^{7}$~{m/s$^2$} which is experimentally feasible. Therefore, an efficient GP measurement inside a cavity is supposed to manifest the non-inertial quantum field theoretic effects more robustly as compared to any other setup proposed so far.
\section{Open quantum system description of the rotating atom and the Geometric phase}
The state of a closed quantum system is represented, up to an overall phase, by a state vector $\ket{\psi} \in \mathcal{H}$, where $\mathcal{H}$ is the Hilbert space of the system and $\ket{\psi}$ is said to be a pure state of the system. The evolution of a closed system is governed by Schroedinger equation~\cite{Sakurai1985}.
In general, a system is found to be interacting with an external environment and pure states are not adequate for the description of such systems. The most general state of such a composite quantum system, called a mixed state, is represented by a density operator $\rho$ such that
\begin{equation}
	\rho = \rho^{\dagger}, \;\;\; \Tr \rho = 1, \;\;\; \expval{\rho}{\phi} > 0 \;\;\; \forall \; \ket{\phi} \in \mathcal{H}_{\text{SE}},
\end{equation} where $\mathcal{H}_{\text{SE}}$ is the Hilbert space of the composite system. This will be helpful in the upcoming calculations where we consider an open system, state of which is described by a density operator.

We start by considering a two-level atom interacting with an electromagnetic field. The atomic ground and excited states are denoted, respectively, by $\ket{g}$ and $\ket{e}$. The proper frequency gap between the two atomic levels is $\Omega_0$ and the atom carries an electric dipole moment four-vector $
\hat{d}'^{\mu}$. Throughout this paper, primed quantities refer to the atom's co-moving frame. In the interaction picture, the dipole moment operator $\hat{\vb{d}}'(\tau)$ is given in terms of its matrix elements (as they will appear in later expressions) by
\begin{equation}\label{eq:dipole_operator}
	\hat{\vb{d}}'(\tau) = \vb{d}' \sigma_{-} \exp(-i \Omega_0 \tau) + \vb{d}'^* \sigma_{+} \exp(i \Omega_0 \tau), 
\end{equation} 
where $\vb{d}' \equiv \bra{g} \hat{\vb{d}}'(\tau = 0) \ket{e}$ and $\sigma_{+} = \sigma^{\dagger}_{-} = \dyad{e}{g}$ is the step-up operator for the atomic states. For simplicity, we assume that $\vb{d}' = (0, d', 0)$.  The electromagnetic field is assumed to be in vacuum state $\ket{0}$. The interaction Hamiltonian between the atom and the electromagnetic field is given by $\hat{H}_I = - \hat{d}^{\mu}E_{\mu}$~\cite{Anandan2000}, where $E_{\mu} \equiv F_{\mu \nu} u^{\nu}$, $F_{\mu\nu}$ is the electromagnetic field strength tensor and $u^{\nu}$ is the four-velocity of the atom. The interaction Hamiltonian takes the form $\hat{H}_I = - \vb{\hat{d}'}\cdot \vb{E'}$ in the rest frame of the atom, where $\vb{E}'$ is the electric field 3-vector as seen by the atom. 
The electric field operator inside a quantization volume $V$ is given by~\cite{Gerry2004}
\begin{equation}
	\vb{E}[x(\tau)] = i \sum_{\vb{k},\lambda} \mathcal{E}_k \epsilon_{\vb{k},\lambda} \left(a_{\vb{k},\lambda} e^{-i\left(\omega_k t(\tau) - \vb{k}.\vb{x}(\tau)\right)}- \text{h.c.}\right),
\end{equation}
where $\mathcal{E}_k \equiv \sqrt{\hbar \omega_k/(2\epsilon_0 V)}$, $\epsilon_{\vb{k},\lambda}$ with $\lambda = 1,2$ are the two orthogonal polarization vectors, and h.c. denotes Hermitian conjugate.

In the rest frame of the atom, the evolution of the atom-field composite system is governed by \cite{Breuer2007} 
\begin{equation}\label{total_time_evolution}
	\derivative{\rho_{\text{AF}}(\tau)}{\tau} = - \frac{i}{\hbar} \comm{H}{\rho_{\text{AF}}(\tau)},
\end{equation} 
where $\rho_{\text{AF}}$ is the density operator of the composite system and $\tau$ is the proper time of the atom. From (\ref{total_time_evolution}), we can obtain the Lindblad evolution of the reduced density operator for the atom $\rho(\tau) \equiv \Tr_{F}{\left(\rho_{\text{AF}}\right)}$, which is given by \cite{Lindblad1976}
\begin{equation}\label{lindblad_eqn}
	\derivative{\rho(\tau)}{\tau} = -\frac{i}{\hbar} \comm{H_{\text{eff}}}{\rho(\tau)} + \mathcal{L}[\rho(\tau)],
\end{equation} 
where 
\begin{equation} \label{eq:lindbladoperator}
	\mathcal{L}[\rho] = \frac{1}{2} \sum_{i,j = 1}^{3} a_{ij} \left(2 \sigma_j \rho \sigma_i - \sigma_i \sigma_j \rho - \rho \sigma_i \sigma_j \right),
\end{equation} 
captures the dissipation and decoherence of the atom induced by its interaction with the electromagnetic field. The $H_{\text{eff}}$ represents the Hamiltonian of the two-level atom with the renormalized atomic level spacing, $ \Omega $, which consists of the Lamb shift~\cite{Breuer2007}. The $\sigma_i$'s are the standard Pauli matrices~\cite{Sakurai2017} and the coefficients $a_{ij}$ are given by \cite{Benatti2005} 
\begin{equation} \label{eq:Kosskowskimatrix}
	a_{ij} = A \delta_{ij} - i B \epsilon_{ijk} \delta_{k3} - A \delta_{i3} \delta_{j3},
\end{equation} 
with $A$ and $B$ defined as 
\begin{align}\label{eq:AB}
	A &= \frac{1}{4} \left[\Gamma_{\downarrow} + \Gamma_{\uparrow}\right], \;\;\; B = \frac{1}{4} \left[\Gamma_{\downarrow} - \Gamma_{\uparrow}\right],
\end{align} 
whereas,
\begin{equation} \label{FTCF}
	\begin{split}
		\Gamma_{\downarrow \uparrow} &= \abs{\bra{\psi_f} \hat{\vb{d}}'(0) \ket{\psi_i}}^2 \int_{-\infty}^{\infty} \dd{\tau_-} e^{\pm i \Omega_0 \tau_-} G'^{+}(x(\tau_-)),
	\end{split}
\end{equation} 
with `$+$' and `$-$' corresponding to $\Gamma_{\downarrow}$ and $\Gamma_{\uparrow}$, i.e., the emission and the absorption rates, respectively. Here, $\tau_- \equiv \tau_2 - \tau_1$; $\ket{\psi_i}, \ket{\psi_f} \in \{\ket{g}, \ket{e} \}$; and $G'^+(x(\tau_-)) \equiv \bra{0}E'^{y}(\tau_1)E'^{y}(\tau_2)\ket{0}$ is the two-point vacuum Wightman function. 
By taking the initial state of the atom to be
\begin{equation}\label{atomic_initial_state}
	\ket{\psi(0)} = \cos(\theta/2) \ket{e} + \sin(\theta/2) \ket{g},
\end{equation} 
and solving Eq.~\eqref{lindblad_eqn}, we get the reduced density operator to be

\begin{equation} \label{eq:reduceddensity}
	\rho(\tau) = \begin{pmatrix}
		e^{-4A\tau}\cos^2(\theta/2) + \frac{B-A}{2A}\left(e^{-4A\tau}-1\right) & \frac{1}{2} e^{-2A\tau-i\Omega \tau}\sin\theta \\ \noalign{\vskip7pt}
		\frac{1}{2} e^{-2A\tau+i\Omega \tau}\sin\theta & 1 - e^{-4A\tau}\cos^2(\theta/2) - \frac{B-A}{2A}\left(e^{-4A\tau}-1\right)\\
	\end{pmatrix},
\end{equation}
in which the effect of the environment is contained in $A$ and $B$.  The GP for an $N$-level quantum system in a mixed state and evolving non-unitarily is given by \cite{Tong2004} 
\begin{align}
	\gamma_g
	=\arg \left( \sum_{k=1}^{N} \sqrt{p_k(0)p_k(T)} \ip*{\phi_k(0)}{\phi_k(T)} e^{-\int_0^{T} \ip*{\phi_k(\tau)}{\dot{\phi}_k(\tau)} d\tau } \right) \label{eq:tong},
\end{align}
where $p_k(\tau)$ and $\ket{\phi_k(\tau)}$ are instantaneous eigenvalues and eigenvectors of the system's density operator $\rho(\tau)$. The eigenvalues of $\rho(\tau)$ are
\begin{equation} \label{eq:eigenvalue}
	p_{\pm}(\tau) = \dfrac{1}{2} \left( 1 \pm \lambda\right),
\end{equation}
where $\lambda = \sqrt{r_3^2 + e^{-4A}\sin^2\theta}$ and $r_3 = e^{-4A\tau}\cos\theta + \tfrac{B}{A}(e^{-4A\tau} -1)$. Since $p_-(0) = 0$, the only eigenvector that contributes to $\gamma_g$  is the one corresponding to $p_+$ which reads
\begin{equation} \label{eq:eigenvector}
	\ket{\phi_+(\tau)} = \sin (\theta_{\tau}/2) \ket{+} + e^{i \Omega_0 \tau} \cos (\theta_{\tau}/2) \ket{-},
\end{equation}
with
\begin{equation}
	\tan (\theta_{\tau}/2) = \sqrt{\dfrac{\lambda + r_3}{\lambda - r_3}}.
\end{equation}
By substituting Eq.~\eqref{eq:eigenvalue} and Eq.~\eqref{eq:eigenvector} in the expression of the GP in Eq.~\eqref{eq:tong}, we get \cite{HuYu2012}
\begin{align} 
	\gamma_g = - \dfrac{\Omega}{2} \int_{0}^{T} \dd{\tau} \left(1 - \dfrac{\mathcal{R} - \mathcal{R} e^{4 A \tau} + \cos \theta}{\sqrt{e^{4 A \tau} \sin^2 \theta + (\mathcal{R} - \mathcal{R} e^{4 A \tau} + \cos \theta)^2}}\right) \label{eq:geophase},
\end{align}
where $\mathcal{R} \equiv B/A$. The above expression is valid for all time $T$. Now, if we take $T = 2 \pi n/\Omega_0$, where $n$ is the number of quasi-cycles, and if $A/\Omega_0 \ll 1$, we can further simplify the expression for $\gamma_g$ and obtain~\cite{HuYu2012}
\begin{equation} \label{eq:HuYuGamma}
	\gamma_g = -\pi n (1 - \cos \theta) - \dfrac{2 \pi^2 n^2}{\Omega_0} \left( 2 B + A \cos \theta \right) \sin^2 \theta,
\end{equation} 
where $n$ is restricted by the demand $\pi n A/\Omega_0 \ll 1$, because to obtain \eqref{eq:HuYuGamma} from \eqref{eq:geophase}, we need $4A\phi/\Omega_0 \ll 1$ and for $n$ cycles $\phi = 2\pi n$. The GP has two contributions: the first term in Eq.~\eqref{eq:HuYuGamma} is due to the unitary evolution of the atom and the second term is coming from the non-unitary evolution of the atom which results from the atom's interaction with the environment. We will shortly see that in case of non-inertial motion of the atom, the non-unitary contribution can be further separated into an inertial and a non-inertial part. Further, using the cavity's resonance structure, the non-inertial contribution to the GP can be significantly enhanced.

\section{GP response of the circularly rotating detector}\label{Sec:GP response of the rotating detector}
In this section, we study the GP acquired by an atom moving on a circular trajectory of radius $R$ and angular frequency $\omega$, inside an electromagnetic cavity. We have seen in Eq.~\eqref{eq:HuYuGamma} that the GP depends directly on the quantities $A$ and $B$ which, in turn, depend on the atomic transition rates. The non-inertial motion of the atom only affects the transition rates and leaves the form of Lindblad master equation [Eq.~\eqref{lindblad_eqn}] unchanged. Therefore, the expression of GP in Eq.~\eqref{eq:geophase} can be used in the case of a rotating atom with the modified transition rates. It is natural to start, therefore, by computing the transition rates and then obtain the GP. 

We study the GP in two different regimes distinguished by the relative magnitudes of the rotational frequency of the atom $(\omega)$ and the atomic frequency gap $(\Omega_0)$. Specifically, we study the $\omega >> \bar{\Omega}_0$ and $\omega << \bar{\Omega}_0$ regimes, where $\bar{\Omega}_0 \equiv \Omega_0 \sqrt{1-\omega^2R^2/c^2}$.
As we shall see, in the $\omega >> \bar{\Omega}_0$ regime, the non-inertial contribution in the spontaneous decay rate dominates over the inertial one when the normal frequency of the cavity is tuned at $\omega+\bar{\Omega}_0$. In the $\omega << \bar{\Omega}_0$ regime, on the other hand, the inertial contribution to the GP overshadows the non-inertial component. However, with a suitable choice of parameters, the non-inertial contribution can be made comparable to the inertial component. Therefore, in both regimes, we can analyze the non-inertial contribution to the GP effectively.

\subsection{Transition rates in the atom's frame}

We get the transition rates in the lab frame for small $\zeta(\omega) \equiv \omega^2 R^2/c^2$, and to the first order in $\zeta(\omega)$ as~\cite{AryaMittal2022}
\begin{equation}\label{transition_rates}
	\begin{split}
		\Gamma^{\text{lab}}_{\downarrow \uparrow} = &\eta \int_0^{\infty} \dd{k}\rho(k)  \omega_k \Bigg[\delta(\pm \bar{\Omega}_0 - \omega_k) +\frac{R^2 \omega^2}{2c^2}  \frac{ 1}{2} \left[\delta(\pm \bar{\Omega}_0  - \omega_k + \omega) + \delta(\pm \bar{\Omega}_0  - \omega_k - \omega)\right] \\  &\qquad\qquad- \frac{\omega_k^2 R^2}{c^2} \frac{2}{5} \left\{ \delta(\pm \bar{\Omega}_0 - \omega_k)  - \frac{1}{2}  \left(\delta(\pm \bar{\Omega}_0 + \omega -\omega_k) + \delta(\pm \bar{\Omega}_0 - \omega -\omega_k)\right) \right\} \Bigg],
	\end{split}
\end{equation}
where $\eta \equiv \abs{\vb{d}}^2/(3 \pi \hbar \epsilon_0 V)$; `$+$' and `$-$' correspond, respectively, to the spontaneous decay rate $(\Gamma_{\downarrow})$ and the excitation rate $(\Gamma_{\uparrow})$ with $\rho(k)$ being the density of the field states. The transition rates in the co-moving frame can be obtained as $ \Gamma_{\downarrow \uparrow} = \gamma \Gamma^{\text{lab}}_{\downarrow \uparrow}$. Inside a cavity, the density of states can be taken to be of Lorentzian form~\cite{Scully_Zubairy1997,Lewenstein1993}
\begin{align}
	\rho(\omega_k) \sim \frac{\left(\omega_c/Q\right)}{\left(\omega_c/Q\right)^2 + \left(\omega_k - \omega_c\right)^2},
\end{align} 
where $\omega_c$ is the normal frequency and $Q$ is the quality factor of the cavity. 
The expression for the transition rates takes a simpler form in the two regimes, namely $\omega \gg \bar{\Omega}_0$ and $\omega \ll \bar{\Omega}_0$. Now, we shall explore the transition rates and the GP response of the atom in these two regimes.

\begin{figure*}
	\centering
	\subfigure[]{
		\includegraphics[height=4.3cm]{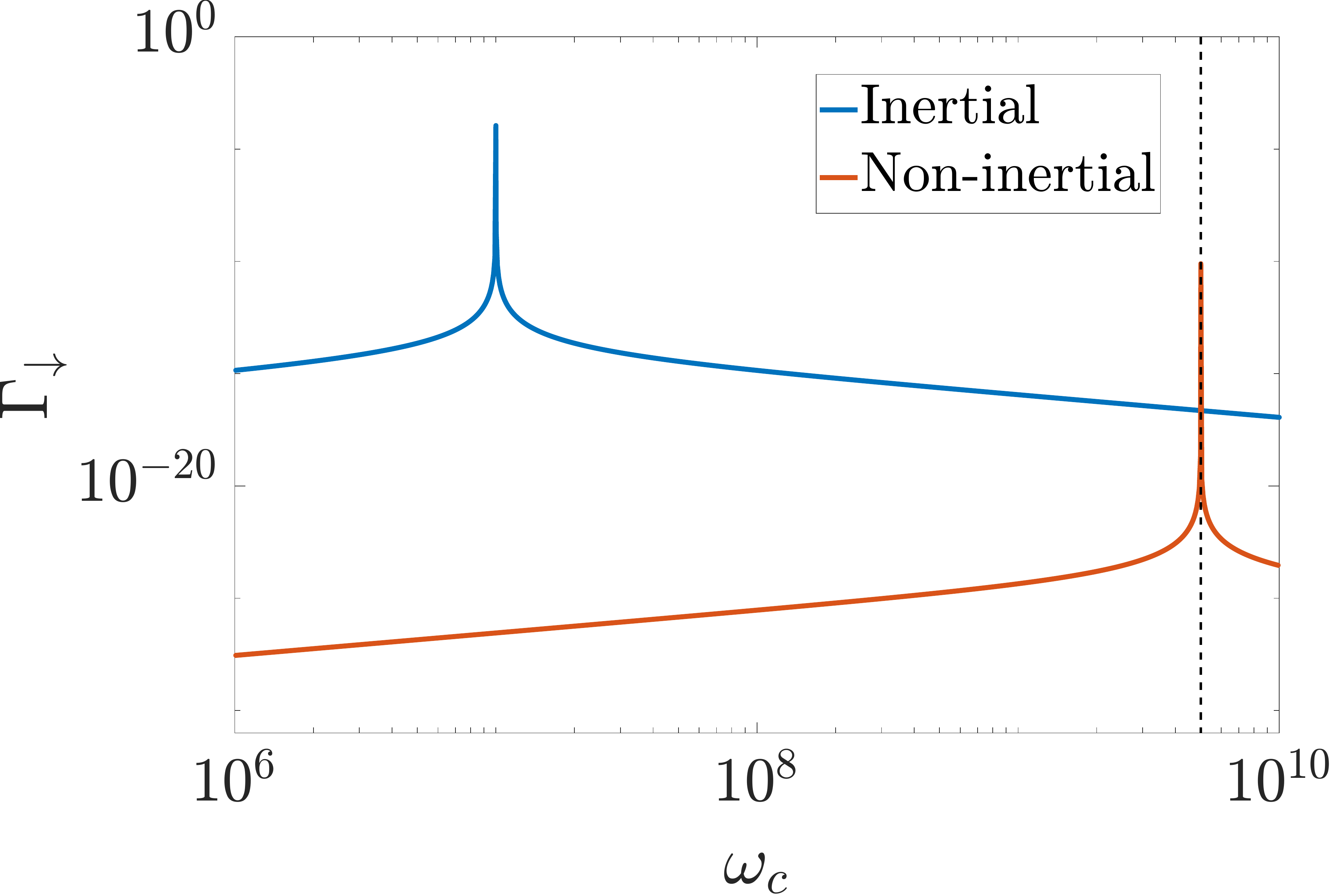}
		\label{fig:decayrate1}}
	\subfigure[]{
		\includegraphics[width=6.5cm]{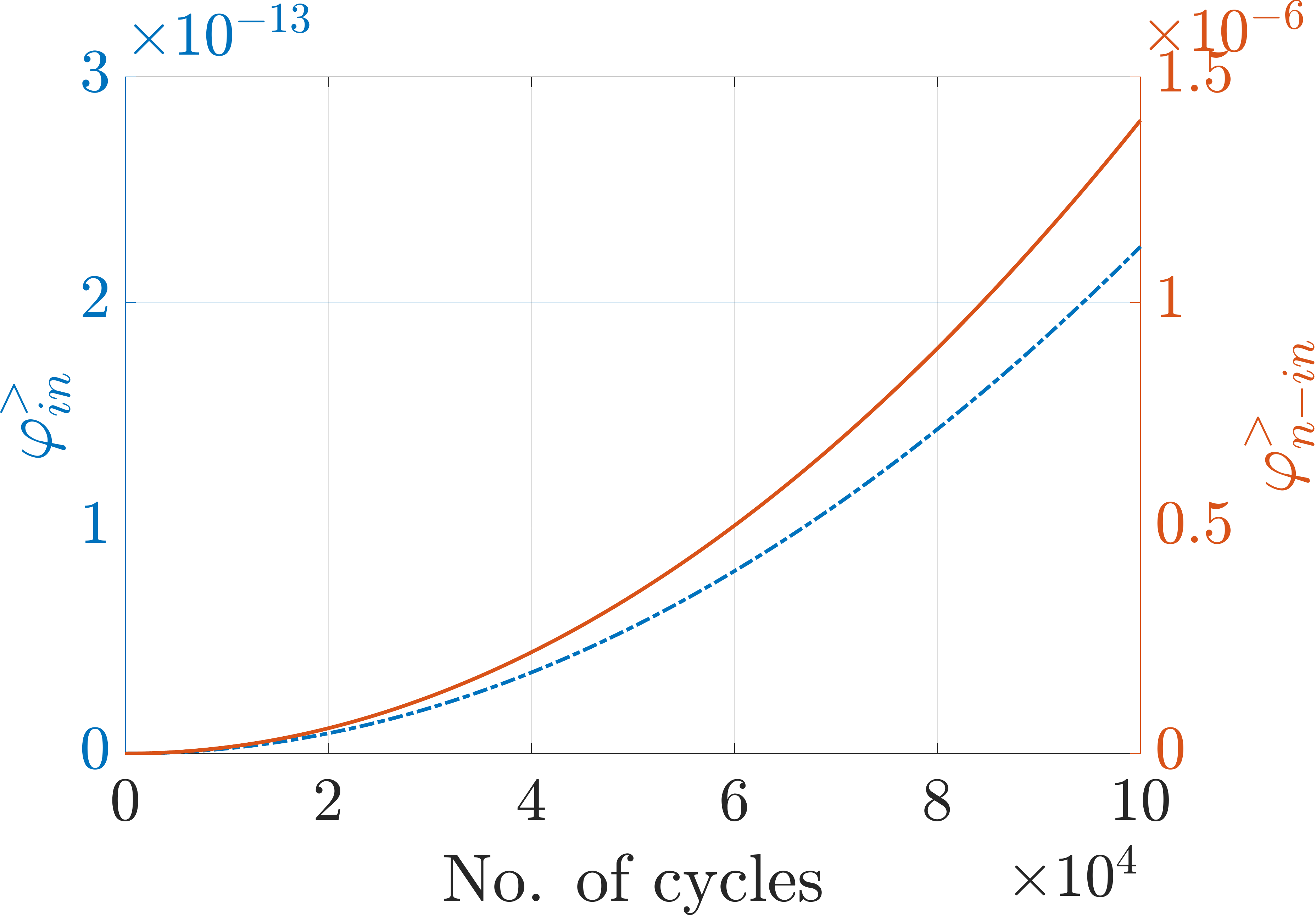}
		\label{fig:gp1}}
	\subfigure[]{
		\includegraphics[height=4.3cm]{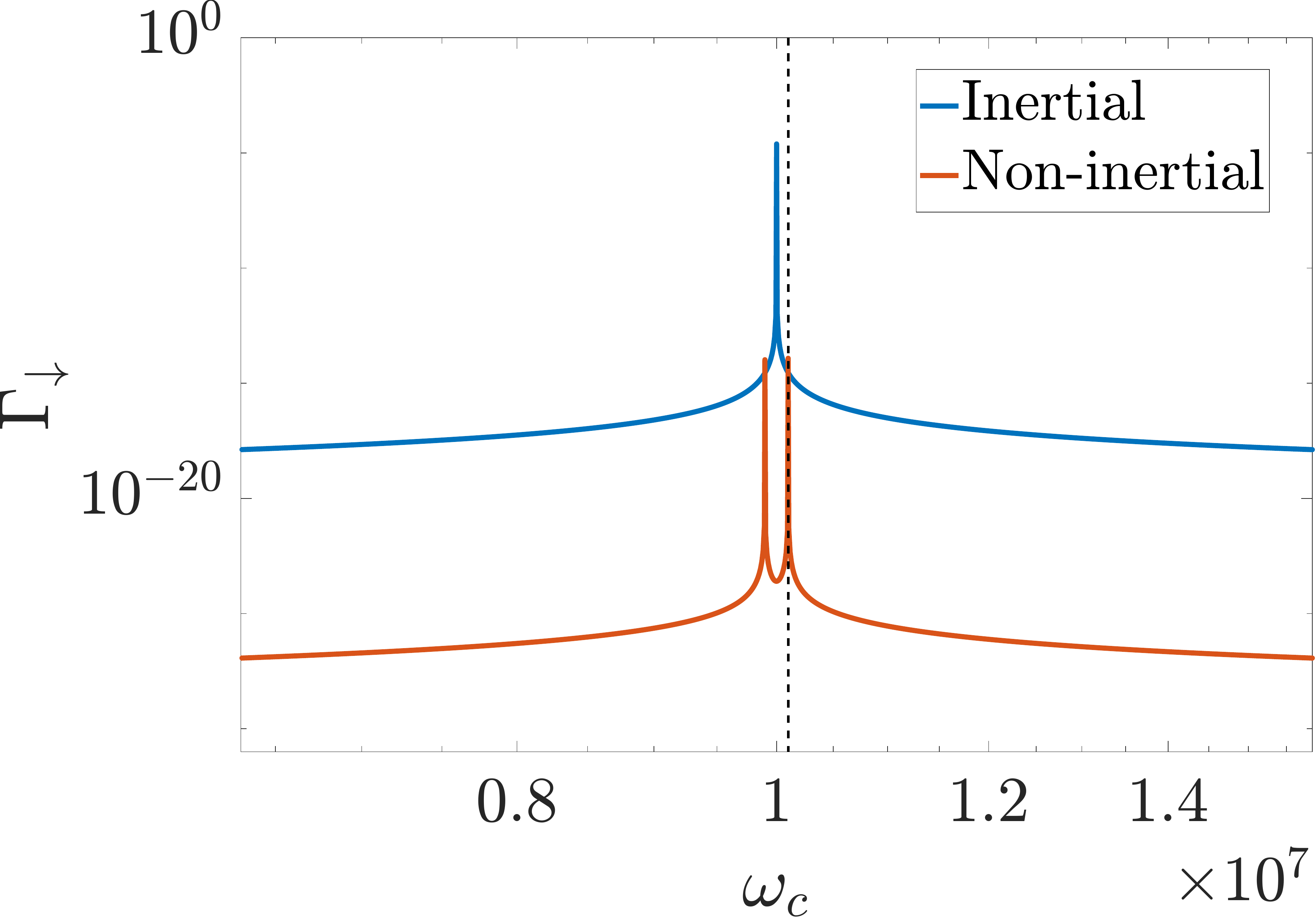}
		\label{fig:decayrate2}} 
	\subfigure[]{
		\includegraphics[width=6.5cm]{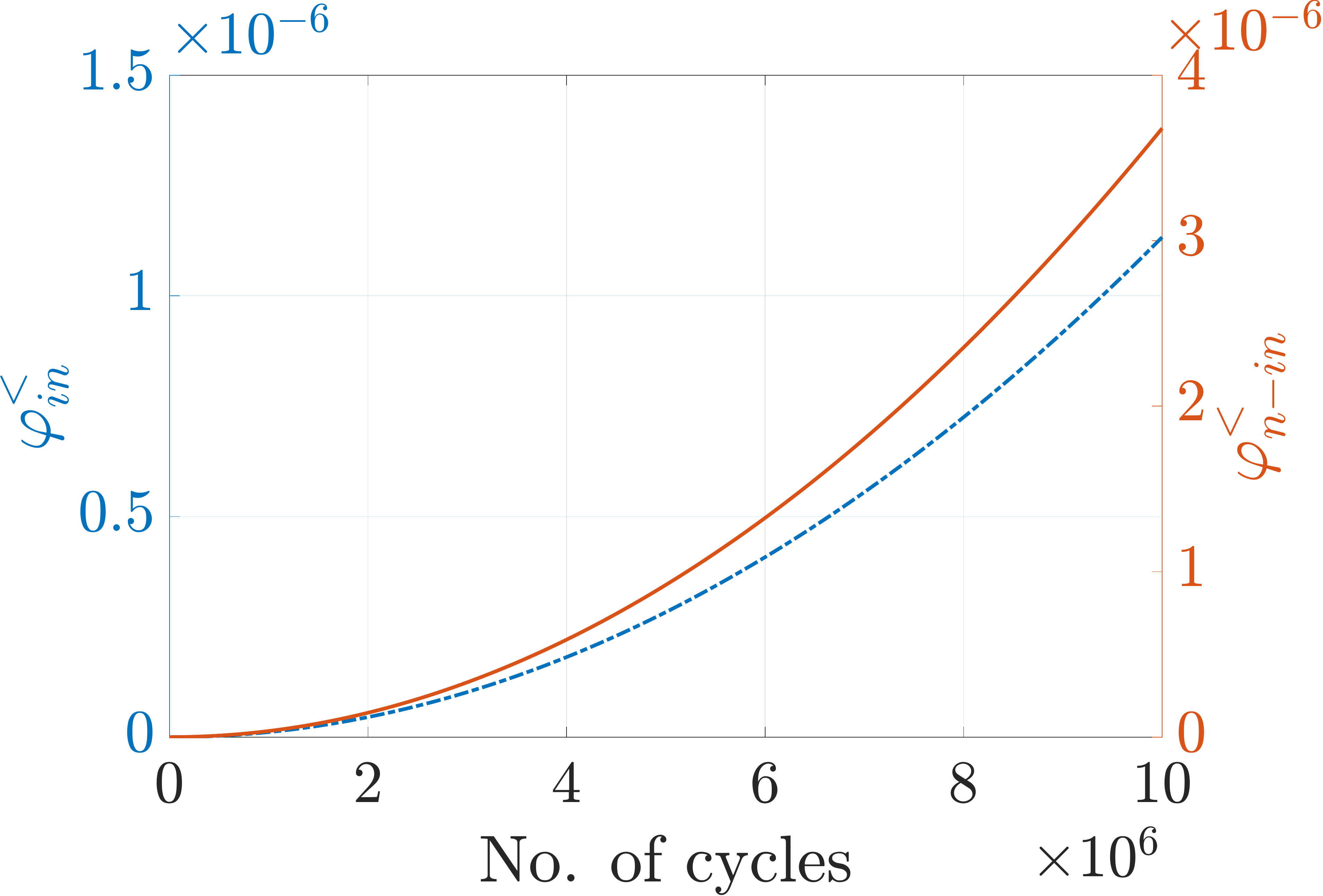}
		\label{fig:gp2}}
	\caption{(Color online) The plot for $\Gamma_{\downarrow}$ vs. $\omega_c$ and the non-unitary GP versus the number of quasi-cycles ($n$) in the two regimes discussed in the paper. We plot the inertial ($\varphi_{in}$) and the non-inertial contributions ($\varphi_{n-in}$) to the non-unitary GP here. \subref{fig:decayrate1},\subref{fig:gp1} $\omega >> \bar{\Omega}_0$ with $\omega = 5$ GHz, $\Omega_0 = $ 10 MHz, $V = 10^{-7}$~{m$^3$}, and $R = 10^{-6}$~m which correspond to an average acceleration $a = \omega^2 R \sim 2.5 \times 10^{13}$~{m/s$^2$}. For this set of parameters, $\pi n A/\Omega_0 \sim 10^{-16} n$. \subref{fig:decayrate2},\subref{fig:gp2} $\omega << \bar{\Omega}_0$ with $\omega = $ 0.1 MHz, $\Omega_0 = $ 10 MHz, $V = 10^{-3}$~{m$^3$}, and $R = 10^{-3}$~m which correspond to an average acceleration $a = \omega^2 R \sim 10^{7}$~{m/s$^2$}.The plots are for $\theta = \pi/2$ in Eq.~\eqref{atomic_initial_state}. The vertical black dashed lines in Figs.~\subref{fig:decayrate1},~\subref{fig:decayrate2} mark the normal frequencies at which the cavity is tuned to obtain the GP versus $n$ plots in Figs.~\subref{fig:gp1},\subref{fig:gp2}.}
	\label{fig:case2}
\end{figure*}

\subsubsection*{Case 1: $\omega \gg \bar{\Omega}_0$}
From Eq.~\eqref{eq:AB}, we know that $A$ is the sum of the spontaneous emission and absorption rates and, therefore, up to the first order in $\zeta(\omega)$, using Eq.~\eqref{transition_rates} is given by
\begin{equation}\label{eq:Acase1}
	A = \frac{\eta}{4}  \Bigg[ \rho(\Omega_0) \Omega_0 - \frac{\zeta(\omega)}{2} \Omega_0^2 \rho'(\Omega_0) + \frac{9}{20} \zeta(\omega) \left\{ \omega_+ \rho(\omega_+) +  \rho(\omega_-) \omega_- \right\}\Bigg].
\end{equation}
where $\rho'(\omega) = \partial \rho / \partial \omega$ and $\omega_{\pm} \equiv \omega \pm \bar{\Omega}_0$.Similarly, $B$ is obtained by replacing the term in curly brackets in Eq.~\eqref{eq:Acase1} by $\omega_+ \rho(\omega_+) -  \rho(\omega_-) \omega_-$. Using \eqref{eq:HuYuGamma}, we obtain the non-unitary GP acquired by the atom in $n$ number of quasi-cycles to be
\begin{multline} \label{eq:Gamma_g_high}
	\varphi^{>} = - \dfrac{2 \pi^2 n^2}{\Omega_0} \frac{\eta}{4} \Bigg[ \left(\rho(\Omega_0) \Omega_0 - \frac{\zeta(\omega)}{2} \Omega_0^2 \rho'(\Omega_0)\right) (2 + \cos\theta) \\ +\frac{9}{20} \zeta(\omega) \Bigg\{ \omega_+ \rho(\omega_+)  (2 + \cos\theta) - \omega_- \rho(\omega_-) (2 - \cos\theta) \Bigg\} \Bigg] \sin^2 \theta,
\end{multline}
where a quasi-cycle consists of a time period of $T = 2\pi/\Omega_0$.
Note that we are using the symbol $\varphi^{>}$ to denote the non-unitary GP in the $\omega \gg \bar{\Omega}_0$ regime. Next, we discuss some numerical estimates of GP for realistic settings~\cite{Ahn2018,Leduc2010,lochan2020}.

When we use $A$ and $B$ obtained by tuning the cavity to $\omega_+$, the non-unitary GP [see Eq.~\eqref{eq:Gamma_g_high}] acquired by the atom is predominantly non-inertial [see Fig.~\ref{fig:gp1}]. In Fig.~\ref{fig:gp1} we plot the inertial and the non-inertial contributions to the GP acquired by the atom as a function of the number of quasi-cycles for $V = 10^{-7}$~{m$^3$} and $R = 10^{-6}$~m. With these values for the parameters, $\pi A n/\Omega_0 \sim 10^{-16}n$ which decides the allowed number of quasi-cycles consistent with the approximation made to obtain Eq.~\eqref{eq:HuYuGamma}. We note from Fig.~\ref{fig:gp1} that the system acquires an experimentally observable~\cite{Wang2018} non-inertial GP $\sim 10^{-6}$~radian (1 second radian) in roughly $10^5$ quasi-cycles, whereas the inertial contribution to the GP is $\sim 10^{-13}$~radian in the same number of quasi-cycles. 

\subsubsection*{Case 2: $\omega \ll \bar{\Omega}_0$}
If $\omega \ll \bar{\Omega}_0$, the spontaneous emission rate can be approximated, to the first order in $\zeta(\omega)$, to be
\begin{multline} \label{eq:sdecayratecase2}
	\Gamma_{\downarrow} = \eta  \Bigg[ \rho(\Omega_0)  \Omega_0 - \frac{\zeta(\omega)}{2} \Omega_0^2  \rho'(\Omega_0) + \frac{\zeta(\omega)}{4} \left( \rho(\bar{\Omega}^+_0)  \bar{\Omega}^+_0  + \rho(\bar{\Omega}^-_0)  \bar{\Omega}^-_0 \right) \\ - \frac{2}{5} \left\{ \zeta(\Omega_0) \rho(\Omega_0)  \Omega_0  - \frac{1}{2}  \left( \zeta(\bar{\Omega}^+_0)\rho(\bar{\Omega}^+_0)  \bar{\Omega}^+_0 +  \zeta(\bar{\Omega}^-_0)\rho(\bar{\Omega}^-_0) \bar{\Omega}^-_0 \right) \right\} \Bigg],
\end{multline}
where $\bar{\Omega}^{\pm}_0 = \bar{\Omega}_0 \pm \omega >0$.
The absorption rate to the first order in $\zeta(\omega)$ vanishes. Therefore, we have 
\begin{equation}
	A = B = \Gamma_{\downarrow}/4,
\end{equation} which leads to a non-unitary GP given by
\begin{equation} \label{eq:Gamma_g_low}
	\varphi^{<} = - \dfrac{\pi^2 n^2}{2\Omega_0} \Gamma_{\downarrow} \left( 2 +  \cos \theta \right) \sin^2 \theta,
\end{equation}
where we have used the symbol $\varphi^{<}$ to denote the non-unitary GP in the $\omega \ll \bar{\Omega}_0$ regime.
By choosing the parameters of the cavity carefully, we can make the inertial and the non-inertial contributions to the transition rate become comparable 
as shown in Fig.~\ref{fig:decayrate2}. Fig.~\ref{fig:gp2} gives the plot of the inertial and the non-inertial contributions to the GP as a function of the number of quasi-cycles. With the parameters taken in Fig.~\ref{fig:gp2}, the allowed number of quasi-cycles consistent with Eq.~\eqref{eq:HuYuGamma} is determined by $\pi A n/\Omega_0 \ll 1$, that is, $10^{-21} n \ll 1$. 

Here, we have studied the GP acquired by a circularly rotating two-level atom, inside an electromagnetic cavity, interacting with the electromagnetic field in the inertial vacuum. The acceleration-induced modifications to the field correlators perceived by the atom depend on the angular frequency of the rotating atom. We have studied GP in two distinct regimes characterized by $\omega \gg \bar{\Omega}_0$ and $\omega \ll \bar{\Omega}_0$. The $\omega \ll \bar{\Omega}_0$ regime is of particular experimental interest because one of the main hindrances to the detection of acceleration-induced modifications to field correlators is that such detection requires very high accelerations.

In the $\omega \gg \bar{\Omega}_0$ regime, for $\omega \sim 10^9$~Hz and $\Omega_0 \sim 10^7$~Hz, we have shown that the atom acquires a non-inertial GP $\sim 10^{-6}$~radian in $10^5$ quasi-cycles, i.e., $\sim 10^{-2}$~s, while the inertial contribution remains insignificant, thereby successfully isolating the non-inertial response to the GP from the inertial one.

In general, in the $\omega \ll \bar{\Omega}_0$ regime the non-inertial GP comes out be much weaker than the inertial GP. However, we show that it is possible to make the two contributions comparable by tuning the cavity to $\bar{\Omega}_0 + \omega$ and taking a larger radius $(R)$. Specifically,  we achieve this by weakening the inertial response by tuning the cavity away from the atomic resonance.  Note that we cannot indiscriminately increase $R$ because an atom rotating on a larger radius requires a bigger cavity to encase it and a larger cavity volume suppresses the overall detector response. By allowing the atom to evolve for $\sim 10^7$ quasi-cycles, a non-inertial GP $\sim 10^{-6}$~radian can be acquired, which is comparable to the inertial GP acquired by the atom. This will enable the possibility of the detection of acceleration-induced modifications to field correlators with much more relaxed parameters compared to previous studies~\cite{Martinez2011,lochan2020}. 

Specifically, we have shown, in the $\omega \ll \bar{\Omega}_0$ regime, that it is possible to detect the acceleration-induced modifications to the field correlators at an acceleration $\sim 10^7$~{m/s$^2$}. Compare this with the accelerations required for the non-inertial effects to be substantial in other proposals. For example, the Unruh effect demands acceleration of the order of $10^{21}$~{m/s$^2$} if the detector transition rates are used as an observable~\cite{unruh1976}, and $10^{17}$~{m/s$^2$} if GP is used as an observable~\cite{Martinez2011}. Similarly, detecting non-inertial effects using a circularly rotating atom inside an electromagnetic cavity, by observing the atomic spontaneous decay rate, requires an acceleration of the order of $10^{12} $~{m/s$^2$}~\cite{lochan2020}. Thus we demonstrate that, aided by the cavity, usage of the GP response for observing weak, but nontrivial, non-inertial effects in quantum field theory is a much sensitive and powerful tool.
\chapter{Conclusion and Future Outlook}
The geometric phase is one of the fundamental aspects of quantum mechanics as it accounts for a wide range of phenomena. It provides a way to classify distinct phases (topological) of matter in condensed matter physics which is a central theme. It further facilitates the precise measurement and capturing of the weak effects. In this thesis, we reviewed the fundamental structure of the geometric phase and studied a number of its applications. 

The geometrical representation of the state space of an $n$-level quantum system is essential in characterizing the system. One possible way to achieve that is to understand the structure of geodesics and null phase curves in the state space. 
In this thesis, we proposed a fresh geometrical perspective of geodesics and null phase curves in higher-dimensional state spaces. This geometrical decomposition provides a consistent way to construct geodesics and a class of null phase curves in $ d $-level systems. We used Majorana star representation which maps a pure quantum state of an $ n $-level system to the symmetric subspace of ($n$-$1$)  two-level systems. This work can be instrumental in studying the topological phases in the systems with three or more band structures. Also, geodesics can be used to design optimal quantum circuits, which is equivalent to finding the shortest path between two points in a certain curved geometry.

The geometric phase is an integral component in characterizing the topological phases. Quantum walks are the quantum analog of classical random walks. They have started gaining popularity among condensed matter physicists in the last decade because one can simulate these topological phases in one (1D) and two-dimensional (2D) discrete-time quantum walks (DTQW). We studied the topological order in 1D and 2D quantum walks. We investigated the behavior of topological phases in the presence of a lossy environment that leads to non-Hermiticity. The non-Hermitian systems exhibit rich topological structure due to the complex-valued spectrum and the presence of exceptional points. We take the non-Hermitian Hamiltonian approach to include the environmental effects and find that the topological phases in quantum walks are robust against moderate losses. We unveiled a persistence of topological order for one-dimensional quantum walks as long as the system respects the parity-time $\mathcal{PT}$-symmetry. We also find the persistent nature of topological phases in two-dimensional quantum walks. However, $\mathcal{PT}$-symmetry does not play a role there. We further showed the bulk-boundary correspondence to support our results. 

We further used the geometric phase to encapsulate the weak effect of the non-inertial motion of a quantum system. We studied the non-inertial effects in a rotating two-level atom placed inside an electromagnetic cavity. The electromagnetic cavity allows us to isolate and strengthen the non-inertial effects, over inertial ones, by appropriately tuning the resonance frequency. The sensitive nature of the geometric phase is instrumental in capturing the impact of the rotation through transition rates. Further, the accumulative nature of the geometric phase facilitates the experimental observation of the weak effects at very low accelerations. We demonstrated that aided by the cavity, usage of the GP response for observing weak but nontrivial, non-inertial effects in quantum field theory is a much more sensitive and powerful tool.

The existing definitions of geometric phases are primarily mathematical. In the future, we plan to introduce an operational definition of the geometric phase, which does not require explicit information about the state. It will lead to a more profound understanding of the subject. Continuing our work on geometric phase-assisted observation of non-inertial cavity-QED effects, we plan to use the geometric phase to observe the weak response in other phenomenons, for example, to observe the gravitational waves.

\chapter*{List of Publications}
\begin{enumerate}
	\item {\textbf{Vikash Mittal}, Aswathy Raj, Sanjib Dey \& Sandeep K. Goyal, ``Persistence of topological phases in non-Hermitian quantum walks'' \href{https://www.nature.com/articles/s41598-021-89441-8}{Scientific Reports \textbf{11}, 10262, (2021)},}
	
	\item \noindent{\textbf{Vikash Mittal}, Akhilesh K.S. \& Sandeep K. Goyal, ``Geometric decomposition of
		geodesics and null phase curves using Majorana star representation'' \href{https://journals.aps.org/pra/abstract/10.1103/PhysRevA.105.052219}{Phys. Rev. A \textbf{105}, 052219 (2022)}},
	
	\item \noindent{\protect\daggerfootnote{\normalsize Partially included in the thesis} Navdeep Arya, \textbf{Vikash Mittal}, Kinjalk Lochan and Sandeep K. Goyal, ``Geometric phase assisted observation of non-inertial cavity-QED effects'' \href{https://journals.aps.org/prd/abstract/10.1103/PhysRevD.106.045011}{Phys. Rev. D \textbf{106,} 045011, Aug (2022)}}. 
\end{enumerate}


\end{document}